\documentclass[]{tADP2e}
\usepackage{color}

\newcommand{\be}{\begin{equation}}
\newcommand{\ee}{\end{equation}}
\newcommand{\bea}{\begin{eqnarray}}
\newcommand{\eea}{\end{eqnarray}}
\newcommand{\ba}{\begin{eqnarray*}}
\newcommand{\ea}{\end{eqnarray*}}
\newcommand{\dagga}{{\phantom{\dagger}}}

\newcommand{\bk}{\mathbf{k}}

\newcommand{\dis}{\displaystyle}

\newcommand{\fract}[2]{\frac{\dis #1}{\dis #2}}

\newcommand{\eqn}[1]{(\ref{#1})}

\newcommand{\bw}{\begin{widetext}}
\newcommand{\ew}{\end{widetext}}

\begin{document}

\markboth{C. Giannetti, M. Capone, D. Fausti, M. Fabrizio, F. Parmigiani, D. Mihailovic}{}

\articletype{REVIEW}

\title{Ultrafast optical spectroscopy of strongly correlated materials and high-temperature superconductors: a non-equilibrium approach}

\author{$^{\ast}$\thanks{$^\ast$Corresponding author. Email: claudio.giannetti@unicatt.it
\vspace{6pt}}Claudio Giannetti$^{\rm a}$, Massimo Capone$^{\rm b}$,  Daniele Fausti$^{\rm c}$, Michele Fabrizio$^{\rm b}$, Fulvio Parmigiani$^{\rm c,d}$\\
\vspace{6pt}
$^{\rm a}${\em{Department of Physics \& i-Lamp, Universit\`a Cattolica del Sacro Cuore, Brescia I-25121, Italy}};
$^{\rm b}${\em{International School for Advanced Studies (SISSA), and CNR-IOM Democritos, via Bonomea 265, I-34136, Trieste, Italy}}; 
$^{\rm c}${\em{Department of Physics, Universit\`a degli Studi di Trieste, Trieste, I-34127, and Sincrotrone Trieste S.C.p.A., Basovizza I-34012, Italy}}; $^{\rm d}${\em{Physikalisches Institut, Universit\"at zu K\"oln, 50937 K\"oln, Germany}}\\
\vspace{6pt}
\&
\\
Dragan Mihailovic$^{\rm e}$\\
\vspace{6pt}
$^{\rm e}${\em{Jozef Stefan Institute \& CENN-Nanocenter, Jamova 39, SI-1000 Ljubljana, Slovenia}}\\
\vspace{6pt}\received{\today}
}

\maketitle

\begin{abstract}

In the last two decades non-equilibrium spectroscopies have evolved from avant-garde studies to crucial tools for expanding our understanding of the physics of strongly correlated materials. The possibility of obtaining simultaneously spectroscopic and temporal information has led to insights that are complementary to (and in several cases beyond) those attainable by studying the matter at equilibrium. From this perspective, multiple phase transitions and new orders arising from competing interactions are benchmark examples where the interplay among electrons, lattice and spin dynamics can be disentangled because of the different timescales that characterize the recovery of the initial ground state. 
For example, the nature of the broken-symmetry phases and of the bosonic excitations that mediate the electronic interactions, eventually leading to superconductivity or other exotic states, can be revealed by observing the sub-picosecond dynamics of impulsively excited states. Furthermore, recent experimental and theoretical developments have made possible to monitor the time-evolution of both the single-particle and collective excitations under extreme conditions, such as those arising from strong and selective photo-stimulation. These developments are opening the road toward new non-equilibrium phenomena that eventually can be induced and manipulated by short laser pulses.

Here, we review the most recent achievements in the experimental and theoretical studies of the non-equilibrium  electronic, optical, structural and magnetic properties of correlated materials. The focus will be mainly on the prototypical case of correlated oxides that exhibit unconventional superconductivity or other exotic phases. The discussion will extend also to other topical systems, such as iron-based and organic superconductors, MgB$_2$ and charge-transfer insulators. Under the light of this review, the dramatically growing demand for novel experimental tools and theoretical methods, models and concepts, will clearly emerge. In particular, the necessity of extending the actual experimental capabilities and the numerical and analytic tools to microscopically treat the non-equilibrium phenomena beyond the simple phenomenological approaches represents one of the most challenging new frontier in physics.
\bigskip  

\newpage
\hbox to \textwidth{\hsize\textwidth\vbox{\hsize20pc
\centerline{\bfseries Contents}
\vspace{5pt}
\hspace*{-12pt} {1.} {\textbf{Introduction}}\\

{2.} {\textbf{Non-equilibrium optical techniques for}}\\
\hspace*{7pt} {\textbf{ correlated electron systems: a survey}}\\
\hspace*{10pt}{2.1.}  {Visible and near-IR pump-}\\
\hspace*{24pt} {probe experiments}\\
\hspace*{10pt}{2.2.} {Multi-pulse techniques}\\
\hspace*{10pt}{2.3.} {Laser heating artifacts}\\
\hspace*{10pt}{2.4.}  {Towards a non-equilibrium optical}\\
\hspace*{24pt} {spectroscopy: supercontinuum}\\
\hspace*{24pt} {white-light generation}\\
\hspace*{10pt}{2.5.}  {Time domain THz}\\
\hspace*{24pt} {spectroscopy}\\
\hspace*{24pt} {2.5.1.}   {Optical pump \&}\\
\hspace*{47pt} {THz probe spectroscopy}\\
\hspace*{10pt}{2.6.}  {Other time-resolved techniques:}\\
\hspace*{24pt} {2.6.1.}   {Time-resolved photoemission}\\
\hspace*{24pt} {2.6.2.}   {Time-resolved electron}\\
\hspace*{47pt} {and X-ray diffraction}\\
\hspace*{10pt}{2.7.}  {The next generation of ultrafast}\\
\hspace*{24pt} {sources}\\

{3.} {\textbf{Quasiparticle dynamics in correlated}}\\
\hspace*{7pt} {\textbf{materials: basic concepts and theoretical}} \\
\hspace*{7pt} {\textbf{background}} \\
\hspace*{10pt}{3.1.}   {The dynamics of quasiparticles}\\
\hspace*{24pt} {3.1.1}   {Quasiparticle scattering in the normal}\\
\hspace*{47pt} {state: electron-phonon coupling}\\
\hspace*{24pt} {3.1.2}   {Quasiparticle scattering including}\\
\hspace*{47pt}  {bosonic degrees of freedom}\\
\hspace*{47pt}  {of different nature}\\
\hspace*{24pt} {3.1.3}   {Magnetic degrees of freedom}\\
\hspace*{24pt} {3.1.4}   {Quasiparticle scattering channels}\\
\hspace*{47pt} {in the gapped phases}\\
\hspace*{10pt}{3.2.}   {Microscopic description of}\\
\hspace*{24pt} {quasiparticles and beyond}\\
\hspace*{24pt} {3.2.1.}{Quasiparticles from strong correlations}\\
\hspace*{47pt} {in the Gutzwiller approximation}\\
\hspace*{24pt} {3.2.2.}{Quasiparticles coexisting with}\\
\hspace*{47pt} {high-energy features in}\\
\hspace*{47pt} {Dynamical Mean-Field Theory}\\
\hspace*{24pt} {3.2.3.}{Quasiparticle evolution}\\
\hspace*{47pt} {approaching the Mott-Hubbard}\\
\hspace*{47pt} {transition}\\
\hspace*{10pt}{3.3.}  {Effective-temperature models}\\
\hspace*{10pt}{3.4.}  {The basic concepts of non-equilibrium}\\
\hspace*{24pt}{superconductivity}\\
\hspace*{24pt} {3.4.1}   {The bottleneck in the dynamics:}\\
\hspace*{47pt} {the Rothwarf-Taylor equations}\\
\hspace*{24pt} {3.4.2}   {The response due to the}\\
\hspace*{47pt} {effective temperature and chemical}\\
\hspace*{47pt} {potential perturbations in}\\
\hspace*{47pt} {the superconducting state}\\
\hspace*{24pt} {3.4.3}   {Sub-gap photo-excitation and}\\
\hspace*{47pt} {and gap enhancement}\\
\hspace*{10pt}{3.5.}   {Dynamics of the order parameters:}\\
\hspace*{24pt} {time-dependent Ginzburg-Landau}\\ 
\hspace*{24pt} {functionals}\\
\hspace*{10pt}{3.6.}   {On the complexity of the}\\
\hspace*{24pt}  {scattering processes which regulate}\\
\hspace*{24pt}  {the quasiparticle dynamics}\\

{4.} {\textbf{Optical properties: a non-equilibrium}}\\ 
\hspace*{7pt} {\textbf{approach}}\\
\hspace*{10pt}{4.1.}   {The fundamental optical features}\\
\hspace*{24pt} {of correlated materials}\\
\hspace*{24pt} {4.1.1.}   {The extended Drude model}\\
\hspace*{24pt} {4.1.2.}   {The electron-boson scattering}\\
\hspace*{47pt} {in optics}\\
\hspace*{24pt} {4.1.3.}   {Infrared-active modes}\\
\hspace*{24pt} {4.1.4.}   {Electrodynamics of the}\\
\hspace*{47pt} {condensate}\\
\hspace*{24pt} {4.1.5.}   {Interband transitions and high-energy}\\ 
\hspace*{47pt} {excitations}\\
\hspace*{24pt} {4.1.6.}   {Optics and Dynamical Mean}\\ 
\hspace*{47pt} {Field Theory}\\
\vspace{20pt}\\}\hspace{-24pt}\vbox{\noindent\hsize20pc
\hspace*{10pt}{4.2.}   {The non-equilibrium response}\\
\hspace*{24pt} {4.2.1.} {Probing the spectral}\\
\hspace*{47pt} {response: dynamics of the}\\
\hspace*{47pt} {dielectric function}\\
\hspace*{24pt} {4.2.2.} {The optical response of}\\
\hspace*{47pt} {a superconductor}\\
\hspace*{24pt} {4.2.3.} {Description of the non-equilibrium}\\
\hspace*{47pt}{experiments in terms of}\\ 
\hspace*{47pt}{stimulated Raman scattering}\\

{5.} {\textbf{Non-equilibrium spectroscopies}}\\
\hspace*{7pt} {\textbf{of high-temperature superconductors}}\\
\hspace*{7pt} {\textbf{and correlated materials}}\\
\hspace*{10pt}{5.1.} {Single-color pump-probe in the}\\
\hspace*{24pt} {near-IR}\\
\hspace*{24pt} {5.1.1.} {Cuprate superconductors}\\
\hspace*{24pt} {5.1.2.} {Iron-based superconductors}\\
\hspace*{24pt} {5.1.3.} {The conventional superconductors}\\
\hspace*{47pt} {MgB$_2$ and NbN}\\
\hspace*{24pt} {5.1.4.} {Organic superconductors}\\
\hspace*{24pt} {5.1.5.} {The basic concepts emerging from}\\
\hspace*{47pt} {single-color experiments}\\
\hspace*{10pt}{5.2.} {Multi-color experiments: main results}\\
\hspace*{24pt} {5.2.1.}   {Electron dynamics of charge}\\
\hspace*{47pt} {-transfer and Mott insulators}\\
\hspace*{24pt} {5.2.2.}   {Electron-phonon coupling in}\\
\hspace*{47pt} {correlated materials}\\
\hspace*{24pt} {5.2.3.}   {Ultrafast electron-boson}\\
\hspace*{47pt} {coupling and the magnetic}\\
\hspace*{47pt} {degrees of freedom}\\
\hspace*{24pt} {5.2.4.}  {The ultrafast dynamics}\\
\hspace*{47pt} {in the pseudogap phase}\\
\hspace*{24pt} {5.2.5.}   {Superconductivity-induced}\\
\hspace*{47pt} {spectral-weight change}\\
\hspace*{24pt} {5.2.6.}   {Fluctuations of the superconducting}\\
\hspace*{47pt} {order parameter}\\
\hspace*{24pt} {5.2.7.}   {The cuprate phase diagram}\\
\hspace*{47pt}  {from non-equilibrium spectroscopies}\\
\hspace*{24pt} {5.2.8.}   {Time-domain competition}\\
\hspace*{47pt} {between different orders}\\
\hspace*{10pt}{5.3.}   {Optical generation of coherent}\\
\hspace*{24pt} {bosonic waves}\\
\hspace*{24pt} {5.3.1.} {Lattice modes}\\
\hspace*{24pt} {5.3.2.} {Charge-Density-Waves}\\
\hspace*{24pt} {5.3.3.} {Superconducting Amplitude Modes}\\

{6.}  {\textbf{Toward the optical manipulation}}\\
\hspace*{7pt} {\textbf{and control of electronic phases}}\\
\hspace*{7pt} {\textbf{in correlated materials}}\\
\hspace*{10pt}{6.1.} {Photo-induced phase transitions}\\
\hspace*{24pt} {(PIPT) between quasi-thermodynamic}\\
\hspace*{24pt} {phases}\\
\hspace*{24pt} {6.1.1.} {Insulator-to-metal transition}\\
\hspace*{47pt}{in transition metal oxides}\\
\hspace*{24pt} {6.1.2.} {Photo-induced metastable}\\
 \hspace*{47pt}{non-thermal states}\\
\hspace*{10pt}{6.2.} {Photo-induced transient non-equilibrium}\\
\hspace*{24pt} {states and photodoping}\\
\hspace*{24pt} {6.2.1.} Transient Mott transition\\
\hspace*{24pt} {6.2.2} {Non-thermal melting}\\
\hspace*{47pt}{of the superconducting condensate}\\ 
\hspace*{24pt} {6.2.3} {Photoinduced superconductivity}\\

{7.} {\textbf{Recent advances and perspectives in the}}\\
\hspace*{7pt} {\textbf{theoretical approaches for non-equilibrium}}\\
\hspace*{7pt} {\textbf{correlated materials}}\\
\hspace*{10pt}{7.1.}   {Non-equilibrium techniques}\\
\hspace*{24pt} {for strongly correlated materials}\\
\hspace*{24pt} {the Gutzwiller variational technique}\\
\hspace*{10pt}{7.2.}  {Non-equilibrium techniques for}\\
\hspace*{24pt} {strongly correlated materials:}\\
\hspace*{24pt} {Dynamical Mean Field Theory}\\
\hspace*{24pt} {7.2.1.}   {DMFT: basic ideas and}\\
\hspace*{47pt} {equations}\\
\hspace*{24pt} {7.2.2.}   {DMFT and high-temperature}\\
\hspace*{47pt} {superconductors}\\
\hspace*{24pt} {7.2.3.}   {Non-equilibrium DMFT:}\\
\hspace*{47pt} {the equations and the}\\
\hspace*{47pt} {algorithm}\\}

\vspace{0pt}\\}\hspace{-24pt}\vbox{\noindent\hsize20pc
\hspace*{10pt}{7.3.}   {Correlated systems in electric fields}\\
\hspace*{24pt} {7.3.1.}   {Correlated electrons in a}\\
\hspace*{47pt} {d.c. electric field}\\
\hspace*{24pt} {7.3.2.}   {Dissipation and approach}\\
\hspace*{47pt} {to steady states}\\
\hspace*{24pt} {7.3.3.}   {Periodic a.c. electric}\\
\hspace*{47pt} {fields}\\
\hspace*{24pt} {7.3.4.}   {Electric fields}\\
\hspace*{47pt} {in correlated heterostructures}\\
\hspace*{24pt} {7.3.5.}   {Electric-field driven Mott}\\
\hspace*{47pt} {transitions and resistive switching}\\
\hspace*{24pt} {7.3.6.}   {Light-induced excitation}\\
\hspace*{47pt} {and photodoping}\\
\hspace*{10pt}{7.4.}  {Quantum quenches and}\\
\hspace*{24pt} {dynamical phase transitions}\\
\hspace*{24pt} {in simple models}\\
\hspace*{10pt}{7.5.}   Melting of antiferromagnetism\\
\hspace*{24pt} and broken-symmetry phases\\
\hspace*{10pt}{7.6.}   Non-equilibrium dynamics\\
\hspace*{24pt} beyond the single-band\\
\hspace*{24pt} Hubbard model\\
\hspace*{10pt}{7.7.}   Electron-phonon interaction beyond\\
\hspace*{24pt} the two-temperature model\\
{8.} \textbf{Experimental perspectives}\vspace{5pt}\\
{9.} \textbf{Acknowledgements}\vspace{5pt}\\
{10.} \textbf{References}\\
}
\end{abstract}
\setcounter{section}{0}
\section{Introduction}
Since a long time, the study of the physical properties of materials is pervaded by non-equilibrium methods and concepts. Any form of transport or tunneling property is inferred from measurements performed under the application of electric fields that drive the motion of the conduction electrons. As well, optical and photoemission spectroscopies require electromagnetic fields for inducing transitions between different energy levels. Although perturbative of an equilibrium state, the common nature of these experiments relies on the fact that the external stimuli are either extremely weak or their application lasts for a time much longer than the interaction time among the internal degrees of freedom of the system. Expressly, a steady state regime, which may or may not correspond to the real ground state, is reached on a timescale much faster than the observation time, thus providing a quasi-equilibrium information which can be treated consistently by conventional statistical approaches. 

The possibility of breaking this paradigm by applying external perturbations faster than the typical relaxation times led to dramatic advances. One of the most important examples is offered by cold atoms, in which the typical lifetimes of the excited states are so long that the transformation from an excited quantum state to a statistical ensamble can be directly followed in real time. Similar approaches in condensed matter have been sought for a long time, for their fundamental interactions are often confined in the sub-nanosecond timescale. A turning point was attained with the advent of ultrafast laser sources. These lasers deliver ultrashort, coherent and tunable light pulses with durations ranging from a few to several hundreds of femtoseconds (1 fs=10$^{-15}$ s). Since then, it was realized that such lasers could be used to create non-equilibrium conditions on timescales faster than the energy exchange with lattice vibrations, magnetic excitations and other degrees of freedom, hence disclosing physical phenomena not observable by equilibrium spectroscopies. 

Among the wealth of systems and physical processes that could be investigated by ultrafast experiments, strongly correlated materials soon attracted a substantial attention. These systems are generally on the verge of multiple phase transitions, because of the strong interactions between electrons occupying the same lattice sites, and hence prone to dramatic changes of the electronic properties upon small changes in the external parameters, like temperature, doping, pressure and applied magnetic fields. As a natural extension of this concept, ultrashort light pulses could be used as an additional knob to control the physical properties  of the transient electronic states in the sub-ps timescale, till to reach nowadays the sub-fs time scale. In some cases, the interplay among different degrees of freedom that are strongly intertwined in the frequency domain, has been disentangled in the time domain by exploiting specific thermalization processes characterized by different timescales. 

Within correlated materials, copper oxide-based compounds are an interesting system for observing ultrafast mechanisms depending on temperature, pressure, doping and external fields, hence unveiling complex phenomena that impact on some of the paradigms of the solid state physics. The information attainable by non equilibrium studies is also relevant for studying the electron-boson interactions, the development of low temperature symmetry-breaking instabilities, the competing or cooperative interactions among different order parameters and the doping evolution of the fundamental electronic excitations from incoherent charge fluctuations to coherent quasi-particles, along with other many-body mechanisms in condensed matter. 

Along with the above mentioned topics, the study of non-equilibrium phenomena in superconductors has become so important that it can be regarded as a fundamental topic for condensed matter physics. 
Since the first observation that intense laser pulses can non-thermally destroy the superconducting state
, laser-based spectroscopies gradually surpassed the junction carrier injection as a tool for studying non-equilibrium phenomena in superconductors. This topic further evolved after the advent of ultrafast lasers and time-resolved experiments have been expanded from superconductivity to other systems for studying the charge-orders, the spin and charge density waves, the electronically-driven phase transitions and other mechanisms resulting from strong correlation effects.

There are important reasons for studying non-equilibrium superconductivity in the time domain. 
As an example, the dynamics of gapped phases is ubiquitously characterized by the presence of bottlenecks in the relaxation dynamics. In this case, part of the system, such as gap-edge electrons and optical phonons, reach a quasi-equilibrium condition whereas the remaining degrees of freedom are almost unaffected. This experimental fact suggests that the single particle kinetics at the gap edges can be observed in some detail. For example, In the \textit{bottleneck regime}, the measured relaxation rates are ruled by the fundamental interactions among the elementary single-particle excitations, thus providing complementary information to those obtained by equilibrium spectroscopic methods. 
The proper treatment of bottlenecks turns out to be crucial for understanding the relaxation of quasiparticles across the superconducting gap, hence suggesting the most relevant bosonic modes that are populated during the recombination process.

More in general, the timescale of the relaxation of high-energy electronic excitations and  the quasi-particle (QP) recombination dynamics, via electron-electron, electron-phonon or electron-spin scattering, can unveil the most important interactions in the compound. By tuning the laser photon energy it is possible to induce resonant transitions between states with energies ranging from $\sim$0.001 eV to $\sim$6 eV, thus allowing to select particular relaxation channels. Finally, time-domain laser spectroscopy can be used to investigate the dynamics of collective states, such as the destruction of the superconducting condensate and its collective recovery or the intrinsic dynamics of the amplitude and phase modes associated to specific order parameters. For example, the study of the destruction and recovery of the superconducting condensate in the time-domain can reveal new processes, such as the formation of vortex-antivortex pairs or the coexistence of superconducting and normal domains, eventually similar to those observed at equilibrium when an external magnetic field is applied. 

The possibility of selectively exciting specific degrees of freedom in correlated materials and superconductors opens the way to create novel metastable states, that cannot be reached via thermodynamic transformations. Even though a comprehensive review of photo-induced phase transitions is beyond the scope of the present work, we will report some important examples of non-equilibrium phenomena that can be induced by strong laser-field excitations. The recent advances in the production of intense THz and mid-infrared pulses led to the possibility of directly coupling the laser field to low-energy degrees of freedom and to directly investigate, for example, the emergence of cooperative phenomena induced by selective lattice excitations. All these results point to achieve a real optical control of macroscopic electronic phases on the sub-picosecond timescale.

Nowadays, time-resolved optical experiments can probe almost entirely the electromagnetic spectrum from THz to the hard X-rays using either conventional ultrafast laser sources or free electron lasers. This impressive development offers a wealth of possibilities to perform non-equilibrium studies based on optical and photoemission spectroscopies or diffraction and inelastic scattering experiments. Conversely, the development of suitable theoretical frameworks to treat non-equilibrium problems is still at its infancy, while significant works are currently under development to describe the many-body non-equilibrium physics in condensed matter.  
For this reason, an entire chapter is focused on the ongoing theoretical efforts to treat strong-electronic correlations in non-equilibrium conditions with the aim of going beyond the phenomenological models currently used to interpret time-resolved experiments in the ultrafast time domain. 
\\
The review is organized as follows.

In Chapter \ref{sec_pp}, we provide a survey of the pump-probe (P-p) techniques for correlated electron systems, with particular focus on table-top experiments. 

Chapter \ref{sec_QPdynamics} reports on the basic models used for describing the quasiparticle dynamics in strongly correlated materials starting from the link between the concepts established by frequency-domain experiments/theories and the novel phenomenology emerging from non-equilibrium approaches. In particular, a significant emphasis is given to the Gutzwiller approximation and Dynamical Mean-Field Theory, that today are among the best methods to describe the quasiparticle evolution in the vicinity of Mott-Hubbard transitions. Finally, we present the most widely used models to treat non-equilibrium superconductivity and extract the microscopic interaction parameters from the measured relaxation time. Specifically, we  discuss the effective-temperature models used to extract the electron-boson coupling in the normal state and the Rothwarf-Taylor and other effective models to describe the dynamics of gapped excitations in the superconducting phase. Finally, we introduce the time-dependent Ginzburg-Landau functionals to phenomenologically describe the dynamics of any perturbed order parameter in a symmetry-broken phase. 

In Chapter \ref{sec_noneqopticalprop} we introduce the basic physics related to the optical properties of strongly correlated materials out of equilibrium. The core of this Chapter is the link between the equilibrium optical properties of correlated materials and the non-equilibrium physics that is probed by time resolved optical experiments. We introduce the extended Drude model and the electron-boson scattering as observed in optical experiments, along with the THz/infrared conductivity. In order to bridge the theory with the optical experimental data, we dedicate a section to the description of the optical conductivity of correlated materials by means of Dynamical Mean Field Theory. The second part of the Chapter is aimed at introducing the concept of differential dielectric function for non-equilibrium measurements and to discuss the possible strategies to extract quantitative information from the experiments. 

Chapter \ref{sec_experiments} constitutes the core of this work and presents an outline of the most important results in the field. In the first part we report a historical overview of the first pioneering single-colour experiments on superconductors and other correlated materials. In the second part of the Chapter we present the outcomes of the most advanced non-equilibrium spectroscopies and we discuss the main ideas and concepts emerging from the whole of the experimental data. In particular, we tackle the electron dynamics of charge-transfer and Mott insulators, the electron-boson coupling in correlated materials, the ultrafast dynamics in the pseudogap phase, the superconductivity-induced spectral-weight change, the fluctuations of the superconducting order parameters. Finally, the possibility of impulsively exciting coherent waves in symmetry-broken phases is discussed.

Chapter \ref{sec_optical_control} overviews some selected topics that are relevant for achieving the all-optical manipulation of the electronic properties of correlated materials. Far from pretending of being comprehensive, we mainly discuss the recent attempts for non-thermally destroying or photoinducing the superconductivity and for optically creating transient metastable states with exotic electronic properties.

Chapter \ref{sec_theory} presents the state-of-the art techniques to theoretically treat the many-body problem of correlated materials driven out of equilibrium. While we mainly focus on the Gutzwiller variational approach and on Dynamical Mean Field Theory, we also discuss the different approaches today available and we introduce the next steps required to achieve a realistic description of time-resolved experiments. Since the development of the theoretical models for non-equilibrium experiments is still under development, some of the presented results should be considered of relevance by themselves, even without a direct connection (for the moment) to the experiments.

\setcounter{section}{1}
\section{Non-equilibrium optical techniques for correlated electron systems: a survey}
\label{sec_pp}
In the 1980s, when the first $\sim$100 fs laser pulses have been generated by colliding-pulse mode- locked (CPM) dye laser \cite{Fork1981}, it was clear that the route for studying the matter out of equilibrium was open. A few years later the pulse length was brought down to $\sim$30 fs pulses \cite{Valdmanis1986} and then further down to $\sim$6 fs \cite{Fork1987} and the direct observation of the charge transfer process in semiconductors and molecules was made possible. A further step ahead was done in the early 1990s when coherent pulses of  $\sim$60 fs infrared (IR) were generated using the self mode-locking technique in Ti:sapphire-based lasers \cite{Spence1991}. 

Nowadays, ultrafast lasers are commonly used for a variety of cutting-edge spectroscopies that share the capability of detecting ultrafast non-equilibrium dynamics in gases and condensed matter systems.

\begin{figure}[t]
\begin{centering}
\includegraphics[width=0.9\textwidth]{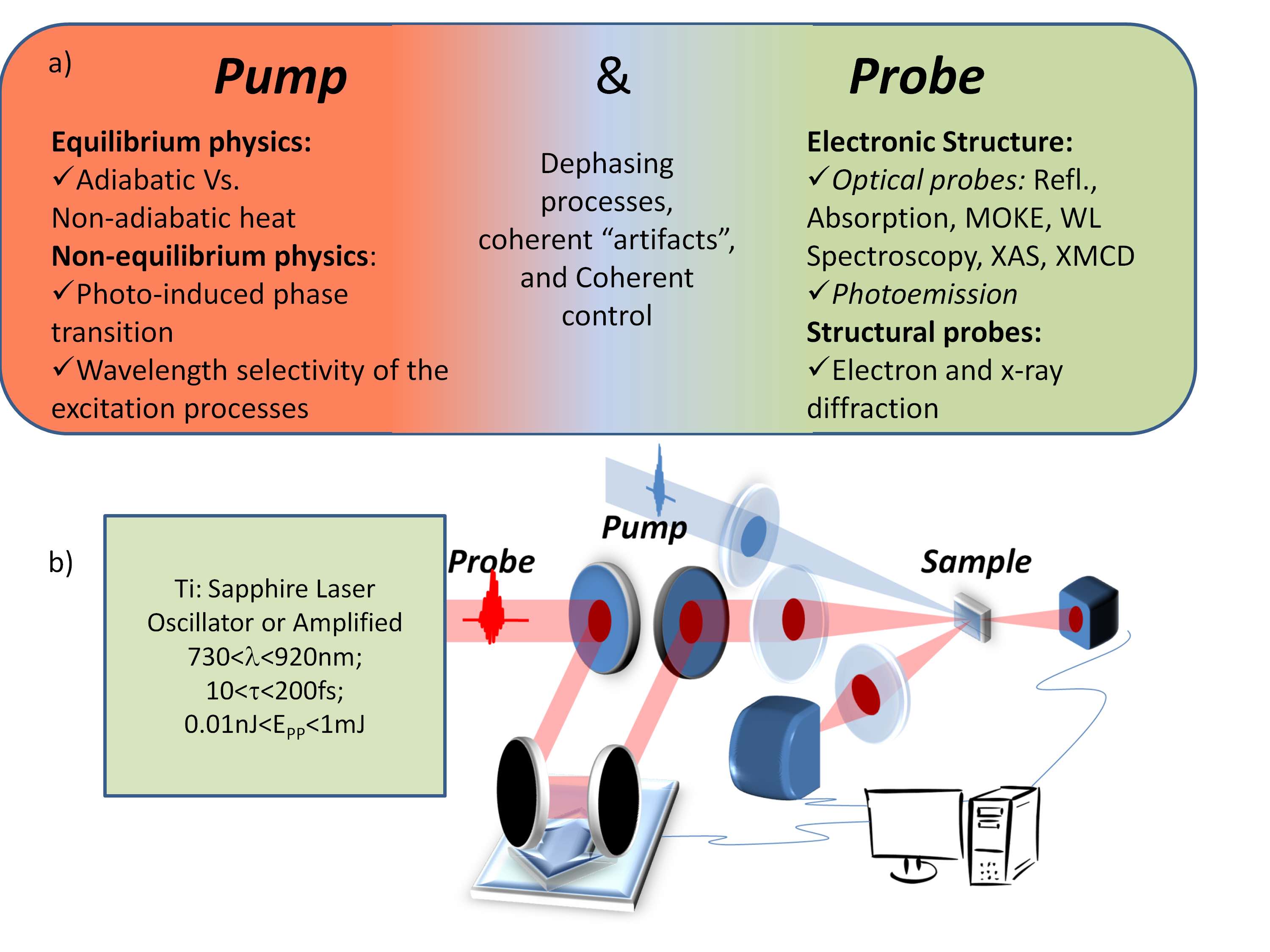}
\caption{a) Pump and Probe experiments. Various excitation schemes (pump) and spectroscopic tools (probes) have been developed over the years (see texts). b) Sketch of the basic setup for optical pump and probe.}
\label{fig_pumpprobe}
\end{centering}
\end{figure}

The great and constantly increasing success of the ultrafast spectroscopies can be attributed to the fact that a single experiment, often naively referred to as Pump-probe (P-p), can be applied to investigate a large variety of phenomena that take place in diverse physical regimes.
As shown in Figure \ref{fig_pumpprobe}, P-p experiments are based on the idea of perturbing the matter with a \textit{pump} (P), i.e., an ultrashort pulsed electromagnetic field, while taking snapshots of the time evolution of some observables in the system using a quasi non-perturbing \textit{probe} (p) pulse. 
This simple arrangement can be adapted to investigate the non-equilibrium physics of solids in substantially different frameworks, that can be grouped into three main categories.
\begin{itemize}
\item[1.] \textbf{Non-equilibrium spectroscopy}. In this case, the experiments aim at using the pump pulse as a mean to induce a non-equilibrium initial condition, often represented as a non-thermal distribution of quasi-particles, with a minor influence on the overall properties of the systems. In the simple potential-landscape picture, the system is instantaneously driven to a non-equilibrium initial condition far from the potential minimum occupied at equilibrium. For sufficiently weak perturbations, the potential is not altered and the system relaxes back to the equilibrium state. In this process, the relaxation dynamics is regulated by the same thermodynamic parameters that determine the equilibrium properties. Microscopically, the initial non-equilibrium distribution of the electronic, vibrational, and magnetic degrees of freedom recovers on timescales that are governed by the same interaction strengths (i.e. electron-phonon coupling) that can be inferred from equilibrium spectroscopies. After a rather fast relaxation process, usually in the fs-ps timescale, the electronic and bosonic (phonons, magnons) excitations recover a quasi-equilibrium distribution that differs from equilibrium only for an increased \textit{effective temperature}. Many examples of this technique, along with the links between the equilibrium quasiparticle dynamics and the P-p experiments, will be extensively provided and discussed in Sections \ref{sec_history}-\ref{sec_results}. 

In general, the details of the effects of the electromagnetic field on the sample are disregarded and treated as a sudden injection of quasi-particles. However, there are many evidences that the initial non-equilibrium condition may have a strongly non-thermal character (e.g., a pronounced anisotropic \textbf{k}-space distribution) that cannot be obtained by adiabatically changing the common thermodynamic variables, such as temperature, pressure and external fields. In this regime, the recovery dynamics can explore pathways different from those that are followed during the relaxation after the sudden change of thermodynamic variables. In this perspective, the \textit{non-equilibrium spectroscopy} goes far beyond the simple time-domain approach, in which the time-domain trace simply represents the Fourier-transform of spectral features detected by the energy-domain spectroscopy. In Secs. \ref{sec_noneqopticalprop}-\ref{sec_theory} we will report different examples of \textit{non-equilibrium spectroscopies} which provide genuine new information as compared to conventional equilibrium techniques.    
\item[2.] \textbf{Excitation of coherent bosonic modes}. Any kind of impulsive excitation coupled to specific bosonic modes, such as lattice vibrations and charge/spin order, can trigger a coherent oscillation at the typical frequency of the mode and with a relaxation time that is related to its de-phasing time. In this case, the P-p technique can be considered as a real time-domain technique, since the Fourier-transform of the time-domain signal provides the frequency and lifetime of the mode. In non-resonant conditions, the lowest order coupling of the light pulse to the mode is given by an inverse Raman process (see Sec. \ref{sec_SRStensor}). The role of the frequency of the pump excitation follows the Raman cross section \cite{Merlin1997,Stevens:2002p6798} that is usually a well-known input from equilibrium Raman spectroscopy. The use of this technique, which will be discussed in Sec. \ref{sec_coherentbosons}, is increasingly expanding, since it provides a new tool to investigate highly-damped modes in transiently non-equilibrium conditions.
\item[3.] \textbf{Optical manipulation}. In the strong excitation regime, the pump pulse can bring the system very far from equilibrium and, eventually, can trigger the creation of "phases" that are not accessible via adiabatic transformations. In this perspective, the research lines pursued in the last years by various groups worldwide aim at driving catastrophic changes in matter to investigate many important phenomena, such as photo-induced phase transitions, ultrafast optical switching and the photo-enhancement of superconductivity. The possibility of using P-p techniques to disclose novel exotic non-equilibrium "phases" and the different excitation schemes which have been used to selectively manipulate the properties of various strongly correlated materials and high-temperature superconductors will be reviewed in Sec. \ref{sec_optical_control}.
\end{itemize}  

The recent advances in ultrafast techniques are opening novel routes in P-p experiments. In state-of-the-art femtosecond experiments, the optical properties can be probed from the THz to the XUV spectral range, with temporal resolutions ranging from a few picoseconds to attoseconds. Ultrafast lasers can be used in time-resolved photoelectron spectroscopy with a time-energy resolution given only by the indeterminacy rule. Time-resolved electron and X-ray diffraction provides a tool to directly investigate the structural and electron density dynamics. Many efforts are currently focused to exploit the new high-energy and high-repetition rate sources to produce trains of light pulses with photon beam parameters complementary to those typical obtained by storage rings. Furthermore, the possibility of combining the direct observation of the ultrafast time evolution with sub-micron spatial resolution is also gaining the attention of multidisciplinary teams extending from physics to biology and nano-science. The recent development of radiation sources capable to generate EUV and X-ray ultrashort pulses starting from suitable relativistic electron bunches, i.e. free electron laser facilities, represents a novel and rapidly expanding area for advanced and unique experiments on matter out-of-equilibrium.

Even though a comprehensive presentation of the technical details of the specific ultrafast techniques is beyond the purpose of this review, we will discuss the main concepts at the base of P-p experiments, with a particular attention to optical techniques for they represent the prototypical and probably the more versatile example of non-equilibrium spectroscopies.

\subsection{Visible and near-IR pump-probe experiments}

In spite of the fact that various early days non-equilibrium studies of high temperature superconductors were performed making use of actively mode locked systems \cite{Matsuda1994, Albrecht1992, Ginder1988, Donaldson1989, Brocklesby1989, Frenkel1989, Chekalin1991, Reitze1992, Semenov1992, Yu1992, Gong1993, Lindgren1994}, most of the recent works have been implemented starting from the Ti:Sapphire laser technology.
The discovery of passive mode locking laser systems, such as Kerr-Lens mode locking \cite{Spence1991, sarukura1989} in Ti:Sapphire based cavities, together with the implementation of chirp pulse amplification techniques \cite{Strickland1985}, boosted the field of ultrafast spectroscopy. The possibility of studying the ultrafast dynamics with relatively cheap table-top laser has stimulated in the last 20 years a dramatic development of the time-resolved version of the traditional "equilibrium" spectroscopies, with a particular ascendancy of optical techniques such as reflectivity and transmissivity. A typical building block for a Ti:Sapphire based time resolved reflectivity (transmissivity) experiment is shown in Figure \ref{fig_pumpprobe}b. The laser pulses are split into two beams. The first and more intense (P) perturbs the sample while the second (p) is used to measures the time evolution of the reflectivity. The time delay between the two pulses is varied by a mechanical translator controlling the length of one of the two arms of the setup (commonly the pump pulse for stability reasons), or, more recently, by exploiting the ferquency detuning between two synchronized cavities (ASOPS: ASynchronous OPtical Sampling) \cite{Bartels2007,Stoica2008}.

The most common acquisition scheme for experiments using high repetition rate lasers ($>$1 kHz) includes a mechanical modulator (chopper), which modulates the pump beam at a known frequency. Being the pump beam modulated, the pump-induced variation of the sample reflectivity (transmissivity) is measured by acquiring the ac-component (at the chopper frequency) of the photo-current produced on a photodiode measuring the reflected (transmitted) probe beam. This "differential" acquisition scheme, allows for a very high sensitivity reaching easily signal to noise (S/N) ratio of 10$^{-4}$ that can be extended to 10$^{-8}$ in experimental setups specifically designed for very low noise measurements.
Alternative acquisition schemes, allowing also single pulse measurements, are based on fast analogue-to-digital converters, that digitize the amplitude of each reflected probe pulse. 
\begin{figure}
\begin{center}
\includegraphics[width=8cm]{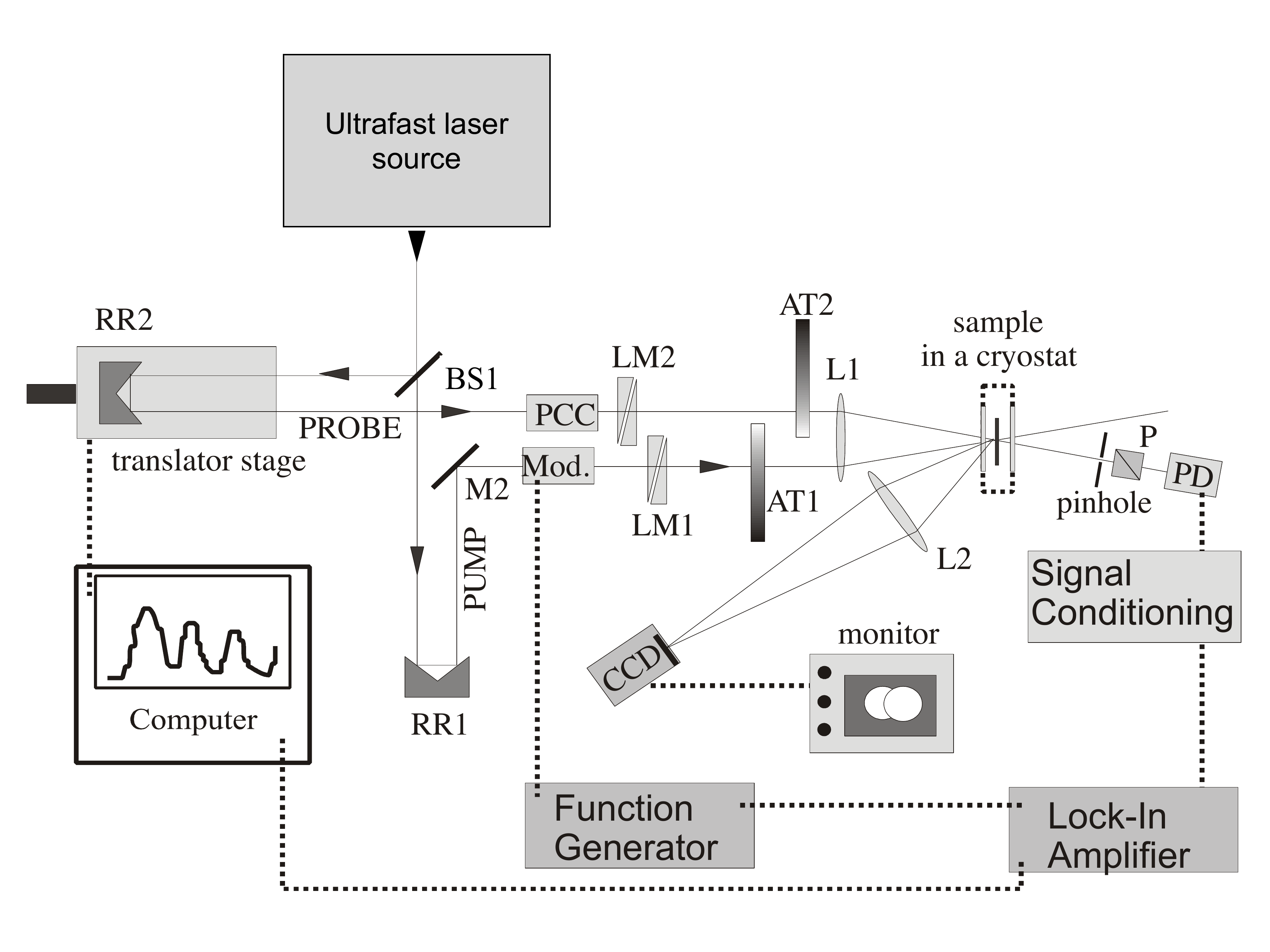}
\caption{\label{fig:A-shematic-Pp-setup}A shematic presentation of a typical
pump-probe experimental setup. M1, M2 are mirrors, BS1: beam
splitter, LM1, LM2: = $\lambda/$2 plates, AT1, AT2: attenuators,
PCC: pre-chirp compensation with the flint glass, which was needed
when acousto-optical modulator was used, Mod.: acousto-optical modulator
or mechanical chopper, L1, L2: lenses, RR1, RR2: retro-reactors, P:
polarizer, PD: photo detector, CCD: CCD monitoring camera. The beams
paths are represented by solid lines and the electrical signal connections
are indicated by dotted lines.}
\end{center}
\end{figure}

More elaborate optical schemes make use of a combination of waveplates (typically l/2 and l/4) and polarizing optics (typically polarizing beam splitter or Wollaston prisms) to measure different properties of the probe pulses such as the polarization state. The most common optical configuration used to this purpose splits ortogonal polarization states on different detectors. This scheme is preferable as it corrects the intensity fluctuations of the laser source. A description of a typical acquisition scheme can be found in Ref. \citenum{Yoneda2004}. The possibility of unraveling the polarization state, in combination with magnetic fields, is commonly used in time-domain magneto optical Kerr effects measurements (or Faraday effect in transmission) \cite{Guidoni2002,Beaurepaire1996}. 

The large majority of the optical experiments reported to date makes use of the fundamental lasing mode of Ti:Sapphire cavities (centered at the energy $\hbar\omega$=1.55 eV) \cite{Stevens1997,Demsar1999,Misochko2002,Segre:2002p9832,Xu2003}, but for a large set of time domain measurements the wavelength tunability of both pump and probe is desirable. This possibility is enabled in time domain studies by the intrinsic advantage of pulsed laser sources allowing to exploit a large set of non-linear optic techniques \cite{Boyd}. The most common approach consists in the doubling (second harmonic), tripling (third harmonic) or quadrupling (fourth harmonic) the laser fundamental frequency by suitable non-linear crystals. This technique allows to easily perform P-p experiments with either pump or probe at different frequency with respect to the fundamental lasing mode \cite{Bloembergen,Boyd}. 

\begin{figure}
\begin{center}
\includegraphics[width=10cm]{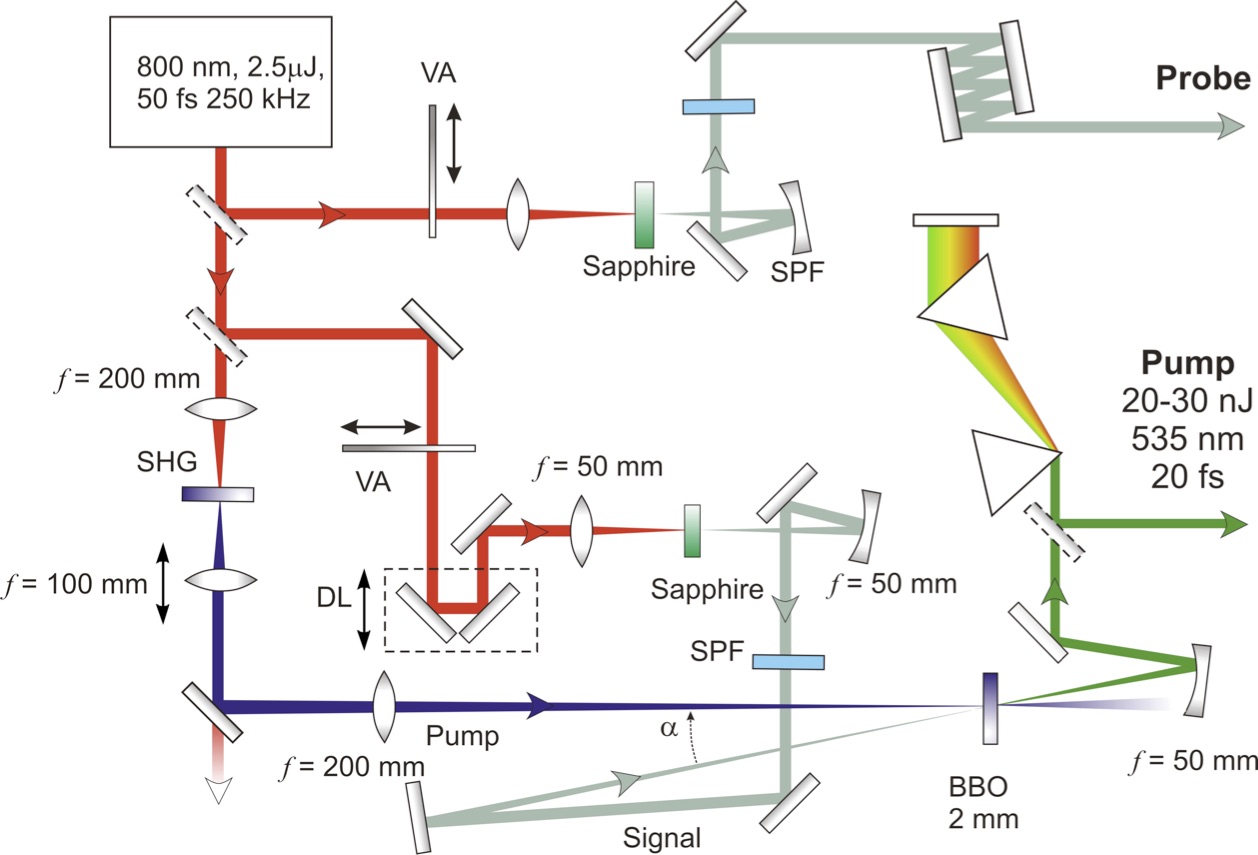}
\caption{\label{fig:The-high-repetition-rate-NOPA} The high-repetition rate non-collinear optical parametric amplifier (NOPA). The $\beta$-barium borate (BBO) crystal is placed in the Rayleigh range of the focussing lens to achieve the high pump intensities. Pump and seed beams overlap in the 2 mm BBO crystal cut for type I phase matching at 3.7$^{\circ}$. At this angle phase matching is ensured over a large bandwidth and the bandwidth of the output is given by the temporal overlap between the pump and the chirped seed pulse. The resulting pulses have a bandwidth of approximately 60 nm and are tunable between centre wavelengths of 500 to 650 nm by changing the pump-seed delay. At 535 nm centre wavelength, the amplified pulse energy is 20 to 30 nJ. A prism compressor compensates the chirp to give pulse durations of $\sim$20 fs.}
\end{center}
\end{figure}

In order to improve the wavelength tunability, a commonly used setup is based on Optical Parametric Amplifiers (OPA). OPAs allow for the generation of light pulses with wavelength continuously tunable in the visible and near-infrared (IR). Most systems use optical parametric amplification in $\beta$-barium borate (BBO) or litium triborate (LBO). The visible region is usually covered by OPA pumped with the Ti:Sapphire second harmonic ($\lambda$=400 nm) that allows tunability from 450 nm (signal) to 2500 nm (idler). The near-infrared (IR) region is commonly covered with OPA pumped with the Ti:sapphire fundamental. The wavelength tunability can be extended up to 5 $\mu$m by means of different non-linear crystals such as  KTiOPO$_{4}$ or its isomorphs, KNbO$_{3}$, or MgO:LiNbO$_{3}$. The range of the optical parameteric amplification can be extended up to 10 $\mu$m wavelength with ZnGeP$_{2}$ and HgGa$_{2}$S$_{4}$ pumped with longer wavelength pulses. 
An alternative to produce mid-IR pulses is provided by Different Frequency Generation between the signal and idler of a standard OPA in AgGa$_{2}$, GaSe or GaAs. This setup  allows a wavelength tunability up to 20 $\mu$m. A detailed review of the non-linear methods for extending the wavelength tunability goes can be found in Refs. \citenum{Cerullo2003,Brida2010}.

\subsection{Multi-pulse techniques}
\label{sec_multipulse}
\begin{figure}
\begin{center}
\includegraphics[width=13cm]{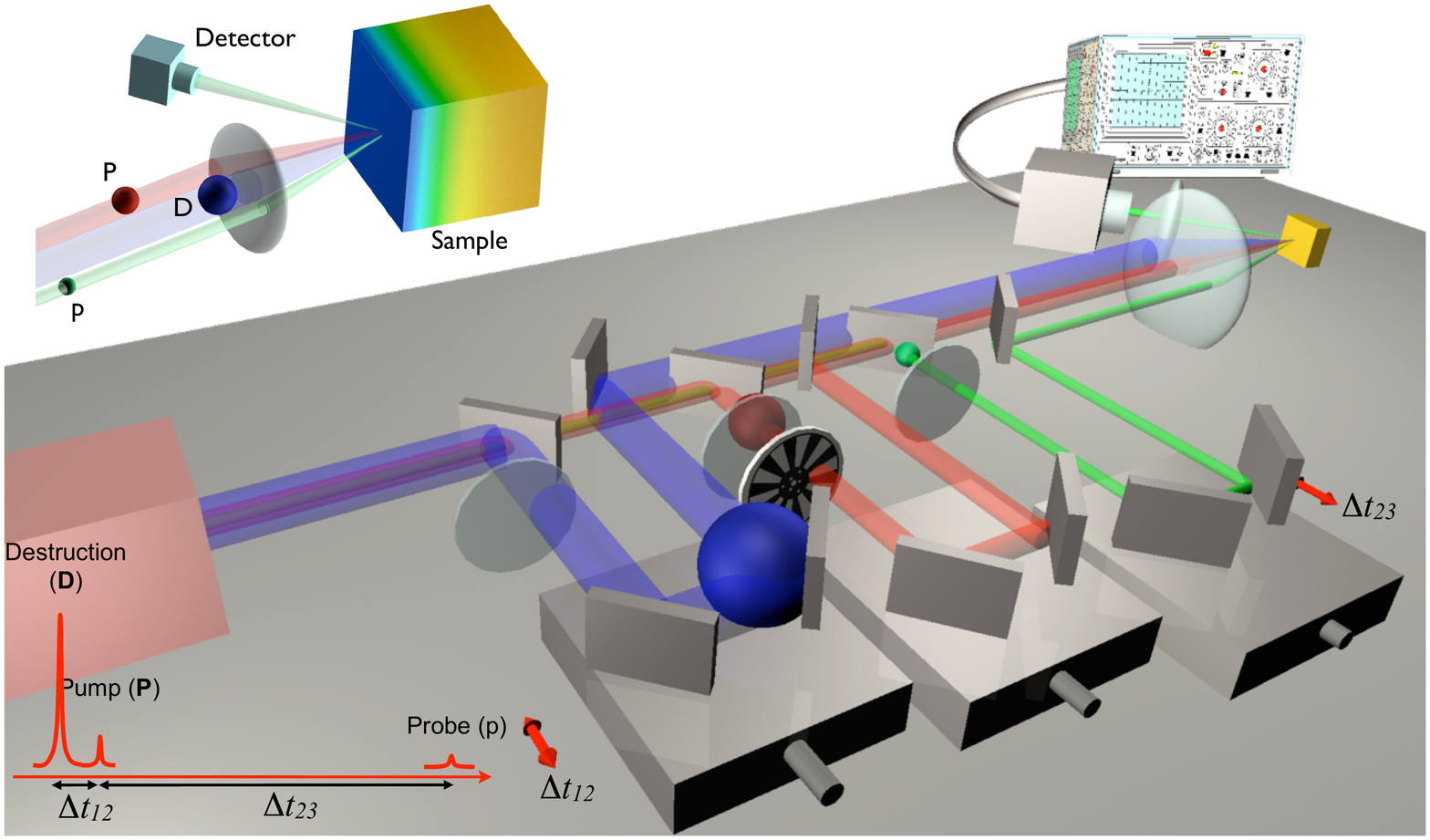}

\protect\caption{\label{fig:3pulse}A schematic diagram of the 3-pulse technique for
studying the temporal evolution of systems freely evolving through
a phase transition. The insert defines the time delays between the
three pulses.}
\end{center}
\end{figure}

Recently, a multi-pulse technique was introduced for studying the coherent
control of collective states of matter, such as the superconducting
state or the charge-density-wave ordered state \cite{Yusupov2010}. 
In its most simple version (see Fig. \ref{fig:3pulse}), an intense laser
pulse $-$ Destruction (D) pulse $-$ is first used to excite the system
into the high-symmetry state. The evolution through
the symmetry-breaking phase is monitored using a weaker standard P-p sequence, where
the P pulse is properly delayed with respect to D.  
The main reason for introducing this two-pumps technique (D-P-p) is that the standard P-p spectroscopy cannot distinguish the order parameter
dynamics from the energy relaxation and single-particle recombination
processes, all of simultaneously present in the response.
In contrast, the D-P-p technique can access the coherent dynamics of the order parameter $\eta$, that can be distinguished from other single-particle and collective mode excitations either by symmetry, relaxation time, temperature dependence or by its behaviour through the transition, particularly if it shows critical behavior.
Notably, this technique can directly provide the wondered information
on the state of the many-body system at any time instant through
the symmetry-breaking transition.

\subsection{Laser heating artefacts}
\label{sub:Laser-heating-effects}

Since in P-p experiments a significant amount of energy is absorbed in a relatively small volume,
this results in a significant and undesired local heating of the sample. The sample temperature can easily increase by many tens of
K or more above the base temperature. In superconducting
samples, this can be checked by cooling a superconducting sample through
$T_{c}$ and recording the temperature difference between the nominal
and actual transition temperature of the sample. 

Laser heating in time-resolved experiments can induce major artifacts, in particular for intensity dependence studies. Surprisingly,
the problem of sample heating is particularly noxious in experiments using
weak laser pulses from the Ti:sapphire oscillators at high-repetition
rates. In this configuration, the time interval between pulses, typically a few ns, is not sufficient to dissipate the small energy released by the single laser pulse, thus leading to a local temperature build-up. On the other hand, high-energy and
low-repetition rate experiments are strongly affected by the impulsive heating, typically a few ps, that may drive the system out of the phase under scrutiny.
The heating problem can be treated starting from the solutions of the thermal diffusion equation:
\begin{equation}
\nabla\centerdot\mathbf{J}(\boldsymbol{r},t)+\rho c\frac{\partial T(\boldsymbol{r}(,t)}{\partial t}=A(\bold{r},t)
\end{equation}
where $T(\mathbf{r},t)$ is the temperature, $J(\mathbf{r},t)$ is
the thermal energy crossing unite area per unit time, $\rho c$ is
the heat capacity per unit volume, $\rho$ is the density, $c$ is
the specific heat capacity and $A(\mathbf{r},t)$ is the net energy
per unit time generatied by the laser. With the additional use of the
Fourier's law $\mathbf{J}(\mathbf{r},t)=-K\nabla T(\mathbf{r},t)$,
where $K$ is the thermal conductivity, analytical solutions for the heating
problem have been
obtained by Bechtel \cite{Bechtel1975} for several experimental cases
encountered in P-p experiments. The long time behavior is typically
diffusive, where the temperature falls off as $T\sim\sqrt{t}$ . A
useful approximation for the temperature evolution after absorption
of light in a surface layer of thickness $z_{0}$ is given in Ref. \citenum{Mertelj:2009p6019}:
\begin{equation}
\text{\ensuremath{\delta}}T_{s}(t)=\text{\ensuremath{\frac{\delta T_{0}}{\sqrt{1+\frac{4Dt}{z_{0}^{2}}}}}}4Dt
\end{equation}
where $D=K/\rho c$ is the diffusivity. For more specific geometries,
the temperature as a function of depth and time must be calculated numerically. To some extent, the effect of laser heating can be experimentally calibrated by comparing
the equilibrium reflectivity at two temperatures, i.e., $R_{eq}(T)-R_{eq}(T+\delta T)$, with the transient
$\delta R(t)/R$, which is more relevant when
dealing with signals on timescales beyond a few ps.

\subsection{Towards a non-equilibrium optical spectroscopy: supercontinuum white-light generation}

The experimental setup dedicated to optical P-p experiments aiming at a time dependent reconstruction of the frequency-dependent dielectric functions are based on the possibility of producing short pulses with a broad wavelength content. Broadband pulses are commonly produced via self-phase modulation in transparent crystals (energy E$>$0.5 $\mu$J/pulse in amplified systems) \cite{Bloembergen} or, more recently, with microstructured photonic-crystal fibers (PCF) operating also with low-energy light pulses (E$<$100 nJ/pulse) \cite{Leonard2007,Dudley2006}. White-light generation in sapphire (of CaF$_{2}$) crystals with 800 nm pumps allows for the generation of a continuous spectra between 450 and 1300 nm. Most commonly, the transient reflectivity (transmissivity) is measured simultaneously on a large spectral range by means of silicon-based multichannel detectors covering the visible and near IR range (400-1000 nm).

It should be noted that the non-linear interactions lead to a broadband non-transform-limited pulse. The residual spectral chirp of the light pulses can be compensated by optomechanical schemes \cite{Wegkamp2011} or with a post-processing procedure of the time resolved data \cite{Cilento2010, Novelli2012}. The advantage of broadband time domain spectroscopy lies in the fact that broadband measurements can be used to develop effective non-equilibrium models for the dielectric function and to disentangle the different physical mechanisms perturbing the optical properties at equilibrium. This issue will be extensively discussed in Sec. \ref{sec_results}. 

\begin{figure}[t]
\begin{centering}
\includegraphics[width=1.1\textwidth]{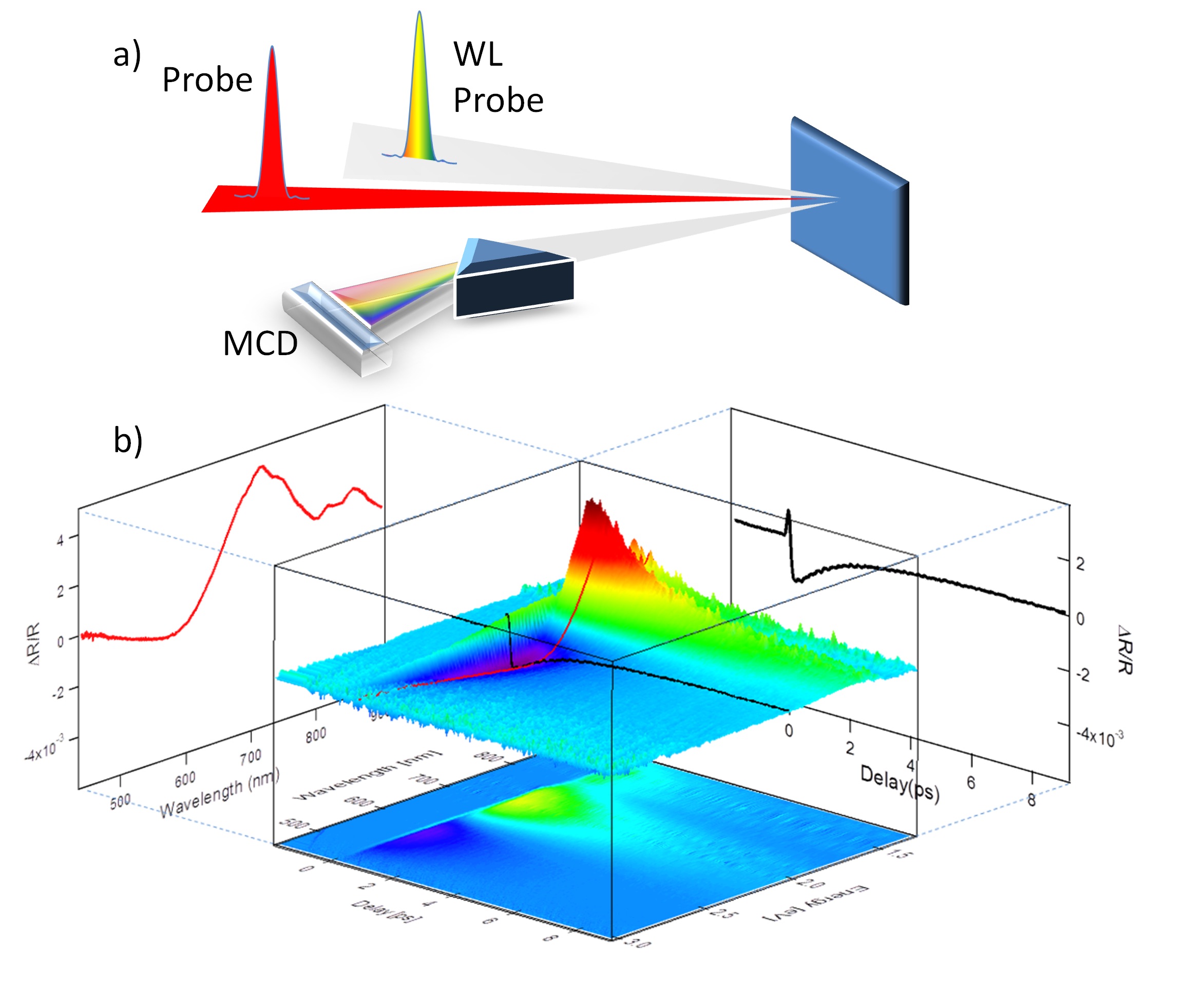}
\caption{a) In ultrafast broadband P-p spectroscopy the white light pulses are generated by non-linear optics (see text) and measured on multichanned detectors allowing for frequency resolution. b) The measurement of the wavelength dependent response on a large energy range for every time delay between pump and probe allows for a differential approach starting from equilibrium optical properties.}
\label{fig_WL}
\end{centering}
\end{figure}

Different approaches can be used to extract the time evolution of the optical functions from time domain broadband reflectivity measurements. Starting from a Drude-Lorentz fit to the static ellipsometry data it is possible to calculate the equilibrium reflectivity ($R_{eq}(\omega)$). From a model describing the static response, it is possible to fit the measured transient reflectivity variation, i.e., $\delta R/R$=$[R(\omega,t)-R_{eq}(\omega)]/R_{eq}(\omega)$, with a differential model for the perturbed reflectivity obtained by the variation of the parameters used to fit the equilibrium data. The values of the oscillator parameters obtained by this fitting procedure for any time $t$ are used to calculate the time-evolution of the different optical constants, and thereby the evolution of the physical parameters relevant for the experiment \cite{Giannetti2011}. 

An alternative approach is to start from the broadband equilibrium reflectivity measurements, to perform a Drude-Lorentz fit over a wide energy range, from few meV up to several tens of eV. In such a case, the real and imaginary parts of the equilibrium dielectric function are calculated in the entire energy range through the Kramers-Kronig (KK) relations. In order to obtain the time-domain changes of the optical functions, it is reasonable to assume that for small reflectivity changes the variations outside the measured energy range are either small or distant in the energy scale so that they do not affect significantly the optical response in the probed range. Starting from these assumptions, the time evolution of the optical quantities can be obtained by the Kramers-Kronig transformations \cite{Novelli2014}. 

\subsection{Time Domain THz spectroscopy}

\begin{figure}[t]
\begin{centering}
\includegraphics[width=1\textwidth]{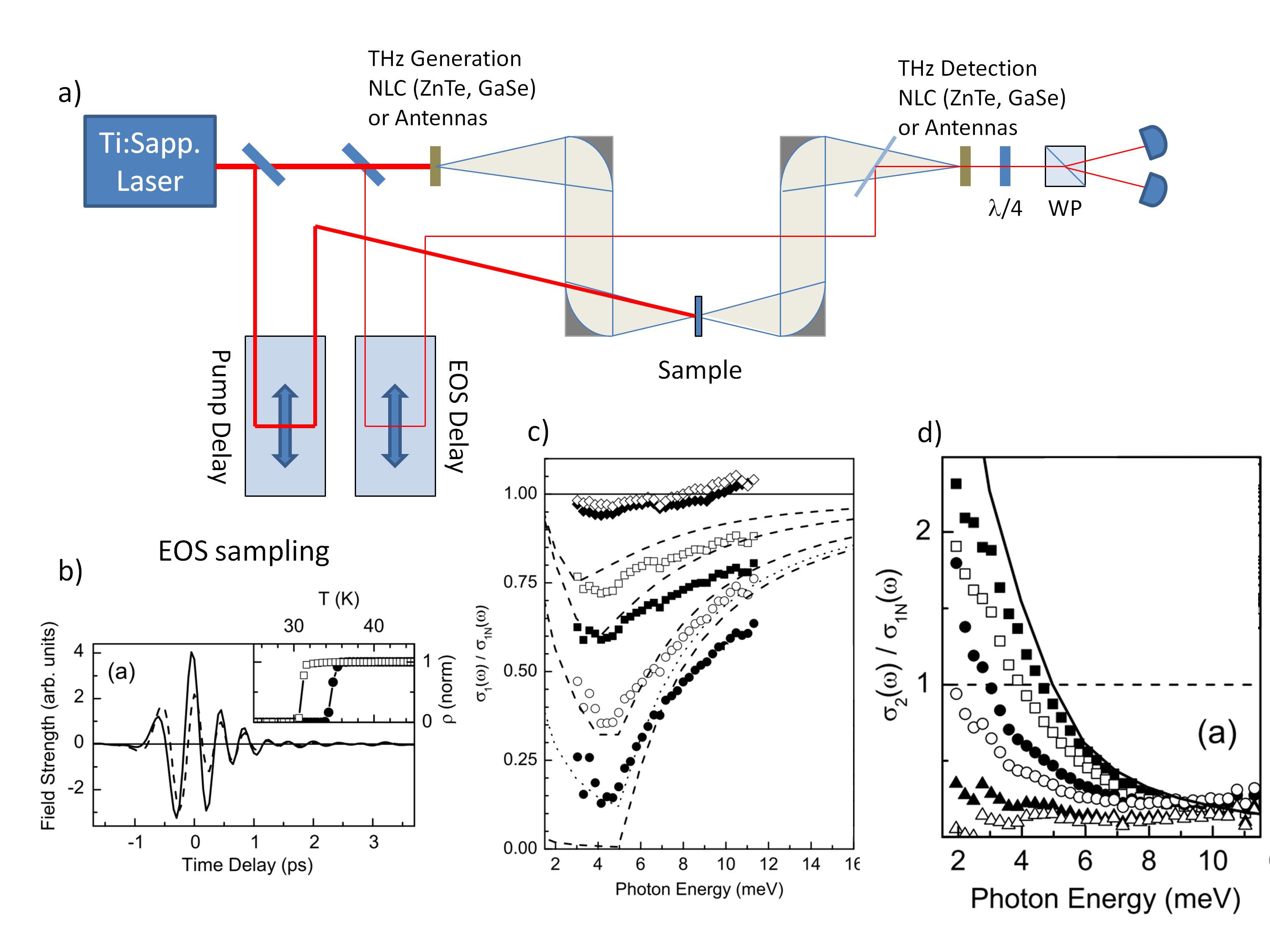}
\caption{a) Sketch of optical pump and THz probe setup. The short light pulses produced by a Ti:Sapphire laser are rectified through optical reptification in non-linear Crystals (NLC) of THz antennas. The produced fields are measured by Electro Optic Sampling (EOS) after the interaction with the sample. Time domain THz spectroscopy can be used to measure either the static response in the far-IR region as well as to study the dynamical response in a pump$\&$probe configuration. b) A typical time domain THz trace measured in transmission ($MgB_{2}$) carries both amplitude and phase information on the transmitted e.m. field allowing for the extraction of both real and imaginary part of the optical conductivity in the THz region. c) and d) depict respectively the temperature dependence of the $\sigma_{1}$ and $\sigma_{2}$ in the THz region for superconducting $MgB_{2}$.\cite{Kaindl2002}}
\label{fig_THz}
\end{centering}
\end{figure}

The field of time domain THz spectroscopy rapidly evolved in the last decades and various configurations making use of ultrashort THz pulses to measure the static optical constants in the THz region have been developed. The interest in investigating and controlling the population dynamics of low energy excitation is driven by the idea of accessing dynamically the excitations that determine the thermodynamic properties of the samples. In particular, light at terahertz frequencies allows for the measurement of excitations at the characteristic energy scales of the wondered physical mechanisms. In this frame, interesting cases are given by phonon, spin or orbital excitations as well as superconducting and charge density waves gaps in transition metal oxides and related compounds. 

We define the THz range as non-ionizing radiation with photon energy ranging from  0.1 THz (0.4 meV or 4 K) to 5 THz (20 meV or 220 K). A standard setup for measuring the optical conductivity in this region is shown in Figure \ref{fig_THz}. Ultrashort light pulses with a frequency content in the THz region are commonly produced by a second order non-linear process known as optical reptification of ultrashort light pulses in the NearIR-visible range \cite{Hangyo2005}. Various non-linear media as ZnTe \cite{Nahata1996}, GaSe \cite{Shi2002}, LiNbO$_{3}$ \cite{Hebling2008} and even organic crystals (DAST \cite{Schneider2006}) can be used to this purpose. The ultrashort light pulses commonly employed have a duration of order 100 fs, so that the optical rectification produces electromagnetic fields with frequency content up to a few THz, depending (inversely) on the length of the generation pulse and the absorption behavior of the non linear crystal. In ZnTe the generation of high THz frequencies (3-4 THz) is limited by the absorption from phonon modes. However, other optical rectification (EOS) schemes, based on thin GaSe, can produce (measure) higher frequencies \cite{Huber2000,Huber2001,Huber2005,Sell2008}.

Another common scheme used for the optical rectification is based on semiconductor THz photoconductive antennas (PA). In PA the free carriers created by the short light pulses illuminating a biased semiconductor lead to a discharge between the electrodes acting as a source for the THz field. The free photo-carriers exhibit a time-dependent behaviour due to both the time-varying exciting pulse and the relaxation dynamics characteristic of the material. This limit the speed of the discharge and therefore the spectral content of the THz radiation emitted \cite{Xi_Cheng2009}. The low damaging threshold of PA makes them ideal for high repetition rate and low power operation, hence they are most commonly used in combination with Ti:sapphire oscillators. 

As shown in Fig. \ref{fig_THz}, THz pulses can be measured with via the so-called Electro Optics Sampling (EOS) \cite{Hangyo2005, Kaindl2006}, that simultaneously accesses both the amplitude and the phase of the electromagnetic field. The intrinsic amplitude and phase sensitivity in EOS measurements allow for the study of the optical constants in the whole terahertz range without using KK transformation \cite{Kaindl2006}.

\subsubsection{Optical pump THz probe spectroscopy}
Time domain THz spectroscopy can be easily implemented in a P-p configuration. The almost single-cycle THz pulse is measured after the interaction with the sample (reflectivity or transmission) perturbed by a second THz pulse. Most commonly, THz P-p spectroscopy combines the excitation with visible or near-IR light pulses with the measurement of the time evolution of the optical properties in the THz region. A typical configuration for pump and probe spectroscopy is presented in Fig. \ref{fig_THz}. In addition to the electro-optical sampling path, a second delay line is used to control the arrival time of the pump with respect to the THz probe. Recent upgrades for the production of ultra-intense THz pulses that can be employed in P-p experiments where THz pulses are used both as a time domain spectroscopic tool as well as the mean to drive matter into transient states (see Sec. \ref{sec_optical_control}).

\subsection{Other time domain domain techniques}

In addition to all-optical P-p experiments other techniques have been implemented for detecting both the time evolution of the electronic structure and the ionic position. Here, we will briefly introduce the basic concepts of time-resolved ARPES and electron diffraction, for they are attracting a fast-growing interest in the scientific community.

\subsubsection{Time resolved photoemission}
\label{sec_TRARPES}
The P-p sequence in time-resolved angle-resolved photoemission
spectroscopy (TR-ARPES) use the same P excitation
as for the optical methods, but the probe pulse is in
the UV spectral range. The interests in this spectroscopy are rapidly expanding for it offers the possibility of directly measuring the non-equilibrium momentum dependence and electronic distribution functions. The kinetic energy and momentum of the photoelectrons are typically measured by time-of-flight spectrometers (TOF) or by state-of-the-art hemispheric analyzers.
TR-ARPES also allows to measure the spin of the photoemitted electrons, when the electron analyzer is equipped with a Mott detector or other spin-filter devices. 

In order to photoemit electrons from a solid the photon energy must exceed the work function, hence for the most common materials ultraviolet laser pulses are required, i.e. typically $\hbar \omega >$6 eV \cite{Perfetti:2006p1775,Perfetti2007,Smallwood2012_b}.  The UV photons can be obtained by frequency-quadrupling the fundamental 
Ti-sapphire frequency ($h\nu$=1.55 eV) with a set of $\beta$-BBO crystals. Unfortunately, because of kinematic constraints, with 6 eV photons 
only a limited part of the Brillouin zone (BZ) can be accessed. This is a serious limit preventing, for example, the investigation of the dynamics of the antinodal regions of high-temperature
superconductors. Other means, mainly based on high harmonic generation (HHG) non-linear processes in gases, have been exploited in these last years and ultrashort pulses in the XUV range ($\sim$10-100 eV) have been generated at a relatively low (few kHz) repetition rate. Although these sources have been proved to be suitable for covering the entire BZ of all the solids, the relatively large pulse bandwidth along with the relatively low repetition rate significantly limit the TR-ARPES momentum and energy resolution along with the signal statistic \cite{Ishizaka:2011dg}.

\subsubsection{Time-resolved electron and X-ray diffraction}
In time domain X-ray diffraction the stroboscopic scheme of P-p experiments combines ultrashort pumps with time domain measurements of the ultrashort diffracted X-ray probe. Various schemes have been implemented using both table top \cite{Murnane1991,Siders1999} and synchrotron radiation \cite{Schoenlein2000, Zholents1996, Lindenberg2000,Cavalleri2001,Larsson1997,Johnson2008,Beaud2007}. More recently hard and soft X-ray diffraction experiments based on free electron laser (FEL) sources have been made possibile \cite{dejong2013,Beye2012,Mankowsky2014}. In particular, the combination of soft X-ray resonant at the Ni and Cu $L_3$ edges ($\sim930$ eV) with tunable pump excitation in the THz/mid-IR/visible range, opened important perspectives in the study of the non-thermal dynamics of charge-order in nickelates \cite{Chuang2013,Kung2013,Forst2015} and cuprates \cite{Forst2014,Forst2014b}.

In the case of time-resolved electron diffraction, the light pulses are used to generate electron bunches by photoemission from a photocathode. Then these ultrashort electron pulses, properly delayed with respect to the pump pulse, can be used to generate diffraction patterns from the material in transmission or in the low angle configuration. Transmission experiments require thin samples for the electrons to cross the sample, hence requiring the use of sample preparation techniques which are standard in electron microscopy, such as slicing, ion milling etc. Apart from direct Bragg diffraction of the incident electrons, more sophisticated techniques are becoming available, such as time-resolved observation of Kikuchi bands, which occur when some electrons, originally diffusely scattered in the sample, satisfy the Bragg condition for a particular plane \cite{Yurtsever2011}.

Details on the different approaches to time resolved structural measurements are beyond the purpose of this review and can be found elsewhere \cite{Pfeifer2006,Hada2013,wang2009,Zewail:2010hn,Sciaini2011}.

\subsection{The next generation of ultrafast sources}
Even though the recent advances in ultrafast techniques determined the success of P-p techniques as an innovative and important tool to investigate the properties of solids, the field is still rapidly growing and is expected to provide in the near future new and advanced tools for ultrafast time-dependent experiments.
A major achievement of this last decade arises from the remarkable development of novel radiation sources capable to generate EUV and ultrashort X-ray pulses. The extension of the time resolved spectroscopies and scattering to the X-ray region is allowing to picture the transient changes of electronic states and lattice structures. A major revolution for time resolved X-ray based experiments started in 2005 when the first VUV free-electron laser (FEL) delivered the first light at the DESY laboratory in Hamburg (SASE) \cite{Ackermann2007}. Today, other facilities operating on the SASE FEL principle include the Linac Coherent Light Source (LCLS) at the SLAC National Accelerator Laboratory \cite{Emma2010} and the SPring-8 Compact SASE Source (SCSS) \cite{Ishikawa2012}, whereas two other FEL facilities are under construction, the European x-ray free electron laser (XFEL) in Hamburg and the SwissFEL at the Paul Scherrer Institute (Switzerland), along with the fully coherent laser seeded FEL operating in the EUV region at Elettra Sincrotrone Trieste (Italy) \cite{Allaria2012}. All this huge technological effort in developing new X-ray sources, delivering fully coherent EUV, soft and hard X-ray pulses with a time structure of few tens of fs, is expected to unlock the gate for the next generation of pump probe experiments. The possibility of directly investigating the electronic core-levels and the lattice structure of materials in a time-resolved fashion is progressively becoming reality.

\setcounter{section}{2}
\section{Quasiparticle dynamics in correlated materials: basic concepts and theoretical background}
\label{sec_QPdynamics}
The first results of P-p experiments on correlated materials and superconductors evidenced a rich phenomenology that boosted, in the first stage, the development of simple models to describe the ultrafast dynamics of quasiparticles and, more in general, of low- and high-energy charge excitations in gapped and correlated states. Nevertheless, it was also soon realized that the microscopic processes which lead to the relaxation of the out-of-equilibrium charge distribution are the same than those regulating the scattering processes in (quasi-)equilibrium conditions and determining the properties of the quasiparticles and of the fundamental excitations. In this chapter we will discuss the most important interaction mechanisms that lead to the dressing of the naked charge excitations and to the development of the concept of "quasiparticles" and incoherent excitations in doped Mott insulators. Even though the use of these concepts is widespread and has been extensively introduced in many works and reviews, here we will touch the most important issues that are of relevance for the understanding of the quasiparticle dynamics and constitute the foundation of the non-equilibrium models that will be presented in Sec. \ref{sec_basicconceptsNES} and Chapter \ref{sec_theory} and extensively used in Chapter \ref{sec_experiments}.

\subsection{The dynamics of quasiparticles}
\subsubsection{Quasiparticle scattering in the normal state: electron-phonon coupling}
\label{sec_ephcoupling}
In metals, the dynamics of a quasiparticle (QP) at frequency $\omega$ and momentum $\bf{k}$ is determined by the scattering processes between the QP itself and all the external degrees of freedom coupled with the QP. Scattering mechanisms include both electron-electron interactions and the coupling with bosons, such as the lattice vibrations, that can be considered as an external reservoir in thermal equilibrium with the QPs population. The scattering processes affect both the QP effective mass $m^*$ and lifetime $\tau$ and $-$ within the effective mass approximation $-$  are accounted for by the complex single particle self-energy, $\Sigma(\textbf{k},\omega,T)$. The inverse of the imaginary part of $\Sigma(\textbf{k},\omega,T)$ provides the finite QP lifetime $\tau$=$\hbar$/Im$\Sigma$.

In the normal state ($T>T_c$) of conventional superconductors, the QP dynamics is strongly affected by the scattering with lattice vibrations, whose dispersion, $\Omega(\textbf{q})$, and high-energy cutoff, $\Omega_c$, determine the frequency region over which the QP scattering time, $\tau(\omega)$, is strongly frequency-dependent. All the allowed QP-phonon scattering processes can be wrapped up in the electron-phonon coupling function (or bosonic function) $\alpha^2F(\bf{k}, \bf{k}$-$\textbf{q}, \Omega)$ that is determined by both the phonon density of states and the matrix element of the electron-phonon interaction. After integration over all the possible scattering wavevectors $\textbf{q}$, the electron self-energy $\Sigma(\omega,T)$ can be calculated as a convolution integral between the bosonic function $\alpha^2F(\bf{k}, \Omega)$ and a kernel function $L(\omega,\Omega,T)$ \cite{Kaufmann1998}:
\begin{equation}\label{eq_SelfEnergy}
\Sigma(\textbf{k},\omega,T)=\int_0^\infty\alpha^2F(\textbf{k}, \Omega)L(\omega,\Omega,T)d\Omega
\end{equation}
The kernel function
\begin{equation}\label{eq_Kernel_EDM2_a}
L(\omega,\Omega,T)=\int \left[ \frac{n(\Omega',T)+f(\Omega,T)}{\Omega-\omega+\Omega'+i\delta} + \frac{1+n(\Omega',T)-f(\Omega,T)}{\Omega-\omega-\Omega'-i\delta} \right ]d\Omega'
\end{equation}
accounts for the distribution of the Fermionic QPs and bosonic excitations through the Fermi-Dirac ($f(\Omega,T)$) and Bose-Einstein ($n(\Omega,T)$) distributions, and can be calculated analytically:
\begin{equation}\label{eq_Kernel_EDM2_b}
L(\omega,\Omega;T_e,T_b)=-2\pi i\left[n(\Omega,T_b)+\frac{1}{2}\right] +\Psi \left(\frac{1}{2}+i\frac{\Omega-\omega}{2\pi T_e}\right) -\Psi \left(\frac{1}{2}-i\frac{\Omega+\omega}{2\pi T_e}\right)
\end{equation}
where $\Psi$ are digamma functions and the dependence of the different terms on the temperatures of the electronic QPs ($T_e$) and bosonic excitations ($T_b$) has been made explicit. 

\begin{figure}[t]
\begin{centering}
\includegraphics[width=0.7\textwidth]{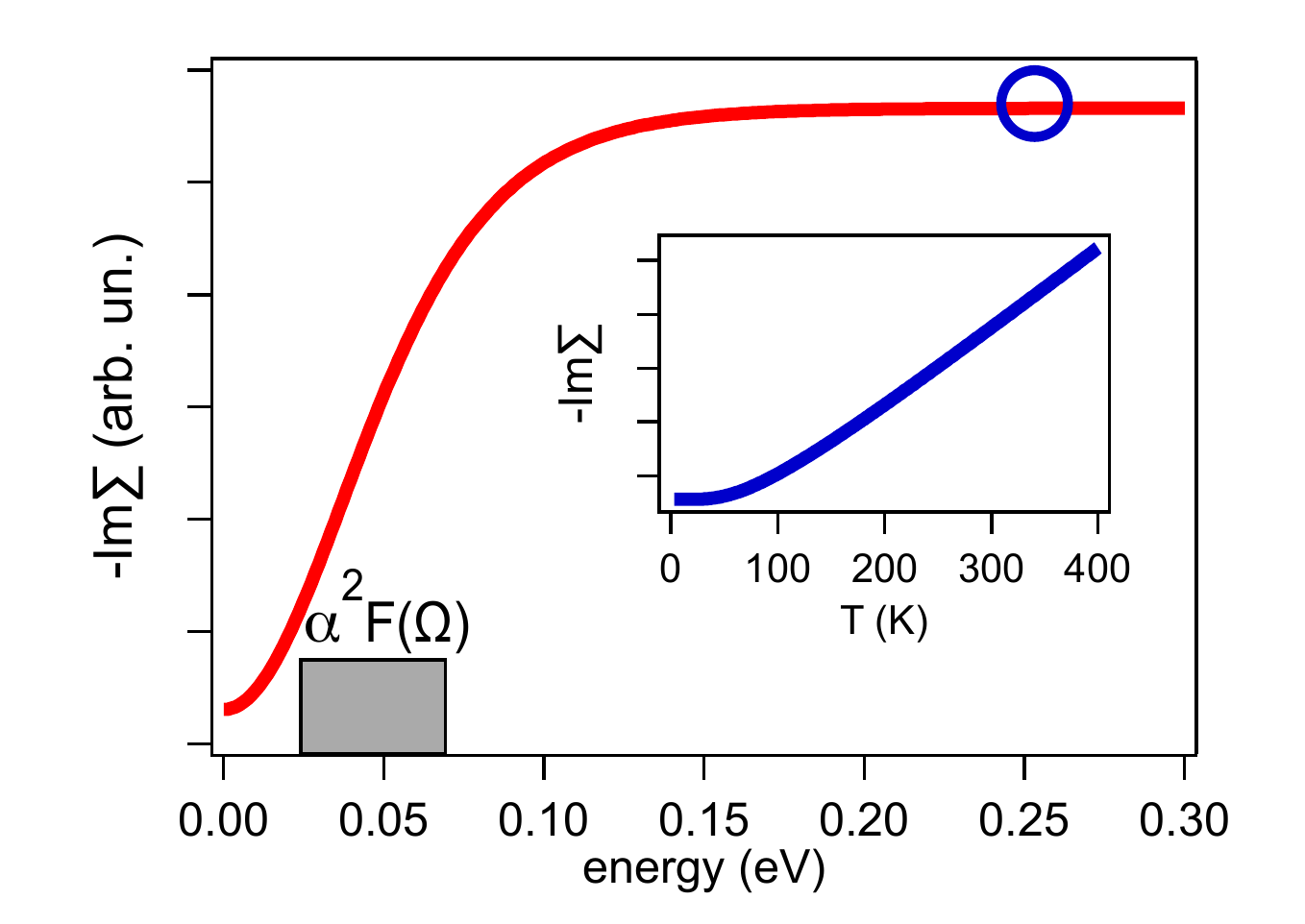}
\caption{The $\textbf{k}$-space integrated Im$\Sigma(\omega,T=300 K)$ is reported as a function of the frequency. For the sake of simplicity we calculated Im$\Sigma(\omega,T)$ by solving Eq. \ref{eq_SelfEnergy}, in which we considered the coupling with a bosonic function localized in the 25-75 meV energy range and represented by the histogram reported in the figure. The inverse QP lifetime reaches a constant asymptotic value on the energy scale corresponding to few times the energy scale of the bosons. In the inset we report the temperature dependence of Im$\Sigma(\omega,T)$ at the fixed value $\hbar\omega$=0.25 eV. The increase of the inverse lifetime (scattering rate) is a consequence of the increase of the boson density as the temperature of the system is increased. The temperature-dependent boson density is accounted for by the kernel function $L(\omega,\Omega;T)$ defined in Eqs. \ref{eq_Kernel_EDM2_a},\ref{eq_Kernel_EDM2_b}.}
\label{fig_selfenergy}
\end{centering}
\end{figure}

In this formalism, the frequency-dependent scattering rate is a consequence of the microscopic interaction of the QPs with a distribution of bosons at temperature $T_b$. As an example, Figure \ref{fig_selfenergy} shows the electronic self-energy for the simple case of a generic coupling represented by a histogram function peaked at 50 meV. On an energy scale larger than the $\alpha^2F(\textbf{k}, \Omega)$ cut-off, the QP scattering rate reaches an asymptotic value that results from all the scattering process with bosons at energies $\Omega$$<$$\Omega_c$. As the temperature is increased, the change in the total number of phonons present in the system causes the increase of Im$\Sigma(\omega,T)$, as shown in the inset of Figure  \ref{fig_selfenergy}. We stress that when the QP scattering processes can be effectively described by the coupling with bosonic degrees of freedom, the temperature of the system and the bosonic function are the only parameters that control $\tau(\omega)$.

\begin{figure}[t]
\includegraphics[bb= 40 170 1000 620, width=1\textwidth]{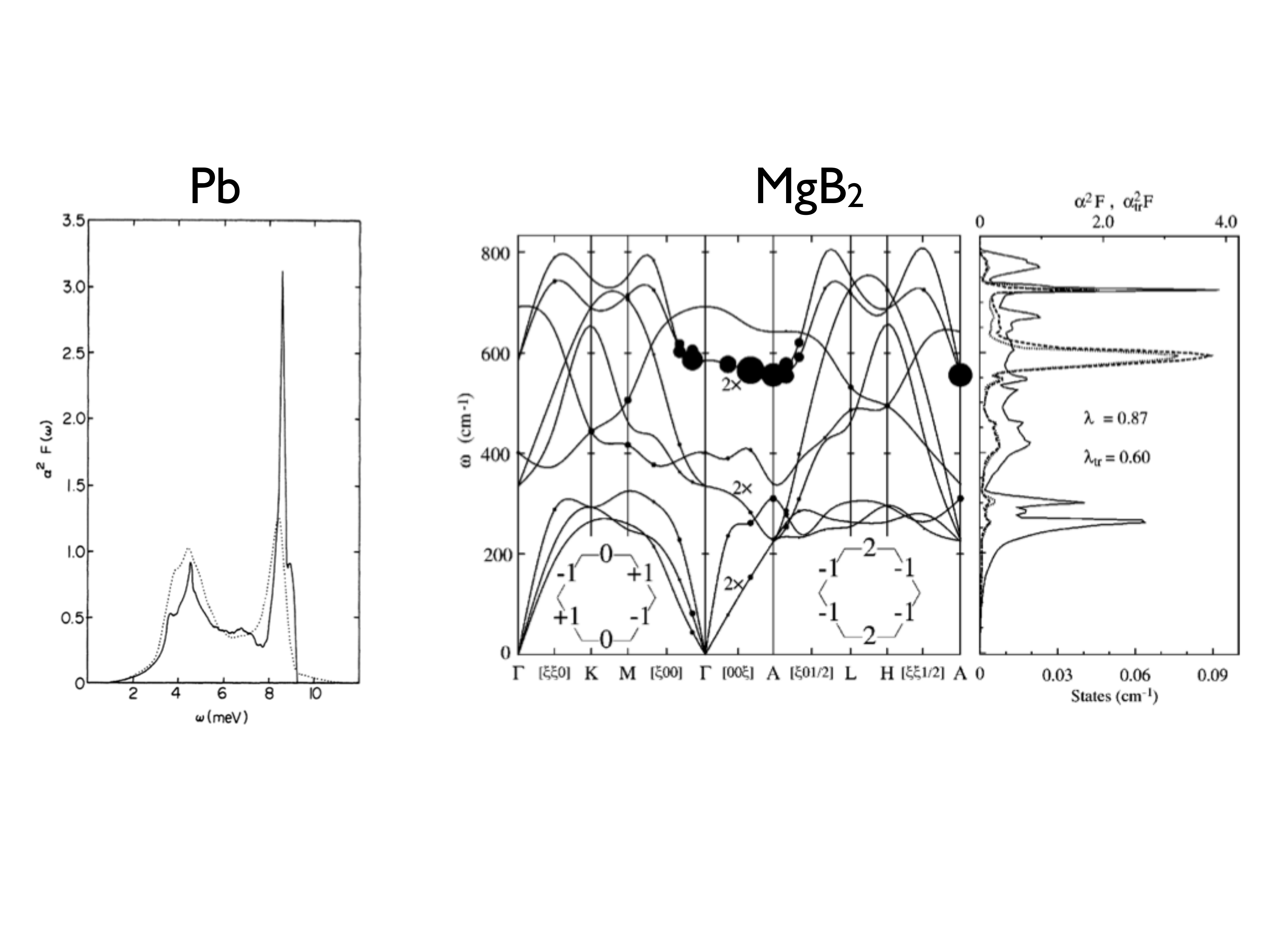}
\caption{Left panel: Bosonic function of Pb as measured by tunneling experiment \cite{Rowell1965} (dotted line) and calculated  \cite{Tomlinson1976} (solid line). Right panel: phonon dispersion and density of states and bosonic function calculated for MgB$_2$ \cite{Kong2001}.}
\label{fig_alfa2F}
\end{figure}

The $\textbf{k}$-space integrated bosonic function is defined as:
\begin{equation}
\label{eq_k_integrated_glue}
\alpha^2F(\Omega)=\frac{\int_{S_F}\frac{dS_k}{\hbar v_k}\alpha^2F(\textbf{k}, \Omega)}{\int_{S_F}\frac{dS_k}{\hbar v_k}}
\end{equation}
$dS_k$ being an infinitesimal surface element of the Fermi surface $S_F$ and $v_k$ the Fermi velocity at momentum $\textbf{k}$. This approach has been successfully used \cite{Carbotte1990} to explain the scattering properties of conventional metals that develop an isotropic superconducting gap in the electronic density of states ($\Delta(\omega$,$T$)), at temperatures smaller than the critical temperature $T_c$. The close correspondence between the $\alpha^2F(\Omega)$ function measured \cite{Rowell1965} by tunneling experiments (see Figure \ref{fig_alfa2F}) and the phonon density of states $F(\Omega)$ reconstructed by inelastic neutron scattering experiments \cite{Brockhouse1962} is considered as one of the most convincing proofs of the phonon-mediated pairing mechanism driving the formation of the Cooper pairs in metals. 

The excited state dynamics in this formalism is expected to be characterized by the instantaneous Coulomb repulsion between charged QPs and a finite timescale attraction corresponding to the retarded interaction with the phonons. In the prototypical case of Pb, the high-energy cutoff in the $\alpha^2F(\Omega)$ function ($\Omega_c\simeq$9 meV) corresponds to $\hbar$/$\Omega_c\simeq$70 fs ($\hbar$=658 meV fs). This defines the fastest timescale of the electron-boson interaction.
The possibility of defining a retarded electron-boson interaction was the key to the success of the BCS theory to explain superconductivity in conventional isotropic metals. 

The critical parameter determining the $T_c$ of the system is the electron-phonon coupling constant, defined as $\lambda_{lat}$=$2\int\alpha^2F(\Omega)/\Omega\;d\Omega$.  In the strong-coupling formalism, the McMillan's formula \cite{McMillan1968}, based on the Eliashberg theory \cite{Eliashberg1960,Eliashberg1961}, can be corrected to calculate \cite{Dynes1972,Allen1975} the critical temperature for pairing in the $s$-wave channel:
\begin{equation}
\label{eq_Tc}
T_{c}=0.83\tilde{\Omega}\exp\left[\frac{-1.04(1+\lambda_{lat})}{\lambda_{lat}-\mu^*(1+0.62\lambda_{lat})}\right]
\end{equation}
where ln$\tilde{\Omega}$=2/$\lambda_{lat}\int_0^{\infty}\alpha^2F(\Omega)\mathrm{ln}\Omega /\Omega d\Omega$ and $\mu^*$ is the non-retarded screened Coulomb pseudopotential that accounts for all the instantaneous electron-electron interactions.

This picture becomes more complex in the case of anisotropic systems. Particularly interesting is the case of MgB$_2$ that becomes superconductor \cite{Nagamatsu2001} at $T_c\sim$40 K, which is the largest critical temperature for what is commonly believed to be a conventional electron-phonon superconductor. MgB$_2$ is a layered system consisting of graphene-like boron planes separated by magnesium planes. In this system, the electronic properties at the Fermi level are mostly determined by boron $p_x$, $p_y$ orbitals that form in-plane $\sigma$-bands and $p_z$ orbitals that form an out-of-plane $\pi$-band. The strong coupling \cite{Kong2001} of the $\sigma$-bands to optical B-B bond-stretching modes is responsible for a strong electron-phonon coupling, that is described by an $\alpha^2F(\Omega)$ with a strong peak at $\Omega_{SCP}\sim$70 meV, as shown in Figure \ref{fig_alfa2F}. Although the phonon density of states is almost flat around 70 meV, the coupling of the $\sigma$-bands to optical B-B stretching modes alone provides $\lambda$=0.62 that represents about the 70$\%$ of the total coupling. The strong anisotropy of the electron-phonon coupling is expected to be reflected in the dynamics of the system. Besides the instantaneous Coulomb repulsion, the characteristic structure of $\alpha^2F(\Omega)$ defines two different timescales: i) $\hbar$/$\Omega_{SCP}\simeq$10 fs related to the coupling with Strongly Coupled optical Phonons (SCP) and ii) $\hbar$/$\Omega_{lat}\gg$10 fs related to the residual coupling with the other optical and acoustic modes.

\subsubsection{Quasiparticle scattering including bosonic degrees of freedom of different nature}
\label{sec_QPscattering}
The physics of the cuprates represents a challenge to a picture of QP interacting with bosons within the effective-mass approximation. The insulating character of the undoped compounds (3$d^9$ electronic configuration, which corresponds to a half-filled band of $d{x^2-y^2}$ character) is the consequence of the strong on-site Coulomb repulsion, $U$, between two holes occupying the same Cu lattice sites. The large $U$ value ($U$$>$$W$, $W$ being the bandwidth) leads to the failure of the single-electron approximation and to the necessity of introducing new models including strong correlation effects and Mott-Hubbard insulators. Although the hybridization between the Cu-3$d_{x^2-y^2}$ and the O-2$p_{x,y}$ orbitals plays a role into a wealth of interesting phenomena, it is commonly accepted that many basic properties of the underdoped compounds are well captured by the single-band Hubbard hamiltonian \cite{Hubbard1963,Anderson1963,Lee2006}:
\begin{equation}
\label{eq_Hubbard}
\hat{H}=-\sum_{i,j,\sigma}(t_{ij}\hat{c}^{\dag}_{i\sigma}\hat{c}_{j\sigma}+c.c.)+U\sum_i \hat{n}_{i,\uparrow} \hat{n}_{i,\downarrow}-\mu\sum_i \hat{n}_{i} 
\end{equation}
where $\hat{c}^{\dag}_{i\sigma}$ ($\hat{c}_{j\sigma}$) create (annihilate) an electron with spin $\sigma$ on the $i$ ($j$) site, $\hat{n}_{i,\sigma}$=$\hat{c}^{\dag}_{i\sigma}\hat{c}_{i\sigma}$ is the number operator, $t_{ij}$ the hopping amplitude from the site $i$ to $j$ and $\mu$ is the chemical potential that controls the average electron density $n$=$\sum_{i\sigma}\langle\hat{n}_{i,\sigma}\rangle/N$ on each of the $N$ lattice sites. This model represents the paradigm of the physics of strong correlations as it features the competition between the kinetic energy term, which naturally leads to metallic states, and the Coulomb energy term, which describes the repulsion between two electrons on the same lattice site.

In the limit $U\rightarrow\infty$ and for $n$=1 (half-filling), the electrons are completely localized by the repulsion and the ground state of the system is an insulator described by atomic-like wavefunctions $\Psi_i$ localized on the lattice site $i$, with infinitely small fluctuations of the average occupation number, i.e., $\delta n \rightarrow$0. 

This is the extreme limit of a Mott-Hubbard insulator with a large gap $U$. On the other hand, for finite $U$ values and for $t \ll U$ --where $t$ is the nearest neighbor hopping-- an antiferromagnetic coupling $J$=4$t^2$/$U$ between neighboring sites arises due to virtual  hopping of holes into already occupied sites (with a $U$ energy cost) through Anderson's superexchange mechanism. If the lattice is not very frustrated, this leads to a magnetic ordering of the spins of the localized electrons. In this framework, the low-energy dynamics is described by a Heisenberg model.

The addition of carriers to the half-doped system by electron ($n$$>$1) or hole doping ($n$$<$1) leads to the so-called t-J model, which allows for the motion of carriers in the presence of superexchange interactions. By increasing the doping the antiferromagnetic ordering is destroyed, the system develops a low-energy spectral weight and it becomes a  ``bad metal", whose nature is still object of a lively debate, while a Fermi-liquid is  progressively recovered at larger dopings. The intertwining of high-energy incoherent excitations at energies of the order of $U$ and low-energy coherent quasiparticles makes the description of the electron dynamics of cuprates an extremely challenging problem.
In the case of the cuprates the above Mott-Hubbard picture does not completely apply as doped holes have been shown to occupy predominantly the oxygen sites, thus calling for a multi band description of the material. This brings into play another energy scale ($\sim$2 eV) associated to the charge transfer process and a coupling with oxygen phonons which can lead to polaronic effect for light doping \cite{Alexandrov:1994wdb,Mihailovic:2001ur,Muller:2007p3138}. 

Despite the overwhelming consensum about the role of strong electron-electron interactions not only in driving the Mott insulating state and antiferromagnetism, but also as the main source of the superconducting pairing, one of the main open questions is related to the very possibility of defining the boson-mediated interaction in doped cuprates \cite{Anderson2007}. The issue is whether the low-energy hole (electron) dynamics can be described through the interaction with a bosonic spectrum with a cutoff of the order of 2$J$ \cite{Scalapino2012}, i.e., the high-energy cutoff of the spin-fluctuation spectrum,  or the non-retarded (sub-fs) $U$ energy scale is the only relevant parameter \cite{Anderson2007}. 

A wealth of techniques have been used to extract the electron-boson coupling in cuprates, as discussed in a recent comprehensive review \cite{Carbotte2011}. Here we report the main experimental outcomes suggesting that at finite hole (electron) concentrations, the $\alpha^2F(\Omega)$ function can be recast into a more general bosonic function:
\begin{equation}
\Pi(\Omega)=\alpha^2F(\Omega)+I^2\chi(\Omega)
\label{eq_Pi}
\end{equation}
that includes electronic correlations through the term $I^2 \chi (\Omega)$. We particularly focus on Bi$_2$Sr$_2$CaCu$_2$O$_{8+\delta}$, that is one of the most extensively studied copper oxides and allows a comparative study with different techniques.

\begin{figure}[t]
\includegraphics[width=1\textwidth]{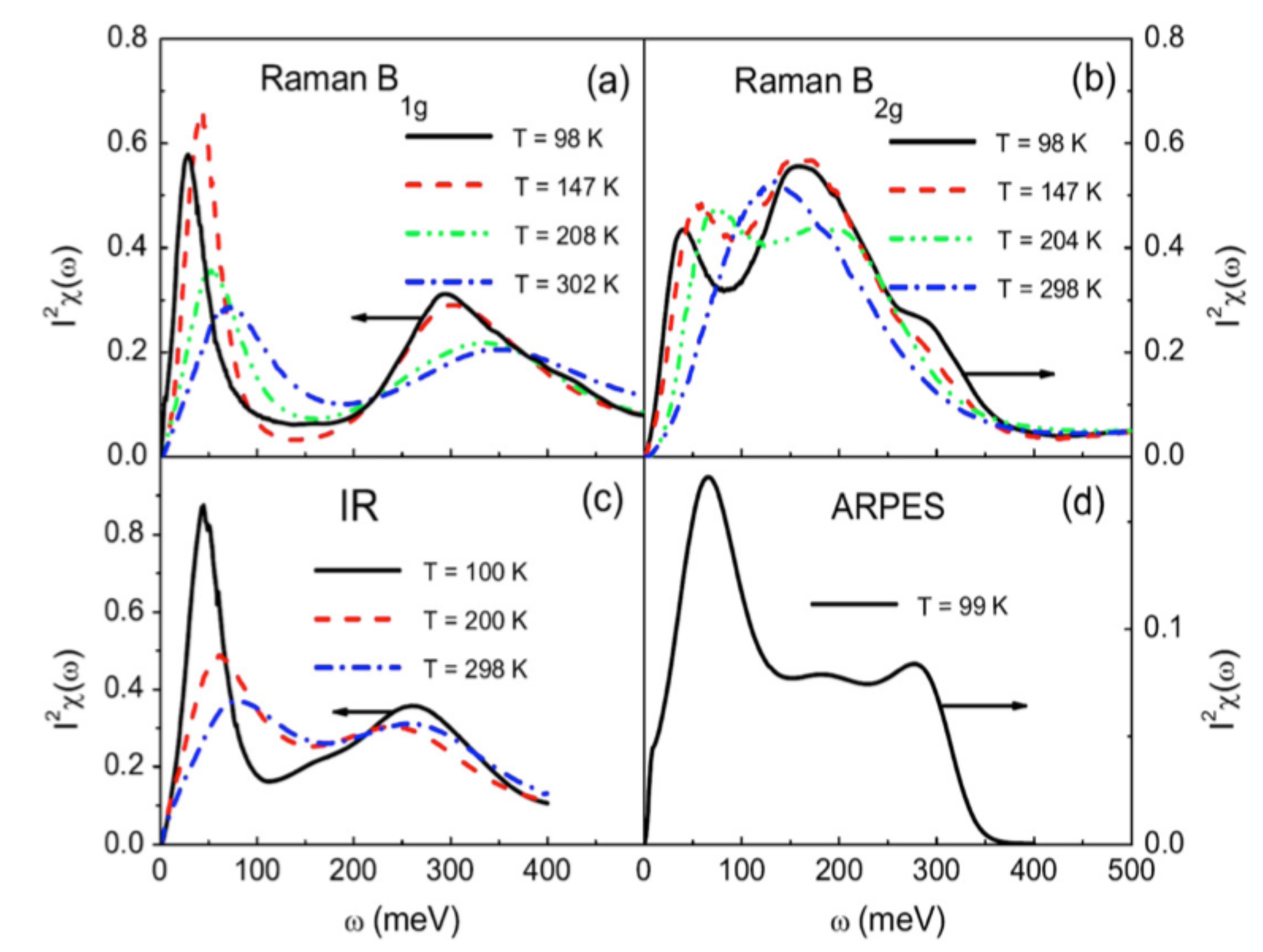}
\caption{Summary of the electron-boson spectral function in Bi2212 extracted from: (a) B$_{1g}$ (a) and (b) B$_{2g}$ Raman data \cite{Muschler2010}; (c) optical conductivity \cite{Schachinger2006,Tu2002}; (d) from nodal direction ARPES \cite{Schachinger2008,Zhang2008}. The Figure has been taken from \cite{Carbotte2011}.}
\label{fig_glueCarbotte}
\end{figure}

Angle Resolved Photoemission Spectroscopy (ARPES) \cite{Damascelli2003} has been widely used to investigate the electron-boson coupling in doped cuprates and other correlated materials. In case of strong electron-boson coupling, the QP dispersion exhibits a deviation from the non-interacting band and a kink at the energy corresponding to the boson mode coupled to the QPs appears. Considering the photoemission intensity profile at constant energy, usually called momentum distribution curve (MDC), the QP peak position ($\textbf{k}$) and width ($\Delta k$) are related to the self-energy by the relations:
\begin{eqnarray}
\mathrm{Re}\Sigma(\omega,\textbf{k})&=&\omega_k-\epsilon^b_k\\
\mathrm{Im}\Sigma(\omega,\textbf{k})&=&-\Delta kv^b_k
\label{eq_ARPES_selfenergy}
\end{eqnarray}
where $\omega_k$ is the QP energy, $\epsilon^b_k$ is the bare band energy at momentum $\textbf{k}$ and $v^b_k$ is the QP velocity. Despite the difficulties in determining $\epsilon^b_k$ and in measuring the QP coupling with broad and featureless bosons at energies $\gtrsim$100 meV, ARPES provides fundamental informations to address the electron-boson coupling problem. Figure \ref{fig_glueCarbotte}(d) shows the bosonic function extracted \cite{Schachinger2008} from ARPES spectra taken \cite{Zhang2008} along the nodal direction of optimally-doped Bi$_2$Sr$_2$CaCu$_2$O$_{8+\delta}$ (Bi2212), through Equation \ref{eq_SelfEnergy}. $\Pi(\Omega)$ exhibits a peak at $\sim$70 meV and a flat continuum with a cutoff at $\sim$350 meV.

Raman spectroscopy is another useful tool to probe the low-energy electron dynamics. In Raman experiments the spectrum of inelastically scattered light is acquired as a function of the light polarization $\sigma$, that determines the symmetry of the scattering processes that can be accessed. It has been demonstrated \cite{Muschler2010} that the Raman Scattering rate is related to the bosonic function through a relation very similar to Equation \ref{eq_SelfEnergy}, provided a weighting that takes into account the symmetry of the modes is introduced. In Figure \ref{fig_glueCarbotte}(a,b), the bosonic function extracted by Raman experiments on optimally-doped Bi2212 is reported. Depending on the polarization of the light, the extracted bosonic function is  weighted by a factor cos$^2$(2$\theta$) ($B_{1g}$ symmetry, sensitive to antinodal directions) and sin$^2$(2$\theta$) ($B_{2g}$ symmetry, sensitive to nodal directions). Considering the average over the whole Brillouin zone, the $\Pi(\Omega)$ obtained by Raman spectroscopy has features very similar to that obtained by ARPES. It consists of a strong peak at low energy ($\sim$40 meV), which is strongly temperature dependent, and a broad continuum up to $\sim$400 meV, which is less sensitive to temperature variations.

The frequency-dependent optical scattering rate ($\tau_{opt}(\omega)$) measured from IR optical spectroscopy can be also used to extract the electron-boson scattering rate, further corroborating the results from ARPES and Raman spectroscopies. The \textbf{k}-space integrated $\Pi(\Omega)$ can be obtained from the optical data through the so-called maximum entropy techniques \cite{Hwang2007}, provided a relation between the single-particle self-energy and the optical scattering rate, that involves particle-hole excitations, is found. In Section \ref{sec_dielfunction} a more comprehensive overview of the optical spectroscopy and of the underlying Extended Drude Model will be given. In Figure \ref{fig_glueCarbotte}(c) the $\Pi(\Omega)$ function of optimally-doped Bi2212 samples is reported \cite{Schachinger2006,Tu2002}. The results are in striking agreement with those obtained by ARPES and Raman spectroscopy, confirming the presence of a temperature dependent peak at $\sim$40 meV and a background with a cutoff at about $\sim$400 meV. More recent results \cite{Hwang2014}, suggests the possibility that the spectrum of bosonic fluctuations may extend up to very high energies of the order of 1-2 eV.

Finally, following the pioneering measurements on conventional superconductors \cite{Rowell1965}, tunnel conductance measurements have been extended to correlated superconductors and revealed the opening of the superconducting gap and the manifestation of the electron-boson coupling. Being the tunnelling current sensitive to the integral of the joint density of states of the two sides of the junction used in the experiment, its derivative (the tunnel conductance) $dI/dV(V,T)
$ can reveal the opening of a gap in the density of states. Meanwhile, electron-boson interactions show up as dip features in $dI/dV(V,T)$. In the particular case of Scanning Tunneling Microscopy (STM), the junction is formed by the sample and the scanning tip. STM technique is strongly complementary to \textbf{k}-space resolved spectroscopies, such as ARPES, since it provides a direct measurement of the real-space dependance of the electronic properties. A recent work \cite{Pasupathy2008} on Bi2212 has demonstrated a strong inhomogeneity of the normal state electronic excitations in the 150-300 meV energy range. These inhomogeneities are spatially related to the opening of the superconducting gap at $T$$<$$T_c$ suggesting a direct relation between the two phenomena. Furthermore, a novel analysis \cite{Ahmadi2011} of data taken with the superconductor-insulator-superconductor (SIS) technique on Bi2212 demonstrated the possibility of extracting an extended bosonic function that exhibit a strong peak at $\sim$40 meV and a weaker broad contribution at higher energies.

Taken all together, the results of ARPES, Raman, optical and tunnelling spectroscopies on Bi$_2$Sr$_2$CaCu$_2$O$_{8+\delta}$, suggest that, close to optimal doping ($\simeq$16$\%$), the QP low-energy dynamics can be effectively described by the interaction with bosonic excitations, whose spectrum is characterized by a strong peak at 40-70 meV and a featureless part extending up to $\sim$400 meV. These observations support a picture in which, the vertex corrections can be neglected for hole concentrations close or larger than the optimal doping concentration. This conclusion is in agreement with the Fermi-liquid based theoretical analysis of the electron-phonon coupling in the presence of strong electronic interactions \cite{Grilli1994,Capone2010}. In particular, in Ref. \citenum{Grilli1994} it is shown that the contribution of the vertex corrections to the effective electron-phonon coupling scales with the ratio $\omega /v_Fq$, $\omega$ and $q$ being the transferred frequency and momentum and $v_F$ the Fermi velocity. The strong correlation-driven renormalization of $v_F$ in underdoped systems thus leads to the enhancement of the bare electron-phonon coupling as the Mott insulating phase is approached.

Far from solving the problem, the experimental results raise the fundamental question of the nature of the bosonic excitations coupled to QPs. While for $\Omega$ larger than the cutoff of the optical phonon branch ($\sim$90 meV)  $\Pi(\Omega)$ is most likely related to the coupling with bosons of electronic nature, the low energy part ($\Omega<$90 meV) is the result of the interplay of electronic interactions ($I^2\chi(\Omega)$) and electron-phonon coupling ($\alpha^2F(\Omega)$). Considering the phonon acoustic and optical branches typical of cuprates, $\alpha^2F(\Omega)$ is expected to be characterized by some universal features:\\
i) the coupling of QPs to acoustic \cite{Johnston2012} and Raman-active optical \cite{Kovaleva2004} phonons in the $<$40 meV energy range;\\
ii) a relatively stronger and anisotropic coupling to either out-of-plane buckling and in-plane breathing Cu-O optical modes \cite{Devereaux2004} at $\sim$60 meV. \\

\subsubsection{Magnetic degrees of freedom}
\label{sec_magnons}
The strong on-site Coulomb repulsion $U$, that is responsible of the insulating phase at zero doping, has also profound consequences in the QP dynamics of doped systems. The first dramatic consequence of the strong $U$ is the intrinsic increase of Im$\Sigma$ related to the electron-electron interactions. The strong repulsion between the charge carriers eventually leads, at low hole concentration, to the loss of the concept of quasiparticle itself \cite{Fournier2010}, driving the metal-to-insulator phase transition. Electronic correlations also give rise to additional bosonic degrees of freedom that should be considered to correctly describe the the low-energy QP dynamics. The $I^2\chi(\Omega)$ term in Eq. \ref{eq_Pi} accounts for all these scattering processes that include the coupling with antiferromagnetic fluctuations and with the fluctuations of the order parameter associated to a possible quantum critical point underneath the superconducting dome \cite{Varma2006,Sachdev}.
\begin{figure}[t]
\begin{centering}
\includegraphics[width=1\textwidth]{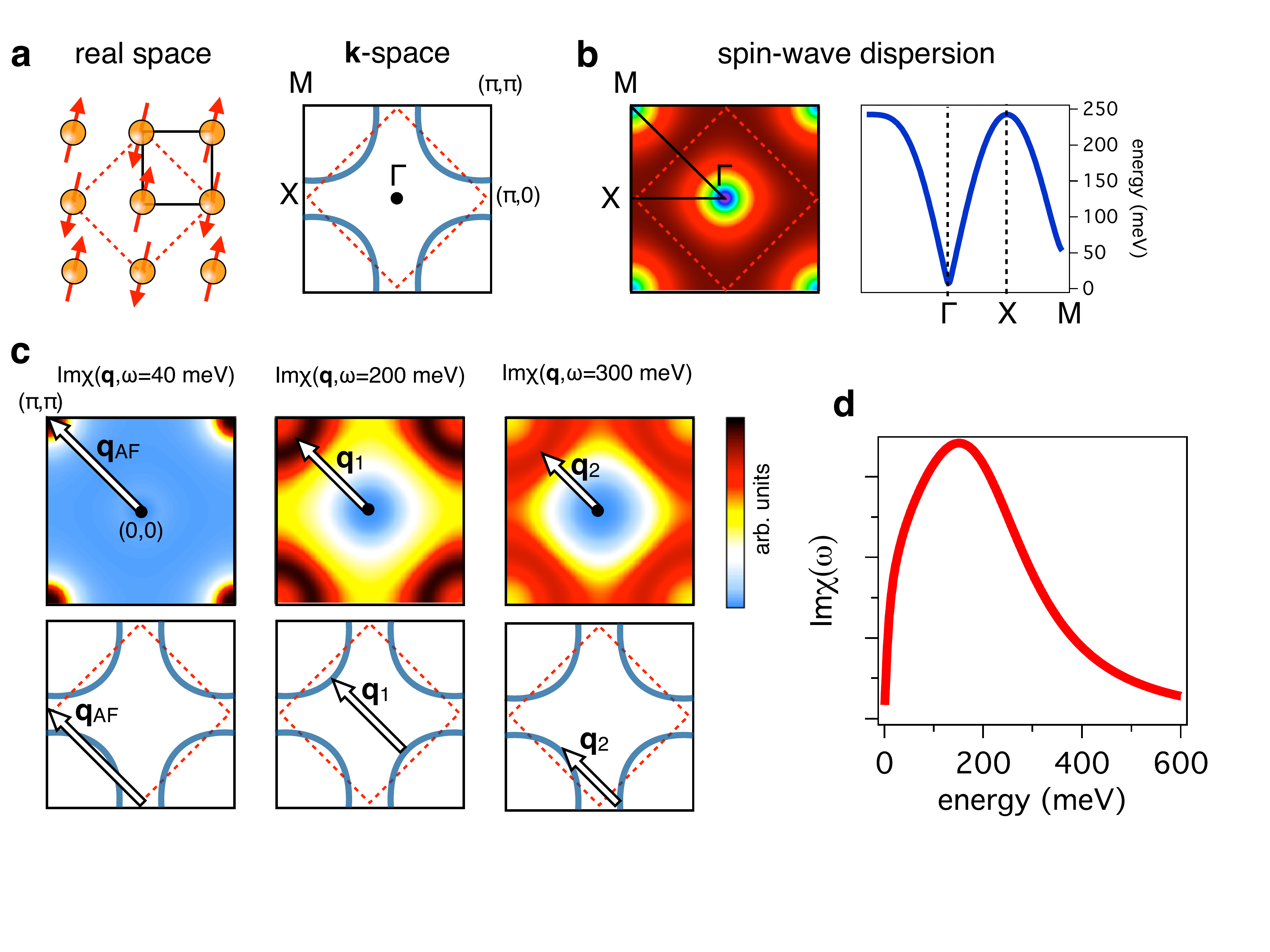}
\caption{Schematic picture of the magnetic excitations of the antiferromagnetic copper sublattice. }
\label{fig_paramagnons}
\end{centering}
\end{figure}

In the undoped compounds the charge-transfer insulating phase is accompanied by long-range antiferromagnetism (AF), that is the consequence of the $J$=4$t_{ij}^2$/$U$ AF coupling constant. The dispersion of the excitations (AF magnons) of the AF lattice can be phenomenologically reproduced by the expression \cite{letacon2011}: 
\begin{equation}
\label{eq_magnon_dispersion}
\omega_{\textbf{q}}=J\sqrt{(2-\mathrm{cos}q_x-\mathrm{cos}q_y)(2+\mathrm{cos}q_x+\mathrm{cos}q_y+\omega^2_{\textbf{q}_{AF}}/4J^2)}
\end{equation}
where $\omega_{\textbf{q}_{AF}}$=50-70 meV is the spin gap that opens up at the AF wavevector $\textbf{q}_{AF}$=($\pi$,$\pi$) (in units of the lattice parameter) in bilayer compounds. In Figure \ref{fig_paramagnons}a we report the electronic unit cell and Brillouin zone (black squares), along with AF ones (red dashed squares). Figure \ref{fig_paramagnons}b displays the $\omega_{\textbf{q}}$ dispersion of AF paramagnons, as calculated by Equation \ref{eq_magnon_dispersion}. The maximum of the magnon dispersion is $\omega_{\textbf{q}}$=$2J\simeq$250 meV (assuming $J$=125 meV) and it is found in correspondence of the border of the AF Brillouin zone (X symmetry point), whereas $\omega_{\textbf{q}}$=$\omega_{\textbf{q}_{AF}}$ at the M point. 

\begin{figure}[t]
\begin{centering}
\includegraphics[bb= 20 200 970 520,width=1\textwidth]{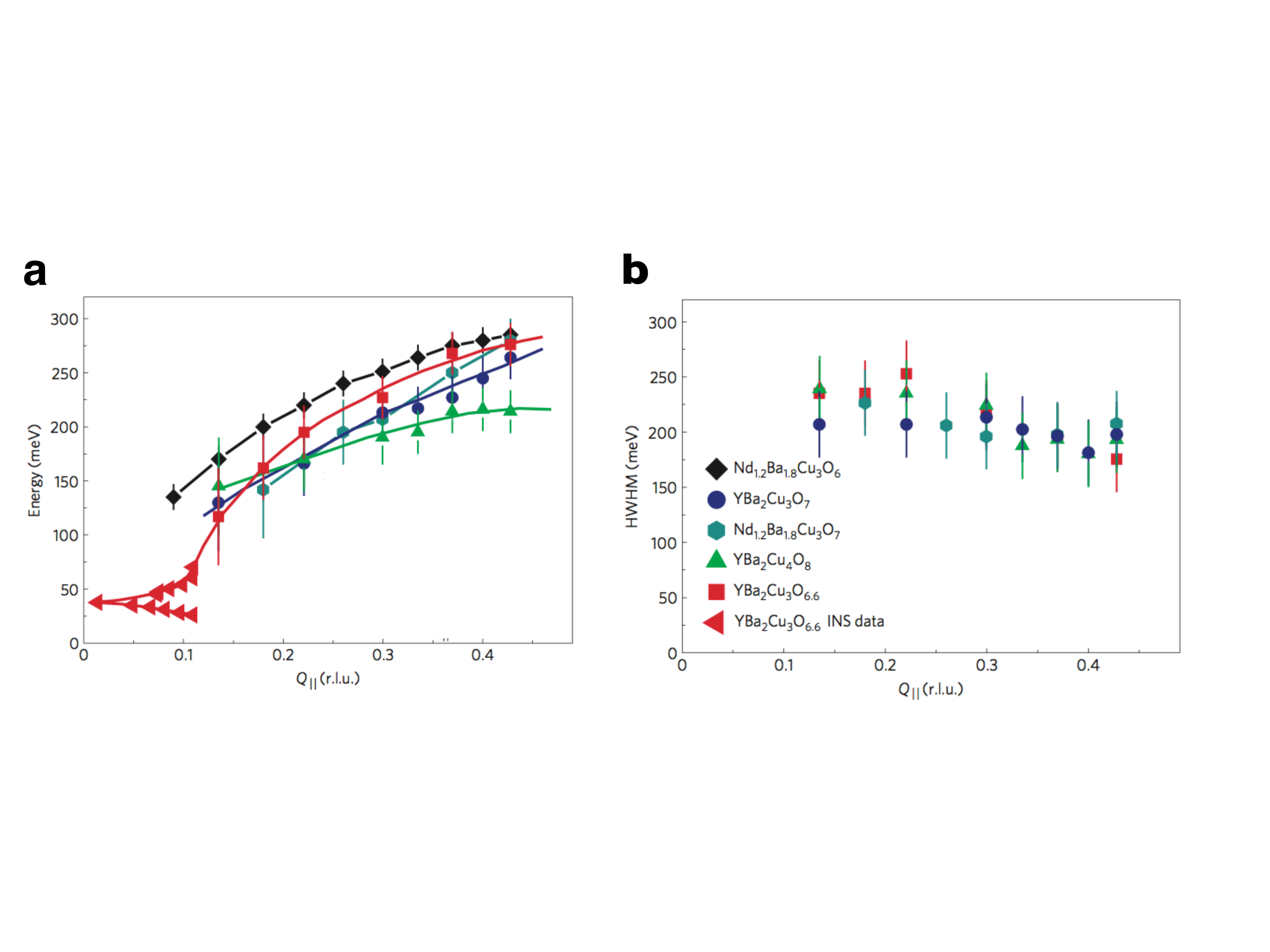}
\caption{a) Dispersion and b) width of the magnetic excitations in the antiferromagnetic Nd$_{1.2}$Ba$_{1.8}$Cu$_{3}$O$_{6}$, the underdoped Nd$_{1.2}$Ba$_{1.8}$Cu$_{3}$O$_{7}$, YBa$_{2}$Cu$_{3}$O$_{6.6}$, YBa$_{2}$Cu$_{4}$O$_{8}$ and YBa$_{2}$Cu$_{3}$O$_{7}$ measured by resonant inelastic X-ray scattering at $T$=15 K. Readapted from Ref. \citenum{letacon2011}.}
\label{fig_RIXS}
\end{centering}
\end{figure}

Recent Resonant Inelastic X-ray Scattering (RIXS)  \cite{Dahm2009,letacon2011,letacon2013,dean2013,dean2013b} measurements and  have demonstrated (see Figure \ref{fig_RIXS}) the persistence of short-range spin-fluctuations far in the highly overdoped region of the phase diagram, challenging the common expectation that AF correlations should significantly weaken when moving away form the insulating phase. 
The tendency of doped systems to develop short range antiferromagnetic correlations is captured by the spin susceptibility, that can be phenomenologically described by the function \cite{letacon2011}:
\begin{equation}
\label{eq_Chi}
\chi(\textbf{q},\omega)=\frac{1-(\mathrm{cos}q_x+\mathrm{cos}q_y)/2}{\omega^2_{\textbf{q}}-\omega^2-i\Gamma\omega}
\end{equation}
where $\Gamma$ is the damping that account for the fluctuating character of short-range AF correlations. The $\chi(\textbf{q},\omega)$ function is shown in Figure \ref{fig_paramagnons}c for the different values of the magnon frequencies and assuming the damping $\Gamma\sim$200 meV, as extracted from the RIXS data reported in Fig. \ref{fig_RIXS}. For $\hbar \omega$=40 meV, $\chi(\textbf{q})$ is strongly peaked at the AF wavector, revealing a very efficient and selective electron-boson scattering mechanism with an exchange of momentum $\textbf{q}_{AF}$=($\pi$,$\pi$). At higher energies the momentum selectiveness of the scattering processes is progressively lost, leading to the coupling with spin fluctuations incommensurate to the AF lattice and with less-defined $\mathrm{\bf{q}}$ vectors, as shown in Fig \ref{fig_paramagnons}. The richness of the magnetic excitations in hole-doped cuprates is also supported by neutron scattering experiments \cite{Fujita2012}, that evidence a universal magnetic spectrum with a bandwidth of $\sim$2$J$.

To summarize, short-range antiferromagnetic correlations lead to a new dissipation channel for the charge carriers. Although the development of microscopic models that could account for the initial and final states available for the scattering with spin fluctuations is still underway, the naive expectation is that the general bosonic function, $\Pi(\Omega)$, extended to include also bosonic fluctuations of electronic origin, should define three different timescales. Besides $\hbar$/$\Omega_{SCP}$$\simeq$10 fs and $\hbar$/$\Omega_{lat}$$\gg$10 fs, electronic correlations introduce an additional timescale, i.e., $\hbar$/2$J<$3 fs, that is related to the $\mathrm{\bf{k}}$-integrated coupling with short-range AF fluctuations. The SCP and magnetic timescales can be thus extremely fast and difficult to be disentangled, which leads to the complexity in these materials and poses a challenge both experimentally and theoretically.

\subsubsection{Quasiparticles scattering channels in the gapped phases}
\label{sec_gappedphases}
\begin{figure}[t]
\begin{centering}
\includegraphics[width=1\textwidth]{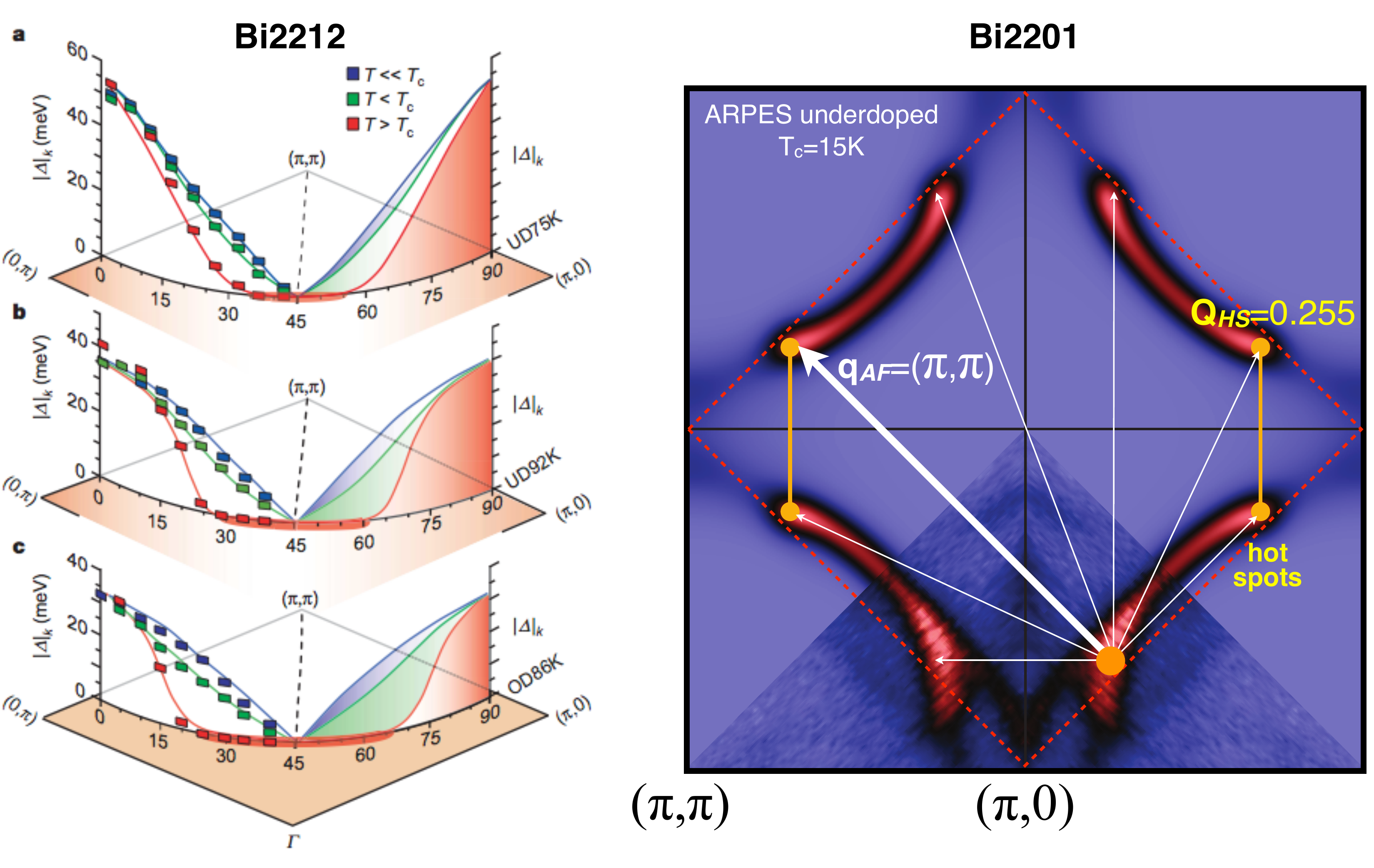}
\caption{Pseudogap in underdoped cuprates. a) The doping- and temperature-dependce of the \textbf{k}-dependent pseudogap, $\Delta_{PG}$(\textbf{k}, is extracted from photoemission data on Bi2212 \cite{Lee2007}. $\Delta_{PG}$(\textbf{k}) is maximum at \textbf{k}$\sim$($\pi$,$\pi$) (antinode), while $\Delta_{PG}$(\textbf{k})=0 at \textbf{k}$\sim$($\pi$/2,$\pi$/2) (node). b) Hot spots of the Fermi arcs in Bi2201 \cite{Comin2014}. The wavevector of the main electron-electron and electron-boson scattering processes are indicated by the coloured arrows.}
\label{fig_PG}
\end{centering}
\end{figure}

The low-energy physics of correlated materials is further complicated by the onset of (pseudo-)gapped phases in the temperature-doping phase diagram. Considering the ubiquitous pseudogap of cuprates ($\Delta_{PG}$(\textbf{k}), see Fig. \ref{fig_PG}) in the low T-low doping region of the phase-diagram, the inherent anisotropy of the electronic density of states can dramatically affect the phase-space available for electron-electron and electron-boson scattering processes. 
A further complication in describing the relaxation processes in the pseudogap phase of correlated materials is the intrinsic instability of these systems towards different kinds of ordered phases. Such phases can be favoured by the reduced kinetic energy of the carriers and they might be associated with a quantum critical point underneath the superconducting dome \cite{Varma2006,Sachdev}. Indeed, a wealth of different broken-symmetries, such as unusual \textbf{q}=0 magnetism \cite{Fauque2006,Li2008,Li2010}, stripes \cite{Tranquada1995}, nematic and smectic phases \cite{Kivelson1998,Mesaros2011,Hinkov2008}, and, more recently, charge density waves (CDW)\cite{Ghiringhelli2012,Chang2012,Comin2014,SilvaNeto2014} have been reported. The fluctuations of these incipient order parameters are expected to provide a wealth of additional scattering channels that are very selective in the momentum space. This can be rationalized by the inspection of the prototypical Fermi surface of underdoped Bi$_2$Sr$_{2-x}$La$_x$CuO$_{6+\delta}$ (Bi2201), shown in Fig. \ref{fig_PG} (right panel). The intersections of the Fermi arcs, characteristic of the pseudogap phase, with the AF Brillouin zone define the so-called hot-spots that are particularly sensitive to a wealth of different scattering processes: i) since the quasiparticle density of states is maximum in the hot-spots ($|\nabla_kE(\textbf{k})|^{-1}$ is maximum), they define an octet of most-probable elastic scattering processes that give rise to the typical QP interference pattern in STM measurements \cite{McElroy2003}; ii) The hot-spots are connected by the antiferromagnetic vector \textbf{q}$_{AF}$=($\pi$,$\pi$), therefore they are subject to a strong scattering with AF fluctuations at $\hbar \Omega\sim$40 meV; iii) Recent experiments \cite{Comin2014} suggest that the CDW instability is driven by a Fermi surface nesting at the wavevector \textbf{q}$_{HS}$=(0.255,0), shown in Fig. \ref{fig_PG} (right panel). As a consequence, the QP scattering with fluctuations of the CDW order is particularly strong at the hot-spots.
On a more general level, the fluctuations of any incipient order parameter with a characteristic wavevector provide additional electron-boson scattering channels that are strongly selective in the momentum space. 

These additional processes should be taken into account when modelling the QP relaxation processes after the impulsive photo-excitation of the pseudo-gapped system. In particular, the photoexcitation process can be roughly reduced to two main steps. In the first step, the pump pulse ($\hbar\omega$ photon energy) is absorbed creating electron-hole excitations extending from -$\hbar\omega$ to $\hbar\omega$ across the Fermi energy ($E_F$) and with a distribution that is regulated by the Joint Density of States (JDOS) of the photo-excitation process. As a consequence of the extremely short scattering time of high-energy excitations, this photoexcited population undergoes a fast energy relaxation related to multiple electron-electron and electron-boson scattering processes that lead to the creation of a large number of low-energy excitations. In the second and slower step, the subset of scattering processes that allow large momentum exchange, while conserving the energy, leads to the recovery of a quasi-equilibrium distribution dominated by nodal QPs at \textbf{k}$\sim$($\pi$/2,$\pi$/2). At this stage, the anisotropic energy gap is expected to provide strong constraints to the large-momentum and small-energy exchange scattering processes necessary for the relaxation of antinodal excitations at \textbf{k}$\sim$($\pi$,$\pi$) and the recovery of the quasi-thermal nodal population. Strictly speaking, the 
phase-space constraints for the scattering processes of the photoinduced non-thermal population can decouple, at least on the picosecond timescale, the populations at different \textbf{k}. This non-thermal QPs distribution cannot be described by a Fermi-Dirac function at any effective temperature and cannot be captured by any effective-temperature model. Nonetheless, the possibility of creating a transient QPs population characterized by an occupation of the empty states that cannot be achieved in equilibrium conditions, opens intriguing perspectives to shed new light into the elusive pseudogap phase in correlated materials.

\subsection{Microscopic Descriptions of quasiparticles and beyond}

In the previous paragraphs we have outlined the main phenomenological
approaches to describe strongly correlated systems at equilibrium along with
their optical properties. Here we briefly show how the quasiparticle properties of a correlated system
can be derived in a microscopic framework, i.e., solving 
models of interacting electrons like the Hubbard model described by Eq. \ref{eq_Hubbard}, in which a band of lattice fermions is influenced by a local Coulomb interaction which acts when two fermions with opposite spin are
present on the same lattice site. Despite its formal simplicity, this
model can not be solved exactly in more than one dimension as a consequence of  the direct competition between the delocalizing effect of
the hopping term and the constraint to the electronic motion imposed
by the local repulsion.
In order to describe realistic
features of actual materials, the simple model above can be enriched
by including more orbitals either on the transition-metal atom and on the ligand oxygen atoms, longer-range electron-electron interaction,
electron-phonon coupling and other terms, but, as long as the Hubbard
term remains the largest, the physics will be dominated by the above
described competition between kinetic energy and electron-electron correlations.

In this section we do not aim at reviewing the huge amount of work which has been devoted to the investigation and the solution
of these prototype modes and we refer the reader to recent reviews, Refs. \cite{Avella2012} and \cite{Avella2013} respectively dedicated to analytical (semianalytical) and numerical approaches.
Instead we focus
on two methods are are well suited for the description of the
quasiparticle dynamics in the strongly correlated regime and to an accurate characterization of the
interaction effects contributing to the the self-energy. The first approach, based on the
Gutzwiller approximation (GA)\cite{Gutzwiller1965}, is a variational method that introduces an
effective quasiparticle description by projecting out the
``high-energy'' configurations from a non-interacting wavefunction
which would -without projection- describe a non-interacting metal. The
second approach, centered on the Dynamical Mean-Field Theory (DMFT)\cite{Georges1996}, goes beyond the QP description and simultaneously accounts for the
low-energy QP particle properties and the higher-energy
excitation (exemplified by the ``Hubbard bands''). This method is particularly suited for the
description of the dynamical response of correlated systems, both in
equilibrium, when linear-response theory holds and out of equilibrium.

\subsubsection{Quasiparticles from strong correlations in the Gutzwiller approximation}

The Gutzwiller Approximation, described in more details in Sec. \ref{Sec:Gutzwiller}, represents one of the most popular
and effective approximations to treat strongly correlated electron
systems. The method is based on the variational principle and on a
wave function that introduces real-space constraints on an
uncorrelated wave function described by a Slater determinant. In the
case of the single-band Hubbard model, the idea is simply that the
Hubbard $U$ makes doubly occupied sites energetically
unfavourable, while the kinetic energy gives rise to delocalized
states, in which any of the local configurations is equally probable. Therefore,
at every value of $U$ there is an optimal value of double
occupation resulting from this balance. The Gutzwiller wave
function describes this process in terms of a variational parameter
(projector). For multi-band systems more projectors with similar
meaning can be introduced. Despite its approximate nature, the
expectation value of a many-body Hamiltonian on the Gutzwiller wave
function can not be computed analytically except for the case of
infinite coordination number, where the the variational energy depends
only on the average double occupation. In finite dimensions this
property becomes an approximation which goes under the name of
Gutzwiller Approximation.

Despite its limitations, the GA provides precious insights on the
physics of strong correlations, including the Mott transition and the
reduction of quasiparticle weight when the critical point is
approached, as pioneered by Brinkman and Rice\cite{Brinkman1970}.

The picture emerging from a GA calculation can be interpreted as a
liquid of quasiparticles with a renormalized hopping amplitude and
bandwidth. In the standard GA, the renormalization is
momentum-independent and the full dispersion is renormalized by a
unique factor, leaving the Fermi surface unaffected by the interactions. In this case, the reduction of the effective bandwidth
is equivalent to an effective mass enhancement, and the Mott
transition is associated to a vanishing kinetic energy and a divergent
effective mass. The main limitation of the GA is that the spectral
weight which is lost at low-energy when the QPs lose their mobility is
not recovered in high-energy spectral features.

One strategy to go beyond this limitation is to consider a {\it
  time-dependent} GA (t-GA), where the Slater determinants and/or the
projectors are assumed to depend on time. In Sec. 7 we will review the
non-equilibrium time-dependent GA which is used to describe the
real-time evolution of correlated systems, but it is important to
mention that the t-GA can be use to extract the linear response
functions which are associated to intrinsic equilibrium properties of
the system\cite{Bunemann2013}. 

The Gutzwiller approximation can be generalized to include the most relevant
realistic features, including a
multi-orbital electronic structure \cite{Bunemann1998}, the coupling
with bosons \cite{Barone2007,Barone2008} and geometry, including the role
of surfaces \cite{Borghi2009,Borghi2010}. Furthermore, this approximation can be
combined with density-functional theory (DFT) to provide a consistent
picture of actual correlated materials\cite{Ho2008,Deng2009,Schickling2012,Lanata2015} that is
complementary to DFT+U methods \cite{Anisimov1991} while less computationally demanding than DFT+DMFT
methods \cite{Kotliar2006}.

\subsubsection{Quasiparticles coexisting with high-energy features in Dynamical Mean-Field Theory}
\label{sec_DMFTscattering}
Dynamical Mean-Field Theory has nowadays a paramount role among the
methods designed to treat strong correlation effects. The reason of
this success is at least twofold: on one hand DMFT allows for the
calculation of a variety of experimentally relevant quantities both
for model systems and for more realistic situations, on the other
hand, DMFT has provided us with a powerful reductionist approach to
strongly correlated electrons on the verge of a metal-insulator
transition. These electrons show the low-energy behavior of a metal,
as expected in the Landau quasiparticle picture, while, at high
energy, they behave as the localized carriers of a Mott insulator.

DMFT can be derived as the quantum version of a static mean-field
theory, where the spatial fluctuations are frozen, but the local
quantum dynamics is instead completely accounted for. In practice,
the theory maps a lattice problem (for example the Hubbard model
\ref{eq_Hubbard}) onto a {\it local theory} in which a single
interacting site is embedded into a non-interacting bath which
describes effectively the interaction of the chosen site with the rest
of the original lattice. The equivalence between the original lattice
model and the effective local theory is enforced by a self-consistency
condition for the local single-particle Green's function that we describe in Sec. \ref{DMFT_section}.
 
The self-consistent solution of the effective local theory is typically realized with a simple
iterative procedure, where a new local theory is generated applying
the self-consistency condition to the output of a previous iteration,
and the iterations proceed until  the input and the output
function coincide. The solution of the effective local theory can not be obtained analytically, but several numerically exact ``impurity solvers" have been identified and, at least for the single-band model, the solution is nowadays completely established and solid. 

It can easily be shown that DMFT becomes exact both in the non-interacting limit $U=0$ and in the atomic limit where all the hoppings go to zero. This highlights that the method is intrinsically nonperturbative and it is therefore perfectly suited to study intermediate regimes, where the different energy scales are comparable. 

\subsubsection{Quasiparticle evolution approaching the Mott-Hubbard transition}

We have discussed in Sec. \ref{sec_QPscattering} that the half-filled Hubbard model describes a Mott-Hubbard insulator in the large $U$ limit. If we neglect the onset of antiferromagnetism, we can follow the evolution from a metal to the Mott insulator as a function of $U$, i.e., the correlation-driven Mott transition, as well as the effect of doping on a Mott insulator.
The Mott-Hubbard transition is arguably the most spectacular direct effect of strong correlations and its understanding
is believed to be essential to understand less obvious effects including high-temperature superconductivity.
The Mott-Hubbard transition is simply the process connecting a metallic state with a Mott-Hubbard insulator, a state in which the electronic conductions is inhibited by strong local Coulomb repulsion despite the partial filling of the band. 
The Mott-Hubbard insulator is indeed easily understood considering a half-filled single-band Hubbard (see Eq. \ref{eq_Hubbard}) with one electron per site. In this case, if $U$ is much larger than the bandwidth, the lowest-energy state is obtained occupying each site with one electron and the electronic motion is inhibited because it would imply the double occupation of a site and hence a higher energy.
Starting from this state, one can then reach a metal either reducing the interaction (or, equivalenty, enhancing the bandwith) or doping carriers.
In this section we describe the theoretical Mott-Hubbard transition which can be driven either by doping 
which takes place in a system described by
the Hubbard model  by increasing the interaction strength $U$. 
Within the GA, Brinkman and Rice\cite{Brinkman1970} have demonstrated an evolution of the
quasiparticles controlled by the progressive reduction of the average number
of doubly occupied sited $d = 1/N_s \sum_i \langle
n_{i\uparrow}n_{i\downarrow}\rangle$ from 0.25 (non interacting limit)
to 0 (Mott insulator), a value which is reached at a critical value of
the interaction.

\begin{figure}[t]
\begin{centering}
\includegraphics[width=1.2\textwidth]{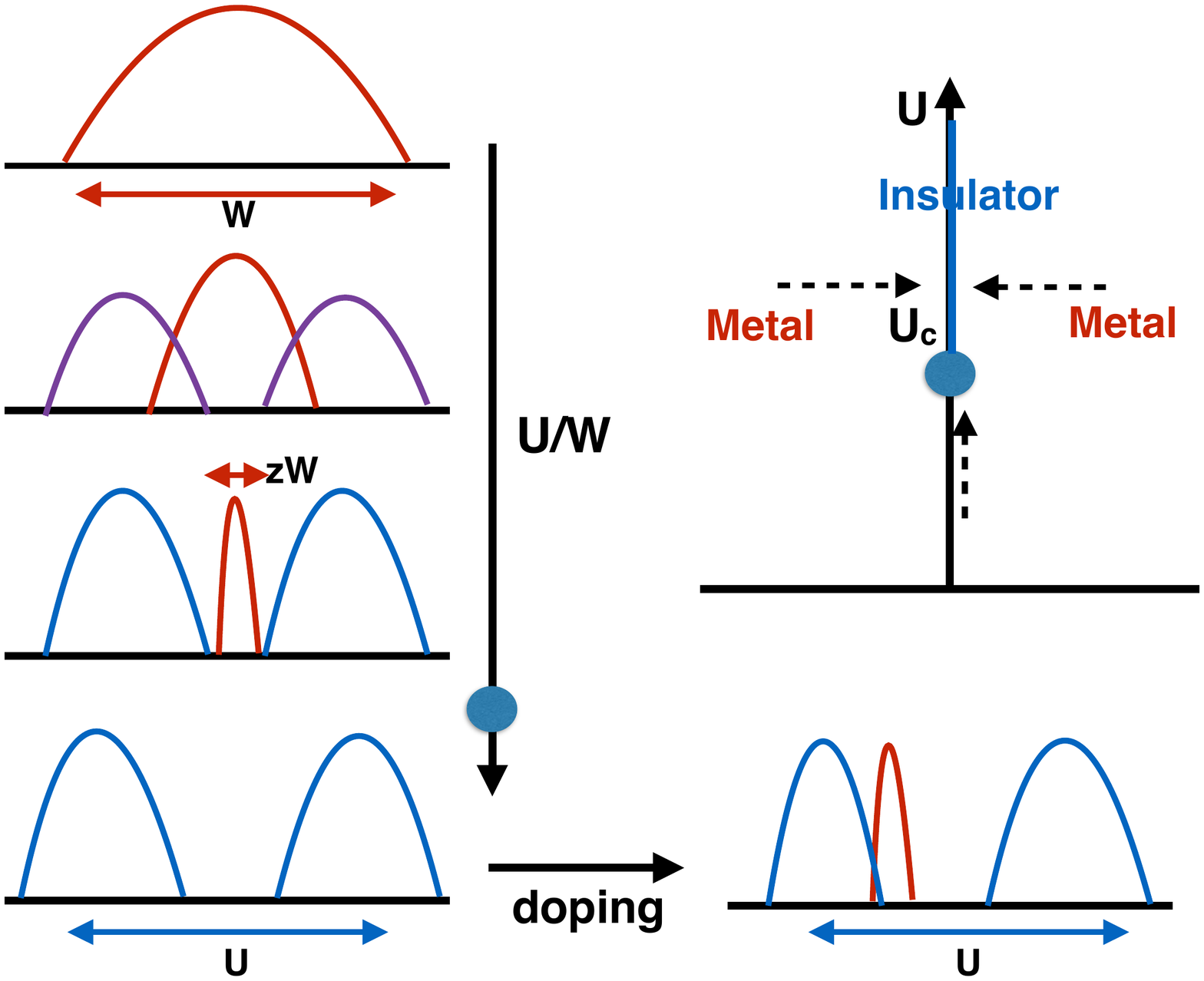}
\caption{Schematic picture of the Mott-Hubbard transition within DMFT. On the left, following the vertical arrow from top to bottom: evolution of the single particle spectral function (local density of states) as a function of $U/W$. Read lines indicate the quasiparticle metallic part of the spectrum, blue lines indicate the insulating high-energy Hubbard bands and purple lines indicate an intermediate situation. Bottom of the plot: following the horizontal arrow we visualize the doping-driven Mott transition from a Mott insulator with a clear gap to a strongly-correlated doped metal with a quasiparticle peak. The figure corresponds to hole doping, while an electron doping would have led to a peak close to the upper Hubbard band. In the top-right part of the picture we report a schematic diagram locating the correlation-driven and doping-driven transitions (dashed arrows). The color notation is the same as in the left panels and the blue dot marks the Mott transition point. }
\label{fig:spectral}
\end{centering}
\end{figure}

The elimination of double occupancy
is mirrored in a reduction of the {\it quasiparticle weight}, which measures the weight of 
coherent Fermi-liquid excitations with respect to incoherent states. For a non-interacting 
system $Z=1$, while the Mott transition is associated to a complete loss of coherence, which implies $Z=0$. This occurs for a finite value of $U = Uc \equiv 8\vert\varepsilon_0\vert$, where $\varepsilon_0$ is the non-interacting kinetic energy per particle. 
This effect can be described also in
term of a self-energy $\Sigma(\omega) = A+ \left( 1-\frac{1}{Z}\right)\omega$, which is momentum-independent and it only depends linearly on frequency. 
 We notice that a similar frequency-dependent can not be captured within standard Hartree-Fock mean-field which can only lead to a constant $\Sigma$. 
Due to the momentum-independent self-energy the effective mass is given by  $m* =1/Z$.
As a consequence, the Mott-Hubbard transition is
associated to a divergence of the effective mass which clearly
highlights the difference with band-like or disorder-driven
metal-insulator transitions. Another consequence of the momentum-independence is that $Z$ also renormalizes the full dispersion, which becomes ${\varepsilon_k}^{\prime} = Z\varepsilon_k$, implying that the Fermi surface shape is not influenced by correlations. Thus one can picture the evolution of the spectral function obtained in GA by following only the red part of the spectrum shown in Fig. \ref{fig:spectral} (left column), where a broad non-interacting spectrum shrinks as $U/W$ increases until a critical point where it simply vanishes.

It is worth to notice that the linear dependence of $\Sigma(\omega)$ can not be obtained within mean-field Hartree Fock, in which the self-energy is unavoidably constant and does not allow to describe loss of coherence and Mott physics in the absence of symmetry breaking.
In Sec. 8 we will present more details about the equilibrium Mott transition within the GA approximation as a starting point to investigate the non-equilibrium dynamics.

The GA can be used also out of half-filling, where the system is always metallic for any doping. 
If we start $U > U_c$, one can study a doping-driven insulator-to-metal transition, in which $Z$ is zero only
in the half-filled case, and it linearly grows as a function of doping
(in the standard Hubbard model with nearest-neighbor hopping only, doping with holes or electrons is
equivalent by particle-hole symmetry). 

The main deficiency of the GA is that it only describes the
quasiparticle component of the correlated electrons and its disappearance as the interaction grows. The Mott transition is thus associated with the destruction of the metallic state and the description of the Mott insulator is completely trivial. One indeed expects
that the reduction of the low-energy quasiparticle
weight close to the Fermi level should be accompanied by a shift of spectral weight towards
high energy. This latter contribution is expected to be incoherent and
to evolve towards the Hubbard bands of the Mott-Hubbard insulator. The
incoherence would be associated to a finite value of the imaginary part of
the self-energy, which determines a finite lifetime for these
high-energy excitation, in contrast with the long-lived
quasiparticles.

While these effects are not accessible by the GA, that starts from a metallic wave function, they
are all present in the DMFT picture of the Mott-Hubbard
transition. Here the spectral function and the self-energy do not take
a simple analytical form, but -at least for the single-band model- a
numerically exact solution can be obtained straightforwardly using
different solvers.

The evolution of the single-particle spectral function obtained within DMFT is reproduced
in Fig. \ref{fig:spectral} (left column). The low-energy part apparently mirrors the
behavior of the Brinkman-Rice transition, with a quasiparticle peak
that gets narrower when the interaction is increased. However, in
DMFT, the quasiparticle peak is flanked by high-energy features,
centered roughly at $\pm U/2$. Increasing the interaction the spectral
weight is shifted toward higher energy. Interestingly, the quasiparticle
peak disappears for a value of $U \simeq 1.5 W$, when the Hubbard
bands are already well separated (bottom-left panels in Fig. \ref{fig:spectral}). As a consequence the Mott-Hubbard
gap opens abruptly when the system becomes an insulator, even if $Z$
vanishes continuously and the zero-temperature transition is of second order. 

When the Mott insulator is doped (bottom-tight panel in Fig. \ref{fig:spectral}) the system immediately becomes metallic. However, the insulator-to-metal transition does not lead to a major reshuffling of spectral weight. Indeed a tiny coherent quasiparticle peak appears close to one of the Hubbard bands (the lower band for hole doping and the upper band for electron doping). As the doping increases the peak becomes larger and larger and eventually merges with the Hubbard band, gradually turning the system into a weakly correlated metal. Roughly speaking, the non-trivial strongly correlated region coincides with parameters such that the spectral function shows three coexistent features: the incoherent high-energy Hubbard bands and the low-energy coherent quasiparticle peak. This is vivid picture of a physical electron which has a partially itinerant and partially localized part, and one of the greatest successes of DMFT is precisely to provide a relatively simple and cheap representation of this non-trivial physics. 

In the top-right part of Fig. \ref{fig:spectral} we report a schematic phase diagram which hold for both GA and DMFT. The dashed arrows indicate the two Mott-Hubbard transitions: the correlation driven transition that occurs at half-filling by increasing the value of $U$ and the doping-driven transition that occurs for $U > U_c$ as a function of the density. 

\subsection{Effective-temperature models}
\label{sec_ETM}
The signature of the electronic coupling with the  different degrees of freedom of the system could be unveiled by time-resolved techniques, under the form of different relaxation dynamics following the pump excitation. The stronger is the coupling to a specific subset of bosonic modes, the faster is the relaxation dynamics. Quantitatively, the link between the bosonic function extracted by conventional techniques and the outcomes of P-p experiments has been set in a seminal work by Allen \cite{Allen1987}. In this work the energy exchange process between an electronic population at the effective temperature $T_e$ and the bosonic fluctuations at the effective temperature $T_b$ is investigated. Although the original work addresses the simplest case of an isotropic coupling with the phonons, in the following we extend the same formalism to the more general case of the QP coupling with bosonic fluctuations of electronic origin at the effective temperature $T_{be}$, with a subset of strongly-coupled phonons (usually buckling and breathing modes involving the Cu-O bonds) at $T_{SCP}$ and with the rest of the lattice at $T_{lat}$.

\begin{figure}[t]
\begin{centering}
\includegraphics[width=0.8\textwidth]{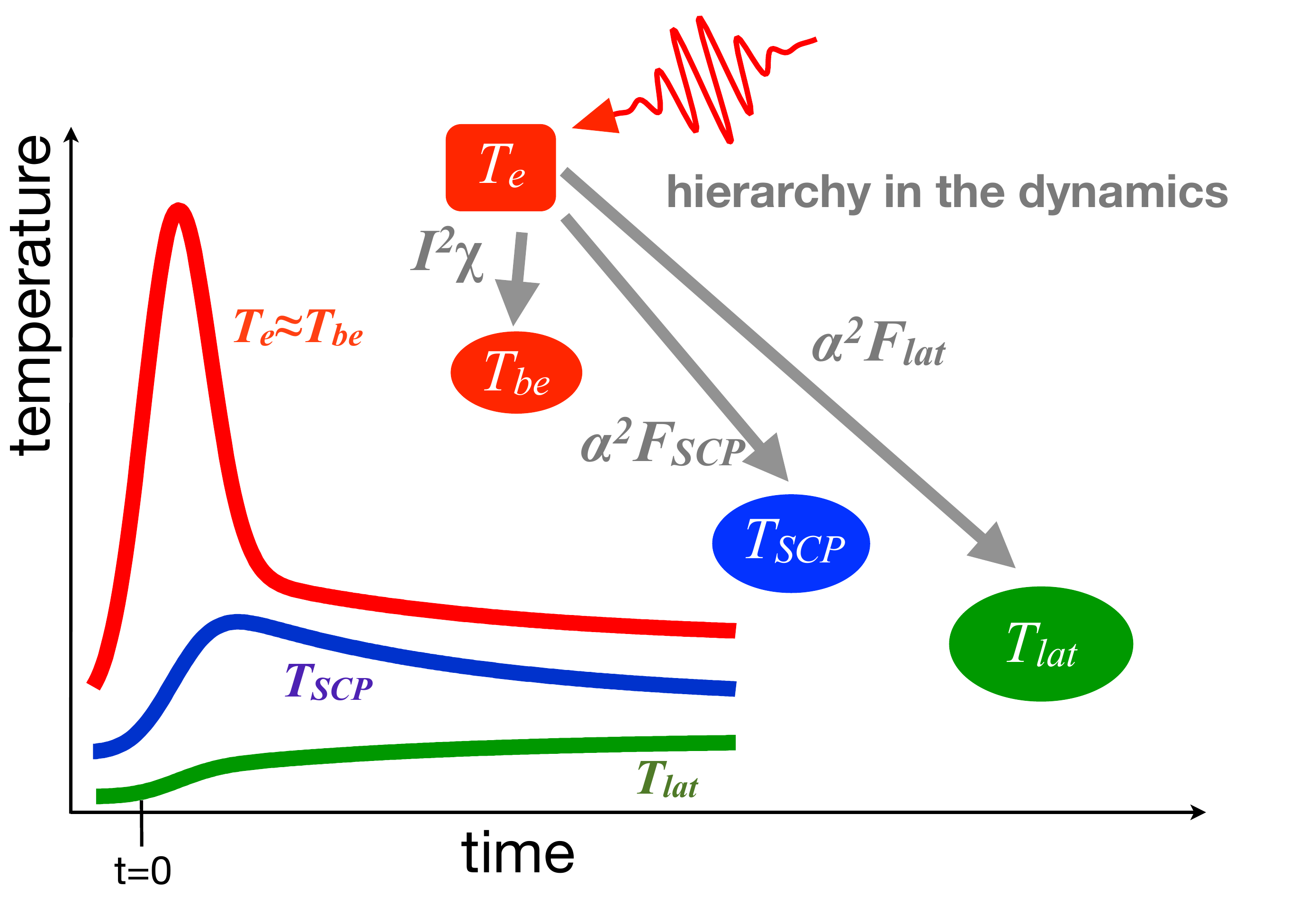}
\caption{Temporal dynamics of the electronic and bosonic effective temperatures, as given by the Effective Temperature Model.}
\label{fig_4TM}
\end{centering}
\end{figure}
A set of four coupled differential equations can be used to represent the following physical processes: a short laser pulse, with power density (absorbed) $p$, impulsively raises the effective electronic temperature of the QPs with a specific heat $C_{e}$=$\gamma_eT_e$ ($\gamma_e$=$\pi^2N_cN(\epsilon_F)k^2_b/3$, $N_c$ being the number of cells in the sample and $N(E_F)$ the density of states of both spins per unit cell). $T_e$ then relaxes through the energy exchange with all the coupled degrees of freedom  that linearly contribute to the total $\Pi(\Omega)$=$I^2 \chi(\Omega)$+$\alpha^2F(\Omega)_{SCP}$+$\alpha^2F(\Omega)_{lat}$.
The rate of the energy exchange among the different populations is given by \cite{Allen1987}:
\begin{eqnarray}
\label{eq_4TMeqs_1}
\frac{\partial T_e}{\partial t}=\frac{G(I^2 \chi,T_{be},T_{e})}{\gamma_eT_e}+\frac{G(\alpha^2F_{SCP},T_{SCP},T_{e})}{\gamma_eT_e}+\frac{G(\alpha^2F_{lat},T_{lat},T_{e})}{\gamma_eT_e}+\frac{p}{\gamma_e T_e}\\
\frac{\partial T_{be}}{\partial t}=-\frac{G(I^2 \chi,T_{be},T_{e})}{C_{be}}\\
\frac{\partial T_{SCP}}{\partial t}=-\frac{G(\alpha^2F_{SCP},T_{SCP},T_{e})}{C_{SCP}}\label{4TM3}\\
\label{eq_4TMeqs_4}
\frac{\partial T_{lat}}{\partial t}=-\frac{G(\alpha^2F_{lat},T_{lat},T_{e})}{C_{lat}}\label{4TM4}
\end{eqnarray}

where 
\begin{equation}
G(\Pi_{i},T_{i},T_{e})=\frac{6\gamma_e}{\pi \hbar k^2_b}\int^{\infty}_0d\Omega \Pi_{i}(\Omega) \Omega^2 [n(\Omega,T_{i})-n(\Omega,T_{e})]
\label{eq_G_ETM}
\end{equation}
with $\Pi_{i}$=$I^2 \chi$, $\alpha^2F_{SCP}$, $\alpha^2F_{lat}$ and $n(\Omega,T_i)$=($e^{\hbar\Omega/k_BT_i}-1$)$^{-1}$ the Bose-Einstein distribution at the temperatures $T_{i}$ ($i$=$be$, $SCP$, $lat$). The specific heat ($C_{SCP}$) of SCPs is proportional to their density of states and is taken as a fraction $f$ of the total specific heat, i.e., $C_{SCP}$=$fC_{lat}$. We underline that the Eqs. \ref{eq_4TMeqs_1}-\ref{eq_4TMeqs_4} constitute a special case of more general equations in which the coupling is not only mediated by the electronic population, but explicit terms that directly link the bosonic populations are considered. In Fig. \ref{fig_4TM} we report the characteristic dynamics of the effective temperatures $T_i$, as calculated from equations \ref{eq_4TMeqs_1}-\ref{eq_4TMeqs_4}. Considering that the specific heat of the spin fluctuations should be a fraction of the electronic specific heat \cite{Singh1998}, that is much smaller than $C_{SCP}$ and $C_{lat}$, the temperatures $T_i$ are expected to decouple very quickly after the excitation with the pump pulse. In particular, while $T_{be}$ promptly follows the electronic temperature, $T_{SCP}$ and  $T_{lat}$ increase on longer timescales, finally leading to the local effective thermalization ($T_e\approx T_{be}\approx T_{SCP}\approx T_{lat}$) on the picosecond timescale. Assuming that the variation of the optical properties is directly proportional to the electronic temperature $T_e$, the relaxation dynamics in the P-p experiments should contain the fingerprint of all the relaxation processes with the different subset of bosonic fluctuations. 

\begin{figure}[t]
\begin{centering}
\includegraphics[width=1\textwidth]{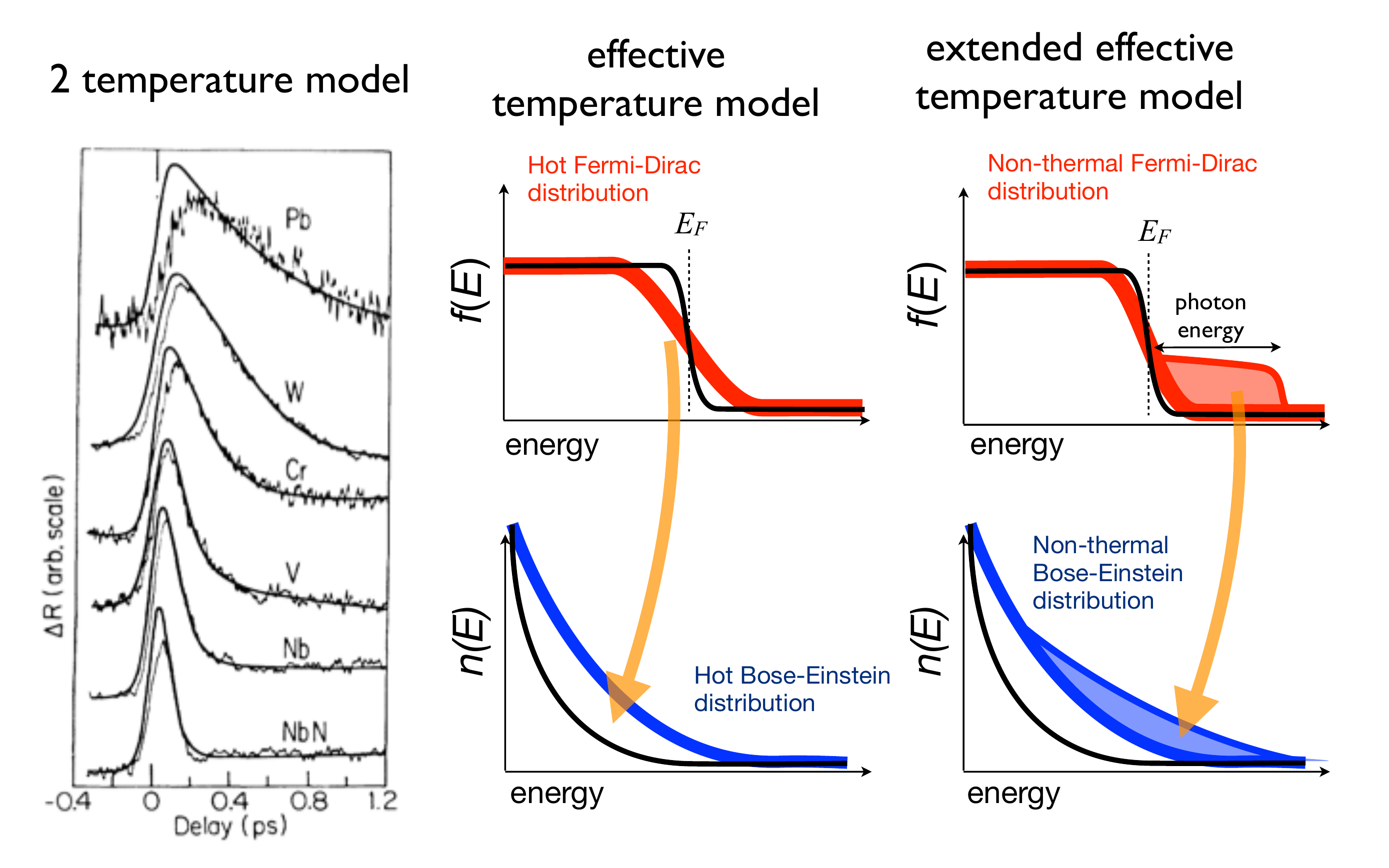}
\caption{Effective temperature model to extract the electron-phonon coupling in metals. The left panel shows time-resolved reflectivity measurements on conventional superconducting metals \cite{Brorson1990}. The fit to the data (black line) is an exponential decay, whose time constant is given by Eq. \ref{eq_tau_e_lat}. The central and left panels represent a sketch of the effective temperature model and of its extension to include a non-thermal distribution of QPs and phnons.}
\label{fig_Brorson}
\end{centering}
\end{figure}

Further insight into this simple model, known as effective-temperature model, is gained by using the Taylor expansion of the Bose-Einstein distribution, $n(\Omega,T_i)$, under the assumption $\hbar\Omega$$\ll$$k_BT_i$. In this case, Eq. \ref{eq_4TMeqs_1} reduces to:
\begin{equation}
\label{eq_4TMeqs_red}
\frac{\partial T_e}{\partial t}=\frac{g_{be}}{\gamma_eT_e}(T_{be}-T_e)+\frac{g_{SCP}}{\gamma_eT_e}(T_{SCP}-T_e)+\frac{g_{lat}}{\gamma_eT_e}(T_{lat}-T_e)\\
\end{equation}
where $g_{i}$=(3$\gamma_e \hbar$/$\pi k_B^2$)$\cdot$($\lambda \langle\Omega^2\rangle_i$) and $\lambda\langle\Omega^2\rangle_i$=$\int_0^{\infty}\Pi_i(\Omega)\Omega d\Omega$ is the second momentum of the Eliashberg coupling with the sub-population $i$. Therefore, Eqs. \ref{eq_4TMeqs_1}-\ref{eq_4TMeqs_4} reduce to a set of linear coupled equations for the functions $T_e(t)$, $T_{be}(t)$, $T_{SCP}(t)$, $T_{lat}(t)$. Considering the simplest case of an isotropic coupling with the lattice (2-temperature model), the relaxation dynamics of $T_e(t)$ further reduces to:
\begin{equation}
\frac{\partial T_e}{\partial t}=-\frac{(T_e-T_{lat})}{\tau_{e-lat}}
\end{equation}
that determines a quasi-exponential decay of $T_e(t)$ with initial slope given by:
\begin{equation}
\label{eq_tau_e_lat}
\tau_{e-lat}(0)=\frac{\pi k_B^2T_e(0)}{3\hbar\lambda\langle\Omega^2\rangle}
\end{equation}
Although useful to quantitatively support the outcomes of the first P-p experiments, the effective-temperature model has some intrinsic limits that mostly arises from the definition of $T_e$ and $T_i$ during the relaxation process of the photoexcited population. 

The basic picture of the effective-temperature model is based on the assumption that the energy delivered by the pump pulse, under the form of a non-thermal occupation of the empty electronic states, rapidly relaxes to a "hot" Fermi-Dirac population before any energy is exchanged with the lattice (or with other bosonic fluctuations). Although the electron-electron interactions can be extremely effective in many materials, the validity of the hypothesis underlying the effective-temperature model could be disputed even in the simplest frame of the Fermi Liquid (FL) theory. As a consequence of the phase-space available for the scattering processes, the FL theory predicts a QP scattering rate ($\gamma$) of the form: $\gamma$=$a$($E$-$E_F$)$^2$+$b$($k^2_BT^2$), where $a$ and $b$ are coefficients, $E$ is the QP energy and $T$ is the temperature. Considering finite temperatures (e.g. $T$=300 K), the QP lifetime (1/$\gamma$) can range from $\sim$10$^{-15}$ s at $E$=1 eV to $\sim$10$^{-10}$ s at $E$=$E_F$, that is much longer than the typical electron-phonon scattering time (typically in the 10$^{-14}$-10$^{-12}$ s range). In a more realistic picture, while the high-energy electronic excitations ($\sim$1 eV) promptly decay towards $E_F$, the low-energy excitations start exchanging energy with the bosonic fluctuations before an effective thermalization is achieved (see Figure \ref{fig_Brorson}). 

From the experimental perspective, the lack of any fluence dependance in the relaxation dynamics, is often considered as the failure of the effective-temperature approach. 
Considering that $\tau_{e-lat}(0)\propto T_e(0)$ (see Eq. \ref{eq_tau_e_lat}) and assuming a direct proportionality between $T_e(0)$ and the laser power density $p$ (that is reasonable for very high excitation fluences, i.e., $T_e(0)\gg T_{sample}$), the effective-temperature model predicts a relaxation dynamics that linearly scales with $p$. This trend can be easily verified though intensity-dependent P-p experiments. Nonetheless, while this is true for the simplest case of an isotropic electron-phonon coupling (2-temperature model), the case of simultaneous energy exchange with different subset of phonons is less straightforward and no characteristic fluence-dependence of $\tau_{e-lat}(0)$ can be predicted \textit{a priori}. This is a consequence of the fact that, while the electronic specific heat linearly increase with $p$, $C_{SCP}$ and $C_{lat}$ are almost independent of the pump fluence. In particular, considering a typical fraction $f$=0.1 of SCPs, the SCP specific heat reduces to 0.1$C_{lat}$ and $C_{SCP}\simeq\gamma_eT_e$ at the typical pump fluences necessary to have $T_e(0)\gg T_{sample}$. Therefore, in the realistic experimental conditions the fastest relaxation dynamics is no longer proportional to $T_e(0)$, as given by Eq. \ref{eq_tau_e_lat}.

As already pointed out by Allen \cite{Allen1987},
the assumption that the distribution functions for the electrons are
thermal (i.e. equilibrium ones) may not be correct. In other words,
the assumption that electron-electron relaxation is instantaneous as compared to
electron-phonon scattering may not always apply. In particular, when dealing
with unconventional metals, such as cuprate superconductors, this
assumption needs to be verified before quantitative values of $\gamma_{eL}$
are discussed.
To explicitly account for the non-thermal nature of the photoinduced electron population, the temporal evolution of the electron distribution should be included in the model through the Boltzmann equation (BE). The simplest way to tackle this problem is to assume a two-fluid picture in which the transient electron distribution, $f(E,t)$, is the sum of a thermal Fermi-Dirac distribution ($f_T$) at the effective temperature $T_e$ and a non equilibrium part that is calculated by BE in the relaxation-time approximation:
\begin{equation}
\label{eq_nonthermal_FD}
f(E,t)=f_T(E,T_e(t))+f_{NT}(E,t)
\end{equation}
Accordingly to this model, usually quoted as Extended Multi Temperature Model (EMTM), the energy provided by the pump pulse is transferred to $f_T(E,T_e(t))$ through the non-thermal population $f_{NT}$. This process can be included in Eqs. \ref{eq_4TMeqs_1}-\ref{eq_4TMeqs_4} by replacing the source term $p(t)$/$\gamma_e T_e$ with $\int p(t')K(t$-$t')dt'$, in which $K(t$-$t')$ is a kernel function determined by the BE \cite{Sun1994,Groeneveld1995,Carpene2006}. Recently, the improvement of the temporal resolution and spectral coverage of time-resolved experiments led to directly measure the dynamics ($\sim$400 fs) of formation of the thermalized electron distribution in gold \cite{DellaValle2012}, that can be exactly reproduced by the EMTM. 

\begin{figure}
\begin{center}
\includegraphics[width=8cm]{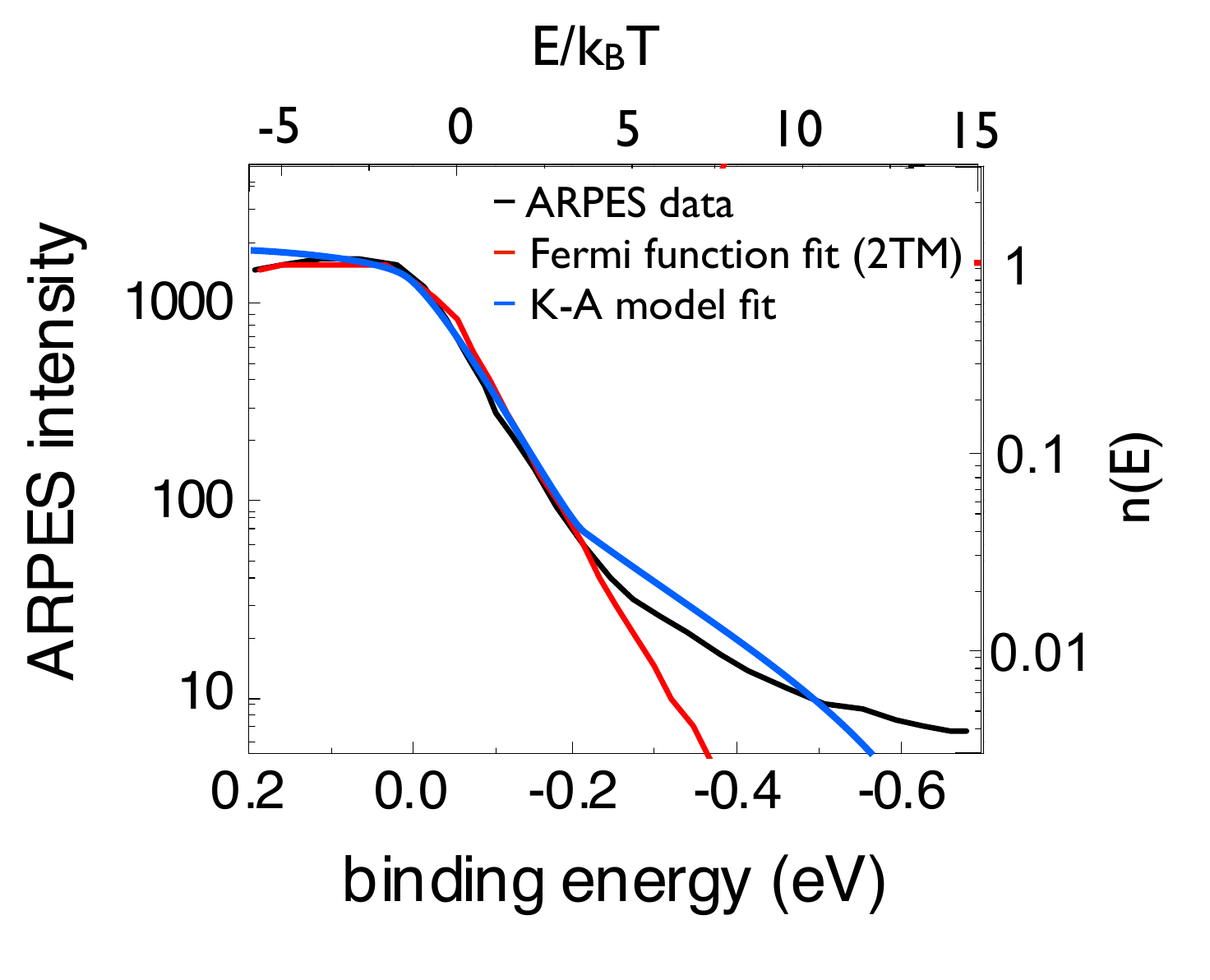}
\protect\caption{\label{fig:Distribution-functions}The ARPES spectrum of Bi$_{2}$Sr$_{2}$CaCu$_{2}$O$_{8}$
immediately after excitation (at zero delay) from Ref. \citenum{Perfetti2007}
(black line), a fit to a Fermi-Dirac function (red line) and the calculated
distribution function from Kabanov and Alexandrov \cite{Kabanov2008} (Blue line)}
\end{center}
\end{figure}

Nonetheless, a complete description of the problem should rely on the calculation of the real QPs distribution function beyond the approximation of the EMTM. 
Kabanov and Alexandrov re-examined the kinetic equations without prior
assumptions regarding electron thermaization, and obtained an exact
solution for the energy relaxation rate \cite{Kabanov2008}.
They analyzed P-p relaxation rates using an analytical approach
to the Boltzmann equation, without assumptions regarding quasi-equilibrium
distributions for the electrons or phonons, and with particular focus
on cuprate superconductors. They applied the Landau-Fokker-Planck
expansion, expanding the electron-phonon collision integral at room or higher
temperatures in powers of the relative electron energy change in a
collision with a phonon, $\hbar\omega/(\pi k_{B}T)\lesssim1$. Then
the integral Boltzmann equation for the non-equilibrium part of the
electron distribution function $\phi(\xi,t)=f(\xi,t)-f_{0}(\xi)$
is reduced to a partial differential equation in time-energy space \cite{Kabanov2008}: 
\begin{equation}
\gamma^{-1}\dot{\phi}(\xi,t)=\frac{\partial}{{\partial\xi}}\left[\tanh(\xi/2)\phi(\xi,t)+\frac{\partial}{{\partial\xi}}\phi(\xi,t)\right],\label{eq:reduced_equation}
\end{equation}
where $f(\xi,t)$ is the non-equilibrium distribution function, and
$\gamma=\pi\hbar\lambda\langle\Omega^{2}\rangle/k_{B}T$. The electron
energy, $\xi$, relative to the equilibrium Fermi energy is measured
in units of $k_{B}T$.
Multiplying Eq. \ref{eq:reduced_equation} by $\xi$ and integrating over all energies yield
the rate of the energy relaxation: 
\begin{equation}
\dot{E}_{e}(t)=-\gamma\int_{-\infty}^{\infty}d\xi\tanh(\xi/2)\phi(\xi,t),
\end{equation}
 where $E_{e}(t)=\int_{-\infty}^{\infty}d\xi\xi\phi(\xi,t)$. Under
appropriate conditions, the characteristic electron-phonon relaxation
rate given by the calculation differs from the rate derived by Allen
in the numerical factor (which is twice larger) and the fact that
the ambient temperature enters the formula instead of the electron
temperature:
\begin{equation}
\label{eq_tau_e_lat_K}
\tau_{e-lat}=\frac{2\pi k_B^2T_{lat}}{3\hbar\lambda\langle\Omega^2\rangle}
\end{equation}
 alleviating the need to determine the initial electron temperature
$T_{e}$. The validity of these approaches requires careful consideration
of electron-electron scattering rates, that crucially dependent on the specific material considered. The estimate that leads to Eq. \ref{eq_tau_e_lat_K} is supported by the experimental
evidence of a non-thermal distribution in ARPES, which shows that
the 2TM fails to describe the high-energy electronic distribution
at very early times on the order of 50-100 fs \cite{Kabanov2008,Gadermaier2010}.
In Fig. \ref{fig:Distribution-functions} we show a comparison of
the Fermi function used by the 2TM and the TR-ARPES data for Bi$_{2}$Sr$_{2}$CaCu$_{2}$O$_{8}$
together with a fit to a Fermi-Dirac distribution used to estimate
$T_{e}$. Compared to the Fermi-Dirac curve, these data show a high-energy
tail very similar to the exact non-equilibrium distribution from Ref. \citenum{Kabanov2008}.
Further confirmation of the validity of the formula \ref{eq_tau_e_lat_K} is
the observation of linear dependence on sample temperature, just as
predicted. The most important conclusion from this work is that the
assumption that either electrons or phonons have equilibrium distribution
functions is very approximate, and it may not hold quantitatively for
the case of cuprates. 

The 2TM can be easily extened to include the effect of thermal diffusion,
in which case an additional term appears for the electronic and bosonic ($i=be,SCP,lat$) temperatures
in the form of Fick's second law $dT_{e,i}/dt=-D\nabla^{2}T_{e,i}$,
where $D=K_{e,i}/\rho_{e,i} C_{e,i}$ is the thermal diffusivity, where $K_{e,i}$ is
the thermal conductivity and $\rho_{e,i}$ is the density.

In all-optical measurements, the energy relaxation rates are inferred
from the transient reflectivity response. The probe is assumed to measure
the relaxation of the dielectric constant (and thus reflectivity)
as a result of the relaxation of the electronic distribution function
for the electrons, assuming electron-hole symmetry. The optical experiments
rely on the existence of excited states for the final states in the
probe transition, in which
case the response is a summation over all the possible vertical transitions
weighted by the matrix elements $M_{if}(E,k)$. Time resolved angular
resolved photoelectron spectroscopy (ARPES) masures the electronic
distribution functions more directly \cite{Perfetti2007,Smallwood2012,Cortes:2011df}.
Optical and ARPES experiments potentially give similar information,
allowing a comparison of the model predictions with experiements.
The values of relaxation time from time-resolved
ARPES experiments \cite{Perfetti2007,Smallwood2012,Cortes:2011df,Graf:2011jw}
agree very well with the optical experiments in BiSCO \cite{Toda:2011cs,Gadermaier:2012vz,DalConte2012}.

All these models rely on the Fermi-liquid description of the electronic properties, of the electron-electron, and of the electron-boson scattering processes. The extension of these models to correlated materials, in which even the applicability of the concept of quasiparticle is disputed, is one of the great challenges of non-equilibrium physics in solid state systems. More broadly, electronic correlations provide a wealth of additional scattering channels, such as spin fluctuations and spin/charge orders, that could accelerate the heat exchange processes and the onset of quasi-thermal distributions.

\subsection{The basic concepts of non-equilibrium superconductivity}
\label{sec_basicconceptsNES}
The outcomes of the first single-colour experiments (see chapter \ref{sec_history}) triggered an intense effort to develop simple models to interpret the novel information coming from the ultrafast dynamics. Although the energy released by the pump pulse would eventually lead to a quasi-thermal state characterized by a local effective temperature, the high temporal resolution of the experiments allows to access the dynamics of the thermalization process itself. On this timescale the thermodynamic state, i.e., the distribution of the charge-carriers and bosonic excitations, cannot be described by a single temperature. The simplest approach is to treat the electrons and the bosonic baths as different coupled systems, characterized by specific effective temperatures. In this picture, the thermal state is achieved via the energy exchange among the different subsystems and the relevant timescale is regulated by the electron-boson coupling function (see chapter \ref{sec_QPdynamics}). 

On the other hand, the study of the relaxation dynamics becomes a more complex problem when the systems undergoes the superconducting phase transition or, more in general, any kind of symmetry-breaking transition. In this case, the opening of a gap in the density of state introduces additional phase-space constraints for the scattering processes, that typically result in a bottlenecked dynamics. This problem will be tackled either from the microscopic point of view, in which the dynamics is the result of the interaction between non-thermal gap-energy electrons and bosons, or from the thermodynamic point of view, through the Ginzburg-Landau functionals.
Even though the simple models presented in this chapter sometimes rely on brutal approximations, they represent the fundamental tools to move the first steps towards the quantitative study of the results of ultrafast experiments and to connect the ultrafast relaxation processes to the quasiparticle dynamics presented in chapter \ref{sec_QPdynamics}. 

\subsubsection{The bottleneck in the dynamics: the Rothwarf-Taylor equations}
\label{sec_RT}
The issue of QPs relaxation in superconductors is relevant for non-equilibrium
superconducting devices, superconducting particle detectors and
for photoexcitation experiments. From QP recombination studies one
can infer the intrinsic relaxation times and their dependence on the experimentally tuneable parameters, such as the excitation intensity. Uniquely, P-p techniques can distinguish between gap states
and pseudogap states on the basis of their lifetimes. This is particularly
important when the system under study is inhomogeneous, either in
real space or in k-space. Real-space inhomogeneity may arise due to
the presence of self-assembled stripes or aggregated textures
\cite{Mertelj:2005p2,Mertelj:2007hn}. In \textbf{k}-space, we can speak of inhomogeneity when states at different
regions in the Brillouin zone (BZ) have different character: for example
some states may have polaronic character, while other states may have
extended state character. This is particularly important in in cuprates, due to the nodal-antinodal dichotomy, and in multi-band
superconductors, such as pnictides, where a number of bands cross
the Fermi level and become gapped below the superconducting transition.
Very similar phenomenology is encountered also in other gapped systems,
particularly CDW systems, where the Fermi surface is gapped in some
regions of the BZ due, for example, to a Peierls instability \cite{Yusupov:2008p5698}.
The main difference is that in superconductors the gap is intended for
single particle excitations out of the ground state, whereas
in CDW systems, there is a gap for electron-hole excitations. 

\begin{figure}
\includegraphics[width=1\columnwidth]{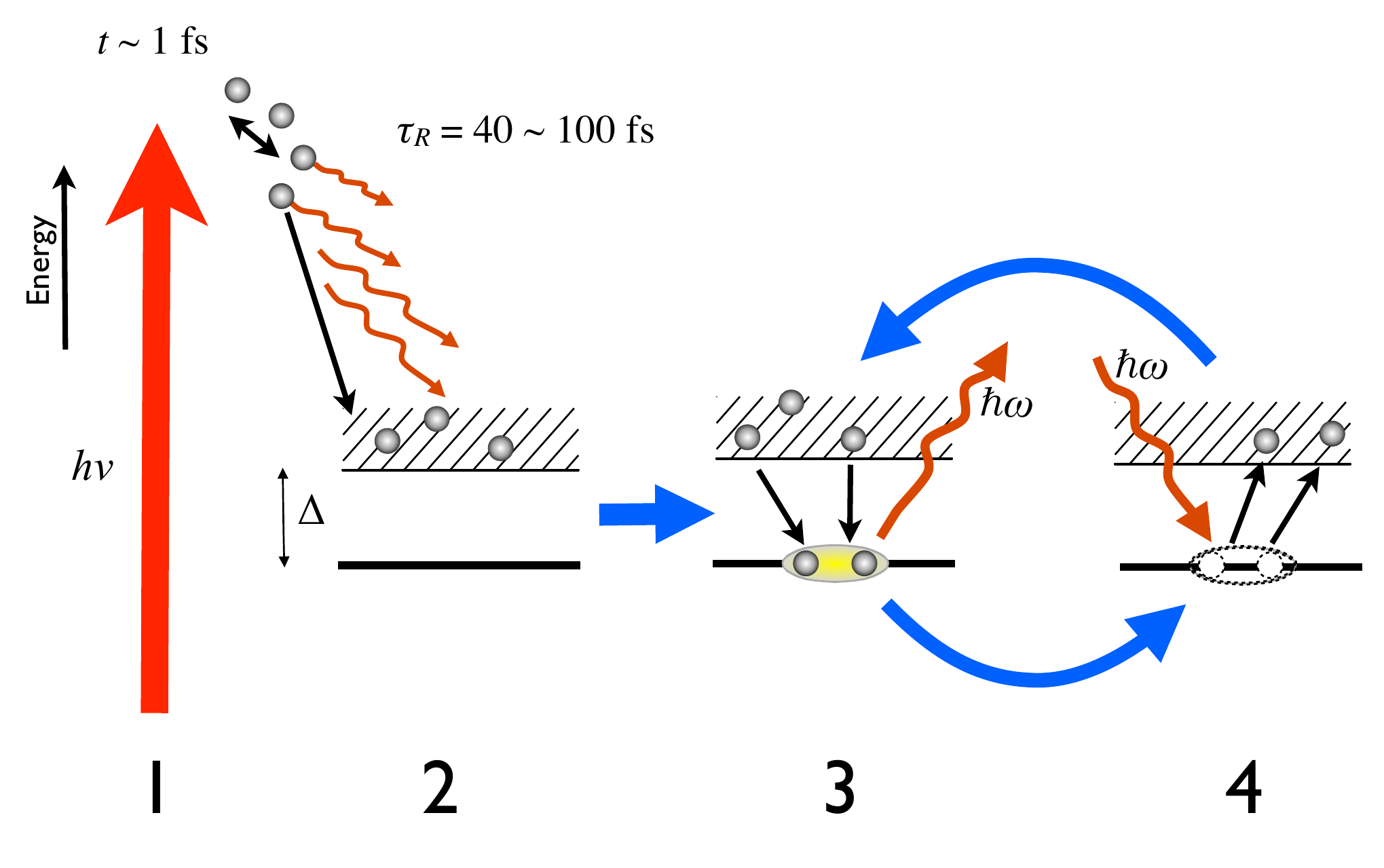}
\protect\caption{\label{fig:The-relaxation-of}The relaxation of photoexcited electrons
takes place in distinct steps (an analogous process takes place for
holes). The inital steps are photoexcitation (1), electron-electron
thermalisation and avalanche energy relaxation (2). The latter takes
place within 40-100 fs. Thereafter quasiparticles recombine into pairs
across the superconducting gap $\Delta$ (3) with the emission of
high frequency phonons (orange wiggly arrows) with an energy $\hbar\omega>2\Delta$.
These GFPs can immediately break pairs (4), leading to a cyclical
process shown by the blue arrows. This process is terminated either
by anharmonic decay of GFPs with energy $\hbar\omega<2\Delta$ or
by phonons escaping from the photoexcited volume\cite{Rothwarf:1967p9182,Mihailovic:2005p57}.}
\end{figure}

In photoexcitation experiments, after the initial rapid energy relaxation
lasting for few hundreds femtoseconds, the electrons (holes) are nearly in equilibrium
with phonons irrespective of whether there is a gap in the electronic
spectum or not. While in a gapless metal, the relaxation proceeds
further towards thermodynamic equilibrium by the QP emission of low-energy
acoustic phonons, in superconductors and CDW systems the final states
are limited by the presence of a gap. In this case the QPs can relax
to the ground state only provided that they emit a phonon with an energy
greater than $2\Delta$, where $\Delta$ is the single particle gap
energy (see Fig. \ref{fig:The-relaxation-of}). In other words, the energy relaxation process
is interrupted when photoexcited particles reach states near the gap
in the energy spectrum. As already discussed in the previous section,
these non-equilibrium quasiparticles can be detected by excited state
absorption or photoemission, measuring the QP density ($n(t)$) at the gap edge. 

It was recognized already in 1979 by Rothwarf and Taylor (R-T) that
the measured QP lifetime for these states is not governed by the intrinsic
recombination rate $R$ because one needs to take into account phonon
reabsorption processes from the ground state, as shown in Fig. \ref{fig:The-relaxation-of}. They proposed a simple model to account for phonon re-absorption, in
which the QP and pairing boson population dynamics are described in
terms of two non-linear differential equations \cite{Rothwarf:1967p9182}
\begin{align}
\frac{dn}{dt} & =I_{0}+\eta N-Rn^{2}\label{eq:RT-1}\\
\frac{dN}{dt} & =-\frac{\eta N}{2}+\frac{Rn^{2}}{2}-\gamma(N-N_{T})\label{eq:RT-2}
\end{align}

where $n$ is the QP population, $N$ is the total gap-frequency
phonon (GFP) population that can be separated into the photoexcited and
thermal contributions: $N=N_{PE}+N_{T}$. $\eta$ is the probability
for pair breaking by the absorption of a phonon, and $R$ is the bare
QP recombination rate with the emission of a phonon and $I_{0}$ describes
the incindent pulse. The factor 2 comes from the fact that two
QPs are annihilated with an emission of a single phonon in a recombination
event. $\gamma$ describes the loss of phonons by mechanisms such
as phonon escape from the superconducting region \cite{Mihailovic:2005p57}
in mesoscopically inhomogeneous systems or in thin films; or anharmonic
decay of the high-frequency phonons ($\hbar\omega>2\Delta$) into
lower frequency phonons ($\hbar\omega<2\Delta$). Phonon escape may
be important in inhomogeneous systems, or in thin films. In such cases,
the rate limiting step is determined by the characteristic length
scale of the superconducting regions or the thickness of the film
\cite{SaiHalasz:1974dq}. 
\begin{figure}
\begin{center}
\includegraphics[width=1\columnwidth]{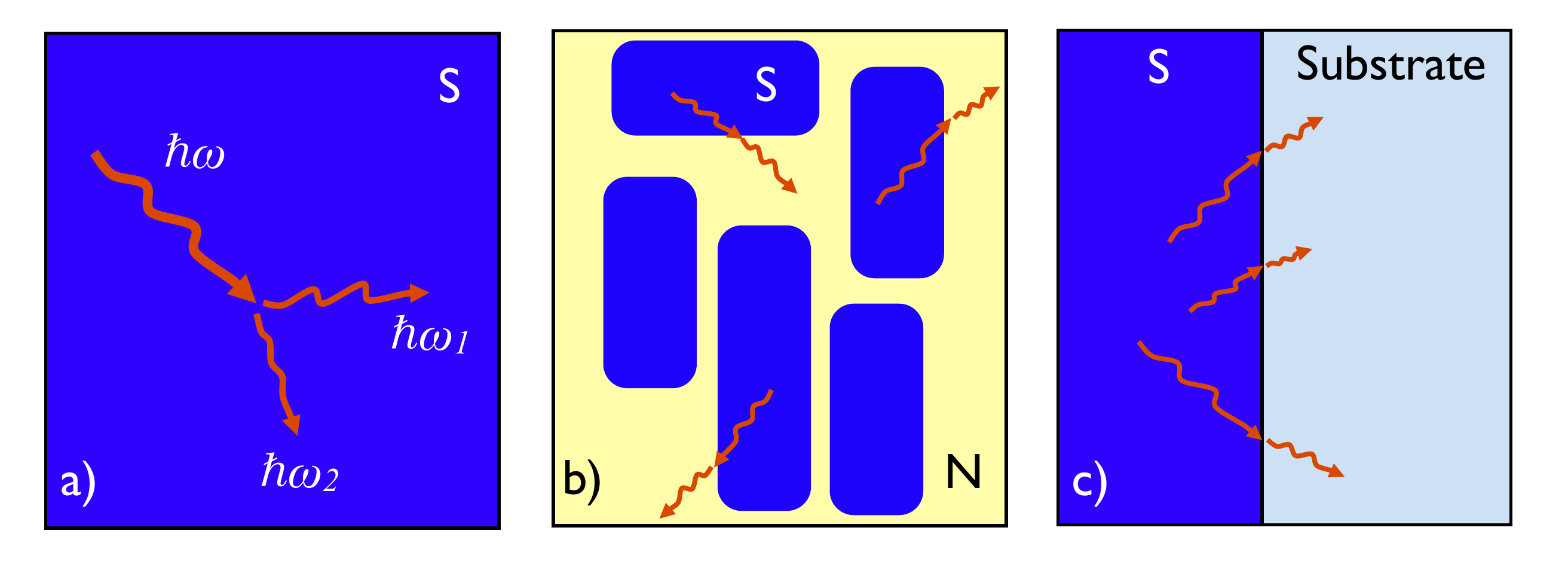}

\protect\caption{\label{fig:Escape}The processes which determine the rate limiting
step, $\gamma$ in the R-T equation \ref{eq:RT-2}. a) Anharmonic
decay of phonons with energy $\hbar\omega>2\Delta$ to lower energy
phonons prevents further pair-breaking. b) Phonon escape from superconducting
regions in an inhomogeneous state, such as a vortex state or an intrinsically
textured state, removes them from the R-T loop\cite{Mihailovic:2005p57}.
c) Phonons escaping from the superconductor to the substrate cannot
be reabsorbed, and present a rate limiting step in thin film samples.}
\end{center} 
\end{figure}

The anharmonic decay rate is known from phonon linewidths and time-resolved
Raman measurements to be in the $0.2\sim1$ ps$^{-1}$ range \cite{Kabanov1999}.
While the model was originally formulated for phonons as the pairing
bosons, the model in its various forms applies equally to phonons and spin
excitations as pairing bosons \cite{Kabanov2005}.
Indeed, apart from classical superconductors \cite{Brorson:1990p6245},
hole \cite{CHWALEK:1991p1951,Demsar:1999p112,Demsar:2001p1518,Gay1999,Gedik2005,HAN:1990p1246,Kabanov1999,Kaindl2005gp,Kusar:2008p5434,Schneider:2004p2385,Giannetti2009b,Coslovich2011,Coslovich2013}
and electron doped \cite{Liu:1993p4053,Cao:2008p1505} cuprates, MgB$_{2}$ \cite{Demsar:2003p7494},
and recently pnictides \cite{Mertelj:2009p6019,Chia:2008p4912,Torchinsky:2009p6282},
the model has been successfully used for optical studies
concerning the QP recombination across the gap in several charge-density wave
systems \cite{Demsar:1999p1520,Toda:2009p6297,Demsar:2002p6773,Perfetti:2006p1775,Yusupov:2008p5698} and
the relaxation of electrons between low-energy electronic states in
Jahn-Teller ordered systems and in heavy electron systems \cite{Demsar:2006p9224,Demsar:2006p9230,Demsar:2009p9187}. 

Although the R-T equations can be solved numerically, analytic solutions
\cite{Kabanov2005} are more useful in performing systematic
analysis as a function of particular parameters, such as excitation
power, carrier density and temperature. In particular, analytical
solutions reveal non-linear dependences of the relaxation time on
pump intensity, as well as bimolecular decay kinetics or simple exponential
behaviour with different excitation densities. 

In thermal equilibrium $N=N_{T}$, $n=n_{T}$ and the equations reduce
to $\eta N_{T}=Rn_{T}^{2}$. The thermal-QP density at temperature
$T$ can be estimated as \cite{Kabanov2005}
\begin{equation}
n_{T}(T)\cong N(0)\sqrt{2\pi\Delta k_{B}T}\exp{(-\Delta/k_{B}T)}\label{EqThermallyExcQuasi}
\end{equation}
Here $N(0)$ is the electronic density of states at Fermi energy,
and $\Delta$ is the energy gap. The estimate of the GFP density is given by: 

\begin{equation}
N_{T}(T)\cong36\frac{\nu'\Delta^{2}k_{B}T}{(\hbar\omega_{D})^{3}}\exp{(-2\Delta/k_{B}T)},
\end{equation}
where $\hbar\omega_{D}$ is the Debye energy and $\nu'$ is the number
of atoms per unit cell. 

As discussed in the previous section, for small perturbations the
optical response is proportional to the excess QP density $n_{p}$.
Substituting $n_{p}=n-n_{T}$ and $N_{p}=N-N_{T}$ into \ref{eq:RT-1}
and \ref{eq:RT-2}we can rewrite the RT equations: 
\begin{equation}
\begin{array}{ccl}
\frac{dn_{p}}{dt} & = & I_{0}+\eta N_{p}-2Rn_{T}n_{p}-Rn_{p}^{2}\\
\\
\frac{dN_{p}}{dt} & = & J_{0}-\eta N_{p}/2+Rn_{T}n_{p}+Rn_{p}^{2}/2-\gamma N_{p},
\end{array}\label{EqRothwarfTaylorTran}
\end{equation}
 here we used the equilibrium condition: $\eta N_{T}=Rn_{T}^{2}$. 

We can now discuss a number of different regimes for QP recombination.
In the so called \emph{weak bottleneck regime} the decay is dominated
by $\gamma$, i.e. $\gamma/\eta\gg1$, the first equation simplifies
to $\frac{dn_{p}}{dt}=-2Rn_{T}n_{p}-Rn_{p}^{2}.$ The relaxation rate
$\tau^{-1}$, defined as $\frac{-dn_{p}/dt}{n_{p}}$ for $t\rightarrow0$,
is temperature and intensity dependent: 
\begin{equation}
\tau^{-1}=R(2n_{T}+n_{p}(0)),\label{EqRRweakBottleneck}
\end{equation}
where $n_{p}(0)$ is $n_{p}$ at time $t=0$. The relaxation time is dominated by two-body processes
for low temperatures when $n_{T}\ll n_{p}$, but it shows a single particular
exponential decay for the low excitation case ($n_{T}\gg n_{p}$). 

More interesting is the opposite limit, namely the \emph{strong bottleneck
regime }where $\gamma/\eta\ll1$. In this regime breaking of pairs
by GFP is dominant, causing an initial fast exchange of energy between
QPs and GFP (compared with $1/\gamma$) leading to a quasi equilibrium
between GFP and QP. 

This can be described as a two temperature regime where GFP and QP
are in thermal equilibrium at a higher temperature than the rest of
the system. This regime was named \emph{prebottleneck dynamics} and
the relaxation following it \emph{superconducting state recovery dynamics}.

We can look at the \emph{superconducting state recovery dynamics}
in two limits. The low temperature limit is for the case when $n_{T}\ll\eta/R$.
In this case $R(n_{T}+n_{p})\ll\gamma$ and the thermal equilibrium
is not established. The relaxation rate for this limit is again temperature
and intensity dependent: 
\begin{equation}
\tau^{-1}=2\frac{R\gamma}{\eta}(2n_{T}+n_{p}(0))\label{EqRRstrongBtlLowT}
\end{equation}

For high temperature limit $n_{T}>\eta/R$ the relaxation rate is:

\begin{equation}
\tau^{-1}=\left\{ \begin{array}{ccl}
\gamma &  & \textrm{ for }\;\; n_{p}\ll n_{T},\\
\gamma/2 &  & \textrm{ for }\;\; n_{p}\gg n_{T}
\end{array}\right.\label{EqRRstrongBtlHighT}
\end{equation}
For this limit it is also interesting to write down the $n_{p}$ as
a function of light excitation density $\varepsilon_{l}$ as quasi
thermal equilibrium is achieved: 
\begin{equation}
n_{p}^{0}=\frac{1}{4}\left(\frac{\eta}{R}-4n_{T}+\sqrt{\frac{\eta^{2}}{R^{2}}+8n_{T}\frac{\eta}{R}+16n_{T}^{2}+8\frac{\eta}{R}\frac{\varepsilon_{l}}{\Delta}}\right)
\end{equation}

The buildup of the signal \emph{prebottleneck dynamics} depends on
the initial processes after photoexcitation in RT equations marked
as $I_{0},J_{0}$. For the case where the first relaxation of highly
excited quasiparticles is predominantly electron-phonon relaxation
and the majority of the energy before thermal equilibrium is in the
phonon system the finite risetime can be described with Eqs. \ref{EqRothwarfTaylorTran}:
\begin{equation}
\frac{dn_{p}}{dt}\approx\eta N_{p},
\end{equation}
 while $n_{p}^{2}$ is negligible.

By use of the two temperature model governed with bottleneck dynamics
and conservation of energy Kabanov  \cite{Kabanov1999} derived
equations for temperature dependence of the density of optically excited
QP in low excitation limit as quasi-thermal equilibrium is reached
between QP and GFP:

\begin{align}
n_{p}= & \frac{\varepsilon_{l}/\Delta_{p}}{1+B\exp{(-\Delta_{p}/k_{B}T)}}\,\label{EqSizeOfTheGapTind}\\
n_{p}= & \frac{\varepsilon_{l}/(\Delta(T)+k_{B}T/2)}{1+B\sqrt{\frac{2k_{B}T}{\pi\Delta_{c}(T)}}\exp{(-\Delta(T)/k_{B}T)}}.\label{EqSizeOfTheGapTdep}
\end{align}

The first equation is for the case of a temperature independent gap
like pseudogap with $\Delta_{p}=constant$ and the second for a temperature
dependent gap $\Delta(T)$ such as the superconducting gap. In the
equations the parameter $B=2\tilde{\nu}/N(0)\hbar\Omega_{c}$ with
$N(0)$ a density of states at Fermi energy, $\tilde{\nu}$ is the
effective number of phonon modes per unit cell participating in the
recombination process and $\Omega_{c}$ is the characteristic phonon
cutoff frequency. It should be pointed out that these equations are
derived for an isotropic energy gap ($s$-wave), whereas the calculations for
anisotropic ($d$-wave) gap predict a different temperature behavior
\cite{Kabanov1999}. Since the simple $s$-wave model works well with several materials (see Sec. \ref{sec_history}), it is argued that the recombination process mainly takes place at anti-nodes, where the
gap magnitude is maximum.

\subsubsection{The response due to the effective temperature and chemical potential perturbations in the superconducting state}
\label{sec_effTmu}
A phenomenological description of the electron dynamics in the superconducting phase should account for the non-thermal modification of the electronic occupation induced by the pump pulse. To address this issue we recall the dependence of the superconducting (isotropic) gap on the excitation distribution function, as given by the BCS gap equation:
\begin{equation}
\label{eq_gapequation}
1=N(0)V_{pair}\int^{\hbar \omega_D}_0\frac{d\epsilon_k}{\sqrt{\epsilon_k^2+\Delta^2}}\left[1-2f_0\left(\sqrt{\epsilon_k^2+\Delta^2}/k_BT\right)\right]
\end{equation}
where $N(0)$ is the electronic density of states at the Fermi level, $\Delta$ the gap and $V_{pair}$ the attractive pairing potential. The main role of the thermal energy $k_BT$ is to control the number and the distribution of excitations in the empty states, $\epsilon_k$, above the Fermi energy. When the temperature is increased, the thermal occupation of the available states $E_k$=$\sqrt{\epsilon_k+\Delta^2}$, given by the Fermi-Dirac distribution $f_0(E_k/k_BT)$, prevents these states from participating to the formation of the superconducting condensate and eventually leading to the superconducting-to-normal state second-order phase transition. When the system is excited by light pulses with photon energy $\hbar\omega$$>$$\Delta$, the thermal distribution is perturbed leading to a transient non-equilibrium population described by:
$f_k$=$f_0(E_k/k_BT)$+$\delta f_k$. The number of excess excitations can be calculated by integrating the non-equilibrium distribution:
\begin{equation}
\label{eq_excess_exc}
\delta n=\frac{1}{\Delta (0)} \int_0^{\infty}\delta f_kd\epsilon_k
\end{equation}

The simplest approach to account for the photoinduced non-equilibrium population is attained by introducing some effective parameters, namely an effective temperature $T_{eff}$ \cite{Parker1975} or chemical potential $\mu_{eff}$ \cite{Owen1972} to describe $\delta f_k$. Although simple, these two effective distributions represent the general classes of perturbation that can be applied to a superconducting system \cite{Tinkham}. In the effective temperature model, $\delta f_k(T_{eff})$ is even with respect to inversion through the Fermi energy and the excess distribuition is the same for electron-like and hole-like excitations. In this case the non-equilibrium population is calculated by a Fermi-Dirac distribution at the effective temperature $T_{eff}$$>$$T$. In contrast, a change in the effective chemical potential describes any odd perturbation of $f_k$, in which the excess electron- and hole-like excitations are unbalanced. In the superconducting state, the possible shift of the chemical potential of "normal" quasiparticles can be balanced by the opposite shift of the chemical potential of superconducting pairs. The $\mu_{eff}$ model can be used when the excitation process is \textit{charged}, in the sense that it intrinsically changes the number of charged carrier, such as the case of electron injection in superconductors. 

The case of photoexcitation in multiband systems is more complex. The photoexcitation process does not alter the total charge distribution of the system, but creates a non-thermal distribution of electron-hole excitations. If the electron and hole density of states are completely symmetric, the chemical potential remains constant during and after the photoexcitation process. However, in realistic multiband systems the density of states and the effective mass of the empty levels filled by the electrons can be dramatically different from those of the occupied states below the Fermi level. As a consequence, an excitation can lead to the simultaneous variation of $T_{eff}$ and $\mu_{eff}$ in order to maintain the charge-neutrality. This picture is expected to represent well the case of underdoped cuprates, that are characterized by the pseudogap and the superconducting gap, and of charge-transfer insulators, in which the excited electron have 3$d$ character, while the excited holes mainly reside in the O-2$p$ orbitals.

Both models predict a decrease of the superconducting gap $\Delta(\delta n,T)$ when increasing $\delta n$ (expressed in units of 4$N(0)$$\Delta(0,0)$) but with important differences \cite{Nicol2003}. In the $T_{eff}$ model the gap dependence can be approximated by $\Delta(\delta n,0)$/$\Delta(0,0)$$\approx$1-32(3$\delta n$)$^{3/2}$/$\pi^3$ and a complete closing is obtained at $\delta n_{cr}$($T_{eff}$)$\simeq$0.33, causing a second-order phase transition to the normal state. In the $\mu_{eff}$ model the gap closing is slower, i.e. $\Delta(\delta n,0)/\Delta(0,0)\approx$1-4$\sqrt{2}\delta n^{3/2}$/3, and, before the complete gap collapse at $\delta n\simeq$0.65, the free energy of the superconducting state equals the normal state one. The free energy difference between the two phases, in units of the condensation energy $U_{c}$=$N(0)\Delta^2(0)/2$, is expressed \cite{Nicol2003} as $\Delta F(\delta n)\approx -1/2+16\sqrt{2}\delta n^{3/2}/3$ and reduces to zero at $\delta n_{cr}$($\mu_{eff}$)$\simeq$0.16. The most important consequence is that, within the $\mu_{eff}$ model, a first-order phase transition from the superconducting to the normal state can take place. This possibility introduces a non-thermal scenario in which superconducting and normal domains can spatially coexist, in contrasts to what expected in equilibrium conditions. We underline that the differences between the predictions of two models are not related to the density of excitation injected into the system but to their energy distribution. 

We stress that the crucial issue to describe the photo-excitation process through the $T_{eff}$ or $\mu_{eff}$ models is the way the pulse energy, $E_p$, is shared among the different degrees of freedom of the system, i.e.$E_p$ =$E_{QP}+E_c+E_{ph}+E_{bos}$, where $E_{QP}$ is the energy directly absorbed by QPs excitation, $E_c$ is the energy necessary to modify the condensation energy of the system, $E_{ph}$ and $E_{bos}$ are the energies released to the phonons and to other bosonic fluctuations, such as short-range AF excitations. While the $T_{eff}$ model directly provides the amount of energy delivered to the bosonic degrees of freedom, that are described by a Bose-Einstein distribution at $T_{eff}$, in the $\mu_{eff}$ model the excess bosons are not accounted for and the absorbed energy is entirely absorbed by the shift of the chemical potential.
Calculating the excess excitations injected by a short light pulse in the limit of weak optical perturbation ($\delta n$$\ll$$n_T$, where $n_T$ is the total number of thermal excitations), it has been shown \cite{Kabanov1999} that:
\begin{equation}
\label{eq_n_photoex_kabanov}
\delta n=\frac{Ep/[\Delta(T)+k_BT/2]}{1+g(k_BT/\Delta)\mathrm{exp}[-\Delta(T)/k_BT]}
\end{equation}
where $g(k_BT/\Delta)$ is a function of the temperature dependent gap. A common assumption is that the photoresponse of a superconducting systems is proportional to the number of excess excitations. This is particularly reasonable in the case of optical measurements in transmission on thin films, in which the relative variation of transmissivity, $\delta T/T$ is directly proportional to the change in absorption, given by the imaginary part of the complex refractive index ($n$+i$k$). In this case, the relation $\delta k/k\simeq\delta n/n_T$  and Eq. \ref{eq_n_photoex_kabanov} suggests that the photoresponse is given by a universal function of $\Delta(T)/k_BT$ and can be used as a powerful probe of the gap amplitude, also in systems that exhibit multiple gaps \cite{Demsar:1999p112,Kabanov1999}.

Once the optical perturbation is halted, the equilibrium distribution at the temperature $T$ is recovered by exchanging energy with the thermal bath constituted by the different degrees of freedom interacting with quasiparticles. In conventional superconductors, the main relaxation channels is provided by the inelastic scattering with phonons, although elastic electron-electron scattering processes can play a role in the case of an anisotropic gap. Although a quantitative estimation of the relaxation time should be based on a microscopic model accounting for the density of states and all the possible couplings with bosonic fluctuations, a simple and general phenomenological expression links the relaxation time, $\tau_{T_{eff}, \mu_{eff}}$ to the gap value, in the limit $\Delta$$\ll$$k_BT$, i.e., for $T$$\simeq$$T_c$:
\begin{equation}
\label{eq_rel_time_Tinkham}
\tau_{T_{eff}, \mu_{eff}}=c_{T_{eff}, \mu_{eff}}\frac{k_BT_c}{\Delta(T)}
\end{equation}
where $c_{T_{eff}, \mu_{eff}}$ is a coefficient that depends on the microscopic parameters of the system and is proportional to the inelastic scattering time of a quasiparticle at the Fermi surface. This relation predicts a characteristic divergence of the relaxation time when $T_c$ is approached, that is related to the progressive decrease of the phase-space available to inelastic scattering processes that involve a fraction $\Delta (T)$/$k_BT_c$ of the thermally occupied states. 

\subsubsection{Sub-gap photo-excitation and gap enhancement}
\label{sec_gapenhancement}
A particularly interesting case is related to the sub-gap ($\hbar \omega<$2$\Delta (T)$) optical excitation of a superconducting or, more in general, of a gapped system at the equilibrium temperature $T$.
As discussed in the previous section, Eq. \ref{eq_gapequation} suggests that the QP distribution plays the fundamental role in determining the gap value. Therefore, by manipulating the equilibrium QP distribution function on the energy scale $k_BT$, it is possible to enhance the gap value \cite{Eliashberg1970,Tinkham}, provided that the total number of excitations is not increased, i.e., $\delta n$=0. During the photoexcitation process, the constraint $\hbar \omega<$2$\Delta (T)$ does not allow breaking Cooper pairs and photoinjecting additional excitations, while thermal quasiparticles are preferentially elevated from the low-lying states at
$\Delta \le E_k\le\Delta$+$\hbar\omega$, to the less populated states at  $E_k>\Delta$+$\hbar\omega$. Intuitively, this process increases the number of states ($\delta f_k<$0) at $E_k\geq\Delta$ available for the formation of the superconducting condensate, thus increasing the value of the gap. From the gap equation, it is possible to define the effective temperature change $\delta T_{eff}$ corresponding to the change of the gap amplitude:
\begin{equation}
\label{eq_deltaTeff}
\frac{\delta T_{eff}}{T}\simeq\int_{-\infty}^{+\infty}\frac{\delta f_k}{\sqrt{\epsilon_k^2+\Delta^2(T)}}d\epsilon_k
\end{equation}
that becomes negative (\textit{effective cooling}) when $\delta f_k<$0. 
This effect is particularly evident at moderate temperatures while at very low temperatures, the absence of thermal excitations prevents the formation of the non-thermal distribution necessary to achieve the gap enhancement. 
  
In conventional superconductors, the gap is of the order of a few meV, that requires excitation with sub-THz electromagnetic radiation to fulfil the relation $\hbar \omega$$<$2$\Delta (T)$. Indeed, a gap enhancement corresponding to $\delta T_{eff}$$\sim$2$\%$ by continuous-wave microwave radiation has been demonstrated in thin metallic films \cite{Klapwijk1977,Kommers1977}. For $T$$>$$T_c$ it has been shown that a new superconducting state with a finite gap is the stable solution of the gap problem upon CW irradiation \cite{Schmid1977}.

In principle, the same approach could be extended to achieve the gap enhancement in high-temperature superconductors. The large value of 2$|\Delta|\sim$40 meV \cite{Huefner2008} permits to design experiments in which short and very intense THz and far-infrared pulses are used to create a transient non-thermal population. However, the $d$-wave character of the superconducting gap prevents from achieving the condition $\hbar \omega<$2$\Delta_k (T)$ for every $k$ value. In particular, a large rate of QP production is expected in the nodal region, i.e., $\mathbf{k}\sim(\pi/2$,$\pi/2)$, in which $\Delta_{(\pi/2,\pi/2)}$=0. Furthermore, the non-thermal population can be sustained on the timescale determined by the relaxation time given by Eq. \ref{eq_rel_time_Tinkham}, that is proportional to the inelastic scattering time of QPs at the Fermi surface. Therefore, the wealth of different scattering channels available in correlated materials (see previous sections) could confine the typical inelastic scattering processes to the femtosecond timescale, making extremely difficult the feed of the non-thermal distribution necessary to achieve the gap enhancement condition. A microscopic description of the out-of-equilibrium superconductivity in correlated materials and the possible $d$-wave gap enhancement via ultrashort THz pulses represents a challenge for condensed matter theory and experiments.          
 
\subsection{Dynamics of the order parameters: time-dependent Ginzburg-Landau functionals}
\label{sec_tdGL}

Any system characterized by a broken symmetry can be described through the Ginzburg-Landau functionals that introduce the concept of (complex) order parameter $\Psi(r)$ \cite{Landau1}. The ground state of the system is obtained by minimization, over the entire volume, of the normalized (and adimensional) Landau free-energy, $\tilde{U}$:
\begin{equation}
\label{eq_GL}
\tilde{U}=\int (\tilde{u}_0+\frac{A_0(P,T)}{2}|\eta|^2+\frac{B(P)}{4}|\eta|^4+\lambda|\nabla \eta|^2)dv
\end{equation}
where $\eta$=$\Psi/\Psi_{eq}$ is the order parameter normalized to its equilibrium value, $A_0(P,T)$ and $B(P)$ are the coefficients that determine the thermodynamics of the symmetry-broken second-order phase transition, and $\lambda$ is the stiffness to the spatial perturbations of $\eta$. The generality of this approach allows treating on the same footing a wealth of different ordering phenomena of relevance to correlated materials, such as 
lattice distortions, magnetic ordering, superconductivity and charge density wave (CDW) instabilities.

\begin{figure}[t]
\begin{centering}
\includegraphics[width=1\textwidth]{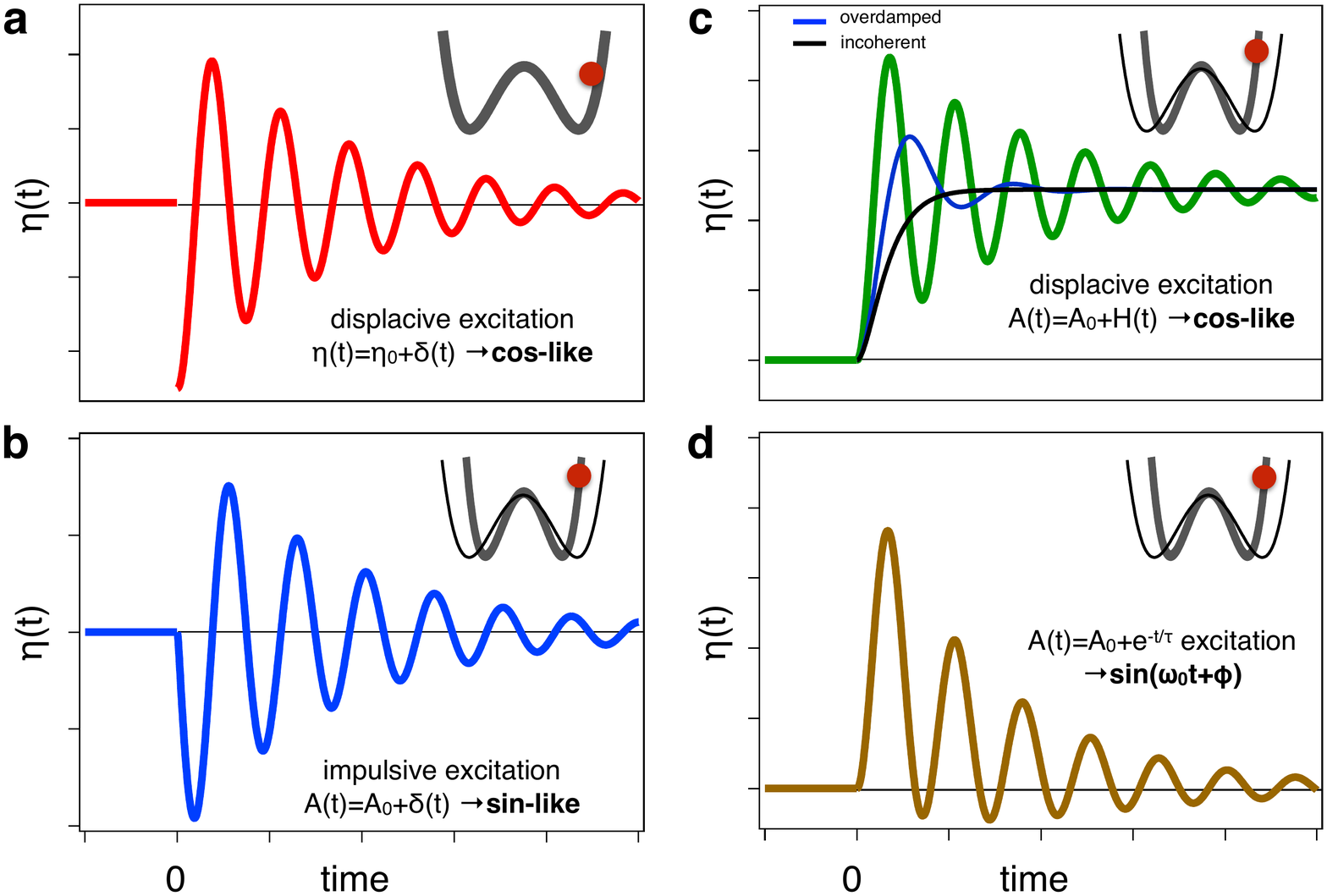}
\caption{Dynamics of the order parameter in a broken-symmetry phase.}
\label{fig_GL}
\end{centering}
\end{figure}

The extension of this approach to describe the dynamics of the order parameter after an impulsive excitation is based on the possibility of replacing the temperature-dependent coefficient $A_0(P,T)$ with a time-dependent function $A(t)$. A general equation of motion can be derived by differentiating $\tilde{U}$ with respect to the order parameter:
\begin{equation}
\label{eq_motion_equation}
\frac{d^2 \eta}{dt^2}+\alpha \omega_0 \frac{d \eta}{d t}=-\omega_0^2\frac{\partial \tilde{U}}{\partial \eta}=-\omega_0^2[A(t)\eta+B\eta^3]
\end{equation}
where $\omega_0$ is the eigenfrequency of the mode, characterized by a well-defined dispersion. For sake of simplicity, we assume a real homogeneous order parameter which describes an optical mode whose dispersion is almost flat. The possible damping of the coherent motion of $\eta$ is introduced through the adimensional coefficient $\alpha$. At equilibrium, $\partial \tilde{U}$/$\partial \eta$=0 and Eq. \ref{eq_motion_equation} does not contain any forcing term. As soon as a perturbation is applied, $\partial \tilde{U}$/$\partial \eta\neq$0 and a coherent motion is triggered, whose dynamics is controlled by the shape of the mexican-hat potential $\tilde{U}$ (see Fig. \ref{fig_GL}). 

This simple formalism can be applied to describe different processes that are relevant to time-resolved experiments. 
The na\"ive picture is that the interaction with the pump pulse prepares a perturbed state of the system in which the value of the order parameter is partially quenched from $\eta_{eq}$=1 to 1-$\delta \eta(t)$. In this case, the potential is unchanged and the evolution of the system is that of an harmonic oscillator in which the initial displacement is maximum, while the initial velocity is null. This typical cos-like oscillatory dynamics is shown in Fig. \ref{fig_GL}a. The damping term $\alpha$ mimic the energy exchange process with the thermal bath that transforms the energy selectively stored in the coherent motion into thermal energy shared among the different degrees of freedom. In other terms, the loss of coherence of the oscillation of $\eta(t)$ is a consequence of the dissipation to the thermal bath that ultimately allows the relaxation to the equilibrium value $\eta$=$\eta_{eq}$=1. 

Nonetheless, this simple picture does not represent most of the excitation processes in which there is no direct dipole-allowed coupling between the light pulse and the order parameters. In contrast, the excitation pathway is usually mediated by some electronic transition, either real or virtual, that impulsively modify the potential landscape of the system. This process can be modelled by introducing the time-dependent coefficient $A(t)$=$A_0$+$f(t)$ which describes the temporal evolution of the perturbation, that can have either an impulsive or displacive character. In the first case, the change of $A(t)$ instantaneously follows the temporal profile of the laser pulse and can be represented by a delta-function, i.e., $f(t)$=$\delta(t)$. This process is analogous to impulsively change the initial velocity of an harmonic oscillator, while leaving the initial position unchanged. As a consequence, a sin-like oscillation of the order parameter is triggered, as shown in Fig. \ref{fig_GL}b, that eventually relaxes back to the equilibrium value. 
Among the processes that can be represented by an impulsive change of teh potential, there is the Impulsive Stimulated Raman Scattering (ISRS) mechanism. Here, a portion of energy is impulsively released to initiate the coherent oscillation of $\eta$ via a virtual electronic transition, following the cross section and the selection rules of the conventional Raman scattering processes. However, in many cases the interaction with the pump pulse involves real electronic transitions that absorb most of the energy that is locally stored in the system and is released to the rest of the bulk on a longer timescale (typically on the order of 0.1-1 ns) that is related with the heat diffusion process. In this case, the excitation process is displacive and the change in the potential is a step-like function given by $f(t)$=$H(t)$, where $H(t)$ is the Heaviside function. The typical oscillatory dynamics, reported in Fig. \ref{fig_GL}c, exhibits a cos-like character but approaches an asymptotic value, $\eta_{\infty}$=$A(t\rightarrow\infty)/B$, that is different from $\eta_{eq}$=1. This kind of dynamics is general to all processes in which the pump excitation induces a long-lived change of the equilibrium potential, such as the case of a photoinduced phase transition or of a temperature increase directly induced by the pump pulse or consequent on the decoherence of $\eta(t)$, as recently shown in the case of the spin dynamics in the antiferromagnetic dielectric KNiF$_3$ \cite{Bossini2014}. Intriguingly, the motion equation of Eq. \ref{eq_GL}) constitutes also the link to thermodynamic description of the relaxation process of a system adiabatically driven out of equilibrium. 

In Fig. \ref{fig_GL}b we show the relaxation dynamics in the case of an overdamped oscillation ($\alpha \gg$1, blue line) and a completely incoherent relaxation ($\alpha \rightarrow \infty$, black line). In this last case, the second-order term can be neglected and the dynamics reduces to the classical kinetic equation \cite{Landau2}:
\begin{equation}
\label{eq_kin_eq}
\frac{d\eta}{dt}=-g\frac{\partial \tilde{U}}{\partial \eta}
\end{equation}
where $g$=$\omega_0$/$\alpha$. This equation can be linearized for small perturbations, $\delta \eta$, obtaining:
\begin{equation}
\label{eq_kin_eq_lin}
\frac{d\delta\eta}{dt}=-\frac{\delta \eta}{\tau_{GL}}
\end{equation}
where $\tau_{GL}$=-2$gA(P,T)$ is the exponential decay that describes the relaxation dynamics of $\delta \eta$. Finally, Fig. \ref{fig_GL}b displays an intermediate scenario in which the time-dependent coefficient $A(t)$=$A_0$+$ae^{-t/\tau}$ relaxes on a timescale of the same order of the damping of the system and of the oscillation period. 

When an inhomogeneous system is considered, the $|\nabla \eta |^2$ term must be included in the functional $\tilde{U}$. This leads to a second-order equation in the time and space coordinates that accounts for the propagation of $\delta \eta (t, \vec{r})$ perturbation. This formalism can be extremely useful to treat thin films, interfaces and devices in realistic geometric configurations, for example the recovery of the superconducting order in LSCO \cite{Mihailovic:2013fpa}. Moreover, it has been recently applied to investigate the space-time propagation of the impulsive quench of a CDW order, in which the inhomogeneity of the process is intrinsically related to the finite penetration length of the electromagnetic field in the material \cite{Yusupov2010}.

The same formalism can be further extended to treat complex order parameters $\eta$=$|\eta|e^{i\phi}$, provided the derivative of the functional $\tilde{U}$ in Eqs. \ref{eq_GL} and \ref{eq_kin_eq} is performed with respect to the function $\eta^*$. Besides the dynamics of the amplitude of the order parameter, additional modes, related to the spatial and temporal variation of $\phi (t)$ emerge from the td-GL.

Finally, the GL functional is extremely useful also in the case of systems, whose properties are the consequence of the interplay between two different order parameters, $\eta_1$ and $\eta2$. The functional $\tilde{U}$ can be extended to include both the order parameters and a quadratic interaction term, i.e., $W|\eta_1|^2|\eta_2|^2$, that can be either repulsive (competing phases, W$<$0) or attractive (cooperative phases, W$>$0):
\begin{equation}
\begin{split}
\label{eq_GLenergy_coupled}
\tilde{U}=&\int \left[\tilde{u}_0+\frac{A_1(P,T)}{2}|\eta_1|^2+\frac{B_1(P)}{4}|\eta_1|^4+\lambda_1|\nabla \eta_1|^2\right]+\\
&\left[\frac{A_2(P,T)}{2}|\eta_2|^2+\frac{B_2(P)}{4}|\eta_2|^4+\lambda_2|\nabla \eta_2|^2+W|\eta_1|^2|\eta_2|^2 \right] dv
\end{split}
\end{equation}
Following the previous steps, it is possible to linearize Eqs. \ref{eq_GLenergy_coupled}, obtaining a set of coupled equations:
\begin{eqnarray}
\label{eq_GL_coupled1}
\frac{d\eta_1}{dt}=-\frac{\delta \eta_1}{\tau_{11}}-\frac{\delta \eta_2}{\tau_{12}}\\
\frac{d\eta_2}{dt}=-\frac{\delta \eta_1}{\tau_{21}}-\frac{\delta \eta_2}{\tau_{22}}
\label{eq_GL_coupled2}
\end{eqnarray}
The dynamics emerging from these equations is characterized by two diagonal time-constants ($\tau_{11}$ and $\tau_{22}$) that describe the intrinsic decay time with non interaction, and two non-diagonal terms ($\tau_{12}$=$\tau_{21}$) that 
depend on the parameter $W$. The possibility of studying the interplay between two competing phases in the time-domain has been recently applied to hole- and electron-doped cuprates \cite{Coslovich2013,Hinton2013}, demonstrating the competitive nature of the interaction between the pseudogap and superconducting phase, and in stripe ordered nickelates, supporting the scenario of strong coupling between the spin and charge sectors \cite{Chuang2013}.

\subsection{On the complexity of the scattering processes which regulate the quasiparticle dynamics}
The brief overview of (quasi-)equilibrium dynamics in strongly correlated systems highlights the inseparable nature of the most relevant lattice and magnetic excitations. This points out the inadequacies of classical theory, particularly for dealing with inhomogeneous and fluctuating mesoscopically ordered phases. The concepts introduced in this chapter also challenge the knowledge coming from conventional equilibrium techniques and pave the way towards the use of non-equilibrium experiments to disentangle the intertwined degrees of freedom. In particular, we highlight the following aspects that constitute the bridge between the equilibrium properties and non-equilibrium dynamics, as discussed in chapter \ref{sec_basicconceptsNES}: 
\begin{itemize}
\item In correlated materials, the charge dynamics is regulated by a rich electron-boson coupling function, which includes the coupling with both phonons and bosons of electronic origin, such as spin fluctuations. These scattering processes can strongly affect the electronic self-energy up to an energy scale of hundreds of milli-electronvolts.
\item The anisotropy of the fundamental interactions in (pseudo-)gapped phases determines a \textbf{k}-dependent renormalization of the electronic self-energy and, therefore, of the most relevant scattering processes. This effect is further enhanced by the intrinsic tendency of correlated materials to develop symmetry-breaking instabilities.
\item The short-range electronic correlations, usually described through the Hubbard model, challenge the notion of quasiparticle in itself. In materials close to the Mott-insulating phase, the fundamental excitations involve incoherent charge fluctuations localized in real space.  
\item Within the effective-temperatures approximation, the timescale of the relaxation dynamics observed in P-p experiments is controlled by the same electron-boson coupling function, $\Pi(\Omega)$, which also regulates the quasiparticle dynamics. In other words, the coupling to the bosonic (eg. phonons, spin fluctuations) baths, which renormalize the single-particle self energy and the quasiparticle effective masses, also provides the relaxation channels necessary for the relaxation of the photoexcited electrons. The electron-boson coupling function can be quantitatively estimated from the sub-ps dynamics in ungapped metallic phases.
\item The opening of a gap in the density of states and the onset of superconductivity call for a microscopic model to describe the relaxation dynamics. While the temporal evolution of the gap can be described by some form of effective-temperature or effective-chemical potential model, the relaxation of the electronic excitations across the gap can be successfully described by the Rothwarf-Taylor model. In this picture, the gap-energy bosons emitted during the recombination process accumulate to form a pair-breaking non-thermal population of bosons, which strongly slows down the relaxation process.
\item The time-dependent Ginzburg-Landau functionals constitute a simple tool to capture the phenomenology of the space-time propagation of the impulsive perturbation in a symmetry-broken phase.
\end{itemize}

\setcounter{section}{3}
\section{Optical properties: a non-equilibrium approach}
\label{sec_noneqopticalprop}
One of the major breakthroughs in the field of ultrafast optical spectroscopies was given by the possibility of tuning both the pump and the probe frequencies over a very wide region, ranging from the THz to the UV. While tuning the pump photon energy allows to selectively excite the system along well-defined pathways, the tunability of the probe can be exploited to explore the dynamics of many different degrees of freedom. Furthermore, the possibility of easily controlling the polarization of both the pump and probe beams makes the table-top tunable ultrafast experiments a novel tool complementary to the conventional equilibrium optical spectroscopies. Although we do not pretend to provide a comprehensive overview of the optical properties of correlated materials, that can be found elsewhere \cite{Basov2005,Basov2011}, here we will present the most important concepts that constitute the bridge between the equilibrium and non-equilibrium optical properties. The possibility of investigating the ultrafast dynamics of different regions of the dielectric function will open the gate to many interesting experiments, whose results will be reported in chapter \ref{sec_results}. 

\subsection{The fundamental optical features of correlated materials}

In this section we overview the general features that constitute the starting point of the differential analysis for time-resolved measurements. In particular, we will focus on copper oxides which can be considered as prototypical materials, since, upon doping and temperature changes, they exhibit some general features that are of relevance for all correlated materials. Among these we cite: i) development of a Drude peak when the carrier concentration is increased; ii) charge-transfer transitions involving transitions from O-2$p$ orbitals to the localized upper Hubbard band of the 3$d$ Cu orbitals; iii) anomalous spectral weight changes involving the Drude peak and the high-energy $U$ energy scale; iv) mid-infrared peaks  (MIP) in the 0.5-0.7 eV energy range; v) signatures of the opening of the pseudogap and of the superconducting gap; vi) inductive response of the condensate.

\label{sec_eqoptprop}
\begin{figure}[t]
\begin{centering}
\includegraphics[width=0.8\textwidth]{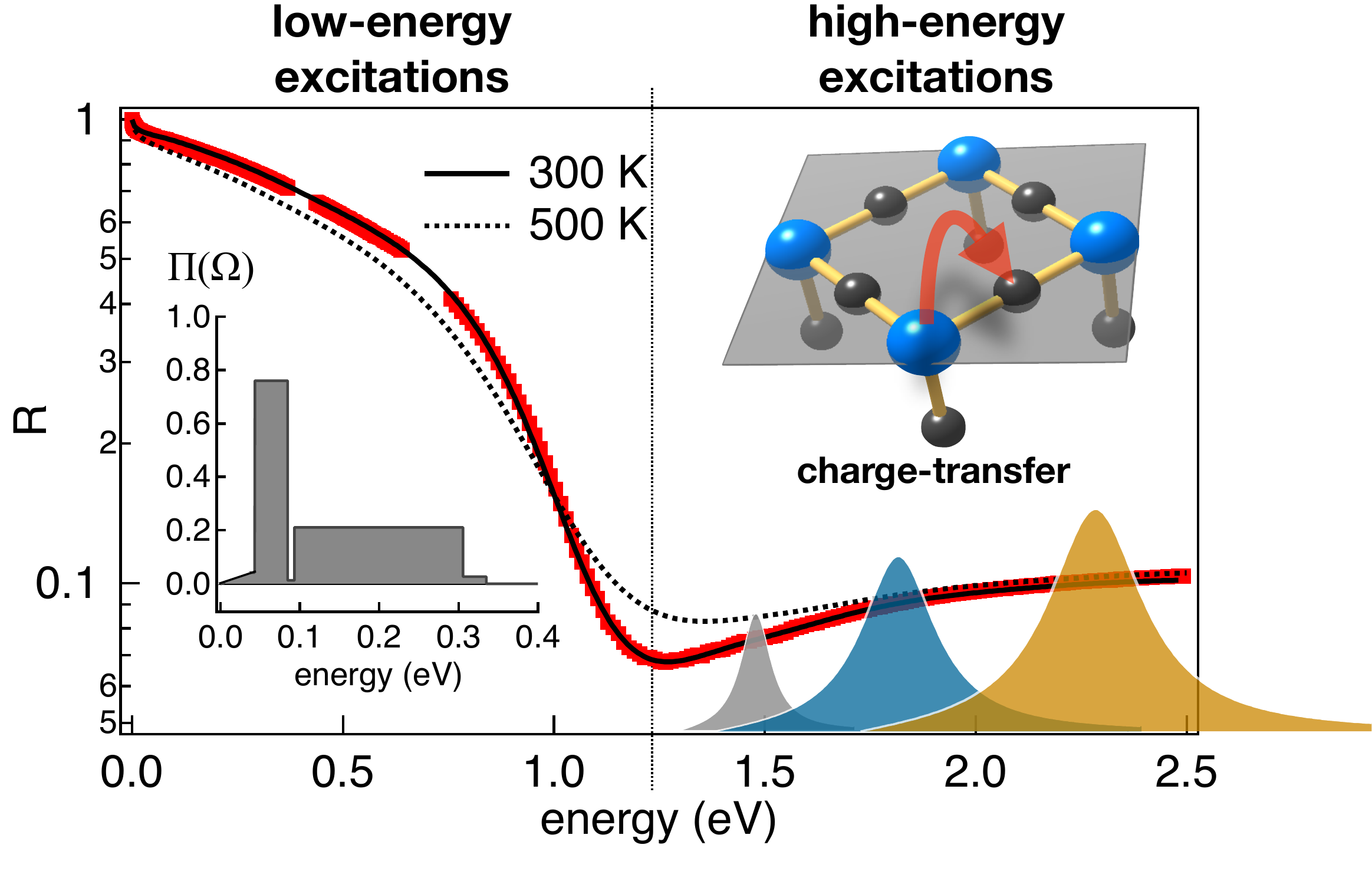}
\caption{Reflectivity of a prototypical cuprate. The red dots are the reflectivity of an optimally-doped Bi$_2$Sr$_2$Y$_{0.08}$Ca$_{0.92}$Cu$_2$O$_{8+\delta}$ sample (Y-Bi2212$_{\mathrm{OP}}$, $p\simeq$0.16, $T_c$=96 K). The black line is the fit of the extended Drude model to the data. The inset displays the electron-boson coupling function $\Pi(\Omega)$ extracted from the EDM. The dashed line is the reflectivity computed in the case of an increase of the effective temperature of $\delta T$=200 K. Taken from \cite{Cilento2013}}
\label{fig_cupratereflectivity}
\end{centering}
\end{figure}
Figure \ref{fig_cupratereflectivity} reports the  \textit{ab}-plane reflectivity of optimally-doped Bi$_{2}$Sr$_{2}$Ca$_{0.92}$Y$_{0.08}$Cu$_2$O$_{8+\delta}$ (Y-Bi2212) crystals ($T_c$=96 K), measured by conventional spectroscopic ellipsometry \cite{vanderMarel2013} at 300 K.
The reflectivity of Y-Bi2212 shows some general features common to most of correlated materials:\\
\begin{enumerate}
\item[i)] below the dressed plasma frequency ($\bar{\omega_p}\sim$1 eV), defined through the relation Re$\{\epsilon(\bar{\omega_p},T)\}$=0, the optical properties are dominated by a broad peak related to the optical response of the low-energy excitations in the conduction band. Interestingly, the dielectric function in this energy range cannot be simply reproduced by a Drude peak, in which a constant scattering rate $\tau$ and effective mass $m^*$ of the QPs is assumed. The strongly frequency-dependent scattering rate and effective mass, resulting from the interaction with bosonic excitations, can be accounted for by the more general extended Drude model \cite{Basov2005}, presented in section \ref{sec_EDM}. When the density of the charge carriers is decreased by chemical doping, the weight of the Drude-like part of the spectrum is progressively suppressed until the insulating phase, exhibiting a charge-transfer gap at 1.5-2 eV (see Fig. \ref{fig_opticscorrelated}), is reached. When moving to the low-doping region of the phase diagram, the optical conductivity of copper oxides is characterized by universal structures at $\sim$0.5 eV, which are usually named as mid-infrared peaks (MIP). Although their nature is often debated, the MIPs constitute a rather universal fingerprint of short-range electronic correlations. As a general picture, the MIPs can be rationalized as interband transitions between the filled lower Hubbard band and the empty quasiparticle states close to $E_F$ or between the filled quasiparticle states at $E_F$ and the empty upper Hubbard band (UHB). In both cases, the transitions involve delocalized QP states and localized states characterized by the doubly-occupancy of the Cu sites. Different interpretations as been also proposed, such as a subtle interplay between the strong coupling of the charge carriers with phonons and spin fluctuations \cite{Mishchenko2008,DeFilippis2009}.\\ 
\item[ii)] above $\bar{\omega_p}$, the high-energy interband transitions dominate. In this energy range, the equilibrium dielectric function can be modeled as a sum of Lorentz oscillators in the $>$1 eV energy range \cite{Basov2005,Basov2011}. The attribution of these interband transitions in cuprates is a subject of intense debate. The ubiquitous charge-transfer (CT) gap edge (hole from the upper Hubbard band with $d_{x^2-y^2}$ symmetry to the O-2$p_{x,y}$ orbitals) in the undoped compounds is about 2 eV \cite{Basov2005}. Upon doping, a structure reminiscent of the CT gap moves to higher energies, while the gap is filled with new transitions. This trend has been reproduced by Dynamical Mean Field Theory (DMFT) calculations of the electron spectral function and of the \textit{ab}-plane optical conductivity for the hole-doped three-band Hubbard model \cite{DeMedici2009}. The structures appearing in the dielectric function at 1-2 eV, that is, below the remnant of the CT gap at 2.5-3 eV, are possibly related to transitions between many-body Cu-O states at binding energies as high as 2 eV (for example singlet states) and states at the Fermi energy. More recently, equilibrium optical spectroscopy measurements \cite{Li2013} on HgBa$_2$O$_{4+\delta}$ have revealed an interband peak at 1.3 eV, which appears at the hole doping $p\sim$0.1 and progressively increases with maximum amplitude at optimal doping.
\end{enumerate}

\begin{figure}[t]
\begin{centering}
\includegraphics[width=0.8\textwidth]{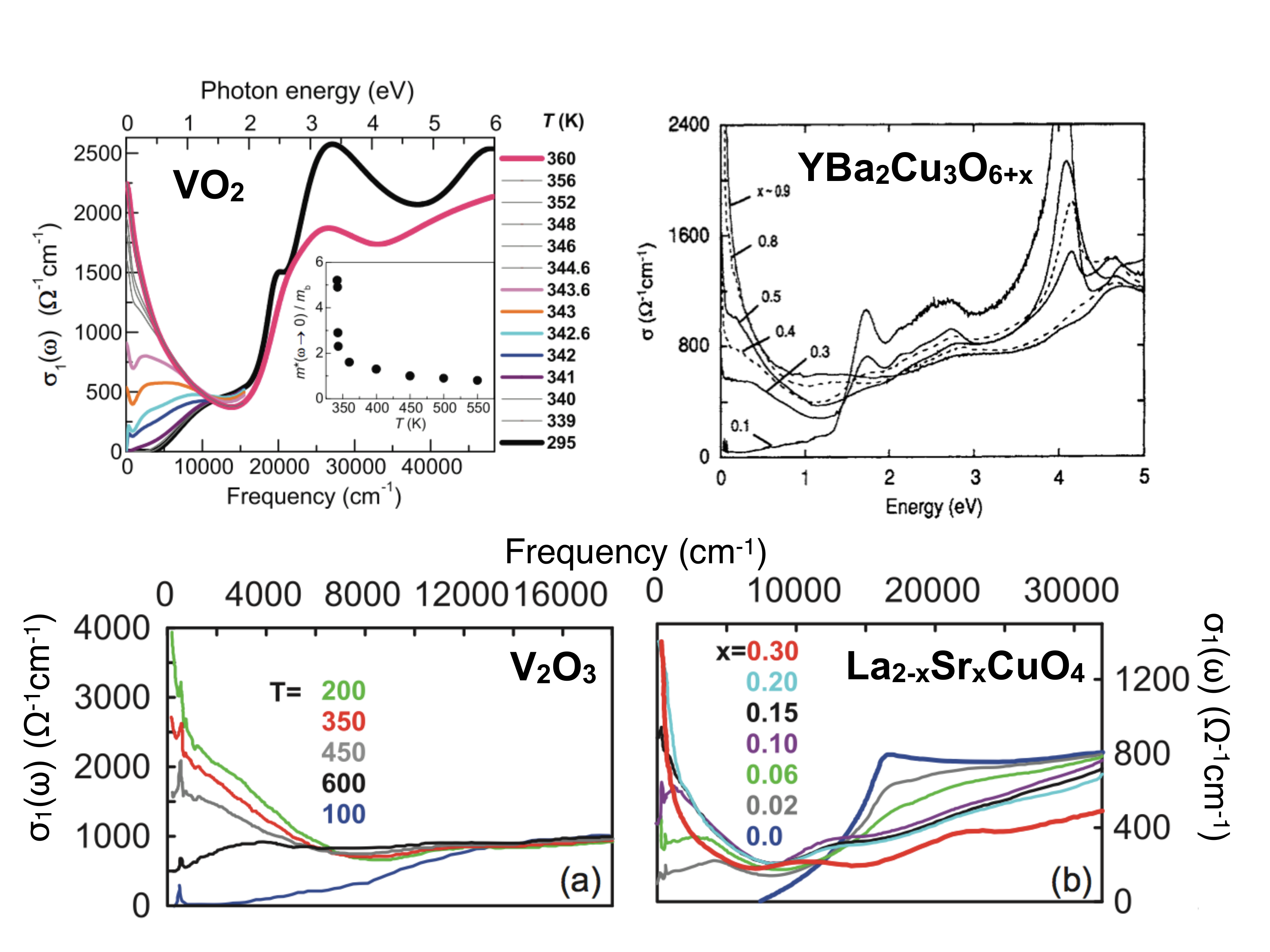}
\caption{Real part of the optical conductivity of prototypical correlated materials, such as VO$_2$  \cite{Baldassarre2008}, YBa$_2$Cu$_3$O$_{6+x}$ \cite{Cooper1993}, V$_2$O$_3$ \cite{Qazilbash2007} and La$_{2-x}$Sr$_x$CuO$_4$ \cite{Uchida1991}.}
\label{fig_opticscorrelated}
\end{centering}
\end{figure}

The spectral evolution of the cuprate optical conductivity when moving from the overdoped to the underdoped region of the $p$-$T$ phase diagram presents many features similar to those characteristic of other correlated materials which undergo a temperature driven insulator-to-metal phase transition. In Fig. \ref{fig_opticscorrelated} we contrast the doping-dependent optical conductivity of YBa$_2$Cu$_3$O$_{6+x}$ and La$_{2-x}$Sr$_x$CuO$_4$ to that of the vanadates VO$_2$ and V$_2$O$_3$ at different temperatures. As the insulating phase is approached, the Drude peak is progressively suppressed concomitantly with the divergence of the optical effective mass (see Section \ref{sec_EDM}) and the increase of the intensity of the interband transitions. As a general behaviour, in correlated materials the changes of the optical conductivity in proximity of a metal-to-insulator transition involve energy scales on the order of the Hubbard $U$. To quantitatively address the temperature- or doping-related changes of the optical conductivity we introduce the optical spectral weight of the Drude-like peak, defined as:
\begin{equation}
\label{eq_SW_def}
SW_{D}=\int_0^{\infty}\mathrm{Re}\sigma_{D}(\omega)d\omega
\end{equation} 
As a consequence of the conservation of the total number of the charge carriers (with density $N$) in the conduction band, $SW_{D}$ is a constant given by $SW_{D}$=$\omega^2_{pl}/8$, where $\omega^2_{pl}$=4$\pi$$N$$q^2$/$m^*$ is the squared plasma frequency and $m^*$ is the effective mass of the free carriers. When the system undergoes a phase transition, e.g., from insulator to metal, the density of states at the Fermi level is expected to dramatically change and a more general sum rule, which accounts for the conservation of the total number of electrons in the conduction and valence bands, holds:
\begin{equation}
SW_{D}+\sum_i SW_i=cost
\end{equation} 
where $SW_i$ is the spectral weight of the interband transitions labeled by $i$.                                                                                                                                                                                        
\\
The normal-state optical properties of copper oxides mostly change, particularly in the infrared region of the spectrum, when the $c$-axis (i.e., perpendicular to the Cu-O planes) conductivity is considered. In particular, the infrared ($\hbar\omega\lesssim$ 100 meV) optical conductivity is strongly suppressed from typical values on the order of 1000-2000 $\Omega^{-1}$cm$^{-1}$ in the $ab$-plane, down to $\ll$400 $\Omega^{-1}$cm$^{-1}$. Further, the optical conductivity along the $c$-axis assumes the characteristics of a strongly incoherent metal. Although the microscopic mechanisms at the origin of this strong scattering are still debated \citep{vanderMarel2000,Basov2005}, the underlying physics of this effect is related to the small interplane transfer matrix elements that are of the order of $t_c\simeq$30 meV for the most conducting systems. Therefore, the $c$-axis coherent is destroyed by the very strong in-plane scattering. Specifically, the timescale of the scattering processes involving the charge carriers is significantly smaller than the the time needed for the interlayer hopping, which is of the order of $\hbar$/$t_c\simeq$20 fs.  

\subsubsection{The extended Drude model}
\label{sec_EDM}
In the conventional Drude model, the optical absorption is mediated by scattering processes that lead to a finite lifetime $\tau$ of the free carriers. However, in most cases a constant value of $\tau$ does not allow to account for the coupling of electrons (or holes) with a rich spectrum of fluctuations and does not reproduce the experimental results. In the extended Drude model (EDM) the physical processes responsible for the renormalization of the lifetime and effective mass  of the QPs are included in a phenomenological way, by replacing the frequency-independent scattering time $\tau$ with a complex temperature- and frequency-dependent scattering time $\tau(\omega,T)$:
\begin{equation}
\tau^{-1} \Rightarrow \tilde{\tau}^{-1}(\omega)=\tau^{-1}(\omega)-i\omega\tilde{\lambda}(\omega)=-iM(\omega,T)
\end{equation}
where $1+\tilde{\lambda}(\omega)=\frac{m^*}{m}(\omega)$ is the mass renormalization of the QPs due to many-body interactions and $M(\omega,T)$ is the memory function.
\\
In the EDM, the optical conductivity $\sigma_D(\omega,T)$=$[1-\epsilon_D(\omega,T)]i\omega/4\pi$ is given by:
\begin{equation}\label{eq_EDM_Drude2}         
\begin{split}                                                                                                                                                                                                                                                                                                                                                                                                                                                                                                                                                                                                                                                                                                                                                                                                                                                                                                                                                                                                                                                                                                                                                                                                                                                                                                                                                                                                                                                                                                                                                                                                                                                                                                                                                                                                                                                                                                                                                                                                                                                                                                                                                                                                                                                       
\sigma_D(\omega,T)=\frac{i}{4\pi} \frac { {\omega_{p}}^{2} } {\omega+M(\omega,T)}
=\frac{1}{4\pi} \frac {{\omega_{p}}^{2}} {1/\tau(\omega,T)-i\omega(1+\tilde{\lambda}(\omega,T))}
\end{split}
\end{equation}

The renormalized scattering rate and effective mass can be directly extracted from the measured Drude optical conductivity, through the relations:
\begin{eqnarray}
1/\tau(\omega,T)&=&\frac{\omega_p^{2}}{4\pi}\mathrm{Re} \left ( \frac{1}{\sigma_D(\omega,T)} \right )\\
1+\tilde{\lambda}(\omega,T)&=&-\frac{\omega_p^{2}}{4\pi} \frac{1}{\omega}\mathrm{Im} \left ( \frac{1}{\sigma_D(\omega,T)} \right)
\end{eqnarray}
\\
Although this phenomenological version of the EDM does not provide any clue about the microscopic mechanisms responsible for the renormalization of the energy dispersion and lifetime of the QPs, it is very useful for directly extracting $\tau(\omega,T)$ and $m^*(\omega,T)/m$ from the optical data. The EDM can be used to demonstrate the correlated nature of the metallic phase of transition-metal oxides, in which the transition to the insulating phase is driven by the progressive localization of the charge carriers. 

In three-dimensions, this Mott-type transition manifests in the divergence \cite{Qazilbash2007} of the scattering rate ($\tau^{-1}(\omega,T)$) and of the effective mass, which unveils a dramatic correlation-driven decrease of the the kinetic energy of the charge carriers. In lower dimensions, which is the case of copper oxides, the spatial correlations become important and the divergence is partially smoothed out. In copper oxides, the EDM has been widely \citep{Basov2005} used to estimate the $ab$-plane scattering time of the charge carriers due to the interaction with any fluctuations present in the system. For example, in the normal state of Bi2212, $\tau^{-1}(\omega,T)$ linearly increases from $\sim$1000 cm$^{-1}$ ($\sim$5 fs) at zero frequency, to $\sim$4000 cm$^{-1}$ ($\sim$1.3 fs) at $\hbar\omega\sim$0.5 eV. Above this frequency, which corresponds to the cut-off of the highest-energy fluctuations, $\tau^{-1}(\omega,T)$ saturates to a constant value \citep{vanderMarel2003}. As soon as the system enters a gapped phase, $\tau^{-1}(\omega,T)$ rapidly drops to zero on the energy scale lower than the energy gap.

In some simple cases, the spectral weight $SW_D$ can be directly related to the average kinetic energy $\langle K \rangle$ of the charge carriers in the conduction band. As an example, in a single-band system in which the tight-binding model with nearest-neighbour hopping is a valid approximation, it is demonstrated \cite{Hirsch2000}:
\begin{equation}
\label{eq_SWk_energy}
SW_D=\frac{\pi^2a^2_{\delta}e^2}{4\hbar^2V_{atom}}\langle -K \rangle
\end{equation}
where $a_{\delta}$ is the lattice spacing projected along the direction determined by the the polarization $\delta$ of the incident light and $V_{atom}$ is the volume per atom. Eq. \ref{eq_SWk_energy} is valid when only the Drude contribution to the optical conductivity is included in the integration \ref{eq_SW_def}, in which the upper bound is infinite. In realistic analysis, however, the Drude contribution can be hardly disentangled from the additional interband transitions which appear at the $>$1 eV energy scale. Therefore, the spectral weight calculations are often performed using a finite-energy cut-off, $\Omega_c$, positioned in-between the plasma frequency and the onset of the interband transitions. In this case, the spectral weight sum rules are not strictly valid since any change of the optical scattering rate leads to the modification of the Drude peak width which is missed when the finite cut-off integral is calculated.  
Quantitatively, this can be easily shown for the simplest Drude model in which the scattering rate is frequency-independent:
\begin{equation}
SW(\Omega_c,T)\simeq \frac{\omega^2_{p}}{8}\left[1-2\frac{\gamma(T)}{\pi \Omega_c}\right]
\label{eq_SW_finite}
\end{equation}
This relation is valid in the limit $\Omega_c\gg\gamma$ and can be extended to a frequency-dependent scattering rate, provided that a suitable frequency-averaged $\langle \gamma(T) \rangle_{\omega}$ is used. Considering the direct relation between the Drude spectral weight and $\gamma(T)$, Eq. \ref{eq_SW_finite} suggests that $SW_D$ can be used to study the temperature-dependence of the scattering rate of the charge carriers, which is strictly related to the scattering with bosonic fluctuations, as it will be discussed in the next section.

\begin{figure}[t]
\begin{centering}
\includegraphics[width=0.8\textwidth]{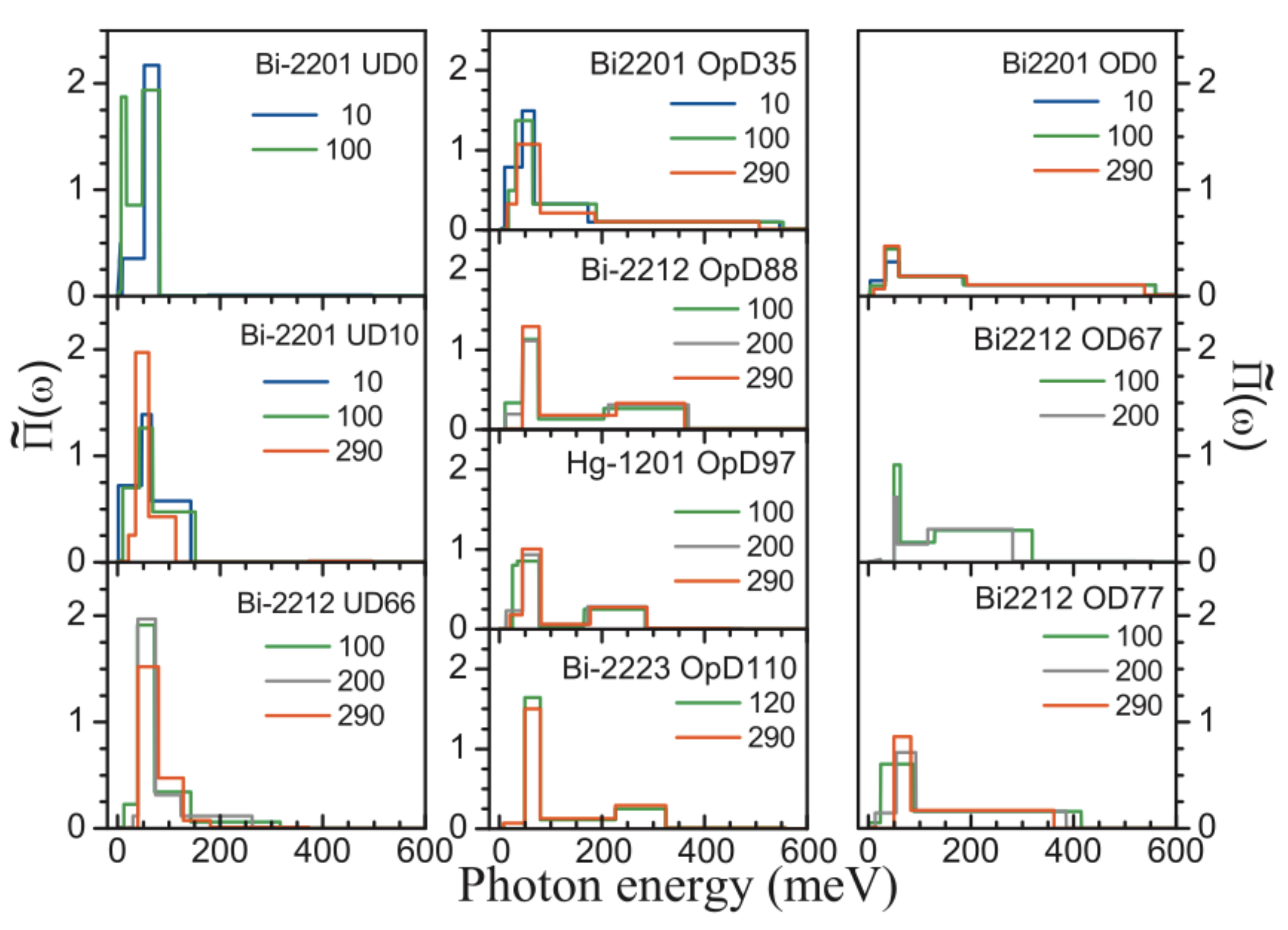}
\caption{Electron-boson coupling function extracted from the Extended Drude analysis of spectroscopic ellipsometry data of different copper-oxides. Taken from \cite{vanHeumen2009}.}
\label{fig_bosonglue}
\end{centering}
\end{figure}

\subsubsection{The electron-boson scattering in optics}
\label{sec_dielfunction}
From the microscopic point of view, the extended Drude formalism can be derived from the Holstein theory for normal metals \cite{Kaufmann1998}. Considering the Kubo formula and using complex diagrammatic techniques to evaluate the electron and boson thermal Green's functions and omitting vertex corrections (Migdal approximation), the memory function $M(\omega,T)$ results:
\begin{equation}\label{eq_Memory_EDM2_b}
M(\omega,T)=\omega\left\{ \int_{-\infty}^{+\infty} \frac{f(\xi,T)- f(\xi+\omega,T)}{\omega+\Sigma^*(\xi,T)-\Sigma(\xi+\omega,T)+i\gamma_{\mathrm{imp}}} d\xi \right\}^{-1}-\omega
\end{equation}
where $f$ is the Fermi-Dirac distribution, $\gamma_{\mathrm{imp}}$ an intrinsic decay rate that accounts for the scattering by impurities and $\Sigma(\omega,T)$ and $\Sigma^*(\omega,T)$ the electron and hole \textbf{k}-space averaged self-energies, which can be calculated by Eq. \ref{eq_SelfEnergy}. We pinpoint that, although the memory function $M(\omega,T)$ has the same analytical properties of the single-particle self-energy $\Sigma(\omega,T)$, it has a conceptually different meaning, since the optical transition at frequency $\omega$ involves a particle-hole excitation of the many-body system and provides information about the joint particle-hole density of states.
Considering the vast amount of experimental indications that, at moderate doping concentrations ($p>$0.16), the fermionic and bosonic degrees of freedom can be separated \cite{Carbotte2011}, we assume that the EDM can be used to extract $\Pi(\Omega)$ from the optical conductivity, as measured at the equilibrium temperature $T$. Although sophisticated maximum entropy techniques \cite{Hwang2007} have been developed to unveil the rich details of the bosonic function \cite{Carbotte2011}, the main features can be evidenced by a simple histogram form for $\Pi(\Omega)$. The inset of Figure \ref{fig_cupratereflectivity} displays the typical $\Pi(\Omega)$ histogram function, extracted by the equilibrium optical conductivity of optimally-doped Y-Bi2212. $\Pi(\Omega)$ is characterized by:\\
i) a low-energy part (up to 40 meV) linearly increasing with the frequency. This part is compatible with either the coupling of QPs to acoustic \cite{Johnston2012} and Raman-active optical \cite{Kovaleva2004} phonons or the linear susceptibility, as expected for a Fermi liquid \cite{Mirzaei2013};\\
ii) a narrow, intense peak centered at $\sim$60 meV, attributed to the anisotropic coupling to either out-of-plane buckling and in-plane breathing Cu-O optical modes \cite{Devereaux2004} or bosonic excitations of electronic origin such as spin fluctuations \cite{Dahm2009,Fujita2012};\\
iii) a broad continuum extending up to 300-400 meV, well above the characteristic phonon cutoff frequency ($\sim$90 meV) that strongly resembles the dispersion of magnetic excitations, evidenced by scattering experiments \cite{Abanov2001,Chubukov2005,Norman2006,letacon2011,Scalapino2012,letacon2013,dean2013,dean2013b}, or loop currents \cite{Varma2006}.

The EDM model allows to treat on equal footing the electron-phonon coupling and the coupling to the additional degrees of freedom, mostly of electronic origin, that are common to all correlated materials. Interestingly, the formalism presented in this section constitutes the link between the single-particle self-energy and the optical self-energy. This fills the gap between the concept of QP scattering, as discussed in chapter \ref{sec_QPdynamics}, and the \textit{optical} scattering rate introduced through the memory function. In this framework, the probability of a scattering event which enables the interaction of a free QP with the incoming photons is ultimately proportional to the density of bosons and to the strength of their coupling with the charge carriers. As extensively reviewed in Ref. \cite{Carbotte2011}, the $\Pi(\Omega)$ extracted by optical spectroscopy in copper oxides is in very good agreement with the electron-boson function measured by other single-particle and two-particle techniques such as ARPES, STM and Raman. In this framework, the temperature-dependence of $\gamma(T)$ naturally emerges from Eq. \ref{eq_Memory_EDM2_b} and is directly related to the change of the number of bosons which can scatter with the carriers. Given the $\Pi(\Omega)$ function with a $\Omega_c$ cutoff, $\gamma(T)$ can be easily calculated. $\gamma(T)$ exhibits a sub-linear increase for $k_BT$$<$$\hbar\Omega_c$, which approaches a linear trend as the temperature becomes larger than the energy scale of the bosonic fluctuations.

Fig. \ref{fig_bosonglue} displays the doping and temperature dependent $\Pi(\Omega)$ on several families of copper oxides. Although the main role of the temperature is to modify the population of the bosonic modes, the histograms exhibit a small temperature-dependence, which becomes stronger in the underdoped regime. This result suggests that the picture of a sharp distinction between fermionic QPs and bosonic fluctuations, which interact though a temperature-independent coupling, is progressively lost when electronic correlations dominate the electronic properties. As a general trend, $\Pi(\Omega)$ is characterized by an intense and peaked low-energy structure, which progressively evolves in a rather flat background as the doping density is increased. This result can be connected to the presence of well-defined AF resonances at $\textbf{q}_{AF}$=($\pi$,$\pi$) in the underdoped compounds, which broaden and evolve in the continuum of the short-range magnetic fluctuations when the density of the charge carriers is increased by doping (see Section \ref{sec_magnons}). 

The EDM has been widely employed to discuss the dynamics of the charge carriers in copper oxides  and its relation with the onset of superconductivity \cite{Hwang2007,Hwang2007b,Hwang2007c,Schachinger2008,vanHeumen2009,Yang2009b,Basov2011,Hwang2011,Vladimirov2012,Hwang2012,Hwang2013,Park2014}. Recent results suggest that the fluctuation spectrum could extend up to several eV \cite{Markiewicz2012,Hwang2014}, hence of the order of the Hubbard energy $U$. The same formalism has been also applied to investigate the coupling in iron-based superconductors \cite{Yang2009} and more conventional ferromagnetic systems \cite{Dordevic2009}.
When the density of the carriers is decreased (underdoping), the application of the EDM to copper oxides and, more in general, to correlated materials becomes more questionable for various reasons: i) the opening of the $\textbf{k}$-dependent pseudogap in the electronic density of states, which will be discussed in Section \ref{sec_DOSnonc}; ii) the loss of well-defined quasiparticles and the impossibility of separating fermionic and bosonic degrees of freedom; iii) the appearance of the MIPs which progressive replace the broad Drude-like peak.

\paragraph*{Non-constant density of states.}
\label{sec_DOSnonc}

\begin{figure}[t]
\begin{centering}
\includegraphics[width=1\textwidth]{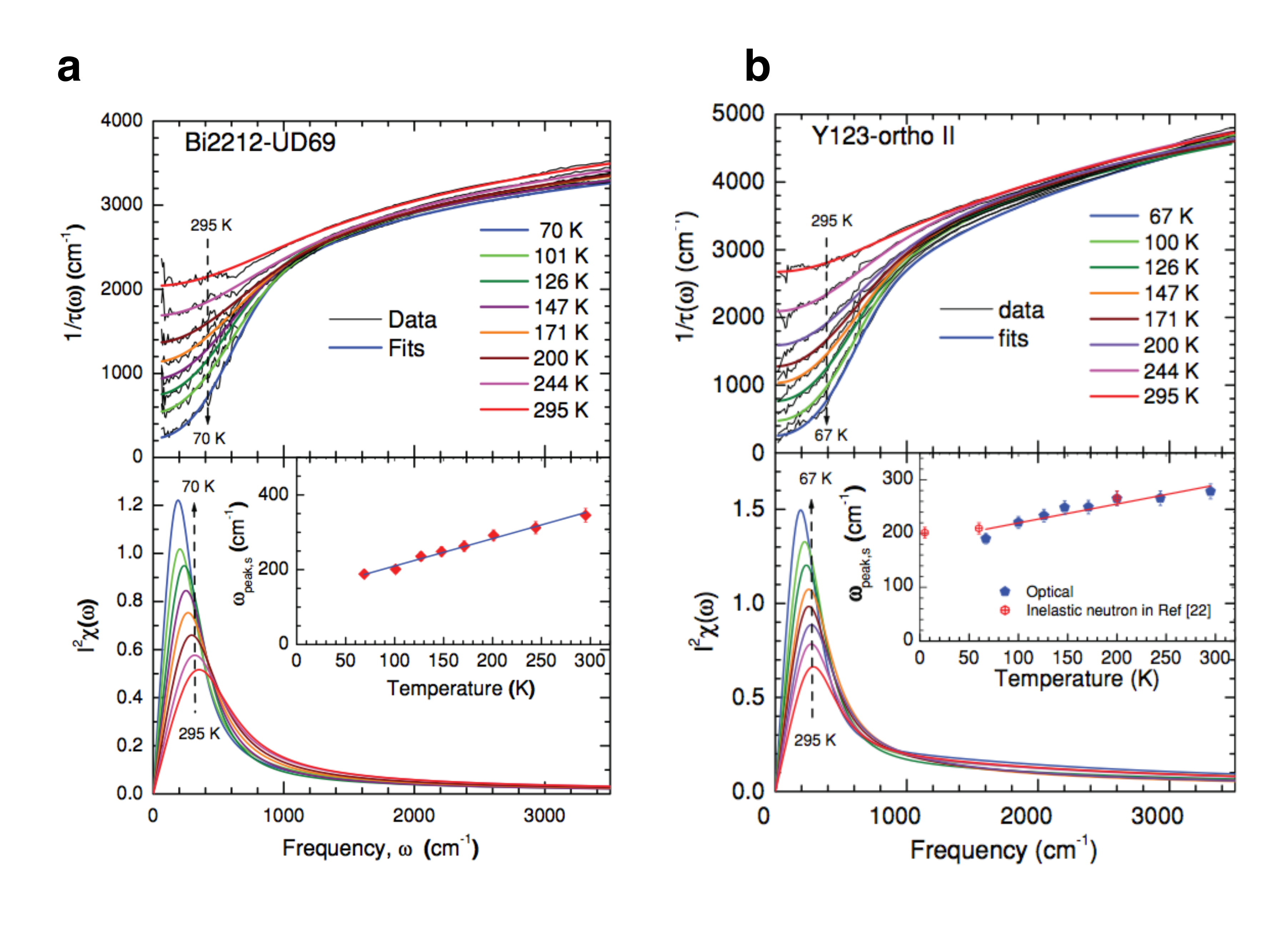}
\caption{Extended Drude model analysis of the frequency-dependent scattering rate in underdoped \textbf{a}) Bi$_2$Sr$_2$CaCu$_2$O$_{8+\delta}$ ($T_c$=67 K) and \textbf{b}) YBa$_2$Cu$_3$O$_{6.50}$ ($T_c$=59 K). The coloured lines are the EDM fit to the experimental data (black lines) at different temperatures. The bottom panels display the electron-boson spectral function at different temperatures. Taken from Ref. \citenum{Hwang2011}.}
\label{fig_pseudogapoptics}
\end{centering}
\end{figure}

In the calculation of the self-energy (see Eq. \ref{eq_SelfEnergy}), a constant density of states at the Fermi level has been considered. This assumption is dramatically broken by the opening of a partial or full gap in the DOS close to $E_F$. Such a gap, lying within the same energy scale of the bosonic fluctuations, hinders a reliable modelling of the electron-boson dynamics.
A further evolution of the EDM, accounting for a non-constant electronic density of states, has been developed by Sharapov and Carbotte \cite{Sharapov2005}, and has been used to analyze spectroscopic data at equilibrium \cite{Hwang2011}.
In this model, the imaginary part of the electronic self energy is given by:
\begin{equation}\label{eq_ESelf_EDM3}
\begin{split}
\mathrm{Im}\Sigma(\omega,T)&=-\pi \int_0^\infty \Pi(\Omega)\{\tilde{N}(\omega+\Omega,T) \left[n(\Omega,T)+f(\omega+\Omega,T)\right]+ \\
&+\tilde{N}(\omega-\Omega,T) \left[1+n(\Omega,T)-f(\omega-\Omega,T)\right]\}         d\Omega
\end{split}
\end{equation}
where $\tilde{N}(\omega,T)$ is the frequency-dependent normalized density of states, which can be modelled through a proper function of $\omega$. Re$\Sigma(\omega,T)$ is calculated from (\ref{eq_ESelf_EDM3}) through the Kramers-Kronig relations. This formalism has been introduced and successfully used to study the electron-boson coupling in the underdoped cuprates, characterized by the onset of the pseudogap $\Delta_{pg}(\textbf{k})$ at $T$$<$$T^*$ \cite{Hwang2011}.
The normalized density of states $\tilde{N}(\omega,T)$ can be modelled by \cite{Hwang2011}:
\\
\begin{equation}\label{eq_N_EDM3}
	\tilde{N}(\omega,T)=\begin{cases} \tilde{N}(0,T)+ [ 1-\tilde{N}(0,T) ] \left( \frac{\omega}{\Delta_{pg}}\right)^2 & for \left| \omega \right| \leqslant \Delta_{pg} \\
1+\frac{2}{3}[ 1-\tilde{N}(0,T)] & for \left| \omega \right| \in (\Delta_{pg}, 2\Delta_{pg}) \\
1 & for \left| \omega \right| \geqslant 2\Delta_{pg} \end{cases}
\end{equation}
\\
where $\Delta_{pg}$$\simeq$40 meV is the maximum energy gap amplitude at the antinodes $\textbf{k}$$\simeq$($\pm\pi$,0) and (0,$\pm\pi$), while 0$<$$\tilde{N}(0,T)$$<$1 is the gap filling. The \textbf{k}-space integrated electronic DOS is completely recovered between $\Delta_{pg}$ and 2$\Delta_{pg}$. This model has been used to extract $\Pi(\Omega)$ as a function of the temperature, demonstrating a significative pseudogap-driven increase of the electron-boson spectral function in the same doping range in which $T_c$ decreases (see Fig. \ref{fig_pseudogapoptics}). This result strongly supports a scenario where the pseudogap and the superconductivity are competing.

\subsubsection{Infrared active modes}
\label{sec_infrared_active_modes}
The low-frequency optical properties contain valuable information about the electronic and structural properties of correlated materials. As discussed in Chapter 1, the THz and mid-IR (2-130 meV) regions of the electromagnetic spectrum can be almost continuously covered by time-resolved experiments. This energy range is particularly sensitive to a wealth of relevant processes such as the infrared active modes, the Josephson plasma resonances, the inductive response of the superconducting condensate and the opening of a full or partial low-energy gap in the density of states. In principle, all these processes can be used as the fingerprints to track the ultrafast transformations of the system, in which the structural and electronic degrees of freedom can be directly disentangled in the time domain. Here, we will review some of the general features that can be accessed by THz or infrared pump-probe experiments. In general, the optical conductivity associated to the infrared active modes is on the order of $\lesssim$500 $\Omega^{-1}\mathrm{cm}^{-1}$. In metallic systems, these values are easily overwhelmed by the intrinsic Drude conductivity of the system. Therefore, the infrared features are usually more visible in insulating systems or in the cases in which the metallic conductivity is suppressed either by the opening of a gap or as a consequence of an extremely large scattering rate of the charge carriers. 
\begin{figure}[t]
\begin{centering}
\includegraphics[width=1\textwidth]{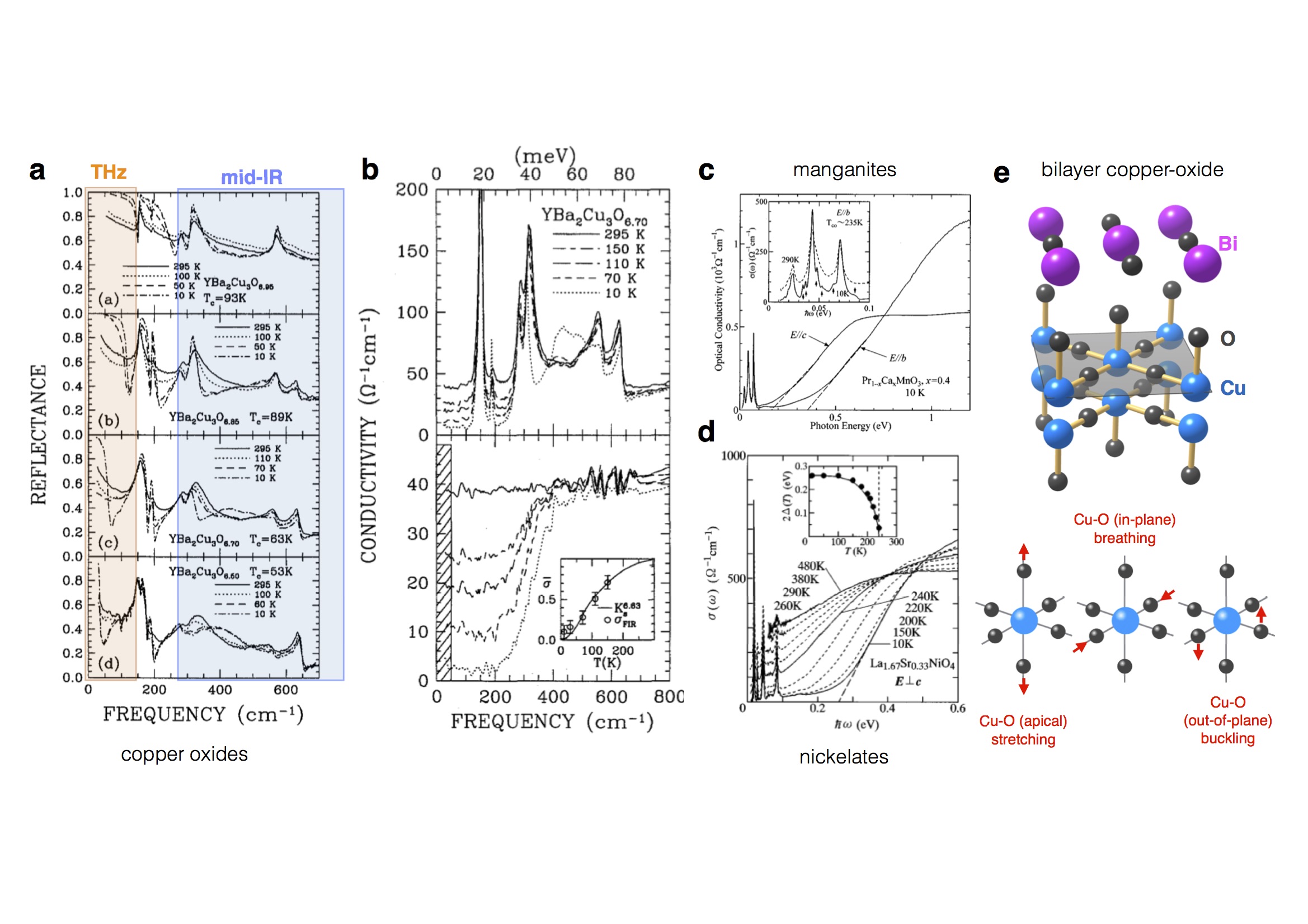}
\caption{a) Reflectivity and b) optical conductivity of YBa$_2$Cu$_3$O$_{6+x}$ at different doping concentrations and temperatures. Taken from \citep{Homes1995}. THz-infrared conductivity of: c) the prototypical doped manganite Pr$_{0.6}$Ca$_{0.4}$MnO$_3$ \citep{Okimoto1999}; d) the prototypical nickel oxide La$_{1.67}$Sr$_{0.33}$NiO$_4$ \citep{Katsufuji1996}. e) Sketch of part of the Bi2212 unit cell and of the typical infrared-active vibrational modes in correlated materials.}
\label{fig_phononmodes}
\end{centering}
\end{figure} 
\paragraph*{Copper oxides}
Figure \ref{fig_phononmodes}a reports the low-frequency $c$-axis reflectivity \citep{Homes1995} of YBa$_2$Cu$_3$O$_{6+x}$ (YBCO) at different hole dopings, ranging from under ($x$=0.5, $T_c$=53 K) to optimal ($x$=0.95, $T_C$=93 K) doping concentrations. Due to the large incoherent nature of the charge carriers along the $c$-axis, a large number of narrow peaks in the 20-80 meV energy range are clearly visible, already at room temperature. These peaks are related to the infrared active phonon modes which are allowed by the dipole optical selection rules, given the symmetry of the crystal structure. Of particular relevance for time-resolved experiments \citep{Rini2007,Kubler2007,Carbone2008,Pashkin2010,Hu2014,Kaiser2014,Coslovich2013} (see Section \ref{sec_results}) are two $B_{1u}$ symmetry modes:
\begin{itemize} 
\item the bond bending mode at $\sim$40 meV (322 cm$^{-1}$), which involves the out-of-plane motion of the oxygens in the Cu-O plane, and 
\item the $\sim$70 meV (570 cm$^{-1}$) mode which modulates the displacement of the apical oxygen atoms (see Fig. \ref{fig_phononmodes}) along the $c$ direction (apex oxygen vibrations). The latter is more clearly visible in the optimally-doped system which is closer to the true orthorhombic phase of YBa$_2$Cu$_3$O$_{7}$. When decreasing the carrier density, this mode progressively weakens and a new mode at $\sim$76 meV (610 cm$^{-1}$), related to a change of the crystal symmetry, appears.
\end{itemize}  
When cooling down (e.g. 100 K), many new players come into the game. While in the most metallic system ($x$=0.95) the THz reflectivity increases, as expected for a metal, in the more underdoped samples the conductivity is suppressed below an energy scale of $\Delta_{PG}\sim$50 meV. This effect is related to the onset of the pseudogap at $T^*$, which leads to the progressive depletion of states at $\hbar\omega<\Delta_{PG}$, as shown in the optical conductivity of the underdoped (($x$=0.7)) sample (see Fig. \ref{fig_phononmodes}b). When the temperature is further decreased and $T_c$ is crossed, an additional plasma edge appears at $\sim$19 meV in the optimally-doped sample. This feature, known as Josephson plasma edge, is related to the coupling between the parallel superconducting Cu-O layers and will be discussed in Sec. \ref{sec_electrodynamics_condensate}.
Furthermore, the temperature also affects the position and lineshape of the infared-active phonon modes. As a general trend, the vibrational modes undergoes a blueshift when the temperature is decreased. This is a consequence of the anharmonicity of the lattice potential which stiffens for smaller vibrations. On the other hand, the bending mode at $\sim$40 meV is subject to an anomalous redshift when $T_c$ is crossed. This shift in the position of the mode is also accompanied by a dramatic change of the lineshape, which becomes asymmetric and can be phenomenologically described by the celebrated Fano resonance \citep{Fano1961}:
\begin{equation}
\label{eq_Fano}
\sigma_1(\omega)=A\frac{(x+q)^2}{1+x^2}
\end{equation}
where $x=\hbar(\omega-\omega_0)/\Gamma$, $A$ is an amplitude factor and $q$ and the asymmetry parameter. The Fano resonance describes any system characterized by the quantum interference between a narrow level and a continuum of states. In the case of YBCO, both the anomalous redshift and the asymmetry of the bending mode have been explained \citep{Munzar1999} as the effect of the change of the local electric field experienced by the ions, which arises from the onset of the Josephson plasma resonance at $T<T_c$. The same effect has been measured in Tl$_2$Ba$_2$Ca$_2$Cu$_3$O$_{10}$ \cite{Zetterer1990}.

The phenomenology of the infrared-active phonon mode in YBCO, is alike for almost all the copper oxides characterized by perovskite-derived structures with orthorhombic (YBCO) or tetrahedral symmetries (e.g. Bi2212, see Fig. \ref{fig_phononmodes}e), where the copper ions sit at the center of an octahedral cell constituted by the oxygen ions. In multi-plane cuprates, the octahedra are replaced by square pyramids with the base lying in the Cu-O plane. Since the electric properties of cuprates are dominated by the conduction in the Cu-O planes, we can single-out three main families of infrared active phonon modes (see Fig. \ref{fig_phononmodes}e) which are expected to be more relevant for the electronic properties:
\begin{itemize}
\item[i)] the modes involving the stretching of the oxygens ions at the apex of the pyramids (or octahedra);
\item[ii)] the stretching modes involving a change of the in-plane Cu-O distance;
\item[iii)] the bending modes related to the out-of-plane movement of the oxygens;
\end{itemize}  

\paragraph*{Other correlated materials}       
Similar infrared features are present in different transition-metal oxides, such as manganites, nickelates and vanadates. As an example, Figure \ref{fig_phononmodes}c reports the optical conductivity of Pr$_{0.6}$Ca$_{0.4}$MnO$_{3}$, which exhibits an anisotropic charge-ordering at $T<$235 K with electronic gap on the order of $\Delta_{CO}\simeq$0.18 eV. The suppression of the optical conductivity below $\Delta_{CO}$ allows observing a wealth of infrared active modes (see inset). The three dominant phonon modes (23, 42 and 71 meV) correspond to the three $F_{2u}$ infrared active vibrational modes of the cubic perovskite. Due to the orthorhombic distortion, a number of weaker resonances are clearly visible. The two highest-frequency vibrations are assigned to the Mn-O-Mn bending mode and the Mn-O stretching mode, respectively \citep{Okimoto1999,Rini2007}.  

Also the charge-ordered nickelate La$_{1.67}$Sr$_{0.33}$NiO$_{4}$ exhibits similar phonon modes that can be clearly visible at low temperature, when the electronic DOS is suppressed by the opening of a gap attributed \cite{Katsufuji1996,Homes1997,Jung2001,Uchida2011} to static charge order, fluctuating stripes or small polarons formation. Also in this case, the Ni-O in-plane stretching mode \cite{Katsufuji1996,Coslovich2013} (84 meV) develops a strongly asymmetric Fano-like lineshape at low temperature, that is attributed to the modification of the physical environment around the ions when the charge carriers progressively localize.    
Phonons in this frequency range are critical for the metal-insulator transition in vanadium oxides. For the VO$_2$, the $A_g$ lattice modes associated with stretching and tilting of V-V dimers are found at $\sim$23 and 27 meV, respectively \cite{Kubler2007,Pashkin2011}. 

Similar cases are given by the layered transition-metal dichalcogenides. For example, 1$T$-TiSe$_2$ shows a transition into a commensurate charge density wave (CDW), when cooled below $T_{CDW}\simeq$200 K. The microscopic mechanism driving the formation of CDW is still under debate, for the electronic correlations and the structural order are strongly intertwined. In the THz-infrared frequency range 1$T$-TiSe$_2$ exhibit features that can be used to disentangle the dynamics of the structure of the system from that of the excitonic correlation \cite{Porer2014}. The crystal symmetry can be tracked by looking at the infrared-active phonons \citep{Holy1977}. The symmetry change related to the CDW yields additional infrared-active in-plane modes at 19 meV and 22 meV, which adds to the 17 meV optical phonon resonance measured at room temperature \cite{Porer2014}. Differently, the pure electronic response of the system can be monitored by measuring the dynamics of the collective plasma resonance of unbound electrons and holes \citep{Li2007}, that manifests as a pole in the dielectric function, which moves from $\sim$145 meV at 300 K to 45 meV at 10 K.

In conclusion, the THz-infrared optical properties of correlated materials can be effectively used to disentangle the dynamics of the lattice, that determines the nature and the position of the infrared-active modes, from that of the electronic degrees of freedom, manifesting themselves in the opening of low-energy gaps, in the change of the plasma frequency and/or scattering rate, in the modification of the local potential experienced by the ions and in the formation of collective states, such as the superconducting condensate, that will be addressed in the next session.

\subsubsection{Electrodynamics of the condensate}
\label{sec_electrodynamics_condensate}
The expression of the optical conductivity of a perfect condensate can be easily obtained starting from the frequency-dependent conductivity of a conductor in which the charge carriers are subject to an average scattering rate 1/$\tau$, i.e., $\sigma$=$\epsilon_0\omega_{pl}^2$/$(1/\tau-i\omega)$, where $\omega_{pl}^2$=$\sqrt{n_qe^2/\epsilon_0m_q}$ is the plasma frequency of the system. In the limit $\tau\rightarrow\infty$ for a perfect superconductor, the optical conductivity becomes purely imaginary with a $\omega^{-1}$ frequency dependence. However, in order to satisfy the causality given by the Kramers-Kronig relations, $\sigma(\omega)$ must also acquire a $\delta$-like real component. Therefore, the optical conductivity of a perfect $s$-wave superconductor reads:

\begin{equation}
\label{eq_sigmaSC}
\sigma(\omega)=\epsilon_0\omega_{pl}^2\left( \pi\delta(\omega)+i\frac{1}{\omega}\right)
\end{equation}
where the $\sigma_1(\omega)$ component satisfies the sum rule:
\begin{equation}
\label{eq_sumruleDrude}
\int_{0}^{+\infty}\sigma_1(\omega)d\omega
=\frac{\pi}{2}\epsilon_0\omega_{pl}^2
\end{equation}
Considering that the amplitude of $\sigma_2(\omega)$ is proportional to the fraction of charge carriers that participates to the condensate, i.e., $n_{SC}$=$n_q-n_N$, near-IR pump/THz-probe measurements in the $ab$-plane have been widely used to investigate the in-plane dynamics of the condensate in high-temperature superconductors \cite{Averitt2001,Carnahan2004,Kaindl2005gp,Perfetti2015}.
When the frequency of the probe is increased up to the energy scale of the superconducting gap, the breaking of the Cooper pairs and the related excitation of above-gap electron-hole pairs provide an additional absorption channel that is manifested under the form of an absorption edge at $\omega\sim 2\Delta_{SC}$ in the $\sigma_1(\omega)$ function. Therefore, non equilibrium optical spectroscopies in the THz/mid-infrared region can be used to directly trace the recovery dynamics of the energy gap in both unconventional \cite{Averitt2001} and conventional \cite{Beck2011,Beyer2011,Beck2013} $s$-wave superconductors.

Considering the case of layered copper oxide superconductors, also the $c$-axis optical properties carry important information about the superconducting properties. While the $c$-axis optical conductivity in the normal state is completely incoherent, the sudden decrease of the in-plane scattering rate drives the coherent coupling between the condensates confined in the Cu-O planes that form an array of Josephson junctions \cite{Basov2005}. In the most common case of bi-layer cuprates, this effect results in the appearance of two longitudinal Josephson plasma resonances \cite{vanderMarel1996,vanderMarel2001} in the reflectivity and an absorption peak in the $\sigma_1(\omega)$ that arises from a transverse mode at the frequency $\omega^2_T$=$d_{in}\omega^2_{in}+d_{ex}\omega^2_{ex}$/$d_{in}+d_{ex}$, where $d_{in}$ is the distance between the two Cu-O planes within the unit cell and $d_{ex}$ is the distance between the double layers in the neighbouring unit cells. The plasma resonances can be modelled by a dielectric function that vanishes at the two proper screened Josephson plasma frequencies, $\omega_{in}$ and $\omega_{out}$, with a pole at $\omega_{T}$:
\begin{equation}
\label{eq_JPF}
\epsilon(\omega)=\epsilon_{\infty}\frac{(\omega^2-\omega^2_{in})(\omega^2-\omega^2_{ex})}{\omega^2(\omega^2-\omega^2_T)}
\end{equation}  
where $\epsilon_{\infty}$ is constant. As an example, YBa$_2$Cu$_3$O$_{6.5}$ displays the two longitudinal plasma resonances at 3.7 meV and 60 meV, whose dynamics can be probed by THz and mid-infrared techniques \cite{Hu2014}.

\subsubsection{Interband transitions and high-energy excitations}
\label{sec_interband}
The high-energy region ($\bar\omega\gtrsim$1.5 eV) of the dielectric function of correlated materials is dominated by interband transitions that correspond to excitations across the Mott or charge-transfer gaps and, at even higher energies, to optical transition between different bands. In general, these transitions are modelled through a high-energy dielectric function ($\epsilon_{HE}{\omega}$) that contains a number of Lorentz oscillators:
\begin{equation}
\label{eq_Lorentz}
\epsilon_{HE}(\omega)=1+\sum_i \frac{\omega^2_{pi}}{\omega^2_{oi}-\omega^2-i\omega\gamma_i}
\end{equation}
where $\omega^2_{pi}$, $\omega^2_{oi}$ and $\gamma_i$ are the intensity, position and frequency of the $i$-oscillator. From the Lorentz dielectric function, the optical conductivity can be directly calculated through the relation:
\begin{equation}
\label{eq_epsilonsigma}
\epsilon(\omega)=1-\frac{i\sigma(\omega)}{\epsilon_0\omega}
\end{equation}
Since the sum of the spectral weight of the high-energy transitions, i.e. $SW_{HE}$=$\int_0^{\infty}\mathrm{Re}\sigma_{HE}(\omega)d\omega$, and of the Drude component is proportional to the total number of charges of the system, this quantity must be conserved. Therefore, any change in the Drude plasma frequency as a function of the change of an external parameter (e.g. temperature, pressure, etc.) or during a phase-transition, must be compensated by the consequent change of the total spectral weight of the interband transition which follows the rule $SW_D+SW_{HE}$=const. 

This conservation rule has some important consequences in the study of the superconducting phase transition. While in conventional superconductors the spectral weight of the condensate zero-frequency $\delta$-function, $SW_{\delta}$, is compensated exclusively by the loss in $SW_D$ in the 0-$\Delta_{SC}$ frequency range, in unconventional superconductors a more general sum rule (Ferrel-Glover-Tinkham) holds \cite{Tinkham}:
\begin{equation}
\label{eq_FGTsumrule}
SW^N_D+SW^N_{HE}=SW^{SC}_D+SW^{SC}_{HE}+SW_{\delta}
\end{equation}
where the superscripts indicate the normal ($N$) and superconducting ($SC$) phases. Considering that the Drude spectral weight is proportional to the total kinetic energy of the charge carriers in the conduction band (see Sec. \ref{sec_EDM}), possible superconductivity-driven changes of $SW^N_{HE}$, in both equilibrium and non-equilibrium conditions, can be used to investigate the nature of the superconducting phase transition as the doping concentration is changed (see Sec. \ref{sec_SWshift}).   

A particular case is given by the cuprate parent insulators ($p$=0), in which the fluctuations of the Cu-3$d^9$ charge-configuration are suppressed by the strong Coulomb repulsion ($U_{dd}\sim10$ eV) between two electrons occupying the same Cu orbital. In this case, the lowest excitation is the charge-transfer (CT) of a localized Cu-3$d_{x^2-y^2}$ hole to its neighbouring O-2$p_{x,y}$ orbitals, with an energy cost $\Delta_{CT}\sim$2 eV$<U_{dd}$. In the optical conductivity, this process is revealed by a typical charge-transfer (CT) edge at $\hbar\omega$=$\Delta_{CT}$, which defines the onset of optical absorption by particle-hole excitations in the complete absence of a Drude response\cite{Uchida1991}. The shape of the charge-transfer edge can be modelled through the Urbach's expression \cite{Urbach1953} $\alpha$=$\alpha_0$exp$s(E-E_0)/k_BT$
where $\alpha_0$ and $E_0$ are temperature-independent parameters, while $s$ is given by $s$=$\sigma_0(2k_BT/\hbar\Omega_P)$tanh$(\hbar\Omega_P/2k_BT)$, in which $\hbar\Omega_P$ is the characteristic phonon energy and $k_B$ the Boltzmann constant. Monitoring the dynamics of the CT edge allows addressing the dynamics of the decoherence of Mott excitons \cite{Wall2010} (holon-doublon pairs) and clarify the nature of the final state of the system after the photoexcitation \cite{Giannetti2009b}. In system in which the electron-phonon coupling is sufficiently strong, the local excitons created by the light absorption can be dressed by the local deformation of the crystal, creating Frenkel polarons. This process accounts for the temperature-driven redshift of the CT edge measured on La$_2$CuO$_4$ \cite{Fan1951,Falck1992,Matsuda1994,Lovenich2001,Okamoto2010}.

\subsubsection{Optics and Dynamical Mean Field Theory}
\label{sec_DMFToptics}
One of the reasons for the success of Dynamical Mean-Field Theory is the direct access to dynamical properties, including both frequency-dependent single-particle spectra and response functions and time-resolved observables. In Sec. \ref{sec_DMFTscattering} we have discussed the basic picture of strongly correlated electrons emerging from DMFT, in which a quasiparticle peak with Fermi liquid properties is flanked by largely incoherent high-energy Hubbard bands. Here we show the consequent picture of the optical conductivity for an equilibrium correlated system before attacking the non-equilibrium properties. 

The optical conductivity is related to the current-current response function. It can be shown on general grounds that, for single-site DMFT, only the local an frequency-independent component of the vertex function contributes to the response functions \cite{Khurana1990}, which implies that we can compute the real part of the optical conductivity as

\begin{equation}
\label{sigma-dmft}
\sigma(\omega) = \frac{\pi e^2}{\omega} \sum_k \int d\varepsilon [f(\varepsilon - \omega) - f(\varepsilon)]Tr[\rho(k,\omega)v_{k\alpha}\rho(k,\omega-\varepsilon)v_{k\beta}],
\end{equation} 
where the sum over k is extended over the whole Brillouin zone, $f(\omega)$ is the Fermi function, $\rho(k,\omega)= -1/\pi Im G(k,\omega)$ is the interacting single-particle spectral function, $v_{k\alpha}$ is the band velocity along the $\alpha$ direction for fermions with momentum $k$. The trace is taken over the (implicit) band index labelling both the spectral functions and the velocity tensor. For more details on the calculation of the optical conductivity in the framework of electronic structure calculations of correlated materials, we refer to Ref. \cite{Tomczak2009}.

\begin{figure}
\begin{centering}
\includegraphics[width=1.1\columnwidth]{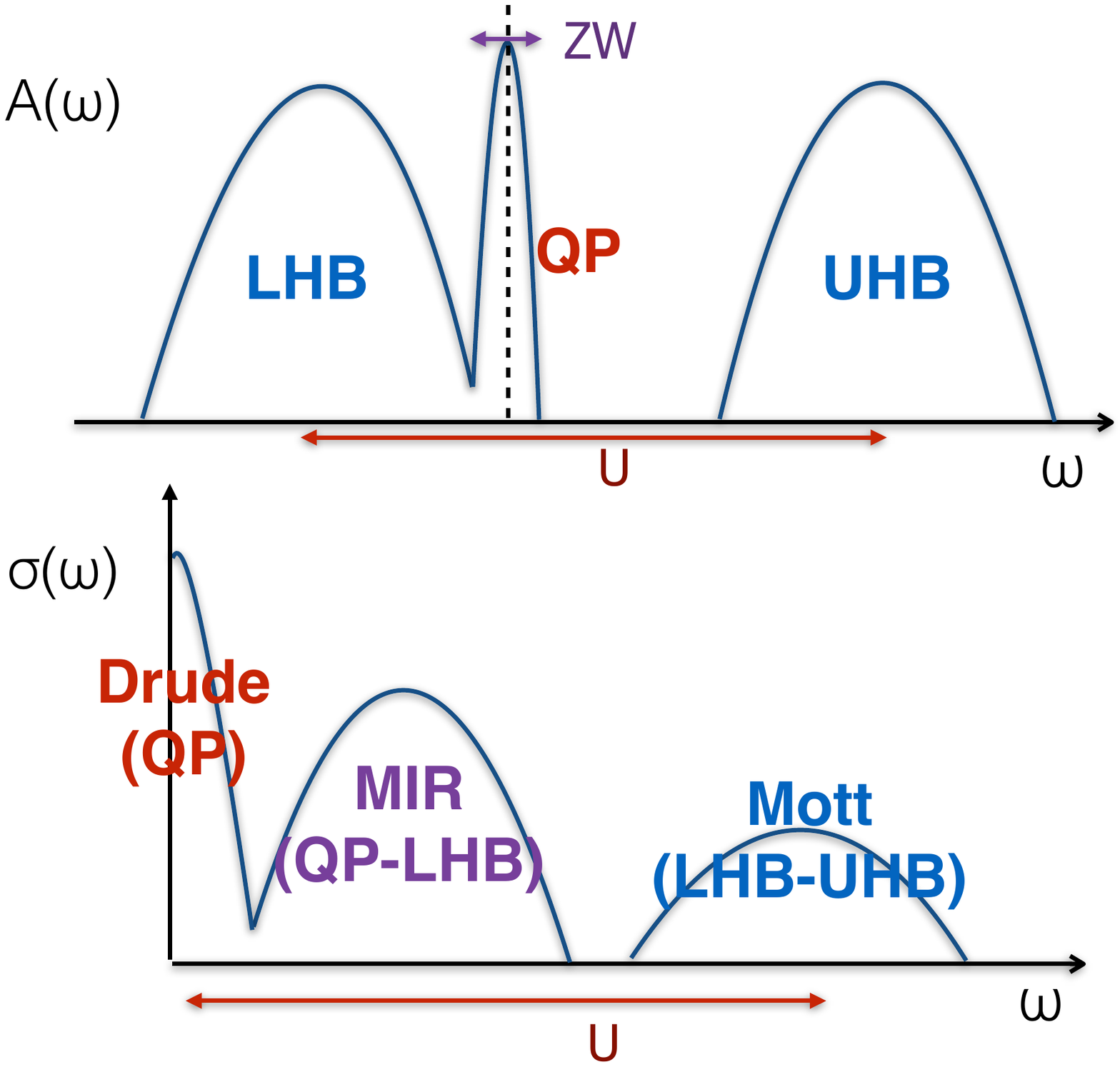}
\caption{\label{fig:optics} Single-particle spectral function $A(\omega)$ and corresponding real part of the optical conductivity $\sigma(\omega)$ for a cartoon doped Mott insulator. Transitions within the narrow quasiparticle peak (QP) of width $ZW$ around the Fermi level (marked by a dashed vertical line) lead to the low-frequency Drude contribution in the optical conductivity. Transitions connecting the lower Hubbard band (LHB) with the quasiparticle peak  lead to the ``MIR" structure in the conductivity, while transition connecting LHB and Upper Hubbard Band (UHB) give rise to the high-energy ``Mott" feature.}
\end{centering}
\end{figure}

For the purpose of the present section we focus however on the basic concepts which stem from Eq. (\ref{sigma-dmft}) and emerged since the first calculations for the single-band Hubbard model\cite{Pruschke1993,Jarrell1995,Rozenberg1995,Rozenberg1996}. As a matter of fact, for a fixed bandstructure, the essential frequency dependence of the optical conductivity is given by the convolution of two spectral functions, which clearly reflects the idea of optical transitions connecting occupied and empty states in the single-particle spectra. Starting from the spectral functions described above, we can simply build a cartoon picture of the optical conductivity for a strongly correlated metal close to a Mott transition. The case of major interest is that of a slightly hole-doped Mott insulator, where the quasiparticle peak is shifted very close to the lower Hubbard band with respect to the half-filled situation, but it retains its separate character, as shown in Fig. \ref{fig:optics}.

The optical conductivity is characterized by three main features: (a) A Drude peak, associated to transition within the quasiparticle peak. This states are essentially coherent, and they give rise to a sharp Drude peak, which can be broadened by interactions of any kind. The weight of the Drude peak is proportional to the quasiparticle weight $Z$; (b) A ``mid-infrared'' peak, which is associated to transitions from the lower Hubbard band to the empty quasiparticle states. As a result, this relatively high-frequency feature is influenced by the low-energy excitations of the system; (c) a very high-energy ($\sim U$) feature which arises from transitions connecting the lower and the upper Hubbard bands. In the case of a charge-transfer insulator, this would coincide with the charge-transfer gap rather than to the Coulomb repulsion.

This simple description represents the backbone of our understanding of optical properties of strongly correlated systems. This picture indeed successfully captures the properties of three-dimensional oxides like V$_2$O$_3$ once the material-dependent aspects are taken into account\cite{Rozenberg1995,Rozenberg1996,Tomczak2009}. The most prominent feature is the evolution of the spectral weight as a function of $U$ and/or the doping, which mirrors the evolution of the quasiparticle peak and the (much less pronounced) evolution of the Hubbard bands in the same process.

While this simple picture remarkably works for three-dimensional materials,  the application of the above picture to the highly anisotropic cuprate superconductors is certainly much more questionable. 
The structure of these materials strongly suggest that the two-dimensional copper-oxygen layers are the main building blocks where the superconducting phenomenon and the other electronic features take place, leading to identify the two-dimensional Hubbard model (or its three-orbital version) as the basic electronic model to describe these materials.
The need to go beyond the DMFT picture with a momentum-independent self-energy is also triggered by the physical properties of the cuprates, starting from the d-wave superconducting phase, which requires a momentum-dependent anomalous self-energy, and the pseudogap regime, in which a clear differentiation between the antinodal and nodal regions takes place. 

If one wants to build on the success of DMFT and include short-ranged correlations, it is therefore necessary to go beyond the "single-site" version of DMFT that we described so far. The most straightforward and popular extensions of DMFT are ``cluster extensions" in which an effective theory is built for a small cluster rather than for a single site\cite{Maier2005a,Kotliar2001,Lichtenstein2000}.

It has been repeatedly shown that even the smallest possible clusters, like a suitably chosen two-site cluster\cite{Ferrero2008,Ferrero2009} or a 2$\times$2 plaquette\cite{Civelli2005,Capone2006,Haule2007,Gull2008,Civelli2008,Kancharla2008,Civelli2009,Sakai2009,Sakai2012,Sordi2012,Sordi2013b} all the main features of the cuprate phase diagram, including antiferromagnetism, d-wave superconductivity and the pseudogap appear and their competition is compatible with the experimental framework. Larger clusters of around 8-16 sites are instead sufficient to make the agreement with experiments stronger and almost quantitative\cite{Maier2005b,Gull2010,Gull2012,Gull2013,Chen2015}.


The present manuscript is not the appropriate forum to discuss all these developments, but we would like to end this session with an important observation. 
Despite the necessity to include at least short-range non-local correlations to properly describe the phase diagram of the cuprates,  even the single-site DMFT with its local self-energy describes accurately the distribution of spectral weight as a function of frequency. In particular, it has been shown that DMFT correctly reproduces both the doping\cite{Comanac2008,Toschi2008} and the temperature\cite{Nicoletti2010} evolution of the integrated optical spectral weight $S = \int_0^{\Omega_c} d\omega \sigma(\omega)$ for different values of the  cut-off $\Omega_c$ chosen in order to include either the pure Drude contribution or the Drude and the mid-infrared structure. 

We stress that this success is not accidental or a minor aspect. The ability of single-site DMFT in capturing the main features of the evolution of the spectral weight  reflects the fact that the separation of the optical spectrum in the three features (Drude, MIR and Mott) described above seems to survive also in low-dimensional materials like the cuprates. 
In this perspective, we can conclude that the inclusion of short-range correlation in cluster methods (or any other suitable extension of DMFT) gives rise essentially to a redistribution of weight {\it within} each main spectral feature, but it does not transfer substantial spectral weight from one feature to the other. In particular, we expect the low-energy manifold in the spectral function to be significantly reorganized in cluster calculations, reflecting in the internal structure of the "Drude" and ``mid-infrared" features. In this sense, we can always view the single-site DMFT picture has the ``backbone" description of the spectral properties of correlated materials even in low dimensions. In the latter case, we have to keep in mind that non-local effects will certainly play a role, but this does not imply a breakdown of the hierarchy of energy scales emergent from DMFT.

\subsection{The non-equilibrium response}
\label{sec_non_eq_response}
In general, the non-equilibrium response of excited materials can be the sum of different contributions, ranging from the plasma
response, the thermomodulation effect, the Fermi surface smearing and the Mathis-Bardeen
(M-B) response in the superconducting state.  

Here we limit the discussion to the case of relatively weak perturbation induced by the photoexcitation, where the photoexcited carrier density $n_{p}$
 is small as compared to the normal
state carrier density $N_{0}$. In this regime, for any specific probed wavelength, it is possible to assume the linear response approximation and
expand the optical constants to the first order in carrier density $n$ or, in the case of phonons, to the displacement coordinate $Q$. In this framework, the temperature- and fluence-dependences of the amplitude and lifetimes are described well by the phenomenological theory presented in the previous sections. However, for longer timescales the thermomodulation response proportional to $d \epsilon/dT$ becomes important. 
In the next subsections we will discuss the transient dynamics of the dielectric function on a very general ground, while some simple examples (e.g. Fermi level smearing) relevant for the results presented in Chapter \ref{sec_experiments} will be provided.
Finally, for a quantitative description of the selection rules related to the polarizations of the pump and probe pulses, we will discuss the transient response in terms of the
stimulated Coherent Raman scattering. 

\subsubsection{Probing the spectral response: dynamics of the dielectric function}

\paragraph*{Fermi surface smearing and interband transitions.}
The possibility of probing the dynamics of the dielectric function of correlated materials over a significantly broad frequency range, represents a breakthrough in the quantitative understanding of the spectrally resolved non equilibrium physics.
The pioneering pump-probe experiments on metals \cite{Brorson1990} and correlated materials were based on the generic assumption of a proportionality between the reflectivity variation at the frequency $\omega$ and the instantaneous "effective" electronic temperature, i.e., $\delta R$/$R$(t)$\propto\delta T_e$/$T_e$(t). In this perspective, the dynamics of the transient reflectivity was reproduced through an effective-temperature model trying to estimate the electron-phonon coupling which ultimately regulates the relaxation dynamics of $T_e$(t) (see Chapter \ref{sec_QPdynamics}). Nonetheless, in many cases the variation of the optical properties is uncorrelated to the effective electronic temperature, whose definition is often questionable, particularly for the sub-ps timescale. 

\begin{figure}[t]
\begin{centering}
\includegraphics[width=1\textwidth]{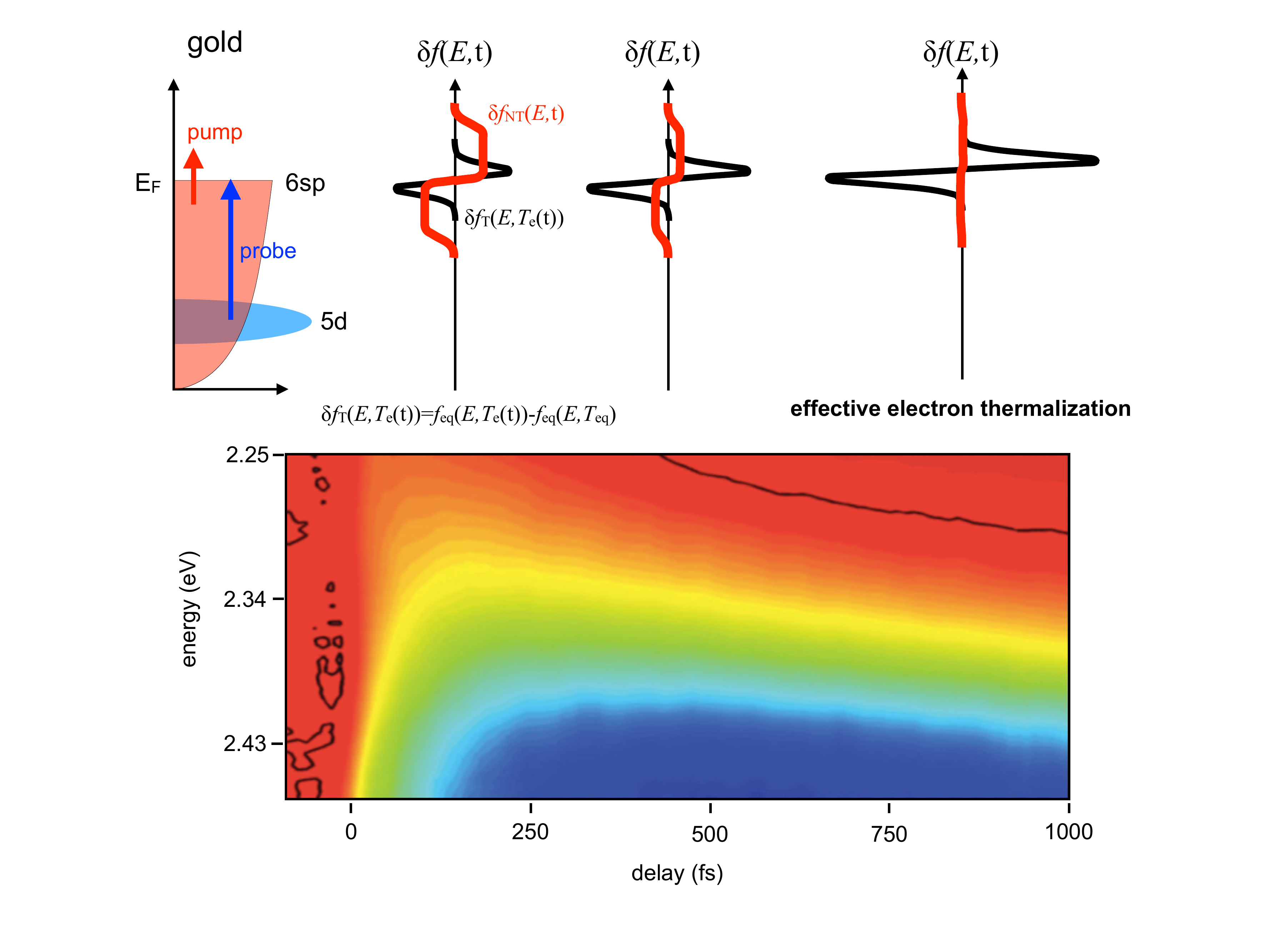}
\caption{Cartoon of the electronic thermalization process in gold. The time-resolved reflectivity experiments in which the 5$d$$-6$$pd$ optical transitions are probed are mostly sensitive to the photoexcited non-thermal distribution $f_{NT}(E,t)$ extending in the energy range $E_F$-$\hbar \omega_{pump}$$-$$E_F$+$\hbar \omega_{pump}$. Taken from \cite{DellaValle2012}}
\label{fig_elthermalizationgold}
\end{centering}
\end{figure}
The simplest example is related to the dynamics of the optical properties in a metal, when a resonant optical transition from a narrow band (e.g. a $d$-band) to the Fermi level is probed, as schematically shown in Fig. \ref{fig_elthermalizationgold}. In this case, the excitation process at $\hbar\omega_{pump}$ creates a step-like non-thermal function, $f_{NT}(E,t)$, which extends from $E_F$-$\hbar\omega_{pump}$ to $E_F$-$\hbar\omega_{pump}$. The relaxation dynamics of $f_{NT}(E,t)$ is regulated by the charge-charge and the charge-phonon interactions, which eventually lead to a thermalized electronic distribution at the effective temperature $T_e(t)$, as discussed in Section \ref{sec_ETM}. Snapshots of this thermalization process can be taken by measuring the dynamics of the complex dielectric function $\delta\epsilon(\omega,t)$. Considering the optical transitions from the $d$-band to the conduction band, we note that $f_{NT}(E,t)$ is expected to increase the phase space available for optical transitions whose final state lies in the ($E_F$-$\hbar\omega_{pump}$)$-E_F$ energy range (see Fig. \ref{fig_elthermalizationgold}), while the photoinduced filling of the bands from $E_F$ to $E_F$+$\hbar\omega_{pump}$ leads to the opposite effect. As a consequence, the absorption of the system is strongly modulated in the ($\omega_0$-$\hbar\omega_{pump}$)$-$($\omega_0$+$\hbar\omega_{pump}$) energy range, where $\omega_0$ is the central frequency of the $d\rightarrow$\textit{conduction band} transition. As far as the $f_{NT}(E,t)$ evolves into a thermal hot distribution $f(E,T_e(t))$, the smearing of the Fermi-Dirac distribution progressively leads to a narrower modification of the optical properties in the 2$k_BT_e$ range across $\omega_0$. The frequency-dependence of the variation of the dielectric function can be easily modelled by assuming \cite{Guerrisi1975,Sun1994} that the absorption change at the probe frequency $\omega$ is given by: 
\begin{equation}
\label{eq_Depsilon_metal}
\delta \epsilon_2(\omega, t)\simeq\frac{1}{(\hbar\omega)^2}\int_{E_{min}}^{E_{max}} D(E, \hbar\omega)\delta f(E,t)dE
\end{equation}
where $D(E,\hbar\omega)$ is the JDOS of the optical transition from $E$-$\hbar\omega$ to $E$ and $\delta f(E,t)$=$f(E,t)$-$f_{eq}(E,T_{eq})$ is the relative change of the electronic occupation with respect to the equilibrium Fermi-Dirac distribution at the temperature $T_{eq}$. While the lower integration limit, $E_{min}$ can be taken as -20$k_BT_e$,  $E_{max}$ depends on the details of the band structure of the material \cite{Guerrisi1975}. Assuming that both the matrix elements of the optical transition and the JDOS remain constant during the excitation and relaxation processes, it possible to model the dynamics of $\delta\epsilon(\omega,t)$ by computing $\epsilon_1$ through the Kramers-Kronig relations and letting $f(E,t)$ evolving within the extended multi-temperature model (see section \ref{sec_ETM}). 

Recent time-resolved measurements with sub-20 fs resolution \cite{DellaValle2012} mapped the dynamics of the electron thermalization process in gold, demonstrating the formation of a fully-thermalized electron distribution on a timescale of the order of 500 fs. The onset of a thermal electron distribution is bottlenecked by the Fermi-liquid behaviour of the low-energy quasiparticles, whose lifetime diverges when $E\rightarrow E_F$ leading to a long-lived low-energy tail in $\delta f(E,t)$. As a further consequence, a significative amount of the energy stored in the electron gas can be exchange with the phonons (or other bosonic fluctuations) before the complete thermalization of the electrons, as pointed out in Ref. \cite{Baranov2014}, making the applicability of the effective temperature model a quite elusive problem.

\paragraph*{Fermi level smearing and transient Drude response.}

In the normal state, the transient change of the electronic occupation, $\delta f(E,t)$, can also lead to important variations of the low-energy optical properties, usually described by the (extended) Drude model. We report here two simple examples, that show how it is possible to relate the complex changes of the intraband optical properties to the variation of the parameters which enter in the equilibrium dielectric function.

\begin{figure}
\begin{centering}
\includegraphics[width=1\columnwidth]{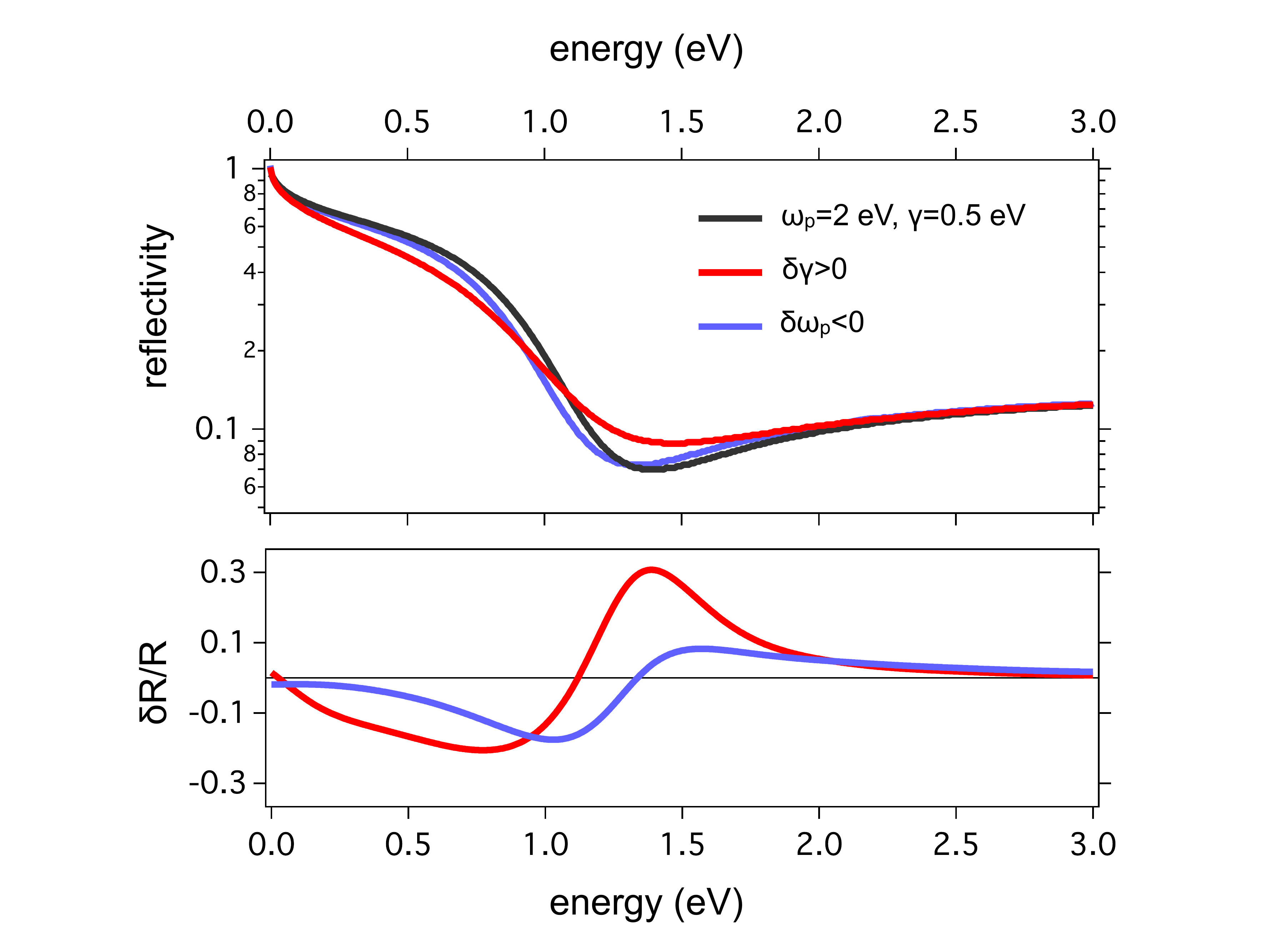}
\caption{\label{fig_plasma-response}The reflectivity (top panel)  and differential
reflectivity $\delta R/R$ (bottom panel) calculated within the simple Drude model. The black line represent the reflectivity for a Drude model with parameters similar to those typical of doped copper oxides: $\omega_{pl}$=2.5 eV and $\gamma$=0.5 eV. The  contribution of the interband transitions is accounted for by setting $\epsilon_{\infty}$=5. The blue line represents the reflectivity variation in the case of a photoinduced decrease of the plasma frequency ($\delta\omega_{pl}\sim$5\%). The red line is the reflectivity variation calculated in the case of an increase of the scattering rate ($\delta\gamma\sim$50\%). The possible photoinduced variations of $\omega_{pl}$ and $\gamma$ have been exaggerated for graphical reasons.}
\end{centering}
\end{figure}

The first simple case is related to a transient change of the density of the free carriers in the conduction band. This can be achieved, for example, by strongly and resonantly pumping a transition from states near $E_{F}$ to high-energy empty bands, whose decay is slower than the temporal resolution of the experiment. The excitation in which the effective number of carriers in the conduction band, $n_q(t)$, is time dependent can lead to a transient change of the
plasma frequency $\omega_{pl}(t)=\sqrt{n_q(t)e^{2}/\epsilon_{0}m_q}$,
which in turn induces a transient change of the reflectivity by the free-carrier plasma. This kind of response is shown in Fig. \ref{fig_plasma-response} (blue line),
for the case in which only the plasma frequency changes. For simplicity we have assumed a simple Drude model, i.e., $\epsilon$=$\epsilon_{\infty}$-$\omega_{pl}^2/(\omega^2$+$\omega\gamma)$ with the parameters typical of doped copper oxides (plasma frequency: $\omega_{pl} \sim $2.5 eV; scattering rate: $\gamma \sim$0.5 eV). The presence of high-energy interband transitions, that are intrinsic to any realistic model, is simulated by using $\epsilon_{\infty}$=5 in the model dielectric function.

The second example is the photoinduced change of the scattering rate of the charge carriers in the conduction band. The reflectivity variation induced by the increase of the electronic scattering rate is shown in Fig. \ref{fig_plasma-response} (red line). This case can represent the simple picture in which the energy released by the pump excitation to the charge carriers is rapidly dissipated into the bosonic bath, whose effective temperature rapidly increases and leads to the rise of the electron-boson scattering rate (see Sec. \ref{sec_ebosresults}). An alternative scenario, implying the photoinduced variation of $\gamma$, is given when carriers have strongly anisotropic
scattering rates, e.g. the nodal-antinodal dichotomy in copper oxides. 
After the photoexcitation, the underlying \textbf{k}-dependent distribution of the thermal excitations can be strongly perturbed and the total scattering rate can be modified as a result of the non-thermal distribution of excitations in the \textbf{k}-space (see Sec. \ref{sec_PGdynamics}). 

While in both cases ($\delta\omega_{pl}<$0 and $\delta\gamma>$0) the non-equilibrium response
is relevant for probe frequencies $\omega\sim\omega_{pl}$, the frequency-dependent relative reflectivity variation, i.e., $\delta R(\omega)/R$, contains the fingerprint of the two phenomena. These two cases constitute the simplest example in which two different physical scenarios can be disentangled by probing the reflectivity variation over a very broad energy range.

\paragraph*{The differential dielectric function.}
Considering the "local" relation between the reflectivity $R(\omega)$ and $\epsilon(\omega)$, the assumption $\delta R$/$R$(t)$\propto\delta T_e$/$T_e$(t) can be invalidated, even in the simple case of weakly interacting electrons in a metallic system. However, the rapid growing of the broadband time-resolved optical spectroscopies has paved the road to the development of quantitative models for the electron dynamics in solids, allowing to address important questions concerning non-equlibrium optical experiments. The foundation of the \textit{differential dielectric function} approach is grounded on the knowledge of the equilibrium dielectric function ($\epsilon_{eq}(\omega)$;$\eta_i$), that is a functional of a number of parameters $\eta_i$ (or functions if they depend on the energy $E$). Starting from this information, it is necessary to identify the smallest subset of parameters necessary to reproduce the measured $\delta\epsilon(\omega,t)$, which can be expressed as:
\begin{equation}
\label{eq_Depsilon}
\delta \epsilon(\omega, t)=\sum_i\frac{\partial \epsilon(\omega; \eta_i)}{\partial \eta_i}\delta \eta_i(E,t)
\end{equation}
As a consequence, the dynamics of the dielectric function is mapped into the dynamics of the parameters $\eta_i(E,t)$ that can provide valuable informations about the relevant physics. Here below we list some physical parameters that are of relevance for the dynamics of correlated materials, along with the physics disclosed by their dynamics:

\begin{enumerate}
\item[i)] \textbf{non-thermal carrier distribution $f(E,t)$} $\rightarrow$ dynamics of the non-thermal distribution of the charge carriers photoexcited by the pump pulse;
\item[ii)] \textbf{plasma frequency $\omega^2_{pl}$} $\rightarrow$ transient modification of the plasma frequency as a consequence of either a change of the carriers density or an ultrafast spectral weight shift;
\item[iii)] \textbf{density of states $N(E,t)$} $\rightarrow$ change of the density of state at the Fermi level;
\item[iv)] \textbf{optical scattering rate $\gamma(\omega,t)$} $\rightarrow$ change of the electron-boson scattering rate as a consequence of the change of either the electron-boson coupling constant or the boson density;
\item[v)] \textbf{optical spectral weight} SW($\Omega_c,t$) $\rightarrow$ ultrafast modification of the spectral weight, energy and damping of interband transitions;
\item[vi)] \textbf{infrared-active modes with frequency $\omega_i$, damping $\Gamma_i$ and oscillator strength $\omega^2_i$} $\rightarrow$ changes of the intensity and shape of infrared-active modes as a consequence of the energy transfer to the lattice;
\item[vii)] \textbf{condensate density $n_{SC}(t)$} $\rightarrow$ transient modification of the inductive response of the superconducting condensate;
\end{enumerate}

The time-domain THz techniques \cite{Neshat2012,Morris2012} provide direct access to the complex optical functions, i.e., the dielectric function $\epsilon(\omega)$, the optical conductivity $\sigma(\omega)$=i/4$\pi$($\epsilon(\omega)$-1) and the refraction index $n(\omega)$=$\sqrt{\epsilon(\omega)}$. Conversely, the non-equilibrium optical spectroscopies in the infrared/visible range usually probe the dynamics of the reflectivity (or transmissivity) that is, at normal incidence, related to the dielectric function by:
\begin{equation}
R(\omega,T)=\left| \frac{1-\sqrt{\epsilon(\omega,T)}}{1+\sqrt{\epsilon(\omega,T)}}\right|^2
\end{equation}

To overcome the limitation in directly probing the complex optical properties, a spectroscopic ellipsometry can be carried out for each probe wavelength by performing measurements at different incident angles or polarizations of the probe beam \cite{Roeser2003} and by numerically inverting the Fresnel relations. However, performing a real time-resolved ellipsometry over a broad frequency range is usually experimentally demanding. More frequently, the optical properties are reconstructed through measurements on thin films, in which the transmissivity variation is directly related to imaginary part of the refraction index, i.e., $T(\omega)$=(1-$R$)exp(-2$\omega n_2(\omega)/c$), when Fabry-Perot internal reflections are neglected. In the case of broadband reflectivity measurements,  the complex dielectric function can be extracted either by performing a differential dielectric function analysis starting from a model of the dielectric function which is intrinsically Kramers-Kronig constraint or by directly performing KK transformations on the $\delta R(\omega,t)$ signal constraining $\delta R(\omega,t)$=0 in the range outside the probe frequency window. This last method, recently introduced \cite{Novelli2014}, basically assumes that the variations of the pump-perturbed reflectivity outside of the probed region are either negligible or too distant from the probed energy scale to significantly change the local optical response.

\subsubsection{The optical response of a superconductor}

\paragraph*{Strong photoexcitation.}
The Mattis-Bardeen (M-B) formula \cite{Tinkham} gives the ratio between the superconducting
and normal state optical conductivity $\sigma_{s}/\sigma_{n}$, (
$=\alpha_{s}/\alpha_{n}$) calculated within a single band model.
While it is large for $\omega\simeq\Delta$, it diminishes with frequency
as $\omega^{-2}$. The amplitude \emph{A}$_{s}$, of the induced change
in reflectivity in the regime of condensate vaporization, is given
by: 
\begin{equation}
\emph{A}_{s}=\left\vert \frac{\delta R}{R}\right\vert _{\mathcal{F}>\mathcal{F}_{T}}=\frac{R_{n}-R_{s}}{R_{s}}\simeq\frac{R_{n}-R_{s}}{R_{n}}\text{\ }
\end{equation}
where $R_{n}$ and $R_{s}$ are the reflectivities in the normal and
in the superconducting states respectively, and $R_{n},R_{s}\gg R_{n}-R_{s}$.
At optical frequencies, the induced change in reflectivity is mainly affected by the imaginary component of the refractive
index, hence proportional to the induced change in the real
part of the optical conductivity $\frac{\delta R}{R}\varpropto\frac{\delta k}{k}\varpropto\frac{\delta\varepsilon_{2}}{\varepsilon_{2}}\varpropto\frac{\delta\sigma_{1}}{\sigma_{1}}$
giving 
\begin{equation}
\emph{A}_{s}=\left\vert \frac{\delta R}{R}\right\vert _{\mathcal{F}>\mathcal{F}_{T}}\varpropto\frac{\sigma_{1}^{n}-\sigma_{1}^{s}}{\sigma_{1}^{n}}\text{\ }
\end{equation}
To determine the temperature dependence of $\emph{A}_{s}$, the ratio 
$\frac{\sigma_{1}^{s}}{\sigma_{1}^{n}}(T)$ is evaluated, being $\frac{\sigma_{1}^{s}}{\sigma_{1}^{n}}$ given by the Mattis-Bardeen relation \cite{Mattis1958} 
\begin{multline}
\frac{\sigma_{1}^{s}}{\sigma_{1}^{n}}\left(\hbar\omega\right)=\frac{2}{\hbar\omega}{\displaystyle \int\nolimits _{\Delta}^{\infty}}\left(f\left(\varepsilon\right)-f\left(\varepsilon+\hbar\omega\right)\right)g\left(\varepsilon\right)d\varepsilon\text{\ }\\
+\frac{1}{\hbar\omega}{\displaystyle \int\nolimits _{\Delta-\hbar\omega}^{-\Delta}}\left(1-2f\left(\varepsilon+\hbar\omega\right)\right)g\left(\varepsilon\right)d\varepsilon\label{MatBarEq}
\end{multline}
Here $f\left(\varepsilon\right)$ is the Fermi-Dirac distribution
function, $\hbar\omega$ is the photon energy, $\Delta$ is the superconducting
gap, and $g\left(\varepsilon\right)$ is 
\begin{equation}
g\left(\varepsilon\right)=\frac{\varepsilon\left(\varepsilon+\hbar\omega\right)+\Delta^{2}}{\sqrt{\varepsilon^{2}-\Delta^{2}}\sqrt{\left(\varepsilon+\hbar\omega\right)^{2}-\Delta^{2}}}\text{\ }
\end{equation}
In the limit of $\Delta\ll\hbar\omega$ \ Eq. \ref{MatBarEq} becomes: 
\begin{multline}
\frac{\sigma_{1}^{s}}{\sigma_{1}^{n}}\left(\hbar\omega\right)\simeq\frac{1}{\hbar\omega}{\displaystyle \int\nolimits _{\Delta}^{\hbar\omega-\Delta}}g\left(\varepsilon-\hbar\omega\right)d\varepsilon\text{\ }\\
+\frac{2}{\hbar\omega}{\displaystyle \int\nolimits _{\Delta}^{\infty}}f\left(\varepsilon\right)\left(g\left(\varepsilon\right)-g\left(\varepsilon-\hbar\omega\right)\right)d\varepsilon\label{I1I2}
\end{multline}
For $2\Delta\ll\hbar\omega$,
the integral \ref{I1I2} can be used to obtain $\emph{A}_{s}(T)$ \cite{Mattis1958,Kusar:2008p5434}:
\begin{equation}
\emph{A}_{s}(T)\varpropto\frac{\Delta(T)^{2}}{(\hbar\omega)^{2}}\ln\left(\frac{4\hbar\omega}{e\Delta(T)}\right)\text{\ .}\label{EqMatissBaardenResult}
\end{equation}
Eq. \ref{EqMatissBaardenResult} can be compared with the temperature
dependence of $\delta R/R$, when a strong P pulse completely destroys the condensate. Note that the response is proportional to $\omega^{-2}$. This fact
can be experimentally tested. 

\paragraph*{Weak photoexcitation.}
If the perturbing pulse is too weak to destroy the ordered state, the derivative
change of reflectivity $\delta R$ with respect to $\Delta$ must be calculated. Using
the two-fluid model, we obtain that $\Delta^{2}=\Delta_{0}^{2}n_{s}=\Delta_{0}^{2}(1-n_{q})$,
where $n_{s}$ is the superfluid density, $n_{q}$ is the quasiparticle
density, and $\Delta_{0}$ is the gap at $T=0$. $\delta R$
can be related to the photoexcited carrier density $n_{p}$ using the fact that $\delta n_{q}=n_{p}$:
\begin{equation}
\delta R=\frac{\partial R}{\partial\Delta}\frac{\partial\Delta}{\partial n_{q}}\delta n_{q}=\frac{A\Delta_{0}^{2}}{2(\hbar\omega)^{2}}\left(1-2ln\left[\frac{1.47\hbar\omega}{\Delta}\right]\right)n_{p}\label{eq:transient reflectivity response}
\end{equation}
If we neglect the derivative of the logarithmic correction with respect
to $\Delta$, we can substitute $\Delta_{0}$ for $\Delta(t$) in the
logarithm and obtain: 
\begin{equation}
R_{s}(t)-R_{n}\simeq A\frac{\Delta\left(t\right){}^{2}}{(\hbar\omega)^{2}}ln\left[\frac{1.47\hbar\omega}{\Delta_{0}}\right]
\end{equation}
Eq. \ref{eq:transient reflectivity response} then further simplifies
to: 
\begin{equation}
\delta R=\frac{\partial R}{\partial\Delta}\frac{\partial\Delta}{\partial n_{q}}\delta n_{q}=\frac{2A\text{\ensuremath{\Delta(t)}}}{(\hbar\omega)^{2}}\delta\Delta\propto-\frac{\text{\ensuremath{\Delta_{0}^{2}}}}{(\hbar\omega)^{2}}n_{p}.\label{eq:simple response}
\end{equation}
This is an intuitive result, which is consistent with an excited
state absorption probe process. This response function,
just as for strong excitation, does not consider resonant processes
and diminishes at higher energies as $\omega^{-2}$.

Under conditions of weak excitation often encountered experimentally
in superconductors, the QPs accumulate at the bandgap edge forming
a bottleneck. Under such conditions, the QPs and phonons are in near
equilibrium, and $n_{p}$ is given by: 
\begin{equation}
n_{p}=\frac{1/(\Delta(T))+T/2)}{1+B\sqrt{2T/(\Delta(T)}\text{Exp}[-\Delta(T)/T]}.\label{eq:epsilon in terms of psi}
\end{equation}
where $\Delta$ is the superconducting gap, $B=\nu/N(0)\hbar\Omega_{c}$,
where $\nu$ is the effective number of phonon modes of frequency
$\hbar\Omega_{c}$ per unit cell participating in the relaxation process.
$B$ can be determined by fitting the temperature dependence of $\delta R/R$
and $N(0)$ is the density of electronic states at the Fermi energy.

\subsubsection{Description of the non-equilibrium experiments in terms of stimulated Raman scattering}
\label{sec_SRStensor}

A vast literature \cite{Mukamel,Merlin1997,Stevens:2002p6798} showed that P-p experiments can be described as a two step stimulated Raman scattering (SRS) process, which applies not only to lattice vibrations, but also to electronic \cite{Mansart2013b,Toda:2014ga} and magnetic excitations. This picture is valid in the limit in which the excitation is shorter than the typical relaxation time of the excited bosonic mode. In this case, the first order in the expansion of the kinetic equations for the electronic and bosonic distributions \cite{Rossi2002} can be approximately considered as the leading one. This model is very useful, since it makes possible to compare quantitatively the polarisation selection rules using the Raman tensor, and intensities of the P-p measurements with theoretically calculated resonant Raman scattering cross-sections. Moreover, it allows detailed analysis of broken symmetries on short timescales. In contrast to ordinary stimulated Raman scattering the process is coherent, so one can extract lifetime information using P-p spectroscopy. The beauty of this approach is that once the dielectric function is known, either experimentally or theoretically, the P-p response can be calculated.

 In the \textbf{first step}, of the pump-probe process, the \emph{pump}
pulse excitation can proceed either by impulsive excitation (IE) or displacive excitation (DE) \cite{Merlin1997,Stevens:2002p6798}. The  two typically apply to non-resonant and resonant conditions respectively. 
In the DE process, photo-excited carriers scatter among themselves
and rapidly release their energy to the lattice within tens of fs,
resulting in a transient non-equilibrium population of electrons, phonons,
magnons, etc. All memory of the initial electron (or hole) momenta
is lost in the process. If the excitation is not too intense, the
process can be considered as a weak coherent perturbation of the
QP density $n_{q}$ near the Fermi level which does not cause a significant
renormalisation of the electronic structure. $n_{q}$ can linearly
couple (only) to symmetric $A_{g}$ excitations, whereas the appropriate Raman tensor $\pi^{R}$ \cite{Stevens:2002p6798,Merlin1997} governs all the excitations (phonons, electronic states, etc.)
in the IE process. In particular, for the pseudo-tetragonal ($D_{4h}$) symmetry, usually
considered as appropriate for the parent structure in cuprates, $A_{1g}$, $A_{2g}$ as well as
$B_{1g}$ and $B_{2g}$ modes can be excited for
polarisations in the $a-b$ plane. 

\begin{figure}[t]
\begin{centering}
\includegraphics[width=100mm]{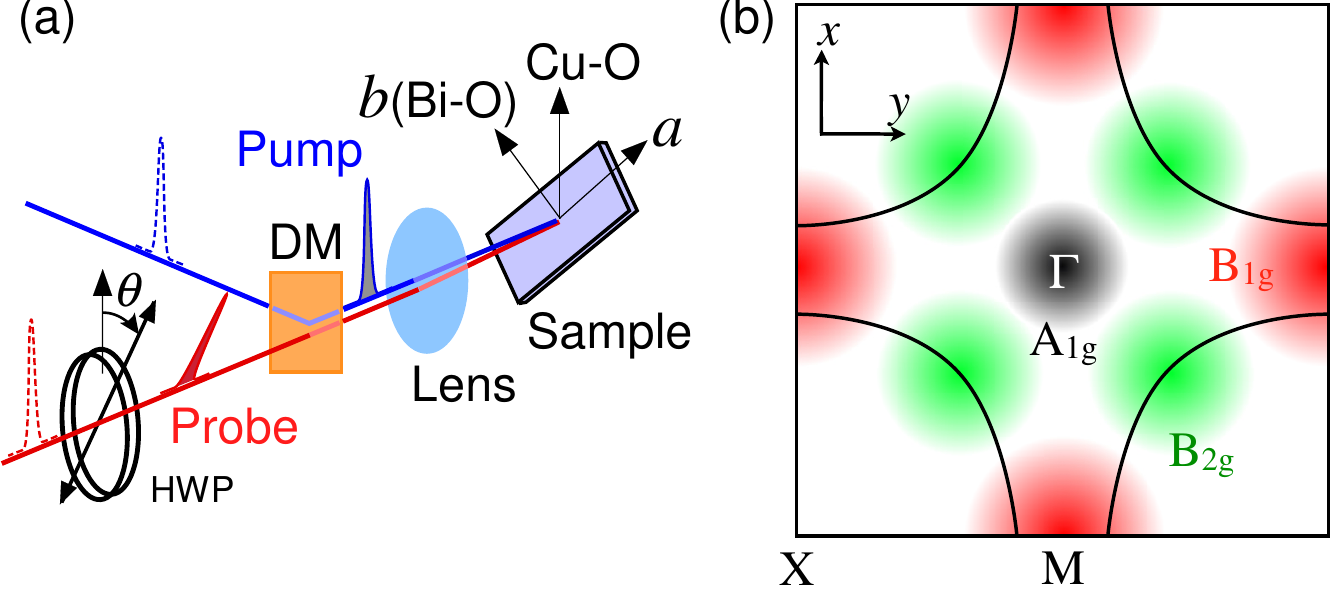}
\caption{ (a) A schematic illustration of the two-color pump-probe setup for polarization-resolved measurements. The probe polarization can be controlled by a half-wave plate (HWP) and is recombined with the pump by a dichroic mirror (DM). The angle $\theta$ is measured relative to the Cu-O bond axes. (b) The \textbf{k}-space selectivity of the probe, according to the Raman-like selection rules is indicated by the colours: A$_{1g}$: black, B$_{1g}$: red; B$_{2g}$: green.}
\label{fig_temp} 
\end{centering}
\end{figure}

In the \textbf{second step}, i.e. the \emph{probe} experiment, the detection of the
transient change of reflectivity can also be described
by a SRS process \cite{Merlin1997,Stevens:2002p6798}. 
The IE and DE components can be distinguished by the phase of the response at $t=0$, since $R^{IE} \propto $sin$(\omega(t-t_0))$, while $R^{DE}\propto $cos$(\omega(t-t_0))$. The coupling of the photon with the crystal excitations (for either the pump or the probe) can be written as: 
\begin{equation}
P_{k}=\Sigma_{l}\chi_{kl}E_{l}+P_{k}^{R}
\end{equation}
The SRS polarisation $P_{k}$
due to a pump-induced change of $n_{q}$ is given by: 
\begin{equation}
P_{k}^{R}=\sum_{l}\mathcal{R}_{kl}E_{l}n_{q},
\end{equation}
where $\mathcal{R}_{kl}=\frac{\partial\epsilon_{kl}}{\partial n_{q}}$
is the Raman tensor, $\epsilon_{kl}$ is the complex dielectric tensor,
and $E_{l}$ is the $l$-th component of the electric field. 

For DE and IE the frequency dependence of the Raman tensor components in terms of the complex dielectric constant are:
\begin{equation}
\xi^{DE}(\omega,\omega+\Omega)\propto [\frac{dRe(\epsilon)}{d\omega}+i\frac{d Im(\epsilon)}{d\omega}]
\end{equation}
and 
\begin{equation}
\xi^{IE}(\omega+\Omega,\omega)\propto [\frac{d \text{Re}(\epsilon)}{d\omega}+2i \text{Im}(\epsilon)/\Omega]
\end{equation}
respectively. Note that the DE response is proportional to the derivative,  $\frac{dIm(\epsilon)}{d\omega}$, while the IE response is proportional to $Im(\epsilon)$. Thus, by measuring the frequency dependence of the P-p response it is possible determine the excitation mechanism. A nice example of such an effect of resonance is  demonstrated by the derivative line shape in the photoinduced transmission experiments of Okamoto et al \cite{Okamoto2011}, signifying a DE mechanism associated with the Cu-O charge transfer excitation around 1.5 eV. 

The case of Bi2212 gives a good example for the analysis of the polarisation response associated with  SRS. Assuming a pseudo-tetragonal structure ($D_{4h}$ point group) for
BiSCO for example, the photoinduced changes of the in-plane dielectric
tensor components can be decomposed, according to symmetry, as: 
\begin{equation}
\delta\mathbf{\epsilon}=\left[
\begin{array}{cc}
\delta\epsilon^{\mathrm{A}_{1\mathrm{g}}}\\
 & \delta\epsilon^{\mathrm{A}_{1\mathrm{g}}}
\end{array}
\right]+\left[\begin{array}{cc}
\delta\epsilon^{\mathrm{B}_{1\mathrm{g}}}\\
 & -\delta\epsilon^{\mathrm{B}_{1\mathrm{g}}}
\end{array}\right]+\left[\begin{array}{cc}
 & \delta\epsilon^{\mathrm{B}_{2\mathrm{g}}}\\
\delta\epsilon^{\mathrm{B}_{2\mathrm{g}}}
\end{array}\right].
\end{equation}
Here we have omitted the $A_{2g}$ symmetry term, because the pseudovector
$A_{2g}$ symmetry components do not appear in low-energy spectra
\cite{Devereaux:2007tg}. However, they may appear at high energies associated
with either $d_{x^{2}-y^{2}}^{9}-d_{xy}^{9}$ intra-Cu transitions
or in $d_{x^{2}-y^{2}}^{9}-p_{\sigma}$ charge transfer transitions \cite{Liu:1993p4053}. 

Considering that the reflected light intensity is proportional to
$P_{k}P_{k}^{*}$, and taking the incident electric field (only the
in-plane components) as $E=E_{0}\bigl(\begin{smallmatrix}\cos\theta\\
\sin\theta
\end{smallmatrix}\bigr)$, then to lowest order the angle-dependence of the photoinduced change
of reflectivity $R$ reduces to:
\begin{eqnarray}
\delta R(\theta) & =\partial R/\partial\epsilon_{1} & \left[\delta\epsilon_{1}^{\mathrm{A}_{1\mathrm{g}}}+\delta\epsilon_{1}^{\mathrm{B}_{1\mathrm{g}}}\cos(2\theta)+\delta\epsilon{}_{1}^{\mathrm{B}_{2\mathrm{g}}}\sin(2\theta)\right]+\nonumber \\
 & +\partial R/\partial\epsilon_{2} & \left[\delta\epsilon_{2}^{\mathrm{A}_{1\mathrm{g}}}+\delta\epsilon_{2}^{\mathrm{B}_{1\mathrm{g}}}\cos(2\theta)+\delta\epsilon{}_{2}^{\mathrm{B}_{2\mathrm{g}}}\sin(2\theta)\right].
\end{eqnarray}
Here $\epsilon=\epsilon_{1}+i\epsilon_{2}$, where $\epsilon_{1}$
and $\epsilon_{2}$ are the real and imaginary parts of the dielectric
constant, respectively. Assuming a linear response $\delta\epsilon_{i}\simeq\delta R_{i}$,
the angle-dependence of the transient reflectivity results:
\begin{equation}
\label{eq_polanisotropy}
\delta R(\theta)\propto\delta R_{\mathrm{A}_{1\mathrm{g}}}+\delta R_{\mathrm{B}_{1\mathrm{g}}}\cos(2\theta)+\delta R{}_{\mathrm{B}_{2\mathrm{g}}}\sin(2\theta).
\end{equation}
Thus by measuring the angle dependence of $\Delta(R)$, we can obtain
the relative magnitudes of the $A_{1g}$, $B_{1g}$ and $B_{2g}$
components. The expression in Eq. \ref{eq_polanisotropy}, together with the temperature dependence and relaxation
time measurements, allows to extract the different components
of the response associated with specific symmetries and, consequently,
to identify the states involved\footnote{This analysis for the probe process applies for the case where the
system does not relax (e.g. by Frank-Condon relaxation) before emitting
the reflected photon, and the perturbation should be weak, which is
easily fulfilled.%
}. 

Although this response was originally derived for collective modes (phonons), it may also hold for electronic excitations, provided the electronic bandwidth is not too large. Therefore, it may apply to superconducting gap excitations,  charge-density-wave gap excitations, single-ion electronic levels in insulators etc. but not for wide band semiconductors, where higher order corrections would need to be included \cite{Rossi2002}. 
Pump-probe experiments and electronic Raman scattering can be thus be viewed as complementary spectroscopies in time domain and spectral domain respectively. Previous analysis of electronic Raman scattering on Bi-based cuprates
have shown that B$_{{\rm 2g}}$ symmetry probes excitations in the
nodal ($\pi$/2,$\pi$/2) direction in $k$-space, while the B$_{{\rm 1g}}$
probes excitations in the antinodal directions ($\pi$/2,0) and (0,$\pi$/2)
\cite{Devereaux:2007tg,Sugai:1989jya,Nemetschek:1997er}. Recent studies suggest a PG in
the nodal direction to have $s$-wave symmetry, in contrast to the
common assumption of a pure $d$-wave PG \cite{Sakai:22tRGgJS}, indicating
that the PG symmetry is still an open issue. A rather similar debate
exists for the SC state symmetry, which entertains the possibility
of different order parameter symmetry on the surface and in the bulk
as indicated by different experimental probes  \cite{BussmannHolder:2007dpa}.
 
\setcounter{section}{4}
\section{Non-equilibrium spectroscopies of high-temperature superconductors and correlated materials}
\label{sec_experiments}
The first pioneering age of pump-probe experiments was characterized by the effort of understanding whether the out-of-equilibrium properties of superconductors and correlated materials could provide a novel piece of information relevant for the general understanding of these materials. This nascent field, mostly based on single-colour experiments, soon brought to light a wealth of interesting phenomena emerging from the possibility of investigating the relaxation processes on a timescale (sub-picosecond) faster than the time necessary to achieve transient quasi-thermodynamic states. 

In this chapter we will provide an overview of the main experimental results ranging from the first single-color experiments to the use of more advanced ultrafast techniques. The aim is to follow the historical path and to collect the main results so far achieved and leading to the foundation of the field. Even though the first experimental outcomes could appear as scattered and unrelated attempts, they led to development of the conceptual framework (see Chapter \ref{sec_basicconceptsNES}) that is still at the base of most of the recent advances (see Section \ref{sec_results}). 

\label{sec_history}
\subsection{Single-color pump-probe in the near-IR}
\subsubsection{Cuprate superconductors}
\label{sec_singlecolour}
Since the early times of ultrafast spectroscopies, single-colour P-p experiments were performed on almost all the families of copper oxides and on many different correlated materials. Only within the cuprate class of high-temperature superconductors, the ultrafast dynamics was measured on the Y-
\cite{Han1990,Chekalin1991,Albrecht1992,Gong1993,Thomas:1996wc,Stevens1997,Kabanov1999,Demsar:1999p112,Demsar:1999wh,Misochko2002,Dvorsek:2002p82,Segre:2002p9832,Gedik:2003p2289,Gedik2004,Luo2003,Luo2006,Gadermaier2010}, La- \cite{Bianchi:2003kj,Schneider2003,Bianchi:2005p3988,Bianchi:2005p3989,Kusar:2005p4778,Kusar:2008p5434,Kusar:2010hz,Gadermaier2010},
Bi- \cite{Gay1999,Smith2000,Murakami2002,Schneider2003,Gedik2005,Liu:2008p4917,Giannetti2009b,Coslovich2011,Toda:2011cs,Coslovich2013,Chia2013}, Tl- \cite{Smith:1999p4056,Easley1990,Chia:2007p1669} and Hg-based \cite{Demsar2000,Demsar:2001p1518} families, as well as on the electron-doped (La/Nd)$_{2-x}$Ce$_{x}$CuO$_{4}$ (LCCO, NCCO) copper oxides \cite{Liu:1993p4053,Cao:2008p1505,Hinton2013b}. Since detailed
systematic studies were performed mainly on YBa$_{2}$Cu$_{3}$O$_{7-\delta}$
 (YBCO), La$_{2-x}$Sr$_{x}$CuO$_{4}$ (LSCO) and Bi$_{2}$Sr$_{2}$CaCu$_{2}$O$_{8+\delta}$ (Bi2212), in this section we will mainly focus on these cuprate families, referring to other materials when illustrating a particular point.

\begin{figure}
\begin{center}
\includegraphics[width=5cm]{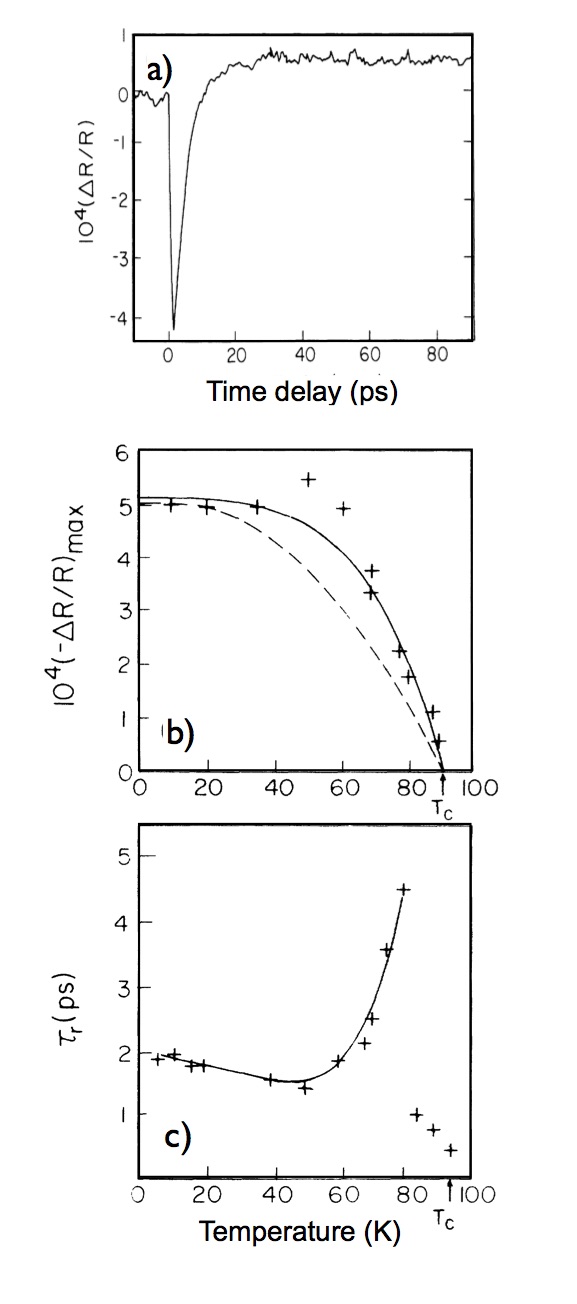}
\caption{\label{fig:Han}Transient reflectivity measurements on YBa$_{2}$Cu$_{3}$O$_{7-\delta}$. a) Typical $\delta R/R(t)$ time trace at $T<T_c$. b) Temperature dependence of the maximum amplitude of the $\delta R/R(t)$ signal. c) Divergence of the relaxation time, $\tau_r$, as $T_c$ is approached. Taken from Ref. \citenum{HAN:1990p1246}}
\end{center}
\end{figure}

\paragraph*{Y-based cuprate superconductors.}
\label{sec_Y-cuprates}
The pioneering work on single particle relaxation in cuprates was
performed on YBa$_{2}$Cu$_{3}$O$_{7-\delta}$ (YBCO) and its related compounds.
Early experiments by Han et al. \cite{HAN:1990p1246} using dye laser
photoexcitation (at 625 nm) on thin film samples of near-optimally
doped YBa$_{2}$Cu$_{3}$O$_{7-\delta}$ detected the appearance of
a transient reflectivity below the superconducting critical temperature
that was unambiguously related to QP dynamics in the superconducting
state (Fig. \ref{fig:Han}). By checking the QP relaxation dynamics
as a function of excitation intensity at 20 K they verified that the dynamics
was consistent with the Rothwarf-Taylor model, (see discussion in sec.
\ref{sec_RT}), in which the gap-energy bosons emitted during the recombination of quasiparticles into Cooper pairs act as a bottleneck for the recombination dynamics. The transient change of reflectivity
was interpreted in terms of a transient change of dielectric constant due to the change in the Drude response arising
from photoinduced QPs. However, the (negative) sign of the QP response compared
to the positive sign for the thermal signal was identified as a potential
problem. The authors also reported a divergence
of the QP lifetime just below $T_{c}$, shown in Fig. \ref{fig:Han}
c), consistent with the closure of the superconducting gap at $T_{c}$. 

\begin{figure}
\begin{center}
\includegraphics[width=0.6\columnwidth]{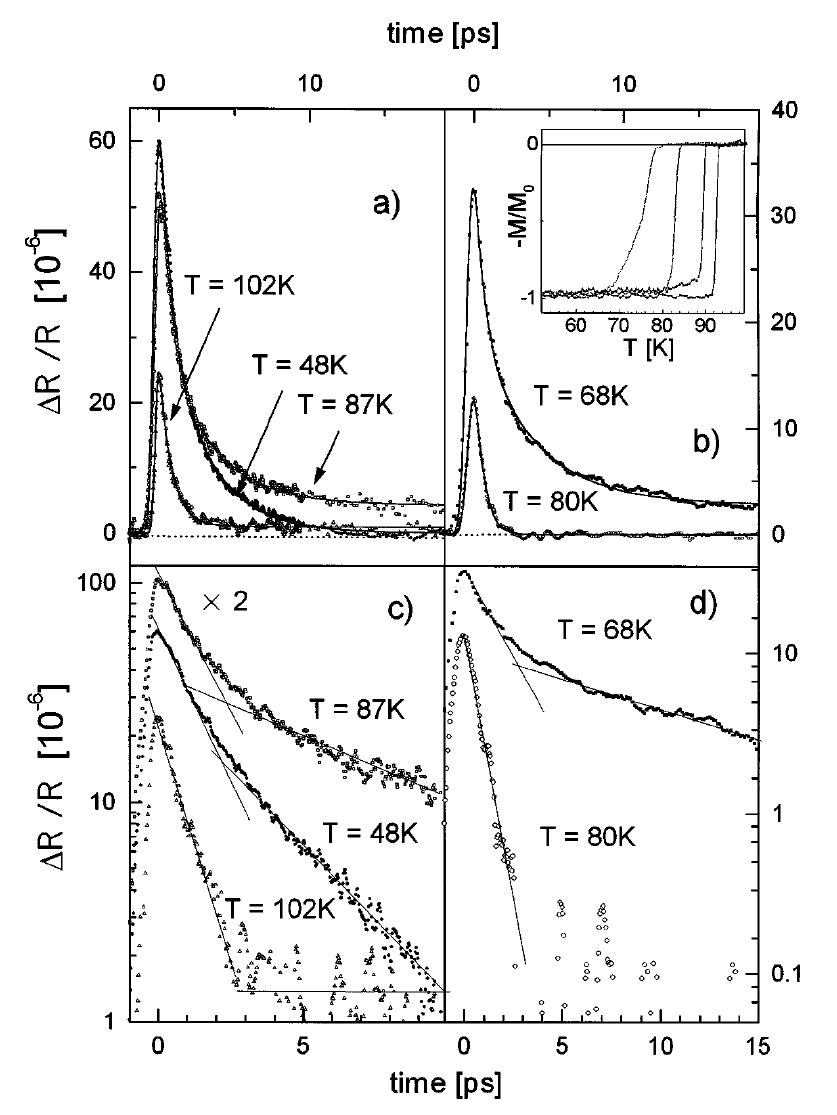}
\includegraphics[width=0.6\columnwidth]{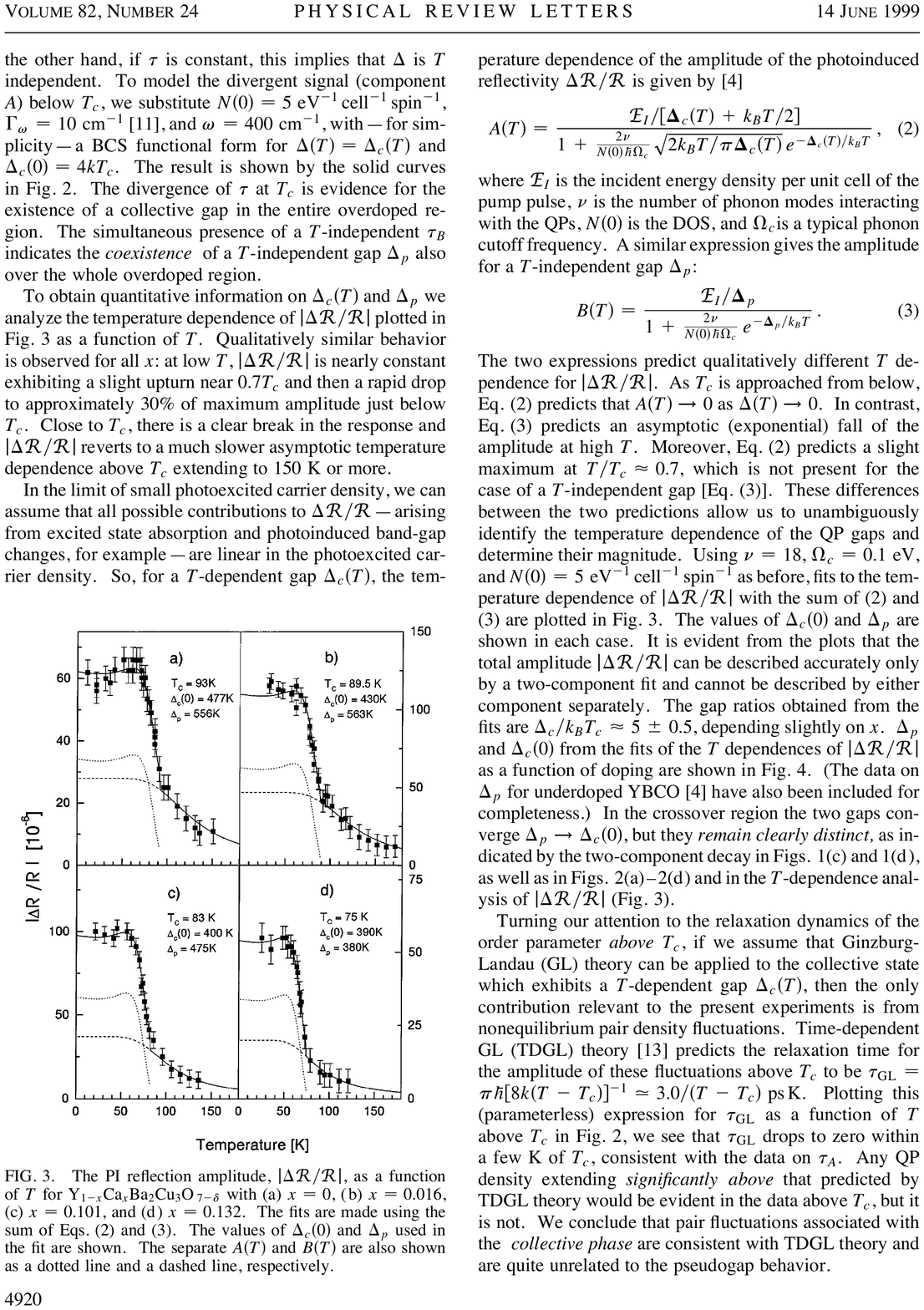}
\caption{\label{fig:The-transient-reflectivity}Top panel: the transient reflectivity
$\delta R/R$ in the superconducting state as a function of time in
optimally doped and overdoped Y$_{1-y}$Ca$_{y}$Ba$_{2}$Cu$_{3}$O$_{7-\delta}$.
The lifetimes of the QP recombination across the superconducting gap
$\Delta_{s}$ and the pseudogap $\Delta_{p}$ differ by nearly an
order of magnitude. Above $T_{c}$ only the PG relaxation is observed. Bottom panel: The temperature dependence
of the amplitude of the transient reflectivity $\delta R/R$ as a
function of temperature in overdoped YBCO. These results suggest the simultaneous presence of two different gaps which exhibit an independent temperature behaviour. Taken from Ref. \citenum{Demsar:1999p112}.}
\end{center}
\end{figure}

\begin{figure}
\begin{center}
\includegraphics[width=0.6\columnwidth]{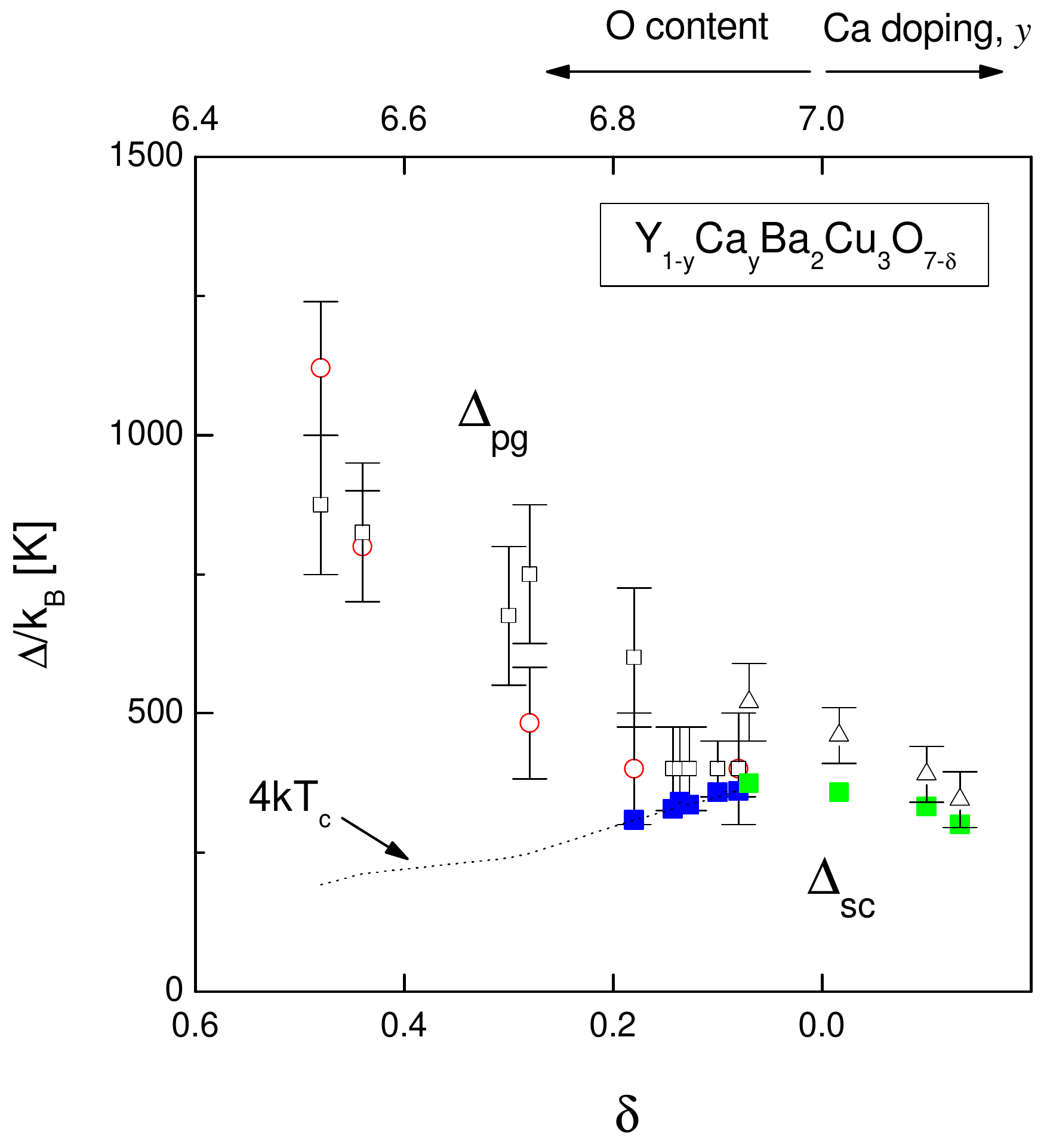}
\caption{\label{fig:The-two-gaps}The two gaps $\Delta_{SC}$ and $\Delta_{PG}$
as a function of doping in Y$_{1-y}$Ca$_{y}$CuO$_{7-\delta}$ are
shown by full and open symbols respectively. A divergence of $\tau_{s}$
in the relaxation time is evident for optimally doped and overdoped
materials \cite{Kabanov1999,Demsar:1999p112}.}
\end{center}
\end{figure}

\begin{figure}
\begin{center}
\includegraphics[width=0.6\columnwidth]{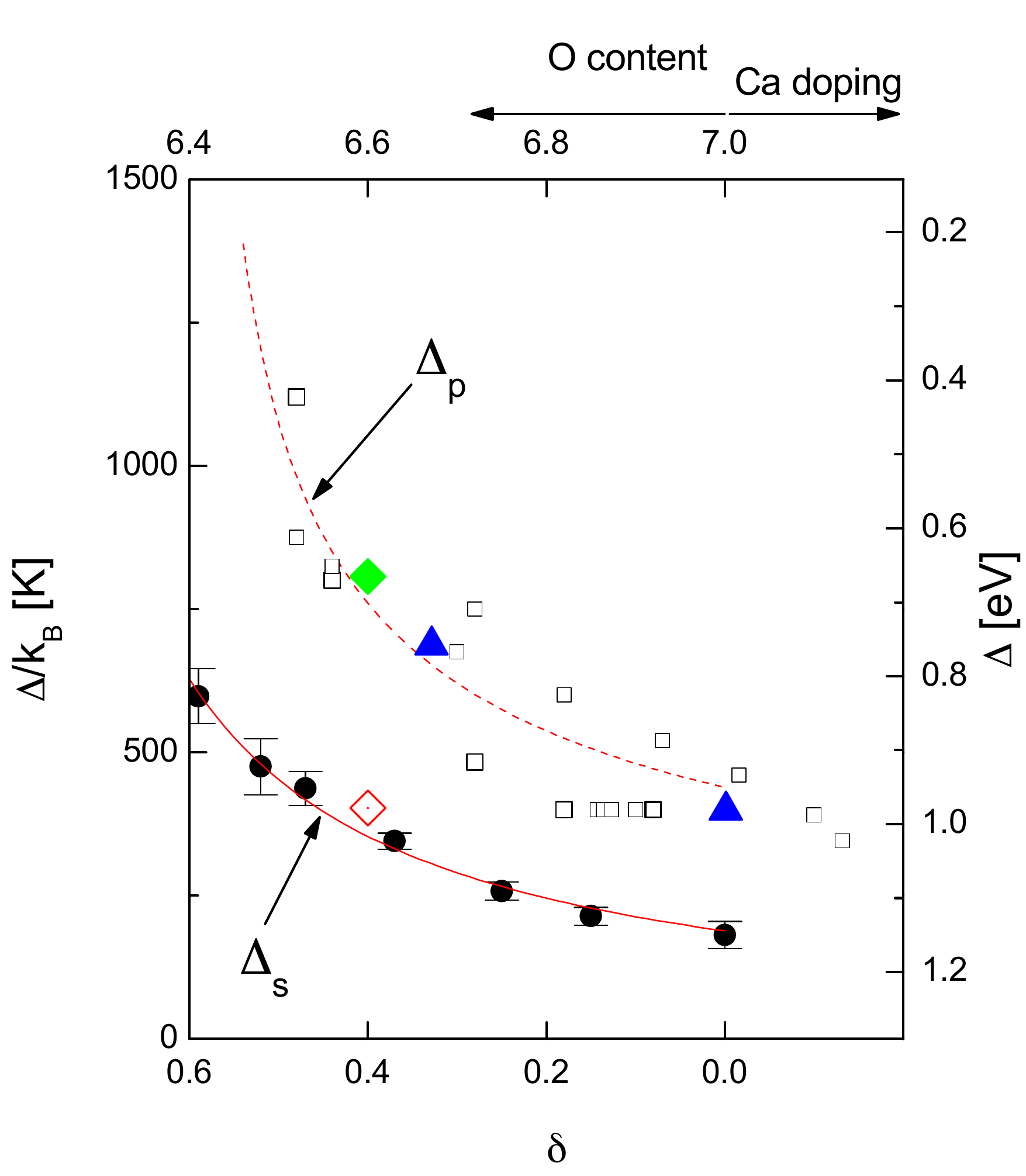}
\caption{\label{fig:The-charge-and-spin}The charge channel pseudogap $\Delta_{PG}$$ $
and spin gap $\Delta_{s}$$ $ in YBa$_{2}$Cu$_{3}$O$_{7-\delta}$
as a function of doping. The full circles are from $^{89}$Y NMR Knight
shift $K_{s}$ . The open squares are from time-resolved QP relaxation
measurements\cite{Kabanov1999,Demsar:1999p112}. The open and
full diamonds are from spin polarised neutron scattering and charge
excitation neutron data, respectively, while the triangles are from
the tunneling data \cite{Deutscher1999,Mihailovic:1999p4786}.}
\end{center}
\end{figure}

Subsequent systematic work on YBa$_{2}$Cu$_{3}$O$_{7-\delta}$ \cite{Stevens1997},
particularly as a function of doping \cite{Thomas:1996wc,Kabanov1999,Demsar:1999p112,Segre:2002p9832,Gedik2004},
showed a very consistent picture of a multi-component response (see Fig. \ref{fig:The-transient-reflectivity}), in which the QP recombination
dynamics across the superconducting gap is quite distinct from the
QP relaxation (recombination) dynamics across the pseudogap. The QP
recombination time across the superconducting gap, $\Delta_{SC}$, vanishes
sharply at $T_{c}$, as a consequence of the closing of the gap itself ($\Delta_{SC}\rightarrow0$), as discussed in Sec. \ref{sec_effTmu}. This divergence is especially pronounced in optimally-doped and overdoped YBCO, but is hardly noticeable in underdoped thin films.
The superconducting QP lifetime remains almost constant for most
samples measured except for the very clean ortho-II phase samples,
where the relaxation rate is reported to decrease with decreasing
temperature. Such behavior appears to be a property of very high quality
single crystals, also of other cuprates, such as LSCO and Hg-1223 \cite{Demsar:2001p1518}.
The temperature dependence of the transient response due to QP recombination, whose amplitude and relaxation time dramatically increase below $T_c$, was shown to fit well to the model
predictions under bottleneck conditions, as described by Eq. \ref{EqSizeOfTheGapTdep}
using a superconducting gap with a BCS-like temperature dependence. 
On the other hand, the PG response can be fit very well
with a model of QP relaxation across a temperature-independent gap
$\Delta_{PG}$ (see Eq. \ref{EqSizeOfTheGapTind}). In the normal state, the PG response gradually diminishes
until it vanishes at a temperature $T^{*}$, which is consistent
with the "pseudogap temperature" measured by other techniques
which measure a charge gap \cite{Mihailovic:1999p4786}. Such behaviour
has been found to be quite generic: all cuprate superconducting materials show similar temperature-dependence for the SC and
PG response. Similar behaviour has also been observed in pnictides,
as we shall see in Sec. \ref{sec_ironbased}. 
The values of $\Delta_{SC}$ and $\Delta_{PG}$ estimated from the
the fitting with the gap-dependent model that will be discussed in Sec. \ref{sec_effTmu} suggested a phase diagram (reported in Fig. \ref{fig:The-two-gaps}) that was remarkably consistent with conventional equilibrium spectroscopies. 
Of particular interest is the comparison between the magnitude of the normal-state pseudogap
$\Delta_{PG}$ and the spin gap measured by NMR. Fig. \ref{fig:The-charge-and-spin} shows that the the charge
gap inferred from the QP recombination in the normal state is approximately
twice the spin gap measured by Knight shift measurements, which was
attributed to the different nature of the charge and spin excitations
\cite{Mihailovic:1999p4786}.

The sign of the SC and PG responses in YBa$_{2}$Cu$_{3}$O$_{7-\delta}$
 was reported to depend
on doping \cite{Thomas:1996wc} and on the polarization of the optical probe
light with respect to the crystal axes \cite{Dvorsek:2002p82}. For
probe parallel to the crystallographic $a$-axis, the PG and SC transient reflectivity responses are
of the same sign, while for probe parallel to the $b$-axis, the two are of opposite
sign. While the SC response changes sign, the PG response is
independent of the probe polarization, consistently
with the band structure anisotropy of YBCO
\cite{Dvorsek:2002p82}. An important issue in analysing the dynamics of the equilibrium-state recovery is the determination of the intrinsic timescale of the quasiparticle decay via reformation of Copper pairs and emission of gap-energy bosons. This process, that has been formalized through the Rothwarf-Taylor equations discussed in Sec. \ref{sec_RT},
is expected to be strongly temperature-dependendent and, therefore, also strongly affected by the average heating accumulated within the laser spot. 

Extremely low-fluence measurements \cite{Segre:2002p9832}, performed on YB$_{2}$Cu$_{3}$O$_{6.5}$ Ortho-II with a high-repetition rate Ti:sapphire oscillator, accessed the low-temperature and low-fluence regime, in which the recovery dynamics can be described by the \textit{weak bottleneck} (see Sec. \ref{sec_RT}). In this limit, the slowest relaxation process is given by the bi-particle recombination and the dynamics is expected to be strongly fluence-dependent, in agreement with the outcome of low-fluence experiments \cite{Segre:2002p9832,Gedik2004,Gedik2005}. 

Measurements of the mid-infrared reflectivity using difference-frequency
generated laser pulses tunable in the range between 60 to 180 meV
(7-21 $\mu$m) on YBa$_{2}$Cu$_{3}$O$_{7}$, confirmed the two-gap
scenario: a picosecond recovery of the superconducting condensate
in underdoped and optimally doped material and, in underdoped YBa$_{2}$Cu$_{3}$O$_{7-\delta}$,
an additional subpicosecond component related to pseudogap correlations \cite{Kaindl2000}. Both the components exhibited a temperature-dependence similar to that previously observed in single-colour P-p experiments.   

Important insights into the diffusion of photoexcited quasiparticles in YBCO were provided by the use of transient-grating spectroscopy \cite{Gedik2003}. In this experimental scheme, the sample is excited by two non-collinear coherent pump pulses, which "write" a transient grating in the dielectric function whose period is of the order of the wavelength. As long as the excitation pattern persists in the sample, the probe beam is partially diffracted with an intensity proportional to the contrast of the grating. By performing a phase-sensitive heterodyne detection, the authors were able to disentangle the in-plane diffusion of the excited quasiparticles, which alters the profile of the grating, from the recombination/relaxation, which only impacts on the contrast of the grating. A diffusion coefficient $D$=20 cm$^2$/s along the \textbf{a}-axis and $D$=24 cm$^2$/s along the \textbf{b}-axis was determined.

\begin{figure}
\begin{center}
\includegraphics[width=0.5\columnwidth]{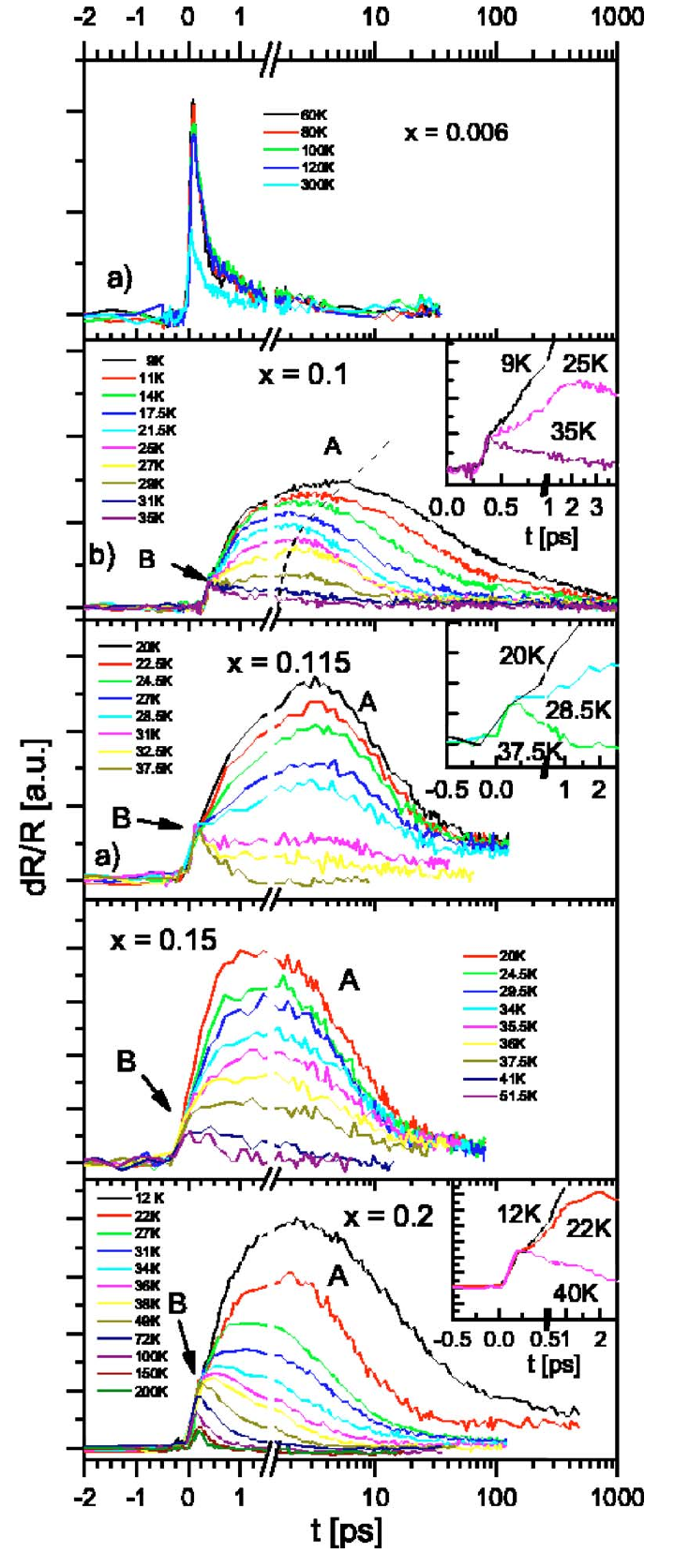}
\caption{\label{fig:The-photoinduced-reflection-lasco}The transient reflection
$\delta R/R$ in La$_{2-x}$Sr$_{x}$CuO$_{4}$ single crystals measured
with photoexcitation intensities $1.7-20\times10^{-7}$ J/cm$^{2}$ for different doping $x$ as a function of temperature.
Two distinct components A and B are clearly resolved (see also insets), which are attributed to superconducting quasiparticle recombination (A) and pseudogap quasiparticle relaxation (B). Taken from Ref. \citenum{Kusar:2005p4778}.}
\end{center}
\end{figure}

\begin{figure}
\begin{center}
\includegraphics[width=0.5\columnwidth]{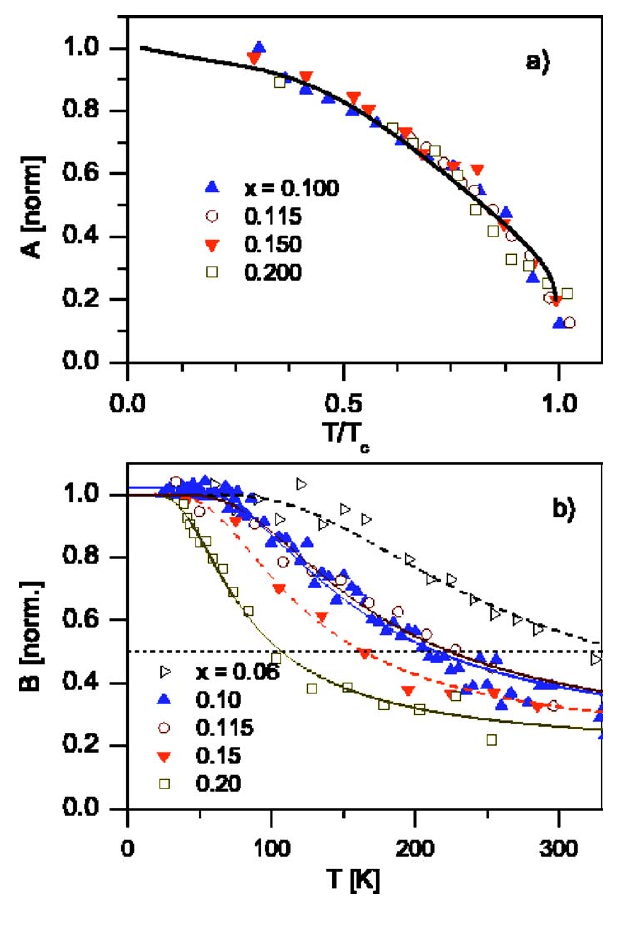}
\caption{\label{fig:Gap-and-PG-LSCO}The temperature dependence of the amplitude of the superconducting (A) and pseudogap (B) components of the $\delta R/R$ signal in
La$_{2-x}$Sr$_{x}$CuO$_{4}$. The lines
are the best fit to the data using the expressions of the temperature-dependent gaps, $\Delta_{SC}$ and $\Delta_{PG}$, reported in Eqs. \ref{EqSizeOfTheGapTdep} and
\ref{EqSizeOfTheGapTind}. The intersection of the fit with
horizontal line in panel b) corresponds to the pseudogap temperature $T^{*}$. Taken from Ref. \citenum{Kusar:2005p4778}.}
\end{center}
\end{figure}

\paragraph*{La-based cuprate superconductors.}
\label{sec_La-cuprates}
The large range of doping concentrations available in La$_{2-x}$Sr$_{x}$CuO$_{4}$ (LSCO)
was of great help to perform systematic measurements which improved the understanding of the phenomena underlying the relaxation dynamics in high-temperature superconductors. Time-resolved experiment as a function of temperature and
doping were reported by Kusar \cite{Kusar:2005p4778}. Similarly as in
YBCO and other cuprates, the experiments on LSCO
revealed a two-component relaxation at $T<T_{c}$ (see Fig. \ref{fig:The-photoinduced-reflection-lasco}), that was attributed to the SC and pseudogap (PG) responses.
This observation turned out to be important, because this material does not contain charge reservoir
layers, thus excluding the possibility that the two-component response
arises as a result of the decoupled charge reservoirs and CuO$_{2}$
planes respectively. Instead, these data suggested that the ultrafast response is strictly related to the electron dynamics within the CuO planes, which is a common feature of all the cuprates. Since in degenerate P-p experiments, the SC and PG signals are of the same sign, particular care is needed to resolve them. However,
the unambiguous temperature dependence of the two signals and
the evolution with doping, turned out to be very helpful in disentangling the two components (see Fig. \ref{fig:Gap-and-PG-LSCO}). 
The temperature dependence of the superconducting signal
follows a markedly different curve than that observed in YBCO, but can
be fit similarly with the bottleneck QP relaxation formula of \ref{EqSizeOfTheGapTdep}.

Bianchi et al. \cite{Bianchi:2003kj} discussed also an alternative possible interpretation to the independent PG and SC responses, which was based on the hypothesis that the observed relaxation traces a 2-step cascade-like process arising from the coexistence of two energy scales ($\Delta_{SC}$ and $\Delta_{PG}$) in the superconducting state. Even though it is not possible to discriminate between a cascade process and the sum of two components, the dependence of the amplitudes of the two signals on P pulse
intensity \cite{Kusar:2005p4778}, the doping dependence and the comparison
of the electronic structure with the outcome of other spectroscopies,
indicate a coexistence of parallel relaxation channels for the two types of
excitations, rather than a cascade. In particular, the phase diagram
obtained from the La$_{2-x}$Sr$_{x}$CuO$_{4}$ data suggests that
the PG and SC gaps coexistence is ubiquitously present in cuprates.


Bianchi et al. also found that in the SC state the response is remarkably
sensitive to a magnetic field \cite{Bianchi:2005p3988,Bianchi:2005p3989}
along the \emph{c} axis of the crystal. This result was attributed  to the
appearance of normal regions surrounding vortex cores (approximately
13 nm diameter) which show characteristic PG relaxation dynamics. 
The temperature dependence of the superconducting quasiparticle recombination
time below $T_{c}$ shows remarkably little variation with
doping.
The expected anomaly is clearly observed near $T_{c}$, but the divergence
of the relaxation time is not as pronounced as in overdoped (Y,Ca)Ba$_{2}$Cu$_{3}$O$_{7-\delta}$ single crystals (see Sec. \ref{sec_Y-cuprates}). More, recently, the physics of the vortices in LSCO has been investigated by broadband time-domain THz spectroscopy \cite{Bilbro2011b}, which suggested an anomalous contribution to the diamagnetic response that is not superconducting in origin, but could be ascribed to a $d$-density wave (DDW) state within the pseudogap phase \cite{Sau2011}.

\paragraph*{Bi-based cuprate superconductors.}
\label{sec_Bi-cuprates}
Because of the easiness in preparing cleaved surfaces, Bi$_{2}$Sr$_{2}$CaCu$_{2}$O$_{8+\delta}$ is one most studied superconducting cuprates through conventional techniques, such as ARPES and STM. Therefore, it was soon recognized as fundamental the comparison between ultrafast optical experiments and the ARPES and
STM results. Unfortunatey, the range of doping in the 2-layer Bi2212 system is typically not as large as that in YBCO or LaSCO, therefore only results near optimum doping can be compared. Furthermore, the weak interlayer coupling leads to a small thermal conductivity perpendicular to the layers, which implies that it is very easy to accumulate local heat with high
repetition rate laser pulse trains. 

\begin{figure}
\begin{center}
\includegraphics[width=1\columnwidth]{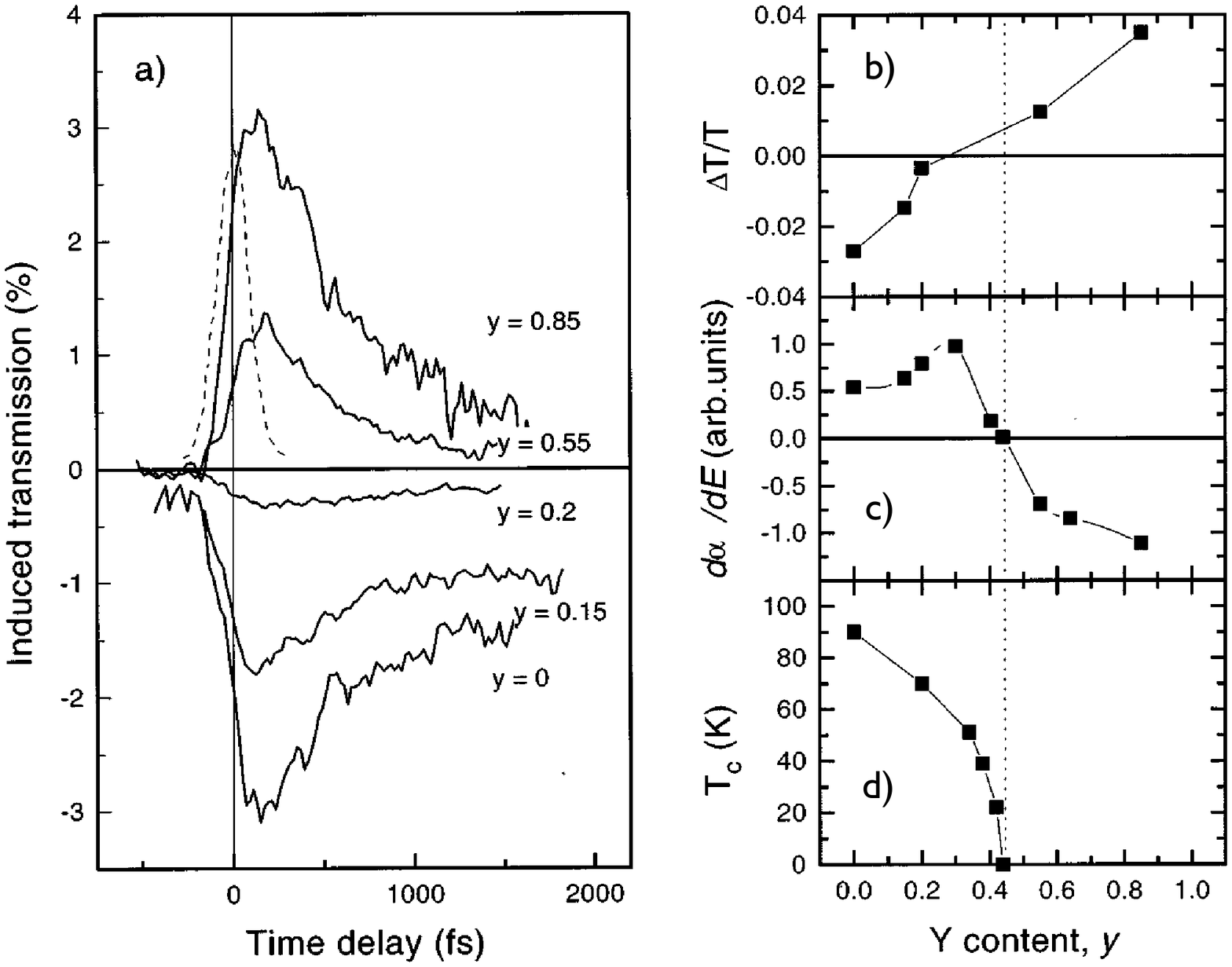}\caption{\label{fig:Time-resolved-BiSCYO}Time-resolved transmission $\delta T/T$=$-\delta\alpha/\alpha$
through thin single crystal films of Bi$_{2}$Sr$_{2}$Ca$_{1-y}$Y$_{y}$Cu$_{2}$O$_{8}$
for different doping $y$. b) the magnitude of $\Delta T/T$ as a
function of $y$. c) The magnitude of the slope $(d\alpha/dE)$ as
a function of $y$. d) The critical temperature $T_{c}$ as a function
of $y$. Taken from Ref. \citenum{Thomas:1996wc}.}
\end{center}
\end{figure}

Thomas et al. \cite{Thomas:1996wc} performed measurements on
Bi$_{2}$Sr$_{2}$Ca$_{1-y}$Y$_{y}$Cu$_{2}$O$_{8}$ (Y-Bi2212) as a function
of $y$, as shown in Fig. \ref{fig:Time-resolved-BiSCYO}a).
The dye laser output was a train of pulses with $\hbar\omega$=2 eV and temporal length of $\sim$150 fs, which prevented high temporal resolution studies. Nevertheless,
the data clearly revealed a low-temperature divergence of the QP lifetime
in the superconducting state, consistently with the results on YBCO and LSCO families, and a systematic cross-over of the sign
of the transient absorption (negative transmission -$\delta T/T$).
The authors correlated the sign of $\delta T/T$ to the sign of the
slope of the absorption curve ($d\alpha/dE)|_{\lambda_{p}}$, where
$\lambda_{p}$ is the probe wavelength (see Fig. \ref{fig:Time-resolved-BiSCYO}b) and c)).
The doping concentration of the crossover was found to be close, but not the same as the point at which the sign changes (see Fig. \ref{fig:Time-resolved-BiSCYO}d)). Furthermore, the magnitude of ($d\alpha/dE)|_{\lambda_{p}}$ showed markedly
different behaviour than $\delta\alpha/\alpha$ as a function of $y$. Remarkably, the cross-over of the slope ($d\alpha/dE)|_{\lambda_{p}}$
coincides with the superconductor-insulator transition as a function
of $y$, although the authors suggested this to be fortuitous. A similar
sign change was reported for Bi$_2$Sr$_2$Ca$_{1-y}$Dy$_y$Cu$_2$O$_{8+\delta}$ and YBa$_{2}$Cu$_{3}$O$_{7-\delta}$, as
a function of the hole-doping concentration \cite{Gedik2005,Thomas:1996wc}. Gedik et al. suggested a deeper significance for this crossover \cite{Gedik2005}, related to sharp transition of the quasiparticle dynamics which takes place precisely at optimal doping. However, since the sign of the response is strongly dependent on the probe wavelength and polarisation, one cannot make more substantial statements on this issue without detailed knowledge of both the initial and final states for the probe transition. This problem, along with the origin of the $\delta R/R$ signal in the infrared-visible region, will be solved thanks to the advent of broadband techniques \cite{Giannetti2011}, as extensively discussed in Sec. \ref{sec_SWshift}. 

Some interesting physics arising from the two-dimensional nature of
the system is discussed by Corson et al. \cite{Corson:1999bo}, who
reported measurements of high-frequency conductivity using time-domain
transmission spectroscopy to capture the linear response in the 100-600 GHz frequency range.
Direct measurement of the electric field, rather than intensity, yields
both the real and imaginary parts of the optical conductivity, without the use of Kramers-Kronig analysis.
The samples were thin (40-65 nm) epitaxial films of underdoped Bi$_{2}$Sr$_{2}$CaCu$_{2}$O$_{8+\delta}$
grown by atomic layer-by-layer molecular-beam epitaxy. The authors tracked
the phase-correlation time, $\tau_{pc}$, in the normal state suggesting
that, just above $T_{c}$, $\tau_{pc}$ reflects the motion of thermally
generated vortices. The phase correlation time is related
to the crossover frequency at which phase-fluctuations become
indistinguishable from the DC superconducting state response.
Corson et al. found that the vortex proliferation reduces $\tau_{pc}$ to
a value indistinguishable from the lifetime of the normal state electrons
at $\sim$100 K, i.e., well below the pseudogap temperature $T^{*}$. As
a consequence, they suggested that while phase correlations indeed persist
above $T_{c}$, they vanish well below $T^{*}$. Similar conclusions have been later drawn on the basis of broadband time-domain THz spectroscopy on LSCO \cite{Bilbro2011}.  Unfortunately,
more direct data on vortex formation, which would confirm the proposed
vortex dynamics, have not been obtained since then. Significant progress on the separation of pairing and phase coherence dynamics above $T_c$ from the pseudogap will be discussed later in section \ref{sec_fluctuations}.

\begin{figure}
\begin{center}
\includegraphics[width=0.8\columnwidth]{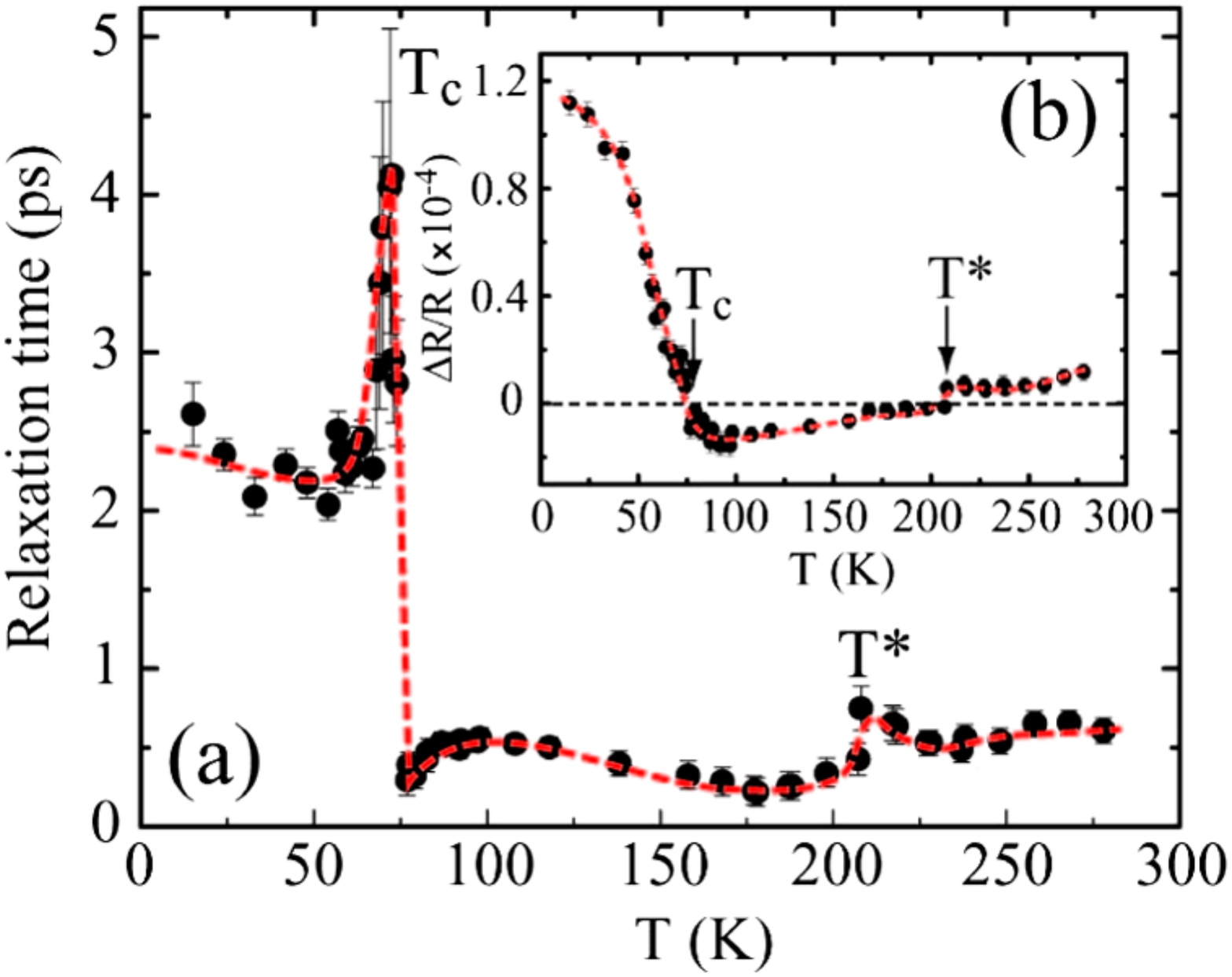}
\caption{\label{fig:T-dep-BiSCO-Toda}Temperature dependence of: a) the QP
relaxation time; b) the measured amplitude of transient
reflectivity signal, $\delta R/R$ . The dashed lines are guides to the eye. Taken from Ref. \citenum{Liu:2008p4917}.}
\end{center}
\end{figure}

Building on the results of Corson
et al. \cite{Corson:1999bo}, systematic transient terahertz conductivity experiments using a high repetition
rate Ti-Sapphire laser source were performed on Bi$_{2}$Sr$_{2}$CaCu$_{2}$O$_{8+\delta}$ by Kaindl at al. \cite{Kaindl2005gp}. After depletion of the superconducting
condensate by optical excitation, Kaindl at al. studied the ensuing dynamics
at various excitation densities and temperatures. Furthermore, they measured the low-energy
spectra and shapes of decay rates that were attributed to the bimolecular
kinetics of the condensate formation, accordingly with the Rothwarf-Taylor model (see Sec. \ref{sec_RT}). 
On the basis of the fact that the kinetics could be fit using bimolecular kinetics, the authors argued that the phonons do not cause breakup of Cooper pairs, in agreement with the prediction that bimolecular kinetics is expected at low temperatures (see Sec. \ref{sec_RT}). Unfortunately, heating is very difficult to control at low temperatures for the vanishing thermal conductivity and heat capacity, thus limiting the experiments to very low fluences \cite{Kaindl2005gp}. More recent experiments, where the heat build-up is smaller, were reported by Liu et al.  \cite{Liu:2008p4917} and Toda et al. \cite{Toda:2011cs}, at various pump and probe wavelengths and temperatures in the bottleneck regime, where an exponential decay is observed.
Liu et al \cite{Liu:2008p4917} reported
an ultrafast optical response of quasiparticles (QPs) in both the
pseudogap (PG) and the superconducting (SC) state of an underdoped Bi$_{2}$Sr$_{2}$CaCu$_{2}$O$_{8+\delta}$
single crystal measured with a probe energy $\hbar\omega$=1.55 eV. Since the $\delta R/R$ signal changes the
sign exactly at $T_{c}$, the direct separation of the PG and SC dynamics was easily achieved. Furthermore, the transient signals associated with the PG and SC dynamics was found to depend on the probe beam energy and polarization. By tuning them below
$T_{c}$, two distinct components could be detected simultaneously,
providing evidence for the coexistence of PG and SC excitations. Remarkably, a clear divergence of the relaxation time at $T_{c}$,
as well as an anomaly at $T\sim$210 K, i.e. approximately at $T^{*}$ was observed (see Fig. \ref{fig:T-dep-BiSCO-Toda}).
The presence of the divergence at $T_{c}$ is in line with the results obtained on the YBCO and LSCO families
and on other materials \cite{Demsar:1999p112}, but the sharp
anomaly in both amplitude of the photoinduced reflectivity and relaxation
time at $T^{*}$ implies the existence of a critical temperature,
which could be associated with the onset of a charge-order in the Bi-O layers, although this behavior is solely for this compound. 
The authors also found pronounced sharp resonant effects with respect
to the probe energy, with a peak near $h \nu_{probe}\simeq$1.17 eV, which
is visible only with polarization parallel with respect
to the crystal \emph{a} axis. This is
clearly a consequence of the interband selection rules, which will be discussed in Chapter \ref{sec_noneqopticalprop}. A similar resonance was found near $h \nu_{probe}\sim$1.5 eV also
in YBa$_{2}$Cu$_{3}$O$_{7-\delta}$ \cite{Stevens1997}. No dependence on pump energy was observed
in either case. 

More recently, Toda et al \cite{Toda:2011cs} presented a
further detailed investigation of the relaxation dynamics in underdoped
Bi$_{2}$Sr$_{2}$CaCu$_{2}$O$_{8+\delta}$ ($T_c$=78 K). By changing
the excitation fluence and the polarization of the probe beam, two different
types of relaxation dynamics, associated with superconducting (SC)
and pseudogap (PG) QPs, were quantitatively analyzed independently.
The amplitudes of the gap were estimated from the temperature dependency, obtaining $\Delta_{SC}$=24 meV and $\Delta_{PG}$=41 meV, in good agreement with the values coming from conventional spectroscopies \cite{Huefner2008,Kurosawa2010}.
\begin{figure}
\begin{center}
\includegraphics[width=1\columnwidth]{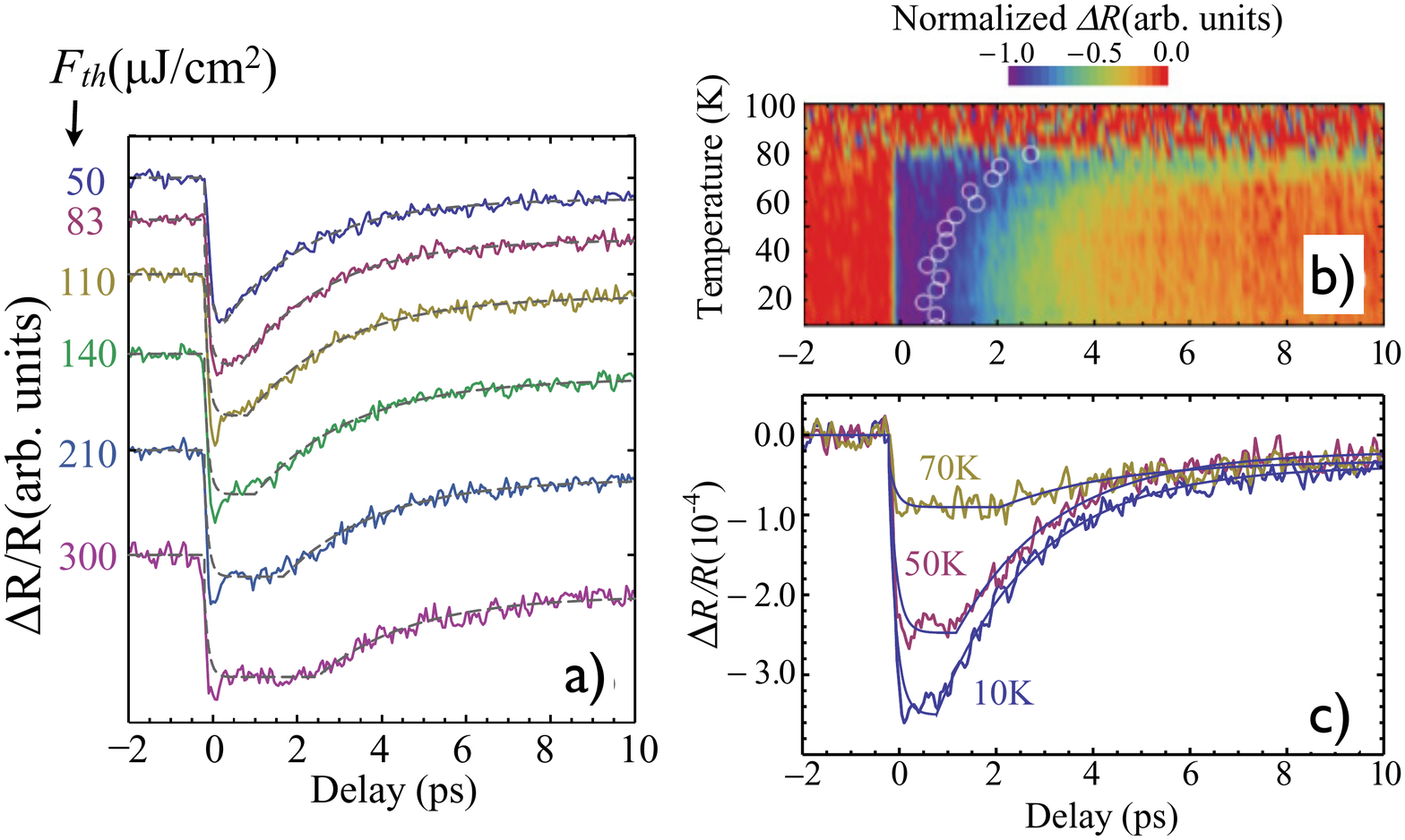}
\caption{(a) $\delta R/R(T=10K)$-$c\delta R/R(90K)$ fitted with a delayed exponential
decay (dashed line), where $c$=$\delta R_{PG}(10K)/\delta R_{PG}(90K)\simeq1.3.$
b) Normalized density plot of $\delta R/R(T)-c(T)\delta R/R(T=90K)$
with $\hbar\omega_{pu}(\hbar\omega_{pr})=0.95(1.55)$ eV and F=70 $\mu$J/cm$^{2}$.
$c(T)=\delta R_{PG}(T)/\delta R_{PG}(90K)$ is the calibration coefficient
for the \emph{T} -dependent amplitude of the PG component. (c) Subtracted
$\delta R/R$ transients at three different temperatures. Each transient
is fitted by an exponential function. White circles show the delay
of each peak obtained from the delayed exponential fits in (a). Taken from Ref. \citenum{Toda:2011cs}.}
\end{center}
\end{figure}

By performing fluence-dependent studies on Bi2212 and on other relevant materials, it was soon realized that the SC response was strongly non-linear, as a consequence of the ultrafast melting of the superconducting condensate \cite{Averitt2001,Carnahan2004,Kaindl2005gp,Kusar:2008p5434,Giannetti2009b,Kusar:2010hz,Stojchevska:2011fz,Coslovich2011,Toda:2011cs,Beyer2011,Coslovich2013}. Phenomenologically, it was shown that, above saturation fluences in the range of $F_{th}$=10-70 $\mu$J/cm$^2$, the dynamics of the $\delta R/R$ signal presented a flat response and a pronounced delay in the build-up time, while a fast component similar to the response of normal QPs appeared on the 0-200 fs timescale \cite{Giannetti2009,Toda:2011cs}. These observations were interpreted as the proof of a non-thermal superconducting-to-normal state phase transition, which will be discussed in Sec. \ref{sec_SCquench}.     
These results opened the possibility for studying the dynamics of the non-thermal melting and recovery of the SC state, which turned out to be a highly-non-thermal process in which the superconducting condensate is vaporized before the complete closing of the gap. This process can lead to a complex transient inhomogeneous state characterized by the spatial coexistence of SC and normal regions \cite{Giannetti2009}. The comprehension of this process is still an open and fascinating aspect of the out-of-equilibrium physics of unconventional superconductors. The new emerging techniques, such as time-resolved ARPES, are expected to shed light on the dynamics of the electronic structure during the non-thermal melting process \cite{Cortes:2011df,Zhang2013,Smallwood2014,Rameau2014,Piovera2015}.
Similarly to the melting of the superconducting condensate, whose threshold has been proved to scale with $\Delta_{SC}^{2}$ \cite{Stojchevska:2011fz}, also the PG signal presents a non-linear behaviour above some threshold fluence. Consistently, the saturation threshold of the PG component was found to be from 4 to 8 times larger than that necessary to melt the condensate \cite{Toda:2011cs,Coslovich2013b,Zhang2013}. This difference opens interesting scenarios where the interplay between the two states can be studied in real time \cite{Coslovich2013b,Zhang2013}.

\paragraph*{Hg- and Tl-based cuprate superconductors.}
\label{sec_Hg-cuprates}
Measurements of the quasiparticle dynamics in Hg$_{1}$Ba$_{2}$Ca$_{2}$Cu$_{3}$O$_{8}$
(Hg-1223) \cite{Demsar:2001p1518} and the one-, two- and three- layer
Tl-based cuprates, evidence the same pattern as
the other cuprate superconductors, i.e., the QP and PG response \cite{Demsar:2001p1518,Smith:1999p4056,Easley1990,Chia:2007p1669} have opposite sign, enabling a clear distinction of the PG  and QP dynamics
in the SC state. As a consequence, it is possible to infer about the existence of two gaps, one temperature-dependent superconducting gap, $\Delta_{SC}(T)$, and another temperature-independent pseudogap, $\Delta_{PG}$, from the temperature dependence of the amplitude
of the photoinduced reflection variation. The zero-temperature magnitudes of $\Delta_{SC}(T)$
and $\Delta_{PG}$ obtained from fits using the bottleneck model \cite{Kabanov1999}
or the analytic solutions to the Rothwarf-Taylor equations \cite{Kabanov2005} were found to be in close agreement with the gaps measured by other techniques, particularly tunneling (STM). Further general features are the rapid increase of QP recombination time in the SC state and
a tendency to diverge of the relaxation time, as $T\rightarrow T_{c}$
from below. Chia et al. \cite{Chia:2007p1669} interpreted the two components as the competition of normal-state and SC-state QPs.
However, this picture is challenged by the fact that
the sign of the response depends on the polarisation and the probe photon
energy, as a consequence of the selection rules for the probe transition
\cite{Dvorsek:2002p82}, as discussed in Sec. \ref{sec_SRStensor}.

\paragraph*{Electron-doped cuprates.}
Ultrafast nonequilibrium carrier relaxation in single-crystal of Nd$_{1.85}$Ce$_{0.15}$CuO$_{4-y}$
was reported very early by Liu et al. \cite{Liu:1993p4053}, who found
that the energy relaxation time in the normal state
increases by more than an order of magnitude at low temperatures, consistently with later reports on other superconducting cuprates. The authors also observed a weak divergence
near $T_{c}=24$ K, and a saturation of the relaxation time below
$T_{c}$, similar to what observed, for example, in thin films of YBa$_{2}$Cu$_{3}$O$_{7-\delta}$ \cite{Mihailovic:1998p4077}. They discussed this phenomenology as the complex interplay among the characteristic relaxation
times, including scattering and recombination of photo-excited quasiparticles  and the order parameter relaxation close to $T_{c}$. 

More recently, Cao et al. \cite{Cao:2008p1505} and Long et al., \cite{Long:2006gp} 
studied the photoexcited carrier dynamics in the electron-doped La$_{2-x}$Ce$_{x}$CuO$_{4}$
(LCCO). They observed a behaviour similar to the results obtained on
hole-doped cuprates (see Secs. \ref{sec_Y-cuprates}-\ref{sec_Hg-cuprates}) consisting in a
general increase of the relaxation time at low temperature, a divergence
at $T$=$T_{c}$, and a mono-exponential (not bimolecular) decay of the SC signal. Furthermore,
they also observed a relatively long (ps) rise time, that was
attributed to the Cooper pairs breaking dynamics. The experimental results
were analyzed through the Rothwarf-Taylor model (see Sec. \ref{sec_RT}), with good agreement.

Single-color P-p spectroscopy has also used for studying NCCO, at a doping concentration near optimal doping, as a function of time, temperature, and laser fluence \cite{Hinton2013b}. The relatively slow decay of $\delta R/R$ above $T_c$ (23$<T<$75 K), compared to the analogous signal in hole doped compounds, allowed to resolve a time-temperature scaling consistent with critical fluctuations. This additional fluctuating order, which could be identified as the charge-order observed on the same compound \cite{daSilvaNeto2015}, was found to compete with superconductivity below the superconducting temperature $T_c$.

\subsubsection{Iron-based superconductors}
\label{sec_ironbased}

\begin{figure}
\begin{center}
\includegraphics[width=0.8\columnwidth]{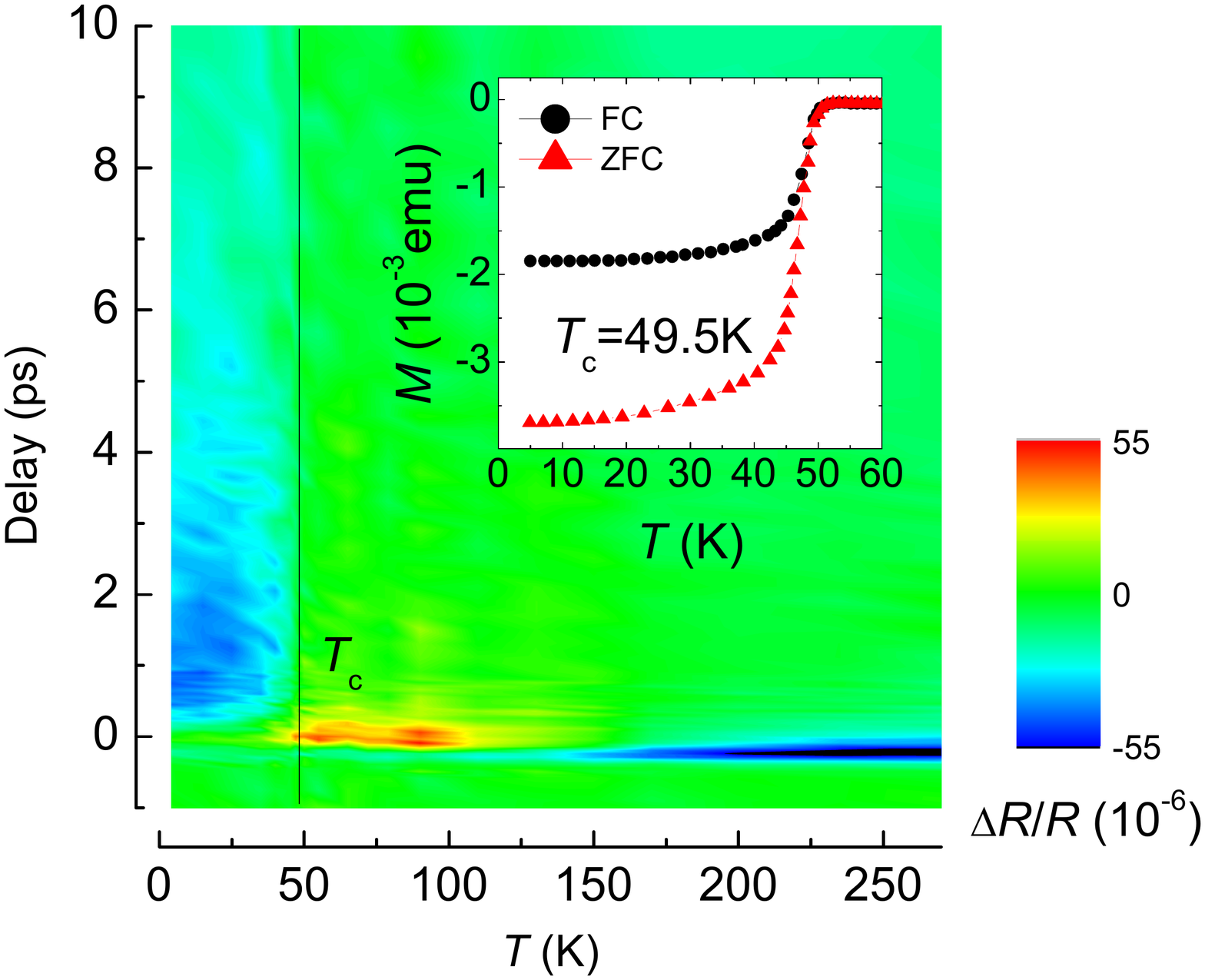}
\caption{\label{fig:Temperature-dep-SmFeAsOF-1}Temperature dependence of the
photoinduced reflectivity $\delta R/R$ at the pump fluence of $3\mu$J/cm$^{2}$ in a SmFeAsO$_{0.8}$F$_{0.2}$ single crystal.  
In the inset, the temperature dependence of the magnetization in the
superconducting state. Taken from Ref. \citenum{Mertelj:2009p6019}.}
\end{center}
\end{figure}

The interplay of well-defined spin or charge-density wave orders
and superconductivity in the iron-based materials introduced a new dimension
in the investigations of the QP dynamics. Hence, a significant number of P-p studies, based on systematic experiments as a function of doping and temperature, have been focused on the competition
between the different ground states. The first experiments on the superconducting
pnictides \cite{Mertelj:2009p6019} revealed multiple QP relaxation
components in the superconducting state, similarly to what observed in cuprate superconductors.
In the nearly optimally-doped SmFeAsO$_{0.8}$F$_{0.2}$
superconductor ($T_{c}=49.5$ K), multiple relaxation processes, such as a SC-like signal following the trend expected for a \emph{T}-dependent superconducting gap and a PG-like signal of opposite sign with an onset above 180 K (see Fig.
\ref{fig:Temperature-dep-SmFeAsOF-1}), have been observed. However, the authors
have interpreted these effects as the sum of the two signals, rather than a competing dynamics. This interpretation was confirmed by doping- and polarisation-dependence studies \cite{Stojchevska:2012ur}. All together these results suggest the existence the existence, above $T_{c}$, of a temperature-independent gap
$\Delta_{PG}$ with a magnitude of 61$\pm$9 meV. Both the superconducting
and pseudogap components showed saturation at pump fluences of $\sim$4 $\mu$J/cm$^{2}$ and $\sim$40 $\mu$J/cm$^{2}$, respectively, associated with the non-thermal melting of the two states.

Later on, Torchinsky et al. \cite{Torchinsky:2009p6282} reported on band-dependent
quasiparticle dynamics in Ba$_{0.6}$K$_{0.4}$Fe$_{2}$As$_{2}$
($T_{c}$=37 K) measured using ultrafast P-p spectroscopy.
In the superconducting state, they observed a fast component, whose decay rate increases linearly with
excitation density, and a slow component with a fluence-independent decay rate. On the basis of these observation they argued that these two components could reflect
the recombination of quasiparticles in the two hole bands by means of
intraband and interband processes. The authors also found that the
thermal recombination rate of quasiparticles increases quadratically
with temperature. The temperature and excitation density dependence
of the decays indicated fully gapped hole bands and nodal or very
anisotropic electron bands.

The doping dependence of quasiparticle relaxation dynamics in (Ba,K)Fe$_{2}$As$_{2}$
was reported by Chia et al. \cite{Chia2010}, in optimally doped,
underdoped, and undoped regimes. In the underdoped sample, spin-density
wave (SDW) order forms at $\sim$85 K, followed by superconductivity
at $\sim$28 K. They found the emergence of a normal-state order which
suppressed SDW at the temperature $T^{*}\sim$ 60 K, arguing that this
normal-state order is a precursor to superconductivity.

Stojchevska et al. \cite{Stojchevska:2012ur} systematically investigated
the photoexcited QP relaxation and the low-energy electronic
structure in electron-doped Ba(Fe$_{1-x}$Co$_{x}$)$_{2}$As$_{2}$
single crystals, as a function of Co doping ($0<x<0.11$). Remarkably,
the evolution of the photoinduced reflectivity transients with doping
proceeds with no abrupt changes as the ground state evolves from a SDW to a superconductor. In the orthorhombic
spin-density-wave (SDW) state, a bottleneck associated with a partial
charge-gap opening is detected, similar to previous reports in different
SDW iron pnictides \cite{Chia:2008p4912}. The
relative charge-gap magnitude 2$\Delta$/k$_{B}$T$_{s}$ decreases
with increasing $x$. In the SC state, an additional relaxation
component appears due to a partial (or complete) destruction of the
SC state on a sub-0.5 picosecond timescale. From the SC component
saturation behavior, the optical SC-state melting energy, $U_{p}/k_{B}$=0.3 K/Fe, is determined near the optimal doping. The subsequent
relatively slow recovery of the SC state indicates clean SC gaps.
The $T$ dependence of the transient reflectivity amplitude in the
normal state was found to be consistent with the presence of a pseudogap in the QP density of states. An interesting feature was the observation of a polarization anisotropy of the $\delta R/R$ signal, suggesting that the pseudogap-like behavior might be associated with a broken fourfold rotational symmetry resulting from nematic electronic fluctuations persisting up to $T
\simeq$ 200 K and occuring at any $x$ in the range 0-0.11. 

Mansart et al. \cite{Mansart2010} investigated the photoexcited electron
energy relaxation in Ba(Fe$_{1-x}$Co$_{x}$)$_{2}$As$_{2}$ focusing
on the possibility of different coupling to specific phonon
baths. Separating the phonon baths into a high-frequency phonon bath
and a low frequency bath, and using a three-temperature model (see Sec. \ref{sec_ETM}),
they attributed the faster relaxation component to
the scattering of the electrons with a subset of strongly-coupled
lattice vibration modes with a second moment of the Eliashberg function
$\lambda\left\langle\omega^{2}\right\rangle\simeq$64 meV$^{2}$ (see Secs. \ref{sec_ephcoupling} and \ref{sec_ephresults}).
Assigning this to a fully symmetric $A_{1g}$ optical phonon leads
to a value of $\lambda\simeq0.12$. The conclusion of the authors
is that this is too weak to account for superconductivity with conventional
theory.

More recently, time-resolved x-ray diffraction has been applied to investigate the structural dynamics of the $A_{1g}$ phonon mode in the parent compound BaFe$_2$As$_2$ \cite{Rettig2015}. The coherent modifications of the Fe-As tetrahedra, indicated a transient increase of the Fe magnetic moments, thus demonstrating the importance of this specific mode for the electron-phonon coupling in these compounds.

Kim et al. \cite{Kim:2012kh} examined the coupling of the spin density
wave gap $\Delta_{SDW}$ to the coherently generated phonon oscillations
of the collective amplitude mode, showing that the $\Delta_{SDW}$
closely follows the phonon oscillations and it is nearly adiabatically
(i.e. without any time lag) modulated by this lattice dynamics. Since the coherent phonon
oscillation modulates the Fermi-surface pockets, thus they also modulate the
nesting wavevector which is at the origin of the SDW instability. Similar
gap oscillations, driven by the same mechanism, were observed in time-resolved ARPES
experiments performed on charge density wave (CDW) compounds, such as TbTe$_{3}$
\cite{Schmitt:2011um}, where the CDW gap is modulated
by the coherent phonon oscillations. The observation of Kim et al.
\cite{Kim:2012kh} demonstrates that this can also occur
for the SDW gap. The implication is that, in spite
of the relatively weak electron-phonon coupling suggested by band
structure calculations and photoelectron energy relaxation\cite{Mansart2010,Gadermaier:2012vz},
there is a strong influence of collective nuclear motion on the spin ordering.

\subsubsection{The conventional superconductors MgB$_{2}$ and NbN}

Pump-probe measurements have been performed also on prototypical conventional superconductors, such as MgB$_{2}$ and NbN.
To some extent, the photoexcited quasiparticle dynamics in MgB$_{2}$, measured through optical-pump/THz-probe experiments \cite{Demsar:2003p7494}, was found to follow relaxation patterns similar to those observed in high-temperature superconductors.
The Cooper pair-breaking dynamics was found to be rather slow, taking place
on a timescale up to 10 ps (compared to typical values of 1 ps in
La$_{2-x}$Sr$_{x}$CuO$_{4}$ \cite{Kusar:2008p5434}) and strongly dependent on the temperature and photoexcitation, in agreement with the Rothwarf-Taylor equations. On the other hand, a careful analysis of the wavelength-dependence of the reflectivity variation (obtained from optical-pump/optical-probe measurements \cite{Demsar:2003p7494}) suggested that in MgB$_{2}$ the photoexcitation is initially followed by energy
relaxation to high frequency phonons instead of the electron-electron thermalization. This process appears to be more pronounced in MgB$_{2}$
than in cuprates, suggesting that the last are characterized by additional relaxation channels with respect to conventional superconductors (see Sec. \ref{sec_ebosresults}).
\begin{figure}
\begin{center}
\includegraphics[width=1\columnwidth]{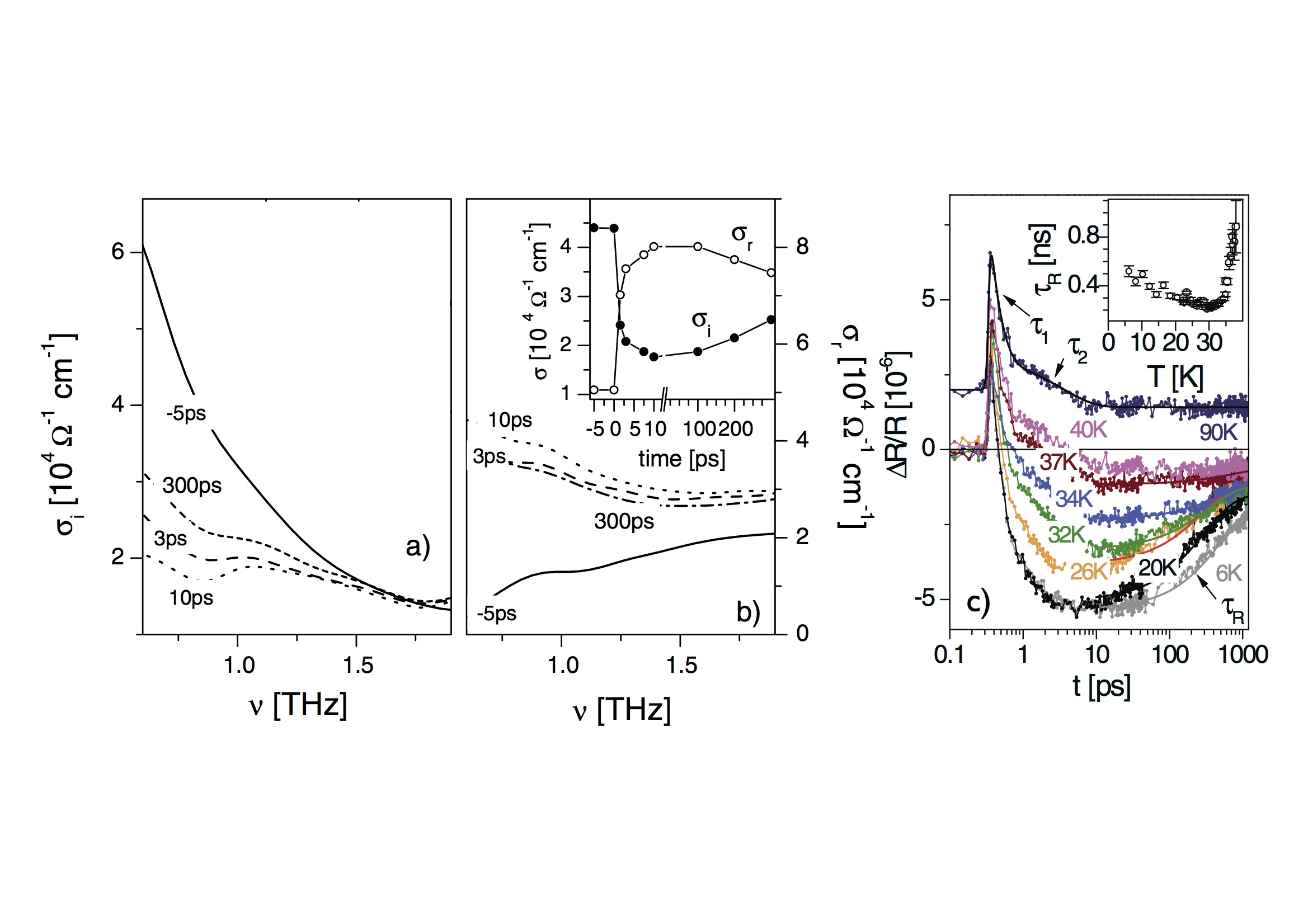}\caption{a) The response of the imaginary part of the optical conductivity,
$\sigma_{2}(\omega)$, in MgB$_{2}$ at different time delays, showing
a slow destruction of the superconducting state taking place over
10 ps or more. b) The time-dynamics of the real part of the conductivity, $\sigma_{1}(\omega)$.
The insert to b) shows the dynamics of $\sigma_{1}$ and $\sigma_{2}$, measured at $\nu=0.8$ THz,
as a function of time delay measured. c) The optical
response at different temperatures shows dynamics on a similar timescale
as the THz response. The insert shows the usual divergence of the
relaxation time $\tau$ as $T\rightarrow T_{c}$ from below. Taken from Ref. \citenum{Demsar:2003p7494}.}
\end{center}
\end{figure}
Additionally, the \emph{T}-dependence of the SC state recovery dynamics was found to be similar to that observed in cuprates. The most significant difference resides in the fact that the recombination rate is two orders of magnitude smaller, as a consequence of the smaller gap value that causes the dynamics to be governed by the lifetime of acoustic phonons instead
of optical phonons as in copper oxides. 

Recent THz experiments on NbN \cite{Beck2011}, showed
remarkable similarity to the MgB$_{2}$. The pair-breaking dynamics
and recovery are on an even longer timescale ($>$10 ps). As in MgB$_{2}$, both the destruction and the recovery are strongly dependent
on the excitation density and temperature. Also in this case, the Rothwarf-Taylor
model applies well, enabling to reliably determine the recombination rate factor. An
electron-phonon coupling constant ($\lambda$=1.1$\pm$0.1) in agreement
with that theoretical estimated for NbN \cite{Haufe1975}, has been extracted from the normal state decay.



\subsubsection{Organic superconductors}
\begin{figure}[t]
\begin{center}
\includegraphics[width=0.8\columnwidth]{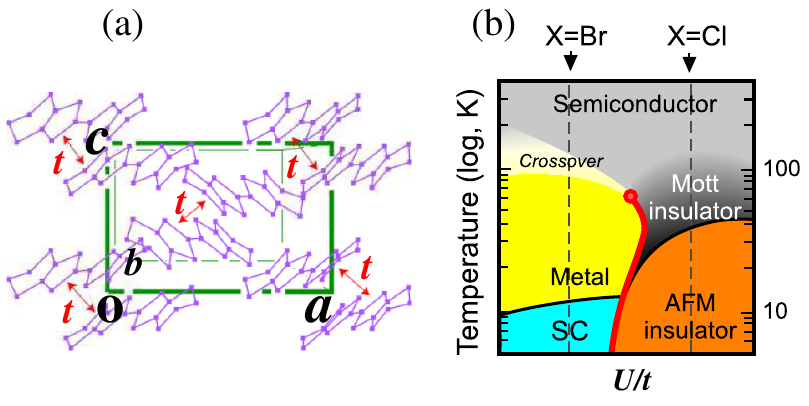}
\caption{\label{fig_BEDT_TTF_diagram}a) The crystal structure of $\kappa$-(BEDT-TTF)$_{2}$X,
b) A schematic phase diagram of $\kappa$-(BEDT-TTF)$_{2}$X as a
function of Coulomb interaction $U/t$. The on-site Coulomb repulsion
U is nearly constant for various anions X, but the substitution of
X changes the transfer integral $t$ between BEDT-TTF dimers. AFM
denotes an antiferromagnetic phase. The black arrows indicate the two salts investigated in Ref. \citenum{Toda:2011gf}.}
\end{center}
\end{figure}

Organic systems were soon considered as promising prototypical systems for the out-of-equilibrium study of the interplay between electronic correlations and complex macroscopic orders (see the prototypical phase diagram of Fig. \ref{fig_BEDT_TTF_diagram}), such as superconductivity and charge-order (CO). Iwai et al. \cite{Iwai:2006vr} reported femtosecond photoinduced
melting of CO in {[}bis(ethylenedithiolo){]}-tetrathiafulvalene
(BEDT-TTF) salts in the non-superconducting state. Ultrafast melting
of the CO demonstrated that the major contribution
to the electronic instability arises from the Coulomb interaction.
A comparative study on two polytypes, exhibiting large - $\theta$-(BEDT-TTF)$_2$RbZn(SCN4)$_4$ - and small - $\alpha$-(BEDT-TTF)$_2$I$_3$ - molecular rearrangements through the
CO transition, were discussed on the basis of low frequency lattice
dynamics, demonstrating that the dynamics of the metallic state are clearly
different in the two systems. The local melting of CO causes ultrafast recovery in the $\theta$-RbZn
salt, whereas the formation of 2D quasi-macroscopic metallic domains
shows a critical slowing down in the $\alpha$-I$_3$ salt. 

More recently, Kawakami et al. \cite{Kawakami:2009p11264}
reported the possibility of optically modulating the effective on-site Coulomb energy
$U$ by approximately 0.4-0.7\% on a dimer, in view of achieving
the Mott insulator-to-metal transition in $\kappa$-(BEDT-TTF)$_{2}$Cu{[}N(CN)$_{2}${]}Br
and $\kappa$-(BEDT-TTF)$_{2}$Cu{[}N(CN)$_{2}${]}Cl. $U$ was optically
modulated by molecular displacements within the dimer as a result
of intradimer excitation. The mechanism of this metallization differs
from the usual photodoping-type mechanism, and a faster transition
via the photodoping mechanism is detected. A metallic-domain-wall
oscillation originating from the modulation of $U$ was also observed
near the critical end-point of the Mott transition line. However,
neither Iwai et al., \cite{Iwai:2006vr} nor Kawakami et al., \cite{Kawakami:2009p11264} investigated the superconducting transition in this system.

The first P-p experiments investigating the superconducting transition
in organic superconductors were performed by Toda et al. \cite{Toda:2011gf}
on $\kappa$-(BEDT-TTF)$_{2}$Cu{[}N(CN)$_{2}${]}Br. The relaxation dynamics of the non-equilibrium carriers in the superconducting system was compared to that measured in the non-superconducting antiferromagnetic $\kappa$-(BEDT-TTF)$_{2}$Cu{[}N(CN)$_{2}${]}Cl (see Fig. \ref{fig_BEDT_TTF_diagram}).
The relaxation
dynamics for both salts showed similar temperature dependences above
$T_{c}$, which was well understood in terms of a carrier relaxation across
a pseudogap (PG) of a magnitude $E_{PG}\sim$16 meV for the Br
salt and $E_{PG}\sim$7.0 meV for the Cl salt.
Below $T_{c}$, the measurements revealed an additional decay component
in the Br salt, which is associated with the SC phase, suggesting
the coexistence of the SC and PG phases. The Br salt showed an abrupt
increase of the decay time at low temperatures, 
associated with the opening of a superconducting (SC) gap below $T_{c}$.
The fluence dependent dynamics at low temperature
confirmed the superposition of the SC and PG components
in the Br salt. The coexistence of the PG and SC phases in the superconducting
state of $\kappa$-(BEDT-TTF)$_{2}$Cu{[}N(CN)$_{2}${]}Br suggested
the presence of phase separation between metallic and insulating phases
in the Br salt, as a consequence of the photoexcitation process.

\subsubsection{Basic concepts emerging from time-resolved measurements}
The exploratory P-p experiments briefly summarised in this chapter have revealed the scope of time-resolved spectroscopy techniques, showing great potential for probing correlated electron systems. It was soon demonstrated that the ultrafast dynamics provides information about the gap(s), the associated quasiparticle relaxation and the electron-boson interactions, opening the way to new more sophisticated spectral- and momentum-resolved techniques (see chapters \ref{sec_noneqopticalprop} and \ref{sec_results}).
The experiments also led to a very useful phenomenological/theoretical description of the elementary phenomena, with analytical expressions for the relaxation times, amplitude of the optical response, and laser fluence dependences. These concepts, which have been widely applied in diverse situations and in different families of superconductors, as well as other gapped systems, will be reviewed in chapter \ref{sec_basicconceptsNES}.
The early experiments in cuprates also pointed out some fundamental observations in cuprates that were not revealed by conventional equilibrium spectroscopies. In particular:
\begin{itemize}
\item the simultaneous presence of the the superconducting gap and the pseudogap in all parts of the phase diagram, with distinct relaxation times and temperature-dependences. This observation could so far be explained only in terms of spatial inhomogeneity with distinct PG and SC regions.
\item the absence of any signature of relaxation from nodal quasiparticles. This evidence strongly suggested that, even in the presence of a strongly anisotropic gap, the dynamics is dominated by the recombination of antinodal excitations. This is in contrast to what expected for a simple $d$-wave gap model at equilibrium, in which the thermally-excited quasiparticle should accumulate in the regions with vanishing gap (nodes).
\item the presence of QP lifetime divergence at $T_c$, particularly in the optimally-doped and overdoped samples, associated with collective ordering. No such divergences were reported for the pseudogap.
\item the presence of multi-component response indicating relaxation processes on very different timescales. In particular, the superconducting dynamics which appears below $T_c$ is universally bottlenecked by some superconductivity-related mechanism which likely involves gap-energy bosons.
\end{itemize}

\subsection{Multi-color experiments: main results}
\label{sec_results}
The recent advances in ultrafast techniques introduced new powerful tools, such as multi-colour experiments and time-resolved ARPES that have been key in bringing the field into a more mature era, in which the experimental results can be quantitatively interpreted and realistic models of the ultrafast dynamics in correlated materials can be developed. In this section we will review the most relevant concepts and problems that have been tackled using the most advanced P-p techniques. Instead of reporting just a list of the main results obtained on different systems, we have organized them by their relevance to the most interesting issues in the physics of correlated materials.  
\subsubsection{Electron dynamics of charge-transfer and Mott insulators}
\label{sec_CTinsulator}
The first and simplest problem that has been tackled by non-equilibrium techniques is the relaxation in Mott-Hubbard or charge-transfer insulators. This issue is related to the many-body effects that regulate the fundamental charge interactions in correlated materials. The interplay between the electronic, magnetic and phononic degrees of freedom can be disentangled by studying the relaxation dynamics, once a non-thermal population of excitations has been created by pumping the system across the correlation gap. 

To better address the problem, we refer to the typical relaxation dynamics of conventional solid state systems. While in ungapped metals the continuum of the electronic levels allows a prompt ($<$1 ps) relaxation of the excited charges through electron-electron and electron-phonon scattering, in semiconductors the gap is usually larger than the spectral width of the bosonic excitations available in the system. Considering the phonons, the cutoff of the maximum energy that can be transferred in a single scattering process is on the order of $\sim$100 meV. As a consequence, a non-thermal population is rapidly accumulated on the bottom of the conduction band, until radiative decays or multi-phonon processes eventually lead to the complete relaxation on a relatively longer timescale (ns). At a first sight, the relaxation process in correlated insulators should be similar to that observed in semiconductors. Since the correlation gap is large and robust, the expectation is that the impulsive photo-excitation would create a long-lived metastable state. The temporal evolution of the lowest-energy electronic excitations, that are those across the Mott-Hubbard or charge-transfer gap ($\Delta_{CT}\gtrsim$1.5 eV), can be easily monitored by performing time-resolved optical experiments in the near-infrared and UV energy ranges.   

The first pioneering measurements on charge-transfer insulators soon evidenced a more complex picture than what was naively expected for conventional systems.
The transient absorption spectra of insulating YBa$_2$Cu$_3$0$_y$ (YBCO) and Nd$_2$Cu0$_4$ (NCO) thin films has been investigated in the 1.3-2.8 eV energy range with a time resolution of $\sim$100 fs \cite{Matsuda1994}. In both cases, an ultrafast photo-induced bleaching of the O2$p$-Cu3$d$ charge-transfer transition at 1.8 eV was observed. Surprisingly, the relaxation of this signal was found to be of the order of 600-900 fs, i.e., extremely faster than what measured in semiconductors with even smaller gaps. The fundamental role of the magnetic excitations in the relaxation process was soon realized. As a consequence of the strong antiferromagnetic coupling ($J$=4$t_h^2/U$) of the electrons (holes) occupying the Cu atoms, the CT insulators are intrinsically characterized by an antiferromagnetic background, whose fluctuation spectrum extends up to 2$J\sim$300 meV. Since the CT process induces a quench of the local magnetic moment on the Cu atom, the photoexcitation corresponds to a local perturbation of the AF ground state, with the immediate consequence of locally increasing the AF fluctuations. In this scenario, the fast decay of the CT bleaching was attributed to nonradiative transitions that involve the emission of high-energy magnetic excitations. 

\begin{figure}[t]
\begin{centering}
\includegraphics[width=1.1\textwidth]{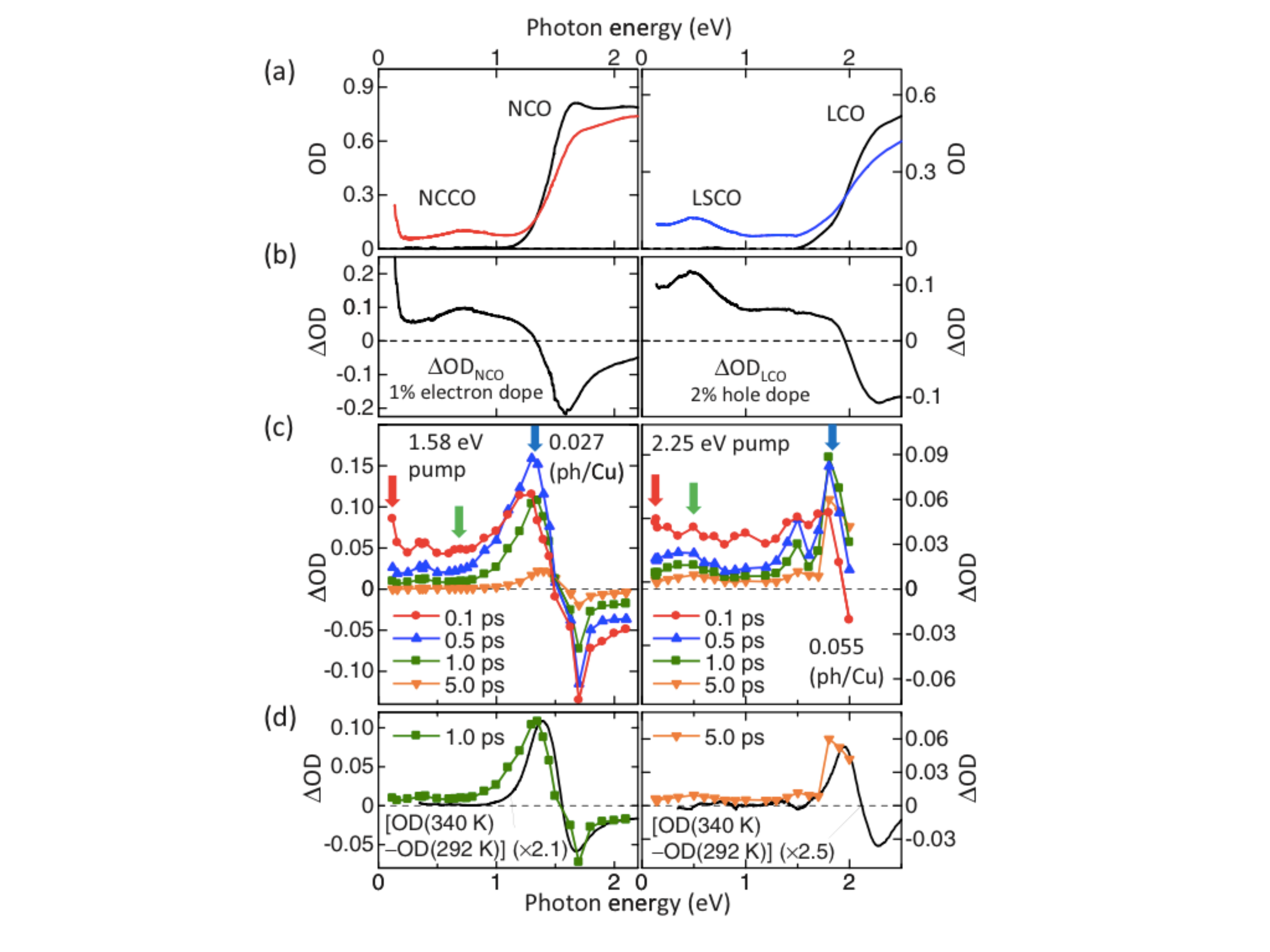}
\label{fig_Okamoto2010}
\caption{Photoexcitation of charge-transfer insulators. \textbf{a}) Optical Density (OD=-log$_{10}$T, where T is the transmittance) OD spectra of electro-doped (Nd$_{2−x}$Ce$_x$CuO$_4$, NCCO) and hole-doped (La$_{2−x}$Sr$_x$CuO$_4$, LSCO) cuprates and their parent compounds (NCO, LCO).  \textbf{b}) The differential OD spectra $\Delta$OD=OD$_{x\neq0}$-OD$_{x=0}$. \textbf{c}) Photoinduced absorption spectra with pump energy of 1.58 eV (NCO) and 2.25 eV (LCO). \textbf{c}) $\Delta$OD spectra at 1 ps of NCO and 5 ps of LCO. Solid lines show the differential OD spectra [OD(340 K)-OD(292 K)]. Taken from Ref. \citenum{Okamoto2010}.}
\end{centering}
\end{figure}

Later on, the probe energy window was extended down to the far infrared (0.1 eV) by performing the frequency difference of two OPAs pumped by an amplified Ti:sapphire oscillator \cite{Okamoto2010}. The dynamics of the CT insulators NCO and La$_2$CuO$_4$ (LCO) was investigated in the 0.1-2.2 eV energy range, after a pump excitation of 1.6 eV and 2.25 eV, that is above the CT gap of NCO and LCO, respectively (see Fig. \ref{fig_Okamoto2010}) \cite{Okamoto2010}. In both systems, the dynamics of the infrared part of the spectrum evidenced the formation of a transient metallic state (see Fig. \ref{fig_Okamoto2010}c) that decays within 200 fs. The delocalized electron-hole excitations subsequently localize because of the charge-spin coupling, giving rise to midgap absorptions at different energies. Interestingly, these features are very similar to those appearing in the spectrum when the hole or electron doping is chemically increased (see Fig. \ref{fig_Okamoto2010}b). On the ps timescale, the variation of the transmittance of the systems is a combination of the increase of the effective doping of the system and of the effective temperature of the charge carriers (see Fig. \ref{fig_Okamoto2010}d). The effective the midgap  absorption has similarities to well-known mid-infrared peak in doped cuprates, which can be attributed to the spin polaron, i.e. charge dressed with spin as well possible phonon excitations. On the other hand, slower recombination in LSCO has predominantly the origin in larger charge gap and smaller exchange interaction $J$. \cite{Lenarcic2012b,Lenarcic2014} 

Similar results have been also obtained in Ca$_2$CuO$_3$, i.e., a prototypical 1D copper-oxide insulator. In this system the photoinduced metallic state decays within $\sim$30 fs, giving rise to the formation of localized polarons via the interplay between charge-phonon and charge-spin coupling \cite{Matsuzaki2015}. 

The role of the spin-charge coupling in the ultrafast dynamics of Mott insulators has been also investigated in other prototypical 1D systems, such as the organic salt bis(ethylenedithio)tetrathiafulvalene-difluorotetracyanoquinodimethane (ET-F$_2$TCNQ), which is characterize by the suppression of the electron-lattice interaction and by a very weak spin-lattice coupling. As a consequence, a transient metallic state is photoinduced and detected as a transient Drude component which relaxes in less than $\sim$200 fs \cite{Okamoto2007}. In contrast to the results obtained on copper oxides, mid-gap states are not expected to form in such 1D systems, where the charge-spin separation suppresses the mechanism which drives the localization of the delocalized charges \cite{Okamoto2007}.  

The increase of the temporal resolution in non-equilibrium optical spectroscopies paved the road to investigate of the decoherence phenomena of the initial many-body excitations. In a Mott insulator, the pumping process drives the formation of Hubbard-excitons (HE), i.e., bound states between double-occupied sites (doublons) and its neighbouring empty sites (holons), which rapidly dephase as a consequence of electron-electron and electron-boson scattering processes. Since the timescale of this process is of the order of few femtoseconds, its observation relies on the development of extremely high-temporal resolution experiments. The physics related to the decoherence of the HEs has been observed in ET-F$_2$TCNQ by exploiting a NOPA seeded by the white-light generated in a sapphire plate and amplified in a BBO crystal, in order to obtain nearly transform-limited 9 fs pulses with a spectrum covering the 0.55-1 eV energy range. After photoexcitation, the delocalization process has been observed to occur through a quantum interference between the HE and ionized holon–doublon pairs, that are created when the holons and doublons move far apart in the 1D lattice. This process is observed as an oscillation at 25 THz (period of $\sim$40 fs) \cite{Wall2010}. The role of HEs in the optical transitions across the Mott-Hubbard gap has been also studied in the orbitally-ordered insulator YVO$_3$. The dynamics of the spectral weight transfer between the HE peak and the single-particle band has been monitored on the fs timescale by means of a white-light probe \cite{Novelli2012}. 

Finally, the role of the excitation process has been also addressed by studying the ultrafast dynamics of quasi-particles in the archetypal strongly correlated charge-transfer insulator LCO. While the above-gap excitation injects electron-hole excitations that subsequently exchange energy with the boson baths (antiferromagnetic fluctuations, phonons), the sub-gap excitation pilots the formation of itinerant quasi-particles, which are suddenly dressed by an ultrafast reaction of the bosonic field \cite{Novelli2014}. This result evidences that, in the case of sub-gap excitation, the interaction between electrons and bosons manifests itself directly in the photo-excitation processes.

The huge experimental effort to investigate the dynamics in Mott-Hubbard and CT insulators was triggered by the possibility of understanding the fundamental relaxation processes in correlated materials and providing the necessary inputs for the development of simple models for the non-equilibrium physics of out-of-equilibrium correlated systems. The problem of the origin of the fast relaxation in a Mott insulator has been tackled by studying a simplified version of the single-band Hubbard model. A generalized $t$-$J$ model for insulators is obtained by performing a canonical transformation of the Hubbard model that, at the lowest order in $t$/$U$, decouples sectors with different numbers of HEs. As compared to the conventional $t$-$J$ model, the transformed Hamiltonian contains also the terms causing recombination of the HEs \cite{Lenarcic2012b,Lenarcic2014}. Within this model, the short picosecond-range lifetimes of photoexcited carriers evidenced by the first P-p experiments is naturally explained as a two step process: i) the formation of an $s$-type HE; ii) the decay of the bound holon-doublon exciton via multimagnon emission. Even though the large gap requires many magnetic excitations to be emitted, leading to an exponentially suppressed recombination rate, the relaxation process is fast as a consequence of the strong charge-spin coupling in 2D systems \cite{Lenarcic2012b,Lenarcic2014}.
In view of reproducing and understanding the dynamics of the entire dielectic function, that has been measured through broadband optical techniques, a linear-response formalism to calculate the time dependent optical conductivity $\sigma(\omega,t)$ within the $t$-$J$ model has been developed \cite{Lenarcic2014b}. The photoexcitation process is accounted for by assuming that the photons absorbed during the interactions with the pump pulse give rise to an instantaneously increase of the kinetic energy of the charge carriers immersed in the spin background. The main features experimentally observed, such as the transient increase of the Drude peak and the formation of midinfrared peaks are reproduced. As an interesting result, the optical sum rules approaches the equilibrium ones extremely fast, even though the time evolution and the final asymptotic behavior of the absorption spectra still reveal the dependence on the type of initial pump excitation \cite{Lenarcic2014b}. 

As a further step in reproducing the ultrafast dynamics in correlated insulators, the electron-phonon coupling is included through the $t$-$J$-Holstein model that includes the coupling with a dispersionless bosonic mode. The formation of spin-lattice polarons after a quantum quench, that simulates absorption of the pump pulse, has been also studied \cite{Golez2012,Kogoj2014}. While in the first stage the kinetic energy of the spin-lattice polarons relax towards its ground-state value, in the second longer stage an energy transfer between lattice and spin degrees of freedom via the charge carriers emerges. This results demonstrate that a direct spin-phonon coupling is not mandatory to quickly achieve the thermalization of the spin and phonon baths \cite{Kogoj2014}. While the $t$-$J$(-Holstein) model correctly describes the coupling to low-energy bosonic excitations, it intrinsically misses the relaxations processes across the correlation gap \cite{Lenarcic2014}. To solve this problem, many efforts are being put forward to solve the full Hubbard(-Holstein) model out of equilibrium, under suitable approximations. Calculations of the relaxation dynamics in a 2D Hubbard-Holstein model after the interaction with an ultrashort powerful light pulse have been recently performed on 8 lattice sites. At non-zero electron-phonon interaction, phonon and spin subsystems are found to oscillate at the period of the phonon mode \cite{Defilippis2012}.

As discussed in Sec. \ref{sec_theory} a breakthrough in this field could be achieved via the development of suitable methods to extend DMFT out-of-equilibrium \cite{Eckstein2008b} and calculate the dynamics of the relaxation process retaining the full Mott physics related to the Coulomb on-site repulsion, as long as the coupling to antiferromagnetic fluctuations \cite{Eckstein2014a} when at least four sites are considered. 

\subsubsection{Electron-phonon coupling in correlated materials}
\label{sec_ephresults}
As a consequence of the development of the BCS theory a huge effort has been put in clarifying whether a strong electron-phonon coupling could lead to high critical temperatures. Although in realistic systems, when the electron-phonon coupling $\lambda_{lat}$ overcomes a threshold value, the crystal structure becomes unstable and the conduction becomes dominated by polaron formation, the Eliashberg equations for the calculation of $T_c$ do not predict any upper bound for the possible critical temperature of metals. Therefore, since the discovery of high-temperature superconductivity in copper oxides, a reliable measurement of the electron-phonon coupling function has been considered crucial for understanding the physical mechanisms responsible for the pair formation. As discussed in Section \ref{sec_QPdynamics}, the state-of-the art spectroscopies, such as ARPES, Raman, optics and tunneling measurements have been applied to measure the electron-phonon coupling in correlated materials and high-$T_c$ superconductors and extract the electron-boson coupling function $\Pi(\Omega)$=$\alpha^2F(\Omega)$+$I^2\chi(\Omega)$. Nonetheless, the interplay of different bosonic degrees of freedom on similar energy scales makes it difficult to single out the electron-phonon constant and to estimate its strength. The application of time-resolved techniques to investigate the electron-phonon coupling was boosted by the seminal work of P. Allen \cite{Allen1987}, in which it is shown that the timescale of the relaxation of the effective electronic temperature $T_e$ is directly related to the frequency-integral of the coupling function $\Pi(\Omega)$. The quantitative estimation of the electron-phonon coupling relies on two major assumptions: i) the transient distribution of the phonon bath and charge carries can be described by the effective temperatures $T_{lat}$ and $T_e$, respectively, larger than the equilibrium temperature $T_0$; ii) the reflectivity-variation at a given wavelength is proportional to the effective electronic temperature, i.e. $\delta R/R$=$\delta T_e/T_0$. 

\begin{figure}[t]
\begin{centering}
\includegraphics[width=0.8\textwidth]{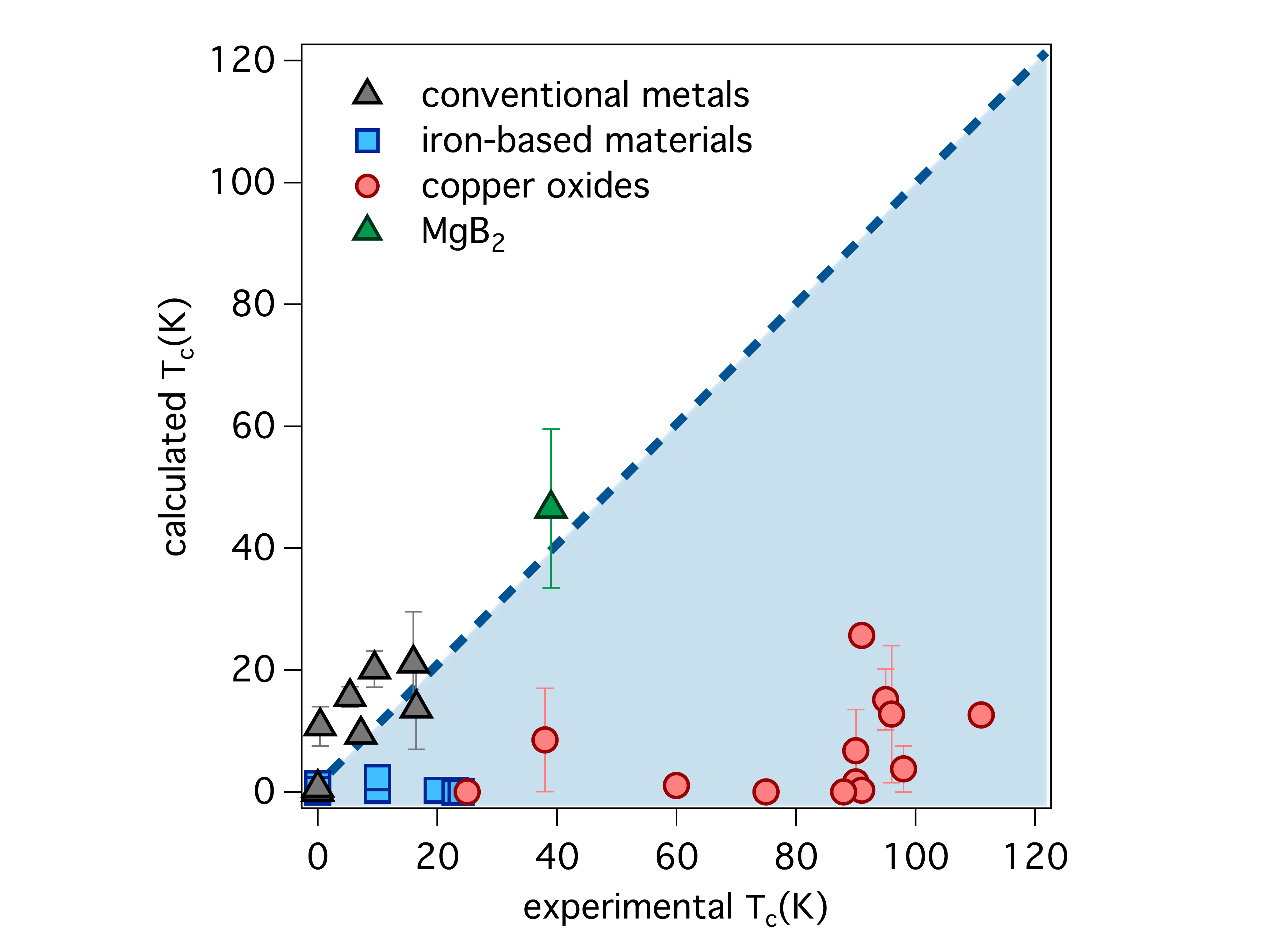}
\caption{The maximum critical temperature, calculated through Eq. \ref{eq_Tc} for different classes of superconducting materials, is reported as a function of the real $T_c$ of the materials. The electron-phonon coupling, $\lambda_{lat}$, determined through time-resolved spectroscopies, has been used to estimate the maximum critical temperatures. The grey triangles represent the conventional metals: Cu ($\lambda_{lat}\simeq$0.08), Au ($\lambda_{lat}\simeq$0.12), Cr ($\lambda_{lat}\simeq$0.11-0.12), W ($T_c$=0.012 K, $\lambda_{lat}\simeq$0.14-0.26), V ($T_c$=5.38 K, $\lambda_{lat}\simeq$0.68-0.79), Nb ($T_c$=9.5 K, $\lambda_{lat}\simeq$0.87-1.16), Ti ($T_c$=0.39 K, $\lambda_{lat}\simeq$0.43-0.58), Pb ($T_c$=7.19 K, $\lambda_{lat}\simeq$1.45-1.51), NbN ($T_c$=16 K, $\lambda_{lat}\simeq$0.53-0.95), V$_3$Ga ($T_c$=1 K, $\lambda_{lat}\simeq$0.45-0.83) \cite{Brorson1990, Gadermaier2010}. The blue squares and red circles indicate the critical temperatures of the iron-based (122 and 1111 families) superconductors and copper oxides listed in Tables \ref{table_Tc_cuprates} and \ref{table_Tc_ironbased}.}
\label{fig_criticaltemperatures}
\end{centering}
\end{figure}

The time-resolved reflectivity measurements (see Fig. \ref{fig_Brorson}) have been promptly employed \cite{Brorson1990} to measure $\lambda_{lat}$=$2\int\alpha^2F(\Omega)/\Omega\;d\Omega$ in many metals and conventional superconductors (Cu, Au, Cr, W, V, Nb, Ti, Pb, NbN, V$_3$Ga). The dynamics of $\delta R/R$ in these materials exhibits a single exponential decay which ranges from 0.1 ps to 1 ps for different systems. Using Eq. \ref{eq_tau_e_lat} and its further evolution (Eq. \ref{eq_tau_e_lat_K}), which accounts for the non-thermal nature of the transient electronic distribution, the second momentum of the Eliashberg coupling ($\lambda_{lat}\langle\Omega^2\rangle$) is directly extracted from the relaxation dynamics. Assuming the values of $\langle\Omega\rangle$ reported in the literature, the following $\lambda_{lat}$ values are extracted: i) 0.1-0.15 for non-superconducting metals (Cu, Au, Cr), ii) 0.15-0.5 for very low-$T_c$ superconductors (W, Ti) and iii) 0.7-1.5 for moderate temperature superconductors (V, Nb, Pb, NbN, V$_3$Ga). In particular, the case of Pb is interesting, for it exhibits the slowest relaxation dynamics that corresponds to the lowest value ($\sim$46 meV$^2$) of $\lambda_{lat}\langle\Omega^2\rangle$. The small value of the second momentum is the consequence of the small energy scale of the phonon modes involved in the relaxation. When the value $\langle\Omega\rangle \sim$5.6 meV is considered, the largest coupling ($\lambda_{lat} \simeq$1.5) among the conventional metals is obtained. The measured value of $\lambda_{lat}$ can be in turn used to estimate the maximum critical temperature attainable. Using the value $\mu^*$=0 and assuming that all the phonon modes couple to the charge carriers in the $s$-wave channel necessary for the pairing, Eq. \ref{eq_Tc} provides the upper bound for $T_c$ which could be reached, in principle, in the most favourable case. Large values of $T_c$ result from the combination of both a strong coupling and a particularly high-energy scale of the phonon modes. As an example, the relatively low energy scale of the phonons in Pb leads to a moderate value of the estimated $T_c$ (6-10 K), despite the large value of the coupling. The plot in Fig. \ref{fig_criticaltemperatures} displays the calculated maximum critical temperatures for phonon-mediated superconductivity as a function of the real $T_c$ of the superconductors. As expected, almost all the BCS materials lye in the upper sector of the plot and evolve along its diagonal, demonstrating the reliability of the values of $\lambda_{lat}$ measured through time-resolved techniques.    

Even more interesting is the case of MgB$_2$, which is the phonon-mediated superconductor with the largest $T_c$ (40 K) known. As a consequence of the intrinsic anisotropy of the layered lattice structure, the charge carriers result strongly-coupled with a stretching mode at $\sim$70 meV which is expected \cite{Kong2001} to provide the largest contribution to the total $\alpha^2F(\Omega)$ function. In fact, recent high-resolution pump-probe measurements \cite{DalConte2015} evidenced a twofold dynamics which is related to the coupling with the 70 meV mode ($\tau_{e-SCP}\sim$90 fs) and, subsequently ($\tau_{e-lat}\sim$500 fs), with the rest of lattice. According to the fit with the 3-temperature model, in which the selective coupling with a subset of phonon modes is introduced in Eqs. \ref{eq_4TMeqs_1}-\ref{eq_4TMeqs_4}, the SCP modes correspond to a small fraction $f$=0.15-0.22 of the total lattice modes but provide a coupling strength $\lambda_{SCP}$=0.53-0.75. Substituting these values in Eq. \ref{eq_Tc}, the range of values 30$<T_c<$60 K is obtained, which is in agreement with the real value of $T_c$. This result further supports the extension of P-p techniques to extract the electron-phonon coupling in intrinsically anisotropic systems.    
\begin{figure}[t]
\begin{centering}	
\includegraphics[width=0.9\textwidth]{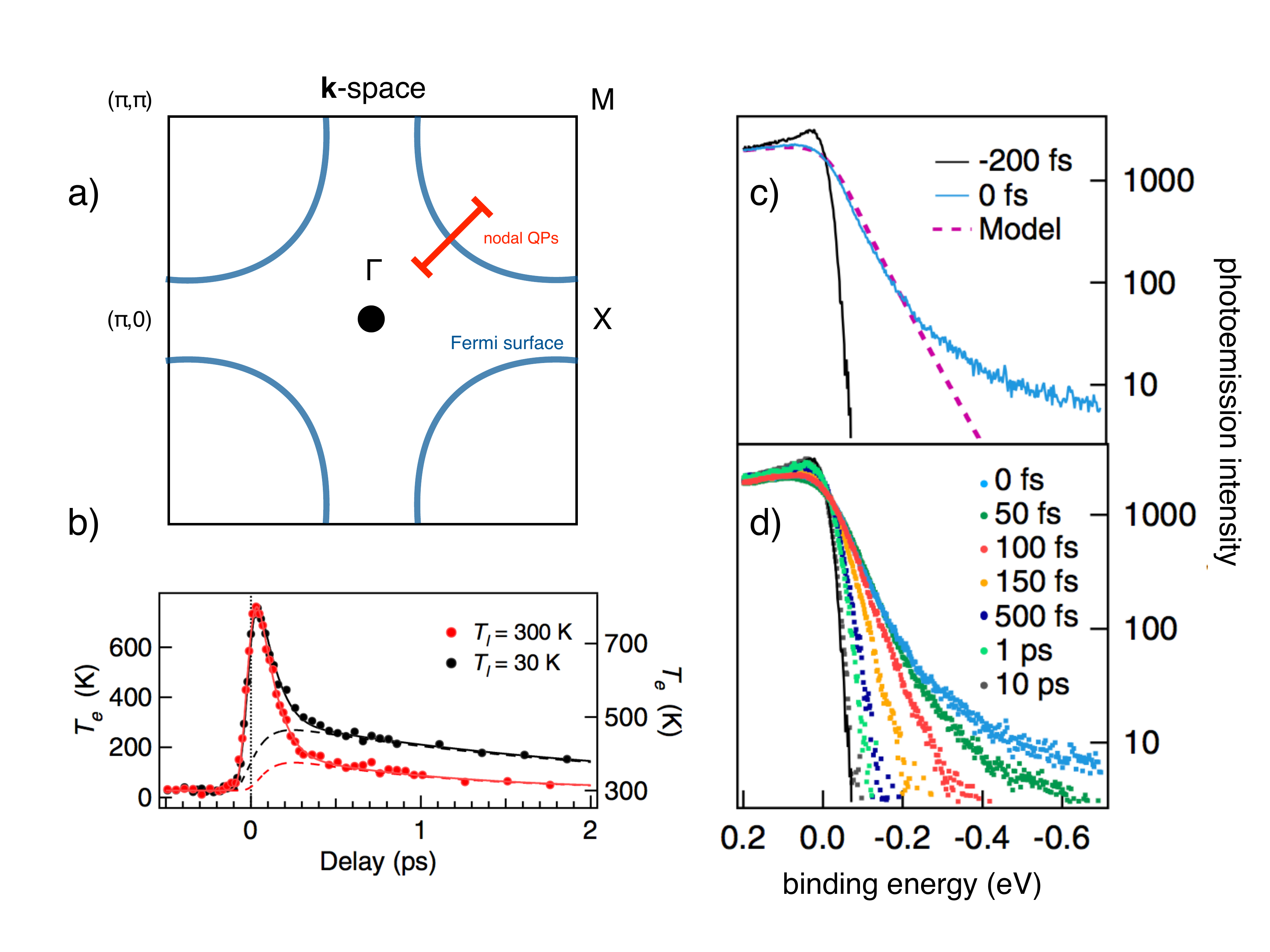}
\caption{Dynamics of the time-resolved ARPES spectra in optimally-doped Bi2212 ($T_c$=91 K). a) Prototypical two-dimensional Brilloiun zone of copper oxides. The time-resolved ARPES experiment probes the transient occupation along a k-space cut perpendicular to the Fermi surface in the nodal region. b) Dynamics of the effective electronic temperature at different equilibrium emperatures before the pump excitation. c) Transient quasiparticle occupation at the maximum overlap between the pump and probe pulses (blue line, $t$=0) contrasted to the equilibrium Fermi-Dirac QP distribution (black line, $t<$0). The dashed red line is the fit to the data of a hot Fermi-Dirac distribution at the effective electronic temperature $T_e$. d) Picosecond evolution of the non-equilibrium nodal QP distribution. Taken from \cite{Perfetti2007}.}
\label{fig_2TMphotoemission}
\end{centering}
\end{figure}  

In principle, the same technique can be extended to correlated materials, provided the electron dynamics can be described as the interaction between a fermionic quasiparticle gas and an external boson bath. Although this picture is clearly inadequate to describe the dynamics in the charge-transfer insulators and in the weakly-doped compounds, a more conventional picture is recovered as the density of the charge carriers is progressively increased \cite{Peli2015,Fournier2010}. Therefore, as far as the underdoped region of the phase diagrams and the pseudogap state are avoided, the concepts underlying the effective temperature model can be extended to interpret the dynamics of moderately-doped cuprates and other correlated materials. In fact, a wealth of non-equilibrium spectroscopies, such as time-resolved optics, photoemission and electron or X-ray diffraction have been applied to investigate the electron-phonon coupling problem in copper oxides. Addressing this issue is of particular relevance, since it is intimately connected to the pairing in unconventional superconductors \cite{Li2014}. Considering the widely studied problem of the charge scattering in cuprates there are many evidences of an important role played by the electron-phonon scattering \cite{Kresin2009,Maksimov2010}. In particular, considering that the conduction is confined in the Cu-O planes, the modes which are typically expected to more effectively couple to the charge carriers are the out-of-plane O buckling modes \cite{Devereaux2004}, which involve small momentum transfers and couple strongly to electronic states near the antinode, the in-plane Cu-O breathing modes \cite{Devereaux2004}, which involve large momentum transfers and couple strongly to nodal electronic states, and the apical oxygen stretching modes, which modulate the low-energy in-plane electric properties through the $p_z$ orbitals \cite{Yin2009}. Furthermore, the electron-phonon coupling is expected to strenghten in the underdoped region of the phase diagram, as a consequence of charge inhomogeneity associated to stripes \cite{Reznik2006} and other possible charge-orders. Recent results also indicated a weak but distinct isotope effect of the $\sim$70 meV kink observed by ARPES on Bi$_2$Sr$_2$CaCu$_2$O$_{8+\delta}$, in which the isotope substitution $^{16}$O$\rightarrow ^{18}$O has been performed.

\begin{table}
\label{table_Tc_cuprates}
\tbl{Electron-phonon coupling in copper oxides from time-resolved techniques. 1L: one-layer; 2L: two-layers; 3L: three-layers. UD: underdoped; OP: optimally doped; OD: overdoped.} {\begin{tabular}{@{}lccccccc}
\toprule
Material
& $T_c$ (K) &$p$ & Ref. & $F_p$ (mJ/cm$^2$) & $\tau_{e-lat}$ (fs)
& $\lambda_{lat}\langle \Omega^2 \rangle$ (meV$^2$) & $\lambda_{lat}$ \\ 
\colrule
1L-UD La$_{1.9}$Sr$_{0.1}$CuO$_4$$^d$ &25 & &\cite{Mansart2013} &5 &- &13$^f$
& 0.04$^n$ \\
& & & &16 &- &24$^f$
& 0.08$^n$ \\
1L-OP La$_{1.85}$Sr$_{0.15}$CuO$_4$$^a$ &38 & &\cite{Gadermaier2010} &0.07-0.5 & 45$\pm$8 &800$\pm$200$^e$
& 0.13-0.5$^g$ \\
1L-OD La$_{1.79}$Sr$_{0.21}$CuO$_4$$^d$ &32 & &\cite{Mansart2013} &16 & - &30$^f$
& 0.09$^n$ \\
1L-OP HgBa$_2$CuO$_{4.1}$$^a$ &98 & &\cite{Gadermaier2014} &0.02 &62  &580$^e$
& 0.09-0.36$^g$ \\

\colrule
2L-UD YBa$_2$Cu$_3$O$_{6.5}$$^a$ &60 & &\cite{Gadermaier2010} &0.07-0.5 &100$\pm$20 &400$\pm$100$^e$
& 0.06-0.25$^g$ \\
2L-UD YBa$_2$Cu$_3$O$_{6.9}$$^a$ &90 & &\cite{Gadermaier2014} &0.02 &77 &450$^e$
& 0.07-0.28$^g$ \\
2L-UD Bi$_2$Sr$_2$CaCu$_2$O$_{8+\delta}$$^c$ &56 & &\cite{Carbone2008} &20 &- &2500$^{f,h}$
& 1$^m$ \\
& & & & &- &300$^{f,l}$
& 0.12$^m$ \\
2L-UD Bi$_2$Sr$_2$CaCu$_2$O$_{8+\delta}$$^a$ &75 & &\cite{Chia2013} &10$^{-4}$ &- &160$^g$
& 0.025-0.1$^g$ \\
2L-OP Bi$_2$Sr$_2$CaCu$_2$O$_{8.14}$$^a$ &90 & &\cite{Gadermaier2014} &0.02 &49 &720$^e$
& 0.11-0.45$^g$ \\
2L-OP Bi$_2$Sr$_2$CaCu$_2$O$_{8+\delta}$$^b$ &91 & &\cite{Perfetti2007} &0.2 &110 &300-380$^f$
& 0.08-0.19$^g$ \\
2L-OP Bi$_2$Sr$_2$CaCu$_2$O$_{8+\delta}$$^c$ &91 & &\cite{Carbone2008} &20 &290$^h$ &1380$^f$
& 0.55$^m$ \\
 & & & & &900$^i$ &450$^f$
& 0.18$^m$\\
 & & & & &2000$^l$ &200$^f$
& 0.08$^m$\\
2L-OP Bi$_2$Sr$_2$CaCu$_2$O$_{8+\delta}$$^a$ &95 & &\cite{Chia2013} &10$^{-4}$ &- &640$^g$
& 0.4$^g$ \\
2L-OD Bi$_2$Sr$_2$CaCu$_2$O$_{8+\delta}$$^a$ &88 & &\cite{Chia2013} &10$^{-4}$ &- &160$^g$
& 0.1$^g$ \\
2L-OP Bi$_2$Sr$_2$Ca$_{0.92}$Y$_{0.08}$Cu$_2$O$_{8+\delta}$$^a$ &96 & &\cite{DalConte2012} &0.01 &200 &1400$\pm$700$^{o}$
& 0.4$\pm$0.2$^o$ \\
\colrule
3L-OP Bi$_2$Sr$_2$Ca$_2$Cu$_3$O$_{10+\delta}$$^c$ &111 & &\cite{Carbone2008} &20 &- &1000$^f$
& 0.4$^m$\\
   \botrule
  \end{tabular}}
\tabnote{$^{\rm a}$Non-equilibrium optical spectroscopy}
\tabnote{$^{\rm b}$Non-equilibrium photoemission spectroscopy}
\tabnote{$^{\rm c}$Time-resolved electron diffraction}
\tabnote{$^{\rm d}$Time-resolved X-ray diffraction}
\tabnote{$^{\rm e}$Evaluated through Eq. \ref{eq_tau_e_lat_K}}
\tabnote{$^{\rm f}$Evaluated through Eq. \ref{eq_tau_e_lat}}
\tabnote{$^{\rm g}$The reported $\lambda$ values correspond to the 80-40 meV energy range of the strongly coupled phonons}
\tabnote{$^{\rm h}$Pump polarization at 0$^{\circ}$ with respect to the Cu-O bond direction}
\tabnote{$^{\rm i}$Pump polarization at 22$^{\circ}$ with respect to the Cu-O bond direction}
\tabnote{$^{\rm l}$Pump polarization at 45$^{\circ}$ with respect to the Cu-O bond direction}
\tabnote{$^{\rm m}$$\lambda$ calculated assuming $\langle \Omega \rangle$=50 meV (B$_{1g}$ buckling mode)}
\tabnote{$^{\rm n}$$\lambda$ calculated assuming $\langle \Omega \rangle$$\simeq$17 meV (E$_{g}$ apical modes)}
\tabnote{$^{\rm o}$$\lambda$ and $\lambda\langle \Omega^2 \rangle$ calculated through the effective temperature model in the integro-differential form (Eqs. \ref{eq_4TMeqs_1}-\ref{eq_G_ETM}) and by the direct integration of the bosonic function $\Pi(\Omega)$}
\label{eph_coupling}
\end{table}

\begin{table}
\label{table_Tc_ironbased}
\tbl{Electron-phonon coupling in iron-based superconductors and parent compounds (PC) from time-resolved techniques.} {\begin{tabular}{@{}lccccccc}
\toprule
Material
& $T_c$ (K) &$p$ & Ref. & $F_p$ (mJ/cm$^2$) & $\tau_{e-lat}$ (fs)
& $\lambda_{lat}\langle \Omega^2 \rangle$ (meV$^2$) & $\lambda_{lat}$ \\ 
\colrule
122-PC EuFe$_2$As$_2$$^b$ &- & &\cite{Rettig2012,Rettig2013,Avigo2013} &0.05-0.8 &- &56-120$^{c,d}$
& 0.1-0.23$^e$ \\
122-PC SrFe$_2$As$_2$$^a$ &- & &\cite{Stojchevska2010} &0.01-0.1 &- &110$^{c}$
& 0.25$^e$ \\
122-PC BaFe$_2$As$_2$$^b$ &- & &\cite{Rettig2013,Avigo2013} &0.05-0.8 &- &28-46$^{c,d}$
& 0.05-0.09$^e$ \\
122-PC BaFe$_2$As$_2$$^a$ &- & &\cite{Gadermaier2014} &0.02 &300 &110$^{c}$
& 0.18$^e$ \\
122 BaFe$_{1.95}$Co$_{0.05}$As$_2$$^a$ &- & &\cite{Gadermaier2014} &0.02 &330 &110$^{c}$
& 0.18$^e$ \\
122 BaFe$_{1.90}$Co$_{0.1}$As$_2$$^a$ &20 & &\cite{Gadermaier2014} &0.02 &320 &110$^{c}$
& 0.18$^e$ \\
122 BaFe$_{1.86}$Co$_{0.14}$As$_2$$^a$ &23 & &\cite{Gadermaier2014} &0.02 &300 &110$^{c}$
& 0.18$^e$ \\
122 BaFe$_{1.85}$Co$_{0.15}$As$_2$$^b$ &23 & &\cite{Rettig2013,Avigo2013} &0.05-0.8 &- &46-55$^{c,d}$
& 0.09-0.1$^e$ \\
122 BaFe$_{1.84}$Co$_{0.16}$As$_2$$^a$ &24 & &\cite{Mansart2010} &1-3 &- &59-66$^{d}$
& 0.11-0.12$^e$ \\
122 BaFe$_{1.78}$Co$_{0.22}$As$_2$$^a$ &10 & &\cite{Gadermaier2014} &0.02 &300                                                                                                                                                                                                                                                                                                                                                                                                                                                                                                                                                                                                                                                                                                                                                                                                                                                                                                                                                                                                                                                                                                                                                                                                                                                                                                                                                                                                                                                                                                                                                                                                                                                                                                                                                                                                                                                                                                                                                                                                                                                                                                                                                                                                                                                                                                                                                                                                                                                                                                                                                                                                                                                                                                                                                                                                                                                                                                                                                                                                                                                              &110$^{c}$
& 0.18$^e$ \\

\colrule  
1111 SmFeAs$_{00.8}$F$_{0.2}$$^a$ &10 & &\cite{Gadermaier2014} &0.02 &190                                                                                                                                                                                                                                                                                                                                                                                                                                                                                                                                                                                                                                                                                                                                                                                                                                                                                                                                                                                                                                                                                                                                                                                                                                                                                                                                                                                                                                                                                                                                                                                                                                                                                                                                                                                                                                                                                                                                                                                                                                                                                                                                                                                                                                                                                                                                                                                                                                                                                                                                                                                                                                                                                                                                                                                                                                                                                                                                                                                                                                                              &180$^{c}$
& 0.29$^e$ \\
1111 SmFeAsO$^a$ &- & &\cite{Mertelj2010} &0.01-0.1 &180                                                                                                                                                                                                                                                                                                                                                                                                                                                                                                                                                                                                                                                                                                                                                                                                                                                                                                                                                                                                                                                                                                                                                                                                                                                                                                                                                                                                                                                                                                                                                                                                                                                                                                                                                                                                                                                                                                                                                                                                                                                                                                                                                                                                                                                                                                                                                                                                                                                                                                                                                                                                                                                                                                                                                                                                                                                                                                                                                                                                                                                              &135$\pm$10$^{c}$
& 0.2$^f$ \\

   \botrule
  \end{tabular}}
\tabnote{$^{\rm a}$Non-equilibrium optical spectroscopy}
\tabnote{$^{\rm b}$Non-equilibrium photoemission spectroscopy}
\tabnote{$^{\rm c}$Evaluated through Eq. \ref{eq_tau_e_lat_K}}
\tabnote{$^{\rm d}$Evaluated through Eq. \ref{eq_tau_e_lat}}
\tabnote{$^{\rm e}\lambda$ calculated assuming $\langle \Omega \rangle$=23 meV (A$_{1g}$ Raman active mode)}
\tabnote{$^{\rm f}$$\lambda$ calculated assuming $\langle \Omega \rangle$=25 meV}
\label{eph_coupling_iron}
\end{table}

Profound insights into the charge relaxation dynamics in cuprates are given by time-resolved ARPES experiments (see Fig. \ref{fig_2TMphotoemission}), in which the transient electronic occupation is probed during the relaxation process triggered by a 1.5 eV ultrafast pump pulse. Due to the energy-momentum conservation in the photoemission process, only the nodal region of the BZ is explored by the 6 eV probe photons. In Fig. \ref{fig_2TMphotoemission}c) and d) the transient occupation of the states in the nodal region of Bi$_2$Sr$_2$CaCu$_2$O$_{8+\delta}$ is shown in the 0.2$<E_F$-$E<$-0.7 energy range. The QP distribution quickly evolves from a non-thermal distribution, characterized by a flat population extending up to 0$<E_F$-$E<E_F$+$\hbar\omega$, to a hot-Fermi-Dirac function at the effective temperature $T_e>T_0$. In Fig. \ref{fig_2TMphotoemission}c) the cooling dynamics of the QP is reported by fitting the hot Fermi-Dirac at the different P-p delays. Similarly to the case of MgB$_2$, the relaxation exhibits two different dynamics. The first ($\sim$100 fs) is related to the fast and effective energy exchange with the strongly-coupled lattice modes, while the second ($\sim$1-2 ps) accounts for the thermalization with the entire phonon spectrum. The 3-temperature model fitting provides a coupling to SCP smaller than 0.2. At the same time, the presence of a weak non-thermal tail for $t<$100 fs triggered a debate about the possibility of defining a single effective electronic temperature which describes the relaxation dynamics. Even if the number of electrons in the high-energy tail is negligible (note the log-scale in Fig. \ref{fig_2TMphotoemission}d)) as compared to the total number of electrons, they can directly and effectively exchange energy with the lattice modes, thus altering the dynamics described the 3TM. This picture has been recently supported by time-resolved measurements in which a broadband midinfrared pulse probes the transient lineshape of the Raman-active mode at 70 meV (see Sec. \ref{sec_infrared_active_modes}) in optimally-doped YBCO \cite{Pashkin2010}. The apex oxygen vibration is strongly excited within 150 fs, demonstrating that the strongly-coupled modes can absorb a portion of the pump energy on the same timescale of the quasiparticles thermalization. However, when the strength of the coupling  is evaluated through Eq. \ref{eq_tau_e_lat_K}, which accounts for the non-thermal nature of the QP and boson occupations during the first stages of the relaxation \cite{Baranov2014}, an even smaller value is obtained. The decrease of the estimated $\lambda_{lat}\langle\Omega^2\rangle$ is a consequence of the fact that the ratio between the values obtained through Eqs. \ref{eq_tau_e_lat_K} and \ref{eq_tau_e_lat} is 2$T_{lat}$/$T_e$ which is, for reasonable values of the experimental parameters, smaller than unity. Even though determining the effective temperature of the non-equilibrium Fermi-Dirac distribution is a difficult task and new methods to accurately extracts both the temperature and the position of the Fermi level for a hot carrier distribution are being developed \cite{Ulstrup2014}, time-resolved photoemission experiments suggest that the value of the electron-phonon coupling in optimally-doped copper oxides is similar to that measured in very low-$T_c$ metals. To support or confute these results, many of the state-of-the art non-equilibrium techniques have been applied to investigate the electron-phonon coupling in doped cuprates. Time-resolved reflectivity \cite{Gadermaier2010} on La$_{1.85}$Sr$_{0.15}$CuO$_4$ and YBa$_2$Cu$_3$O$_{6.5}$ supported similar values, i.e., $\lambda_{lat}\langle \Omega^2 \rangle$=800$\pm$200 meV$^2$ and 400$\pm$100 meV$^2$, respectively. Besides, time-resolved electron diffraction revealed a possible anisotropy in the electron-phonon coupling                                                                                                                                                                                                         \cite{Carbone2008} of optimally-doped Bi2212. In particular, when the pump polarization is set parallel to the direction of the Cu-O bonds (corresponding to the nodal direction in \textbf{k}-space) the largest coupling ($\lambda_{lat}$=0.55) is measured. The coupling progressively decreases as the angle between the polarization and the Cu-O bonds increases, reaching the minimum ($\lambda_{lat}$=0.08) at 45$^{\circ}$, which corresponds to the antinodal direction with the maximum superconducting gap. The average value of the coupling ($\lambda_{lat}\sim$0.3) is consistent with the \textbf{k}-space integrated one (see Eq. \ref{eq_k_integrated_glue}), as probed by non-equilibrium optical spectroscopy. Slightly larger values for the coupling ($\lambda_{lat}\sim$0.4), with no significant anisotropy, have been measured \cite{Carbone2008} in double-layer compounds (Bi2223).
Furthermore, the e-phonon coupling in overdoped LSCO has been investigated by X-ray diffraction experiments \cite{Mansart2013}, in which the fluence is varied from 5 to 27 mJ/cm$^2$. The results evidenced a $\lambda\langle \Omega^2 \rangle$ ranging from 13 to 56 meV$^2$, as a consequence of the temperature-induced modification of the QP distribution and of the density of states at the Fermi level, in agreement with DFT calculations.  

Taken all together, the outcomes of the non-equilibrium spectroscopies (collected in table \ref{table_Tc_cuprates}) provide a comprehensive picture of the electron-phonon coupling in copper oxides. As for conventional superconductors, Eq. \ref{eq_Tc} can be used to estimate the upper bound of the critical temperature achievable when considering the measured $\lambda_{lat}$ and assuming that the whole coupling is due to the interaction with a single mode at the frequency $\langle \Omega \rangle$. Despite the many optimistic approximations, the maximum $T_c$ estimated for cuprates never exceeds their real critical temperature, being almost always confined in the $<$20 K range. As shown in Figure \ref{fig_criticaltemperatures}, this result remarks a dramatic difference in respect to conventional superconducting metals, in which $\lambda_{lat}\sim$ accounts for the critical temperature of the system. The same techniques have been also extensively applied to iron-based superconductors and parent compounds. The results, reported in Table \ref{table_Tc_ironbased}, evidence values of the electron-phonon coupling very similar to those measured in cuprates.

Even though the precise estimation of the total electron-phonon coupling relies on the approximations contained in the effective-temperature models and their further developments, non-equilibrium spectroscopies demonstrate that the timescale of the energy-exchange between the electrons and the lattice in unconventional high-$T_c$ systems is similar to that measured in conventional superconducting metals. By calculating the maximum $T_c$ attainable within the Eliashberg theory (see Eq. \ref{eq_Tc}) and using as input the measured $\lambda_{lat}$, we observe that copper oxides and iron-based materials cluster in the bottom-right corner of the plot in Figure \ref{fig_criticaltemperatures}, thus demonstrating that a simple phonon-mediated mechanism cannot account for the superconducting pairing.

\begin{figure}
\includegraphics[bb= 0 150 850 520, keepaspectratio, width=1\textwidth]{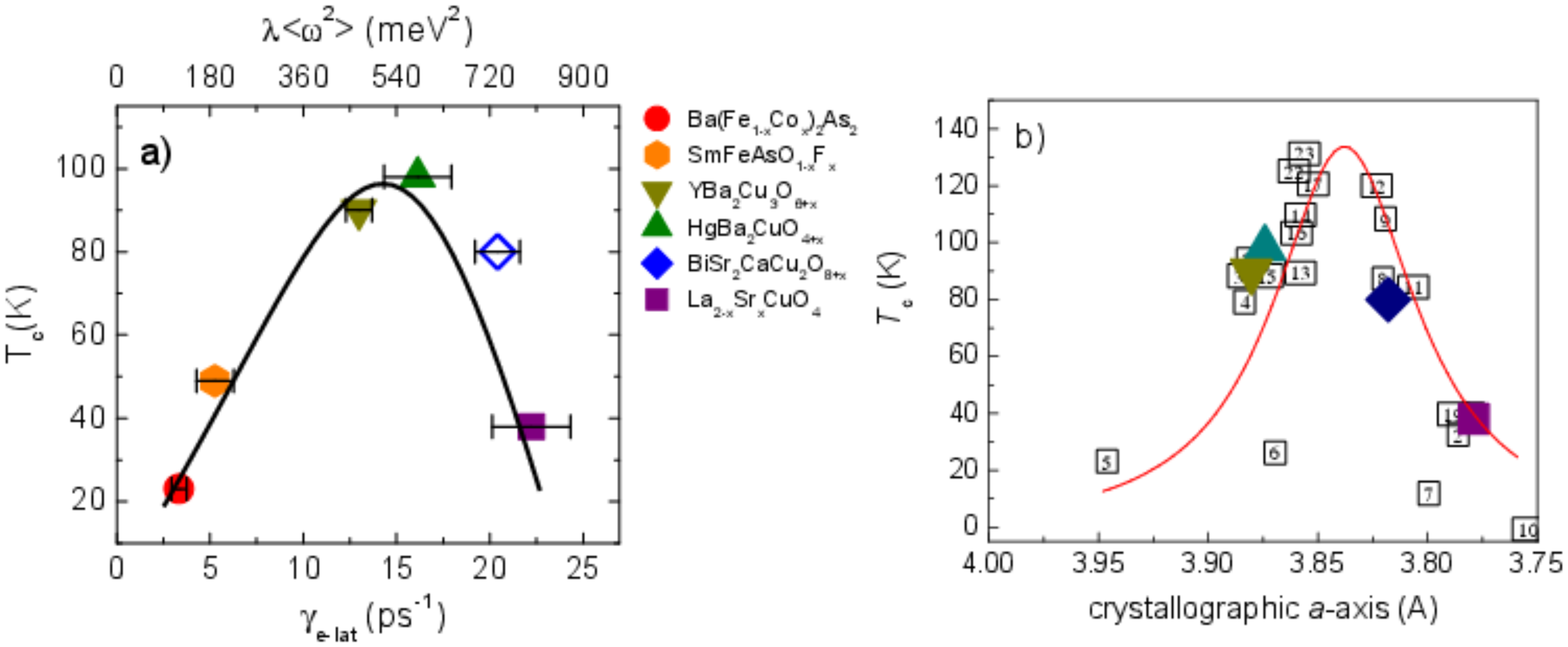}
\label{lambda_gadermaier}
\protect\caption{a) The maximum superconducting transition temperature $T_{c}$ as
a function of the relaxation rate $\gamma_{e-lat}$(left). b) $T_{c}$
as a function of a lattice constant $a$ for cuprates (Note the correspondence
of the symbols (and compounds) in the two panels) \cite{Gadermaier2014}}
\end{figure}
Surprisingly, in cuprates there appears to be a systematic, but non-monotonic
dependence of $T_{c}$ on the relaxation rate $\gamma_{e-lat}=1/\tau_{e-lat}$
as shown in Fig. \ref{lambda_gadermaier}a). The behaviour is clearly well beyond BCS
or Eliashberg electron-phonon coupling theory which predict a monotonically
increasing $T_{c}$ with increasing coupling constant. The behaviour
has been discussed in terms of polaronic effects \cite{Gadermaier2014},
which lead to the entanglement of electronic and lattice degrees of
freedom, as will be discussed in the next section. Insight into the
role of electronic excitations on the superconducting critical temperature
is provided by the remarkable correspondence between the dependence
of $T_{c}$ on $\gamma_{e-lat}=1/\tau_{e-lat}$ and the
in-plane lattice constant $a$, as shown in Fig. \ref{lambda_gadermaier}b). Since $a$ is directly related to the local strain, the correspondence,
and particularly the apparent singular behaviour of $T_{c}$ with
lattice constant in Fig. \ref{lambda_gadermaier}b), implies a critical strain value associated
with an instability at a Cu-Cu distance of 1.92 \AA. 

\subsubsection{Ultrafast electron-boson coupling and the magnetic degrees of freedom}
\label{sec_ebosresults}
Electron-phonon coupling constitutes only a subset of the decay channels for the photoexcited charge-carriers in correlated materials. As discussed in Sections \ref{sec_QPscattering} and \ref{sec_magnons}, the electron-boson coupling function is characterized by a strong peak at $\sim$60 meV and a continuum extending up to 350 meV. This rather featureless part of the coupling function is compatible with the coupling to short-range anti-ferromagnetic fluctuations, that persist far in the over-doped region of the phase diagram\cite{Dahm2009,letacon2011,letacon2013,dean2013,dean2013b}.

Simple arguments can be used to estimate the timescale of the coupling to bosonic fluctuations of magnetic origin in correlated materials. Considering the Hubbard model, the instantaneous virtual hopping of holes into already occupied sites favours an antiferromagnetic coupling $J$=4$t_h^2/U$ between neighbouring sites, where $t_h$ is the nearest-neighbour hopping energy. The spectrum of the antiferromagnetic fluctuations has a high-energy cutoff of 2$J\sim$200-300 meV which naturally brings in a characteristic retarded timescale of the order of $\hbar$/2$J$=2-3 fs. From another point of view, the elementary time ($\hbar$/$t_h \sim$ 2fs) associated with the Cu$-$O$-$Cu hole-hopping process inevitably leads to the creation of local AF excitations. All these evidences point to an extremely fast dynamics of the electron-magnetic fluctuations coupling that, until recently, was concealed by the temporal resolution of pump-probe techniques.

For this reason, the direct experimental study of the coupling to magnetic fluctuations has been anticipated by theoretical works that modeled the coupling of out-of-equilibrium charge carriers to local AF fluctuations using an approach similar to that discussed in Sec. \ref{sec_CTinsulator} for the CT insulators. For this purpose, the starting point is the simplified $t$-$J$ model on a square lattice that retains the complete dynamics at the energy scale $J$. Considering that in the $t$-$J$ model the double occupation of Cu sites is projected out, the photoexcitation process can be simplified as a quantum quench that instantaneously rises the kinetic energy of the holes (electrons) already present in the system. It was readily pointed out\cite{Golez2014} that also the magnetically-mediated relaxation dynamics is a two-step process. The initial ultrafast relaxation is regulated by the hopping integral $t_0$ and the exchange interaction $J$ and can be considered as a local process that involves many antiferromagnetic excitations, each absorbing a fraction of $J$ of energy. The timescale of this process is estimated to be of the order of $\tau\sim (\hbar/t_0)(J/t_0)^{-2/3}\simeq$4 fs for realistic values of $t_0\sim$360 meV and $J\sim$120 meV. The second step is slower and involves the coupling to collective magnetic excitations (magnons) that carry away the local excess energy. The rate of this magnon-mediated cooling depends on the magnon group velocity that is in turn regulated by the details of the magnon dispersion. Even though this model is strictly valid for a single hole in an infinite AF background, a first step to mimic the consequences of the chemical doping is to limit the average number of antiferromagnetic bonds that are available for the relaxation process. This can be achieved by limiting the propagation of the photo-excited hole to a finite cluster, whose size is chosen so as to reproduce the actual density of the chemically-doped charge carriers \cite{DalConte2015}.

Similar results have been obtained by considering the solution of the full Hubbard model with a non-equilibrium version \cite{Eckstein2014} of the dynamical cluster approximation (DCA) for capturing the effects of short-range correlations. The relaxation rate of the high-energy photo-doped charge carriers (i.e. those that are close to the upper edge of the Hubbard band) in the paramagnetic phase is of the order of 10-20 fs and it scales with the strength of the nearest-neighbour spin correlations, further supporting the direct role played by local AF excitations. The relaxation process becomes more complex at finite doping, since the direct charge-charge interactions open additional scattering channels that lead to a faster relaxation of the photo-excited holes (electrons). However, the direct interactions among the carriers are not expected to dramatically influence the time necessary to exchange energy with the boson bath \cite{DalConte2015}, that is ultimately regulated by the electron-boson coupling (see Secs. \ref{sec_QPdynamics} and \ref{sec_basicconceptsNES}).

The coupling to dispersionless phonons can be accounted for by the Holstein model (see Sec. \ref{sec_theory}), thus opening the way to the study of the competition between the magnetic and phononic decay channels. As a first step, this problem has been tackled in a finite one-dimensional chain by diagonalization in a limited functional space (LFS)\cite{Dorfner2015}. The results of time evolution of the electron and phonon distribution have been benchmarked against the relaxation dynamics obtained from the Boltzmann equation. Among the different important results, we report the observation that, in the weak-coupling regime, the natural time unit to measure the relaxation with the bosons is 1/$\omega_0$, where $\omega_0$ is the typical frequency of the bosons. The relaxation dynamics $\tau$ is thus regulated by the relation $\tau\omega_0$=(8/$\pi$)($t_0$/$\lambda\omega_0$), $t_0$ and $\lambda$ being the hopping amplitude and the electron-boson coupling. Considering that in the weak-coupling $\lambda\leq$0.5, the $\tau\omega_0$ product is always larger than one, i.e., the relaxation dynamics is always slower than the inverse energy-scale of the boson involved in the process. As an example, considering a realistic value for the copper oxides ($t_0\sim$350 meV), we obtain for the electron-phonon coupling $\tau\omega_0\sim$30, where we have considered the upper bound $\lambda$=0.4, as determined by pump-probe measurements (see Sec. \ref{sec_ephresults}). This value corresponds to a relaxation dynamics of several times the inverse phonon energy ($(\hbar\omega_0)^{-1}\sim$8 fs, for $(\hbar\omega_0)\sim$80 meV), that is of the order of $\sim$250 fs. This value can dramatically decrease when AF excitations are considered. Since the spectrum of magnetic excitations extends up to about 300 meV, i.e. about 4 times the phonon energy scale, the relaxation towards the magnetic degrees of freedom is expected to be at least 4 times faster assuming the same coupling constant.    

\begin{figure}[t]
\begin{centering}
\includegraphics[width=1\textwidth]{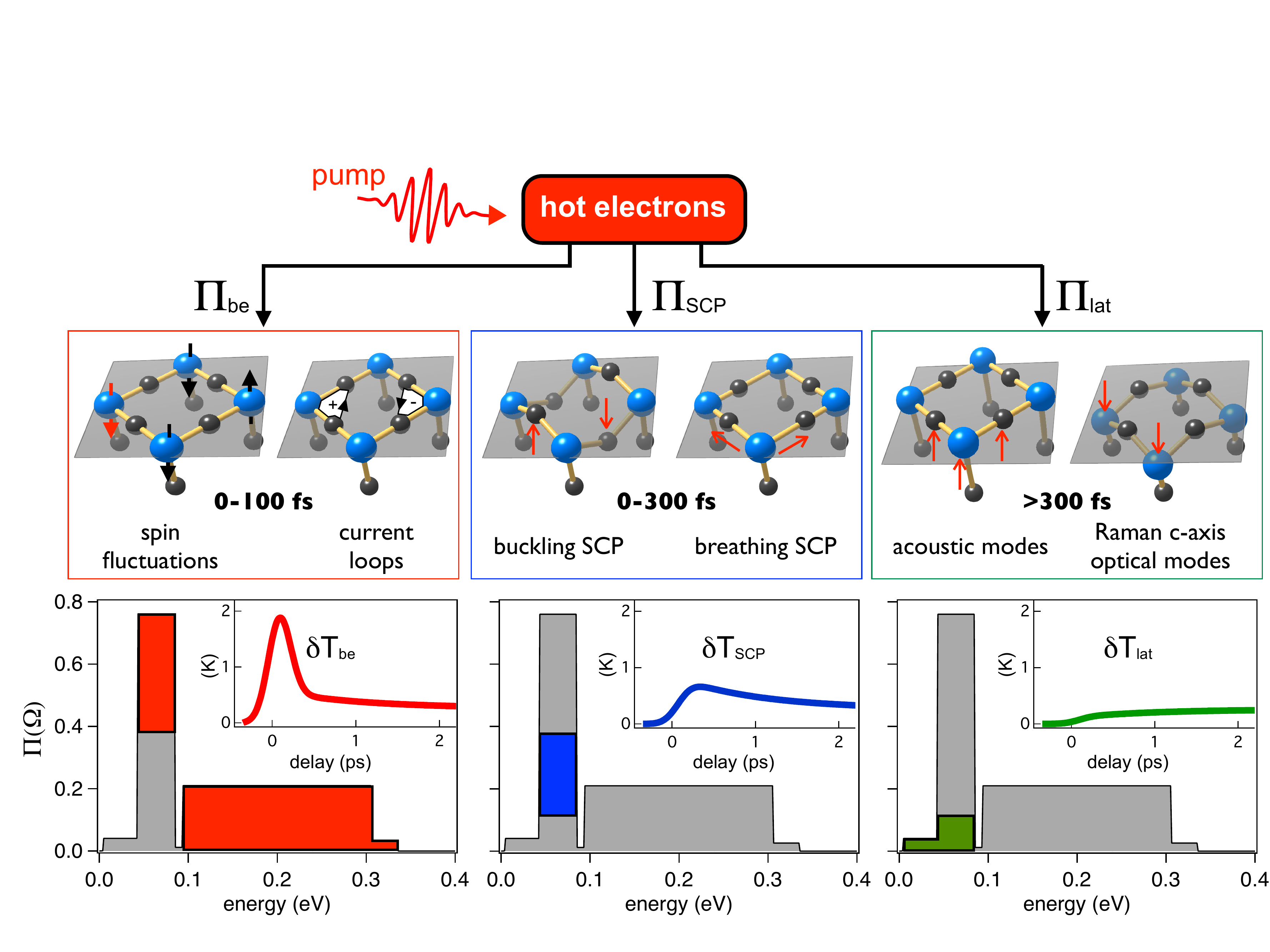}
\caption{Different contributions to the total bosonic glue $\Pi(\Omega)$, extracted from non-equilibrium broadband spectroscopy. The red area indicates the coupling function related to the coupling with bosonic fluctuations of electronic origin. The blue and green areas represent the coupling with the subset of strongly coupled phonons and with the rest of lattice, respectively. Taken from Ref. \citenum{DalConte2012}.}
\label{fig_electronbosoncoupling}
\end{centering}
\end{figure}

From the experimental standpoint, the experimental resolution didn't allow, until recently \cite{DalConte2015}, to directly follow in the time domain the coupling with bosons of electronic origin. Nevertheless, the combination of the ultrafast time-resolution with the broad spectral window accessible by the supercontinuum-based time-resolved spectroscopies has been a turning-point in the study of the electron-boson coupling in copper oxides. By measuring the dynamics of a broad part of the reflectivity around the dressed plasma frequency of doped Bi2212, it has been suggested that the reflectivity variation ($\delta R(\omega,t)/R$) is directly proportional to the electron-boson scattering rate \cite{DalConte2012}. Therefore, by monitoring $\delta R(\omega,t)/R$ it is possible to reconstruct the dynamics of the average boson density and estimate the electron-boson coupling within the Effective Temperature Model (ETM) described in Sec. \ref{sec_ETM}. This technique evidenced that on a timescale faster than the electron-phonon coupling, that is of the order of 100 fs, the charge carriers are already effectively coupled with another type of bosons, characterized by the coupling function $\Pi_{be}(\Omega)$. The spectrum of the coupling with this bosonic excitations, shown in Fig. \ref{fig_electronbosoncoupling} is extracted by considering as a constraint that the total electron-boson coupling function ($\Pi_{tot}(\Omega)$=$I^2 \chi(\Omega)$+$\alpha^2F_{SCP}(\Omega)$+$\alpha^2F_{lat}(\Omega)$, see Sec. \ref{sec_ETM}) should coincide with that extracted from the equilibrium optical conductivity (see Secs. \ref{sec_QPscattering}, \ref{sec_magnons}, \ref{sec_EDM} and \ref{sec_ETM}). The spectral extension of $\Pi_{be}(\Omega)$ (see the spectra reported in Fig. \ref{fig_electronbosoncoupling}) and the strength of the coupling itself ($\lambda_{be}\simeq$1.1) strongly support its magnetic origin \cite{DalConte2012}. Furthermore, when using $\Pi_{be}(\Omega)$ and $\lambda_{be}$ as inputs for the calculation of the critical temperature through the McMillan's formula \cite{McMillan1968}, reported in Eq. \ref{eq_Tc}, a maximal $T_C$=105-135 K is obtained for optimally doped Bi2212. Even though the fraction of the $\Pi_{be}(\Omega)$ that can give rise to superconductivity in the $d$-wave channel is still subject of debate, these results demonstrate that the coupling to AF excitations can, in principle, provide the glue necessary to achieve critical temperatures exceeding 100 K. This mechanism is likely to constitute the "missing" glue that has been evidenced in Fig. \ref{fig_criticaltemperatures}. More recently,
cluster dynamical mean-field calculations have been used to investigate the superconducting gap within the two-dimensional Hubbard model \cite{Gull2014}. The results strongly suggest that superconductivity arises from exchange of spin fluctuations and the inferred pairing glue function is in remarkable qualitative consistency with the pairing function, $\Pi_{be}(\Omega)$, inferred from time-resolved optical spectroscopy \cite{DalConte2012}.

The coupling with two different types of bosons has been also indirectly inferred by the relaxation dynamics of single-colour pump-probe measurements on Bi$_2$Sr$_2$CaCu$_2$O$_{8+\delta}$ single crystals \cite{Chia2013}, whose doping concentration ranged from the underdoped to overdoped regime. The different doping dependences of the electron-boson coupling strengths led the authors to identify them as strongly-coupled phonons and spin fluctuations. While the electron-phonon coupling ($\lambda_{SCP}$) peaks at optimal doping with a maximum value of 0.4, the coupling to AF excitations decreases monotonically with doping, starting from values $\lambda_{be}\simeq$1 at the hole concentration $p$=0.1 \cite{Chia2013}.
Recent time-resolved ARPES experiments on optimally-doped Bi2212 \cite{Rameau2014} also evidenced, above $T_c$, a marked change in the effective mass of the nodal quasiparticles related to the change of 70-meV kink in the $E(\textbf{k})$ dispersion. This ultrafast modification of the kink is found \cite{Rameau2014} to occur during the experiment's 100-fs temporal resolution and is compatible with the ultrafast coupling to AF excitations. 

In conclusion, time-resolved optical spectroscopies opened new scenarios in the study of the electron-boson coupling in copper oxides. The theoretical efforts are currently pointing to address the limits of applicability of the ETM \cite{Kogoj2015} and to develop models for the extraction of electron-boson coupling from time-resolved experiments without assuming an effective temperature model \cite{Sentef2013}. On the other hand, recent advances in ultrafast optical spectroscopy \cite{DalConte2015} succeeded in achieving a temporal resolution of $\sim$10 fs, that is of the order of $\hbar/2J$, i.e., the timescale of the direct coupling to magnetic fluctuations. This extremely high temporal resolution enabled the direct observation the $\sim$16 fs build-up of the effective electron–boson interaction in different families of hole-doped copper oxides \cite{DalConte2015}. This extremely fast timescale has been compared to numerical calculations based on the $t$-$J$ model, in which the relaxation of the photo-excited charges is achieved via inelastic scattering with short-range antiferromagnetic excitations \cite{DalConte2015}. The joint theoretical-experimental effort is expected to provide novel tools to further clarify the issues related to the electron-boson dynamics in correlated materials and to the problem of the retarded interactions and the superconducting glue in copper oxides.

\subsubsection{The ultrafast dynamics in the pseudogap state}
\label{sec_PGdynamics}
\begin{figure}[t]
\begin{centering}
\includegraphics[width=1\textwidth]{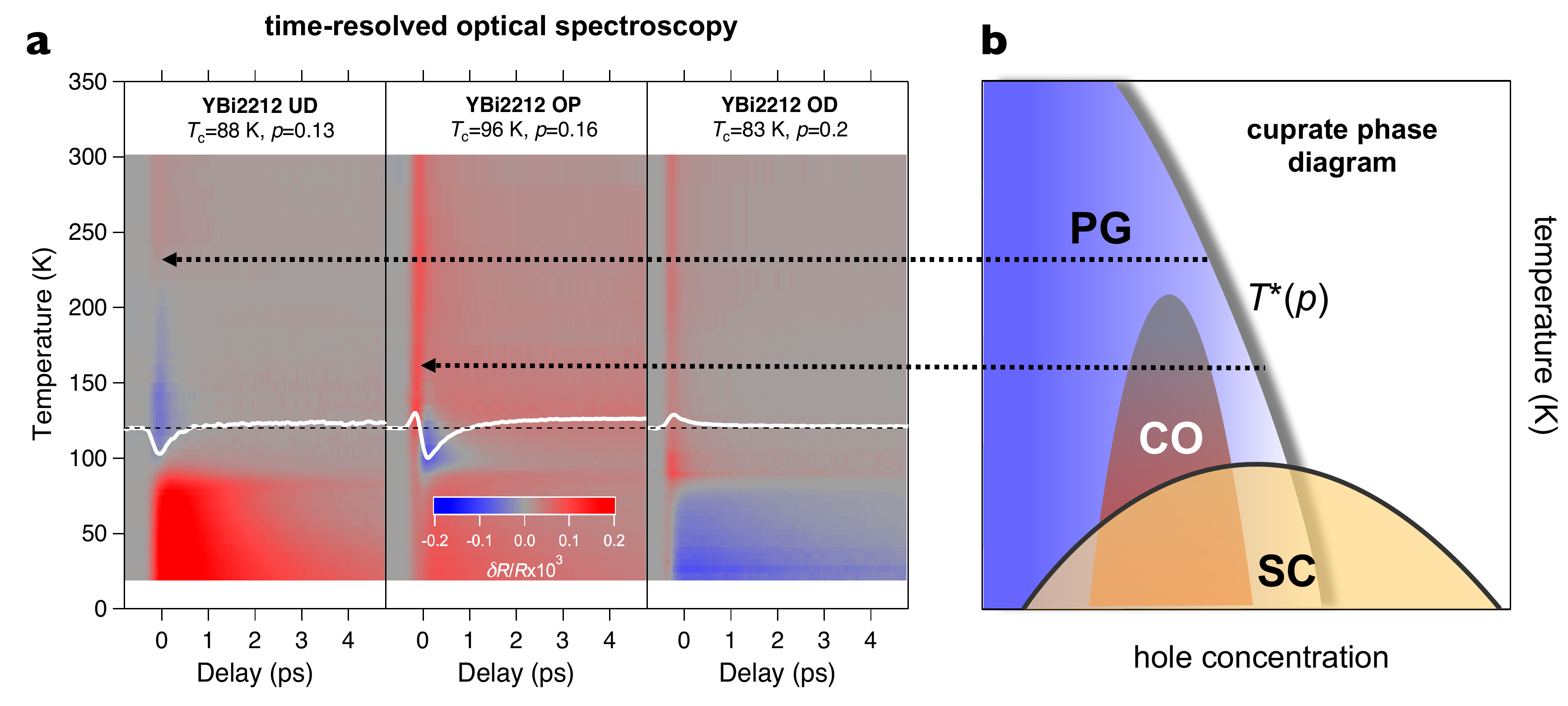}
\caption{The pseudogap state unveiled by non-equilibrium techniques. \textbf{a}) The relative reflectivity variation $\delta R/R(t)$ measured in the single-colour configuration (1.55 eV) is reported as a function of the temperature for three YBi2212 samples with different hole concentrations. The white lines are the $\delta R/R(t)$ time traces at 110 K. \textbf{a}) The generic phase diagram of cuprates is sketched. The pseudogap boundary $T^*(p)$ (grey curve) is determined reporting the temperature at which a negative component in the $\delta R/R(t)$ signal appears. Taken from Ref. \citenum{Cilento2014}.}
\label{fig_PGphasediagram}
\end{centering}
\end{figure}

As extensively discussed in Sec. \ref{sec_singlecolour}, the single-colour P-p experiments are particularly sensitive to the pseudogap (PG) state of copper oxides and, more in general, of correlated materials. For the sake of convincing the reader of the effectiveness of the ultrafast techniques in investigating the PG, we report in Fig. \ref{fig_PGphasediagram}a the temperature- and doping-dependent time-traces measured on Bi$_2$Sr$_2$Y$_{0.08}$Ca$_{0.92}$Cu$_2$O$_{8+\delta}$ (YBi2212) at the probe energy of 1.55 eV \cite{Cilento2014}. A negative component with a fast relaxation dynamics ($<$1 ps) progressively appears at the temperature $T_c<T<T^*$ and dominates the time response, until the onset of the superconducting phase transition (see Sec. \ref{sec_SWshift}). The plot of $T^*(p)$ as a function of the hole doping $p$ (see Fig. \ref{fig_PGphasediagram}) evidences a clear pseudogap-line that intersects the superconducting dome at $p>0.16$. Despite the dramatic evidence of the PG physics at play, the origin and the nature of this component in the transient reflectivity remained unexplained for a long time, until the development of more sophisticated versions of the first single-colour experiments became available. The possibility of changing the polarization of the pump and probe beams, of exploiting the capabilities of multicolour techniques, which extend from the midinfrared to the visible, and of probing the electron dynamics through time-resolved photoemission spectroscopy provided new insights into the nature of the PG state and opened new intriguing scenarios that will be reviewed in this section.
   
The first evidence that the time-resolved signal in the pseudogap state of copper oxides is constituted by two components characterized by different symmetries and different temperature-dependence came from polarization-dependent measurements on (110)- and (100)-YBCO thin films \cite{Luo2006}. Considering the direct relation between the gaps in the electronic DOS and the dynamics in the time-domain, the authors directly investigated the symmetries of the different components by combining the direction of the beam polarization and the orientation of the crystal axes. While the low-energy gap appears below $T_c$ with the same $d$-wave symmetry regardless of the hole concentration, signatures of the high-energy gap, whose symmetry strongly depends on doping, were observed also along the $a$-$b$ diagonal. The authors suggested a strong dichotomy between nodal and antinodal quasiparticles and argued that the two differnt gaps should have different physical origins and compete with each other \cite{Luo2006}. 

Similar polarization-dependent dynamics were lately observed on Bi$_2$Sr$_2$CaCu$_2$O$_{8+\delta}$ (Bi2212), which became the best system to disentangle the superconducting and pseudogap signals and to study the symmetry of the pseudogap through the polarization-dependence of the time-domain response \cite{Liu:2008p4917,Toda:2011cs,Toda:2014ga}, as already discussed in Sec. \ref{sec_singlecolour}. All these results converge on a picture in which the low-temperature dynamics is characterized by a fast pseudogap-like component, which corresponds to the relaxation of antinodal QP, and a slower component that is related to the opening of the superconducting gap and is dominated by the near-node contributions. Furthermore, the observation that the pseudogap dynamics depends only on the probe polarization led to the conclusion that the PG state is characterized by a spontaneous spatial symmetry breaking that is not limited to the sample surface \cite{Toda:2014ga}.

The results obtained by time-resolved optical spectroscopy unveil a picture that is fully consistent with the outcomes of the most advanced \textit{equilibrium} techniques. Recently, a comparative study of the pseudogap in Bi2201 has been carried out by ARPES, polar Kerr effect and time-resolved reflectivity \cite{He2011}. The three experiments revealed a coincident and abrupt onset at $T^*$ of an antinodal gap, a Kerr rotation signal and a change on the relaxation dynamics, making the link between the ultrafast dynamics and the equilibrium concepts that are usually invoked to explain the PG a robust experimental fact. 

\begin{figure}[t]
\begin{centering}
\includegraphics[width=1\textwidth]{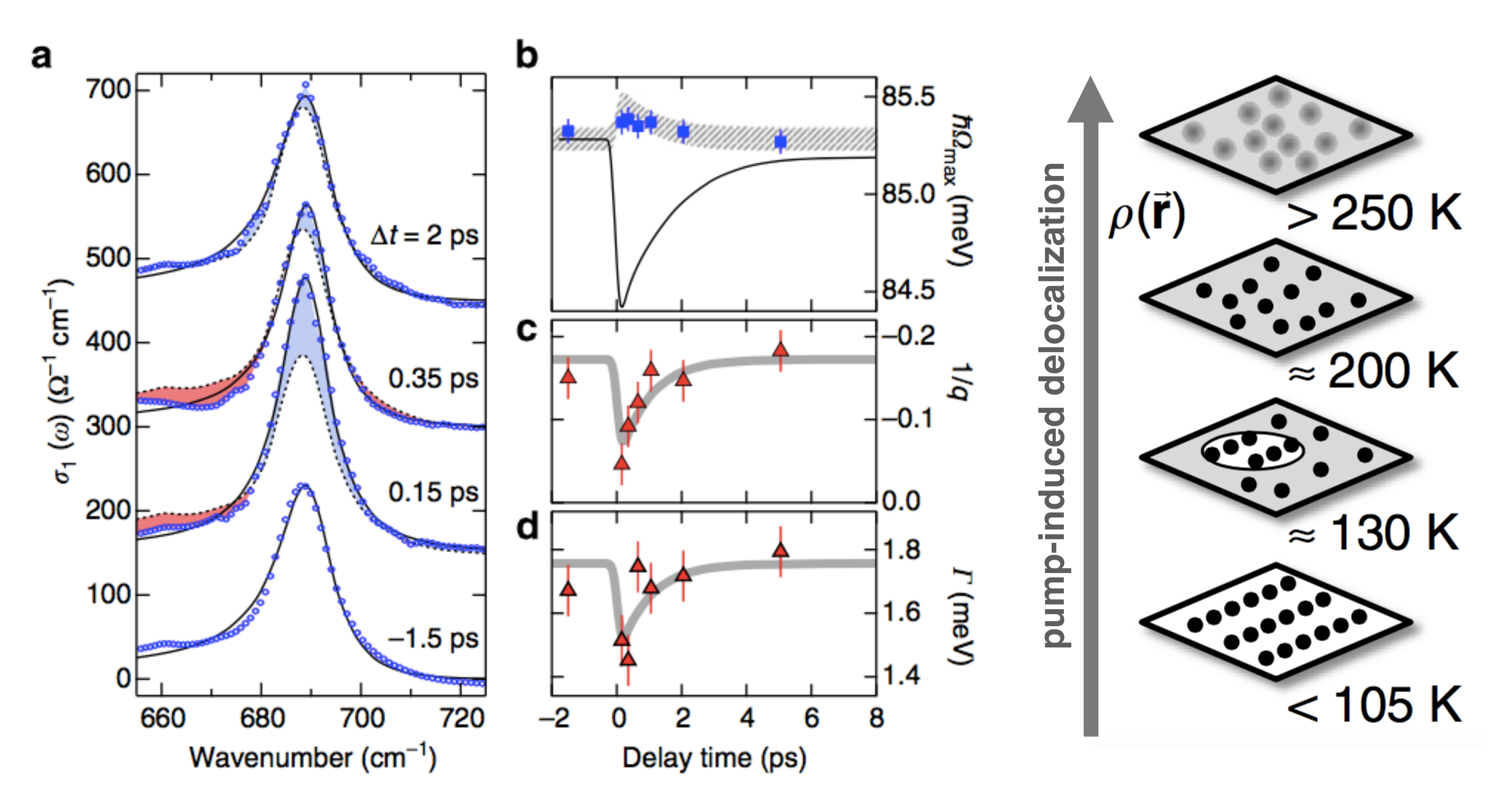}
\caption{Ultrafast dynamics in the pseudogap state of the La$_{1.75}$Sr$_{0.25}$NiO$_4$ nickelate. \textbf{a}) Non-equilibrium mid-infrared optical conductivity around the Ni-O stretching mode (circles) at specific delay times after the 1.5-eV photoexcitation at 30 K. Dashed lines: equilibrium conductivity, solid lines: Fano model. \textbf{b}) Dynamics of the Ni-O mode peak energy $\omega_0$ \textbf{c,d}) Dynamics of the Fano asymmetry parameter 1/$q$ and of the phonon linewidth $\Gamma$, as defined in Eq. \ref{eq_Fano}.
Taken from \cite{Coslovich2013}.}
\label{fig_Fanonickelates}
\end{centering}
\end{figure} 

A further step towards the comprehension of the PG dynamics in correlated materials has been recently done by studying the model compound La$_{1.75}$Sr$_{0.25}$NiO$_4$ \cite{Coslovich2013}. Similarly to the case of copper oxides, the optical conductivity of this prototypical nickelate exhibits a pseudogap-like loss of infrared spectral weight, whose onset temperature ($T^*$) is far above the long-range stripe formation ($T\sim$100 K). By using mid-infrared probe pulses, it has been possible to directly study the dynamics of the Ni-O stretch vibration at 85 meV. As discussed in Sec. \ref{sec_infrared_active_modes}, the spectral shape of the infrared-active phonon modes is usually characterized by a strong asymmetry that is described by a Fano lineshape and originates from the quantum interference between the narrow intrinsic phonon lineshape and the continuum of the electronic levels. The strength of the coupling regulates the asymmetry of the peak through the factor $q$, introduced in Eq. \ref{eq_Fano}. In the experiment, the phonon lineshape is monitored as a function of the delay from a near-infrared pump pulse. The data, reported in Fig. \ref{fig_Fanonickelates}, show a clear decrease of the 1/$q$ factor, while the position of the resonance remains unchanged. This result is interpreted as the transient decrease of the electron-phonon coupling in consequence of the delocalization of the charges induced by the impulsive excitation \cite{Coslovich2013}. This experiment suggests that the real-space charge localization is the key to understand the physics of the PG and that it can be precursory to the onset of charge-ordering at lower temperatures. We stress that, even though the role of the pump is to increase the average energy of the charge carriers, similarly to the case of an adiabatic increase of the temperature, the transient decrease of the electron-phonon coupling has no counterpart in equilibrium measurements. This difference is inherently related to the ultrafast excitation process, during which the electronic degrees of freedom (responsible for the charge localization) are perturbed before the heating the lattice. On a longer timescale, the temperature-related broadening of the phonon linewidth completely washes out the effect, thus demonstrating that the non-equilibrium approach provides a genuinely new information as compared to conventional spectroscopies. Similar conclusions have been derived from three-pulses experiments, that unveiled the tendency to carrier localization in the pseudogap state of Bi2212 \cite{Madan2015}.

The possibility of experimentally decoupling the charge localization dynamics from the lattice heating, also suggests that the charge localization, which characterizes the pseudogap state of nickelates, is primarily driven by the electronic correlations. Similarly, the role of the short range Coulomb repulsion in the PG physics of copper oxides (Bi2212 and Hg1201) has been recently demonstrated by combining broadband time-resolved optical experiments to time-resolved ARPES \cite{Cilento2014}. Ultrashort light pulses are used to prepare a non-thermal electron  distribution that is dominated by antinodal excitations. The dynamics of the dielectric function, simultaneously probed in the 0.5–2 eV range and analyzed within the extended Drude model (see Sec. \ref{sec_EDM}), unveils an anomalous decrease in the scattering rate of the charge carriers. In the pseudogap-like region of the phase diagram, delimited by a well-defined $T^*_{neq}(p)$ line, the photoexcitation process transforms the nature of antinodal excitations from that typical of a strongly correlated metal to that of a less-correlated metal, characterized by delocalized quasiparticles with a longer lifetime. This effect is naturally explained within the single-band Hubbard model, in which the short-range Coulomb repulsion leads to a $k$-space differentiation between nodal quasiparticles and antinodal excitations that are strongly localized in real space \cite{Cilento2014}. A similar picture has been confirmed by three-pulses experiments on Bi2212 samples at different dopings and temperatures \cite{Madan2015}. The pseudogap photodestruction and recovery reveal a marked absence of critical behaviour at $T^*$, which implies an absence of collective electronic ordering beyond a few coherence lengths on short timescales. This result suggests that the pseudogap is characterized by the localization of the charge carriers into a textured polaronic state arising from a competing Coulomb interaction and lattice strain \cite{Madan2015}. Taken all together, the outcomes of non-equilibrium optical spectroscopies strongly support a scenario in which the PG physics is dominated by the \textbf{k}-dependent effect of the short-range Coulomb repulsion that drives the real-space localization of the charge carriers and that can evolve into charge-ordered states upon further cooling.

\begin{figure}[t]
\begin{centering}
\includegraphics{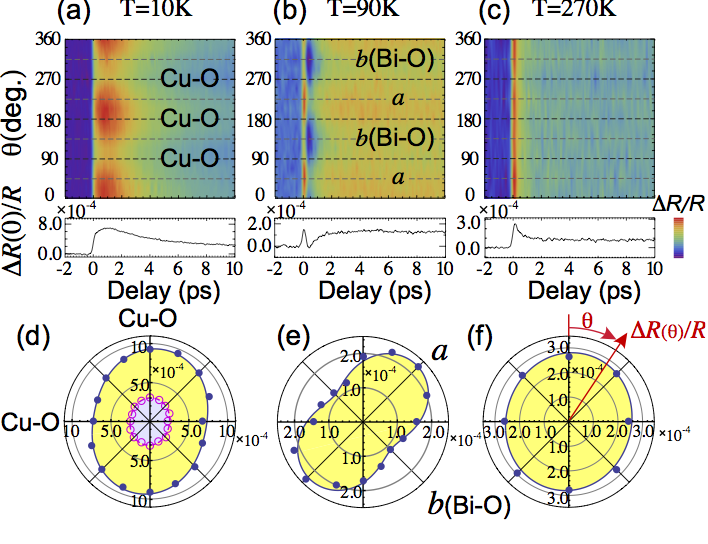}
\label{fig_polarizationanisotropy}\caption{(a-c) $\delta R(\theta)/R$ transients at typical temperatures for
overdoped Bi2212 samples together with polar plots of the maximum
values of $\delta R(\theta)/R$. The solid lines indicate fits using
Eq. \ref{eq_polanisotropy}. $\delta R(\theta)$/$\delta R$ at delay time of 10 ps
is also shown (open circles). Note that the Cu-O bonds directions
are drawn horizontal and vertical, while the crystalline axes are
along the Bi-O bonds, and are rotated nearly 45$^{\circ}$ from the
Cu-O bonds. Taken from Ref. \citenum{Toda:2014ga}.}
\end{centering}
\end{figure}
Recently, polarisation sensitive P-p measurements were used
to ascertain the presence of rotational symmetry breaking associated
with the PG state in Bi$_{2}$Sr$_{2}$CaCu$_{2}$O$_{{8+\delta}}$
compounds at various dopings \cite{Toda:2014ga}. Fig. \ref{fig_polarizationanisotropy} shows the experimental
polarization dependence, where the probe polarization was rotated
by an angle $\theta$ with respect to the crystal $x$ direction.
(Fig. \ref{fig_polarizationanisotropy}a)). Here we use notations in which $x$ and $y$ point
in directions along the Cu-O bond, while $x'$ and $y'$ are directions
rotated by 45$^{\circ}$ in the plane (Fig. \ref{fig_polarizationanisotropy}b)). 
When either global or local symmetry is broken (either dynamically or statically), an additional
possibility exists, where asymmetric states can be excited also by higher
order terms via DE. Without such breaking of symmetry, higher order
coupling cannot coherently excite $B$-symmetry modes, because the phase
of the excitation averages out to zero (for second order coupling).
However, symmetry breaking can provide a coupling mechanism for the excitation of $B$-symmetry modes.
Measurements by Toda et al \cite{Toda:2014ga} on Bi$_{2}$Sr$_{2}$CaCu$_{2}$O$_{{8+\delta}}$ reveal a remarkable $independence$ of the Pp response on $pump$ polarisation, which rules out IE as the excitation mechanism, and indicates the presence of symmetry breaking. The polarisation data shown in Fig. \ref{fig_polarizationanisotropy} show that not only $A_{1g}$, but also B-symmetry modes are excited. The pump must therefore excite $B_{1g}$ and $B_{2g}$ in second order, which is only possible in the presence of breaking of tetragonal symmetry. The T-dependences of the $A_{1g}$, $B_{1g}$  and $B_{1g}$  components in Fig. \ref{fig_polarizationanisotropy} nicely show the onset of the underlying symmetry-breaking to occur near $T^{*}$. Within this mechanism it is necessary that all three components show broken symmetry at the same time, which is indeed the case. Thus it is possible to unambiguously conclude that tetragonal symmetry breaking occurs at $T^{*}=140$ K \cite{Toda:2014ga}.
 
Finally, interesting results have been recently obtained by time-resolved ARPES. The initial pump-induced modification of the occupation of the ungapped nodal states is affected by the onset of the pseudogap state at $T<T^*$ \cite{Zhang2013}. These results allows to directly connect the transient density of the nodal quasiparticles to the global phase diagram of cuprates (see Sec. \ref{sec_cupratePD}), demonstrating an unexpected interplay between the antinodal pseudogap an the pump-induced population at different \textbf{k}-vectors in the Brillouin zone.

\subsubsection{Superconductivity-induced spectral-weight change}
\label{sec_SWshift}

As extensively discussed in Sec. \ref{sec_singlecolour}, single-colour optical techniques unveil a dramatic change in the dynamics, as soon at the system undergoes the superconducting phase transition at $T<T_c$. At the most general level, all the time-resolved optical measurements unveil a slowing down of the relaxation dynamics of the optical signal, whose time constant rapidly changes from 100-200 fs when $T>T_c$, characteristic of the electron-phonon cooling process, to several picosends at $T<T_c$. Simultaneously, the optical constants in the near-infrared range increases of about one order of magnitude. The detailed discussion about the origin of this bottleneck, which is ascribed to the formation of a non-equilibrium population of pair-breaking bosons at the gap energy, has been presented in Sec. \ref{sec_singlecolour} and discussed within the frame of the Rothwarf-Taylor model (see Sec. \ref{sec_RT}). Despite some early attempts of directly measuring the dynamics of the superconducting gap, $\Delta_{SC}(T)$, via time-resolved Raman spectroscopy \cite{Saichu2008}, it was soon realized that the gap dynamics could be mapped into the high-energy ($>$1 eV) optical properties. In this section we will focus on the origin of the variation of the near-infrared optical properties that, for some unexpected mechanisms, results to be directly proportional to the density of the superconducting condensate. To formulate the problem more directly, we can consider the conventional BCS superconductors, in which the onset of superconductivity corresponds to the opening of the superconducting gap, $\Delta_{SC}(T)$, in the electron density of states. As a consequence, the optical conductivity (see Sec. \ref{sec_electrodynamics_condensate}) is strongly affected only for below-gap radiation ($\hbar\omega<\Delta_{SC}(T)$), while the optical properties are almost unaffected at energy scales larger than about 10$\Delta_{SC}(T)$. Even considering an interband transition at the frequency $\Omega_0$, the simple energy-gap model for conventional superconductors \cite{Dresselhaus1962} predicts small changes of the interband transitions over a narrow frequency range of the order of $\Omega_0\pm\Delta_{SC}(T)$, which does not explain the relatively large variation of the optical properties in the entire infrared-visible spectrum. 

The direct relation between the dynamics of the infrared optical properties and the superconducting condensate was directly demonstrated by the first time-resolved experiments in the THz \cite{Averitt2001,Carnahan2004,Kaindl2005gp} and mid-infrared \cite{Kaindl2000} energy range that directly accessed the dynamics of the condensate and of the superconducting gap. Optical (1.5 eV) pump-THz probe experiments were performed on superconducting YBCO crystals \cite{Averitt2001}. At $T<T_c$, the recovery time for long-range phase-coherent pairing increases up to a value of about 3.5 ps, that is in agreement with the relaxation time measured through all optical P-p experiments. Furthermore, the relaxation time exhibits a clear tendency to diverge when $T_c$ is approached from below. This result, which was lately confirmed by infrared probe measurements \cite{Coslovich2011}, suggests that the relaxation time and the temperature-dependent gap are linearly related, i.e., $\tau\propto 1/\Delta_{SC}(T)$. Similar conclusions can be reached by studying the ultrafast mid-infrared reflectivity of YBCO at different hole-doping. In particular, after the optical excitation, the $ab$-plane gap in the optical conductivity undergoes an ultrafast filling, while the equilibrium condition is recovered on the picosecond timescale \cite{Kaindl2000}. Overall, THz and mid-infrared data clearly confirm the scenario suggested by all-optical measurements, i.e., the ultrafast excitation triggers a strong decrease of the superconducting fraction, that progressively recovers on the picosecond time scale.  

\begin{figure}[t]
\begin{centering}
\includegraphics[width=1\textwidth]{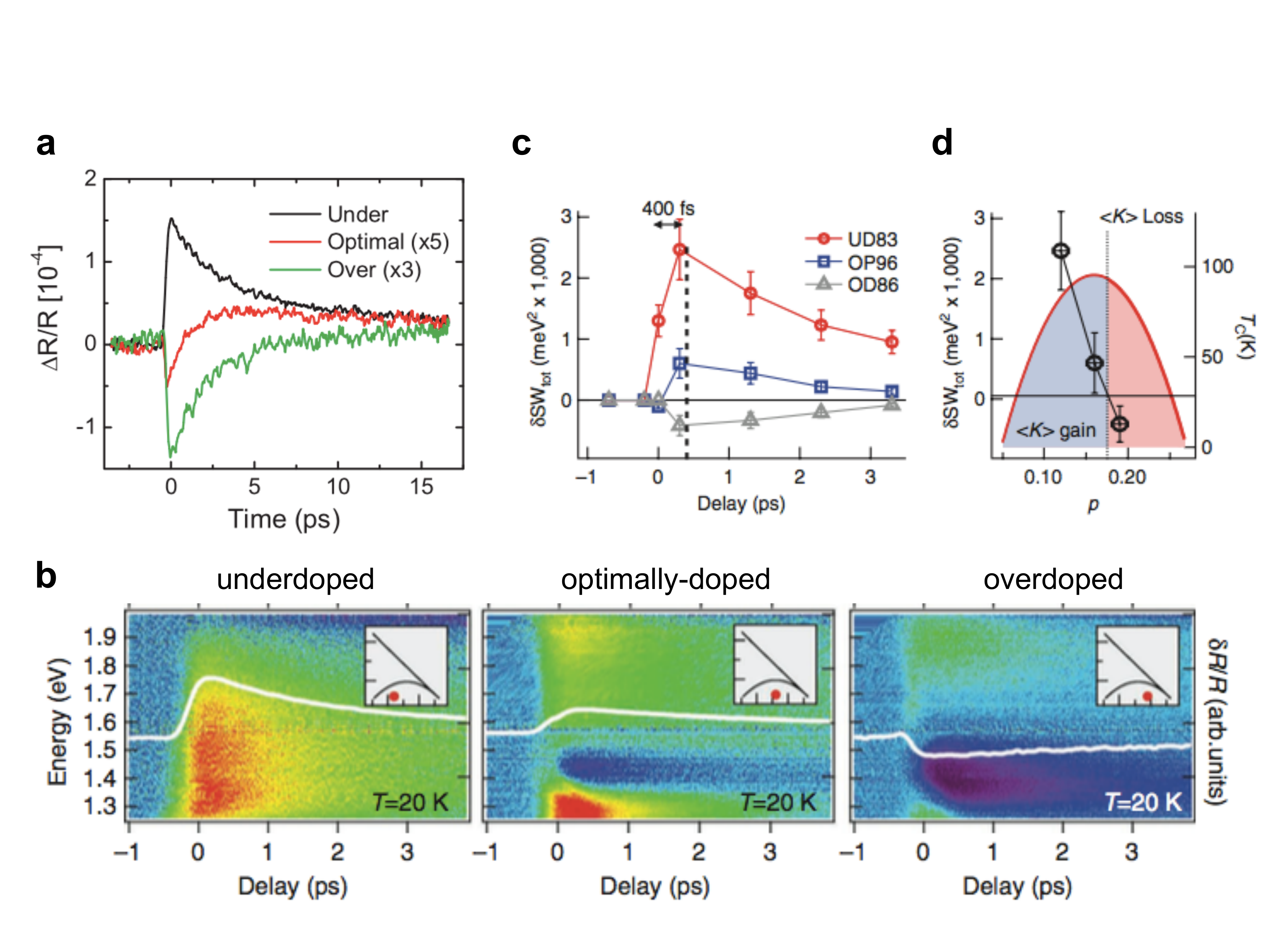}
\caption{Time-resolved optics in superconducting cuprates. \textbf{a}) The single-colour relative reflectivity variation, $\delta R/R(t)$, measured in the superconducting phase of Bi$_2$Sr$_2$Ca$_{1-y}$Dy$_y$Cu$_2$O$_{8+\delta}$ crystals with different hole concentrations. The superconducting temperatures are $T_c$=71 K (underdoped), $T_c$=94.5 K (optimally doped) and $T_c$=77 K (overdoped). Taken from \cite{Gedik2005}. \textbf{b}) The broadband relative reflectivity variation, $\delta R/R(\omega,t)$, measured in the superconducting phase of Y-Bi2212 crystals with different hole concentrations. The superconducting temperatures are $T_c$=83 K (underdoped), $T_c$=96 K (optimally doped) and $T_c$=86 K (overdoped). The insets display schematically the position of each measurement in the $T$-$p$ phase diagram of Y-Bi2212. The white lines are the time traces at 1.5 eV photon energy. Taken from \cite{Giannetti2011}. \textbf{c}) The spectral weight variation, $\delta SW_{tot}$ is reported at different delays for the three dopings. \textbf{d}) The black circles represent the maximum $\delta SW_{tot}$, i.e. the $\delta SW_{tot}$ measured at 400 fs, as a function of the doping level. Taken from Ref. \citenum{Giannetti2011}.}
\label{fig_pumpprobeSC}
\end{centering}
\end{figure}

A possible explanation of the relation between the optical properties in the visible and the superconducting condensate was suggested by time-resolved reflectivity measurements on Bi$_2$Sr$_2$Ca$_{1-y}$Dy$_y$Cu$_2$O$_{8+\delta}$ (BSCCO) crystals with a hole-doping concentration that spans a significant range of values across optimal doping ($p_{opt}$=0.16) \cite{Gedik2005}. These experiments demonstrated that both the kinetics of the quasiparticle decay and the sign of the $\delta R(t)/R$ signal change abruptly at optimal doping, as shown in Fig. \ref{fig_pumpprobeSC}a. As a consequence, the $p=p_{opt}$ value discriminates between two regions characterized by a different nature of the superconducting phase transition. Taking inspiration from the results coming from equilibrium optical spectroscopies \cite{Molegraaf2002,Deutscher2005,Carbone2006}, it was suggested that the change of sign of $\delta R(t)/R$ could be the fingerprint of a spectral weight transfer from the condensate $\delta(\omega)$ function (see Sec. \ref{sec_electrodynamics_condensate}) to higher frequencies. 

This problem was unambiguously addressed after the advent of multicolour experiments that opened the possibility of probing the dynamics of the dielectric function over a broad energy range. In order to perform low-fluence experiments in which the pump energy does not overcome the threshold for the impulsive vaporization of the superconducting condensate (see Secs. \ref{sec_singlecolour} and  \ref{sec_SCquench}), a pump probe technique based on the supercontinuum white light produced by a photonic fibre seeded by a high-rep. rate Ti:sapphire oscillator was developed \cite{Cilento2010}. This novel optical pump-supercontinuum-probe technique was used to investigate the dynamics of the optical properties in the 1-2.5 eV energy range in Bi$_2$Sr$_2$Ca$_{0.92}$Y$_{0.08}$Cu$_2$O$_{8+\delta}$ (Y-Bi2212) crystals at different doping, unveiling a superconductivity-driven change of an optical transition at 1.5 eV (see Fig. \ref{fig_pumpprobeSC}b) \cite{Giannetti2011}. This result unveils an unconventional mechanism at the base of the superconducting transition, both below and above $p_{opt}$, whose nature can be inferred by calculating the spectral weight variation of the 1.5 eV transition, i.e., $\delta SW_{1.5 eV}$=$SW^N_{1.5 eV}$-$SW^{SC}_{1.5 eV}$, as defined in Sec. \ref{sec_interband}. As shown in Fig. \ref{fig_pumpprobeSC}c, the sign change at $p=p_{opt}$ of the $\delta R/R(t)$ signal is just the consequence of a transition from positive to negative $\delta SW_{1.5 eV}$ values. This observation entails important consequences on the nature of the superconducting phase transition, since the spectral weight change of the interband transitions can be related to the change of the average kinetic energy of the charge carriers. By assuming that $\delta SW_{1.5 eV}$ is compensated by an opposite change of the Drude spectral weight, the kinetic energy variation ($\delta\langle K \rangle$) of the charge carriers is obtained by \cite{Giannetti2011}:

\begin{equation}
\label{eq_K_SW}
\delta\langle K \rangle=8\hbar^2\delta SW_{1.5 eV} \cdot\mathrm{(83 meV/eV^2)}
\end{equation}     

Using Eq. \ref{eq_K_SW}, the estimated superconductivity-induced kinetic energy decrease per Cu atom. For the underdoped sample is estimated to be $\sim$1–2 meV, which is very close to the superconductivity-induced kinetic energy gain predicted by several unconventional models \cite{Hirsch2000,Norman2002}.
In this picture, the $p=p_{opt}$ value delimits two regions that correspond to the crossing from a kinetic energy gain- to a potential energy gain-driven superconducting transition (see Fig. \ref{fig_pumpprobeSC}b), consistent with predictions of the 2D Hubbard model \cite{Gull2012}. These results clarify the origin of the doping- and temperature-dependent $\delta R/R(t)$ signal measured in single-colour optical experiments and demonstrate that the dynamics of $\Delta_{SC}$ can be reconstructed from the dynamics at optical frequencies.

The interplay between the low-energy physics and the high-energy optical transition has been further supported by other experiments based on the use of transient broad-band reflectivity. In particular signatures of the dynamics of $\Delta_{SC}$ in La$_{1.85}$Sr$_{0.15}$CuO$_4$ (LSCO) have been found at the typical scale of the Mott physics (2.6 eV) \cite{Mansart2013b}. More recently, a direct link between the $c$-axis phonon modes and the high-energy in-plane optical properties in optimally doped YBCO was made by a similar technique \cite{Fausti2014}.

\begin{figure}[t]
\begin{centering}
\includegraphics[width=0.6\textwidth]{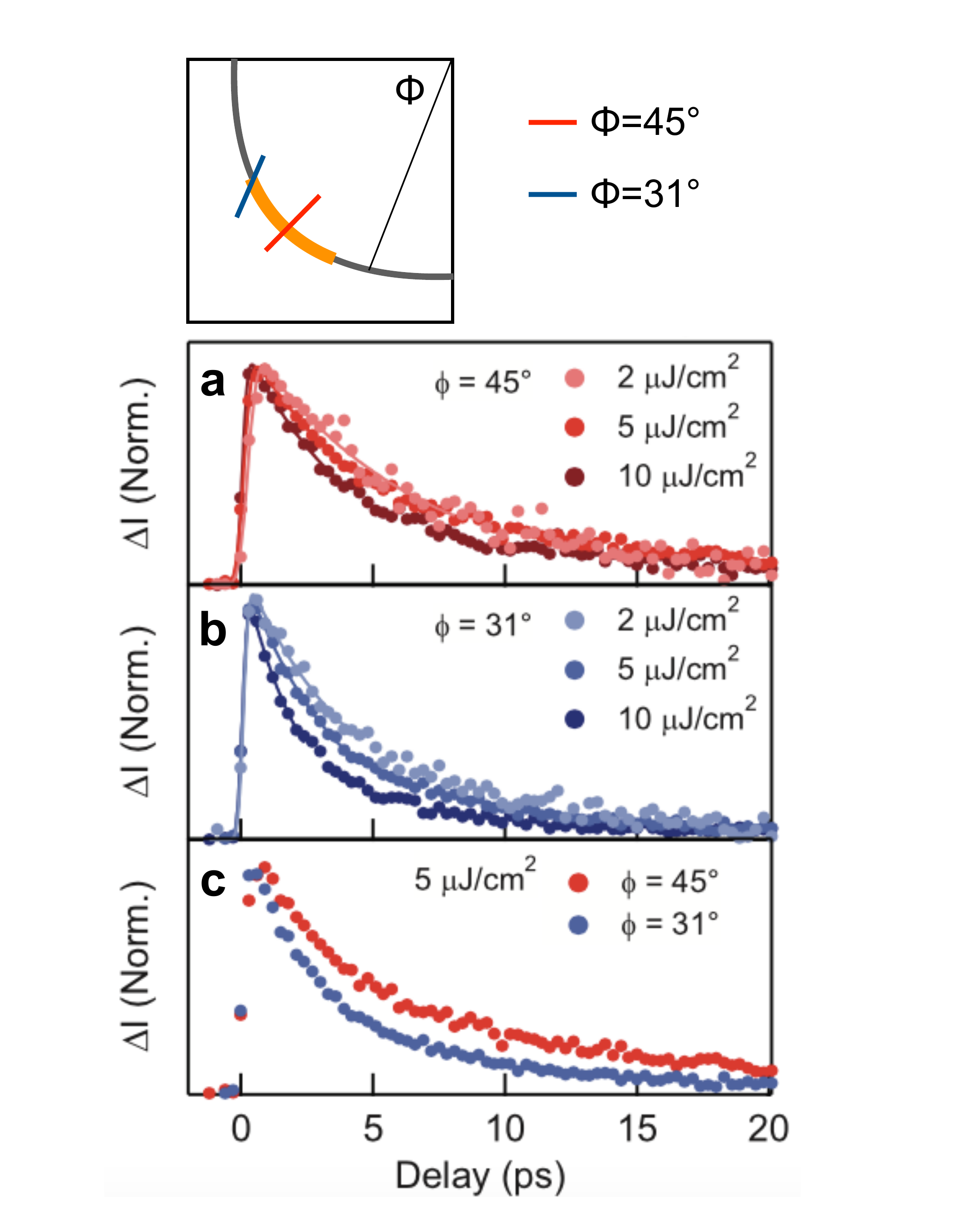}
\caption{Relaxation dynamics of the ARPES intensity change at different pump fluences and \textbf{k}-directions, measured by the azimuthal angle $\Phi$. \textbf{a}) Nodal decay curves at different fluences normalized to the same amplitude. \textbf{b}) Off-nodal decay curves at different fluences normalized to the same amplitude. \textbf{c}) Nodal and off-nodal curves at the same fluence.  Taken from Ref. \citenum{Smallwood2012}.}
\label{fig_smallwooddynamics}
\end{centering}
\end{figure} 

A new window on the ultrafast dynamics of high-$T_c$ superconductors was opened by the recent development of high-resolution TR-ARPES experiments \cite{Smallwood2012_b}. The high repetition rate of the ultrafast sources that can be currently employed for photoemission experiments guarantees the statistics necessary for low-fluence experiments in which the band dispersions and the electron self-energies are only slightly modified by the optical excitation. 
Considering the case of superconducting copper oxides, TR-ARPES experiments substantially confirmed the picture emerging from time-resolved optical technique, beside adding new valuable information related to the excitation and relaxation processes in the crystal momentum space. Unfortunately, the photon energy ($\hbar\omega\sim$ 6 eV) of the fourth harmonics of the Ti:sapphire laser, that is commonly employed as the source of ultrashort pulses of UV photons, limits the accessibility window of this technique in copper oxides to the states that lie within $\sim$1 eV from the Fermi energy and with a momentum that corresponds to a range of $\sim$27$^{\circ}$ of azimuthal angle \cite{Cortes:2011df} from the nodal direction (45$^{\circ}$, see the top panel in Fig. \ref{fig_smallwooddynamics}). As expected, the measurements of the pump-induced dynamics of the nodal quasiparticles in Bi2212 revealed a suppression of the nodal quasiparticle spectral weight, that is greatly enhanced in the superconducting state\cite{Cortes:2011df,Graf2011}. As shown in Fig. \ref{fig_smallwooddynamics}a, the recovery dynamics of the QP spectral weight depends on the laser fluence and ranges from $\sim$4 to $\sim$7 ps, in agreement with the optical results that evidenced the strong superconductivity-induced bottleneck in the dynamics. Interestingly, the evidence of a superconductivity-induced change of the dynamics of the nodal QPs challenges the conventional picture dominated by the nodal-antinodal dichotomy, in which the nodal QPs should be almost unaffected by the onset of superconductivity and the opening of a $d$-wave gap \cite{Graf2011}. Further results \cite{Zhang2013}, also demonstrated the unexpected possibility of establishing a link between the nodal quasiparticles and the cuprate phase diagram, as can be inferred from equilibrium and out-of-equilibrium optical  (see Sec. \ref{sec_cupratePD}). Even though the issue of the \text{k}-dependence of the relaxation dynamics has not been completely clarified \cite{Cortes:2011df,Smallwood2012}, the TR-ARPES measurements far from the node seem to suggest that the decay of hot QPs towards the node is blocked by phase-space restrictions, while the recovery of the equilibrium superconducting state is dominated by the Cooper pair recombination in a boson bottleneck limit \cite{Cortes:2011df}.

Current efforts in the field of TR-ARPES measurements are directed to quantitatively extract the dynamics of the band dispersion and of the electron self-energy and to address the nature of the boson modes which drive the formation of a non-thermal electron/hole distribution in the "boson"-energy window $E_F$-$\hbar\Omega<\hbar\omega<E_F+\hbar\Omega$ \cite{Rameau2015} and its successive relaxation through electron-boson coupling \cite{Yang2015}. Recent data on optimally-doped Bi2212, evidenced a transient shrinking of the Fermi surface on the 200-500 fs timescale, which suggests a possible transient hole-photodoping as a consequence of the electron-hole asymmetry in the relaxation dynamics \cite{Rameau2014}. On the other hand, the analysis of the TR-ARPES experiments on Bi2212 and Pb-Bi2201 evidenced a transient decrease of the electron self-energy, which corresponds to a weakening of the electron-boson coupling \cite{Zhang2014}. The tight relation between the dynamics of the electron-boson coupling and that of the superconducting gap is also supported by the fact that the electron–boson coupling is unresponsive to the ultrafast excitation above the superconducting critical temperature \cite{Zhang2014}. These recent results, along with the current efforts to increase the probe photon energy through high-harmonics generation from high-repetition rate ultrafast sources, are expected to open new perspectives for the physics of correlated systems and high-temperature superconductors.

\subsubsection{Fluctuations of the superconducting order parameter}
\label{sec_fluctuations}
\begin{figure}[t]
\begin{centering}
\includegraphics[width=0.7\textwidth]{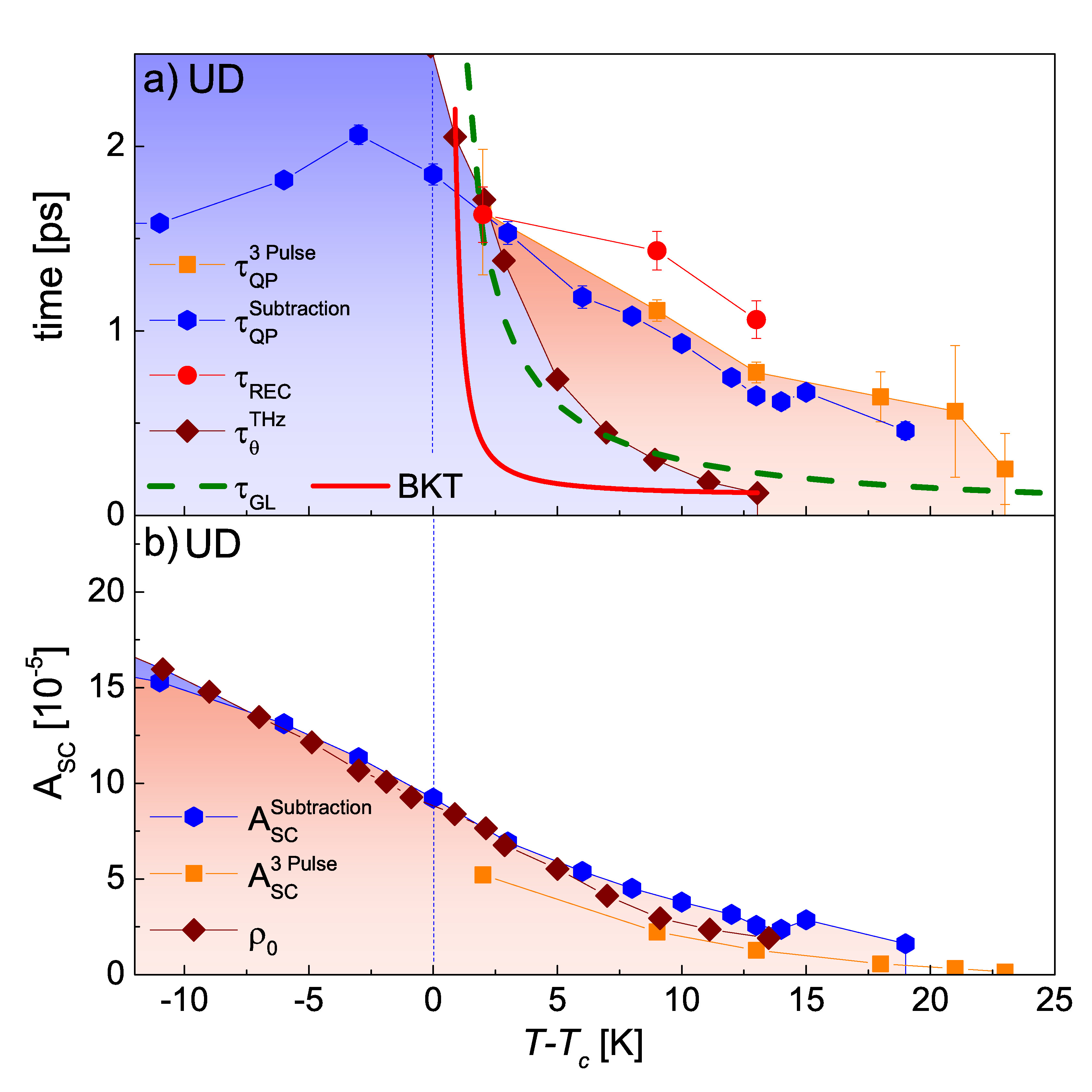}
\caption{A summary of the pairing amplitude and phase coherence dynamics near $T_{c}$ measured with different methods in Bi2212: (a) The recovery time of the optical superconducting signal ($\tau_{REC}$), the QP recombination time $\tau_{QP}^{3Pulse}$ measured by the three pulse technique and the QP recombination time $\tau_{QP}^{Subtraction}$  from
Pump-probe measurements obtained by subtraction of the pseudogap signal. A fit to the data using a Berezinsky-Kosterlitz-Thouless model \cite{Corson:1999bo} is shown by the solid red line. The green dashed line shows the fluctuation lifetime $\tau_{GL}$ given by time-dependent Ginzburg-Landau theory. The phase correlation time $\tau_{THz}$ from THz conductivity measurements is shown for comparison. (b) The amplitude of the SC signal $A_{SC}^{3Pulse}$ and $A_{SC}^{Pp}$ obtained by three pulse and the Pump-probe measurements respectively. The THz response is also shown. As expected on the basis of the response functions discussed in Sections \ref{sec_electrodynamics_condensate} and \ref{sec_non_eq_response}, the superfluid density $\rho_{0}$ from THz measurements shows remarkably close agreement with the amplitude of the optical response $A_{SC}$, above $T_c$.}
\label{Fig:AboveTc}
\end{centering} 
\end{figure}
The non-equilibrium spectroscopies offer also a new tool to investigate the role of 
the fluctuations in the superconducting transition. While in conventional superconductors the opening of the energy gap is simultaneous to the onset of a macroscopic phase coherence, in low-density and two-dimensional systems the phase fluctuations of the order parameter can prevent the formation of a macroscopic condensate, even though a pairing gap is already formed. In this regime the system is governed by a small coherence length, $\xi$, which tends to diverge at $T_c$ following the power law $\xi\propto|T-T_c|^{-\nu}$. Similarly, the relaxation time $\tau$, of the system is expected to diverge as $\tau\propto\xi^z$, where $z$ and $\nu$ are the critical exponents \cite{Tinkham}.

THz spectroscopy directly probes the superfluid response and is thus a useful tool for measuring the temporal fluctuations of the superconducting order parameter. Broadband time-domain THz spectroscopy in Bi2212 \cite{Corson:1999bo} and La$_{2−x}$Sr$_x$CuO$_4$ \cite{Bilbro2011} unveiled the persistence of superconducting correlations up to 16 K above $T_c$. The further development of the techniques also led to the development of more elaborate techniques in optical P-p experiments aimed at investigating the relaxation dynamics with high temporal resolution in the regime that is expected to be dominated by the phase fluctuations of the superconducting condensate. Indeed, using an all-optical three pulse technique (see Sec. \ref{sec_multipulse}) on Bi2212, the fluctuation dynamics of the superconducting pairing amplitude was separated from phase relaxation above the critical transition temperature. Both were shown to be distinct from the contribution related to the pseudogap phase \cite{Madan2014} on the basis of their temperature dependence, and more recently on the basis of symmetry \cite{Toda:2014ga}. The results are summarized in Fig. \ref{Fig:AboveTc}z. Similar critical behavior near $T_{c}$ have been recently obtained on thin Bi2212 films with optical pump-THz probe measurements, directly monitoring the dynamics of the mid-infrared conductivity \cite{Perfetti2015}. Finally, also recent time-resolved ARPES measurements \cite{Zhang2013} suggested that the pump-induced modification of the nodal QP population is affected by the onset of superconducting pairing fluctuations at $T_{fl}>T_c$. Overall, these results strongly support the scenario in which the pairing gap amplitude of underdoped cuprates extends a few tens of degrees beyond $T_c$, while the phase coherence shows a  power-law divergence as $T\rightarrow T_c$.

\subsubsection{The cuprate phase diagram from non-equilibrium spectroscopies}
\label{sec_cupratePD}
\begin{figure}[t]
\begin{centering}
\includegraphics[width=1\textwidth]{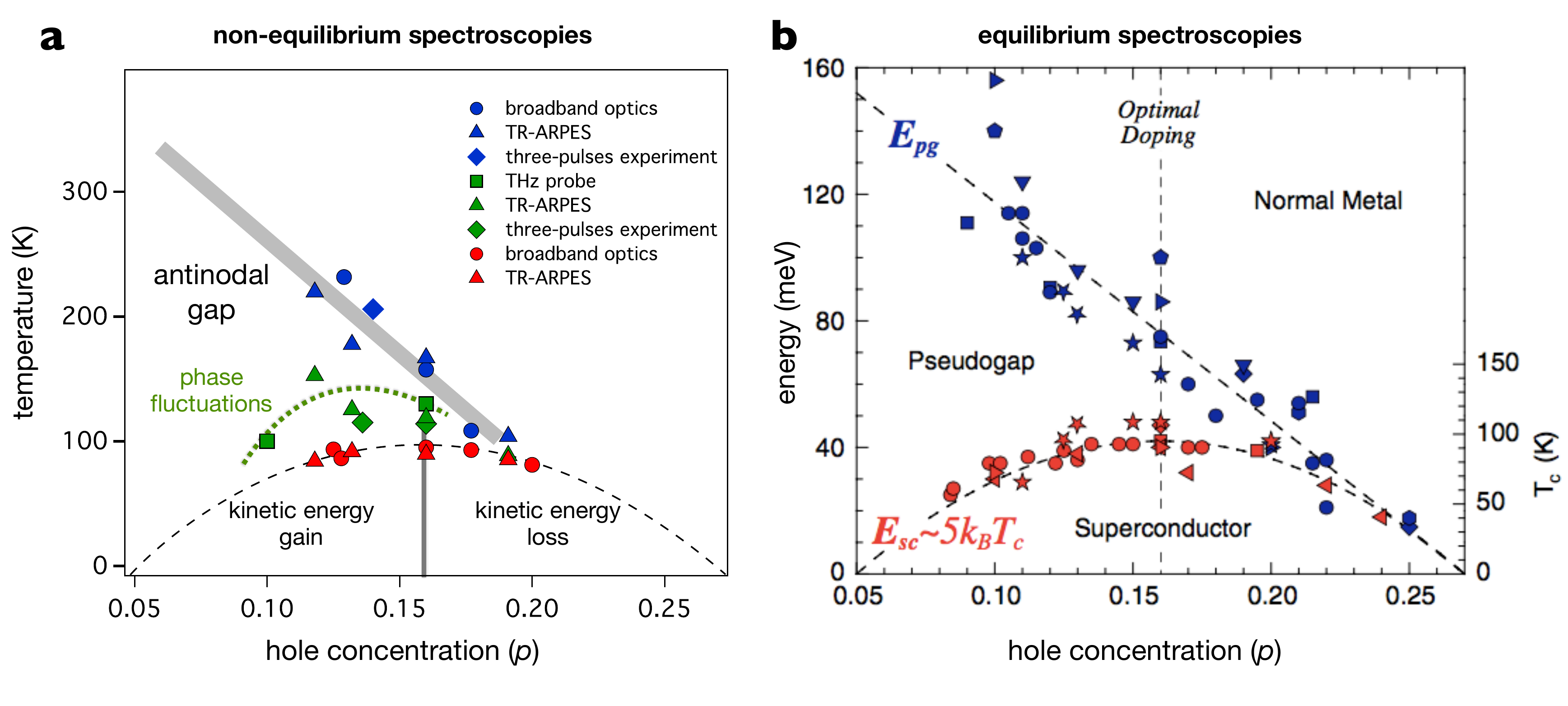}
\caption{The phase diagram of superconducting copper oxides from non-equilibrium spectroscopies. \textbf{a}) The phase diagram of Bi2212 obtained by   Non-equilibrium. \textbf{b}) Pseudogap and superconducting energy scales as obtained from equilibrium spectroscopies for various families (Bi2212, Y123, Tl2201 and Hg1201) of superconducting copper oxides with $T_{c,max}$=95 K. Taken from Ref. \cite{Huefner2008}.}
\label{fig_cuprate_phase_diagram}
\end{centering}
\end{figure} 
Bringing together the results presented in the previous sections, we can easily realize that the non-equilibrium spectroscopies, which combine temporal and spectral information, constitute a unique tool to investigate the phase diagram of copper oxides and, more in general, of correlated materials. In contrast with conventional techniques, such as ARPES, Raman scattering, tunneling, specific-heat measurements and scattering experiments, which usually probe a specific aspect, non-equilibrium technique are sensitive, within a single experiment, to a variety of phenomena ranging from the pseudogap to the onset of superconductivity. In the case of hole-doped copper oxides, the main non-equilibrium results collected in Fig. \ref{fig_cuprate_phase_diagram}a, provide a comprehensive $p$-$T$ phase diagram whose main properties can be summarized as follows:
\begin{enumerate}
\item[] \textbf{the normal state}$-$The high-temperature and large-doping concentration region of the phase diagram is characterized by a relaxation dynamics similar to that observed in conventional metals. Two recovery times of the order of $\sim$100 fs and $\sim$1-2 ps are usually observed and attributed to the coupling to the optical buckling and breathing phonons and, subsequently, to the rest of the lattice vibrations. The coupling of the charge carriers to short-range antiferromagnetic fluctuations is expected to be effective on the $\sim$10 fs timescale (see Secs. \ref{sec_ephresults} and \ref{sec_ebosresults}).
\item[] \textbf{the pseudogap state}$-$When the doping concentration and temperatures are decreased, the ultrafast dynamics exhibits a sharp change in the sign of the relative reflectivity variation, which defines a $T_{neq}^*(p)$ pseudogap-like line that approaches the superconducting dome in the overdoped region. This drastic change in the dynamics has been attributed to a photoinduced decrease of the scattering rate related to the photoinjection of antinodal excitations. Notably, TR-ARPES measurements showed that the opening of the antinodal pseudogap also affects the relaxation dynamics of \textit{nodal} quasiparticles, challenging the concept of complete nodal-antinodal dichotomy (see Sec. \ref{sec_PGdynamics}). Furthermore, polarization-sensitive ultrafast measurements suggested an underlying $s$-wave symmetry of the pseudogap-related signal, in contrast to the
common assumption of a pure $d$-wave pseudogap (see Sec. \ref{sec_SRStensor}). Interestingly, the region defined by the $T_{neq}^*(p)$ corresponds to the region of the $T$-$p$ phase diagram where an anomalous increase of the conductivity, interpreted as a transient superconducting-like state, has been reported after mid-infrared excitation (see Sec. \ref{sec_PIsuperconductivity}).
\item[] \textbf{the superconducting dome}$-$When the systems is cooled down below the critical temperature, the photoinduced relaxation dynamics is dramatically modified, as clearly shown by all the different non-equilibrium techniques. On a very general level, the recovery dynamics suddenly increases from the $\sim$100 fs to the $\gg$1 ps timescale, as soon as $T<T_c$. This drastic increase of the recovery time is attributed to a bottleneck process, in which the pair-breaking excitations photoinjecetd by the pump pulse are trapped on the upper edge of the superconducting gap as a consequence of the interaction with the gap-energy bosons emitted in the recombination process and of the phase-space constraints for the scattering processes. Even though a comprehensive model of the microscopic processes responsible for the relaxation dynamics are still missing, many experimental and theoretical efforts have been recently made to clarify this fundamental aspect in the hope of finding the signature of the elusive superconducting bosonic glue. Furthermore, non-equilibrium optical spectroscopies have demonstrated that the superconducting transition is accompanied by a change of the spectral weight of the optical transitions at the energy scale (1.5-2.5 eV) of the charge-transfer process, which suggest the transition from a kinetic energy-driven superconducting transition in the underdoped region to a more conventional potential energy-driven process in the overdoped region (see Sec. \ref{sec_SWshift}).        
\item[] \textbf{the phase-fluctuation region}$-$In the underdoped region of the phase diagram, the superconducting signal persists well above the critical temperature $T_c$. This has been interpreted as the signature of an underdoped region dominated by the phase-fluctuations of the superconducting order parameter which destroy the long-range order (see Sec. \ref{sec_fluctuations}).
\end{enumerate}

Overall, the non-equilibrium spectroscopies demonstrated an enormous potential in directly probing the rich physics of the copper oxides phase diagram. Current efforts mainly focus on experimentally accessing the dynamics of the antinodal excitations, on investigating the interplay between the charge-order phenomenon and the superconductivity and finally unveiling which are the boson modes involved in the pairing mechanism. 

\subsubsection{Time-domain competition between different orders}
Time-resolved techniques constitute a natural tool to investigate the interplay among multiple orders that can cooperate or compete to determine the actual ground state of complex materials. Recalling the Ginzburg-Landau equations (see Eqs. \ref{eq_GLenergy_coupled}, \ref{eq_GL_coupled1} and \ref{eq_GL_coupled2}) for two coupled order parameters ($\eta_1$ and $\eta_2$), it is easily to show that the recovery dynamics of the two orders are interdependent. This observation has important consequences that can be exploited in time-domain experiments: i) when the intrinsic lifetimes ($\tau_{11}$ and $\tau_{22}$) of the order parameters are significantly different, it is possible to partially quench one of the two and track in the time-domain their subsequent coupling during the restoring of the ground state; b) if $\tau_{11}\sim\tau_{22}$, then the optical pump can be used to completely quench one of the phases and probe the intrinsic dynamics of the residual order parameter.

\begin{figure}[t]
\begin{centering}
\includegraphics[width=1\textwidth]{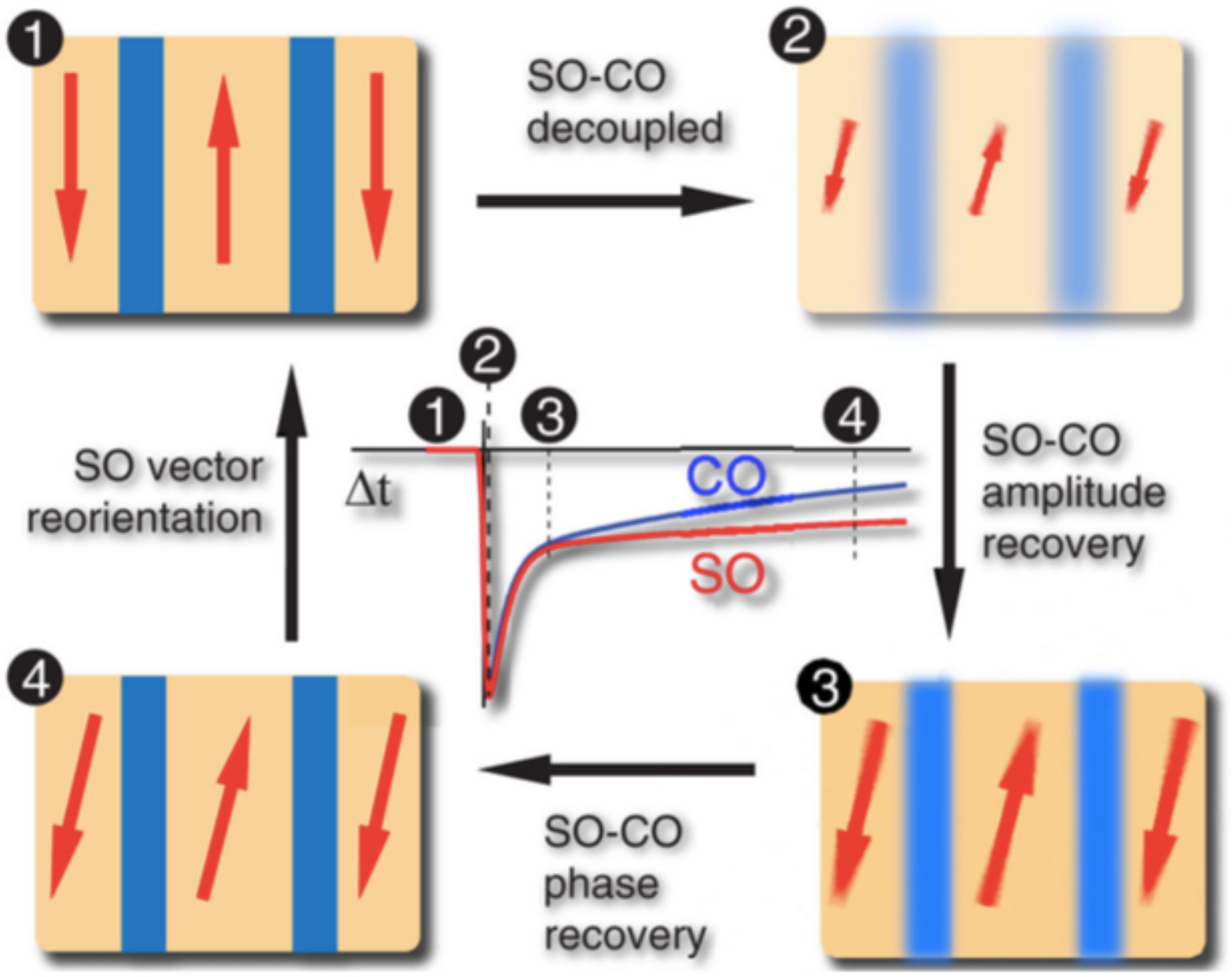}
\caption{Cartoon of the dynamics of CDW (blue areas) and SO (yellow areas) orders in La$_{1.75}$Sr$_{0.25}$NiO$_4$ at four different stages. The intensity of the colours is proportional to teh amplitude of the CDW and SO orders. Taken from Ref. \citenum{Chuang2013}.}
\label{fig_competing_nickelates}
\end{centering}
\end{figure}

\paragraph*{Coupled dynamics of multiple orders.}
The coupled dynamics of different order parameters has been studied in a variety of systems, such as nickel and copper oxides.
One the simplest examples is the La$_{1.75}$Sr$_{0.25}$NiO$_4$ nickelate that exhibits simultaneous spin and charge orders in the stripy phase. In particular, while the formation of the one dimensional charge density waves (CDW) breaks the translational symmetry, the antiferromagnetic spin order (SO), which requires the pre-existing CDW, breaks both translational and rotational symmetries. By using time-resolved femtosecond resonant X-ray diffraction at the Ni-$L_3$ edge \cite{Chuang2013}, which is sensitive to the modulation of the Ni valence electrons at the CDW and SO wavevectors, it has been shown that the dynamics of the two symmetry-broken phases is strongly coupled in the time-domain (see Fig. \ref{fig_competing_nickelates}), i.e. $\tau_{12}\geq \tau_{11},\tau_{22}$, despite the fact that the energy scales for the orders differ by at least an order of magnitude. This work stimulated the theoretical investigation of the role of the amplitude and phase fluctuations in the coupled CDW and SO dynamics, demonstrating that the amplitude dynamics dominates the initial time scale ($\sim$2 ps), while the phase recovery controls the behaviour on the long-timescale ($>$20 ps) \cite{Kung2013}. 

More recently, femtosecond soft X-ray diffraction at the Cu-$L_3$ edge resonance was combined with mid-infrared excitation to investigate the melting of the charge order and its interplay with superconductivity in copper oxides \cite{Forst2014,Forst2014b}. In particular, upon resonant excitation of the in-plane Cu-O stretching mode in La$_{1.875}$Ba$_{0.125}$CuO$_4$, the charge stripe order was found to melt on a sub-picosecond time scale, while the low-temperature tetragonal distortion was only weakly perturbed (and on longer timescales). Similar measurements were also performed on YBa$_2$Cu$_3$O$_{6.6}$ \cite{Forst2014}. Interestingly, the melting of the in-plane charge-density wave was shown to be directly connected to the photo-enhancement of the coherent interlayer transport \cite{Kaiser2014,Hu2014}, suggesting that the charge-order phenomenon competes with the superconductivity. Above $T_c$, superconductivity remains a hidden phase, which can be accessed by nonlinear phonon excitation \cite{Forst2011,Mankowsky2014}.

Time-resolved optical techniques have been also employed to investigate possible phases that compete with superconductivity in Tl$_2$Ba$_2$Ca$_2$Cu$_3$O$_y$ \cite{Chia2007} and (Ba,K)Fe$_2$As$_2$ \cite{Chia2010}, in which spin-density wave (SDW) order anticipates the superconducting transition at $T$=28 K. 

Quite naturally, much effort has been dedicated to clarify the nature of the interplay between the pseudogap (PG) and superconductivity (SC) in doped copper oxides. Single-colour P-p experiments promptly revealed a coexistence of the the PG and SC signals below $T_c$, which allows the direct separation of the charge dynamics of PG and SC excitations \cite{Liu:2008p4917,Giannetti2009b}. The advent of broadband time-resolved experiments constituted a breakthrough in the understanding of the relation between the PG and SC phases. The broadband probe provides spectral fingerprints of the different phases that can be exploited to unambiguously disentangle their dynamics, even in the regime in which one of the two orders is vanishing. This technique has been applied to investigate Bi$_2$Sr$_2$Ca$_{0.92}$Y$_{0.08}$Cu$_2$O$_{8+\delta}$, demonstrating that once superconductivity is established, the relaxation of the pseudogap proceeds two times faster than in the normal state. The sign and strength of the coupling term suggest a weak competition between the two phases, allowing their coexistence \cite{Coslovich2013b}. Similar results, which suggest a repulsive interaction between superconductivity and another fluctuating order, have been also obtained on the Nd$_{2-x}$Ce$_x$CuO$_{4+\delta}$ electron-doped cuprate \cite{Hinton2013b}. 

\paragraph*{Ultrafast quench of the superconducting phase.}
The possibility of using an ultrashort light pulse to completely quench the superconducting phase constitutes a further development in the time-domain study of the PG-SC interplay. More specifically, above a fluence-threshold of the order of few $\mu$J/cm$^2$, the Cooper pairs with long-range order are fully destroyed within the photoexcited volume (see discussion in Sec. \ref{sec_SCquench}). Above this threshold, it is possible to isolate the native SC signal and to study how the PG signal is modified during the recovery of the long-range superconducting order \cite{Toda:2011cs}. Time-resolved broadband spectroscopy on Bi$_2$Sr$_2$Ca$_{0.92}$Y$_{0.08}$Cu$_2$O$_{8+\delta}$ demonstrated that the relaxation of the pseudogap becomes significantly slower once the SC phase is removed through the impulsive optical excitation \cite{Coslovich2013b}. This result suggests a dynamically competing relation between SC and PG, in agreement with the results obtained as a function of the temperature \cite{Coslovich2013b}.
The presence of multiple competing ordering tendencies has been also argued by TR-ARPES measurements on Bi2212 \cite{Smallwood2014}, which evidenced a marked anisotropy in the fluence-dependent response of the gap.

\subsection{Optical generation of coherent bosonic waves}
\label{sec_coherentbosons}
The excitation with an external stimulus that is shorter than the typical lifetime of a specific mode is able to induce a macroscopic coherent oscillation of the mode itself, which can be detected by the probe pulse. Microscopically, the impulsive excitation mechanism can be described as an inverse Raman scattering process (see Sec. \ref{sec_SRStensor}), in which the modes with the symmetry determined by the Raman tensor \cite{Stevens:2002p6798,Merlin1997} can be excited. In the particular case of symmetry-broken phases, the oscillation of the mode at the proper frequency $\omega_0$ can be viewed as the dynamics around the equilibrium value of the order parameter, as determined by the \textit{mexican-hat} potential landscape described in Sec. \ref{sec_tdGL}. The possibility of tracking in the time-domain the oscillation of the order parameter, constitutes the time-domain counterpart of the conventional energy-domain Raman spectroscopy and it has been applied for the investigation of a wealth of different systems. Historically, most of the work has been focused on optical coherent phonon modes that can be easily excited and detected in many materials. More recently, also the possibility of investigating the coherent oscillations of charge-density wave modes and the massive amplitude modes (Higgs modes) in $s$-wave superconductors has opened interesting perspectives. Even though the discussion of the huge amount of literature about the coherent excitation of optical phonons and CDW modes goes beyond the scope of the present review and it would require a separate work, we will here present some recent examples to show how it is possible to exploit coherently excited bosonic modes for understanding the properties of unconventional superconductors and other correlated materials.  

\begin{figure}[t]
\begin{centering}
\includegraphics[width=1\textwidth]{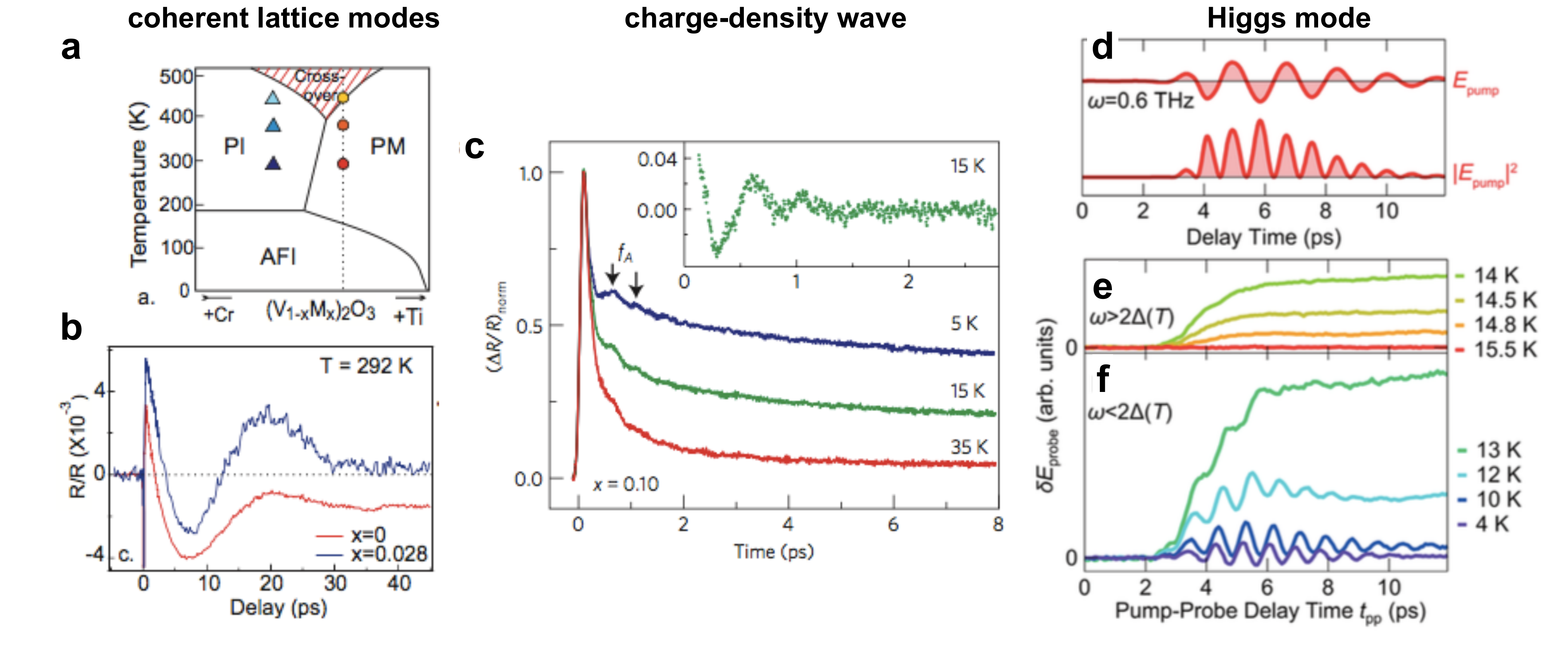}
\caption{Optical generation of coherent bosonic waves. \textbf{a}) Phase diagram of doped vanadium sesquioxide (V$_2$O$_3$). \textbf{b}) Relative reflectivity variation measured in a single-colour P-p experiment on V$_2$O$_3$ and V$_{1.944}$Ti$_{0.056}$O$_3$. Taken from Ref. \citenum{Mansart2010b}. \textbf{c}) Relative reflectivity variation measured in a single-colour P-p experiment on La$_{1.9}$Sr$_{0.1}$CuO$_4$ at different temperatures. Taken from Ref. \citenum{Torchinsky2013}. The change in the probe THz electric field, $\delta E_{probe}$, as a function of the P-p delay time $t_{pp}$, measured on superconducting NbN. Taken from Ref. \citenum{Matsunaga2014}.}
\label{fig_coherent_bosons}
\end{centering}
\end{figure} 

\subsubsection{Lattice modes}
\label{sec_latticemodes}
A major advances in the field was achieved after the observation of coherent optical phonons under the form of distinct oscillations in the single-colour P-p signal on YBa$_2$Cu$_3$O$_7$. Surprisingly, both the amplitude and the dephasing time of a specific mode at $\sim$15 meV, which involves the motion of the Ba atoms, were found to drastically increase below $T_c$ \cite{Albrecht1992}. This result underscored, for the first time, a direct relation between coherent phonon modes and the superconducting condensate. As a possible explanation, \textit{ab-initio} calculations suggested that the superconductivity-induced change of the density of states at the Fermi level induces a small displacement in the equilibrium positions of the ions. As a consequence of the partial quench of the superconducting condensate, induced by the pump pulse, the ions are pulled back to their normal equilibrium positions, thus exciting coherent phonons \cite{Mazin1994}. This picture has been later completed by means of broadband time-resolved measurements which showed that the coherent movement of the Ba atoms resonantly modulates the optical properties at a very high energy-scale (2-2.4 eV) \cite{Fausti2014}. DMFT calculations demonstrated that the specific atomic displacement of the Ba atoms strongly couples to the density of states at the Fermi energy, thus affecting the high-energy optical transitions from the lower Hubbard band to the Fermi level \cite{Fausti2014}.

The possibility of inducing lattice oscillations to coherently modulate the density of states at the Fermi level and, in turn, the chemical potential has been also widely explored in iron-based superconductors. While P-p reflectivity measurements on Ba(Fe$_{1−x}$Co$_x$)$_2$As$_2$ did not evidence any difference below and above the critical temperature \cite{Mansart2010}, TR-ARPES on Ba/Eu(Fe$_{1−x}$Co$_x$)$_2$As$_2$ showed a pronounced oscillation of the Fermi level at the frequency of the A$_{1g}$ phonon mode \cite{Yang2014}. These results have been interpreted as the fingerprint of the modulation of the effective chemical potential in the photoexcited surface region \cite{Yang2014,Avigo2013}.
Furthermore, the use of few-cycle multi-THz pulses enabled to directly track the evolution of the spin-density-wave (SDW) gap of BaFe$_2$As$_2$ after the ultrafast excitation. Interestingly, it was found that the SDW gap is induced upon excitation of the normal state above the transition temperature and that the magnetic order adiabatically follows the coherent lattice oscillation at a frequency of 22 meV (5.5 THz). These results attested a strong spin-phonon coupling that supports the rapid development of a macroscopic order on small vibrational displacement even without breaking the symmetry of the crystal \cite{Kim:2012kh}.

Finally, the excitation of coherent acoustic waves has been exploited to investigate the interplay between the lattice and the electrons in proximity of the insulator-to-metal phase transition in Cr(Ti)-doped V$_2$O$_3$ (see Fig. \ref{fig_coherent_bosons}a,b), which constitutes the paradigm of a Mott-insulator \cite{Mansart2010b}.

\subsubsection{Charge-Density-Waves}
The spontaneous development of charge-order patterns, which break the translational symmetry, is a common property of many correlated materials. The paradigmatic case is the charge-density wave (CDW) instability that emerges from the Fermi-surface nesting at the wavevector \textbf{Q}$_{CDW}$, as a consequence of the coupling between the charge density and the lattice (Peierls instability). Besides this basic mechanism, the CDW instability can also be caused or enhanced by the electron-electron interactions, which can eventually lead to the formation of Mott or excitonic insulators, and by disorder (Anderson localization mechanism). It was recently realized that time-resolved optical spectroscopy can be used to quench the CDW order and to measure the recovery dynamics of the equilibrium state. This technique provided a new way for disentangling the role of the elementary electronic and structural processes and thus identifying the dominant interactions responsible for the insulating state and for the CDW instability \cite{Mansart2012,Hellmann2012}. A comprehensive picture of the melting and reconstruction of the CDW order has been obtained by extending the time-domain investigation to the lattice degrees of freedom, through the use of time-resolved electron diffraction on TaS$_2$ \cite{Eichberger:2010p12340}. 

Besides the clarification of the fundamental interactions responsible for the CDW instability, ultrafast technique can be also used to generate coherent oscillations of the CDW mode. Since the CDW instability is one of the most simple and robust examples of second-order phase transition, in which a complex order parameter $\Delta$exp(i\textbf{Q}$\cdot$\textbf{r}+$\phi$) emerges at $T<T_c$, the CDW systems can be used as a platform to study the dynamics of symmetry-broken phases and the decay of the amplitude/phase modes. Coherent oscillations at the frequency corresponding to the lattice mode which drives the Peierls instability have been observed in various rare-earth tritellurides, in TaSe$_2$, in K$_{0.3}$MoO$_3$, Rb$_{0.3}$MoO$_3$ and in TiSe$_2$ by conventional optical pump-probe \cite{Schafer2010,Schafer2014}, three-pulses optical experiments \cite{Yusupov2010}, TR-ARPES \cite{Rettig2014,Schmitt:2011um}, THz spectroscopy \cite{Porer2014} and time-resolved X-ray diffraction \cite{Huber2014}. 

In P-p experiments the initial quench of the order parameter, $\delta \eta_0(t, \overrightarrow{r})$, is confined to the thin superficial layer corresponding to the pump penetration length. The intrinsic inhomogeneity of the excitation process leads to a complex spatio-temporal dynamics of $\delta \eta(t, \overrightarrow{r})$ that can be reproduced by a second-order differential equation that is derived by the Ginzburg-Landau functionals (see Sec. \ref{sec_tdGL})  \cite{Yusupov2010}. The results of time-resolved spectroscopies on CDW materials triggered a theoretical effort, that is still at its infancy, to develop non-equilibrium microscopic models that can reproduce the temporal evolution of the CDW gap, the charge distribution and the density of states at the Fermi level \cite{Shen2014, Shen2014b}. 
More in general, the study of the temporal evolution of systems undergoing symmetry breaking phase transitions is relevant for the Kibble-Zurek mechanism that has an impact that goes beyond condensed-matter physics, extending to cosmology and finance.   
    
As discussed in Sec. \ref{sec_gappedphases} the physics of the CDW instability is also relevant for copper oxides superconductors. The universal tendency to develop short-ranged incommensurate CDWs that break the translational symmetry has been recently reported in both hole- and electron-doped copper oxides \cite{Ghiringhelli2012,Achkar2012,Chang2012,Comin2014,daSilvaNeto2014,BlancoCanosa2014,Tabis2014,daSilvaNeto2015,Comin2015}.  
These finding opened new intriguing questions about the origin of the CDW instability (Fermi surface nesting \textit{vs} electronic mechanism) and its relation with the pseudogap and superconducting phases. In general, the CDW order is considered as an instability that competes with superconductivity, drastically reducing the $T_c$ in the under-doped region of the phase diagram of copper oxides. The possibility of impulsively removing the CDW order and of achieving its control by optical means, would open interesting, though challenging, ways to increase the critical temperature of copper oxides. Recently, the observation of a coherent oscillation at 2 THz in a single-colour P-p experiment has been reported in underdoped La$_{1.9}$Sr$_{0.1}$CuO$_4$ films ($T_c$= 26 K) up to a temperature of $\sim$100 K (see Fig. \ref{fig_coherent_bosons}a,c) \cite{Torchinsky2013}. This oscillating signal has been attributed to a highly-damped collective mode of the fluctuating CDW. The observation that the CDW mode vanishes in the optimally-doped La$_{1.84}$Sr${0.16}$CuO$_4$ compound ($T_c$=38.5 K), supports the scenario of a competing interaction between the CDW order and superconductivity. Furthermore, the extremely short lifetime of this mode ($\sim$300 fs) demonstrates the capability of ultrafast spectroscopies to investigate fluctuating phenomena that would remain inaccessible by frequency-domain spectroscopies. A similar result was also reported for underdoped YBa$_2$Cu$_3$O$_{6+x}$, where a coherent oscillation with frequency $\nu$=1.87 THz and doping-dependent onset temperature ($T_{CDW}$=105(130) K for $x$=0.67(0.75)) \cite{Hinton2013} has been detected via P-p optical spectroscopy. Interestingly, the characteristics of the CDW mode change when the the system is cooled below the superconducting temperature. For $T<T_c$, the CDW oscillation amplitude increases, the phase shifts by $\pi$, and the frequency decreases by $\delta \nu/\nu\sim$7$\%$. The analysis through the coupled Ginzburg-Landau equations (see Eqs. \ref{eq_GLenergy_coupled}, \ref{eq_GL_coupled1} and \ref{eq_GL_coupled2} in Sec. \ref{sec_tdGL}) also supports the competing scenario between superconducting and CDW orders \cite{Hinton2013}. Finally, recent femtosecond X-ray scattering experiments on underdoped YBCO, in which the vibrational modes involving the apical oxygen are excited though a mid-IR ultrashort pulse, have been performed \cite{Forst2014}. The observation that the CDW is significantly quenched when the $c$-axis trasport is photo-enhanced (see Sec. \ref{sec_PIsuperconductivity}), further confirms the CDW-SC competing scenario as the most plausible.

\subsubsection{Superconducting amplitude modes}
A very interesting topic that is attracting great attention concerns the possibility of directly observing the collective amplitude mode (AM) of the superconducting condensate. This mode corresponds to the the oscillation of the order parameter $\eta(t)$ about the equilibrium value $\eta_{eq}$ (see Sec. \ref{sec_tdGL}) at the proper frequency 2$\Delta_{SC}$. The existence of an intrinsic oscillation frequency, which is derived from the microscopic BCS theory, implies that the collective mode has acquired a \textit{mass}, in analogy with the Higgs mechanism in the Standard Model. For an historical review of the subject, see Refs. \citenum{Anderson2015} and \citenum{Varma2001}. In principle, two main obstacles should impede the excitation and the observation of the AM in a $s$-wave superconductor: i) its frequency resonates with the energy necessary to break Cooper pairs, resulting in a strong damping of the mode itself; ii) it does not couple directly to electromagnetic fields in the linear response regime. However, it was soon realized that when  the superconductivity coexists with the CDW order, the competing CDW gap pushes the total effective energy necessary to inject electron-hole excitations up to a value larger than 2$\Delta_{SC}$. Therefore, the simultaneous presence of the CDW order avoids the strong overdamping of the AM \cite{Littlewood1982} and, furthermore, if there is any coupling between CDW and superconductivity, the AM becomes Raman active \cite{Littlewood1982,Cea2014}. As the most promising candidate, the CDW superconductor NbSe$_2$ attracted many attentions. In this dichalcogenide, the onset of the superconducting phase, characterized by a $T_c$=7.1 K and $\Delta_{SC}\sim$1.2 meV, is anticipated by the onset of the CDW order at $T_{CDW}$=33 K and $\Delta_{CDW}\sim$5 meV. Indeed, a collective mode attributed to the Higgs AM was recently measured in NbSe$_2$ through Raman scattering \cite{Measson2014}. The attribution to the AM was corroborated by the comparison of the results on NbSe$_2$ with those on its isostructural partner, NbS$_2$, which lacks the charge density wave order and, accordingly, does not exhibit the signature of the AM.

These results triggered an effort toward the time-domain investigation of the elusive AM in conventional ($s$-wave) superconductors. It was argued that the hallmark of the AM should be an oscillation of the gap edge at twice the gap frequency \cite{Kemper2015}. The direct detection of this mode could be achieved via either TR-ARPES or optical spectroscopy in the gap energy range (THz) \cite{Matsunaga2012,Krull2015}. In the case of optical P-p measurements, the collective oscillatory response in the optical conductivity can be resonantly enhanced by tuning the frequency of the order-parameter oscillations to the energy of an optical phonon mode that is coupled to the superconducting AM \cite{Krull2014}. 

Unfortunately, the main problem in designing such a time-domain experiment is related to the excitation process. While, in principle, the inverse Raman process in a CDW-SC system could be be exploited to launch coherent AM, the excitation with photon energy $\hbar\omega\gg 2\Delta_{SC}$ would result in the strong photoexcitation of the electron-hole continuum which would probably destroy the superconducting phase. This problem was recently solved by employing a sub-gap THz excitation scheme, in which the intensity of the THz pump pulse is strong enough to \textit{non-linearly} couple to the superconducting condensate, thus exciting the AM beyond the linear response regime \cite{Matsunaga2013}. Time-domain oscillations at the frequency 2$\Delta_{SC}$ were reported on Nb$_{1-x}$Ti$_{x}$N films via THz pump-THz probe experiments, in which the probe directly monitors the dynamics of the absorption edge associated to the superconducting gap. As a further development, it was demonstrated the resonant excitation of collective modes in NbN via non-linear excitation with a coherent THz field. The superconducting nature of the resonance has been demonstrated by the occurrence of a large terahertz third-harmonic generation when the frequency of the THz pulse matches the value of the superconducting gap \cite{Matsunaga2014}. While it was originally proposed that this effect could be ascribed to the resonant excitation of the AM mode \cite{Matsunaga2014}, it has been recently argued that the contribution of the AM is largely subdominant with respect to the collective excitation of Cooper pairs \cite{Cea2015}. The reason is that even if the AM oscillates three-time faster than the incoming light, it weakly couples to the optical probe, that induces a collective density fluctuation. As shown in Ref. \cite{Cea2015} the non-linear optical process behind the third-harmonic generation shares interesting analogies with equilibrium Raman spectroscopy, including a polarization dependence that could be tested, e.g. in cuprates, to selectively excite the AM.

The problem becomes even more complex in the case of unconventional superconductors, which exhibit a $d$-wave order parameter. In these systems, it was theoretically shown the existence of a rich assortment of Higgs bosons, each in a different irreducible representation of the point-group symmetry of the lattice \cite{Barlas2013}. Recently, non-equilibrium broadband optical spectroscopy on copper oxides showed coherent oscillations at the gap frequency, which were attributed to the oscillations of the Cooper pair condensate and were discussed by an NMR/electron spin resonance formalism \cite{Mansart2013b}. Interestingly, the coherent oscillations affect the high-energy optical properties at the typical scale of Mott physics (2.6 eV). This finding further supports the scenario of a strong interplay between the high- and low-energy scale physics in unconventional superconductors (see Secs. \ref{sec_SWshift} and \ref{sec_latticemodes}).

The route towards the optical control of AM modes in unconventional superconductors is still long, but the recent advances in ultrafast techniques provide interesting routes to investigate the character of the order parameter and its coupling to other degrees of freedom \cite{Schnyder2011}, such as phonons, magnons or charge-density oscillations, and to develop novel schemes toward non-linear quantum optics in superconductors.

\setcounter{section}{5}
\section{Toward the optical manipulation and control of electronic phases in correlated materials}
\label{sec_optical_control}
The main body of this review is dedicated to time domain studies performed in the limit of small perturbation. The light pulses trigger a small redistribution of energy among the different degrees of freedom and the time evolution of the observables measured is used to unveil the basic interactions determining equilibrium properties in complex materials. 
A different regime of time domain spectroscopy explores the limits where light pulses can perturb the overall physical properties of materials on picoseconds timescales, enabling, in principle, an ultra-fast control over the material properties. The possibility of controlling materials properties on sub-picosecond timescales led, in the last decades, to a large number of experiments exploring photo-induced \textit{phase transformations}\footnote{Here we use the wording \textit{phase transformations} to address the field which commonly goes under such name. Nevertheless we will argue here that the use of the wording \textit{non-equilibrium phase} is often misleading and we propose in the following a new classification of the phenomena leading from photo-excitation to the onset new material functionalities}  "impulsively" injecting a large number of excitations in a suitable material. The basic idea of these experiments is to drive the transformations, by means of ultrashort light pulses "impulsively" injecting a large number of excitations. The photo-excitation within time windows shorter than the characteristic times of the relaxation processes drives the matter into highly non-equilibrium transient regimes characterized by anomalous energy distribution between electrons, ions, and spins, hence resulting in a sudden changes of the overall material properties.
The non-equilibrium pathways leading to such effects depends on the different physics scenario specific of the material investigated. Unravelling such pathways is recognized as an important challenge for the community in the years to come and a conference series is dedicated to this topic since 1998, while a number of focused issues \cite{PIPT1-5} and books \cite{Nasu2004} have been published.  
\begin{figure}[t]
\begin{centering}
\includegraphics[width=1\textwidth]{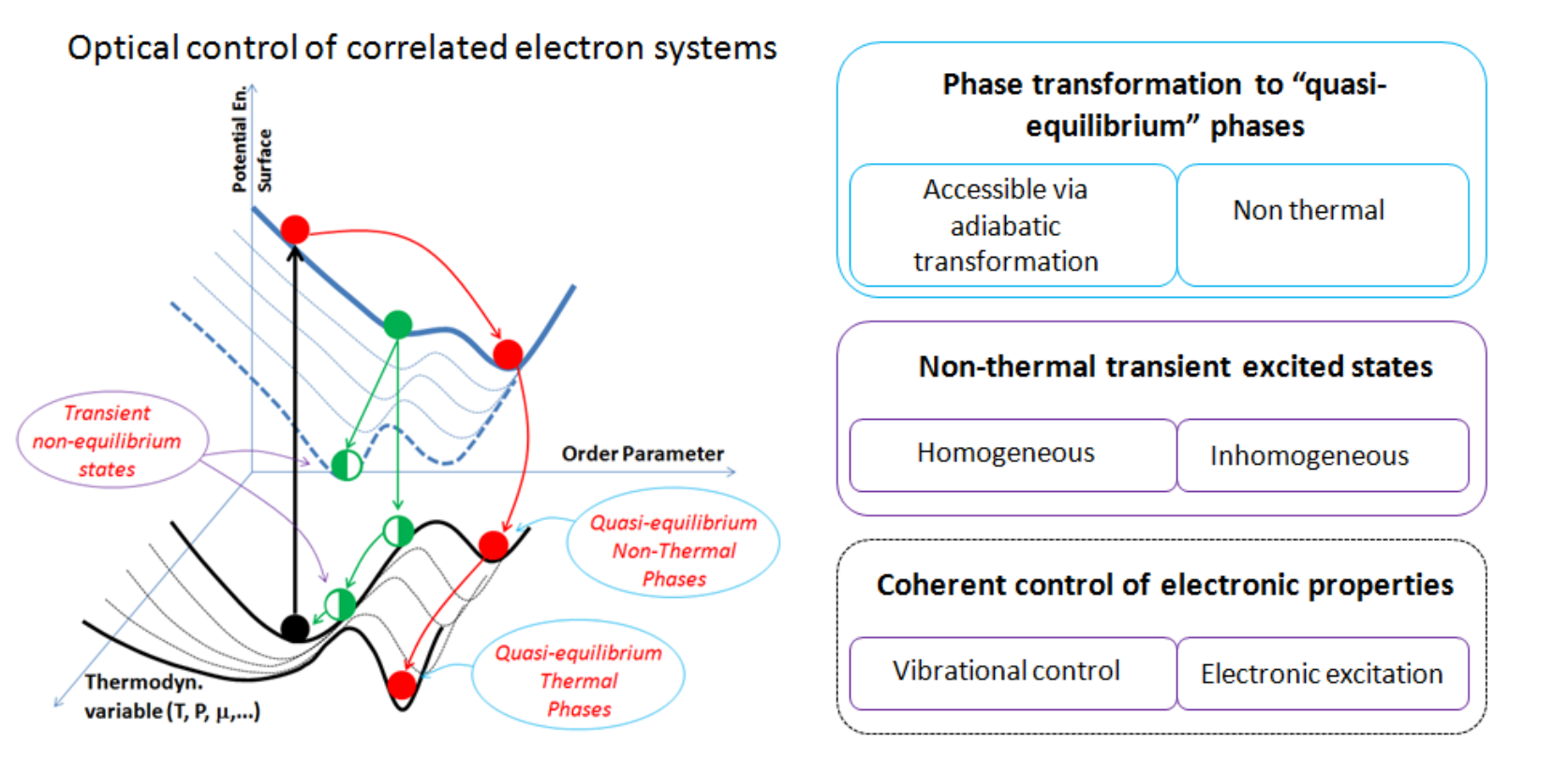}
\caption{Scketch for optical control of correlated electron systems. Photo-excitation can be used to produce transient macroscopic material's functionalities not present at equilibrium. The onset of such functionalities is typically the result of the transient non-equilibrium pathways triggered by photo-excitation (left sketch). The main advantages of this approach is twofold. On one hand the new functionality can be obtained on timescales which are not limited by thermodynamic constraints (sub ps), while on the other hand, the non-adiabatic nature of the processes can produce transient states not attainable at equilibrium. We propose here to classify the different evidences of transient light driven macroscopic change of physical properties as: phase transformation between quasi equilibrium phases, non-thermal transient excited states, and coherent control of electronic properties.}
\label{fig_optical_control}
\end{centering}
\end{figure}

Very often the light driven onset of a transient property in complex materials is reported as Photo-Induced-Phase-Transition (PIPT). Despite the common use of the wording PIPT, strictly speaking it is very rare that the inset of a new functionality is effectively associated to a new thermodynamic phase. 
Therefore, we propose here to classify the most common approaches to the ultrafast control of material properties under the following categories: 

\begin{enumerate}
\item[i)] \textbf{Photo-induced phase transition between quasi-equilibrium phases (PIPT) }; i.e. light driven transient states characterized by new macroscopic functionalities which are associated to a well defined thermodynamics potential.
\item[ii)] \textbf{Photo-Induced transient Non-Equilibrium States (PINES)}; i.e. transient light driven state with a new functionality but undefined thermodynamics potential.
\item[iii)] \textbf{Coherent control of transient matter states (COHCO)}; i.e. transient states associated to the presence of the perturbing fields.
\end{enumerate}

In this chapter we will proceed following the classification proposed and we will limit the discussion to some of the most intriguing experimental evidences and promising theoretical approaches to describe such regimes in correlated electron systems. Furthermore, we will limit the discussion to photo-induced phase transition (i) and transient non thermal states (ii). We stress that what reported here is not a comprehensive review on photo-induced phase transition and it should be rather seen as a selection of evidences reported to date. Actually the scope is to mainly provide a set of references related to this fast growing field along with a tentative framework to classify the most important results so far achieved.

\subsection{Photo-induced phase transitions (PIPT) between quasi-thermodynamic phases}
At equilibrium, the free energy that describes the phase of a material is at the global minimum given the value of the control parameters such as temperature and pressure. The crossing of the boundary between two different phases is commonly called phase transition and it occurs as the consequence of the variation of the control parameters.
The concept beyond photo-induced phase transition to quasi-equilibrium phases is to use ultrashort light pulses as an additional control parameter. The intrinsic advantage of photo-excitation with respect to quasi-static changes of temperature or pressure lies in the fact that the photo-excitation creates a far from equilibrium distribution of the energy among the different degrees of freedom. This can trigger the formation of \textsl{thermal} and sometimes \textsl{non-thermal} metastable phases. The \textsl{thermal} phases are those that can be attained via a quasi-equilibrium changes of temperature and pressure. The \textsl{non-thermal} phases are metastable states accessible only via a pathway including transient non-equilibrium distribution of energy between the different degrees of freedom; i.e. non-thermal phases are those phases not observable in equilibrium phase diagrams.

A central role in the effort of achieving an ultrafast control of electronic properties has been played by transition metal oxides (TMOs). The rich phase diagram of many TMOs is the result of the intricate interplay between electrons, phonons, and magnons that often makes TMOs very susceptible to the fine tuning of control parameters such as the pressure, the magnetic field, and the temperature. The same responsiveness makes TMOs an ideal playground to design experiments where the interaction between ultra-short light pulses and matter can trigger the formation of transient phases with specific, sometime exotic, physical properties.

\subsubsection{Insulator-to-metal transition in transition metal oxides}

The technological drive for increasing data processing speed and storage density led to a significant effort dedicated to photo-induced insulator-to-metal transition in 3$d$ transition metal oxides. In this framework, a large set of evidences revealing the transient changes of electronic properties have been discussed as photo-induced phase transitions. Various reports revealed that photo-excitation can lead to metallicity on picoseconds timescales in different transition metal oxides families ranging from manganites \cite{Chanduri2010, Rini2009, Takubo2008, Ichikawa2011}, iron oxides \cite{dejong2013}, and the vanadate families (VO$_2$ and V$_2$O$_3$) \cite{Cavalleri2004,Liu2011}. These findings anticipated the possible realization of a new generation of metal-oxide-based ultrafast electronic devices, whose electronic properties can be controlled on sub-picosecond timescales, i.e. three or four orders of magnitude faster than current semiconductor-based devices. 
In the following we will shortly review the studies on VO$_2$, allegedly the most widespread photo-induced phase transformation in inorganic compounds. The high electron density of the insulating phase (10$^{22}$ electrons/cm$^3$) as well as the vicinity of the equilibrium insulator-to-metal transition to ambient temperatures led to large number of experiments with different time-domain techniques, such as time-resolved X-ray \cite{Cavalleri2001,Hada2010} and electron diffraction \cite{Baum2007,Morrison2014}, P-p spectroscopies in the THz/mid-IR/visible ranges \cite{Cavalleri2004,Hilton2007,Kubler2007,Pashkin2011,Wall2012,Cocker2012,Hsieh2014} and time-resolved photoemission spectroscopy \cite{Wegkamp2014}.

\begin{figure}[t]
\begin{centering}
\includegraphics[width=0.9\textwidth]{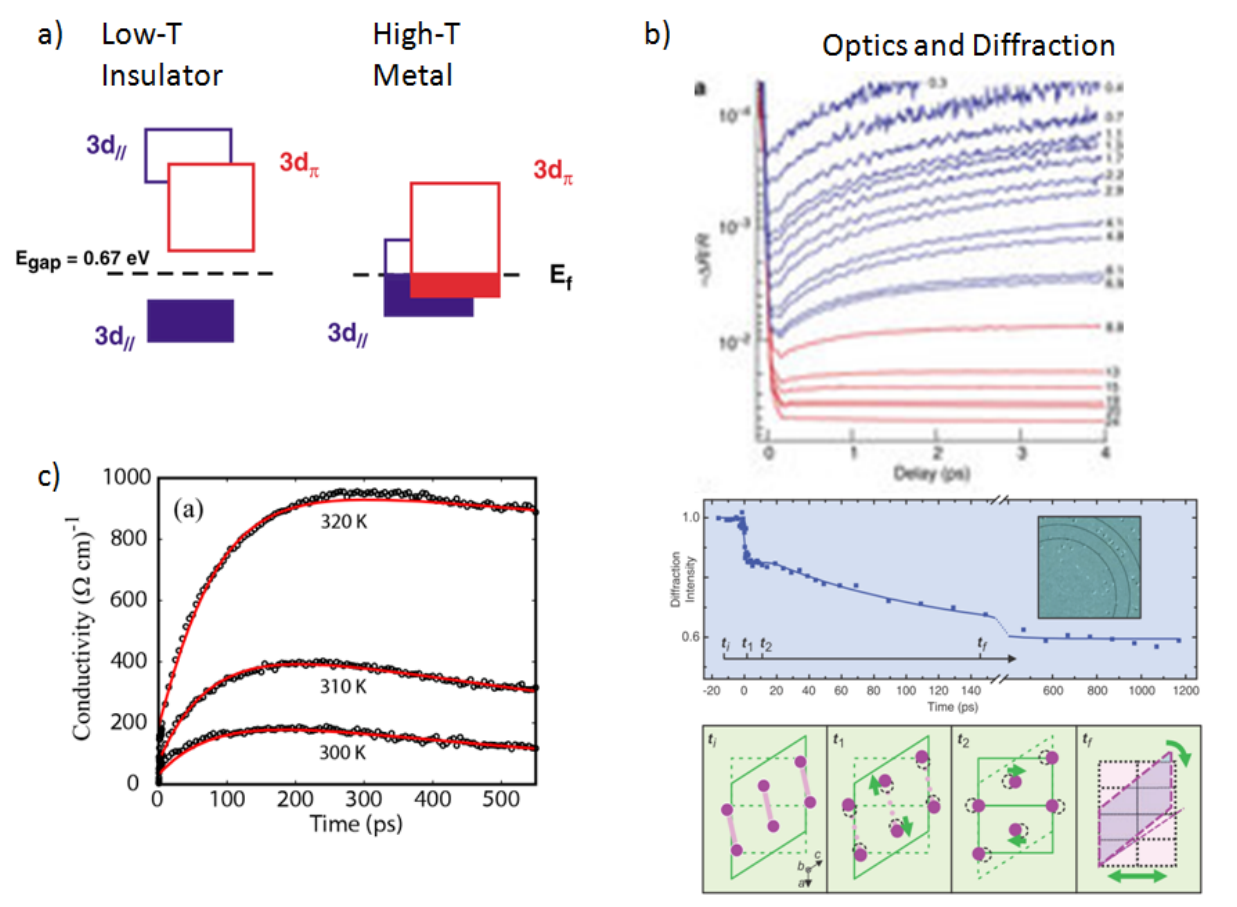}
\caption{a) Schematic of the insulator-metal transition in VO$_2$. While the optical properties (reprint from \cite{Wall2012}) and diffraction (d, \cite{Baum2007}) reveal a quasi-instantaneous response, the far IR conductivity (c, \cite{Hilton2007}) changes on much longer timescales.}
\label{fig_PIPT_VO2}
\end{centering}
\end{figure}

Below the critical temperature $T_c\simeq$340 K, VO$_2$ undergoes a transition from a metallic rutile (R) to an insulating monoclinic (M$_1$) phase. The M$_1$ phase can be obtained from R by V-V dimerization along the rutile c-axis, which thus leads to the doubling of unit cell along this direction. Nowadays, there is general consensus that VO$_2$ can be considered as a \textit{strongly correlated Peierls insulator} in which the short range Coulomb repulsion drives the formation of the dynamical V-V pairs necessary to initiate the structural phase transition \cite{Biermann2005}. 

At ambient temperature, the photo-excitation with visible and near-IR pulses results in a sudden ($<$ps) change of the electronic properties. The large increase of conductivity occurring in the first picoseconds is followed by a slow response revealing that highly photo-excited VO$_2$ reaches a metallic state immediately after photo-excitation. The high energy spectral response is changed within one picosecond \cite{Rini2008,Cilento2010} as well as the local vanadium environment \cite{Baum2007,Wall2012}. On the other hand, in highly photo-excited VO$_2$ single crystal the low energy properties evolve for longer times up to hundreds of picoseconds \cite{Hilton2007,Liu2011}. 

Besides the obvious technological implications, the possibility of photoinducing the insulator-to-metal phase transition in VO$_2$ attracted many interests for disentangling in the time-domain the role of the structural phase transition from the pure electronic mechanisms responsible for the insulating properties.  
The first pioneering works with time-resolved X-ray diffraction and near-IR optical spectroscopy \cite{Cavalleri2001,Cavalleri2004} demonstrated that the time necessary to photoinduce the switching from the M$_1$ to the R phase is $\sim$80 fs, which is exactly half period of the phonon connecting the two crystallographic phases. These results were obtained with a pump excitation larger than $\sim$10 mJ/cm$^2$, which exceeded the energy needed for a quasi-adiabatic transformation. The presence of the 80 fs \textit{bottleneck} suggested that the structural rearrangement is fundamental for achieving the complete phase transformation.

Later on, the dramatic advances in multi-THz spectroscopy led to the possibility of directly disentangling the electronic and lattice dynamics with a time resolution of 40 fs \cite{Kubler2007,Pashkin2011}. These measurements unveiled an instantaneous (within 40 fs) photoinduced metallicity and a coherent response of the V-V structural motion, which were interpreted as the result of the local photoexcitation of the vanadium dimers from the insulating state. A clear threshold of 2-4 mJ/cm$^2$ at 340-300 K was reported. The absence of a \textit{bottleneck} to photoinduce the transient metallic state, led to reconsider the role of the local electronic correlations in the stabilization of the insulating M$_1$ phase. 

All this rich (and to some extent contrasting) phenomenology was recently reconciled by time-resolve electron diffraction measurements \cite{Morrison2014}. The possibility of directly tracking the position of the V atoms allowed the authors to identify two different excitation regimes: i) above $\sim$9 mJ/cm$^2$ the pump excitations drives the non-thermal melting of the M$_1$ phase in a fraction of crystallites; ii) at intermediate fluences (2-9 mJ/cm$^2$), the pump excitation is not sufficient to initiate the structural phase transition, but drives a \textit{non-thermal} metastable metallic state in the M$_1$ crystallographic structure. This transient metallic state, which is not accessible in equilibrium conditions, is the direct consequence of the instantaneous collapse of the Mott gap after the strong photoexcitation of carriers from the localized V-3$d$ valence states. This picture has been confirmed by time-resolved photoemission spectroscopy \cite{Yoshida2014,Wegkamp2014}, which directly showed the instantaneous collapse of the Mott gap in the intermediated 2-9 mJ/cm$^2$ pump-fluence regime.

A similar phenomenology has been also observed on V$_2$O$_3$ by time-resolved far-infrared spectroscopy \cite{Liu2011}. The possibility of directly probing the dynamics of the Drude conductivity allowed the authors to disentangle two different pathways for the photoinduced phase transition: i) a photothermal transition from antiferromagnetic insulator to paramagnetic metal in $\sim$20 ps; ii) an incipient strain-generated paramagnetic metal to paramagnetic insulator transition on a timescale of $\sim$100 ps. 

\subsubsection{Photo-induced metastable non-thermal phases}
\begin{figure}[t]
\begin{centering}
\includegraphics[width=1.2\textwidth]{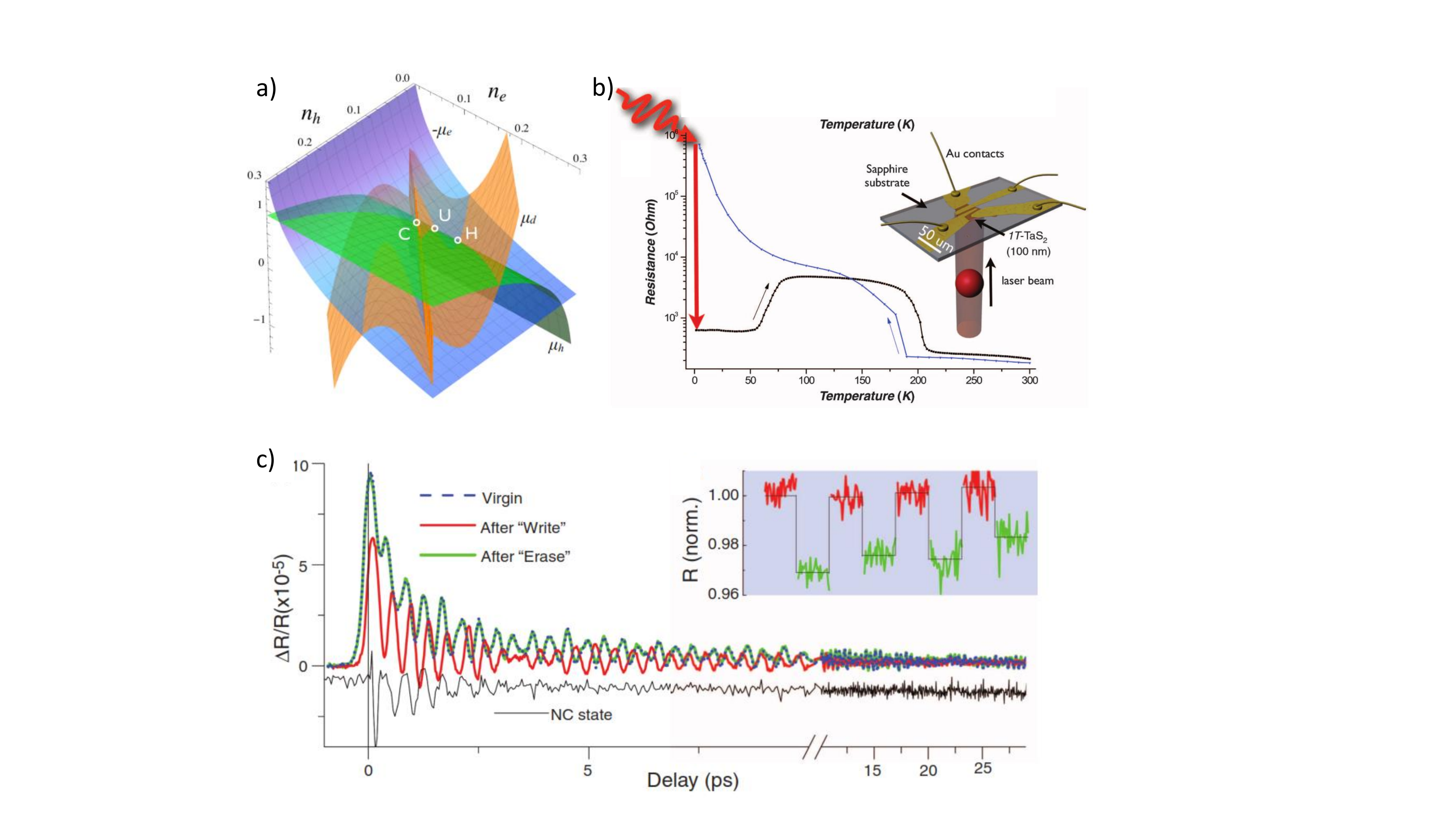}
\caption{Exploring a non-equilibrium phase diagram by photo-excitation in TaS2. A transformation towards a persistent non-thermal phase can be triggered by ultrashort light pulses in TaS2. The transient electron-hole asymmetry photo-excited can push the material into different electronic free energy minima (a), leading to quasi-stable phase characterized by a conductance (b) and structural caracteristic (c) different from any thermodynamically accessible state. Taken from Ref. \citenum{Stojchevska2014}.}
\label{fig_TaS2}
\end{centering}
\end{figure}

The phase transitions towards metastable phases that are not accessible through quasi-adiabatic transformations are quite attractive for both fundamental physics and technological applications. Specifically, it is possible to access non-thermal minima in the free energy by resonantly exciting specific degrees of freedom (electron, phonon, or spin excitation). 
These quasi-equilibrium states of the matter are reached via a transformation requiring a non adiabatic distribution of energy between electron, lattice or spins. One seminal example of such a mechanism has been recently reported in TaS$_2$ \cite{Stojchevska2014}, where a transition towards a stable hidden phase in the material can be obtained through the sudden quench driven by $\sim$35 fs light pulses. In TaS$_2$ the interplay between electron correlation and the coupling of electrons with phonons makes the free energy dependence on different degrees of freedom quite complex. In particular, the transient photo-excited electron-hole asymmetry allows for a transient non-adiabatic doping. In figure \ref{fig_TaS2}a, a plot of the chemical potential surface as function of the number of electrons and holes is represented. The three surfaces represents the chemical potential surfaces for free electrons (blue), holes (green) and localized electrons (orange). Photo-excitation under the condition of electron-hole asymmetry can create electronic distributions not achievable via quasi-equilibrium transformation. As reported in Ref. \citenum{Stojchevska2014}, this mechanism can be exploited to drive the system between different \textit{quasi-ground state} electronic configurations, leading to different sample conductance (figure \ref{fig_TaS2}b) and vibrational characteristic (figure \ref{fig_TaS2}c).

\subsection{Photo-induced transient non-equilibrium states  and photodoping}
Photo-excitation in matter often result in transient non-linear responses that are associated to a non-thermal change of is physical properties. oppositely to the previous paragraph where we have identified photo induced phase transformation as a response associated to a novel quasi-equilibrium phase, transient responses displaying functionalities different from those at equilibrium and not associated to the quasi-equilibrium thermodynamic potentials can be often produced by light irradiation in complex material. In such states the onset of the new functionality is not associated to a phase transition, but rather to a transient non-equilibrium responses with undefined thermodynamic potentials. The dynamical onset and offset of a new functionality is, in this case, determined by the dissipation processes characteristic of the material. 

Many studies have used this non-linear response to address the nature of the interactions leading to the pairs formation in high temperature superconductors. The rationale of those studies have been to use photo-excitations to trigger a partial (or full) quench of the superconducting gaps in order to address via the dynamical response, the leading mechanisms underneath high temperature superconductivity.

\subsubsection{Non-thermal melting of the superconducting condensate}
\label{sec_SCquench}
The destruction of the superconducting state in very thin Pb films was first proved by L.R.
Testardi \cite{Testardi:1971p1592}, by using 40 microsecond pulses at 2 W e 6 microsecond pulses at 5 W, generated by an Ar laser operating at  a wavelength of 5145 \AA. Testardi argued that a non-thermal process was responsible for the destruction of the SC state. Later on, the availability of ultrashort laser pulses enabled more
detailed studies on this topic.
In conventional superconductors, phase coherence and pairing occurs
at the same temperature, so the evolution order parameter coincides
with the appearance of the gap. Thus one can discuss the process of
the destruction of the SC state in terms of pair-breaking using the
Rothwarf-Taylor model (see Sec. \ref{sec_RT}). 

\begin{figure}
\begin{center}
\includegraphics[width=0.5\columnwidth]{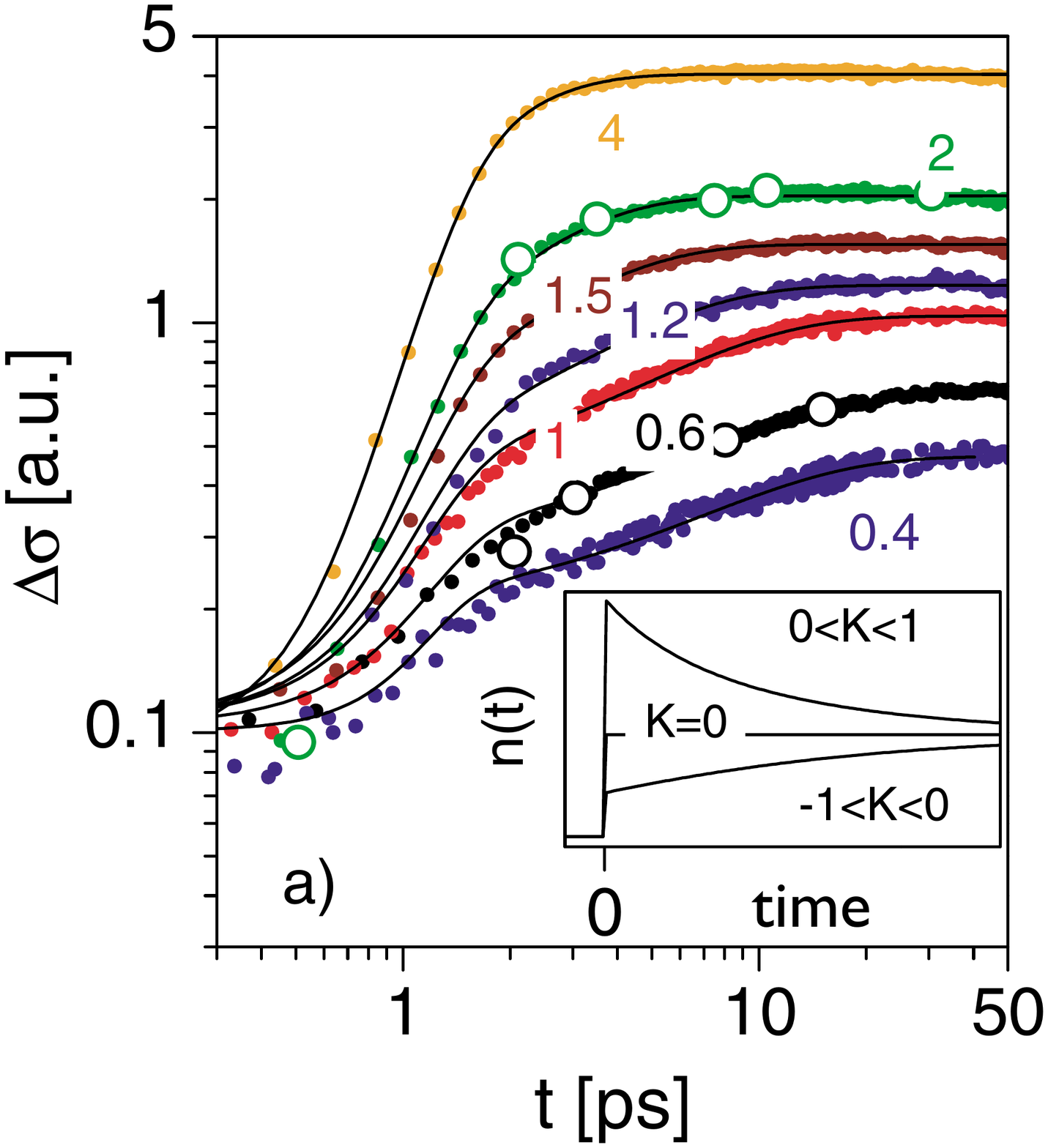}\protect\caption{\label{fig:The-pair-breaking-dynamics_MgB2}The pair-breaking dynamics
in MgB$_{2}$ at 7K exhibited by the transient conductivity $\sigma_{1}(t)$
measured by optical pump/THz probe reflectivity measurements with
different pump fluences $F=0.4-4\mu$J/cm$^{2}$. The insert shows
the time-dependence of \emph{n(t) }for different \emph{K}. Taken from \citenum{Demsar:2003p7494}.}
\end{center}
\end{figure}

Indeed, the conventional superconductors MgB$_{2}$ and NbN
have been discussed very successfully in terms of the R-T equations
taking into account pair-breaking by phonons generated in the photoexcited electron thermalization process. The so-called pre-bottleneck scenario gives and expression for the time-evolution of the QP dynamics within
the R-T model. Exact solutions for the QP density $n$ are given in
\cite{Kabanov2005,Demsar2003}:
\begin{equation}
n(t)=\frac{\beta}{R}\left[-\frac{1}{4}-\frac{1}{2\tau}+\frac{1}{\tau}\left(\frac{1}{1-Ke^{-\beta t/\tau}}\right)\right]\label{eq:pairbreaking_dominant}
\end{equation}
where \emph{R} and $\beta$ are the usual bare QP recombination rate
and pair-breaking rates respectively, while K and $\tau$ are dimensionless
parameters determined from the initial conditions: $K$=$\frac{\tau\left(\xi+1\right)/2-1}{\tau\left(\xi+1\right)/2+1}$,
$\tau^{-1}=\sqrt{\frac{1}{4}+\frac{2R}{\beta}\left(n_{0}+2N_{0}\right)}$,
where $\xi=4R_{0}/\beta$. $n_{0}$ and $N_{0}$ are the initial QP
and phonon populations. Hence, when the number of phonons $N$ is large after the 
photoexcitation, a regime where $-1\le K<0$, i.e., the initial
phonon temperature is higher than the equilibrium value, is established. Fig. \ref{fig:The-pair-breaking-dynamics_MgB2}
shows a good fit to Eq. \ref{eq:pairbreaking_dominant} \cite{Demsar2003}.
Interestingly, a very similar pair-breaking dynamics was discussed also for NbN \cite{Beck2011}. 

\begin{figure}
\begin{center}
\includegraphics[width=0.6\columnwidth]{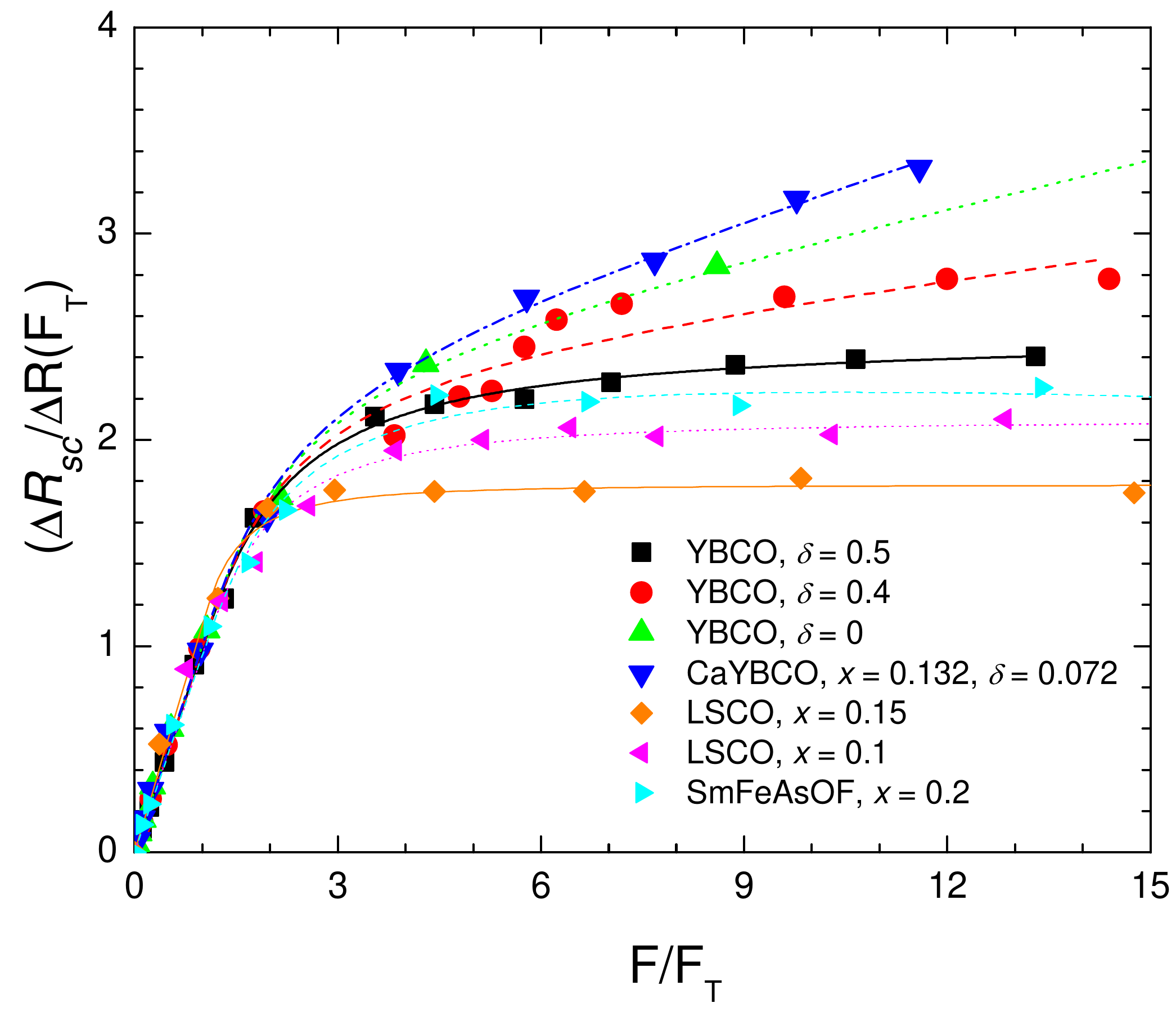}

\includegraphics[width=0.6\columnwidth]{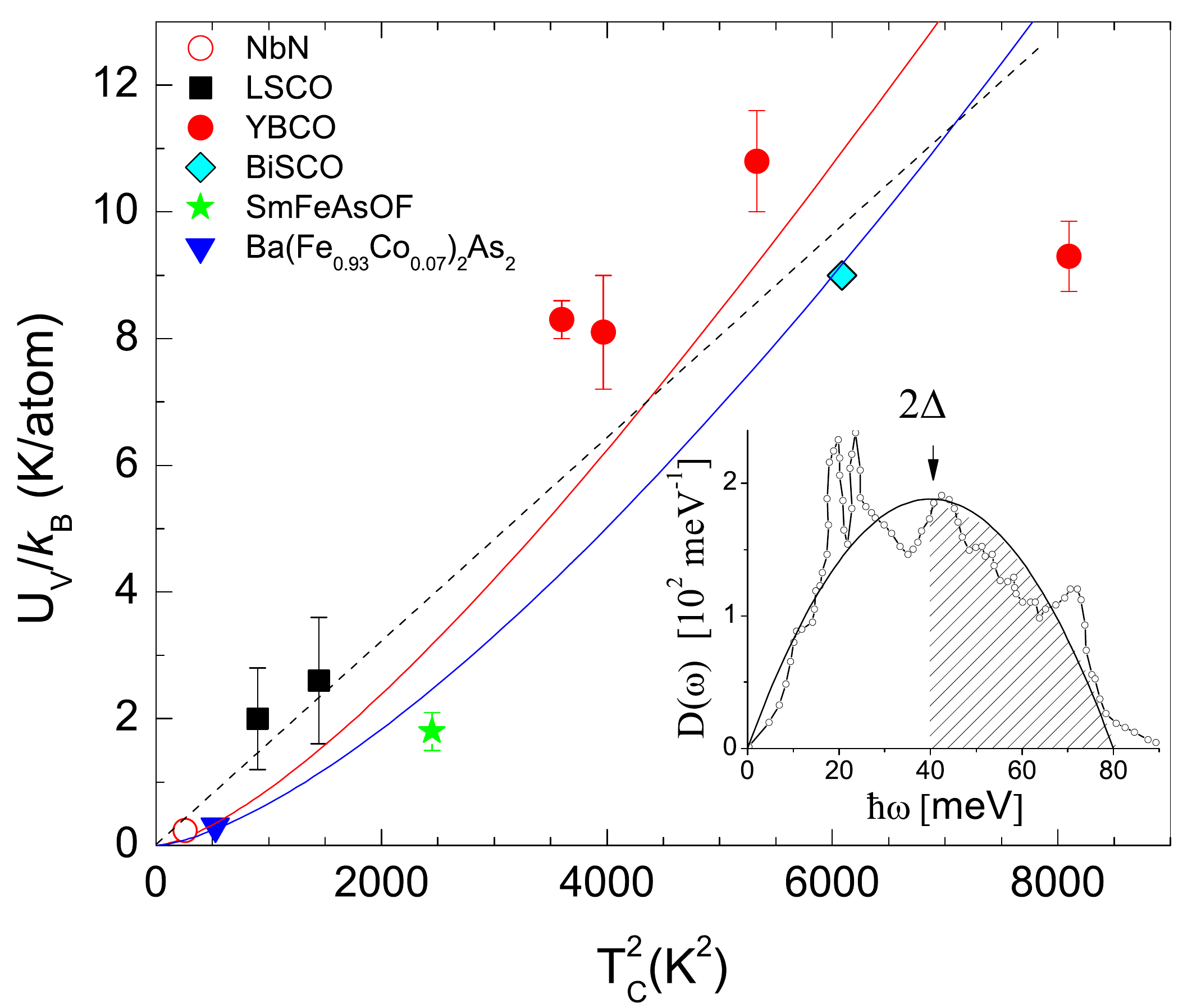}

\protect\caption{
\label{fig:Vaporisation-energy}a) The transient reflectivity $\delta R/R$
as a function of fluence $F$, normalised to the threshold value $F_{T}$
for a number of different superconductors. b) Vaporisation energy
$U_{v}$ , as a function of $T_{c}^{2}$ for the cuprates (expressed
in Kelvin per planar Cu), NbN (in \emph{K}/Nb), and for the pnictides
SmFeAsO$_{0.8}$F$_{0.2}$ and Ba(Fe$_{0.93}$Co$_{0.07}$)$_{2}$As$_{2}$
(in K/Fe). The solid curve is a plot of $U_{lost}$ using $D(\omega$$_{ph}$$)$
from Eq. \ref{eq:U_lost_2} for YBaCuO (red) and the iron oxy-pnictide
(blue). The dashed line is a square law $U_{v}=\eta T_{c}^{2}$. The
inset shows the phonons with $\hbar\omega>2\Delta$ (shaded) participate
in the relaxation. The measured phonon density of states $D(\omega_{ph})$
for YBCO (Ref. 20) is approximated by a parabola (line) for the purposes
of the calculation. Taken from Ref. \cite{Stojchevska:2011fz}.%
}
\end{center}
\end{figure}

As already discussed in Sec. \ref{sec_Bi-cuprates}, a saturation of the P-p signal was universally observed \cite{Averitt2001,Carnahan2004,Kaindl2005gp,Kusar:2008p5434,Giannetti2009b,Kusar:2010hz,Stojchevska:2011fz,Coslovich2011,Toda:2011cs,Beyer2011,Coslovich2013} also in copper oxides at relatively small pump fluences ($F_{th}$=10-70 $\mu$J/cm$^2$). This phenomenon was attributed to the non-thermal superconducting-to-normal state phase transition induced by the pump pulse. Intriguingly, it was argued that this transition is driven by the loss of phase coherence, before the complete closing of the superconducting gap is achieved \cite{Giannetti2009b,Coslovich2011,Piovera2015}. This process can lead, in analogy to the case of a magnetic-field quench of type I superconductivity, to a non-thermal first-order phase transition where superconducting and normal-state domains can coexist at the nanoscale. In this picture, from the saturation of the $\delta R/R$ response it is possible to extract the vaporisation
energy $U_{v}$ necessary to destroy the condensate. The data for a number of
diverse materials are shown in Fig. \ref{fig:Vaporisation-energy}a) \ref{Stojchevska:2011fz}.
A systematic investigation of the processes leading to nonthermal
condensate vaporization shows that the process exhibits a strong systematic
dependence of the vaporization energy on $T_{c},$ varying approximately
as $U_{v}\sim T_{c}^{2}.$ The behaviour can be shown to be described
by a phonon-mediated quasiparticle (QP) bottleneck mechanism within
the RT model, where the ratio between $\Delta_{SC}$ in
relation to the phonon spectrum determines the dependence of $U_{v}$
on $T_{c}$ (see Fig. \ref{fig:Vaporisation-energy}b).
To model the dependence of $U_{v}$ on $T_{c},$ let us consider how
the energy transfer between photoexcited (PE) carriers and the condensate
depends on the size of the SC gap. After initial thermalization
(which is over in 100 fs) a large non-equilibrium phonon population
is created. The phonons whose energy exceeds the gap $\hbar\omega_{\mathrm{ph}}>2\Delta_{SC}$
can subsequently excite QPs from the condensate, but the \textit{low
frequency} phonons with $\hbar\omega<2\Delta_{SC}$ cannot, so they do not contribute
to the vaporisation process. The lost energy to low-energy phonons
is: 
\begin{equation}
U_{\mathrm{lost}}=\int_{0}^{2\Delta_{SC}}\delta f(\omega)D(\omega)\hbar\omega d\omega\label{eq:U_lost}
\end{equation}
where $D(\omega)$ is the phonon density of states and $\delta f(\omega)=f_{NE}(\omega)-f_{E}(\omega)$
is the difference between the non-equilibrium and equilibrium phonon
distribution functions $f_{NE}$ and $f_{E}$ respectively. To estimate the dependence of
$U_{lost}$ on $T_{c}$ from Eq. (1), we assume that $\delta f(\omega)$
is constant, and approximate the experimental $D(\omega)$ by an inverted
parabola $D(\omega)=\frac{\alpha}{8}\omega(2\omega_{0}-\omega)$, where $\alpha$ is a constant. Integrating Eq. \ref{eq:U_lost}, we obtain: 
\begin{equation}
U_{\mathrm{lost}}=\alpha\Delta_{SC}^{3}[\frac{2\hbar\omega_{0}}{3}-\frac{\Delta_{SC}}{2}].\label{eq:U_lost_2}
\end{equation}
Assuming a constant gap ratio $2\Delta_{SC}/k_{\mathrm{B}}T_{\mathrm{c}}=R$,
with $R=4$, for YBCO, $\hbar\omega_{0}=40$ meV extending $D(\omega)$
to $80$ meV, as shown in the insert to Fig. \ref{fig:Vaporisation-energy}, Eq. \ref{eq:U_lost_2}
gives the curve shown in Fig. \ref{fig:Vaporisation-energy}b. For
iron pnictides the phonon frequencies are lower than in oxides, but
taking $\hbar\omega_{0}=20$ meV, the predicted variation
of $U_{lost}$ on $T_{c}$ is not significantly different (Fig. \ref{fig:Vaporisation-energy}b)).
For the superconductor series Eq. \ref{eq:U_lost_2} predicts the
dependence of $U_{v}$ on $T_{c}$ quite well, which is not surprising,
considering that the gross features of $D(\omega)$ do not vary significantly
from material to material.

The total vaporisation energy is the sum of the condensation energy
and the lost energy, $U_{\mathrm{v}}=U_{\mathrm{c}}+U_{\mathrm{lost}}$.
Since for large gap systems $U_{\mathrm{v}}\gg U_{\mathrm{c}}^{\mathrm{exp}}$, then $U_{v}\simeq U_{\mathrm{lost}}$. As a consequence, $U_{c}^{exp}$ is
only a small contribution to $U_{v}$, hence explaining why the anomalous
doping dependence of $U_{c}^{exp}$ observed in the cuprates
is not displayed by $U_{v}$. Comparing $U_{v}$ for NbN measured
using the same technique \cite{Beck2011}, $U_{\mathrm{v}}/U_{\mathrm{c}}^{\mathrm{exp}}=1.7$
is considerably smaller than in the cuprates, which can now be understood
in terms of Eq. \ref{eq:U_lost_2} and the insert to Fig. \ref{fig:Vaporisation-energy}b.
When $2\Delta_{SC}<<\hbar\omega_{\mathrm{Debye}}$ almost all phonons can
excite QPs, so $U_{lost}\rightarrow0$, and $U_{v}\simeq U_{c}$,
in agreement with the obvserved ratio of $U_{\mathrm{v}}/U_{\mathrm{c}}^{\mathrm{exp}}=1.7$.
Thus for small-gap superconductors, the optical method can be used
to give a reasonable estimate of the superconducting condensation energy.

\subsubsection{Photoinduced superconductivity}
\label{sec_PIsuperconductivity}
The conjecture that the superconducting state might emerge, or be
enhanced as a result of an appropriate photoexcitation, goes back many decades\cite{Kumar1968}.
In particular, many interests were focused on materials where the appearance of
superconductivity is governed by the doping. Starting with a material
set in a region of the phase diagram close to the superconducting
transition, a small additional photodoping might result in a metastable superconducting state. The main problem is that photoexcitation -
that creates equal numbers of electrons and holes - does not increase the carrier density unless an unbalanced population is
created near the Fermi level. Moreover, if such a condition can
be achieved, it must hold long enough for establishing the macrosopic
phase coherence. Unfortunately, achieving the required
$e-h$ asymmetry is difficult, and attempts of photoinduced states
with CW excitation lead to heating, and possibly a "photoinduced"
pyroelectric effect \cite{Mihailovic:1990p1474}. 

In recent years, a few examples of doping-induced metastable photoinduced
states have been discovered. Here we will limit the discussion
to materials that exhibit photoinduced superconductivity.

\paragraph*{Persistent photoconductivity.}

A possible mechanism for creating a metastable state is to utilise
the trapping of electrons on defects and vacancies \cite{Fugol1990}.
Such a trapping can occur in insulating layers of a structurally heterogeneous
superconductors, but also on impurities or within twin or
grain boundaries\cite{Tanabe:1994kp}. Indeed, these
effects are quite common in nonstoichiometric complex oxides \cite{Hayes2004,Bridges1990}.
For example, the trapping of electrons in colour centers of the Cu-O chains in YBa$_{2}$Cu$_{3}$O$_{7-\delta}$
leads to hole doping in the CuO$_{2}$ planes, hence mimicking
the chemical doping. Samples close to the insulator-superconductor
phase boundary can then show metastable superconductivity. The effect,
originally observed by Kudinov \cite{Kudinov1993,Hoffmann1997},
was suggested to be intrinsically related to the ordering of the oxygen atoms into
chain fragments subsequent to photoexcitation \cite{Kudinov1993}.
The process is correlated with the appearance of luminescence \cite{Federici1995},
only recently it was clarified by novel experiments and theory \cite{Bahrs2004,Bahrs2005,Bruchhausen:2004hda}.

An interesting arises from the large transient photoinduced enhancement
of the superconducting critical temperature from 43 K to 67
K in epitaxial YBa$_{2}$Cu$_{3}$O$_{6.7}$/La$_{0.7}$Ca$_{0.3}$MnO$_{3}$
bilayers upon visible light illumination \cite{Pena2006}. In contrast
with the case of persistent photoconductivity, here the photoinduced
enhancement of the critical temperature is transient and it relaxes slowly
on a timescale of 100 s. While in the first persistent photoconductivity experiments the oxygen ion defects are involved,
in these layered samples a charge transfer between the layers is suggested to be
responsible for the photodoping. This demonstrates the possibility of
light induced charge transfer through interfaces, by a process similar
to carrier injection though semiconductor junctions.

\begin{figure}
\includegraphics[scale=0.7]{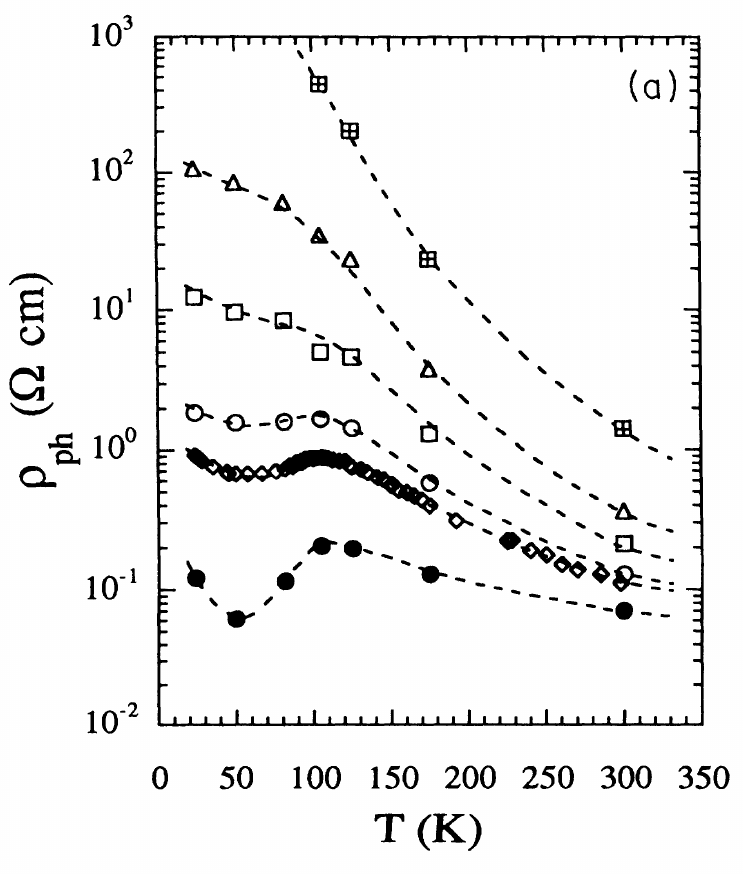}\includegraphics{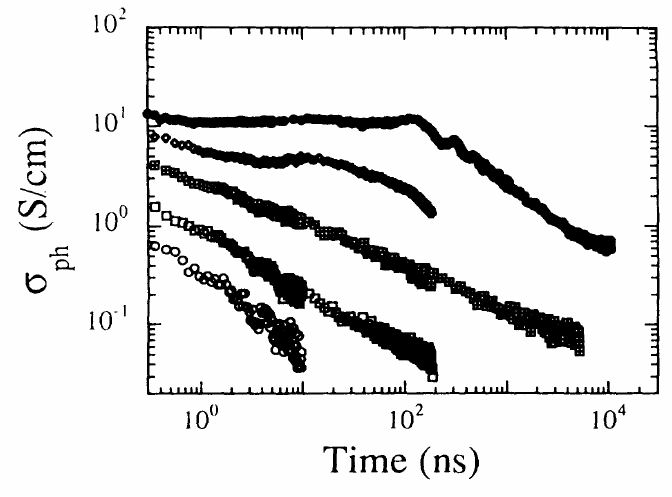}
\label{fig_Yu}
\protect\caption{The transient photocurrents recorded with different laser fluences
in insulating YBa$_{2}$Cu$_{3}$O$_{6.3}$. a) At a critical fluence
of $\sim7\times10^{15}$ photons/cm$^{2}$, the transient photoinduced
resistivity$\rho_{PC}(T)$ shows a dip near 90 K, close to the superconducting
$T_{c}=93$ K of the optimally doped compound. b) $\sigma_{PC}$ as
a function of time shows an increase of lifetime $\tau_{PC}$ with
increasing fluence, again with a remarkable crossover to metastability
at a fluence near $\sim10^{16}$photons/cm$^{2}$. Taken form Ref. \citenum{Yu1992}.}

\end{figure}

\paragraph*{Transient photoconductivity.}
Yu et al. have pioneered ultrafast transient photoconductivity experiments \cite{Yu1992,Yu1991} on insulating YBa$_{2}$Cu$_{3}$O$_{6.3}$
crystals. Their studies show an intriguing transient drop of the resistance on the picosecond
timescale appearing at temperatures near 90 K. Interestingly, this finding reflects the superconducting
critical temperatures in chemically doped YBa$_{2}$Cu$_{3}$O$_{6.9}$.
The crystals were mounted in a stripline circuit, with a resolution
of 50 ps at 2.6 eV photon energy, or 600 ps at 3.7 eV, and the transient
photocurrent was recorded with a bandwidth $>200$ GHz. Interestingly, a similar
effect was also observed in La$_{2}$CuO$_{4+\delta}$ \cite{Yu1992}.
In these experiments, the lifetime of the photocurrent response was found to
be strongly dependent on photoexcitation fluence, displaying a clear
crossover around $7\times10^{15}$ photons/cm$^{2}$ (which corresponds
to $2\sim3$ mJ/cm$^{2}$ for 2.6 eV photons), where a dip appears
in the photoinduced resistivity $\rho_{ph}(T)$ curve at 90 K (Fig.
 \ref{fig_Yu}a). A the same laser fluence, a kink appears in the photoinduced
conductivity $\sigma_{ph}(t)$ (Fig. \ref{fig_Yu}b), which indicates the existence
of a metastable photoexcited state. Yu et al \cite{Yu1992}
attribute this to the formation of metastable metallic droplets embedded in
the insulating matrix and that become superconducting at low temperatures.
The critical fluence is quite large in comparison with the vaporisation
energy required to melt the superconducting condensate, but is close
to the fluence ($\sim$0.75 mJ/cm$^{2}$) required to manipulate the
pseudogap in La$_{1.9}$Sr$_{0.1}$CuO$_{4}$ \cite{Kusar:2010hz},
and it is lower than the fluence that causes photoinduced structural
instabilities ($>$10 mJ/cm$^{2}$) \cite{Gedik2007}. 

The microscopic mechanism for the observed metastability is unknown yet, although it is clear that the process is relatively inefficient.
The majority of photoexcited $e-h$ pairs are presumed to annihilate
very rapidly, and only a small number of electrons remain trapped
in defects, allowing their hole partners to dope the CuO$_{2}$ planes,
forming superconducting droplets. As we shall see in the next section,
signatures may appear at higher frequencies in the optical response.

\begin{figure}
\includegraphics[scale=0.6]{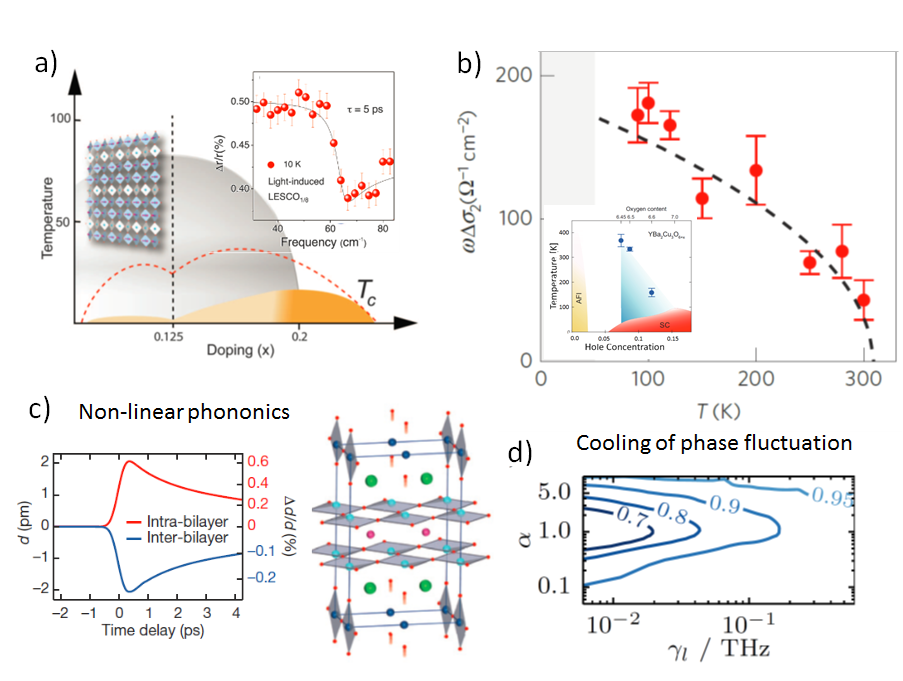}\label{Picture1}
\protect\caption{Various experiments have rekindled the possibility of using Mid-IR pulses to transiently enhance superconductivity. a) The first evidence was reported in La$_{1.675}$Eu$_{0.2}$Sr$_{0.125}$CuO$_{4}$ where irradiation with Mid-IR pulses triggered the onset of a low frequency Josephson plasma resonance (insert) indicative of 3D quantum coherent transport. b)The THz response to similar excitation condition in underdoped YBCO revealed a light-induced inductive response which is indicative of the formation of a condensate whose density vanishes approaching the pseudogap onset temperature. Different scenarios have been proposed for this experiemntal evidences. c) LDA calculation have shown that the resonant excitation of CuO stretching mode could lead to a modulation of the interplanar distances through non-linear phononic coupling. d) An alternative scenario is that the resonant excitation in the mid-IR could lead to an effective cooling of the interlayer phase fluctuation thereby leading to the onset of 3D coherent transport. The figures have been reproduced by Refs. \citenum{Fausti2011,Hu2014,Kaiser2014,Denny2015,Hoppner2015}.}
\end{figure}

\paragraph*{Photo-induced superconductivity based on mid-IR excitation.}
As extensively discussed in Sec. \ref{sec_gapenhancement}, sub-gap photons can be used to manipulate the distribution of the thermal excitations already present in the system, leading to the effective increase of $\Delta_{SC}$, as given by Eq. \ref{eq_deltaTeff}. This effect has been recently observed in NbN, under strong excitation by intense and narrow-band picosecond THz pulses \cite{Beck2013}. Although this effect is interesting \textit{per se}, it can be hardly extended to unconventional systems with anisotropic gaps. In copper oxides, even though the large antinodal gap ($\sim$ 40 meV) would allow to achieve th sub-gap condition by strong mid-IR excitation, the presence of nodal regions with zero or vanishing gap prevents from reaching the full \textit{effective} \textit{cooling} regime. 
Nonetheless, the effort for achieving by photoexcitation superconductivity in condition where a superconducting state is thermodynamically unattainable rekindled thanks to recent discoveries revealing that a superconducting phase may be attainable in photo-excited underdoped cuprates \cite{Fausti2011,Hu2014}. The physical mechanisms leading to transient superconductivity in such systems is still debated but the set of evidences and theoretical proposals is growing by the day.
The first reports proposed to drive by light pulses the formation of superconducting phases in non-superconducting stripe ordered underdoped cuprates. For example, in La$_{1.675}$Eu$_{0.2}$Sr$_{0.125}$CuO$_4$ and La$_{2-x}$Ba$_{x}$CuO$_4$ superconductivity at equilibrium is inhibited by the presence of a competing charge ordered phase  \cite{Fausti2011,Nicoletti2014}. The main idea of these works is to use near-IR pulses to melt the charge ordered phase in underdoped cuprates via vibrational excitation. The qualitative argument reported by Fausti et al. is that the resonant excitation of vibrational modes could melt the charge order \cite{Forst2014b} without a significant heating of the electrons, hence leading to a superconducting state not reachable by adiabatic transitions. In these works, the onset of out-of-equilibrium quantum coherent transport is measured by the light-induced appearance of a Josephson plasma resonance(insert fig. \ref{Picture1}a), that at equilibrium is linked to the onset of 3D coherent transport \cite{Tamasaku1992,Basov2005}.

The evidence that the resonant excitation of vibrational modes may lead to a transient quantum coherent state motivated various recent reports using similar experimental settings to study the effects of resonant excitation of the off-plane CuO modes in underdoped YBCO. A systematic study of the effect as a function of the doping revealed that mid-IR excitations result in a transient response possibly related to the quantum coherent transport in all the pseudogap phase of underdoped YBCO\cite{Hu2014,Kaiser2014}(Fig. \ref{Picture1}b). Most interestingly, a signature of a transient superconducting state have been revealed up to ambient temperature in Y$_2$Ba$_2$CuO$_{6.45}$ \cite{kaiser2014,Hu2014}.

A comprehensive theory describing the light-induced quantum coherent transport at high temperatures is still lacking but some experimental evidences and theoretical speculations provide some clues. Under the excitation condition that leads to the onset of possible superconducting states \cite{Hu2014,Kaiser2014}, the distances between the CuO layers are modified by the photo-excitation process \cite{Mankowsky2014}. This is rationalized at a qualitative level by the presence of a non-linear coupling between the ac-driven CuO phonon mode and the inter- and intra-bilayer distances \cite{Subedi2014,Forst2014}(Fig. \ref{Picture1}c). The resonant excitation of the c-axis CuO mode leads to an increase (reduction) of the intra(inter)-bilayer distance. LDA calculations suggest that such modified intra- inter-layer distances may drive a reduction of the hybridization between the oxygen (in the oxygen-deficient chains) with the Cu orbitals. This is likely to increase the doping and the $d_{x^2-y^2}$ character of the Fermi surface. A doping increase would push the system towards optimal doping while the change of the Fermi surface shape may help suppressing the charge-density wave order, which inhibits superconductivity \cite{Mankowsky2014}. Both effects may favour the formation of a transient superconductive phase, but further investigations are needed to consolidate this scenario.

The debate over the mechanism governing the possible transient superconducting state is lively and partly the same authors have proposed alternative scenarios. An alternated stack of insulating and superconducting layer, which are coupled by the Josephson interaction, can describe the electrodynamical response of the layered superconductors. In particular, in bilayer YBCO two different Josephson coupling are present between intra- and inter-layer planes. This peculiar structure give rise to two different plasma resonances associate to inter- and intra-layer coupling. Recent calculations suggest that an ac-excitation of a by layer superconductor resonant to the intra-layer plasmon can drive a cooling of the interlayer fluctuations, which are mostly connected to 3D transport properties \cite{Denny2015,Hoppner2015} (Fig. \ref{Picture1}d). The optimal suppression of phase fluctuations is observed for electric field resonant with the energy difference between inter- and intra-layer plasmon, while the experimental reports suggested a resonance with the CuO mode \cite{Hu2014,Kaiser2014}. Furthermore, it has been suggested that Dicke superradiance can cause the formation of transient superconducting states at temperature much higher than the equilibrium critical temperature \cite{Baskaran2012}. 

Various question remain open and the debate is lively on both the possibility of observing the onset of quantum coherent transport on ultrashort time scales \cite{Orenstein2015,Nicoletti2015} as well as on the theoretical scenarios disclosing the possibility of driving by ultrashort electric field the formation of a superconducting phase. 

The recent advances in optical and Free Electron Laser schemes to generate intense THz light pulses allowed also for the exploration of various non-linear transport response of superconductors to electromagnetic waves. In particular, the possibility of manipulating the superconducting order parameter along the $c$-axis allowed for the observation of a field driven quench of the Josephson coupling between the superconducting planes and revealed in such a condition the onset of an Ohmic response \cite{Dienst2011}. Further studies combining table top and FEL THz sources could disclose a regime where solitonic solutions for the propagation of low photon energy e. m. field could be observed \cite{Dienst2013}.
More recently in-plane THz excitation has been used to measure non-linear transmission of the condensate \cite{Glossner2012} and to trigger collective excitations associated the charge-density-wave-ordered in underdoped cuprate \cite{Dakovski2015}. The disappearance of the CDW oscillation at Tc revealed a strong coupling between the possibility of launching coherent excitation in the CDW and the onset of the SC order parameter.

\setcounter{section}{6}
\section{Recent advances and perspectives in the theoretical approaches for non-equilibrium correlated materials}
\label{sec_theory}
Almost every attempt to review our theoretical understanding of high-temperature superconductivity features the admission that, despite the
huge effort of the community, these materials still remain puzzling in
many aspects, including the most important question, namely the origin
of superconducting properties. One of the main reasons behind the lack of a complete understanding of these materials is indeed the inherent difficulty to treat the strong correlations that dominate their phase diagram, even in simplified models such as the two-dimensional single-band Hubbard model \cite{Hubbard1963}, (see Sec. \ref{sec_QPscattering}) or the related t-J model, which can be derived as the strong coupling limit of (Eq. \ref{eq_Hubbard}) for $U \gg t$, but it is sometimes considered as a more general model because it can also be obtained also from a more realistic three-band model which includes also oxygen $p_x$ and $p_y$ orbitals\cite{Zhang1988}:
\begin{equation}
\label{equation_tJ}
H = -\sum_{i,j,\sigma} t_{ij} c^{\dagger}_{i\sigma}c_{j\sigma} + J\sum_{ij} \vec{S}_i \cdot \vec{S}_j,
\end{equation}
where $\vec{S}_i = \sum_{\alpha\beta} c^{\dagger}_i\alpha \vec{\sigma}_{\alpha\beta} c_{i\beta}$ and  $\vec{\sigma}$ are the Pauli matrices. 

Despite the huge effort triggered by the idea that these models should feature the main properties of high-temperature superconductors, no exact solution has been found except for one dimension and in the limit of infinite coordination, and there seems to be still some controversy even about the qualitative phase diagram of the two-dimensional version of these models.
In this view, the path towards a theory of non-equilibrium
high-temperature superconductivity may seem long and
winding. On the other hand, one can reverse the perspective and consider
the non-equilibrium dynamics as an extra ``knob'' that we can use  to
explore the intrinsic properties of high-temperature superconductors
and strongly correlated materials. As discussed in the previous
chapters, pump-and-probe spectroscopies can be indeed used to drive
the system in metastable states which are not accessible in
equilibrium and/or to tune the properties of the system, similarly to
what can be done in ultracold atom systems, where we can control the
relevant parameters of a quantum system with a freedom which is not possible in solid
state\cite{Chu2002}. 

The tuning of the physical properties via photoexcitation is clearly not, at
least in the present stage, as simple, direct and controlled as in cold atoms. Nonetheless a joint experimental and theoretical effort can 
help to disentangle the various effects which determine the time-evolution of photoexcited materials, thereby identifying protocols to
control the properties of the materials, a strategy with a huge potential both for technological applications and for basic research.

Of course this process is at in an early stage, even if the experimental advances are setting a very fast pace. As a matter of fact, the present landscape of theoretical investigation of non-equilibrium correlated systems is rather complex and only a part of these studies are directly relevant to strongly correlated materials, while a large body of research is devoted to open quantum systems and to general concepts related to thermalization and lack thereof.
As a consequence, the aim of this section is not devoted a complete a comprehensive review of the 
diverse approaches to study the non-equilibrium physics  strongly correlated models, but it rather presents some of the main ideas which can help to get oriented in the evolving landscape of time-resolved experiments on high-temperature superconductors and correlated materials. In particular, we focus our attention on  the non-equilibrium extensions of the Gutzwiller approximation and the Dynamical Mean-Field Theory,  and we will briefly review their application to study some basic problems in which a correlated materials is driven out of equilibrium.

\paragraph*{Methods for non-equilibrium correlated systems}

Even if the research on the time evolution of non-equilibrium quantum systems is a relatively young field, a variety of methods have been introduced or adapted from their equilibrium counterparts. 

In this work we focus mainly on the time-dependent generalizations of  two of the most popular approaches for
equilibrium correlated systems, the Gutzwiller variational approach\cite{Gutzwiller1964}
and the Dynamical Mean-Field Theory\cite{Georges1996}. The choice of these two approaches is based on their success in the description of strongly correlated materials in equilibrium, especially in three dimensions and for homogeneous phases. Moreover, the two methods systematically improve on independent-particle methods like the Hartree-Fock scheme, and DMFT can be seen in turn as an improvement on the Gutzwiller approximation, thereby defining a hierarchy of theoretical methods which guides us to identify the microscopic mechanism behind the observed phenomena.
On the other hand, we will not discuss in any details many other approaches and ideas
that have been proposed to study strongly correlated systems out of
equilibrium, including for example exact diagonalization and density-matrix renormalization group (DMRG)\cite{Vidal2004,Feiguin2004}, quantum versions of classical methods to explore the dynamics, like the master equation\cite{Breuer2002} or the kinetic equation\cite{Rammer1998}, variational methods\cite{McLachlan1964} and Quantum Monte Carlo\cite{Goth2012}, even if we will comment on  some of the main results in relation with specific physical problems. In the spirit of the present work, we will not discuss issues connected to the thermalization in closed quantum systems and its relation with integrability. We will also avoid discussion of
one-dimensional systems, which display peculiar phenomena which are not directly relevant for high-temperature superconductors or three-dimensional Mott insulating oxides. In the following section we will first discuss the basic information we can gain from equilibrium calculations, then we will present analytical insight based on the Gutzwiller approximation, and we will finally briefly comment on the DMFT results. We remind that this section is not meant to review the huge body of work on non-equilibrium DMFT, for which we refer the refer the reader to Ref. \cite{Aoki2014}.

\subsection{Non-equilibrium techniques for strongly correlated materials: the Gutzwiller variational technique}
\label{Sec:Gutzwiller}
The Gutzwiller variational approach represents one of the simplest
yet effective tools to deal with strongly correlated electron
systems. The method is based on the Rayleigh-Ritz variational principle and it is devised to find the optimal wavefunction within a class of
wavefunctions where a linear operator which gives different weights to different local configurations (the ``Gutzwiller projector"), is applied to a Slater determinant.
 The Gutzwiller approximation is therefore a systematic improvement over the Hartree-Fock method, which in turns spans the space of Slater determinants, which is obviously a subspace of the Gutzwiller variational space.  
The similarity between the two methods suggests that the Gutzwiller approach can be generalized to non-equilibrium situations in analogy with the time-dependent Hartree-Fock scheme. For the same reasons mentioned above, the time-dependent Gutzwiller approximation will be a systematic improvement with respect to the time-dependent Hartree-Fock, which can be recovered in a the limit in which the Gutzwiller projector becomes the unit operator.

Starting from any lattice Hamiltonian with only local interaction terms
\begin{equation}
\mathcal{H} = \sum_{i,j}\sum_{a,b=1}^N\,\Big(
t^{ab}_{ij}\, c^\dagger_{ia}c^{\phantom{\dagger}}_{jb}+\text{H.c.}\Big) + 
\sum_i \, \mathcal{H}_i,
\label{sec-8.1:Ham}
\end{equation}
where $c^\dagger_{ia}$ creates an electron at site $i$ in orbital $a=1,\dots,N$, the index $a$ spans all the local degrees of freedom including the spin, and the orbital index,  and $\mathcal{H}_i$ contains all on-site terms, like the Hubbard $U$ interaction, the Hund's exchange coupling and the crystal-field potential.  
The Gutzwiller wavefunction\cite{Gutzwiller1964,Gutzwiller1965} is defined as
\begin{equation}
\mid\Psi\rangle = \mathcal{P}\mid\Psi_0\rangle = \prod_i\,\mathcal{P}_i\mid\Psi_0\rangle,
\label{sec-8.1:Psi}
\end{equation}
where $\mid\Psi_0\rangle$ is a Slater determinant 
and $\mathcal{P}_i$ a linear operator that acts on the Hilbert space of site-$i$. Following a popular terminology we shall denote $\mathcal{P}_i$ as the {\sl Gutzwiller projector}. The role of  $\mathcal{P}_i$ is to the change the weights of the local electronic configurations with respect to the Slater determinant, mimicking the effect of the interactions, which select some local configurations. For example, in the case of the single-band repulsive Hubbard model, the interaction unfavors double occupancy and the Gutzwiller projector has to ``project out"  configurations rich of doubly occupied sites.
  
At equilibrium, both $\mid \Psi_0\rangle$ and $\mathcal{P}_i$ can be determined variationally by minimizing the expectation value of the Hamiltonian over the variational wavefunction
\be
E = \fract{\langle \Psi\mid \mathcal{H}\mid \Psi\rangle}
{\langle \Psi\mid\Psi\rangle}.\label{sec-8.1:total-E}
\ee

If we want to describe the non-equilibrium dynamics of a correlated system within the same approximation, we can introduce a  time-dependent trial wavefunction~\cite{Schiro2010,Schiro2011} in which both the Slater determinant and the projector depend on time
\begin{equation}
\mid \Psi(t)\rangle = \mathcal{P}(t)\mid \Psi_0(t)\rangle =
\prod_i \mathcal{P}_i(t)\mid \Psi_0(t)\rangle.\label{tGA-Psi}
\end{equation}
and require that $\mid \Psi_0(t)\rangle$ and $\mathcal{P}_i(t)$ extremize the action $\mathcal{S}(t)$, i.e. 
\be 
\delta\mathcal{S}(t) = \delta \int_0^t\,d\tau\,\mathcal{L}(\tau)
= 0,\label{sec-8.1:action}
\ee
where ($\hbar=1$) 
\begin{equation}
\mathcal{L}(t) = \langle \Psi(t)\mid i\partial_t - \mathcal{H}\mid \Psi(t)\rangle.\label{sec-8.1:lagrangian}
\end{equation}

In general situations  the expectation values in Eq.~(\ref{sec-8.1:total-E}) and Eq.~(\ref{sec-8.1:lagrangian}) cannot be evaluated analytically, and the exact values can only be obtained numerically, for instance by means of variational Monte Carlo.~\cite{Sorella2005} 
An exception is the case of models defined on lattices with infinite coordination number. In this case, the expectation values can be computed analytically provided the following two constraints are satisfied:~\cite{Bunemann1998,Fabrizio2007,Fabrizio2013} 
\bea
\langle \Psi_0(t)\mid \mathcal{P}_{i}(t)^\dagger\, \mathcal{P}_{i}(t)^\dagga\mid \Psi_0(t)\rangle &=& 1,\label{sec-8.1:1}\\
\langle \Psi_0(t)\mid \mathcal{P}_{i}(t)^\dagger\, \mathcal{P}_{i}(t)^\dagga
\,c^\dagger_{ia}c^\dagga_{ib}\mid \Psi_0(t)\rangle &=& \langle \Psi_0(t)\mid 
c^\dagger_{ia} c^\dagga_{ib}\mid \Psi_0(t)\rangle, \qquad \forall a,b.\label{sec-8.1:2}
\eea

At equilibrium the wavefunction is time independent which means that enforcing the constraint simply requires to perform a constrained minimization of the expectation value $E$ over the variational parameters.

Away from equilibrium, one should in principle enforce the two constraints Eq.~\eqn{sec-8.1:1} and Eq.~\eqn{sec-8.1:2} at any time $t$. However, it was shown~\cite{Fabrizio2013} that, if Eq.~(\ref{sec-8.1:1}) 
and Eq.~(\ref{sec-8.1:2}), are satisfied at the initial time $t=0$, the equalities will hold during the whole time-evolution $t>0$ that solves  
Eq.~\eqn{sec-8.1:action}, which is a noteworthy simplification for practical implementations.

In the following we parameterize the Gutzwiller projector $\mathcal{P}_i(t)$ at site 
$i$ by means of a variational matrix $\hat{\Phi}_i(t)$, with elements 
$\Big(\hat{\Phi}_i(t)\Big)_{\alpha\beta}$ where $\mid i;\alpha\rangle$ and $\mid i;\beta\rangle$ span a basis set in the many-body Hilbert space of site $i$. For a single band Hubbard model $\vert i;\alpha\rangle$ can assume four values, namely$ \vert 0\rangle, \vert\uparrow\rangle, \vert\downarrow\rangle, \vert \uparrow\downarrow\rangle$. 

By definition~\cite{Fabrizio2013} 
\be
\Big(\hat{\Phi}_i(t)^\dagga\,\hat{\Phi}_i(t)^\dagger\Big)_{\alpha\beta}
= \langle \Psi(t)\mid i;\beta\rangle\langle i\; \alpha\mid\Psi(t)\rangle, 
\ee
is the matrix measuring the probability distribution of the local configurations in the 
variational wavefunction. In terms of $\hat{\Phi}_i(t)$, 
Eq.~(\ref{sec-8.1:1}) and Eq.~(\ref{sec-8.1:2}) read, respectively,
\bea
\text{Tr}\Big(\hat{\Phi}_i(t)^\dagger\,\hat{\Phi}_i(t)^\dagga\Big) &=& 1,
\label{sec-8.1:1-bis}\\
\sum_{\alpha\beta}\,
\Big(\hat{\Phi}_i(t)^\dagger\,\hat{\Phi}_i(t)^\dagga\Big)_{\alpha\beta}\,
\langle i;\beta\mid c^\dagger_{ia}c^\dagga_{ib}\mid i;\alpha\rangle 
&\equiv& \text{Tr}\Big(\hat{\Phi}_i(t)^\dagger\,\hat{\Phi}_i(t)^\dagga\,c^\dagger_{ia}c^\dagga_{ib}\Big) \nonumber\\
&=& \langle \Psi_0(t)\mid c^\dagger_{ia}c^\dagga_{ib}
\mid\Psi_0(t)\rangle,\label{sec-8.1:2-bis}
\eea
where the fermionic operators inside the trace must be interpreted hereafter as their matrix representation in the local basis set
$\big\{ \mid i;\alpha\rangle\big\}$.

If Eq.~\eqn{sec-8.1:1-bis} and Eq.~\eqn{sec-8.1:2-bis} are both satisfied, the variational principle \eqn{sec-8.1:action} can be calculated explicitly and it  leads to the equations of motion for the Slater determinant and the projector, respectively~\cite{Schiro2010,Schiro2011,Fabrizio2013} 
\bea
i\partial_t\mid\Psi_0(t)\rangle &=& \mathcal{H}_*(t)\mid\Psi_0(t)\rangle,\label{sec-8.1:dot-Psi0}\\
i\partial_t\,\hat{\Phi}_i(t)^\dagga &=& 
\fract{\delta E(t)}{\delta \hat{\Phi}_i^\dagger(t)},\label{sec-8.1:dot-Phi}
\eea
where
\be
E(t) = \sum_i\,
\text{Tr}\Big(\hat{\Phi}_i(t)^\dagger\,\hat{\Phi}_i(t)^\dagga\mathcal{H}_i\Big)
+ \langle\Psi_0(t)\mid \mathcal{H}_*(t)\mid\Psi_0(t)\rangle.
\label{sec-8.1:E(t)}
\ee
We have conveniently introduced a non-interacting effective time-dependent Hamiltonian $\mathcal{H}_*(t)$ which is written as
\be
\mathcal{H}_*(t) = \sum_{i,j}\,\sum_{a,b,d,f=1}^N\,
c^\dagger_{ia}\;R_{i,ad}(t)^\dagger\;t^{df}_{ij}\;
R_{j,fb}(t)^\dagga\;c^\dagga_{jb},\label{sec-8.1:H*}
\ee
which is nothing but the non-interacting kinetic terms renormalized by time-dependent renormalization factors $R_{i,ad}(t)$ and $R_{j,fb}(t)$, which generalize the static renormalization of the kinetic energy which characterizes the equilibrium Gutzwiller approximation. The renormalization parameters $R_{i,ab}(t)$ are defined through 
the solution of the linear set of equations 
\be
\text{Tr}\Big(
c^\dagger_{ia}\,\hat{\Phi}_i(t)^\dagger\,
c^\dagga_{ib}\hat{\Phi}_i(t)^\dagga\Big) 
= \sum_d\, R_{i,bd}(t)\, \text{Tr}\Big(
\hat{\Phi}_i(t)^\dagger\,\hat{\Phi}_i(t)^\dagga\,
c^\dagger_{ia}\,c^\dagga_{id}\Big),\qquad \forall \,a,b\,.\label{sec-8.1:R}
\ee

The set of equations (14-18) defines the time-dependent Gutzwiller approximation. In the following sections we will describe the main applications which have been developed so far. Here we briefly analyze the physical content of these equations in order to anticipate what are the main improvement with respect to Hartree-Fock and the limitations of this approach.

It is indeed natural and common to interpret the time-dependent Slater determinant $\mid\Psi_0(t)\rangle$ as describing the dynamics of quasiparticles, whereas the matrices $\hat{\Phi}_i(t)$ encode the dynamics of the Mott-Hubbard side-bands. 
Within the above variational scheme the two distinct degrees of freedom are mutually coupled, but only in a mean-field fashion; each of them evolves in an effective time-dependent potential provided by the other. 

In addition, the equation of motion for $\hat{\Phi}_i(t)$, Eq. (\ref{sec-8.1:dot-Phi}), is essentially semiclassical; there is no quantum entanglement between the matrices at the same and at different sites. 
Therefore, while the feedback between quasiparticles and Hubbard bands represents a fundamental improvement over time-dependent Hartree-Fock, where the systems is approximated by non-interacting quasiparticle states,  Eq.~\eqn{sec-8.1:dot-Psi0} and Eq.~\eqn{sec-8.1:dot-Phi} can not account for all dissipative processes which can take place in the real dynamics. 
This limitation is expected to be particularly severe in the description of the relaxation to a stationary state. 
However,  as we discuss in the following, comparison with exact time-dependent dynamical mean-field theory results, whenever they are available, shows that time-averages of observables turn out to be well captured by the Gutzwiller variational approach.~\cite{Schiro2010,Sandri2012,Sandri2013b}

The dynamical equations \eqn{sec-8.1:dot-Psi0} and \eqn{sec-8.1:dot-Phi}, with the expressions in Eq.~\eqn{sec-8.1:E(t)}, 
\eqn{sec-8.1:H*} and \eqn{sec-8.1:R} can be derived from the variational principle (\ref{sec-8.1:action}), and they have a rigorous variational interpretation, only when the lattice coordination is infinite. Nevertheless, 
the same formulas can also be used when the lattice coordination is finite, where they go under the name of Gutzwiller approximation (GA). 
In fact, several basic concepts in the field of strongly-correlated electron systems in equilibrium have been originally uncovered by the Gutzwiller approximation, or its equivalent slave-boson mean field theory,~\cite{Kotliar1986,Lechermann2007} like for instance the Brinkman-Rice scenario for the Mott transition in V$_2$O$_3$,~\cite{Brinkman1970} or the Resonating Valence Bond mechanism for high-temperature superconductivity.~\cite{Anderson1987} 

Before briefly reviewing some results that have been obtained by the time-dependent variational scheme based on the Gutzwiller wavefunction and approximation, we end mentioning that Eq.~\eqn{sec-8.1:dot-Psi0} and Eq.~\eqn{sec-8.1:dot-Phi} expanded at linear order in the deviations from equilibrium coincide with the linear response theory within the Gutzwiller approximation originally developed in Refs.~\cite{Seibold2001,Bunemann2013}
 
\subsection{Non-equilibrium techniques for strongly correlated materials: Dynamical Mean Field Theory}
  \label{DMFT_section}

Dynamical Mean-Field Theory (DMFT)\cite{Georges1996} has established in the last decades as one of
the most reliable and powerful methods to treat strongly correlated
electron systems.  DMFT is a non-perturbative, yet computationally affordable method
that can be used either to solve simple models (Hubbard and related
models) or it can be combined with
density-functional theory to describe accurately actual
solids\cite{Kotliar2006}. It can treat different interactions (electron-electron,
electron-phonon) and aspects of the electronic structure of a material
without assuming any hierarchy of energy scales.

As the name implies, DMFT can be viewed as a quantum (dynamical) generalization of
classical mean-field theory. The method is based on the neglect of spatial fluctuations, which are treated as in a static mean-field theory. 
On the other hand, all the remaining quantum and thermal fluctuations are
instead treated exactly without any further assumption. This approximation becomes exact in the limit of infinite coordination or infinite dimensionality\cite{Metzner1989},  where the interaction of each site with an infinite number of partners can be described exactly by an exact bath. 
Indeed the DMFT approximation  turns out to be remarkably good for many three-dimensional strongly correlated materials, where it can be used both 
to clarify longstanding problems like the Mott-Hubbard transition\cite{Zhang1993,Georges1996} or in combination with density-functional theory to address 
realistic aspects of correlated solids\cite{Kotliar2006} including successful descriptions of photoemission
and optical spectra as well as many other observables. This success
emphasizes the crucial role of local quantum fluctuations in strongly
correlated materials, but it can not be extended straightforwardly to
two-dimensional or highly anisotropic materials such as the
high-temperature superconducting cuprates. In the latter case, it is
necessary to consider cluster extensions of DMFT\cite{Kotliar2001,Maier2005}, in which the
short-ranged dynamics inside a chosen cluster is treated explicitly. 

The success of DMFT to treat the equilibrium properties of strongly
correlated materials makes the method an ideal candidate for the
investigation of the time-dependent properties of systems driven out
of equilibrium. Indeed  DMFT is naturally extended to the non-equilibrium by
combining it with the Keldysh formalism. The first general formulation of
nonequilibrium DMFT is due to Ref. \cite{Freericks2006} where the
Baym-Kadanoff formalism is used in the case of a system driven by an
electric field. Triggered by the experimental advances in
time-resolved spectroscopies and in the detection of real-time
non-equilibrium phenomena, the nonequilibrium DMFT has been boosted
thanks to the activity of several groups, as reported in the recent
review by Aoki {\it et al.}\cite{Aoki2014}. In this manuscript we
focus mainly on the aspects which are particularly relevant for the
description on non-equilibrium phenomena in high-temperature
superconductors and in Mott insulators.

\subsubsection{Dynamical Mean-Field Theory: Basic Ideas and Equations}
In this section we briefly review the basic concepts of equilibrium and
non-equilibrium DMFT. We refer to previous reviews for all the
technical details, and we limit the present discussion to the basic
information required to appreciate the potential of the method and its
applications to investigate strongly correlated systems out of
equilibrium and their dynamical response.

As we anticipated above, the Dynamical Mean-Field Theory is an
approximated solution of a quantum model in which every lattice site
is assumed to be equivalent, or, in other words, spatial fluctuations
are frozen, just like in a static mean-field theory.
On the other hand, in contrast with classical mean-field methods, the local quantum fluctuations are fully taken into account and their
effect defines the  ``quantum'' or ``dynamical'' character that
appears in the name of the method. The inclusion of quantum
fluctuations is a highly non-trivial improvement over static
mean-field theories. As a matter of fact, the quantum nature of the
method significantly complicates the solution of the effective
approximate theory, but, on the positive side, it remarkably extends
the validity of the method, allowing for a complete characterization
of nonperturbative phenomena such as the Mott-Hubbard
transition\cite{Georges1996}, or the crossover between
Bardeen-Cooper-Schrieffer superconductivity and Bose-Einstein condensation\cite{Toschi2005a,Toschi2005b}.

The route to construct the DMFT is to build an effective theory for one
arbitrary representative site, usually labeled as ``0''. From a formal point of view, this is
realized by means, e.g., of a cavity construction, formally
integrating out all the degrees of freedom of a lattice model
retaining only those defined on the chosen site.
This formal procedure defines an effective local theory which is
however parameterized by all the high-order Green's functions of the
rest of the lattice. In the limit of large lattice coordination, using
a scaling of the hopping that insures a finite expectation energy for
the kinetic energy, only the single-particle Green's function survives as all the higher-order functions vanish.
As a consequence, the local effective theory only depends on a single
function of two time variables. We can write down the form of the
effective theory in terms of a local action, in the specific case of
the single-band Hubbard model, as
\begin{equation}
\label{equation_Seff}
S_{eff} = \int_{-\infty}^{\infty} dt\int_{-\infty}^{\infty} dt' c^{\dagger}_{0\sigma}(t){\cal{G}}_0^{-1} (t,t') c_{0\sigma}(t') + \int_{-\infty}^{\infty} dt U(t) n_{0\uparrow}(t) n_{0\downarrow} (t).
\end{equation}
The second term is the on-site Hubbard repulsion, and it would be
replaced by any other local interaction for different models. The
first term describes precisely the local quantum fluctuations on any
site, and it is measured by the single function ${\cal{G}}_0^{-1}
(t,t')$. This function contains, within DMFT, the effect of the rest
of the lattice on any arbitrary site, just like the Weiss effective
field describes the effect of the other spins onto any ``chosen'' spin
in a classical mean-field theory. For this reason, it is usually
referred to as a ``dynamical Weiss field''. The time dependence of the
Weiss field is the key element of the dynamical mean-field approach. 

In order to have a practical mean-field theory, we need to determine
the Weiss field ${\cal{G}}_0^{-1}$. This is done by imposing a
self-consistency condition, which is derived in the infinite
coordination limit, where the method becomes exact. For the sake of simplicity, we first present the
self-consistency equation in equilibrium, postponing its
non-equilibrium counterpart to the end of the present section.

The main simplification in equilibrium is that ${\cal{G}}_0^{-1}$ is not
an independent function of $t$ and $t'$, but it depends only on the time difference
$t-t'$. As a consequence, it can be Fourier transformed in the
frequency space.
The crucial observables that encodes for the local quantum dynamics of the effective theory is the Green's function 
$G(t,t') = -i\langle T
c_{0\sigma}(t)c^{\dagger}_{0\sigma}(t')\rangle_{S_{eff}}$, where the
expectation value is taken over the effective action
Eq. (\ref{equation_Seff}) and $T$ indicates the time-ordered product of the subsequent operators. In equilibrium also $G(t,t')$ becomes time-translation invariant,
and a function of the time difference alone. We can
define the self-energy of the effective theory 
\begin{equation}
\Sigma(\omega) = {\cal{G}}_0^{-1} (\omega) - G^{-1}(\omega),
\label{eq:sigma}
\end{equation}
and write the self-consistency as 

\begin{equation}
\label{equation_selfconsistency}
G(\omega) = \int_{-\infty}^{\infty} d\varepsilon
\frac{D(\varepsilon)}{\omega+\mu -\varepsilon -\Sigma(\omega)},
\end{equation}
where $D(\varepsilon)$ is the bare density of states for the chosen
lattice. Eq. (\ref{equation_selfconsistency}) implies that the
Green's function of the effective theory $G(\omega)$ coincides with the local component of the lattice Green's function computed using  $\Sigma(\omega)$ as a local self-energy for the lattice model. The right-hand side of Eq. \label{equation_selfconsistency} is indeed equivalent to the sum over all the momenta of the Brillouin-zone of the lattice Green's function $G(k,\omega)$ is the self-energy is local and coincides with $\Sigma(\omega)$ given by Eq. (\ref{eq:sigma})
As a matter of fact, one can indeed  view DMFT as a theory in which the
exact Green's function is obtained provided that the
momentum-dependent lattice self-energy $\Sigma(k,\omega)$ is replaced by a local quantity
which is ideally obtained averaging $\Sigma(k,\omega)$ over all the
momenta of the Brillouin zone.

\begin{figure}
\begin{center}
\resizebox*{15cm}{!}{\includegraphics{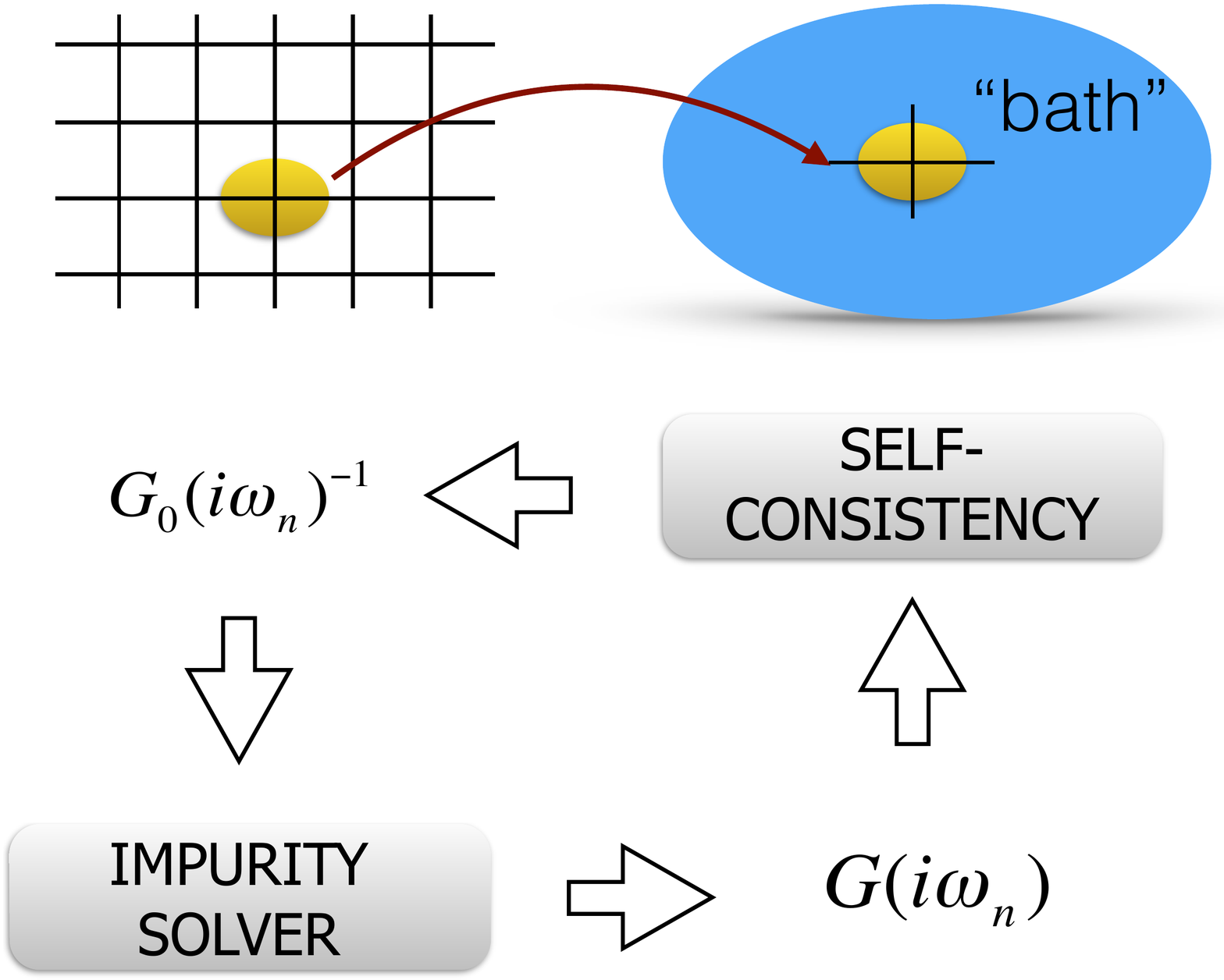}}%
\caption{Schematic representation of the DMFT idea (upper half) in which a lattice model is mapped onto an impurity model and of the DMFT self-consistency loop (lower half)}
\end{center}
\label{Keldysh-contour}
\end{figure}

A practical implementation of DMFT requires to solve the local effective theory and to impose the self-consistency condition. The local effective theory can be mapped onto a quantum impurity model, in which a single interacting site is hybridized with a non-interacting bath, which is the Hamiltonian representation of the Weiss field. Impurity models, albeit much simpler than the original lattice models, are non trivial and do not allow for analytical solutions. There are however different numerically exact approaches that allow to solve the model and compute the Green's function, among which we mention Continuous-Time-Quantum Monte Carlo, Numerical Renormalization Group and Exact Diagonalization. In practice a DMFT calculation is an iterative procedure in which, starting from a given Weiss field, the impurity model is solved, and the new Green's function and self-energy are used to compute a new Weiss field through the self-consistency condition. The procedure is then iterated until convergence is achieved. The numerical solution of the impurity model is computationally extremely cheap for a single-band Hubbard model, but it can become expensive for multi-orbital models. In the next section we briefly discuss the ``impurity solvers'' that are presently available out of equilibrium.

One of the most remarkable features of DMFT is that it gives a direct
access to dynamic observables which include the spectral function
$A(k,\omega) = -1/\pi ImG(k,\omega)$, which retains a momentum dependence despite the local self-energy, and response functions including
the optical conductivity $\sigma(\omega)$, the dynamical spin
susceptibility $\chi(\omega)$.

\subsubsection{Dynamical Mean-Field Theory and High-Temperature Superconductors}

The above discussion of the DMFT should cast some doubts on its
application to high-temperature superconductors. Indeed the
phenomenology we described in the introductory sections does not seem
to be accessible by means of a theoretical approach in which spatial
fluctuations are frozen. The most fundamental limitation is that a
momentum-independent self-energy is unable to describe a d-wave
superconducting order parameter and a momentum-dependent
pseudogap. Charge-density wave are also unaccessible unless for
specific commensurate cases, in which DMFT can be straightforwardly
generalized. The simplest strategy to overcome the limitations of DMFT
is to include the quantum short-range correlations within the range of
a finite cluster which requires to consider one of the methods that go
under the name of cluster-extensions of DMFT. We do not need to enter
the details of these methodologies here, and we limit to mention two
main approaches, the Dynamical Cluster Approximation (DCA) and the
Cellular DMFT (CDMFT). These methods employ a very similar philosophy,
approximating a lattice model with a finite cluster embedded in an
effective medium which is self-consistently calculated analogously to DMFT.
On the other hand the two methods differ in the way they treat the lattice translation
symmetry. Within DCA an artificial translational symmetry within the
cluster is implemented in order to make the model diagonal in the
reciprocal vectors of the cluster, while in CDMFT the translation
symmetry is broken.

\subsubsection{Non-equilibrium Dynamical Mean-Field Theory: the equations and the algorithm}

As mentioned above, when the system is driven out of equilibrium, the
Weiss field ${\cal{G}}_0^{-1} (t,t')$ is no longer a function of $t-t'$,
but it depends on both time variables. Apart from this important
difference, DMFT can be formulated in a way that closely mirrors the equilibrium version.
The key ingredient is the use of the non-equilibrium Green's function theory based on the seminal papers by Schwinger\cite{Schwinger1961}, Kadanoff and Baym\cite{Kadanoff1962} and Keldysh\cite{Keldysh1964}. As a matter of fact the Keldysh formalism allows to generalize the equilibrium Green's function method in a straightforward way and it does not require any assumption or knowledge on the properties of the system other than the initial state and the time-evolution of the Hamiltonian. We refer the reader to previous publications\cite{Kamenev2011,Aoki2014} for detailed reviews of the derivation of the Kadanoff-Baym and Keldysh formalisms. The basic idea is the time evolution of the density matrix in the presence of a time-dependent Hamiltonian ${\cal H}(t)$ can be formally written as 
\begin{equation}
\rho(t) = \left[U(t)\right]^{\dagger} \rho(t_{min}) U(t) \equiv U(t,t_{min})\rho(t_{min})U(t_{min},t),
\end{equation} 
where $U(t) \equiv U(t_{min},t) = Texp(-i\int_{t_{min}}^t dt' {\cal H}(t')$ is the standard time-evolution operator and $T$ is the time-ordering operator, while we define a reversed time evolution operator $U(t,t_{min}) = T^{-}exp(-i\int_{t_{min}}^t dt' {\cal H}(t')$, where $T^{-}$ is an anti-time-ordering operator. Then the calculation of the expectation value of any operator $A$ becomes
\begin{eqnarray}
  \langle A(t)\rangle &=& \frac{1}{Z}Tr[e^{-\beta H} U(t_{min},t)A U(t,t_{tmin})] =\nonumber\\
  &=&\frac{1}{Z} Tr[U(-i\beta+t_{min},t_{min})U(t_{min},t)AU(t,t_{min})]
\end{eqnarray} 
where we have used the cyclic property of the trace and we recast the ``Boltzmann" exponential as an imaginary-time evolution. The trace can be computed following first evolving from time t$_{min}$ to time t and measuring the operator A, then evolving from t back to t$_{min}$ and then from t$_{min}$ to t$_{min}-i\beta$. This leads to define the contour represented in Fig. \ref{Keldysh-contour} and a time-ordering operator along the contour $_C$, represented by the arrows. With this notation, the above expectation value can be written as
\begin{equation}
\langle A(t)\rangle = \frac{Tr[T_C e^{-i\int_C A(t) dt'H(t')]}}{Tr[T_C e^{-i\int_C dt'H(t')]}}.
\end{equation}
\begin{figure}
\begin{center}
\resizebox*{12cm}{!}{\includegraphics{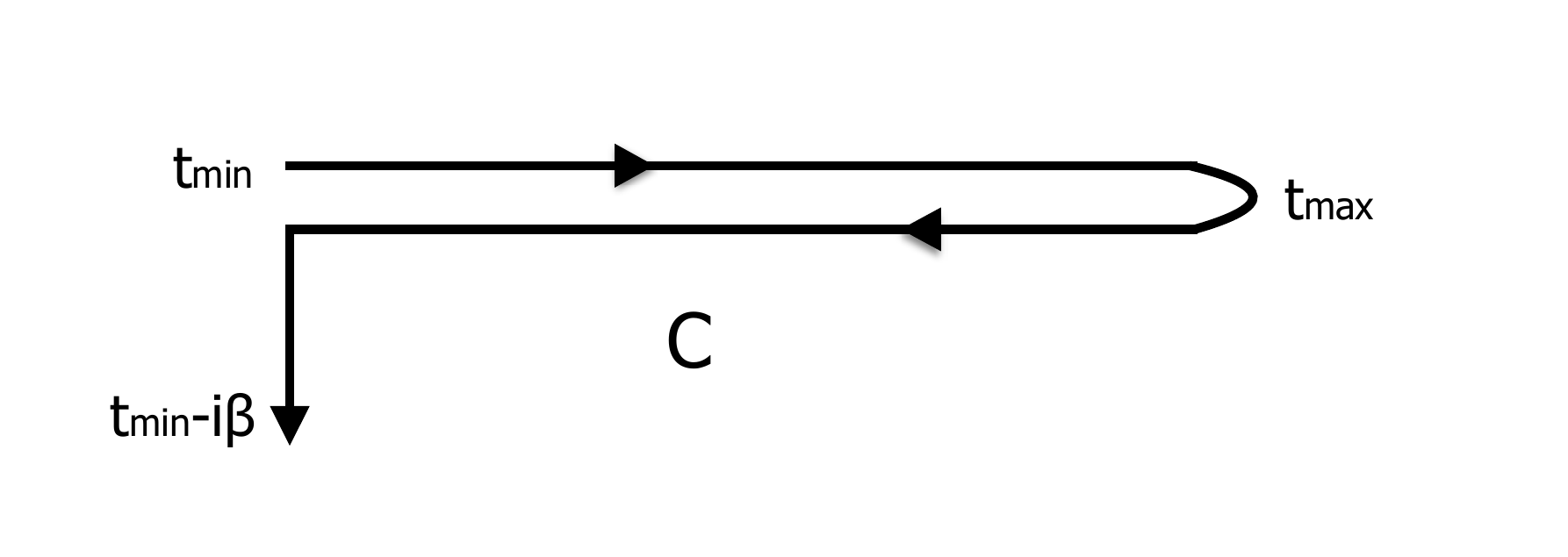}}%
\caption{The ordered contour {\cal C} used to compute non-equilibrium Green's function}
\end{center}
\label{Keldysh-contour}
\end{figure}
The use of the Keldysh contour defined above is particularly useful when dealing with the evaluation of expectation values of products of operators at different times, like the Green's functions which can be simply defined as 
\begin{equation}
G(t,t') \equiv \langle T_C c(t) c^{\dagger}(t')\rangle,
\label{G_contour}
\end{equation}
where we did not specify any label (position, spin, orbital momentum) for the fermionic operators to maintain the generality. Since the contour ${\cal C}$ has three branches, we can unfold Eq. (\ref{G_contour}) into a 3$\times$3 matrix representation where the three branches of the contour correspond to the three rows and columns (1 is the upper real-time branch, 2 is the second real-time branch and 3 is the imaginary-time branch)\cite{Wagner1991}. We refer the reader to the literature on the non-equilibrium Green's functions or its application to DMFT\cite{Aoki2014} for more details. Here we just recall that the components $G_{12} \equiv G^{<}$ and $G_{21} \equiv G^{>}$ correspond to the so-called lesser and greater Green's function and that the expressions of the retarded Green's function and the so-called Keldysh component are given by
\begin{equation}
G^R(t,t') = -i\theta(t-t')\langle{c(t),c^{\dagger}(t')}\rangle = \frac{1}{2}(G_{11}-G_{12}+G_{21}-G_{22})
\end{equation}
\begin{equation}
G^K(t,t') = -i\theta(t-t')\langle[c(t),c^{\dagger}(t')]\rangle = \frac{1}{2}(G_{11}+G_{12}+G_{21}+G_{22}).
\end{equation}

Using this contour, one can essentially construct the same theory of the equilibrium Green's functions for time-dependent problems and perform, at least in principle the perturbation expansion, which leads to the definition of the self-energy $\Sigma(t,t')$ which will now have the same components that we discussed for the Green's function, and to a Dyson equation.

Once the formalism for non-equilibrium Green's function is given, one can prove that in the limit of infinite dimensionality the self-energy becomes local also for time-dependent non-equilibrium problems using the same scaling of the hopping which is required to obtain a sensible infinite-coordination limit in equilibrium. This implies that
\begin{equation}
\hat{\Sigma}_{ij}(t,t') = \delta_{ij}\hat{\Sigma}_i(t,t'),
\end{equation}
where the hat reminds the matrix nature of the self-energy and now i and j are spatial indices referring to lattice sites. Then the Dyson equation which allows to compute the Green's function is
\begin{equation}
(G^{-1})_{ij}(t,t') = [\delta_{ij}(i\partial_t + \mu) - J_{ij}(t)]\delta_C(t,t') - \delta_{ij}\Sigma_i(t,t'),
\label{dyson_timedependent}
\end{equation}
where $J_{ij}$ are the hopping amplitudes, which we relabel to avoid confusion with the time variable t, and $\delta_C(t,t')$ is the delta function on the contour. The equation is easily understood observing that the first term of the right-hand side is nothing but the inverse of the non-interacting lattice Green's function. Exactly like in equilibrium, in infinite coordination the local component of the lattice self-energy coincides with the self-energy of an auxiliary single-site effective theory with the local action what we write for the site 0
\begin{equation}
S_0 = -i\int_{C} dt H_0(t) - i\sum_{\sigma} \int_C dtdt' c_{0\sigma}^{\dagger}(t){\cal G}_0^{-1}(t,t')c_{0\sigma}(t')
\label{action_timedependent}
\end{equation}
Exactly as in equilibrium the Weiss field has to be chosen in such a way that the Green's function of model (\ref{action_timedependent}) $G = -i\langle T_c c_{0\sigma}(t)c^{\dagger}_{0\sigma}(t')\rangle_{S_0}$ coincides with the local component of the lattice Green's function computed from Eq. (\ref{dyson_timedependent}) with 
\begin{equation}
G^{-1}(t,t') = (i\partial_t + \mu)\delta_C(t,t') - \Sigma_i(t,t') - {\cal G}_0^{-1}(t,t').
\end{equation}

\subsection{Correlated systems in electric fields}

In this section we review the theoretical investigations of strongly correlated
systems driven by an electric field. In particular we consider the cases of static or
periodically oscillating uniform fields. 

\subsubsection{Correlated Electrons in a d.c. electric field}
An ideal metal, in which no scattering process disturbs the electronic
motion, is known to react to a static electric field with a current
which oscillates periodically in time displaying the so-called Bloch
oscillations\cite{Bloch1929}. 
The Bloch oscillations are the consequence of a time-dependent shift
of the occupied momenta in the Brillouin zone and their existence
relies on the delocalized character of the electronic states. The
period of the oscillations is given by $T_B = 2\pi\hbar/ae\vert
E\vert$, where a is the lattice spacing, e is the electron charge and
$E$ is the electric field. 

In solid state systems this period is indeed too large, even for the
largest accessible electric fields, and the Bloch oscillations are
hard to observe also in the best metals with very long scattering
times. The detection of the oscillation has become possible only in  
artificially designed systems like semiconducting heterostructures and
it may be realized in cold-atom systems.

From a theoretical point of view, a natural question is the destiny of
the Bloch oscillations in the presence of electron-electron
interactions which lead to scattering processes and limit the coherent
motion of the electrons. Pioneering works on time-dependent DMFT have
shown the damping of the Bloch oscillations as the interaction is
increased in the Falicov-Kimball
model\cite{Freericks2006,Turkowski2007,Freericks2008}, a simplified
Hubbard model in which one of the two spin species is localized. A
similar behavior is found in the Hubbard model\cite{Eckstein2011b,Amaricci2012}.
The many-body spectrum of the system is strongly reminiscent of the Wannier-Stark ladder resonances, with a positive feedback effect between the Wannier-Stark localization and the effects of the interactions. As a matter of fact the correlation properties are enhanced by the electric field\cite{Tsuji2008,Joura2008,Freericks2008,Eckstein2011b}. When the electric field is directed along one of the crystallographic directions of the lattice and it is so large to make the potential drop between two lattice sites the largest energy scale of the problem, a dimensional crossover to a system of dimensionality $d$-1 has been reported\cite{Aron2012a}.

\subsubsection{Dissipation and the approach to steady states}

The results we briefly discussed in the previous section apply to
isolated correlated systems driven out of equilibrium by a static
electric field. In the context of actual materials they can only
provide a starting point to understand the role of electronic
correlation, while their relevance is limited by the neglect of dissipation due to the coupling with phonons and to the other degrees
of freedom that are excluded from the simplified models. 

It should be clear that some kind of dissipation must be included in
order to absorb the energy which
is lumped into the system by the electric field and it is a necessary
condition to reach a stationary states at long times. 

From a general point of view, the role of the dissipative bath is
essentially to allow for energy and momentum relaxation, depleting the
high-energy and high-momentum state which are populated by the driving
field. Only the competition between the two effects can lead to a
stationary population of the different quantum states. 

A simple strategy to study the general effect of a
dissipative bath on the dynamics of a driven system is to include an
artificial thermostat, although one can also consider models in which   some models one can avoid to
introduce an explicit bath, because the dissipative degrees of freedom
are included in the model\cite{Mierzejewski2011a,Golez2014}. 

In a practical implementation the dissipation can be realized either via a
bosonic bath\cite{Eckstein2013c,Eckstein2013a,Kemper2013} (which may describe the coupling with phonons or other bosons) or via a
fermionic bath. The bosonic bath is realized via an infinite set of harmonic oscillators which can be assumed coupled with the electrons
with a simple Holstein coupling. To avoid a feedback of the system on the bath the coupling is treated in linear order with a free-boson
distribution function. 

On the other hand, at least within DMFT it is particularly convenient to implement a fermionic
local bath, which appears in the DFMT equations as a further
thermalized bath which is added to the self-consistent bath discussed
in Sec 8.2. This approach has been successfully implemented in\cite{Amaricci2012}  where the full
relaxation dynamics leading to the non-equilibrium stationary state
was studied and characterized, and in Ref. \cite{Aron2012a}, where the
formalism was limited to study the asymptotic stationary state assumed
to establish at long times. In Ref. \cite{Werner2021z} the same approach has been applied to the Kondo-lattice model.

As we anticipated, the effects of the
thermostat on the non-equilibrium dynamics can be taken into account
by means of an additional self-energy $\Sigma_{bath}$, which can be
written as:
\begin{equation}
\Sigma_{bath}(t-t')= V^2g(t-t')
\end{equation}
where $g(t-t')$ is the Green's function of the bath, which is assumed
to be unaffected by the interaction with the actual system and time-translation invariant. This can be taken as an operative
definition of a bath, an object that can exchange energy and particles
with the actual system without changing its temperature and
thermodynamic state. In this way the bath can absorb the extra energy
pumped into the system by the electric field. Naturally this choice of
bath is not limited to the case of a constant electric field, but it
can be used for any time dependence of the external perturbation,
including impulsive fields such as those used in pump-probe
experiments.  

The thermostated system obeys the following Dyson equation on the Keldysh contour\cite{Joura2008}
\begin{equation}
G_{k}(t,t')={\cal{G}}_{0k}(t,t') + [{\cal{G}}_{0k}\cdot \Sigma_k \cdot G_k](t,t')
\label{eq1}
\end{equation}
where all quantities represent continuous functions of two time variables $(t,t')\in\cal{C}$,  the symbol $\cdot$ denotes the convolution product 
$$
\left[f\cdot g\right](t,t')=\int_{\cal{C}} dz f(t,z)g(z,t')
$$ 
and $\Sigma_k(t,t')$ denotes the Keldysh self-energy function.  Eq.\ref{eq1} is expressed in terms of the ``renormalized" non-interacting lattice Green's function ${\cal{G}}_{0k}$:
\begin{equation}
{\cal{G}}_{0k}^{-1}(t,t')=\left[G^{-1}_{0k}(t,t') - \Sigma_{bath}(t-t')\right]
\label{eq2a}
\end{equation}
which is obtained from the ``bare" non-interacting lattice Green's function:
\begin{equation}
G_{0k}^{-1}(t,t') = [i\overrightarrow{\partial_t} -\epsilon(k)]\cdot\delta_{\cal{C}}(t,t')
\label{eq2b}
\end{equation}
by integrating out locally the electronic degrees of freedom of the external thermostat.
The symbol $\delta_{\cal{C}}$ indicates the delta function on the contour $\cal{C}$.

We can define
$\Lambda= V^2/W$ as an effective coupling of the physical  electrons with the
thermal bath (This would be an exact result for a flat bath density of
states of amplitude $W$, but it gives the order of magnitude for any
kind of DOS).
 
The approach to the stationary non-equilibrium state can be
characterized following the real-time dynamics of the local current
$\mathbf{J}(t)=-ie/\pi \sum_k {\mathbf v}_k G^<_k(t,t)$, where
${\mathbf v}_k=\nabla_k \epsilon(k)$ is the electronic velocity.

\begin{figure}
\begin{center}
\begin{minipage}{120mm}
\subfigure[]{
\resizebox*{6cm}{!}{\includegraphics{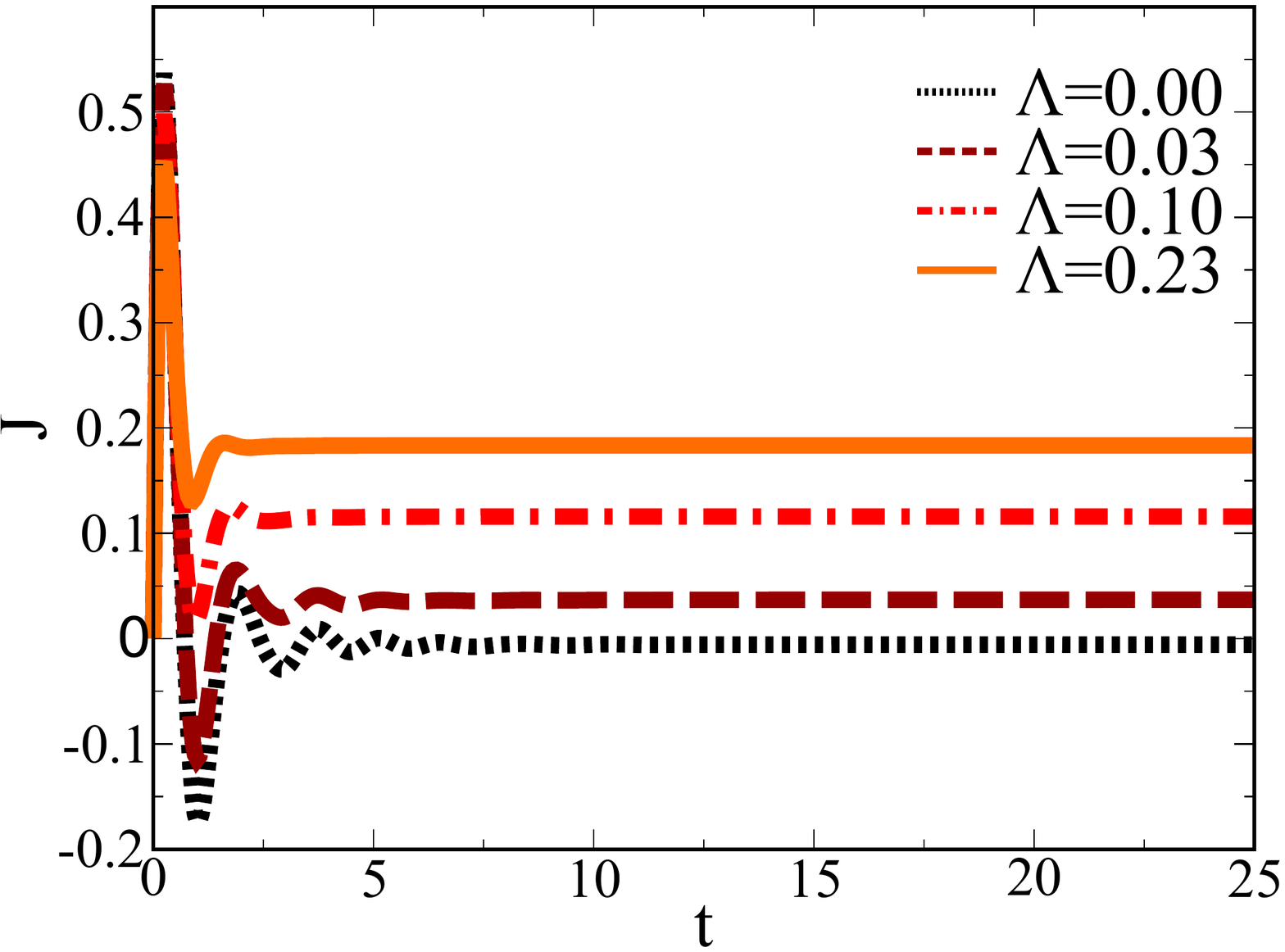}}}%
\subfigure[]{
\resizebox*{6cm}{!}{\includegraphics{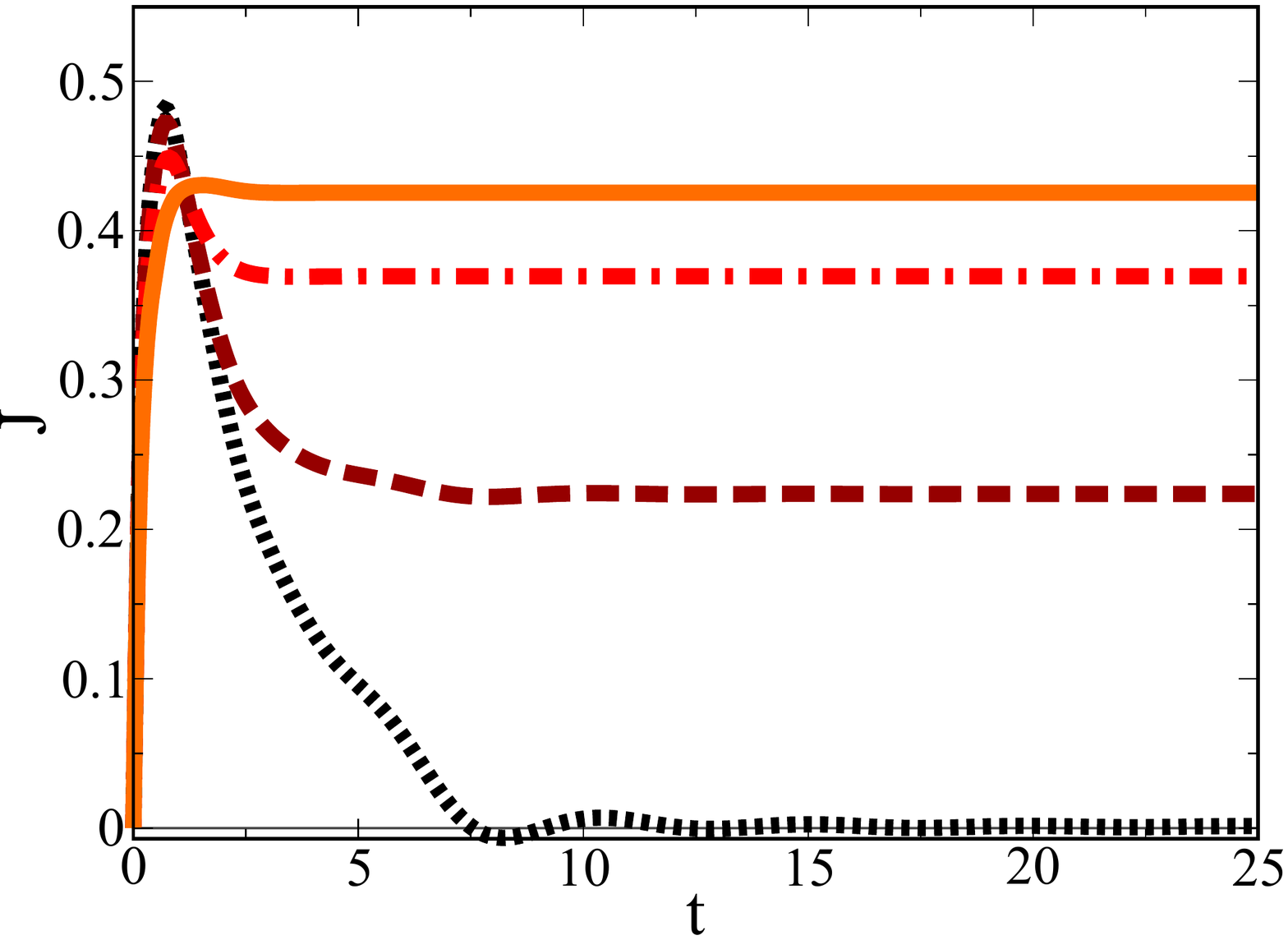}}}%
\caption{Time evolution of the current for two values of the electric
  field (E=4.7 in the top panel and   E=1.26 in the bottom panel) for different values
of the coupling to the dissipative bath. In the absence of a bath the
current goes to zero at long-times, signaling the inability to
establish a real stationary state }
\label{fig8.1}
\end{minipage}
\end{center}
\end{figure}

\begin{figure}
\begin{center}
\resizebox*{12cm}{!}{\includegraphics{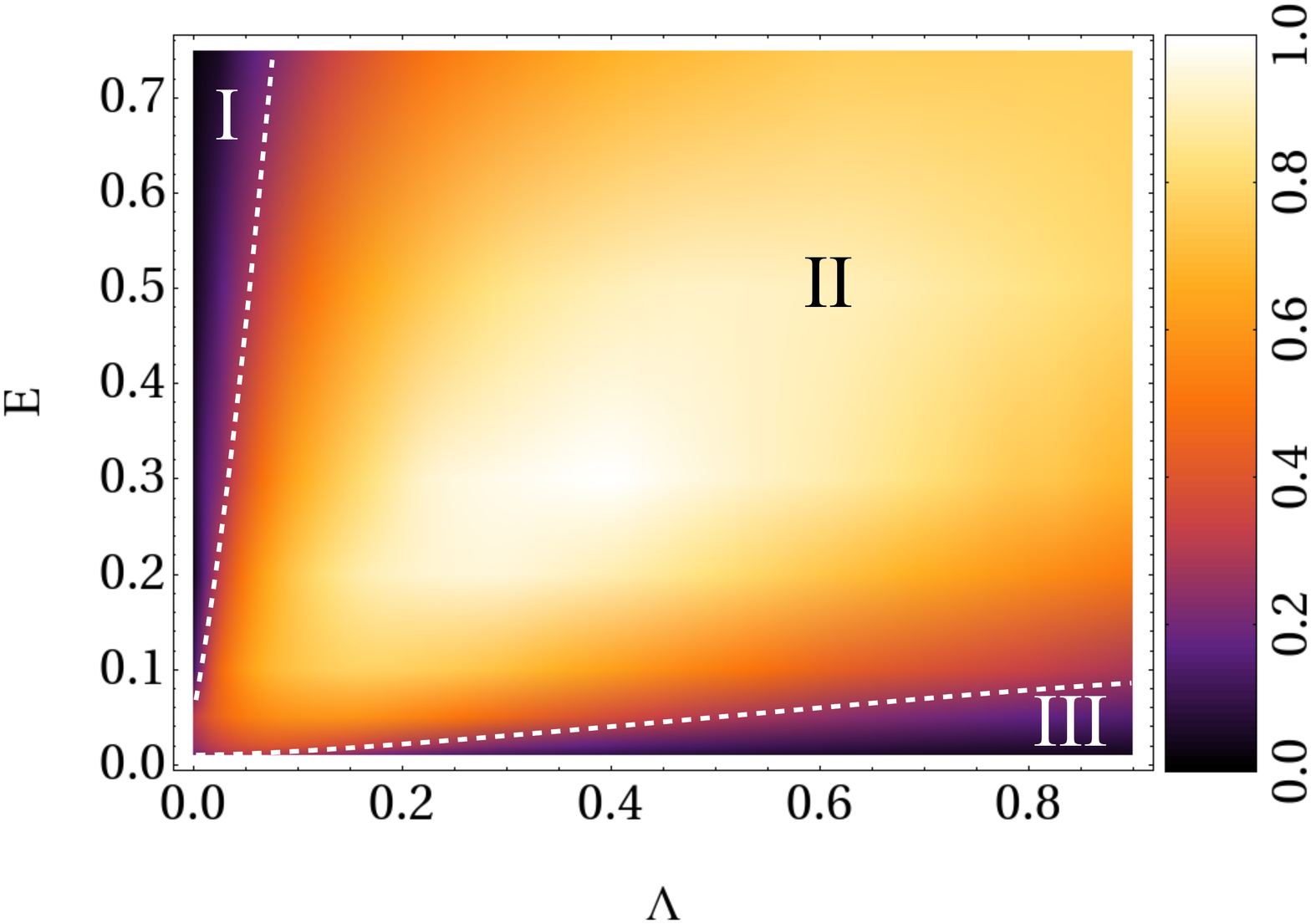}}%
\caption{Phase diagram of the driven Hubbard model as a function of the static electric field $E$ and the dissipation constant $\Lambda$. The colorscale represents the asymptotic value of the current for long times. The current vanishes both for very large values of $E/\Lambda$, for which the system is unable dissipate the extra energy pumped into the system, and it evolves towards a state with infinite effective temperature, and for small values of $E/\Lambda$, in which the coupling with the dissipative bath is so large and it overshadows the effect of the driving field. However, a large region of finite current is found for intermediate values of $E/\Lambda$, and the value of the current does not depend dramatically on the parameters.}
\label{fig8.2}
\end{center}
\end{figure}

 The results for the local current are presented in \ref{fig8.1}. The
 relaxation dynamics is depends naturally on the value of the applied
 electric field and on the coupling with the dissipative
 bath. However, as long as the coupling is finite and not incredibly
 large, a finite value of the asymptotic current at long times is
 obtained and the value does not depend strongly on the value of the
 coupling $\Lambda$. This information is summarized in the phase
 diagram of Fig. \ref{fig8.2}, which shows a wide central region
 (labelled II) in which the competition between the driving field and
 the dissipation channel lead to a non-trivial state with a finite
 current, whose value is largely independent on the detals, while only
 for exceedingly large $E$ (region I) or $\Lambda$ (region III)
 trivial states with no stationary current are obtained. This
 demonstrates the effectiveness of the fermionic bath to drive
 non-trivial stationary states.
 
 Han has later shown that the inclusion of a fermionic bath is sufficient to reproduce a semiclassical Ohm-expression in the conductivity despite the lack of explicit momentum scattering\cite{Han2013a}, confirming that this bath can lead to a proper dissipation despite its intrinsic simplifications. Only in the limit of small dissipation the steady state displays a divergent effective temperature as long as the Bloch oscillation frequency remains finite\cite{Han2013b}.

\subsubsection{Periodic a.c. electric fields}
In the presence of a periodic in time external field, a quantum system can be driven into a nonequilibrium steady state in which the system follows the periodic time dependence of the external stimulus. In principle, such an a.c. field could be used to dynamically control the properties of the system. On the other hand in present solid-state experiments, pulsed lasers are necessary to reach sufficiently large intensities. However, if a sufficiently large number of cycles takes place in the duration of the pulse, we can reasonably describe the external stimulus as an oscillating field. 

When we can approximate the external field with a periodic pulse, we can use Floquet method\cite{Shirley1965,Zeldovich1967}, which is based on Floquet theorem\cite{Floquet1883}, which can be seen as an analog of Bloch's theorem for periodic time-dependent problems. As a consequence of the periodic dependence of time, the time-dependent problem can be mapped onto a time-independent eigenvalue problems introducing quasienergies which are defined up to integer multiples of $2\pi/T$, where $T$ is the oscillation period. The approach is expected to reproduces accurately the spectral properties, while the distribution depends on the initial conditions and requires ad hoc schemes to adequately describe the effect of generic fields. 

Recently, the Floquet method has been employed in combination with DMFT\cite{Schmidt2002,Joura2008,Freericks2008p,Tsuji2008,Lubatsch2009,Tsuji2009}.
In the Floquet DMFT formalism\cite{Schmidt2002,Tsuji2008} a dissipative correlated system is continuously driven by a time-periodic perturbation, and it is postulated that a non-equilibrium stationary state is reached in the long-time limit, when the dissipative processes is expected to eliminate any dependence from the initial conditions. We notice in passing that this assumption has been numerically proved in Ref. \cite{Amaricci2012} in the case of static electric fields and a fermionic dissipative bath. The non-equilibrium periodic state can be directly calculated within Floquet DMFT mapping the problem onto a non-equilibrium state of an impurity model. Therefore this specific version of non-equilibrium DMFT does not require to actually compute the time evolution of the system, leading to a substantially reduced computational weight. 

The method was first applied to the Falicov-Kimball model, where the metallization of a Mott insulator has been shown through the formation of photoinduced midgap states emerge from strong a.c. fields\cite{Tsuji2008}.  The field-induced metallic state has been further characterized, demonstrating how a standard Drude-like peak coexists with intrinsic non-equilibrium features, including dip and kink structures in the optical conductivity exhibits around the frequency of the external field, a midgap absorption arising from photoinduced Floquet subbands, and a negative attenuation due to a population inversion\cite{Tsuji2009}.

In Ref. \cite{Tsuji2011,Tsuji2012} a sudden application of a properly tuned periodic field flips the electronic structure, effectively turning the repulsion into an attraction in the absence of energy dissipation. The
driven system is characterized by a a no-adiabatic shift of the electronic population in momentum space. When the momentum shift reaches $\pi$, the shifted population relaxes to a negative-temperature state, which leads to the interaction switching.

\subsubsection{Electric fields in correlated heterostructures}
The theoretical description of a system subject to an external electric
field naturally leads to address the properties of an actual device
contacted to two metallic leads which enforce the potential bias. From
a technical point of view, this requires to consider spatially
inhomogeneous systems and in layered
heterostructures. The difficulties in solving an inhomogeneous system
out of equilibrium suggested to address the problem starting from the
stationary state\cite{Okamoto2007,Okamoto2008,Heary2009,Li2015,Amaricci2014}.

Okamoto has studied the steady state of a correlated slab sandwiched between two non-interacting metallic leads using DMFT and the Keldysh formalism \cite{Okamoto2007,Okamoto2008}.The current-bias characteristic are clearly nonlinear and this is connected with the evolution of the spectral functions inside the correlated slab as a function of the external bias, which show important deformations with respect to equilibrium. 

Using a simplified approach in which the physical carriers are divided into left and right movers,  Heary and Han\cite{Heary2009} have proposed the existence --in the presence of a bias-- of a second critical value of the interaction $U_d$ smaller than the critical $U$ for the Mott transition where the Landau-Fermi liquid quasiparticles are lost while the system is not Mott localized. In Ref.~\cite{Amaricci2014} the same approach has been applied to an s-wave superconductor described by the attractive Hubbard model and a ``bad superconducting" state was found in an analogous critical region adjacent to a superconductor-insulator transition, where now the insulator is a paired state of preformed pairs without phase coherence\cite{Keller2001,Capone2002,Toschi2005a}.
In a steady-state DMFT calculation for the repulsive model a spatially inhomogeneous state with metallic and insulating islands has been later proposed as the bridge between the field-induced driven metal and the Mott insulator\cite{Li2015}.

In Ref. \cite{Eckstein2013b}  the real-space inhomogeneous DMFT for layered systems has been extended to the Keldysh formalism and solved for a Mott insulator using strong-coupling perturbation theory. In Ref. \cite{Eckstein2014a} the same approach has been used to monitor the spreading of photoexcited carriers, demonstrating the crucial effect of antiferromagnetic correlations, which allow for fast transport between layers and spatial separation of holes and electrons, as opposed to paramagnetic Mott insulators.

\subsubsection{Electric-field driven Mott transitions and resistive switching}

When a band insulator is subject to a very large electric field, a resistive transition to a metallic state can be induced, as proposed in pioneering work by Landau\cite{Landau1932} and Zener\cite{Zener1932}. This mechanism, which goes under the name of Landau-Zener dielectric breakdown is based on the simple idea that the electric field changes the wave vector of the band electrons from $k$ to $k-eEt/\hbar$, therefore it can induce the quantum tunneling of carriers from the conduction band to an otherwise empty valence band. The Landau-Zener tunneling requires a threshold electric field which is clearly of the order of the band gap.
An alternative mechanism for the "dielectric breakdown" in band insulators and semiconductors is the avalanche effect, where electrons accelerated by the external field are able to excite new electrons leading to an exponential rise of the conducting carriers. Also in this case the threshold field for the dielectric breakdown is necessarily controlled by the amplitude of the single-particle gap and the switching of the metallic state is triggered by promoting carriers to the empty valence band.

The extension of this mechanism to Mott insulators is not merely an academic one. While in a band insulator the electric field can reasonably excite only a small number of electrons, in a Mott insulator the number of carriers is large and, if some mechanism is able to destroy the Mott locking, it can release a much larger number of electrons. Therefore the control control of the conduction properties of materials by means of electrc fields would accelerate the development of the "Mottronics", in which the properties of correlated materials would be exploited to devise and engineer novel devices which could overcome the limitations of silicon-based electronics in terms of mininiaturization, large voltages and long reaction times. As a matter of fact, the idea is related to control with external knobs the insulator-to-metal transition, leading to a new kind of dielectric breakdown. The first steps in this direction have been already taken, and a number of strongly correlated insulator have been shown to undergo a resistive transition which does not seem to follow the Landau-Zener paradigm\cite{Cario2010,Cario2013a,Cario2013b}. The list includes prototypical three-dimensional oxides and chalcogenides like V$_2$O$_3$, NiS$_{2-x}$Se$_x$, GaTa$_4$Se$_8$. 

The theoretical investigation of the dielectric breakdown of Mott insulators has focused mainly on the single-band Hubbard model. Solving numerically a time-dependent Schr\"odinger equation in one dimension Oka et al.\cite{Oka2003,Oka2010} have revealed that even in the case of a Mott insulator a  Landau-Zener quantum tunneling leads to the dielectric breakdown of the Hubbard model through quantum tunneling of many-body states across the Mott gap.\cite{Oka2012}

The Mott version of the Landau-Zener mechanism involves the creation of pairs of doubly occupied sites (doublon) and holes (holons) leading to a quasistationary state which carries with a threshold behavior and an exponential behavior\cite{Oka2003,Okamoto2007,Oka2010,Lenarcic2012a}
\begin{equation}
\label{stationary_j}
j(t) \to \gamma e^{-\frac{V_{th}}{V}}.
\end{equation}
\begin{figure}
\begin{center}
\resizebox*{12cm}{!}{\includegraphics{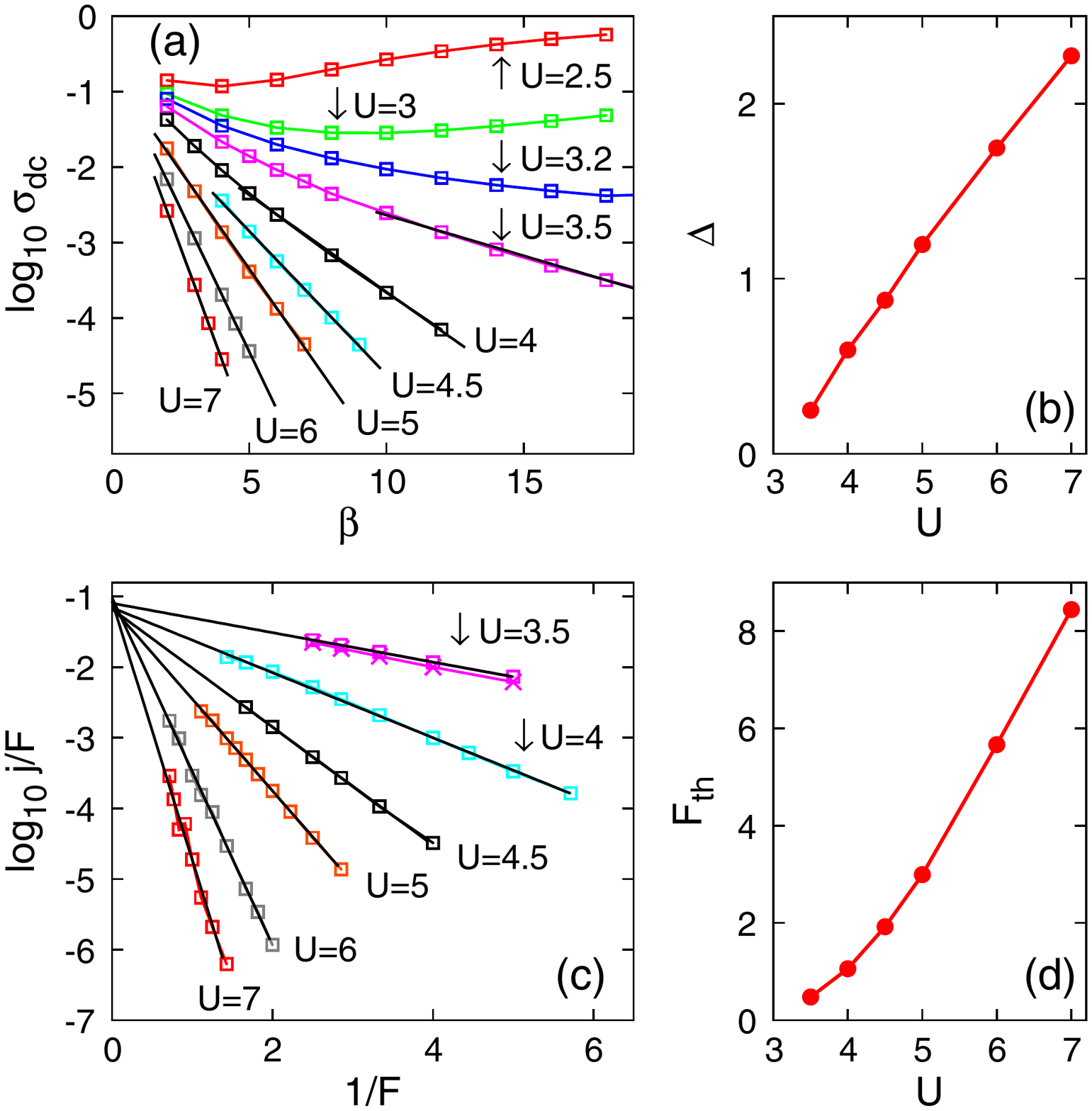}}%
\caption{Asymptotic behavior of the current $\bar{j}$ in the Hubbard model obtained in DMFT. The exponential behavior is made clear by the logarithmic plot in panel (c). Panel (a) shows the linear response conductivity $\sigma_{dc} = \bar{j}/F$, where $F$ here denotes the bias ($V$ in the text). Panel (b) and (d) show respectively the gap $\Delta$ and the threshold bias $F_{th}$, showing that the latter is cotnrolled by the former.}
\label{From_Eckstein2010_dielectric}
\end{center}
\end{figure}
The threshold is found to vanish at the Mott transition and to scale with the Hubbard $U$, confirming that $V_{th}$ is controlled by the Mott gap\cite{Eckstein2010b,Eckstein2013c}. Interestingly, the system displays an asymptotic finite value of the current even in the absence of dissipative mechanisms (which are not included in the studies mentioned above). The result has been interpreted in terms of an infinite effective temperature for the excited carriers which do not contribute to the coherent motion and to the current, which remains associated to the pure quantum tunneling effect. 

In Ref. \cite{Aron2012b} a dissipative mechanism has been introduced and some deviations from a pure Zener picture have been reported. In particular the current at weak field is found to be controlled by the dissipation. In this case the effective temperature of the excited carriers becomes finite leading to a contribution to the asymptotic current. Moreover, in connection with the electric-field-driven dimensional crossover, the dielectric breakdown occurs when the field strength is on the order of the Mott gap of the corresponding lower-dimensional system.
\begin{figure}
\begin{center}
\resizebox*{12cm}{!}{\includegraphics{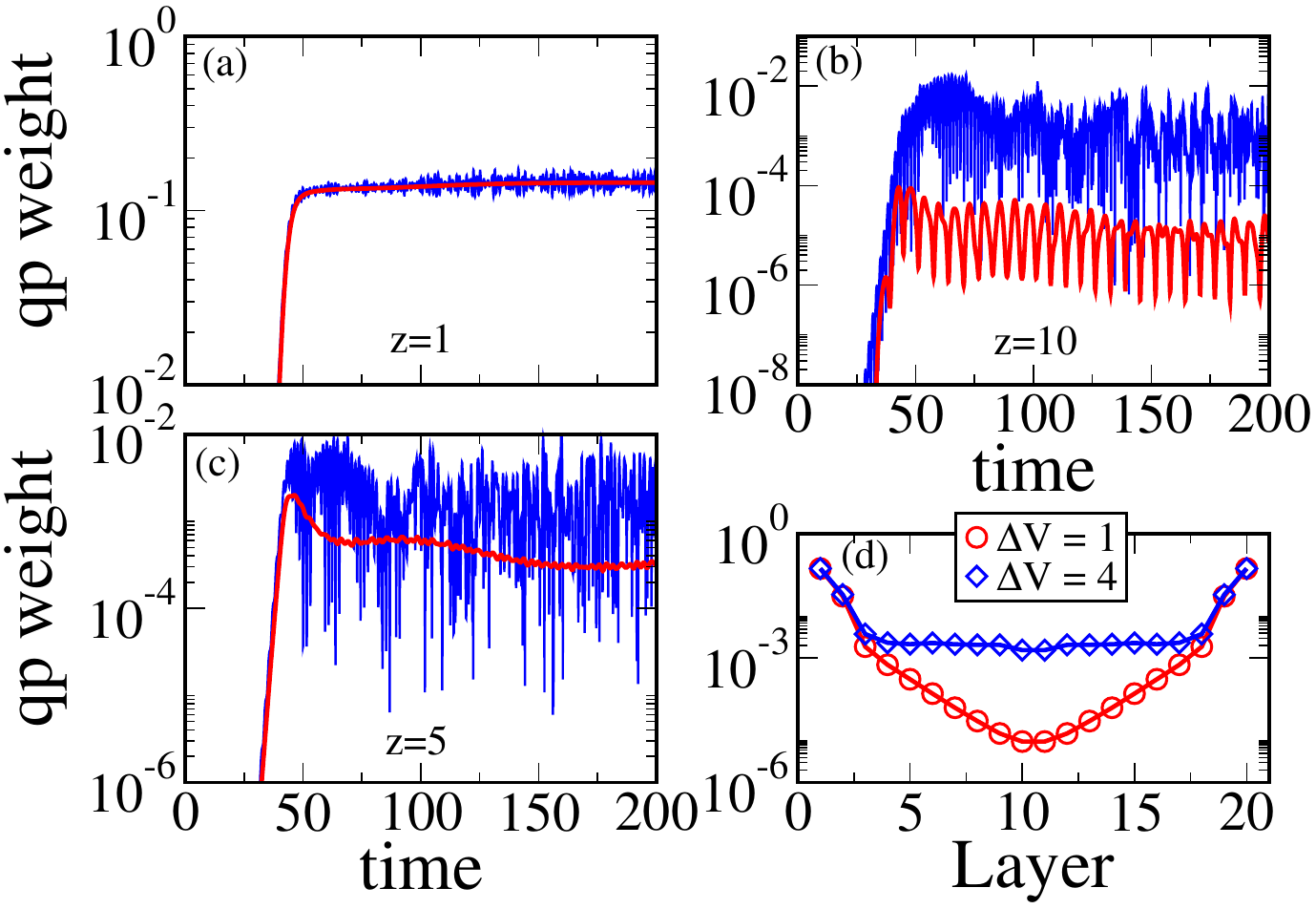}}%
\caption{Time evolution of the quasiparticle weight 
    for layers 1 (a, edge), 5(b) and 10(c, middle) of a 20-layer slab with $U = 16.5$ (larger than the critical $U$ for the Mott transition) and an hybridization between the slab and the leads $v=1.0$ Energies are in units of the hopping amplitude along each later.
   We compare  two values of the applied bias $\Delta V=1.0$(red lines) and $\Delta V=4.0$
    (blue lines). (d): time-averaged stationary quasiparticle weight  spatial profile across the slab.}
\label{Mazza2015fig14}
\end{center}
\end{figure}Using  a inhomogeneous time-dependent Gutzwiller approximation it is possible to monitor the appearance of field-induced evanescent quasiparticles starting from the Mott insulator\cite{Mazza2015} and their spatial distribution. The quasiparticles appear exponentially fast as soon as the system is attached to the leads, with a characteristic time which diverges when $U$ is tuned towards the Mott transition, suggesting an avalanche effect triggered by electron-electron interaction. We notice that the feedback between the quasiparticles and the Hubbard bands characteristic of the time-dependent Gutzwiller scheme is fundamental to obtain this effect.
In Fig. \ref{Mazza2015fig14} we report the time evolution of the quasiparticle weight for different layers ($z=1$ corresponds to one edge, while $z=10$ is the middle layer). While the dynamics displays oscillations reminiscent of the incoherent dynamics of the metallic state, the time-averaged values shown in Panel (d) have a well defined limit, which shows how increasing the bias the quasiparticles penetrate from the outer layers to the bulk region. The penetration of the quasiparticles leads finite values of the stationary current which well fit the expected behavior (\ref{stationary_j}) for a Landau-Zener dielectric breakdown.
A series of different behaviors in the relaxation dynamics in the presence of an electric field has  been reported also within DMFT in Ref. \cite{Fotso2014} for the Hubbard and Falicov-Kimball models highlighting that the integrability of the unperturbed model does not influence the relaxation dynamics.

The effect of a dissipative bath has been explored in \cite{Li2015} in a steady-state formalism for DMFT. A nonmonotonic threshold field for the resistive transition s found as a function of the interaction strength  due to an interplay  quasiparticle renormalization and  field-driven effective temperature. Hysteretic I-V curves suggest that the nonequilibrium current is carried through a spatially inhomogeneous metal-insulator mixed state.

All the studies of the single-band Hubbard model strongly suggest that the Landau-Zener picture survives in the essential aspects also in the case of a Mott insulator described by the Hubbard model. This may seem a surprising and disappointing result, as the many-body character of the insulating Mott state is expected to lead to a more adaptive gap with respect to a band insulator, where the gap is fixed by the structure, the chemistry and the external conditions, but it does not depend on the electronic configuration. Indeed a recent study\cite{Mazza2015b} shows that the picture changes qualitatively as soon as we consider more than one single orbital per site (or one single band). Solving a two-orbital Hubbard model with a crystal field splitting, it has been shown that a novel mechanism for the dielectric breakdown takes place when the equilibrium Mott-Hubbard transition is strongly first-order and the metallic solution obtained either by changing control parameters or via the electric field is substantially different from the metal. As a result, the threshold field is not controlled by the gap but rather by the energy difference between the metallic and the insulating solutions, as proposed at a more phenomenological level in Ref.~\cite{Camjayi2014}.

Finally, it has been recently shown that the noninteracting lesser Green's function can be determined in terms of the Wannier-Stark ladder eigenstates, which are thermalized via the standard canonical ensemble according to the Markovian quantum master equation. As a result, the interplay between strong correlation and large electric fields can generate a sequence of two dielectric breakdowns with the first induced by a coherent reconstruction of the midgap state within the Mott gap and the second by an incoherent tunneling through the biased Hubbard bands. It is predicted that the reconstructed midgap state generates its own emergent Wannier-Stark ladder structure with a reduced effective electric field. The two dielectric breakdowns are mediated by a reentrant insulating phase, which is characterized by the population inversion, causing instability toward inhomogeneous current density states at weak electron-impurity scattering.
\cite{Lee2014}

\subsubsection{Light-induced excitation and photodoping}

One of the main goals of the research on non-equilibrium correlated materials is to describe the excitation process induced by a laser pulse, as realized in pump-probe spectroscopies. This subfield is however one of the most complicated targets, as it requires the inclusion of all the different effects leading to relaxation after the impulsive excitation. This section is therefore mainly devoted to DMFT calculations in which the relaxation effects associated to electron-electron interactions are better accounted for.
\begin{figure}
\begin{center}
\centerline{\includegraphics[width=14cm]{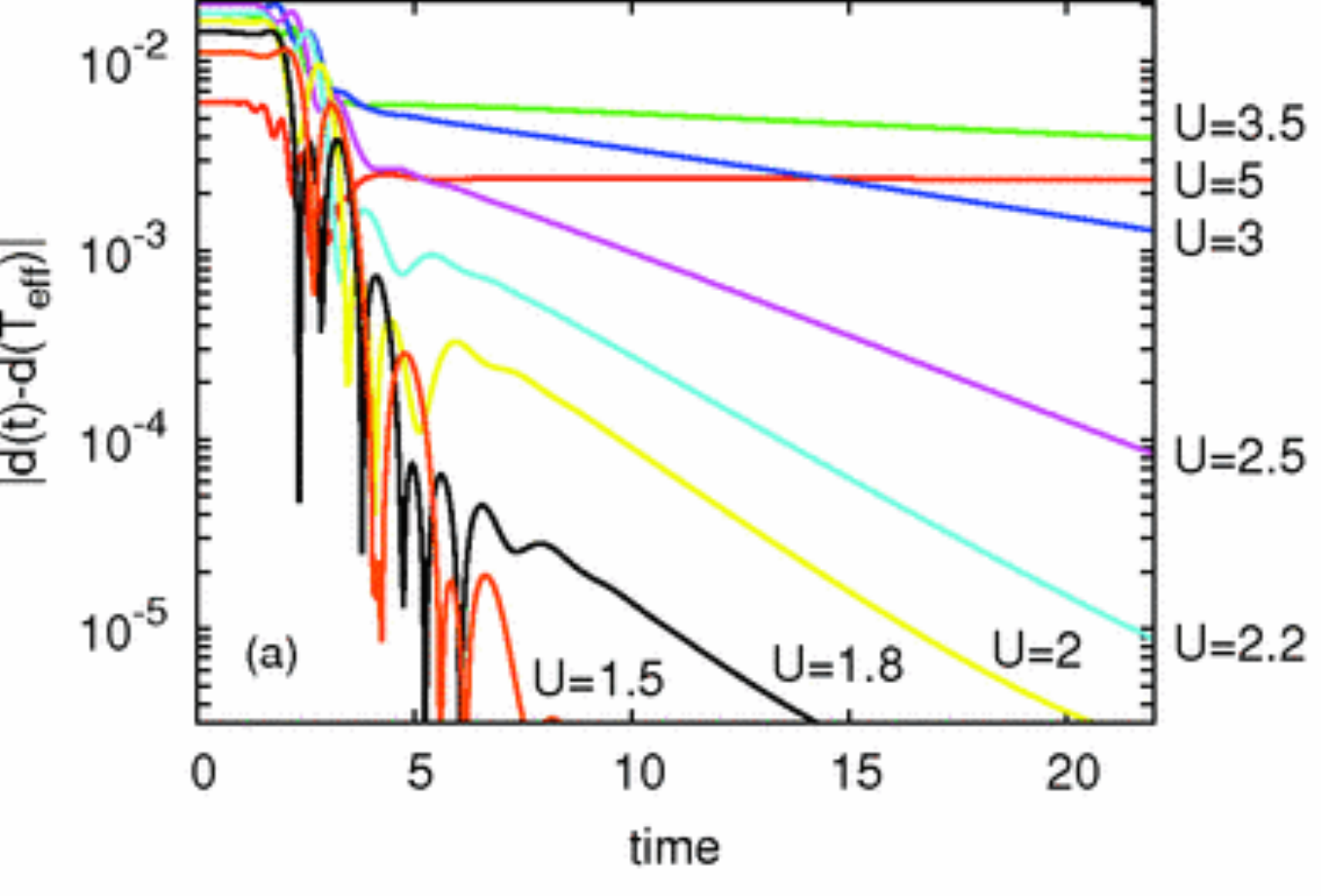}}
\caption{Relaxation dynamics of the difference between the double occupancy $d(t)$ and the effective thermal value calculated in equilibrium with $T=T_{Jeff}$.  For different values of $U$ for a single-band Hubbard model solved within time-dependent DMFT. The linear behavior in the semi-log plot highlights the exponential decay for intermediate and large values of $U$ and the deviation when the Mott transition point is approached. Energies are in units of the bandwidth. Taken from Ref.~\cite{Eckstein2011a}.}
\label{figEckstein2011}
\end{center}
\end{figure}
As discussed at length in Sec. 5.1,  the simple two-temperature model\cite{Allen1987}, in which the photo excitation excites ``hot" electrons to a temperature $T_e$, while the lattice remains at a lower effective temperature $T_{lat}$, has been successful in reproducing the main features of the photo excitation of semiconductors and weakly correlated metals. 
This scheme can be useful only when electron-electron interaction processes drive the system to a quasi-equilibrium state on a timescale which is smaller and clearly separated from the time associated with the electron-lattice dynamics.  DMFT studies have indeed shown that this is the case when the correlated metallic state of the Hubbard model is impulsively excited\cite{Eckstein2011a} and, rather surprisingly, this essentially happens also in the simplified Falicov-Kimball model\cite{Moritz2010,Moritz2013} where the exact solution demonstrates the lack of actual thermalization for quasiparticle properties\cite{Eckstein2008b} which do not coincide with the corresponding thermal values\cite{Freericks2009}.

The qualitative success of the  two-temperature scheme is however mostly limited to the metallic state. When we perturb the Mott insulator the system is unable to thermalize\cite{Moritz2010,Eckstein2011a,Moritz2013} and the dynamics depends on the details of the excitation process\cite{Eckstein2011a}. The dynamics is characterized by a fast transient with strong damping, rapidly followed by an exponential relaxation. For large values of $U$, the relaxation time grows with $U$ as a consequence of the long lifetime of doubly occupied sites (doublons). When we approach the Mott transition from above, the dynamics becomes faster and faster and the system relaxes to an asymptotic state in times of the order of the inverse of the hopping matrix element, until a point in which the relaxation appears even faster than an exponential and the decay time loses significance.

Indeed even in the presence of an explicit dissipation channel the photodoped carriers give rise to a bad metal which  does not follow the Landau Fermi-liquid paradigm, despite their relatively high kinetic energy\cite{Eckstein2013a}. The lifetime of the photo excited holes and electrons is indeed finite at the longest times accessible in the DMFT simulations of Ref.~\cite{Eckstein2013a} so that the bands observed in both the single-particle and the optical spectra are much broader than the corresponding equilibrium counterparts. The limited simulation time does not however rule out the possibility that on a longer timescale the system actually relaxes to a conventional Fermi liquid, which would signal a similarity between the behavior of the photo excited Mott insulator and systems with broken symmetry\cite{Heyl2013,Tsuji2013}. 

In Ref.~\cite{Werner2014} it has been shown that the thermalization dynamics can depend on the energy of the photo-doped carriers, which can be controlled by changing the laser frequency and fluence. In particular, if the Mott gap is smaller than the kinetic energy of the carriers in the Hubbard bands, the excited carriers can produce additional doublon-hole pairs through a kind of impact ionization. This phenomenon can take place on ultrafast timescales of the order of 10 fs, well before the standard relaxation processes take place. Finally it has been recently shown --using extended DMFT to introduce low-energy bosonic modes which are not present in DMFT-- that the photo excitation of carriers rapidly opens new screening channels which are not present in the Mott insulator and therefore reduce the Mott gap. The low-energy bosonic excitations also open new relaxation channels which further increase the speed of the thermalization\cite{Golez2015}. The opposite scenario can be realized in a strongly correlated metal, where the weak quasiparticles can be destroyed by the impulsive excitation, thereby reducing the screening.

Using non-equilibrium inhomogeneous DMFT\cite{Eckstein2013b}, to study a photo excited Mott heterostructure\cite{Eckstein2014a} it has been shown that AFM correlations strongly influence the carrier dynamics. An antiferromagnetic state can in fact exchange energy with the excited carriers on an ultrafast timescale via spin-flip processes, which allows for transport across the heterostructure leading to a spatial separation between excited holes and electrons.

\subsection{Quantum quenches and dynamical phase transitions in simple models}
One of the most popular and paradigmatic non-equilibrium problems is
the quantum quench, a protocol in which a system is driven out of
equilibrium by a sudden change of a relevant parameter (for example
the interaction strength in the Hubbard model). From a theoretical
point of view, this is perhaps the simplest non-equilibrium protocol,
and it attracted a lot of interested in the case of open quantum
systems in connection to the concept of thermalization\cite{Polkovnikov2011}. 

Even if the quantum quench problem is not directly connected to the experiments that represent the core of the present review, where the system is pushed out of equilibrium by an electromagnetic field, it still represents a useful reference problem which allows to reverse engineer the complex dynamics of a correlated material driven out of
equilibrium and to disentangle the effect that stem from the change in the state of the system (for example a change in the double occupancy in the Hubbard model) from the intrinsic effects of the electromagnetic field. Hence in the following we review the main results for quantum quenches  obtained by means of the Gutzwiller approximation and DMFT.

\begin{figure}[t]
\centerline{\includegraphics[width=10cm]{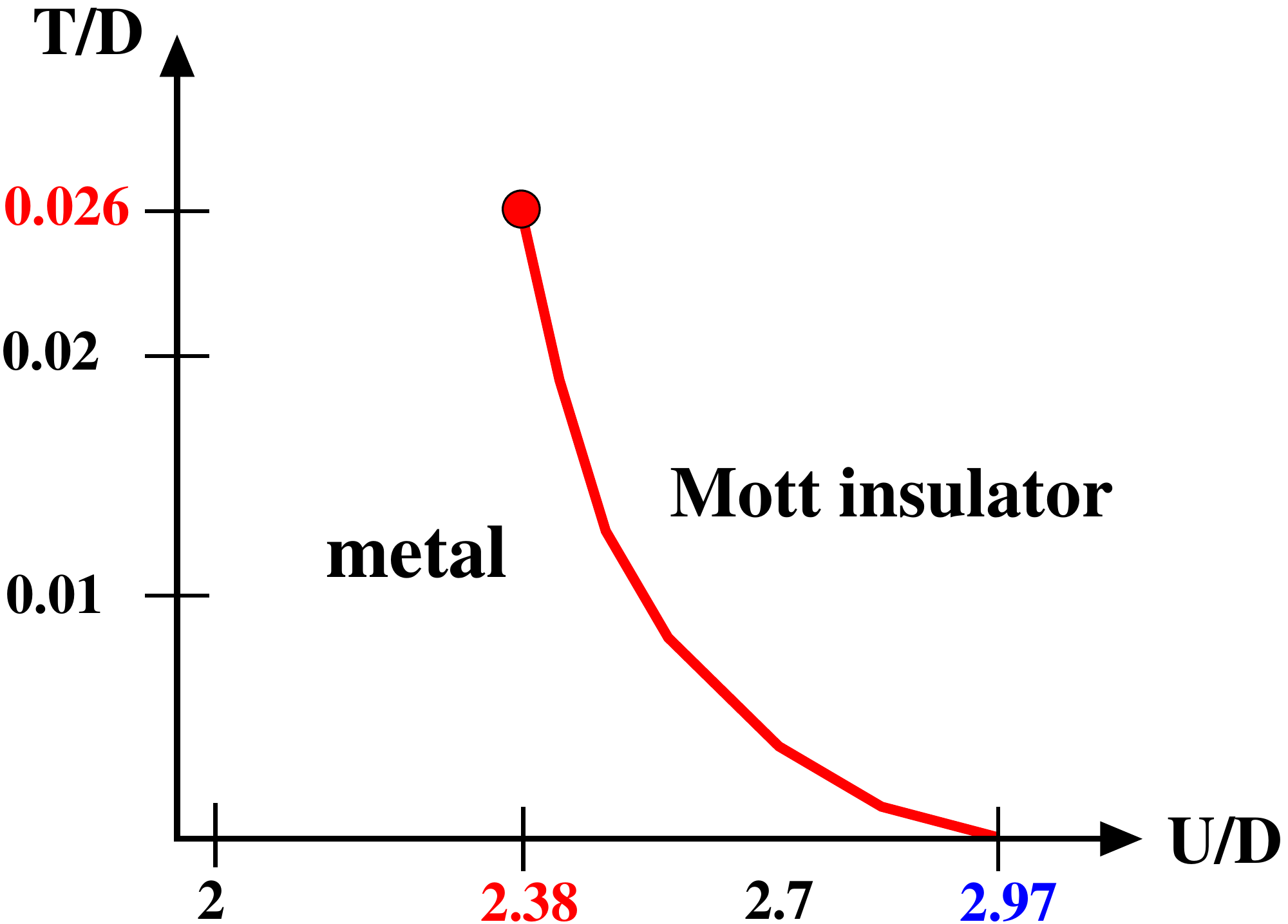}}
\caption{Sketch the DMFT phase diagram of the half-filled Hubbard model on a Bethe lattice and in the paramagnetic sector. The transition line at $T\not=0$ is first order and ends up into a second order critical point at $T\simeq 0.026~D$.  }
\label{fig-8.1-1}
\end{figure}

In this section we mainly focus on the paradigmatic single-band Hubbard model. We can gain intuition on the effect of a change of the interaction parameter exploiting the knowledge on the equilibrium phase diagram of the model within DMFT which has been firmly established after a remarkable collective effort using a combination of different approaches. In particular we consider the half-filled model in the paramagnetic sector, where the antiferromagnetic ordering is neglected (or it assumed to be frustrated).
The system undergoes a Mott-Hubbard metal-insulator transition increasing the interaction $U$. 
The schematic phase diagram in the interaction-temperature plane is reproduced in Fig. ~\ref{fig-8.1-1}, where both $T$ and $U$ are measured in units of the half-bandwidth $D$ of the bare lattice. Here we consider an infinite-coordination Bethe lattice with a semicircular density of states. The transition occurs when the line $U_\text{MIT}$ is crossed and it is continuous at zero temperature and it becomes of first order at finite temperatures below a critical temperature $T_c$, where the line ends in a critical point\cite{Zhang1993,Georges1996}.

\begin{figure}
\begin{center}
\resizebox*{10cm}{!}{\includegraphics{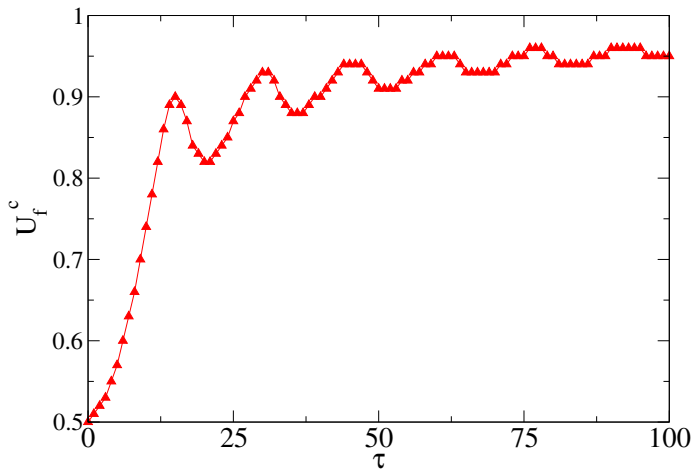}}%
\caption{The critical value for the dynamical Mott transition in  the single-band Hubbard model, $U^c_f$ in units of $U_\text{MIT}(T=0)$,  found by the time-dependent Gutzwiller wavefunction as function of the switching time $\tau$. (From Ref.~\cite{Sandri2012}.)}
\label{fig-8.1-2}
\end{center}
\end{figure}

Let us assume that the system is initially in equilibrium in the groundstate of the non-interacting system ($U=0$). At some given time we turn on the interaction reaching a finite value $\bar{U}$. If the interaction switch is infinitely slow (adiabatic) the system will be able to follow the equilibrium phase diagram which implies that at $\bar{U}=U_\text{MIT}(T=0)\simeq 2.97 D$ it will cross the transition line and turn into a Mott insulator. If the interaction is instead switched on in a finite time $\tau$, the system will no longer be able to follow adiabatically the instantaneous groundstate, and it will instead evolve into a thermal state with a finite effective temperature $T_*\not=0$. The smaller is $\tau$, the larger is expected to be $T_*$. As a consequence, since the the critical interaction decreases as a function of temperature $U_\text{MIT}(T\not=0) \leq U_\text{MIT}(T=0)$ [See Fig.~\ref{fig-8.1-1}] the dynamical phase transition to a Mott insulator is expected to occur at smaller value of the interaction $U_*(\tau) \simeq U_\text{MIT}(T_*) \leq U_\text{MIT}(T=0)$, the equality holding only in the adiabatic limit when the switching-time $\tau\to\infty$. This expectation has been actually confirmed by explicit calculations within the time-dependent Gutzwiller approximation in Ref.~\cite{Sandri2012}, where a dynamical Mott transition is obtained for any value of $\tau$ with a critical value which increases with $\tau$ as shown in the plot of reported in  Fig.  ~\ref{fig-8.1-2}.

It is much less obvious to predict the behavior for of a sudden quench of the interaction, which corresponds to the $\tau\to 0$ limit of the previously described protocol.
Indeed this problem has been the first application of the time-dependent Gutzwiller approximation~\cite{Schiro2010,Schiro2011}. The result is that a dynamical Mott transition occurs also in this limit at a critical $U_*(\tau\to 0) \simeq U_\text{MIT}(T=0)/2$, a value of the interaction for which no Mott transition occurs as a function of temperature within DMFT (See Fig. \ref{fig-8.1-1}). One might be tempted to blame this surprising and counterintuitive result to the limitations of the Gutzwiller approximation which does not account for all the dissipative channels therefore inhibiting a proper thermalization.

This suspicion is completely ruled out by the DMFT studies of the same problem that in fact appeared before the Gutzwiller analysis in two pioneering papers\cite{Eckstein2009,Eckstein2010a}. Here the authors used Continuous-Time Quantum Monte Carlo as the impurity solver, which leads to numerically exact results, but severely limits the simulation time. Despite this limitation, the results clearly show a sharp change in the dynamics at a critical value of the interaction $U_c^{dyn} \simeq 2 D$ which can be associated to a possible dynamical phase transition.  
\begin{figure}
\begin{center}
\resizebox*{12cm}{!}{\includegraphics{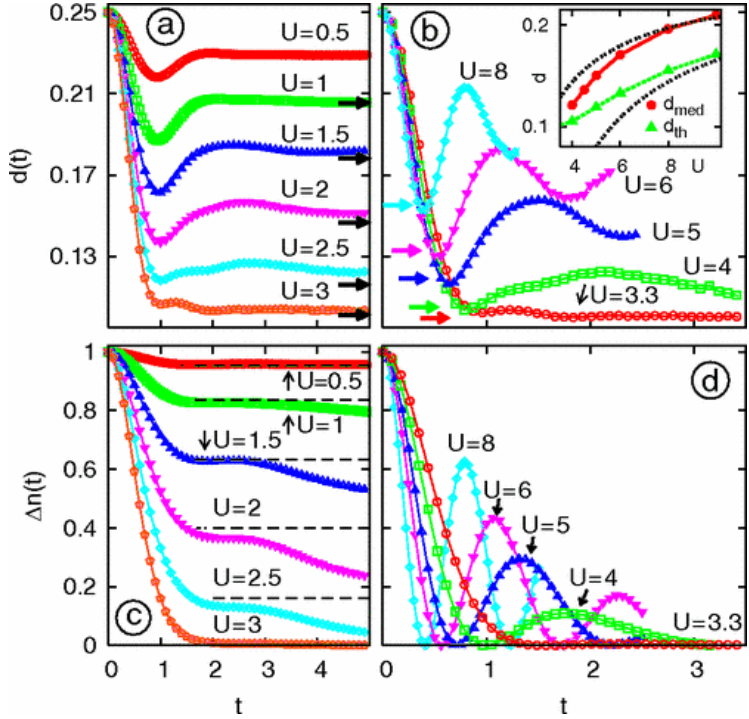}}%
\caption{Fermi surface discontinuity $\Delta n$ and double occupation $d(t)$ after quenches to $U<3$ (left panels) and $U>3.3$ (right panels). Inset: thermal value $d_{th}$ and $d_{med}$, the average of the first maximum and the second minimum of $d(t)$, which provides an estimate of the stationary value $d_{stat}$; black dashed lines are the respective results from the strong- coupling expansion. taken from \citenum{Eckstein2009}.}
\label{fig_EKW2}
\end{center}
\end{figure}
In Fig. \ref{fig_EKW2}, we reproduce the time evolution of the double occupancy $d(t) =\langle \sum_i n_{i\uparrow}n_{i\downarrow}\rangle$ and of the jump of the momentum distribution function at the Fermi level $\Delta n(t)$ from Ref. \cite{Eckstein2009} for different values of the final interaction $\bar{U}$. Panels (a) and (c) refer to small and intermediate values of $\bar{U}$, panels (b) and (d) to large values. In all cases the interaction quench reduces both $d(t)$ and $\Delta n(t)$ from the non-interacting value, but the dynamics is clearly different. For $U < U_c^{dyn}$ the system is trapped for very long times into a quasi stationary state characterized by a double occupancy which differs from the thermal equilibrium value by a quantity of order 1, while for $U > U_c^{dyn}$ the dynamics is characterized by oscillations with a frequency of the order $2\pi/U$ in the observables. The two behaviors are compatible with metallic and Mott-insulating responses, respectively. This might suggest that the DMFT dynamical phase transition is continuously connected with the equilibrium phase diagram we discussed above. On the other hand the critical value of the interaction  $U_c^{dyn} \simeq 2 D$ is smaller than the minimum value of $U_\text{MIT}$, which coincides with the finite-temperature critical point.

The possible connection between the results for finite switching time and the sudden quench has not been studied within time-dependent DMFT yet. We notice in conclusion that if the scenario which we have drawn on the basis of the present result would be confirmed, the existence of such dynamical transition implies the absence of thermalization. 

We finally mention that the time-dependent Gutzwiller approximation 
(t-GA), i.e. the above variational scheme applied to lattices with finite coordination number, also predicts the possibility of a dynamical surface-Mott transition. Specifically, in Ref.~\cite{Andre2012} a correlated metal slab was studied whose surface is suddenly excited into a non-equilibrium state. Above a threshold, the supplied excitation energy 
dynamically drives the surface layer into a Mott insulating phase while the bulk is still metallic. In addition, if a realistic  electron-phonon coupling is included in the calculation, the  dynamical surface Mott-transition is found to be accompanied by a lattice deformation at the surface layers.~\cite{Andre2012} 
Therefore, should the prediction of t-GA be correct, such a phenomenon could be detected by monitoring either the conducting or the structural properties of the surface layers.  

\subsection{Melting of antiferromagnetism and broken-symmetry phases}
\begin{figure}[t]
\centerline{\includegraphics[width=12cm]{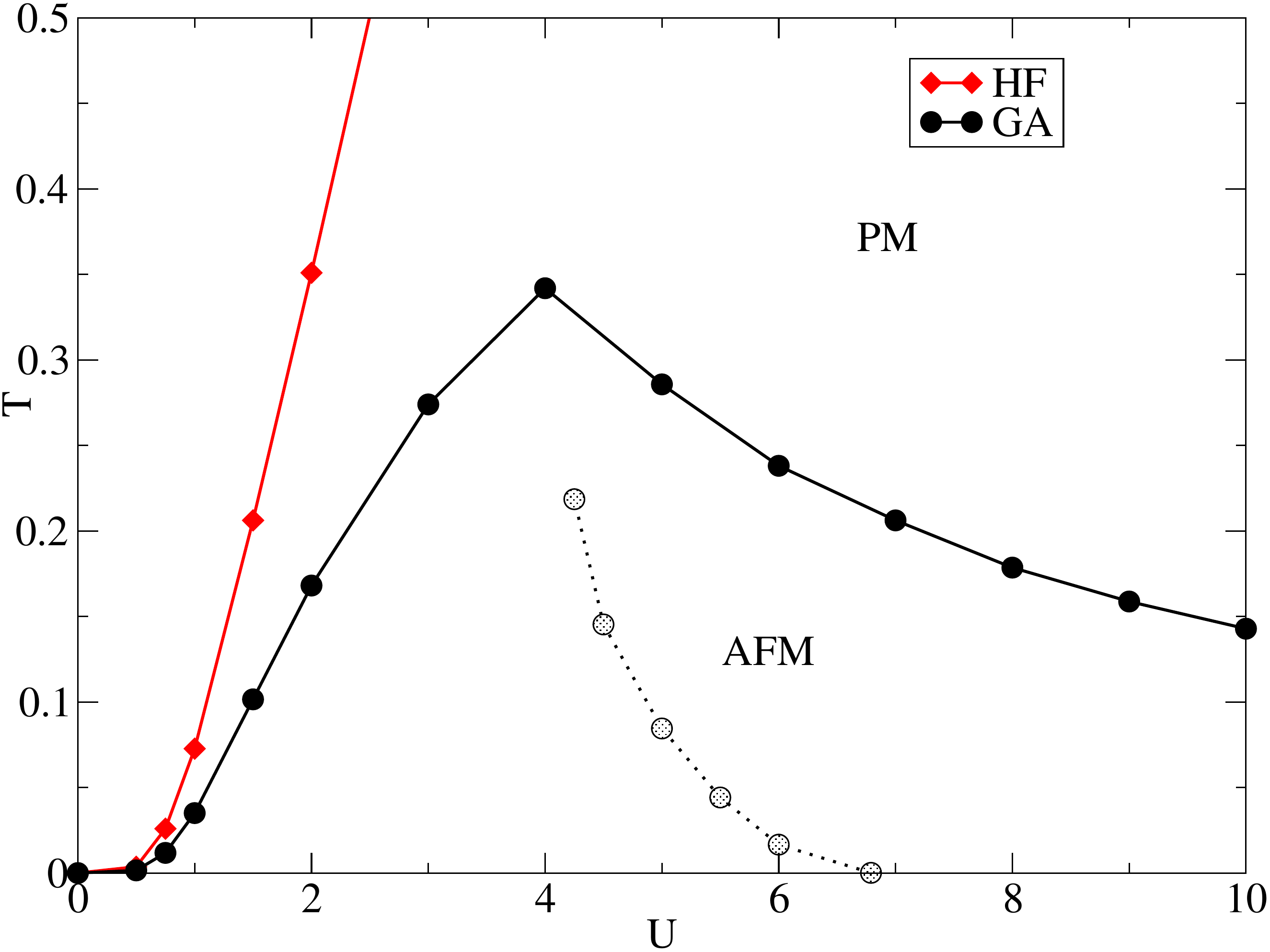}}
\caption{Phase diagram of the half-filled Hubbard model obtained within the Gutzwiller approximation for a model with a semicircular density of states (Bethe lattice) comparing the magnetic and paramagnetic solutions AFM marks the antiferromagnetic solution, which is always insulating, PM indicates the paramagnetic metallic phase. The first-order critical line between a PM and and a paramagnetic Mott insulator which is obtained quenching magnetism is the dashed line with open symbols (See Fig.~\ref{fig-8.1-3}). The red line reproduces the  N\'eel temperature within the Hartree-Fock approximation. Both $U$ and 
$T$ are in units of a quarter of the bandwidth (From Ref.~\cite{Sandri2013b}.)}
\label{fig-8.1-3}
\end{figure}
In the previous section we focused on the paramagnetic phase of the Hubbard model, where no magnetic (or other) symmetry breaking is allowed. Under this assumption a dynamical transition has been identified which has a direct connection with the equilibrium Mott transition between a paramagnetic metal and a paramagnetic insulator. 

However the ground state of the half-filled Hubbard model with nearest-neighbor hopping on a bipartite lattice is an antiferromagnetic (AFM)  insulator for any coupling $U$, and a magnetic state is expected to be stable at low temperature for most values of $U$ also in the presence of realistic lattices and hopping structures (for example in the presence of a next-neighbor hopping) as long as the degree of frustration is not extreme. 

In Fig. ~\ref{fig-8.1-3} we report a phase diagram in the $U$-temperature plane which shows the stability of the AFM state as obtained by means of the   finite-temperature Gutzwiller approximation~\cite{Sandri2013a}. The data are from Ref.~\cite{Sandri2013b}. A qualitatively similar phase diagram is obtained using DMFT.

The N\'eel temperature $T_N$, below which the system is an antiferromagnet increases exponentially at weak-coupling as predicted in a simple Hartree-Fock approximation, then reaches a maximum when $U$ is of the order of the bandwidth $2D$, and then decreases. At large $U$, $T_N$ is proportional to  $J\sim D^2/U$, the superexchange coupling. In this regime the Hubbard model can indeed be mapped onto a Heisenberg model with coupling constant $J$, which is nothing but the t-J model where no hopping is possible due to the double occupancy constraint. It is remarkable that the finite-temperature Gutzwiller variational approach can capture the correct non-monotonic behavior of $T_N$ as function of $U$, which is completely out of the reach of the standard Hartree-Fock approximation, which predicts incorrectly that $T_N$ always increases with $U$, see Fig.~\ref{fig-8.1-3}. We observe that the line marking the transition between the paramagnetic metal and the paramagnetic insulator lies below the $T_N$. Therefore, above $T_N$ there only a crossover connects the metallic to the insulating state.

The above equilibrium phase diagram suggests an interesting perspective for the non-equilibrium behavior~\cite{Sandri2013b}. 
We consider a quantum quench of the interaction starting from an intermediate value of $U_i$.  The wave function at $t=0$ therefore coincides with the variational wavefunction Eq.~\eqn{sec-8.1:Psi} that  minimizes the internal energy at $U_i=4$, see Fig.~\ref{fig-8.1-3}. 
Then the interaction is quenched to $U_f\not = U_i$ defining a final Hamiltonian which governs the time evolution through the dynamical equations \eqn{sec-8.1:dot-Psi0} and \eqn{sec-8.1:dot-Phi}. 
Since the initial state is a superposition of an infinite number of excited states of the final Hamiltonian, the system is expected to thermalize to an asymptotic state with an effective temperature $T^*$. $T^*$ is expected to be larger as the strength of the quench  $\left|U_f-U_i\right|$ increases. Since $T_N$ is nonmonotonic and it vanishes both for $U\to 0$ and $U\to\infty$, thermalization would imply the existence of two critical values of the final interaction such that, for $U^>_f > U_i$ a $U^<_f < U_i$ =$T^*$ exceeds $T_N$ leading to a final state without magnetic ordering.
In other words, we would obtain a field-induced dynamical melting of antiferromagnetism above a critical $\left|U_f-U_i\right|= \left|U_f-4\right|$. In particular, for $U_f\leq U^<_f$, the state would  evolve into a paramagnetic metal at finite temperature, while for $U_f\geq U^>_f$ into a finite-temperature paramagnetic insulator. 
\begin{figure}[t]
\centerline{\includegraphics[width=13cm]{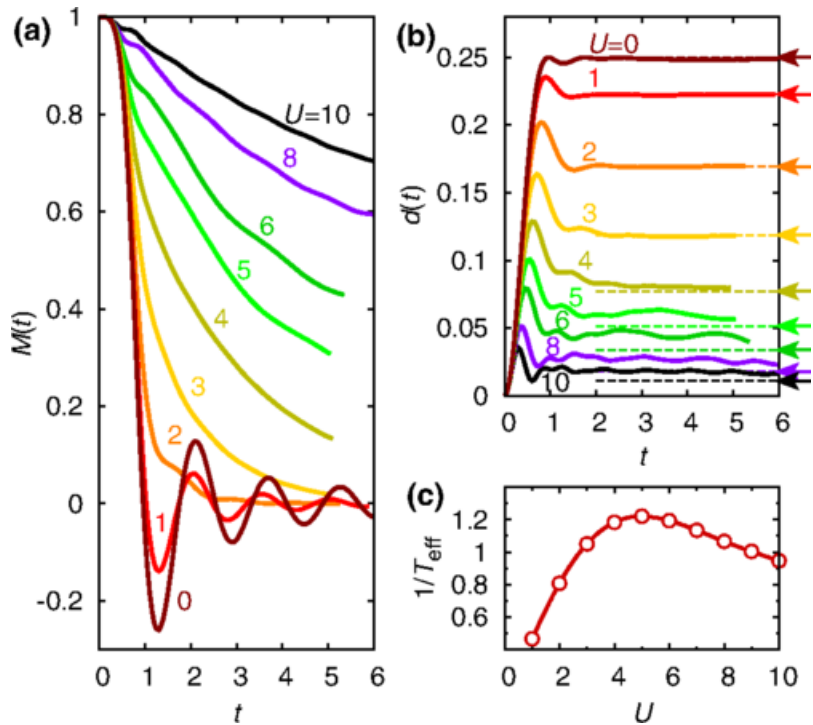}}
\caption{Time evolution of the double occupancy $d(t)$ (Panel (b))  and of the order parameter $M(t) = M/(1-2d(t))$ (Panel (a)) for different values of the interaction in a quench in which the initial state is the N\'eel antiferromagnet. In Panel (c) the inverse of the effective temperature reached after thermalization is plotted as a function of $U$. The arrows in Panel (b) are the value of double occupancy computed in equilibrium for the corresponding effective temperatures. Taken from Ref.~\cite{Balzer2015}}
\label{figBalzer2015}
\end{figure}
This qualitative expectation is indeed confirmed by explicit calculations within the Gutzwiller approximation in Ref.~\cite{Sandri2013b}, where, using $U_i = 4$,  the critical values $U^<_f\simeq 1.7$ and $U^>_f\simeq 21$ were found (here we use a quarter of the bandwidth as energy unit to facilitate the comparison with literature). It must be noted however that such critical values correspond to an effective temperature higher than the corresponding equilibrium $T_N$, which suggests that magnetism is more resilient to the quantum quench than expected on on the basis of thermalization. It must be reminded that the Gutzwiller approximation does not describe the actual dissipation processes, and we can expect that the exact dynamics leads to actual thermalization correcting the non-thermal Gutzwiller asymptotic state.

The same quench protocol has been investigated  within DMFT by Tsuji, Eckstein and Werner\cite{Werner2012,Tsuji2013}.  In particular, Ref.~\cite{Tsuji2013} is dedicated to the weak-coupling regime, with  $U_i$ ranging between 2 and 2.7 and $U_f<U_i$, where the persistence of magnetic ordering below the threshold expected by thermalization is observed in a finite range of parameters. Analyzing  the decay of the  magnetic order parameter for different values of $U_i$, the authors argue that the non-thermal state is only a transient regime in which the dynamics is trapped before reaching, at longer times, the final relaxation to a thermal state. The lifetime of the metastable non-thermal state is shown to increase as $U_i$ is reduced.
We finally turn to the properties of the nonthermal state obtained for $U_f>U_i$. In this region we expect an antiferromagnetic insulator for long times. It was found in Ref.~\cite{Sandri2013b} that although the magnetic ordered phase survives more than expected, for quenches 
$U_* \simeq 8.2\leq U_f\leq U^>_f$ the non-equilibrium antiferromagnet shows incoherent excitations at the gap edges, in contrast to the equilibrium antiferromagnet which has coherent excitations.
A similar trapping into a non-thermal state has been observed using DMFT in Ref.~\cite{Werner2012} for the case of a quench from $U_i =4$ to slightly larger values of $U$. Here the transient state is found to be reminiscent of a doped AFM insulator, and its persistence has been attributed to the slow decay of doubly occupied sites. 

In the previous section we focused on the paramagnetic phase of the Hubbard model, where no magnetic (or other) symmetry breaking is allowed. Under this assumption a dynamical transition has been identified which has a direct connection with the equilibrium Mott transition between a paramagnetic metal and a paramagnetic insulator. 
In Ref.~\cite{Balzer2015} the melting of the N\'eel state has been studied using DMFT with the accurate DMRG solver. In this study the system is prepared in a classical N\'eel state and let evolve under the actual Hubbard Hamiltonian for different values of $U$ and the results are summarized in Fig. \ref{figBalzer2015}. A crossover between two different relaxation dynamics occurs for $U \simeq 2.4$ (following the same notations of this whole section). In strong coupling the local magnetic moment survive the melting of the ordered AFM state. In this case the magnetic state is destroyed by the motion of the photoexcited carriers, similarly to what happens for doped models of the cuprates\cite{DalConte2015,Mierzejewski2011a,Lenarcic2012b,Moritz2013,Eckstein2014a}. The destruction of the magnetic ordering in weak coupling is instead controlled by the quasiparticle excitations. 

Besides the melting of the full AFM state, it is also possible to influence and control the strength of the exchange interaction $J$ in a Mott antiferromagnet. In particular, it has been shown by explicitly simulating the spin precession in an antiferromagnet canted by a perpendicular magnetic field\cite{Mentink2014} that photo doped  excitation can quench the exchange coupling on an ultrafast timescale almost as effectively as chemical doping. Using a Floquet approach, it has been proposed to exploit the same mechanism to control in an ultrafast and reversible way the exchange coupling\cite{Mentink2015}. In particular $J$ can be enhanced and reduced by means of periodic modulations with a frequency  smaller and larger than the Mott gap, and strong perturbations can even lead to a change of sign.

\subsection{Non-equilibrium dynamics beyond the single-band Hubbard model}

In this section we discuss how the non-equilibrium dynamics can become reacher and more appealing if we relax the single-band approximation and introduce other orbitals in the low-energy description of the correlated material. This is not merely a theoretical complication, but it reflects the electronic structure of most correlated materials, while single-band materials like the cuprate superconductors are the exception rather than the rule.

If we assume that the orbitals are completely degenerate and we can neglect the Hund's coupling, we can start from a 
 $SU(N)$  symmetric Hubbard model  
\be
\mathcal{H} = \sum_{i,j,\sigma}\,\sum_{a=1}^N\,t_{ij}\,c^\dagger_{ia}c^\dagga_{ja} 
+ \fract{U}{2}\sum_i \big(n_i - n\big)^2,\label{sec-8.1:SU(N)}
\ee
where $a=1,\dots,N$ labels one of the $N$ orbitals, including the spin multiplicity of each atom (lattice site) and $n\leq N$ is the average number of electrons per site. Mott-Hubbard transitions occur in general at every integer value of $n$ at some $U_c(n,N)$ such that a Mott insulator is stable for $U > U(n,N)$, while for any non-integer $n$ the paramagnetic solution is always metallic.
As in the case of the single-band Hubbard model, we expect some kind of spatial ordering to take place at low temperature to quench the orbital degeneracy corresponding to the number of ways to arrange $n$ fermions into $N$ orbitals, i.e. to the binomial coefficient $C(N,n)$. However, within mean-field theories like Gutzwiller and DMFT we can neglect any kind of ordering and study the fully symmetric solution, what for $SU(2)$ implies enforcing paramagnetism. In this case it is well established within DMFT~\cite{Georges1996} that two critical value of the interaction $U_{c1}$ and $U_{c2}$ exist, with $U_{c1}<U_{c2}$. 
Below $U_{c1}$ only a metallic solution can be found, which means that no insulating minimum of the energy exists, while  above $U_{c2}$ 
there is no metallic minimum, the model is insulating. 
For $U_{c1}\leq U\leq U_{c2}$ both metallic and insulating solutions exist as local minima of an energy functional. In principle a first-order transition is expected at the value of $U$ for which the energies of the two solutions cross. However, for the fully symmetric model  the point at which the crossing of the energy curves coincides with $U_{c2}$, the point at which the metal becomes unstable, so that in the whole coexistence region between $U_{c1}$ and $U_{c2}$ the lowest 
energy state is always metallic and the zero-temperature Mott transition is continuous (of second-order). At any finite temperature the delicate balance between the two internal energies is broken by the different entropy of the two solutions (the metal has a smaller entropy) and the transition becomes of first-order. As a matter of fact, the fully symmetric case has the same physics of the single-band Hubbard model.

Turning back to zero temperature we can now add a symmetry-lowering field 
$\Delta$ to the Hamiltonian 
\be
\delta\mathcal{H} = -\Delta\sum_i\,\mathcal{M}_i = -\Delta\sum_i\,\sum_{a,b=1}^N\,
c^\dagger_{ia}\,M_{ab}\,c^\dagga_{ib},\label{sec-8.1:delta-SU(N)}
\ee
where $\mathcal{M}_i$ can be one of the $SU(N)$ generators, which splits the orbital degeneracy. To visualize the effect we can consider a two orbital model and  $\mathcal{M}_i  = n_{i1} - n_{i2}$ which manifestly makes one of the two orbitals energetically more convenient. 
If we aim at a realistic description of solids, we should also include interaction terms that lower the $SU(N)$ symmetry, specifically the Coulomb exchange splitting responsible of the Hund's rules, whose interplay with the single-particle field $\Delta$ can lead to a variety of phenomena including both cooperation and competition between the two symmetry-breaking terms.~\cite {Werner2007,Werner2009,de'Medici2011a,de'Medici2011b,Georges2013,de'Medici2014} When $n=1$ or $n=N-1$ the role of the Hund's interaction is indeed much more limited and it can be neglected. For the sake of simplicity, we limit the present discussion to such simpler situations and we simply consider the fully symmetric form of the Coulomb interaction.

\begin{figure}[h]
\centerline{\includegraphics[width=10cm]{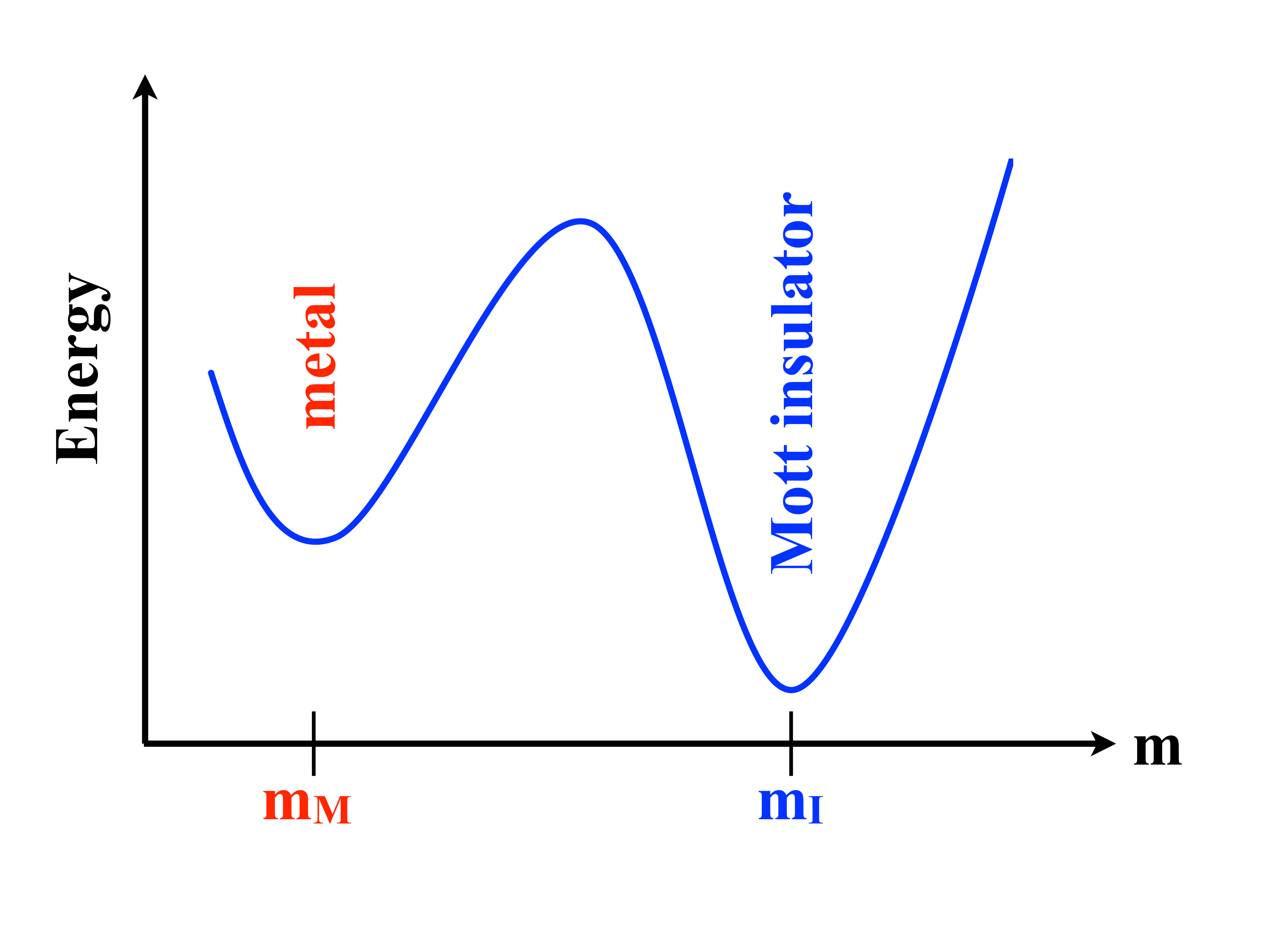}}
\caption{Sketch of the total energy as a function of the orbital polarization $m= \langle \mathcal{M}_i\rangle$ 
on the insulating side of the coexistence region, $U_c<U<U_{c2}$. The insulator, characterized by a larger $\mathcal{M}$ is the absolute energy minimum, while the metal has a lower orbital polarization and higher energy.}
\label{fig-8.1-4}
\end{figure}

When  $\Delta$ is finite the energetic balance between the two solutions in the coexistence region is not obvious and we can safely discard the possibility in which the energies of the two solutions coincide exactly at $U_{c2}$, which remains a marginal case, realized only in the fully symmetric model at zero temperature. 
We hence expect that for any fine $\Delta> 0$, even much smaller than $U$ and the bandwidth $W$, 
there will be a genuine first order phase transition at some $U_c$  within the coexistence region 
$U_{c1}< U_c  <U_{c2}$. Both the strongly correlated metal near the transition and the Mott insulator are expected to have a very large susceptibility to the symmetry-lowering field  
\eqn{sec-8.1:delta-SU(N)}, even if the insulator is expected to display a larger response because of the localized character of the carriers. That implies that the site-independent average $m= \langle \mathcal{M}_i\rangle$, which we shall denote as "orbital polarization", is greater in the insulator than in the metal. Therefore, we expect that on the insulating side of the coexistence region, i.e. for $U_c<U<U_{c2}$, 
even though the ground state is a Mott insulator characterized a larger value $\langle \mathcal{M}_i\rangle = m_\text{I}$, a metastable metal phase with higher energy and $\langle \mathcal{M}_i\rangle = m_\text{M} < m_\text{I}$ exists, as schematically shown in Fig.~\ref{fig-8.1-4}.

The above physical scenario has been explicitly uncovered  in a two-orbital Hubbard model, which corresponds to 
$N=4$ including spin degeneracy at quarter filling $n=1$ (one electron per site), in the presence of a crystal field,~\cite{Poteryaev2008,Sandri2013a}. This simple model has also been proposed to capture the essential physics  of  V$_2$O$_3$. The Hamiltonian reads 
\bea
\mathcal{H} &=& \sum_{a=1}^2\sum_{\bk\sigma}\, 
\epsilon_{\bk}\, c^\dagger_{a\bk\sigma}c^\dagga_{a\bk\sigma} 
+ \sum_{\bk\sigma}\,\gamma_\bk\, \big(c^\dagger_{1\bk\sigma}c^\dagga_{2\bk\sigma}+H.c.\big)
\nonumber\\
&& + \sum_i\bigg[
-\Delta\,\big(n_{1i}-n_{2i}\big) + \frac{U}{2}\,\big(n_{1i}+n_{2i}-1\big)^2\bigg],\label{sec-8.1:Ham2}
\eea
where $a=1,2$ labels the two orbitals, $U$ parametrizes the on-site repulsion and 
$\Delta>0$ the crystal field splitting. We include a small inter-orbital hopping 
$\gamma_\bk $, with a symmetry such that the 
local single-particle density matrix remains diagonal in the orbital indices 1 and 2. This choice assures that the occupation of each orbital is not a conserved quantity and yet that 
both orbitals are irreducible representations of the crystal field symmetry. 

When $\Delta$ is much smaller than the bandwidth $W$ and $U=0$, the Hamiltonian \eqn{sec-8.1:Ham2} describes a metal with two overlapping partially occupied bands. Even though the two orbitals are
hybridized, the lower energy band has mainly orbital character "1" while the upper energy one mostly orbital character "2". Therefore we shall refer to them as band "1" and "2", respectively. 

A finite $U$ brings two main effects. One is the standard reduction of the  coherent quasiparticle bandwidth $W\to W_* < W$ that we described in Sec. 4. In addition, $U$ generates an internal field $\Delta_\text{int}(m)$ that depends on the orbital polarization and it sums up to external field $\Delta$,  i.e. 
\be
\Delta_\text{eff}(m) = \Delta + \Delta_\text{int}(m) >  \Delta. \label{sec-8.1:effective}
\ee
In contrast with the quasiparticle bandwidth reduction, which requires to use the Gutzwiller scheme or DMFT, the Stoner-like enhancement of the crystal-field splitting  can be described also within an independent particle picture. Within Hartree-Fock one immediately finds a simple expression that can guide our interpretation
\be
\Delta_\text{int}(m) = \fract{U}{4}\,m.\label{sec-8.1:effective-HF}
\ee
As a result, when $U$ increases, the reduction of the effective bandwidth and the enhancement of the effective splitting cooperate to gradually deplete the band "2".  At some value of $Z$ the upper band is therefore completely depleted, which implies that lower band becomes half-filled.
If $\Delta$ is small enough, the full polarization leading to half-filling of the occupied band takes place for an interaction value such that the Hubbard bands are already formed and the system becomes an half-filled Mott insulator. An interesting features revealed both by DMFT~\cite{Poteryaev2008} and 
by the GA~\cite{Sandri2013a} is that the correlation effects are gradually washed out as the band "2" becomes less and less populated, so that the 
Mott-Hubbard sub-bands have mostly orbital character "1". This evolution from a two-band metal into 
a Mott insulator is schematically shown in Fig.~{fig-8.1-5}. 

\begin{figure}[th]
\centerline{\includegraphics[width=10cm]{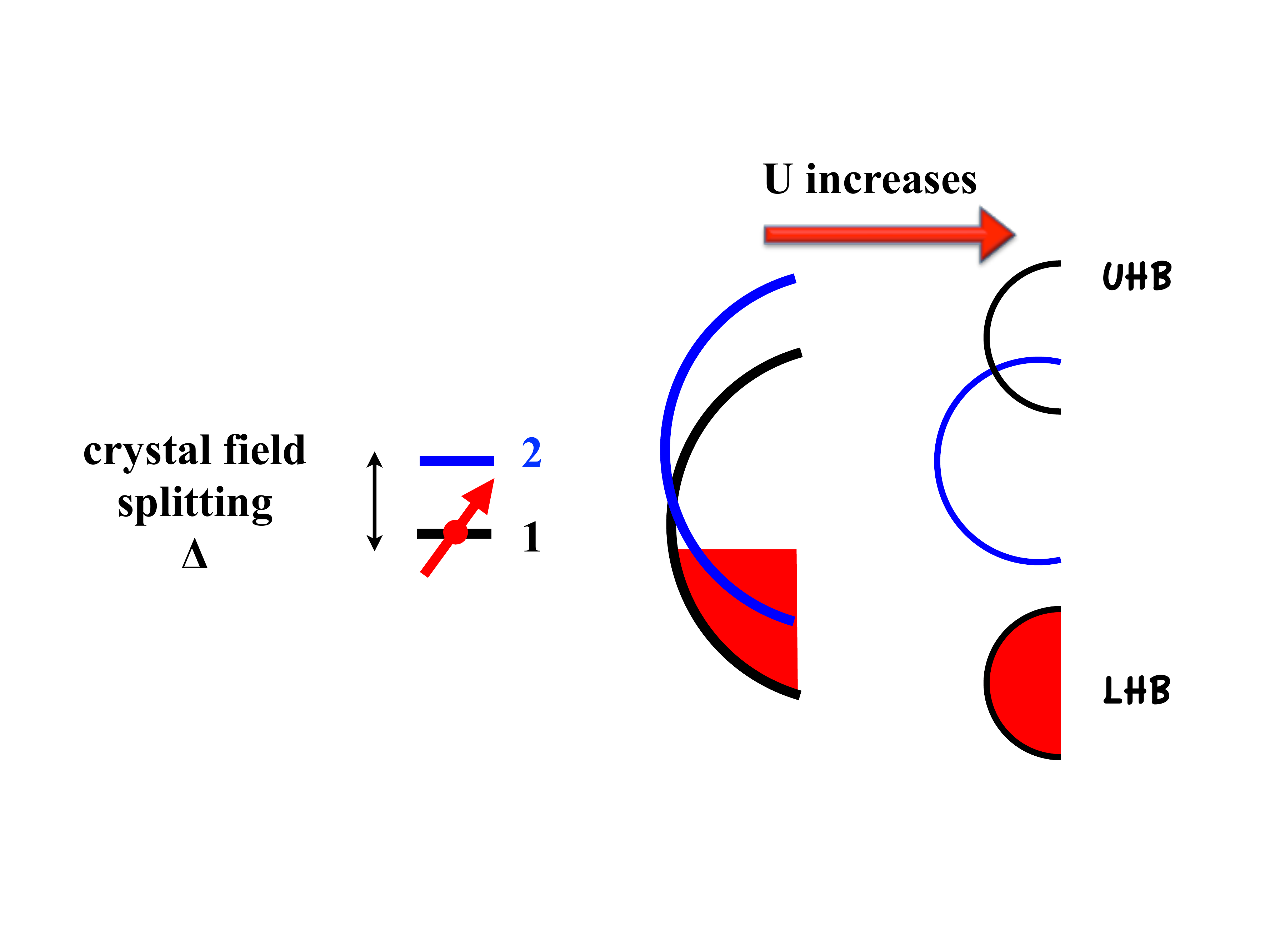}}
\caption{Schematic DOS as the repulsion $U$ increases. On the left the isolated site is shown 
with the two crystal-field split orbitals, the lower one being occupied by the available electron. 
In the center, we sketch the metal DOS when $U=0$. Finally, one the right the DOS in the Mott insulator 
is drawn, with occupied lower Hubbard band (LHB), unoccupied upper Hubbard band (UHB), both with character "1", and, in between, the empty band "2" undressed by correlations. }
\label{fig-8.1-5}
\end{figure}

Within the paramagnetic sector, i.e. preventing magnetic ordering, the physics is exactly the same we previously described in a generic $SU(N)$ model. Therefore we expect the transition from the two-band metal into the Mott insulator to be of first order, which is indeed clearly found in the Gutzwiller approximation~\cite{Sandri2013a}, while DMFT~\cite{Poteryaev2008} does not completely rule out a second order transition.
When we allow for magnetism, in the present bipartite lattice the Mott insulator is expected to order in an AFM way, since the Mott localized band "1" is half-filled. This is indeed what is found,~\cite{Sandri2013a} with a clear evidence, both by DMFT and GA, of a first-order transition from the two-band metal to the antiferromagnetic Mott insulator. The temperature $T$ versus $U$ phase diagram 
obtained in Ref.~\cite{Sandri2013a} by the GA is shown in Fig.~{fig-8.1-6}, and agree well 
with DMFT, also worked out in the same reference.  It is remarkable that, in constrast with the single-band Hubbard model, the first order 
line that separates paramagnetic metal from paramagnetic Mott insulator at finite temperature is not completely covered by the antiferromagnetic dome and a first-order Mott transition is indeed present in the actual solution of the model. Indeed, such a phase diagram closely resembles the experimental data on V$_2$O$_3$ confirming that the present model contains the basic physics of this prototypical material.

\begin{figure}[bh]
\centerline{\includegraphics[width=14cm]{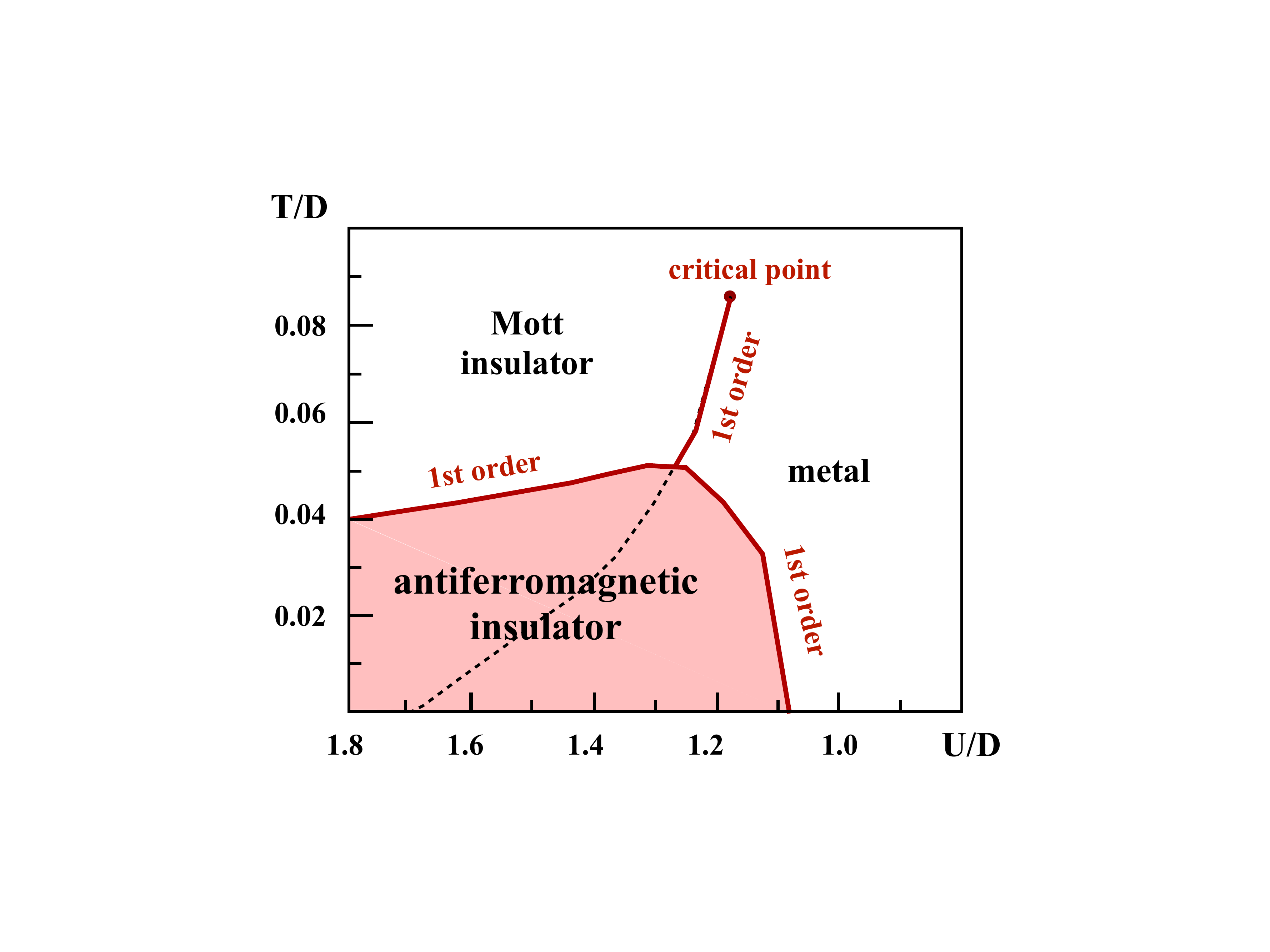}}
\caption{Phase diagram of the model Eq.~\eqn{sec-8.1:Ham2} in the $U$-$T$ plane at $\Delta=0.025W = 0.05 D$ taken from the 
data of  Ref.~\cite{Sandri2013a} and obtained  within the GA.}
\label{fig-8.1-6}
\end{figure}

\begin{figure}[th]
\centerline{\includegraphics[width=10cm]{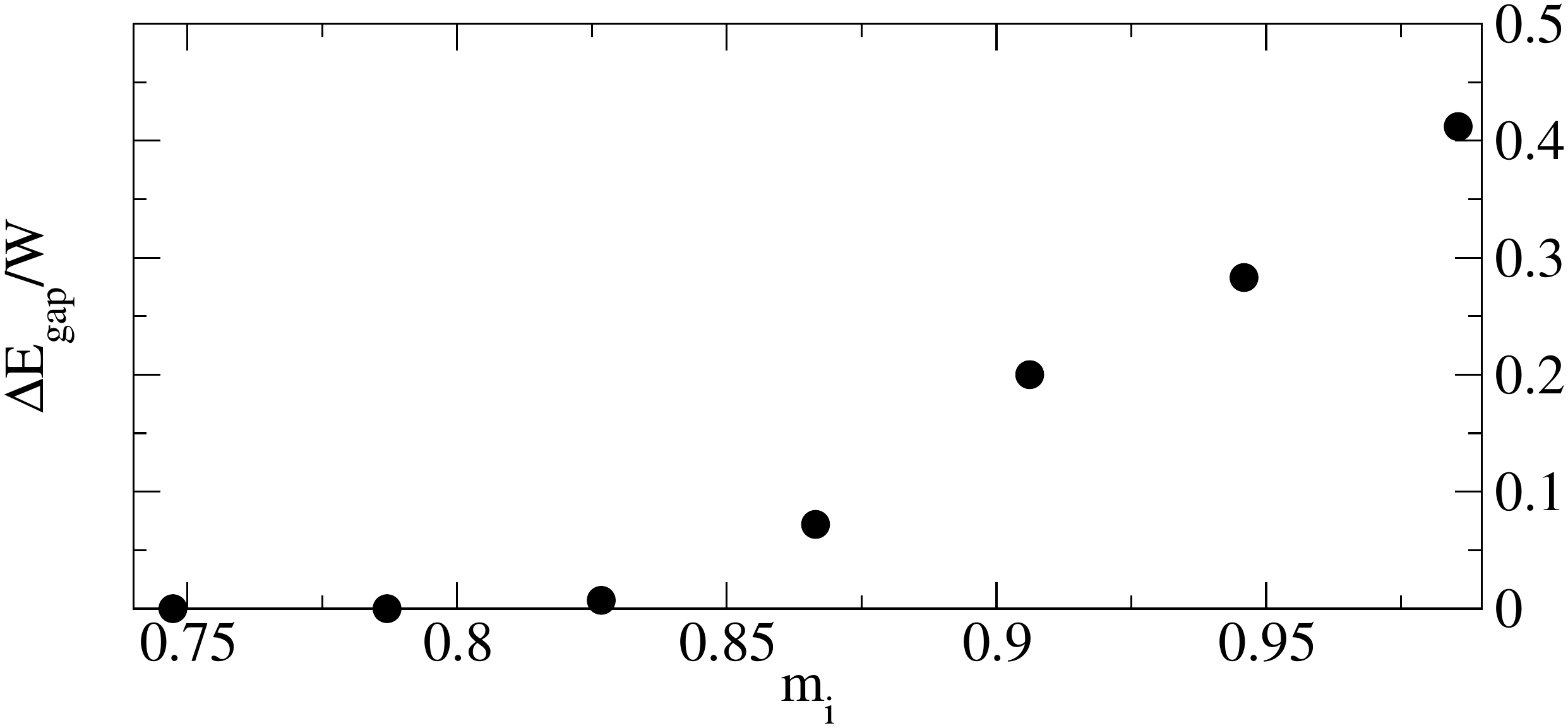}}
\caption{Long time value of the gap 
$\Delta E_\text{GAP}$ between band "1" and "2" as function of the non-equilibrium 
orbital polarization $m_i$ that the system acquires after an infinitely short laser pulse.
[From Ref.~ ~\cite{Sandri2014}.]}
\label{fig-8.1-7}
\end{figure}

The physical scenario exemplified by Fig.~\ref{fig-8.1-4} and realized in the model Eq.~\eqn{sec-8.1:Ham2}, 
assumed to mimic vanadium sesquioxide, suggests in fact an appealing non-equilibrium pathway to drive metallic a Mott insulator. Let us assume that the on-site single-particle excitation from orbital "1" to "2" is optically allowed. It follows that a short laser pulse might excite a certain fraction $\Delta n$ 
of electrons from orbital "1" to "2", thus lowering the orbital polarization by $\Delta m = - 2\Delta n$, where 
$m=n_1-n_2$ in this case.  
When $\left|\Delta m\right|$ is large enough, the system may cross the potential barrier between the two minima and fall into the  attraction range  of 
the relative metallic minimum, see Fig.~\ref{fig-8.1-4}, hence stay trapped in the metastable metal phase 
where the gap $\Delta E_\text{GAP}$ between the valence band "1" and the conduction band "2" has collapsed. This scenario clearly emerges from the time-dependent Gutzwiller approximation in Ref.~\cite{Sandri2014}. In Fig.~\ref{fig-8.1-7} we show the long-time value of the gap 
$\Delta E_\text{GAP}$ that is reached as function of the non-equilibrium 
orbital polarization $m_i$ that the system acquires after the laser pulse, assumed here to last infinitely short time, smaller then the equilibrium value $m_\text{eq}\simeq 1$. 
If we want to compare these results with experiments, we have to consider that the data of 
Fig.~\ref{fig-8.1-7} refer to a simulation where translational symmetry is enforced. In this case, since the 
metallic solution is anyway a minimum of the energy, the system can in principle remain trapped
 in the metastable phase for a very long or even infinite time. 
 In a real solid, however, the coexistence of solution will not lead to a competition between alternative homogenous states, but it will rather give rise to the nucleation of droplets of the metastable phase which would eventually evaporate restoring the Mott insulator, with a dynamics which is expected to be slower and slower as the system approaches the Mott transition. 
Such a non-equilibrium gap collapse driven by an ultrashort laser pulse has been actually observed in VO$_2$~\cite{Wegkamp2014} and in V$_2$O$_3$~\cite{Lantz2014}.

\subsection{Electron-phonon interaction beyond the two-temperature model}

As discussed in Sec. \ref{sec_ephresults}, pump-probe techniques have been widely used to investigate the dynamics of electron-boson coupling in many different materials. The analysis of time-domain data is usually based on the assumption that the ultrafast relaxation dynamics is regulated by the same electron-boson matrix elements that determine the electronic self-energy at equilibrium (see Chap. \ref{sec_QPdynamics}). Furthermore, the quantitative determination of the electron-boson interaction strength often relies on the use of simplified models (see Sec. \ref{sec_ETM}) that do not include quantum coherence and correlation effects. The experimental efforts thus naturally triggered the development of microscopic models of the electron-phonon interaction out of equilibrium aimed at benchmarking the validity of the assumptions behind the standard analysis of ultrafast experiments. 

A rather general result arising from microscopic models is that, when the electron-boson coupling is relatively small, the relaxation dynamics is largely controlled by the equilibrium physics, and the relaxation times characterizing the time evolution turn out to be dictated by the equilibrium scattering rates, computed from the low-frequency limit of the imaginary part of the self-energy. This observation is crucial to use out-of-equilibrium spectroscopy to extract intrinsic properties of the materials from the real-time dynamics without assumptions of specific modeling.

Most of microscopic studies are based on the simplest model for electron-phonon interaction, namely the Holstein model
\begin{equation}
\label{eq_Holstein}
\hat{H}=g\sum_{i}n_i(a_i + a^{\dag}_{i}) + \omega_0 \sum_i a^{\dag}_{i} a_{i},
\end{equation}
where $n_i = \sum_{\sigma} c^{\dag}_{i\sigma}c_{i\sigma}$ is the total fermionic occupation of site $i$, $a_i$ and $a^{\dag}_{i}$ are the annihilation and creation operator for a local bosonic mode on site $i$. The phonons are dispersionless Einstein modes and they are coupled with the total charge. The dimensionless coupling $\lambda$ can be defined as $\lambda = g^2/(2\omega_0t_0)$.
The groundstate of the model generally undergoes a crossover from a metal weakly coupled to phonons to a polaronic regime, where the electrons are almost localized by the strong coupling with the lattice. 
The crossover occurs for $\lambda \sim$1, essentially for every electron density, when the phonon frequency is smaller than the hopping (adiabatic regime). A different criterion, i.e., $g/\omega_0 >1$, is required for large phonon frequencies (antiadiabatic regime) \cite{Ciuchi1997,Capone1997,Bonca1999}. In the absence of Hubbard repulsion polarons bind into bipolaronic pairs which may eventually lead to an bipolaronic insulating state \cite{Capone2003p,Capone2006p}.

Using exact diagonalization Golez et al. \cite{Golez2012,Golez2012b} have studied the effect of an impulsive excitation in a one-dimensional Holstein model and otained  the scattering time as a function of the model parameters finding different regimes. For very weak coupling ($\lambda > > 0.1 t_0/\omega_0$) $1/\tau$ is linear in the coupling in agreement with the Fermi golden rule, while  it becomes sublinear for $\lambda > 0.1 t_0/\omega_0$. As mentioned above, the actual relaxation time is closely connected with the imaginary part of the self-energy, i.e. with the equilibrium scattering rate. In Ref. \citenum{Dorfner2015} a dissipative behavior is recovered for a single fermion coupled with phonons in one dimension in the limit of weak coupling and small phonon frequency.

Using the simple Migdal approximation for which the self-energy can be easily computed \cite{Sentef2013,Kemper2013,Kemper2014} it has been shown that the relaxation dynamics can largely be understood in terms of equilibrium physics, confirming that, at least within this approximation, the equilibrium scattering rate determines the actual relaxation dynamics when the system is driven out of equilibrium. The results have been applied to the understanding of different time-resolved spectroscopies and led the authors to propose a time-resolved Compton scattering to characterize the unoccupied states of materials \cite{Kemper2013}.

The relaxation of electron-phonon systems has been explored recently by Murakami et al. \cite{Murakami2015} using Dynamical Mean-Field Theory and describing the excitation process as a simple quantum quench, where the electron-phonon interaction is suddenly switched on from zero to a finite value. Also this study has shown that different dynamics take place even within the weak-coupling regime, when the system is still far from polaronic localization and one might expect that dissipative effects prevail over intrinsic interactions. For very weak coupling a fast damping of the oscillations associated to phonon excitation is observed, while the electron dynamics is much slower. In the second regime of coupling the electron dynamics becomes faster than the phonon damping, leading to a sort of thermalization. The behavior is explained in terms of a different dependence on the coupling of the phononic and electronic self-energies.

The strong-coupling limit has been instead explored by Sayyad and Eckstein \cite{Sayyad2015} by means of the exact solution of the single-polaron problem which generalized the equilibrium result of Ref. \citenum{Ciuchi1997}. The authors focused on the adiabatic strong coupling regime (small phonon frequency, large coupling) where they observed a regime of coexistence between excited polarons and more itinerant states. Interestingly, the phonon cloud seems to be described by a Fermi golden rule even in this strong coupling regime. 

In the presence of strong Hubbard-like repulsion the scenario for electron-phonon assisted relaxation can change according to the parameters range. Indeed equilibrium studies already suggested that strong correlations can strongly affect the signature of electron-phonon interaction. In Ref. \citenum{Sangiovanni2005} it has been shown that for $U$ larger than the phononic scales and the kinetic energy, phonon effects are strong only for high-energy processes (i.e. on short timescales), while the electron-phonon interaction can be described as an effective non-retarded interaction at low-energy (long timescales). This is particularly strong at half filling, where Mott localization takes place, but it also affects strongly correlated metallic states \cite{Sangiovanni2006}.

In Ref. \citenum{Yonemitsu2009} the transfer of energy from electrons to a phononic environment has been discussed solving the Schr\"odinger equation for many-electron wavefunctions coupled with different kind of vibrations. Werner and Eckstein have shown that electron-phonon coupling can effectively dissipate the excess of doubly excited states following a quench of the Coulomb interaction \cite{Werner2013final}.
An electric-field driven effective enhancement of the electron-phonon coupling in Mott insulators subject to a DC field with potential implications for photoemission spectra has been also observed \cite{Werner2015final}.

\section{Experimental perspectives}
Here we have reviewed the most recent experimental results and theoretical models for studying the non-equilibrium electronic, optical, structural and magnetic properties of correlated materials in the sub-picosecond time domain. Although the main focus was on the prototypical correlated oxides exhibiting the superconducting and other exotic phases at low temperature,
other topical materials, such as iron-based and organic superconductors, MgB$_2$, charge-transfer insulators and many others have been mentioned and discussed to some extent. 
The main conclusion of this extended review is quite clear for it shows that the gate toward novel experiments, concepts and theories is unlocked. As a matter of fact the impressive amount of techniques and experimental data that need to be coherently understood calls for the development of new theoretical concepts and treatises. This is particularly pressing for a major revolution has started recently when the first free-electron laser sources have been used for studying the non-equilibrium electronic, magnetic and structural properties of the matter.  Today it is quite clear that the possibility of generating ultra-bright and ultra-short light pulses, extending from the VUV to the hard X-rays region, requires to expand beyond the present paradigm the numerical and analytic techniques for microscopically treating the non-equilibrium phenomena. From the present review is undoubtedly emerging that the ultrafast optical spectroscopy of correlated materials and, more in general, the non-equilibrium physics in condensed matter represents a challenging new frontier allowing to picture the transient changes of electronic states and lattice structures, along with the dynamics of single-particle excitations and collective phenomena. However, even more challenging is the future possibility to extending to the X-ray region the coherent and quantum optics studies, at the present possible only for the spectral regions available by the conventional laser sources. This is quite clear now, since externally seeded FEL like FERMI have shown the possibility of generating fully coherent EUV, soft X-ray pulses with a time structure of few tens of fs.

From the time of the exploratory pump-probe experiments, the time-resolved techniques have revealed in the past thirty years a great potentiality for probing correlated electron systems, proving that the ultrafast experiments can provide unique information about the gap(s) dynamics, the associated quasi-particle relaxation and the electron-boson interactions. These findings are opening the road to new more sophisticated spectral- and momentum-resolved techniques. Non-equilibrium 
experiments, as shown here, also led to novel and exotic phenomenological theoretical descriptions of the elementary phenomena and  to analytical expressions for evaluating the relaxation times, the amplitude of the optical response, and the laser fluence dependences. These experiments have been particularly critical for studying the non-equilibrium properties in cuprates and for triggering significant theoretical efforts to determine the electron-boson coupling from time resolved experiments without assuming effective temperature models. 

Although the main body of this review is focused to study the non-equilibrium properties of strongly correlated materials in the limit of small perturbations, we also bring to the attention of the reader the fact that different excitations can explore the limits where light pulses can perturb the overall physical properties of materials unlocking the gate for the ultra-fast control of macroscopic properties of the material. The fundamental idea is to induce controlled photo-excitations within time windows shorter than the characteristic times of the relaxation processes. This will bring the matter into highly off-equilibrium transient regimes. These regimes are found to have an anomalous energy distribution between electrons, ions, and spins, possibly resulting in fast changes of the material properties. In other words the photo-excitation can lead to a highly non-thermal distribution of energy among the different degrees of freedom and such a transient non-equilibrium state, eventually enabling the ultrafast light-based control of material properties.

Finally, the recent advances in the production of increasingly short pulses in the sub-10 fs/attosecond range \cite{Cerullo2003,Brida2010,Sansone2011} are expected to open the route to a completely unexplored physics, in which the temporal resolution is shorter than the dephasing time of the excited many-body wavefunctions. In this limit, multi-pulse combinations can be used to implement multi-dimensional spectroscopic techniques \cite{Cundiff2013} in order to study the fundamental optical dephasing processes in correlated materials and optically manipulate the macroscopic polarization by creating quantum superposition of many body wavefunctions. At the same time, the possibility of performing photon-number statistics \cite{Esposito2015} in pump-probe experiments could unlock the gate to novel approaches to track the fluctuations of the atomic positions or of other degrees of freedom that can couple with the electro-magnetic field.

\section{Acknowledgements}
We would like to acknowledge the many useful discussions with A. Amaricci, Avella, R. Averitt, F. Banfi, L. Benfatto, L. Boeri, J. Bon\v{c}a, U. Bovensiepen, F. Carbone, A. Cavalleri, G. Cerullo, F. Cilento, R. Comin, G. Coslovich, S. Dal Conte, A. Damascelli, L. de' Medici, J. Demsar, M. Eckstein, G. Ferrini, J. Freericks, N. Gedik, M. Grilli, V. Kabanov, R. Kaindl, B. Keimer, A. Kemper, G. Kotliar, A. Lanzara, M. Le Tacon, A. Leitenstorfer, Z. Lenar\v{c}i\v{c}, J. Lorenzana, D. Manske, M. Marsi, T. Mertelj, M. Mierzejewski, J. Orenstein, L. Perfetti, F. Peronaci, P. Prelov\v{s}ek, M. Sandri, G. Sangiovanni, G. Sawatzky, D. Scalapino, M. Schir\`o, A.J. Taylor, A. Toschi, L. Vidmar, C. Weber, P. Werner. 
Part of the research activities which the review is based on received funding from the European Union, Seventh Framework Programme (FP7 2007-2013), under Grant No. 280555 (GO FAST). FP and DF acknowledge the FERMI project at Elettra Sincrotrone Trieste S.c.p.A., partially supported by the Ministry of University and Research (Grants No. FIRBRBAP045JF2 and No. FIRBRBAP06AWK3).

\bibliographystyle{tADP} 
\bibliography{biblio,localbib,localbib2,localbib3,localbib4,phonons}

\begin{thebibliography}{637}
\providecommand{\natexlab}[1]{#1}

\bibitem[1]{Fork1981}
R.L. Fork, B.I. Greene, and C.V. Shank, {\itshape Generation of optical pulses
  shorter than 0.1 psec by colliding pulse mode locking}, Applied Physics
  Letters 38 (1981), pp. 671--672.

\bibitem[2]{Valdmanis1986}
J. Valdmanis and R. Fork, {\itshape Design considerations for a femtosecond
  pulse laser balancing self phase modulation, group velocity dispersion,
  saturable absorption, and saturable gain}, Quantum Electronics, IEEE Journal
  of 22 (1986), pp. 112--118.

\bibitem[3]{Fork1987}
R.L. Fork, C.H.B. Cruz, P.C. Becker, and C.V. Shank, {\itshape Compression of
  optical pulses to six femtoseconds by using cubic phase compensation}, Opt.
  Lett. 12 (1987), pp. 483--485.

\bibitem[4]{Spence1991}
D. Spence, P. Kean, and W. Sibbett, {\itshape 60-fsec pulse generation from a
  self-mode-locked Ti:sapphire laser}, Opt. Lett. 16 (1991), pp. 42--44.

\bibitem[5]{Merlin1997}
R. Merlin, {\itshape Generating coherent THz phonons with light pulses}, Solid
  State Communications 102 (1997), pp. 207--220 Highlights in Condensed Matter
  Physics and Materials Science.

\bibitem[6]{Stevens:2002p6798}
T.E. Stevens, J. Kuhl, and R. Merlin, {\itshape {Coherent phonon generation and
  the two stimulated Raman tensors}}, Physical Reviw B 65 (2002), pp. 1--4.

\bibitem[7]{Matsuda1994}
K. Matsuda, I. Hirabayashi, K. Kawamoto, T. Nabatame, T. Tokizaki, and A.
  Nakamura, {\itshape Femtosecond spectroscopic studies of the ultrafast
  relaxation process in the charge-transfer state of insulating cuprates},
  Phys. Rev. B 50 (1994), pp. 4097--4101.

\bibitem[8]{Albrecht1992}
W. Albrecht, T. Kruse, and H. Kurz, {\itshape Time-resolved observation of
  coherent phonons in superconducting YBa$_2$Cu$_3$O$_{7-\delta}$ thin films},
  Phys. Rev. Lett. 69 (1992), pp. 1451--1454.

\bibitem[9]{Ginder1988}
J.M. Ginder, M.G. Roe, Y. Song, R.P. McCall, J.R. Gaines, E. Ehrenfreund, and
  A.J. Epstein, {\itshape Photoexcitations in La$_2$CuO$_4$: 2-eV energy gap
  and long-lived defect states}, Phys. Rev. B 37 (1988), pp. 7506--7509.

\bibitem[10]{Donaldson1989}
W.R. Donaldson, a.M. Kadin, P.H. Ballentine, and R. Sobolewski, {\itshape
  {Interaction of picosecond optical pulses with high Tc superconducting
  films}}, Applied Physics Letters 54 (1989), p. 2470.

\bibitem[11]{Brocklesby1989}
W.S. Brocklesby, D. Monroe, a.F.J. Levi, M. Hong, S.H. Liou, J. Kwo, C.E. Rice,
  P.M. Mankiewich, and R.E. Howard, {\itshape {Electrical response of
  superconducting YBa2Cu3O7−$\delta$ to light}}, Applied Physics Letters 54
  (1989), p. 1175.

\bibitem[12]{Frenkel1989}
A. Frenkel, M.A. Saifi, T. Venkatesan, C. Lin, X.D. Wu, and a. Inam, {\itshape
  {Observation of fast nonbolometric optical response of nongranular high-$T_c$
  Y$_1$Ba$_2$Cu$_3$O$_{7-x}$ superconducting thin films}}, Applied Physics
  Letters 54 (1989), p. 1594.

\bibitem[13]{Chekalin1991}
S.V. Chekalin, V.M. Farztdinov, V.V. Golovlyov, V.S. Letokhov, Y.E. Lozovik,
  Y.A. Matveets, and A.G. Stepanov, {\itshape Femtosecond spectroscopy of
  YBa$_2$Cu$_3$O$_{7-\delta}$: Electron-phonon-interaction measurement and
  energy-gap observation}, Phys. Rev. Lett. 67 (1991), pp. 3860--3863.

\bibitem[14]{Reitze1992}
D.H. Reitze, A.M. Weiner, A. Inam, and S. Etemad, {\itshape Fermi-level
  dependence of femtosecond response in nonequilibrium high-$T_c$
  superconductors}, Phys. Rev. B 46 (1992), pp. 14309--14312.

\bibitem[15]{Semenov1992}
 {Semenov, A. D. and Gol'tsman, G. N. and Gogidze, I. G. and Sergeev, A. V. and
  Gershenzon, E. M. and Lang, P. T. and Renk, K. F.}, {\itshape {Subnanosecond
  photoresponse of a YBaCuO thin film to infrared and visible radiation by
  quasiparticle induced suppression of superconductivity}}, Applied Physics
  Letters 60 (1992), p. 903.

\bibitem[16]{Yu1992}
G. Yu, C.H. Lee, A.J. Heeger, N. Herron, E.M. McCarron, L. Cong, G.C. Spalding,
  C.A. Nordman, and A.M. Goldman, {\itshape Phase separation of photogenerated
  carriers and photoinduced superconductivity in high-$T_c$ materials}, Phys.
  Rev. B 45 (1992), pp. 4964--4977.

\bibitem[17]{Gong1993}
T. Gong, L.X. Zheng, W. Xiong, W. Kula, Y. Kostoulas, R. Sobolewski, and P.M.
  Fauchet, {\itshape Femtosecond optical response of Y-Ba-Cu-O thin films: The
  dependence on optical frequency, excitation intensity, and electric current},
  Phys. Rev. B 47 (1993), pp. 14495--14502.

\bibitem[18]{Lindgren1994}
M. Lindgren, V. Trifonov, M. Zorin, M. Danerud, D. Winkler, B.S. Karasik, G.N.
  Gol'tsman, and E.M. Gershenzon, {\itshape {Transient resistive
  photoresponse of YBa2Cu3O7−$\delta$ films using low power 0.8 and 10.6
  $\mu$m laser radiation}}, Applied Physics Letters 64 (1994), p. 3036.

\bibitem[19]{sarukura1989}
N. Sarukura, Y. Ishida, H. Nakano, and Y. Yamamoto, {\itshape cw passive mode
  locking of a Ti:sapphire laser}, Applied Physics Letters 56 (1990).

\bibitem[20]{Strickland1985}
D. Strickland and G. Mourou, {\itshape Compression of amplified chirped optical
  pulses}, Optics Communications 56 (1985), pp. 219 -- 221.

\bibitem[21]{Bartels2007}
A. Bartels, R. Cerna, C. Kistner, A. Thoma, F. Hudert, C. Janke, and T.
  Dekorsy, {\itshape Ultrafast time-domain spectroscopy based on high-speed
  asynchronous optical sampling}, Review of Scientific Instruments 78 (2007),
  035107.

\bibitem[22]{Stoica2008}
V. Stoica, Y.M. Sheu, D. Reis, and R. Clarke, {\itshape {Wideband detection of
  transient solid-state dynamics using ultrafast fiber lasers and asynchronous
  optical sampling}}, Optics express 16 (2008), pp. 2322--35.

\bibitem[23]{Yoneda2004}
H. Yoneda, H. Morikami, K. Ueda, and R.M. More, {\itshape Ultrashort-Pulse
  Laser Ellipsometric Pump-Probe Experiments on Gold Targets}, Phys. Rev. Lett.
  91 (2003), p. 075004.

\bibitem[24]{Guidoni2002}
L. Guidoni, E. Beaurepaire, and J.Y. Bigot, {\itshape Magneto-optics in the
  Ultrafast Regime: Thermalization of Spin Populations in Ferromagnetic Films},
  Phys. Rev. Lett. 89 (2002), p. 017401.

\bibitem[25]{Beaurepaire1996}
E. Beaurepaire, J.C. Merle, A. Daunois, and J.Y. Bigot, {\itshape Ultrafast
  Spin Dynamics in Ferromagnetic Nickel}, Phys. Rev. Lett. 76 (1996), pp.
  4250--4253.

\bibitem[26]{Stevens1997}
C.J. Stevens, D. Smith, C. Chen, and J.F. Ryan, {\itshape {Evidence for
  Two-Component High-Temperature Superconductivity in the Femtosecond Optical
  Response of YBa$_2$Cu$_3$O$_{7-\delta}$}}, Phys. Rev. Lett. 78 (1997), pp.
  2212--2215.

\bibitem[27]{Demsar1999}
J. Demsar, B. Podobnik, V.V. Kabanov, T. Wolf, and D. Mihailovic, {\itshape
  {Superconducting gap $\Delta_c$, the pseudogap $\Delta_p$, and pair
  fluctuations above $T_c$ in overdoped
  Y$_{1-x}$Ca$_x$Ba$_2$Cu$_3$O$_{7-\delta}$ from femtosecond time-domain
  spectroscopy}}, Physical Review Letters 82 (1999), pp. 4918--4921.

\bibitem[28]{Misochko2002}
O. Misochko, N. Georgiev, T. Dekorsy, and M. Helm, {\itshape {Two Crossovers in
  the Pseudogap Regime of YBa$_2$Cu$_3$O$_{7-\delta}$ Superconductors Observed
  by Ultrafast Spectroscopy}}, Physical Review Letters 89 (2002), p. 067002.

\bibitem[29]{Segre:2002p9832}
G. Segre, N. Gedik, J. Orenstein, D. Bonn, R. Liang, and W.N. Hardy, {\itshape
  {Photoinduced Changes of Reflectivity in Single Crystals of
  YBa$_{2}$Cu$_{3}$O$_{6.5}$(Ortho II)}}, Physical Review Letters 88 (2002), p.
  137001.

\bibitem[30]{Xu2003}
Y. Xu, M. Khafizov, L. Satrapinsky, P. K\'{u}\v{s}, a. Plecenik, and R.
  Sobolewski, {\itshape {Time-Resolved Photoexcitation of the Superconducting
  Two-Gap State in MgB$_2$ Thin Films}}, Physical Review Letters 91 (2003), p.
  197004.

\bibitem[31]{Boyd}
R. Boyd {\itshape {Nonlinear Optics}},    Academic Press, 2003.

\bibitem[32]{Bloembergen}
N. Bloembergen {\itshape {Nonlinear Optics}},    World Scientific, 1996.

\bibitem[33]{Cerullo2003}
G. Cerullo and S. {De Silvestri}, {\itshape {Ultrafast optical parametric
  amplifiers}}, Review of Scientific Instruments 74 (2003), pp. 1--18.

\bibitem[34]{Brida2010}
D. Brida, C. Manzoni, G. Cirmi, M. Marangoni, S. Bonora, P. Villoresi, S.
  De~Silvestri, and G. Cerullo, {\itshape {Few-optical-cycle pulses tunable
  from the visible to the mid-infrared by optical parametric amplifiers}},
  Journal of Optics 12 (2010), p. 013001.

\bibitem[35]{Yusupov2010}
R. Yusupov, T. Mertelj, V.V. Kabanov, S. Brazovskii, P. Kusar, J.h. Chu, I.R.
  Fisher, and D. Mihailovic, {\itshape {Coherent dynamics of macroscopic
  electronic order through a symmetry breaking transition}}, Nature Physics 6
  (2010), pp. 681--684.

\bibitem[36]{Bechtel1975}
J.H. Bechtel, {\itshape Heating of solid targets with laser pulses}, Journal of
  Applied Physics 46 (1975), pp. 1585--1593.

\bibitem[37]{Mertelj:2009p6019}
T. Mertelj, V. Kabanov, and C. Gadermaier, {\itshape {Distinct Pseudogap and
  Quasiparticle Relaxation Dynamics in the Superconducting State of Nearly
  Optimally Doped SmFeAsO$_{0.8}$F$_{0.2}$ Single Crystals}}, Physical Review
  Letters 102 (2009), p. 117002.

\bibitem[38]{Leonard2007}
J. L\'{e}onard, N. Lecong, J.P. Likforman, O. Cr\'{e}gut, S. Haacke, P. Viale,
  P. Leproux, and V. Couderc, {\itshape Broadband ultrafast spectroscopy usinga
  photonic crystal fiber: application tothe photophysics of malachite green},
  Opt. Express 15 (2007), pp. 16124--16129.

\bibitem[39]{Dudley2006}
J.M. Dudley, G. Genty, and S. Coen, {\itshape Supercontinuum generation in
  photonic crystal fiber}, Rev. Mod. Phys. 78 (2006), pp. 1135--1184.

\bibitem[40]{Wegkamp2011}
D. Wegkamp, D. Brida, S. Bonora, G. Cerullo, J. Stähler, M. Wolf, and S. Wall,
  {\itshape Phase retrieval and compression of low-power white-light pulses},
  Applied Physics Letters 99 (2011), 101101, pp.~--.

\bibitem[41]{Cilento2010}
F. Cilento, C. Giannetti, G. Ferrini, S. {Dal Conte}, T. Sala, G. Coslovich, M.
  Rini, A. Cavalleri, F. Parmigiani, and S. {Dal Conte}, {\itshape {Ultrafast
  insulator-to-metal phase transition as a switch to measure the spectrogram of
  a supercontinuum light pulse}}, Applied Physics Letters 96 (2010), p. 021102.

\bibitem[42]{Novelli2012}
F. Novelli, D. Fausti, J. Reul, F. Cilento, P.H.M. Loosdrechtvan , A.A.
  Nugroho, T.T.M. Palstra, M. Gr\"{u}ninger, and F. Parmigiani, {\itshape
  {Ultrafast optical spectroscopy of the lowest energy excitations in the Mott
  insulator compound YVO$_{3}$: Evidence for Hubbard-type excitons}}, Physical
  Review B 86 (2012), p. 165135.

\bibitem[43]{Giannetti2011}
C. Giannetti et~al., {\itshape {Revealing the high-energy electronic
  excitations underlying the onset of high-temperature superconductivity in
  cuprates.}}, Nat. Commun. 2 (2011), p. 353.

\bibitem[44]{Novelli2014}
F. Novelli et~al., {\itshape {Witnessing the formation and relaxation of
  dressed quasi-particles in a strongly correlated electron system}}, Nat.
  Commun. 5 (2014), p. 5112.

\bibitem[45]{Kaindl2002}
R.A. Kaindl, M.A. Carnahan, J. Orenstein, D.S. Chemla, H.M. Christen, H.Y.
  Zhai, M. Paranthaman, and D.H. Lowndes, {\itshape Far-Infrared Optical
  Conductivity Gap in Superconducting MgB$_2$ Films}, Phys. Rev. Lett. 88
  (2001), p. 027003.

\bibitem[46]{Hangyo2005}
M. Hangyo, M. Tani, and T. Nagashima, {\itshape Terahertz Time-Domain
  Spectroscopy of Solids: A Review}, International Journal of Infrared and
  Millimeter Waves 26 (2005), pp. 1661--1690.

\bibitem[47]{Nahata1996}
A. Nahata, A.S. Weling, and T.F. Heinz, {\itshape A wideband coherent terahertz
  spectroscopy system using optical rectification and electro-optic sampling},
  Applied Physics Letters 69 (1996), pp. 2321--2323.

\bibitem[48]{Shi2002}
W. Shi, Y.J. Ding, N. Fernelius, and K. Vodopyanov, {\itshape Efficient,
  tunable, and coherent 0.18--5.27-THz source based on GaSe crystal}, Opt.
  Lett. 27 (2002), pp. 1454--1456.

\bibitem[49]{Hebling2008}
J. Hebling, K.L. Yeh, M.C. Hoffmann, B. Bartal, and K.A. Nelson, {\itshape
  Generation of high-power terahertz pulses by tilted-pulse-front excitation
  and their application possibilities}, J. Opt. Soc. Am. B 25 (2008), pp.
  B6--B19.

\bibitem[50]{Schneider2006}
A. Schneider, M. Neis, M. Stillhart, B. Ruiz, R.U.A. Khan, and P. G\"{u}nter,
  {\itshape Generation of terahertz pulses through optical rectification in
  organic DAST crystals: theory and experiment}, J. Opt. Soc. Am. B 23 (2006),
  pp. 1822--1835.

\bibitem[51]{Huber2000}
R. Huber, A. Brodschelm, F. Tauser, and A. Leitenstorfer, {\itshape Generation
  and field-resolved detection of femtosecond electromagnetic pulses tunable up
  to 41 THz}, Applied Physics Letters 76 (2000), pp. 3191--3193.

\bibitem[52]{Huber2001}
R. Huber, F. Tauser, A. Brodschelm, M. Bichler, G. Abstreiter, and A.
  Leitenstorfer, {\itshape {How many-particle interactions develop after
  ultrafast excitation of an electron-hole plasma}}, Nature 414 (2001), pp.
  286--289.

\bibitem[53]{Huber2005}
R. Huber, C. K\"ubler, S. T\"ubel, A. Leitenstorfer, Q.T. Vu, H. Haug, F.
  K\"ohler, and M.C. Amann, {\itshape Femtosecond Formation of Coupled
  Phonon-Plasmon Modes in InP: Ultrabroadband THz Experiment and Quantum
  Kinetic Theory}, Phys. Rev. Lett. 94 (2005), p. 027401.

\bibitem[54]{Sell2008}
A. Sell, S. R\"udiger, A. Leitenstorfer, and R. Huber, {\itshape
  {Field-resolved detection of phase-locked infrared transients from a compact
  Er:fiber system tunable between 55 and 107 THz}}, Applied Physics Letters 93
  (2008), 251107.

\bibitem[55]{Xi_Cheng2009}
X.C. Zhang and J. Xu {\itshape {Introduction to THz Wave Photonics}},
  Springer, 2009.

\bibitem[56]{Kaindl2006}
R.A. Kaindl, R. Huber, B.A. Schmid, M.A. Carnahan, D. H\"agele, and D.S.
  Chemla, {\itshape Ultrafast THz spectroscopy of correlated electrons: from
  excitons to Cooper pairs}, Physica Status Solidi (b) 243 (2006), pp.
  2414--2422.

\bibitem[57]{Perfetti:2006p1775}
L. Perfetti, P. Loukakos, M. Lisowski, U. Bovensiepen, H. Berger, S. Biermann,
  P. Cornaglia, A. Georges, and M. Wolf, {\itshape {Time Evolution of the
  Electronic Structure of 1T-TaS$_{2}$ through the Insulator-Metal
  Transition}}, Physical Review Letters 97 (2006), p. 67402.

\bibitem[58]{Perfetti2007}
L. Perfetti, P.A. Loukakos, M. Lisowski, U. Bovensiepen, H. Eisaki, and M.
  Wolf, {\itshape {Ultrafast Electron Relaxation in Superconducting
  Bi$_2$Sr$_2$CaCu$_2$O$_{8+\delta}$ by Time-Resolved Photoelectron
  Spectroscopy}}, Physical Review Letters 99 (2007), p. 197001.

\bibitem[59]{Smallwood2012_b}
C.L. Smallwood, C. Jozwiak, W. Zhang, and A. Lanzara, {\itshape An ultrafast
  angle-resolved photoemission apparatus for measuring complex materials},
  Review of Scientific Instruments 83 (2012), p. 123904.

\bibitem[60]{Ishizaka:2011dg}
K. Ishizaka et~al., {\itshape {Femtosecond core-level photoemision spectroscopy
  on 1T-TaS$_{2}$ using a 60 eV laser source}}, Physical Review B 83 (2011).

\bibitem[61]{Murnane1991}
M. Murnane, H. Kapteyn, M. Rosen, and R. Falcone, {\itshape Ultrafast X-ray
  Pulses from Laser-Produced Plasmas}, Science 251 (1991), pp. 531--536.

\bibitem[62]{Siders1999}
C.W. Siders, A. Cavalleri, K. Sokolowski-Tinten, C. Toth, T. Guo, M. Kammler,
  M.v. Hoegen, K.R. Wilson, D. {von der Linde}, and C.P.J. Barty, {\itshape
  Detection of Nonthermal Melting by Ultrafast X-ray Diffraction}, Science 286
  (1999), pp. 1340--1342.

\bibitem[63]{Schoenlein2000}
R.W. Schoenlein, S. Chattopadhyay, H.H.W. Chong, T.E. Glover, P.A. Heimann,
  C.V. Shank, A.A. Zholents, and M.S. Zolotorev, {\itshape Generation of
  Femtosecond Pulses of Synchrotron Radiation}, Science 287 (2000), pp.
  2237--2240.

\bibitem[64]{Zholents1996}
A.A. Zholents and M.S. Zolotorev, {\itshape Femtosecond X-Ray Pulses of
  Synchrotron Radiation}, Phys. Rev. Lett. 76 (1996), pp. 912--915.

\bibitem[65]{Lindenberg2000}
A.M. Lindenberg et~al., {\itshape Time-Resolved X-Ray Diffraction from Coherent
  Phonons during a Laser-Induced Phase Transition}, Phys. Rev. Lett. 84 (2000),
  pp. 111--114.

\bibitem[66]{Cavalleri2001}
A. Cavalleri, C. T\'oth, C.W. Siders, J.A. Squier, F. R\'aksi, P. Forget, and
  J.C. Kieffer, {\itshape {Femtosecond Structural Dynamics in VO$_2$ during an
  Ultrafast Solid-Solid Phase Transition}}, Phys. Rev. Lett. 87 (2001), p.
  237401.

\bibitem[67]{Larsson1997}
J. Larsson et~al., {\itshape Ultrafast x-ray diffraction using a streak-camera
  detector in averaging mode}, Opt. Lett. 22 (1997), pp. 1012--1014.

\bibitem[68]{Johnson2008}
S.L. Johnson, P. Beaud, C.J. Milne, F.S. Krasniqi, E.S. Zijlstra, M.E. Garcia,
  M. Kaiser, D. Grolimund, R. Abela, and G. Ingold, {\itshape Nanoscale
  Depth-Resolved Coherent Femtosecond Motion in Laser-Excited Bismuth}, Phys.
  Rev. Lett. 100 (2008), p. 155501.

\bibitem[69]{Beaud2007}
P. Beaud, S.L. Johnson, A. Streun, R. Abela, D. Abramsohn, D. Grolimund, F.
  Krasniqi, T. Schmidt, V. Schlott, and G. Ingold, {\itshape Spatiotemporal
  Stability of a Femtosecond Hard\char21{}X-Ray Undulator Source Studied by
  Control of Coherent Optical Phonons}, Phys. Rev. Lett. 99 (2007), p. 174801.

\bibitem[70]{dejong2013}
S. Jongde~ et~al., {\itshape {Speed limit of the insulator-metal transition in
  magnetite}}, Nat Mater 12 (2013), pp. 882--886.

\bibitem[71]{Beye2012}
M. Beye et~al., {\itshape X-ray pulse preserving single-shot optical
  cross-correlation method for improved experimental temporal resolution},
  Applied Physics Letters 100 (2012), 121108.

\bibitem[72]{Mankowsky2014}
R. Mankowsky et~al., {\itshape Nonlinear lattice dynamics as a basis for
  enhanced superconductivity in YBa$_2$Cu$_3$O$_{6.5}$}, Nature 516 (2014), pp.
  71--73.

\bibitem[73]{Chuang2013}
Y.D. Chuang et~al., {\itshape Real-Time Manifestation of Strongly Coupled Spin
  and Charge Order Parameters in Stripe-Ordered La$_{1.75}$Sr$_{0.25}$NiO$_4$
  Nickelate Crystals Using Time-Resolved Resonant X-Ray Diffraction}, Phys.
  Rev. Lett. 110 (2013), p. 127404.

\bibitem[74]{Kung2013}
Y.F. Kung, W.S. Lee, C.C. Chen, A.F. Kemper, A.P. Sorini, B. Moritz, and T.P.
  Devereaux, {\itshape Time-dependent charge-order and spin-order recovery in
  striped systems}, Phys. Rev. B 88 (2013), p. 125114.

\bibitem[75]{Forst2015}
M. Forst et~al., {\itshape {Spatially resolved ultrafast magnetic dynamics
  initiated at a complex oxide heterointerface}}, Nat. Mater. 14 (2015), pp.
  883--888.

\bibitem[76]{Forst2014}
M. F\"orst et~al., {\itshape {Melting of Charge Stripes in Vibrationally Driven
  La$_{1.875}$Ba$_{0.125}$CuO$_4$: Assessing the Respective Roles of Electronic
  and Lattice Order in Frustrated Superconductors}}, Phys. Rev. Lett. 112
  (2014), p. 157002.

\bibitem[77]{Forst2014b}
---{}---{}---, {\itshape Femtosecond x rays link melting of charge-density wave
  correlations and light-enhanced coherent transport in
  YBa$_2$Cu$_3$O$_{6.6}$}, Phys. Rev. B 90 (2014), p. 184514.

\bibitem[78]{Yurtsever2011}
A. Yurtsever and A. Zewail, {\itshape {Kikuchi ultrafast nanodiffraction in
  four-dimensional electron microscopy}}, Proc. Natl. Acad. Sci. USA 108
  (2011), pp. 3152--3156.

\bibitem[79]{Pfeifer2006}
T. Pfeifer, C. Spielmann, and G. Gerber, {\itshape Femtosecond x-ray science},
  Reports on Progress in Physics 69 (2006), p. 443.

\bibitem[80]{Hada2013}
M. Hada, K. Pichugin, and G. Sciaini, {\itshape Ultrafast structural dynamics
  with table top femtosecond hard X-ray and electron diffraction setups}, The
  European Physical Journal Special Topics 222 (2013), pp. 1093--1123.

\bibitem[81]{wang2009}
J. Wang and U. Rochesterof~ {\itshape Ultrafast Electronic and Structural
  Dynamics in Solids Using Femtosecond Laser Techniques},    University of
  Rochester, 2009.

\bibitem[82]{Zewail:2010hn}
A.H. Zewail, {\itshape {Four-Dimensional Electron Microscopy}}, Science 328
  (2010), pp. 187--193.

\bibitem[83]{Sciaini2011}
G. Sciaini and R. Miller, {\itshape Femtosecond electron diffraction: heralding
  the era of atomically resolved dynamics}, Reports on Progress in Physics 74
  (2011), p. 096101.

\bibitem[84]{Ackermann2007}
 {Ackermann W.} et~al., {\itshape {Operation of a free-electron laser from the
  extreme ultraviolet to the water window}}, Nat. Photon. 1 (2007), pp.
  336--342.

\bibitem[85]{Emma2010}
 {Emma P.} et~al., {\itshape {First lasing and operation of an
  angstrom-wavelength free-electron laser}}, Nat. Photon. 4 (2010), pp.
  641--647.

\bibitem[86]{Ishikawa2012}
T. Ishikawa et~al., {\itshape {A compact X-ray free-electron laser emitting in
  the sub-angstrom region}}, Nat. Photon. 6 (2012), pp. 540--544.

\bibitem[87]{Allaria2012}
 {Allaria E.} et~al., {\itshape {Highly coherent and stable pulses from the
  FERMI seeded free-electron laser in the extreme ultraviolet}}, Nat. Photon. 6
  (2012), pp. 699--704.

\bibitem[88]{Kaufmann1998}
H.J. Kaufmann, E.G. Maksimov, and E.K.H. Salje, {\itshape {Electron-Phonon
  Interaction and Optical Spectra of Metals}},  11 (1998), pp. 755--768.

\bibitem[89]{Carbotte1990}
J.P. Carbotte and C. Lmf, {\itshape {Properties of boson-exchange}},   (1990).

\bibitem[90]{Rowell1965}
W. McMillan and J. Rowell, {\itshape {Lead phonon spectrum calculated from
  superconducting density of states}}, Physical Review Letters 14 (1965), pp.
  108--112.

\bibitem[91]{Brockhouse1962}
B. Brockhouse, T. Arase, G. Caglioti, K. Rao, and A. Woods, Physical Review 128
  (1962), p. 1099.

\bibitem[92]{McMillan1968}
W. McMillan, {\itshape {Transition Temperature of Strong-Coupled
  Superconductors}}, Physical Review 167 (1968), p. 331.

\bibitem[93]{Eliashberg1960}
 {Eliashberg, G.M.}, Sov. Phys.-JETP 11 (1960), p. 696.

\bibitem[94]{Eliashberg1961}
---{}---{}---, Sov. Phys.-JETP 12 (1961), p. 1000.

\bibitem[95]{Dynes1972}
R. Dynes, {\itshape {McMillan's equation and the $T_c$ of superconductors}},
  Solid State Communications 10 (1972), pp. 615--618.

\bibitem[96]{Allen1975}
P. Allen and R. Dynes, {\itshape Transition temperature of strong-coupled
  superconductors reanalyzed}, Physical Review B 12 (1975), p. 905.

\bibitem[97]{Tomlinson1976}
P. Tomlinson and J. Carbotte, {\itshape {Anisotropic superconducting energy gap
  in Pb}}, Physical Review B 13 (1976), p. 4738.

\bibitem[98]{Kong2001}
Y. Kong, O. Dolgov, O. Jepsen, and O. Andersen, {\itshape {Electron-phonon
  interaction in the normal and superconducting states of MgB2}}, Physical
  Review B 64 (2001), p. 020501.

\bibitem[99]{Nagamatsu2001}
J. Nagamatsu, N. Nakagawa, T. Muranaka, Y. Zenitani, and J. Akimitsu, {\itshape
  {Superconductivity at 39 K in magnesium diboride}}, Nature 410 (2001), pp.
  63--64.

\bibitem[100]{Hubbard1963}
J. Hubbard, {\itshape {Electron Correlations in Narrow Energy Bands}}, Proc. R.
  Soc. Lond. 276 (1963), p. 238.

\bibitem[101]{Anderson1963}
P. Anderson, {\itshape {Theory of Magnetic Exchange Interactions: Exchange in
  Insulators and Semiconductors.}}, Solid State Physics 14 (1963), p.~99.

\bibitem[102]{Lee2006}
P. Lee, N. Nagaosa, and X.G. Wen, {\itshape {Doping a Mott insulator: Physics
  of high-temperature superconductivity}}, Reviews of Modern Physics 78 (2006),
  pp. 17--85.

\bibitem[103]{Alexandrov:1994wdb}
A. Alexandrov and N. Mott, {\itshape {Bipolarons}}, Reports On Progress In
  Physics 57 (1994), pp. 1197--1288.

\bibitem[104]{Mihailovic:2001ur}
D. Mihailovic and V. Kabanov, {\itshape {Finite wave vector Jahn-Teller pairing
  and superconductivity in the cuprates}}, Physical Review B 63 (2001), p.
  054505.

\bibitem[105]{Muller:2007p3138}
K.A. M{\"u}ller, {\itshape {On the superconductivity in hole doped cuprates}},
  Journal Of Physics-Condensed Matter 19 (2007), p. 251002.

\bibitem[106]{Anderson2007}
P.W. Anderson, {\itshape {Is there glue in cuprate superconductors?}}, Science
  316 (2007), pp. 1705--7.

\bibitem[107]{Scalapino2012}
D. Scalapino, {\itshape {A common thread: The pairing interaction for
  unconventional superconductors}}, Review of Modern Physics 84 (2012), p.
  1383.

\bibitem[108]{Carbotte2011}
J.P. Carbotte, T. Timusk, and J. Hwang, {\itshape {Bosons in high-temperature
  superconductors: an experimental survey}}, Reports on Progress in Physics 74
  (2011), p. 066501.

\bibitem[109]{Damascelli2003}
A. Damascelli, Z. Hussain, and Z. Shen, {\itshape {Angle-resolved photoemission
  studies of the cuprate superconductors}}, Reviews of Modern Physics 75
  (2003).

\bibitem[110]{Schachinger2008}
E. Schachinger and J.P. Carbotte, {\itshape {Finite band inversion of
  angular-resolved photoemission in Bi$_{2}$Sr$_{2}$CaCu$_{2}$O$_{8+\delta}$
  and comparison with optics}}, Physical Review B 77 (2008), p. 094524.

\bibitem[111]{Zhang2008}
W. Zhang, G. Liu, L. Zhao, H. Liu, J. Meng, and X. Dong, {\itshape
  {Identification of a New Form of Electron Coupling in the
  Bi$_2$Sr$_2$CaCu$_2$O$_8$ Superconductor by Laser-Based Angle-Resolved
  Photoemission Spectroscopy}}, Physical Review Letters 100 (2008), p. 107002.

\bibitem[112]{Muschler2010}
B. Muschler, W. Prestel, E. Schachinger, J.P. Carbotte, R. Hackl, S. Ono, and
  Y. Ando, {\itshape {An electron-boson glue function derived from electronic
  Raman scattering.}}, Journal of physics. Condensed matter : an Institute of
  Physics journal 22 (2010), p. 375702.

\bibitem[113]{Schachinger2006}
E. Schachinger, D. Neuber, and J.P. Carbotte, {\itshape Inversion techniques
  for optical conductivity data}, Phys. Rev. B 73 (2006), p. 184507.

\bibitem[114]{Tu2002}
J.J. Tu, C.C. Homes, G.D. Gu, D.N. Basov, and M. Strongin, {\itshape Optical
  studies of charge dynamics in optimally doped
  Bi$_2$Sr$_2$CaCu$_2$O$_{8+\delta}$}, Phys. Rev. B 66 (2002), p. 144514.

\bibitem[115]{Hwang2007}
J. Hwang, T. Timusk, E. Schachinger, and J.P. Carbotte, {\itshape Evolution of
  the bosonic spectral density of the high-temperature superconductor
  Bi$_2$Sr$_2$CaCu$_2$O$_{8+\delta}$}, Phys. Rev. B 75 (2007), p. 144508.

\bibitem[116]{Hwang2014}
J. Hwang and J.P. Carbotte, {\itshape High-energy fluctuation spectra in
  cuprates from infrared optical spectroscopy}, Phys. Rev. B 89 (2014), p.
  024502.

\bibitem[117]{Pasupathy2008}
A.N. Pasupathy, A. Pushp, K.K. Gomes, C.V. Parker, J. Wen, Z. Xu, G. Gu, S.
  Ono, Y. Ando, and A. Yazdani, {\itshape Electronic Origin of the
  Inhomogeneous Pairing Interaction in the High-$T_c$ Superconductor
  Bi$_2$Sr$_2$CaCu$_2$O$_{8+\delta}$}, Science 320 (2008), pp. 196--201.

\bibitem[118]{Ahmadi2011}
O. Ahmadi, L. Coffey, J. Zasadzinski, N. Miyakawa, and L. Ozyuzer, {\itshape
  {Eliashberg Analysis of Tunneling Experiments: Support for the Pairing Glue
  Hypothesis in Cuprate Superconductors}}, Physical Review Letters 106 (2011),
  pp. 1--4.

\bibitem[119]{Grilli1994}
M. Grilli and C. Castellani, {\itshape {Electron-phonon interactions in the
  presence of strong correlations}}, Phys. Rev. B 50 (1994), pp. 16880--16898.

\bibitem[120]{Capone2010}
M. Capone, C. Castellani, and M. Grilli, {\itshape {Electron-Phonon Interaction
  in Strongly Correlated Systems}}, Advances in Condensed Matter Physics 2010
  (2010), p. 920860.

\bibitem[121]{Johnston2012}
S. Johnston, I.M. Vishik, W.S. Lee, F. Schmitt, S. Uchida, K. Fujita, S.
  Ishida, N. Nagaosa, Z.X. Shen, and T.P. Devereaux, {\itshape Evidence for the
  Importance of Extended Coulomb Interactions and Forward Scattering in Cuprate
  Superconductors}, Phys. Rev. Lett. 108 (2012), p. 166404.

\bibitem[122]{Kovaleva2004}
N.N. Kovaleva et~al., {\itshape $c$-axis lattice dynamics in Bi-based cuprate
  superconductors}, Phys. Rev. B 69 (2004), p. 054511.

\bibitem[123]{Devereaux2004}
T.P. Devereaux, T. Cuk, Z.X. Shen, and N. Nagaosa, {\itshape Anisotropic
  Electron-Phonon Interaction in the Cuprates}, Phys. Rev. Lett. 93 (2004), p.
  117004.

\bibitem[124]{Fournier2010}
D. Fournier et~al., {\itshape {Loss of nodal quasiparticle integrity in
  underdoped YBa2Cu3O6+x}}, Nature Physics 6 (2010), pp. 905--911.

\bibitem[125]{Varma2006}
C. Varma, {\itshape {Theory of the pseudogap state of the cuprates}}, Physical
  Review B 73 (2006), p. 155113.

\bibitem[126]{Sachdev}
S. Sachdev {\itshape Quantum Phase Transitions},    Cambridge University Press,
  2 edition, 2011.

\bibitem[127]{letacon2011}
M. Le~Tacon et~al., {\itshape Intense paramagnon excitations in a large family
  of high-temperature superconductors}, Nature Phys. 7 (2011), p. 725.

\bibitem[128]{Dahm2009}
T. Dahm, V. Hinkov, S.V. Borisenko, A.A. Kordyuk, V.B. Zabolotnyy, J. Fink, B.
  B\"{u}chner, D.J. Scalapino, W. Hanke, and B. Keimer, {\itshape {Strength of
  the spin-fluctuation-mediated pairing interaction in a high-temperature
  superconductor}}, Nature Physics 5 (2009), pp. 217--221.

\bibitem[129]{letacon2013}
M. Le~Tacon et~al., {\itshape Dispersive spin excitations in highly overdoped
  cuprates revealed by resonant inelastic x-ray scattering}, Phys. Rev. B 88
  (2013), p. 020501.

\bibitem[130]{dean2013}
M.P.M. Dean et~al., {\itshape {High-Energy Magnetic Excitations in the Cuprate
  Superconductor Bi$_2$Sr$_2$CaCu$_2$O$_8$: Towards a Unified Description of
  Its Electronic and Magnetic Degrees of Freedom}}, Phys. Rev. Lett. 110
  (2013), p. 147001.

\bibitem[131]{dean2013b}
 {Dean, M. P. M. and Dellea, G. and Springell, R. S. and Yakhou-Harris, F. and
  Kummer, K. and Brookes, N.B. and Liu, X. and Sun, Y-J. and Strle, J. and
  Schmitt, T. and Braicovich, L. and Ghiringhelli, G. and Bo\u{z}ovi\'c, I. and
  Hill, J. P.}, {\itshape {Persistence of magnetic excitations in
  La$_{2-x}$Sr$_x$CuO$_4$ from the undoped insulator to the heavily overdoped
  non-superconducting metal}}, Nature Materials 12 (2013), p. 1019.

\bibitem[132]{Fujita2012}
M. Fujita, H. Hiraka, M. Matsuda, M. Matsuura, J. Tranquada, S. Wakimoto, G.
  Xu, and K. Yamada, {\itshape {Progress in Neutron Scattering Studies of Spin
  Excitations in High-$T_c$ Cuprates}}, J. Phys. Soc. Jpn. 81 (2012), p.
  011007.

\bibitem[133]{Lee2007}
W. Lee, I.M. Vishik, K. Tanaka, D.H. Lu, T. Sasagawa, N. Nagaosa, Z. Devereaux
  T. P.~Hussain, and Z.X. Shen, {\itshape {Abrupt onset of a second energy gap
  at the superconducting transition of underdoped Bi2212}}, Nature 450 (2007),
  p.~81.

\bibitem[134]{Comin2014}
R. Comin et~al., {\itshape Charge Order Driven by Fermi-Arc Instability in
  Bi$_2$S$_{2-x}$La$_x$CuO$_{6+\delta}$}, Science 343 (2013), p. 391.

\bibitem[135]{Fauque2006}
B. Fauqu\'e, Y. Sidis, V. Hinkov, S. Pailh\`es, C.T. Lin, X. Chaud, and P.
  Bourges, {\itshape Magnetic Order in the Pseudogap Phase of High-$T_c$
  Superconductors}, Phys. Rev. Lett. 96 (2006), p. 197001.

\bibitem[136]{Li2008}
Y. Li, V. Bal\'{e}dent, N. Barisi\'{c}, Y. Cho, B. Fauqu\'{e}, Y. Sidis, G. Yu,
  X. Zhao, P. Bourges, and M. Greven, {\itshape {Unusual magnetic order in the
  pseudogap region of the superconductor HgBa$_2$CuO$_{4+\delta}$}}, Nature 455
  (2008), pp. 372--5.

\bibitem[137]{Li2010}
Y. Li et~al., {\itshape {Hidden magnetic excitation in the pseudogap phase of a
  high-$T_c$ superconductor}}, Nature 468 (2010), pp. 283--5.

\bibitem[138]{Tranquada1995}
J.M. Tranquada, B.J. Sternlieb, J.D. Axe, Y. Nakamura, and S. Uchida, {\itshape
  {Evidence for stripe correlations of spins and holes in copper oxide
  superconductors}}, Nature  (1995), p. 561.

\bibitem[139]{Kivelson1998}
S.A. Kivelson, E. Fradkin, and V.J. Emery, {\itshape {Electronic liquid-crystal
  phases of a doped Mott insulator}}, Nature 393 (1998), p. 550.

\bibitem[140]{Mesaros2011}
A. Mesaros, K. Fujita, H. Eisaki, S. Uchida, J.C. Davis, S. Sachdev, J. Zaanen,
  M.J. Lawler, and E.A. Kim, {\itshape {Topological Defects Coupling Smectic
  Modulations to Intra-Unit-Cell Nematicity in Cuprates}}, Science (New York,
  N.Y.) 333 (2011), p. 426.

\bibitem[141]{Hinkov2008}
V. Hinkov, D. Haug, B. Fauqu\'{e}, P. Bourges, Y. Sidis, a. Ivanov, C.
  Bernhard, C.T. Lin, and B. Keimer, {\itshape {Electronic liquid crystal state
  in the high-temperature superconductor YBa$_2$Cu$_3$O$_{6.45}$}}, Science
  (New York, N.Y.) 319 (2008), pp. 597--600.

\bibitem[142]{Ghiringhelli2012}
G. Ghiringhelli et~al., {\itshape Long-Range Incommensurate Charge Fluctuations
  in (Y,Nd)Ba$_2$Cu$_3$O$_{6+x}$}, Science 337 (2012), pp. 821--825.

\bibitem[143]{Chang2012}
J. Chang et~al., {\itshape {Direct observation of competition between
  superconductivity and charge density wave order in YBa$_2$Cu$_3$O$_{6.67}$}},
  Nature Physics 8 (2012), pp. 871--876.

\bibitem[144]{SilvaNeto2014}
E.H. {da Silva Neto} et~al., {\itshape Ubiquitous Interplay between Charge
  Ordering and High-Temperature Superconductivity in Cuprates}, Science 343
  (2013), p. 6169.

\bibitem[145]{McElroy2003}
K. McElroy, R. Simmonds, J. Hoffman, D.H. Lee, J. Orenstein, H. Eisaki, S.
  Uchida, and J. Davis, {\itshape {Relating atomic-scale electronic phenomena
  to wave-like quasiparticle states in superconducting
  Bi$_2$Sr$_2$CaCu$_2$O$_{8+\delta}$}}, Nature 422 (2003), p. 592.

\bibitem[146]{Avella2012}
A. Avella and F. Mancini {\itshape Stronlgy Correlated Systems. Theoretical
  Methods},  Springer Series in Solid-State Sciences  Springer-Verlag Berlin
  Heidelberg, 2012.

\bibitem[147]{Avella2013}
A. Avella and F. Mancini {\itshape Stronlgy Correlated Systems. Numerical
  Methods},  Springer Series in Solid-State Sciences  Springer-Verlag Berlin
  Heidelberg, 2013.

\bibitem[148]{Gutzwiller1965}
M.C. Gutzwiller, {\itshape Correlation of Electrons in a Narrow $s$ Band},
  Phys. Rev. 137 (1965), pp. A1726--A1735.

\bibitem[149]{Georges1996}
A. Georges, P. Cedex, W. Krauth, and M.J. Rozenberg, {\itshape {Dynamical
  mean-field theory of strongly correlated fermion systems and the limit of
  infinite dimensions}}, Rev. Mod. Phys. 68 (1996).

\bibitem[150]{Brinkman1970}
W.F. Brinkman and T.M. Rice, {\itshape Application of Gutzwiller's Variational
  Method to the Metal-Insulator Transition}, Phys. Rev. B 2 (1970), pp.
  4302--4304.

\bibitem[151]{Bunemann2013}
J. B\"unemann, M. Capone, J. Lorenzana, and G. Seibold, {\itshape
  Linear-response dynamics from the time-dependent Gutzwiller approximation},
  New Journal of Physics 15 (2013), p. 053050.

\bibitem[152]{Bunemann1998}
J. B\"unemann, W. Weber, and F. Gebhard, {\itshape {Multiband Gutzwiller wave
  functions for general on-site interactions}}, Phys. Rev. B 57 (1998), p.
  6896.

\bibitem[153]{Barone2007}
P. Barone, R. Raimondi, M. Capone, C. Castellani, and M. Fabrizio, {\itshape
  Extended Gutzwiller wave function for the Hubbard-Holstein model}, EPL
  (Europhysics Letters) 79 (2007), p. 47003.

\bibitem[154]{Barone2008}
 {Barone, P. and Raimondi, R. and Capone, M. and Castellani, C. and Fabrizio,
  M.}, {\itshape Gutzwiller scheme for electrons and phonons: The half-filled
  Hubbard-Holstein model}, Phys. Rev. B 77 (2008), p. 235115.

\bibitem[155]{Borghi2009}
G. Borghi, M. Fabrizio, and E. Tosatti, {\itshape {Surface Dead Layer for
  Quasiparticles Near a {Mott} Transition}}, Phys. Rev. Lett. 102 (2009), p.
  066806.

\bibitem[156]{Borghi2010}
---{}---{}---, {\itshape {Strongly correlated metal interfaces in the
  Gutzwiller approximation}}, Phys. Rev. B 81 (2010), p. 115134.

\bibitem[157]{Ho2008}
K.M. Ho, J. Schmalian, and C.Z. Wang, {\itshape Gutzwiller density functional
  theory for correlated electron systems}, Phys. Rev. B 77 (2008), p. 073101.

\bibitem[158]{Deng2009}
X. Deng, L. Wang, X. Dai, and Z. Fang, {\itshape Local density approximation
  combined with Gutzwiller method for correlated electron systems: Formalism
  and applications}, Phys. Rev. B 79 (2009), p. 075114.

\bibitem[159]{Schickling2012}
T. Schickling, F. Gebhard, J. B\"unemann, L. Boeri, O.K. Andersen, and W.
  Weber, {\itshape Gutzwiller Theory of Band Magnetism in LaOFeAs}, Phys. Rev.
  Lett. 108 (2012), p. 036406.

\bibitem[160]{Lanata2015}
N. Lanat\`a, Y. Yao, C.Z. Wang, K.M. Ho, and G. Kotliar, {\itshape Phase
  Diagram and Electronic Structure of Praseodymium and Plutonium}, Phys. Rev. X
  5 (2015), p. 011008.

\bibitem[161]{Anisimov1991}
V.I. Anisimov, J. Zaanen, and O.K. Andersen, {\itshape Band theory and Mott
  insulators: Hubbard \textit{U} instead of Stoner \textit{I}}, Phys. Rev. B 44
  (1991), pp. 943--954.

\bibitem[162]{Kotliar2006}
G. Kotliar, S.Y. Savrasov, K. Haule, V.S. Oudovenko, O. Parcollet, and C.A.
  Marianetti, {\itshape Electronic structure calculations with dynamical
  mean-field theory}, Rev. Mod. Phys. 78 (2006), pp. 865--951.

\bibitem[163]{Allen1987}
P. Allen, {\itshape Theory of Thermal Relaxation of Electrons in Metals},
  Physical Review Letters 59 (1987), p. 1460.

\bibitem[164]{Singh1998}
P. Singh and S. Barrie, {\itshape Magnon Specific Heat in Quasi-Two-Dimensional
  Antiferromagnetic Materials: Application to YBa$_2$Cu$_3$O$_6$ and
  La$_2$CuO$_4$ Oxide Superconductors}, Phys. Status Solidi B 205 (1998), p.
  611.

\bibitem[165]{Brorson1990}
S.D. Brorson, A. Kazeroonian, J.S. Moodera, D.W. Face, T.K. Cheng, E.P. Ippen,
  M.S. Dresselhaus, and G. Dresselhaus, {\itshape Femtosecond room-temperature
  measurement of the electron-phonon coupling constant in metallic
  superconductors}, Phys. Rev. Lett. 64 (1990), pp. 2172--2175.

\bibitem[166]{Sun1994}
C.K. Sun, F. Vall\'ee, L.H. Acioli, E.P. Ippen, and J.G. Fujimoto, {\itshape
  Femtosecond-tunable measurement of electron thermalization in gold}, Phys.
  Rev. B 50 (1994), pp. 15337--15348.

\bibitem[167]{Groeneveld1995}
R.H.M. Groeneveld, R. Sprik, and A. Lagendijk, {\itshape Femtosecond
  spectroscopy of electron-electron and electron-phonon energy relaxation in Ag
  and Au}, Phys. Rev. B 51 (1995), pp. 11433--11445.

\bibitem[168]{Carpene2006}
E. Carpene, {\itshape Ultrafast laser irradiation of metals: Beyond the
  two-temperature model}, Phys. Rev. B 74 (2006), p. 024301.

\bibitem[169]{DellaValle2012}
G. {Della Valle}, M. Conforti, S. Longhi, G. Cerullo, and D. Brida, {\itshape
  {Real-time optical mapping of the dynamics of nonthermal electrons in thin
  gold films}}, Physical Review B 86 (2012), p. 155139.

\bibitem[170]{Kabanov2008}
V.V. Kabanov and A.S. Alexandrov, {\itshape Electron relaxation in metals:
  Theory and exact analytical solutions}, Phys. Rev. B 78 (2008), p. 174514.

\bibitem[171]{Gadermaier2010}
C. Gadermaier, a. Alexandrov, V. Kabanov, P. Kusar, T. Mertelj, X. Yao, C.
  Manzoni, D. Brida, G. Cerullo, and D. Mihailovic, {\itshape {Electron-Phonon
  Coupling in High-Temperature Cuprate Superconductors Determined from Electron
  Relaxation Rates}}, Physical Review Letters 105 (2010), pp. 1--4.

\bibitem[172]{Smallwood2012}
C.L. Smallwood, J.P. Hinton, C. Jozwiak, W. Zhang, J.D. Koralek, H. Eisaki,
  D.H. Lee, J. Orenstein, and A. Lanzara, {\itshape Tracking Cooper Pairs in a
  Cuprate Superconductor by Ultrafast Angle-Resolved Photoemission}, Science
  336 (2012), pp. 1137--1139.

\bibitem[173]{Cortes:2011df}
R. Cortes, L. Rettig, Y. Yoshida, H. Eisaki, M. Wolf, and U. Bovensiepen,
  {\itshape {Momentum-Resolved Ultrafast Electron Dynamics in Superconducting
  Bi$_2$Sr$_2$CaCu$_2$O$_{8+\delta}$}}, Physical Review Letters 107 (2011), p.
  097002.

\bibitem[174]{Graf:2011jw}
J. Graf, C. Jozwiak, C.L. Smallwood, H. Eisaki, R.A. Kaindl, D.H. Lee, and A.
  Lanzara, {\itshape {Nodal quasiparticle meltdown in ultrahigh-resolution
  pump--probe angle-resolved photoemission}}, Nature Physics 7 (2011), pp.
  805--809.

\bibitem[175]{Toda:2011cs}
Y. Toda, T. Mertelj, P. Kusar, T. Kurosawa, M. Oda, M. Ido, and D. Mihailovic,
  {\itshape {Quasiparticle relaxation dynamics in underdoped
  Bi$_2$Sr$_2$CaCu$_2$O$_{8+\delta}$ by two-color pump-probe spectroscopy}},
  Physical Review B 84 (2011), p. 174516.

\bibitem[176]{Gadermaier:2012vz}
 C. Gadermaier at al. \textit{High superconducting critical temperatures depend
  universally on the electron-phonon interaction strength}. arXiv:1205.4978v1
  (2012).

\bibitem[177]{DalConte2012}
S. Dal~Conte et~al., {\itshape Disentangling the Electronic and Phononic Glue
  in a High-$T_c$ Superconductor}, Science 335 (2012), pp. 1600--1603.

\bibitem[178]{Mertelj:2005p2}
T. Mertelj, V. Kabanov, and D. Mihailovic, {\itshape {Charged particles on a
  two-dimensional lattice subject to anisotropic Jahn-Teller interactions}},
  94 (2005), p. 147003.

\bibitem[179]{Mertelj:2007hn}
T. Mertelj, V. Kabanov, J.M. Mena, and D. Mihailovic, {\itshape
  {Self-organization of charged particles on a two-dimensional lattice subject
  to anisotropic Jahn-Teller-type interaction and three-dimensional Coulomb
  repulsion}}, Physical Review B 76 (2007), p.~9.

\bibitem[180]{Yusupov:2008p5698}
R.V. Yusupov, T. Mertelj, J.H. Chu, I.R. Fisher, and D. Mihailovic, {\itshape
  {Single-Particle and Collective Mode Couplings Associated with 1-and
  2-Directional Electronic Ordering in Metallic RTe$_3$ (R=Ho,Dy,Tb)}},
  Physical Review Letters 101 (2008), p. 246402.

\bibitem[181]{Rothwarf:1967p9182}
A. Rothwarf and B. Taylor, {\itshape {Measurement of recombination lifetimes in
  superconductors}}, Physical Review Letters 19 (1967), p.~27.

\bibitem[182]{Mihailovic:2005p57}
D. Mihailovic, {\itshape Optical Experimental Evidence for a Universal Length
  Scale for the Dynamic Charge Inhomogeneity of Cuprate Superconductors}, Phys.
  Rev. Lett. 94 (2005), p. 207001.

\bibitem[183]{SaiHalasz:1974dq}
G.A. Sai-Halasz, C. Chi, A. Denenstein, and D. Langenberg, {\itshape {Effects
  of Dynamic External Pair Breaking in Superconducting Films}}, Physiscal
  Review Letters 33 (1974), pp. 215--219.

\bibitem[184]{Kabanov1999}
V.V. Kabanov, J. Demsar, B. Podobnik, and D. Mihailovic, {\itshape
  {Quasiparticle relaxation dynamics in superconductors with different gap
  structures: Theory and experiments on YBa$_2$Cu$_3$O$_{7-\delta}$}}, Physical
  Review B 59 (1999), pp. 1497--1506.

\bibitem[185]{Kabanov2005}
V.V. Kabanov, J. Demsar, and D. Mihailovic, {\itshape {Kinetics of a
  Superconductor Excited with a Femtosecond Optical Pulse}}, Physical Review
  Letters 95 (2005), p. 147002.

\bibitem[186]{Brorson:1990p6245}
S. Brorson, A. Kazeroonian, J. Moodera, D. Face, T. Cheng, E. Ippen, M.
  Dresselhaus, and G. Dresselhaus, {\itshape {Femtosecond room-temperature
  measurement of the electron-phonon coupling constant gamma in metallic
  superconductors}}, Physical Review Letters 64 (1990), pp. 2172--2175.

\bibitem[187]{CHWALEK:1991p1951}
J. Chwalek, C. Uher, J. Whitaker, G. Mourou, and J. Agostinelli, {\itshape
  {Subpicosecond time-resolved studies of coherent phonon oscillations in
  thin-film YBa$_2$Cu$_3$O$_{6+x}$ ($x<0.4$)}}, Applied Physics Letters 58
  (1991), pp. 980--982.

\bibitem[188]{Demsar:1999p112}
J. Demsar, B. Podobnik, V. Kabanov, T. Wolf, and D. Mihailovic, {\itshape
  {Superconducting gap $\Delta_c$, the pseudogap $\Delta_p$, and pair
  fluctuations above $T_c$ in overdoped
  Y$_{1-x}$Ca$_x$Ba$_2$Cu$_3$O$_{7-\delta}$ from femtosecond time-domain
  spectroscopy}}, Physical Review Letters 82 (1999), pp. 4918--4921.

\bibitem[189]{Demsar:2001p1518}
J. Demsar, R. Hudej, J. Karpinski, V. Kabanov, and D. Mihailovic, {\itshape
  {Quasiparticle dynamics and gap structure in
  HgBa$_2$Ca$_2$Cu$_3$O$_{8+\delta}$ investigated with femtosecond
  spectroscopy}}, Physical Review B 63 (2001), p. 054519.

\bibitem[190]{Gay1999}
P. Gay, D.C. Smith, C.J. Stevens, C. Chen, and J.F. Ryan, {\itshape
  {Femtosecond Dynamics of BSCCO-2212}}, J. Low Temp. Phys. 117 (1999), pp.
  1025--1029.

\bibitem[191]{Gedik2005}
N. Gedik, M. Langner, J. Orenstein, S. Ono, Y. Abe, and Y. Ando, {\itshape
  {Abrupt Transition in Quasiparticle Dynamics at Optimal Doping in a Cuprate
  Superconductor System}}, Physical Review Letters 95 (2005), p. 117005.

\bibitem[192]{HAN:1990p1246}
S. Han, Z. Vardeny, K. Wong, O. Symko, and G. Koren, {\itshape {Femtosecond
  optical-detection of quasi-particle dynamics in high $T_c$
  YBa$_2$Cu$_3$O$_{7-\delta}$ superconducting thin films}}, Physical Review
  Letters 65 (1990), pp. 2708--2711.

\bibitem[193]{Kaindl2005gp}
R.A. Kaindl, M.A. Carnahan, D.S. Chemla, S. Oh, and J.N. Eckstein, {\itshape
  {Dynamics of Cooper pair formation in
  Bi$_{2}$Sr$_{2}$CaCu$_{2}$O$_{8+\delta}$}}, Phys. Rev. B 72 (2005), p.
  060510.

\bibitem[194]{Kusar:2008p5434}
P. Kusar, V. Kabanov, J. Demsar, and T. Mertelj, {\itshape {Controlled
  Vaporization of the Superconducting Condensate in Cuprate Superconductors by
  Femtosecond Photoexcitation}}, Physical Review Letters 101 (2008), p. 227001.

\bibitem[195]{Schneider:2004p2385}
M. Schneider, M. Onellion, X. Xi, X. Zeng, A. Soukiassian, P. Omernik, and G.
  Taft, {\itshape {Electron dynamics in metallic and spin-glass cuprates}},
  Physical Review B 70 (2004), p. 012504.

\bibitem[196]{Giannetti2009b}
C. Giannetti, G. Coslovich, F. Cilento, G. Ferrini, H. Eisaki, N. Kaneko, M.
  Greven, and F. Parmigiani, {\itshape {Discontinuity of the ultrafast
  electronic response of underdoped superconducting
  Bi$_2$Sr$_2$CaCu$_2$O$_{8+\delta}$ strongly excited by ultrashort light
  pulses}}, Physical Review B 79 (2009), p. 224502.

\bibitem[197]{Coslovich2011}
G. Coslovich et~al., {\itshape Evidence for a photoinduced nonthermal
  superconducting-to-normal-state phase transition in overdoped
  Bi$_2$Sr$_2$Ca$_{0.92}$Y$_{0.08}$Cu$_{2}$O$_{8+\delta}$}, Phys. Rev. B 83
  (2011), p. 064519.

\bibitem[198]{Coslovich2013}
---{}---{}---, {\itshape {Ultrafast charge localization in a stripe-phase
  nickelate}}, Nat Commun 4 (2013), p. 2643.

\bibitem[199]{Liu:1993p4053}
Y. Liu, J. Whitaker, C. Uher, J. Peng, Z. Li, and R. GReene, {\itshape
  {Ultrafast Nonequilibrium Carrier Relaxation in Single-Crystal
  Nd1.85ce0.15cuo4-Y}}, Applied Physics Letters 63 (1993), pp. 979--981.

\bibitem[200]{Cao:2008p1505}
N. Cao, Y.B. Long, Z.G. Zhang, J. Yuan, L.J. Gao, B.R. Zhao, S.P. Zhao, Q.S.
  Yang, J. Zhao, and P. Fu, {\itshape {Quasiparticle relaxation dynamics in
  n-type superconductor La$_{2-x}$Ce$_x$CuO$_4$}}, Physica C-Superconductivity
  And Its Applications 468 (2008), pp. 894--897.

\bibitem[201]{Demsar:2003p7494}
J. Demsar, R.D. Averitt, A.J. Taylor, V.V. Kabanov, W.N. Kang, H.J. Kim, E.M.
  Choi, and S.I. Lee, {\itshape {Pair-Breaking and Superconducting State
  Recovery Dynamics in MgB$_2$}}, Physical Review Letters 91 (2003), p. 267002.

\bibitem[202]{Chia:2008p4912}
E.E.M. Chia, D. Talbayev, J.X. Zhu, H.Q. Yuan, T. Park, J.D. Thompson, G.F.
  Chen, J.L. Luo, N.L. Wang, and A.J. Taylor, {\itshape {Competing energy
  scales in iron-based layered (Ba,K)Fe2As2 superconductors}}, arXiv
  cond-mat.supr-con (2008).

\bibitem[203]{Torchinsky:2009p6282}
D.H. Torchinsky, G.F. Chen, J.L. Luo, N.L. Wang, and N. Gedik, {\itshape
  {Nonequilibrium quasiparticle dynamics in single crystals of
  Ba$_{0.6}$K$_{0.4}$Fe$_{2}$As$_{2}$.}}, Phys. Rev. Lett. 105 (2009), p.
  027005.

\bibitem[204]{Demsar:1999p1520}
J. Demsar, K. Biljakovic, and D. Mihailovic, {\itshape {Single particle and
  collective excitations in the one-dimensional charge density wave solid
  K$_{0.3}$MoO$_3$ probed in real time by femtosecond spectroscopy}}, Phys.
  Rev. Lett. 83 (1999), pp. 800--803.

\bibitem[205]{Toda:2009p6297}
Y. Toda, R. Onozaki, M. Tsubota, K. Inagaki, and S. Tanda, {\itshape Optical
  selection of a multiple phase order in the charge density wave condensate
  $o$-TaS$_3$ using a spectrally resolved nonequilibrium measurement}, Phys.
  Rev. B 80 (2009), p. 121103.

\bibitem[206]{Demsar:2002p6773}
J. Demsar, L. Forr{\'o}, H. Berger, and D. Mihailovic, {\itshape {Femtosecond
  "snapshots" of gap--forming charge-density-wave correlations in
  quasi-two-dimensional dichalcogenides 1T-TaS$_2$ and 2H-TaSe$_2$}}, Phys Rev
  B 66 (2002), pp. 041101--041101.

\bibitem[207]{Demsar:2006p9224}
J. Demsar, V. Thorsmolle, J. Sarrao, and A. Taylor, {\itshape {Photoexcited
  electron dynamics in Kondo insulators and heavy fermions}}, Physical Review
  Letters 96 (2006), p. 037401.

\bibitem[208]{Demsar:2006p9230}
J. Demsar, J. Sarrao, and A. Taylor, {\itshape {Dynamics of photoexcited
  quasiparticles in heavy electron compounds}}, Journal Of Physics-Condensed
  Matter 18 (2006), pp. R281--R314.

\bibitem[209]{Demsar:2009p9187}
J. Demsar, V.V. Kabanov, A.S. Alexandrov, H.J. Lee, E.D. Bauer, J.L. Sarrao,
  and A.J. Taylor, {\itshape {Hot electron relaxation in the heavy-fermion
  Yb1-xLuxAl3 compound using femtosecond optical pump-probe spectroscopy}},
  Physical Review B 80 (2009), p. 085121.

\bibitem[210]{Parker1975}
W. Parker, {\itshape {Modified heating theory of nonequilibrium
  superconductors}}, Physical Review B 12 (1975), p. 3667.

\bibitem[211]{Owen1972}
C. Owe and D. Scalapino, {\itshape {Superconducting State under the Influence
  of External Dynamic Pair Breaking}}, Physical Review Letters 28 (1972), p.
  1559.

\bibitem[212]{Tinkham}
M. Tinkham {\itshape {Introduction to Superconductivity}},    Dover
  Publications, 2004.

\bibitem[213]{Nicol2003}
E.J. Nicol and J.P. Carbotte, {\itshape {Comparison of s- and d-wave gap
  symmetry in nonequilibrium superconductivity}}, Physical Review B 67 (2003),
  p. 214506.

\bibitem[214]{Eliashberg1970}
G.M. Eliashberg, {\itshape Film Superconductivity Stimulated by a
  High-frequency Field}, JETP Letters 11 (1970), p. 114.

\bibitem[215]{Klapwijk1977}
T. Klapwijk, J. Bergh, and J. Mooij, {\itshape Radiation-stimulated
  superconductivity}, Journal of Low Temperature Physics 26 (1977), pp.
  385--405.

\bibitem[216]{Kommers1977}
T. Kommers and J. Clarke, {\itshape Measurement of Microwave-Enhanced Energy
  Gap in Superconducting Aluminum by Tunneling}, Phys. Rev. Lett. 38 (1977),
  pp. 1091--1094.

\bibitem[217]{Schmid1977}
A. Schmid, {\itshape Stability of Radiation-Stimulated Superconductivity},
  Phys. Rev. Lett. 38 (1977), pp. 922--925.

\bibitem[218]{Huefner2008}
S. H\"ufner, M.a. Hossain, A. Damascelli, and G. Sawatzky, {\itshape {Two Gaps
  Make a High Temperature Superconductor?}}, Rep. Prog. Phys. 71 (2008), p.
  062501 (9pp).

\bibitem[219]{Landau1}
 Landau and  Lifshitz {\itshape {Statistical Physics, Part 1}},    Elsevier,
  Butterworth Heinemann, Third Edition 1980.

\bibitem[220]{Bossini2014}
D. Bossini, A.M. Kalashnikova, R.V. Pisarev, T. Rasing, and A.V. Kimel,
  {\itshape Controlling coherent and incoherent spin dynamics by steering the
  photoinduced energy flow}, Phys. Rev. B 89 (2014), p. 060405.

\bibitem[221]{Landau2}
 Lifshitz and  Pitaevskii {\itshape {Physical Kinetics}},    Elsevier,
  Butterworth Heinemann, 1981.

\bibitem[222]{Mihailovic:2013fpa}
D. Mihailovic, T. Mertelj, V.V. Kabanov, and S. Brazovskii, {\itshape {Coherent
  topological defect dynamics and collective modes in superconductors and
  electronic crystals}}, Journal Of Physics-Condensed Matter 25 (2013), p.
  404206.

\bibitem[223]{Hinton2013}
J.P. Hinton, J.D. Koralek, Y.M. Lu, A. Vishwanath, J. Orenstein, D.A. Bonn,
  W.N. Hardy, and R. Liang, {\itshape New collective mode in
  YBa$_{2}$Cu$_{3}$O$_{6+x}$ observed by time-domain reflectometry}, Phys. Rev.
  B 88 (2013), p. 060508.

\bibitem[224]{Basov2005}
D.N. Basov and T. Timusk, {\itshape Electrodynamics of high-${T}_{c}$
  superconductors}, Rev. Mod. Phys. 77 (2005), pp. 721--779.

\bibitem[225]{Basov2011}
D.N. Basov, R.D. Averitt, D. Marelvan~der , M. Dressel, and K. Haule, {\itshape
  {Electrodynamics of correlated electron materials}}, Reviews of Modern
  Physics 83 (2011), pp. 471--541.

\bibitem[226]{Cilento2013}
F. Cilento et~al., {\itshape In search for the pairing glue in cuprates by
  non-equilibrium optical spectroscopy}, Journal of Physics: Conference Series
  449 (2013), p. 012003.

\bibitem[227]{vanderMarel2013}
D. Marelvan~der , H.J.A. Molegraaf, J. Zaanen, Z. Nussinov, F. Carbone, A.
  Damascelli, H. Eisaki, M. Greven, P.H. Kes, and M. Li, {\itshape Quantum
  critical behaviour in a high-$T_c$ superconductor}, Nature 425 (2003), pp.
  {271--274}.

\bibitem[228]{Mishchenko2008}
A.S. Mishchenko, N. Nagaosa, Z.X. Shen, G. {De Filippis}, V. Cataudella, T.P.
  Devereaux, C. Bernhard, K.W. Kim, and J. Zaanen, {\itshape {Charge Dynamics
  of Doped Holes in High $T_c$ Cuprate Superconductors: A Clue from Optical
  Conductivity}}, Physical Review Letters 100 (2008), p. 166401.

\bibitem[229]{DeFilippis2009}
G. {De Filippis}, V. Cataudella, a. Mishchenko, C. Perroni, and N. Nagaosa,
  {\itshape {Optical conductivity of a doped Mott insulator: The interplay
  between correlation and electron-phonon interaction}}, Physical Review B 80
  (2009), p. 195104.

\bibitem[230]{DeMedici2009}
L. {de' Medici}, X. Wang, M. Capone, and A.J. Millis, {\itshape Correlation
  strength, gaps, and particle-hole asymmetry in high-${T}_{c}$ cuprates: A
  dynamical mean field study of the three-band copper-oxide model}, Phys. Rev.
  B 80 (2009), p. 054501.

\bibitem[231]{Li2013}
Y. Li et~al., {\itshape {Doping-Dependent Raman Resonance in the Model
  High-Temperature Superconductor HgBa$_2$CuO$_{4+\delta}$}}, Physical Review
  Letters 111 (2013), p. 187001.

\bibitem[232]{Baldassarre2008}
L. Baldassarre et~al., {\itshape {Quasiparticle evolution and pseudogap
  formation in V$_{2}$O$_{3}$: An infrared spectroscopy study}}, Phys. Rev. B
  77 (2008), p. 113107.

\bibitem[233]{Cooper1993}
S.L. Cooper et~al., {\itshape {Optical studies of the a-, b-, and c-axis charge
  dynamics in YBa$_{2}$Cu$_{3}$O$_{6+x}$}}, Phys. Rev. B 47 (1993), pp.
  8233--8248.

\bibitem[234]{Qazilbash2007}
M.M. Qazilbash et~al., {\itshape Mott Transition in VO2 Revealed by Infrared
  Spectroscopy and Nano-Imaging}, Science 318 (2007), pp. 1750--1753.

\bibitem[235]{Uchida1991}
S. Uchida, T. Ido, H. Takagi, T. Arima, Y. Tokura, and S. Tajima, {\itshape
  {Optical spectra of La$_{2-x}$Sr$_{x}$CuO$_{4}$: Effect of carrier doping on
  the electronic structure of the CuO$_{2}$ plane}}, Phys. Rev. B 43 (1991),
  pp. 7942--7954.

\bibitem[236]{vanderMarel2000}
D. {van der Marel}, A. Tsvetkov, M. Grueninger, D. Dulic, and H. Molegraaf,
  {\itshape C-axis optical properties of high Tc cuprates}, Physica C:
  Superconductivity {341-348} (2000), pp. {1531--1534}.

\bibitem[237]{vanderMarel2003}
D. {van der Marel}, H.J.A. Molegraaf, J. Zaanen, Z. Nussinov, F. Carbone, A.
  Damascelli, H. Eisaki, M. Greven, P.H. Kes, and M. Li, {\itshape {Quantum
  critical behaviour in a high-Tc superconductor}}, Nature 425 (2003), pp.
  271--274.

\bibitem[238]{Hirsch2000}
J.E. Hirsch and F. Marsiglio, {\itshape Optical sum rule violation, superfluid
  weight, and condensation energy in the cuprates}, Phys. Rev. B 62 (2000), pp.
  15131--15150.

\bibitem[239]{vanHeumen2009}
E. {van Heumen}, E. Muhlethaler, A.B. Kuzmenko, H. Eisaki, W. Meevasana, M.
  Greven, and D. {van der Marel}, {\itshape Optical determination of the
  relation between the electron-boson coupling function and the critical
  temperature in high-${T}_{c}$ cuprates}, Phys. Rev. B 79 (2009), p. 184512.

\bibitem[240]{Mirzaei2013}
S.I. Mirzaei et~al., {\itshape Spectroscopic evidence for Fermi liquid-like
  energy and temperature dependence of the relaxation rate in the pseudogap
  phase of the cuprates}, Proceedings of the National Academy of Sciences 110
  (2013), pp. 5774--5778.

\bibitem[241]{Abanov2001}
A. Abanov, A.V. Chubukov, and J. Schmalian, {\itshape Fingerprints of spin
  mediated pairing in cuprates}, Journal of Electron Spectroscopy and Related
  Phenomena {117-118} (2001), pp. {129--151} Strongly correlated systems.

\bibitem[242]{Chubukov2005}
A.V. Chubukov and J. Schmalian, {\itshape Superconductivity due to massless
  boson exchange in the strong-coupling limit}, Phys. Rev. B 72 (2005), p.
  174520.

\bibitem[243]{Norman2006}
M.R. Norman and A.V. Chubukov, {\itshape High-frequency behavior of the
  infrared conductivity of cuprates}, Phys. Rev. B 73 (2006), p. 140501.

\bibitem[244]{Hwang2007b}
J. Hwang, T. Timusk, and G.D. Gu, {\itshape {Doping dependent optical
  properties of Bi$_2$Sr$_2$CaCu$_2$O$_{8+\delta}$}}, J. Phys.: Condens. Matter
  19 (2007), p. 125208.

\bibitem[245]{Hwang2007c}
J. Hwang, E.J. Nicol, T. Timusk, A. Knigavko, and J.P. Carbotte, {\itshape
  {High Energy Scales in the Optical Self-Energy of the Cuprate
  Superconductors}}, Physical Review Letters  (2007), p. 207002.

\bibitem[246]{Yang2009b}
J. Yang, J. Hwang, E. Schachinger, J.P. Carbotte, R.P.S.M. Lobo, D. Colson, A.
  Forget, and T. Timusk, {\itshape {Exchange Boson Dynamics in Cuprates :
  Optical Conductivity of HgBa$_2$CuO$_{4+\delta}$}}, Physical Review Letters
  (2009), p. 027003.

\bibitem[247]{Hwang2011}
J. Hwang, {\itshape Electron-boson spectral density function of underdoped
  ${\mathrm{Bi}}_{2}{\mathrm{Sr}}_{2}{\mathrm{CaCu}}_{2}{\mathrm{O}}_{8+\delta}$
  and ${\mathrm{YBa}}_{2}{\mathrm{Cu}}_{3}{\mathrm{O}}_{6.50}$}, Phys. Rev. B
  83 (2011), p. 014507.

\bibitem[248]{Vladimirov2012}
A.A. Vladimirov, D. Ihle, and N.M. Plakida, {\itshape Optical and dc
  conductivities of cuprates: Spin fluctuation scattering in the $t$-$J$
  model}, Phys. Rev. B 85 (2012), p. 224536.

\bibitem[249]{Hwang2012}
J. Hwang and J. Carbotte, {\itshape {Determination of boson spectrum from
  optical data in pseudogap phase of underdoped cuprates}}, Physical Review B
  86 (2012), p. 094502.

\bibitem[250]{Hwang2013}
---{}---{}---, {\itshape Evolution of electron-boson spectral density in the
  underdoped region of Bi$_2$Sr$_{2-x}$La$_x$CuO$_6$}, Journal of Physics:
  Condensed Matter 25 (2013), p. 165703.

\bibitem[251]{Park2014}
S.R. Park, T. Fukuda, A. Hamann, D. Lamago, L. Pintschovius, M. Fujita, K.
  Yamada, and D. Reznik, {\itshape Evidence for a charge collective mode
  associated with superconductivity in copper oxides from neutron and x-ray
  scattering measurements of La$_2$-$x$Sr$_x$CuO$_4$}, Phys. Rev. B 89 (2014),
  p. 020506.

\bibitem[252]{Markiewicz2012}
R.S. Markiewicz, T. Das, and A. Bansil, {\itshape Self-energy and fluctuation
  spectra in cuprates: Comparing optical and photoemission results}, Phys. Rev.
  B 86 (2012), p. 024511.

\bibitem[253]{Yang2009}
J. Yang, D. H\"{u}vonen, U. Nagel, T. R\~{o}\~{o}m, N. Ni, P.C. Canfield, S.L.
  Bud'ko, J.P. Carbotte, and T. Timusk, {\itshape {Optical Spectroscopy of
  Superconducting Ba$_{0.55}$K$_{0.45}$Fe$_{2}$As$_{2}$: Evidence for Strong
  Coupling to Low-Energy Bosons}}, Physical Review Letters 102 (2009), p.
  187003.

\bibitem[254]{Dordevic2009}
S.V. Dordevic, L.W. Kohlman, N. Stojilovic, R. Hu, and C. Petrovic, {\itshape
  Signatures of electron-boson coupling in the half-metallic ferromagnet
  Mn$_5$Ge$_3$: Study of electron self-energy $\Sigma$($\omega$) obtained from
  infrared spectroscopy}, Phys. Rev. B 80 (2009), p. 115114.

\bibitem[255]{Sharapov2005}
S. Sharapov and J. Carbotte, {\itshape {Effects of energy dependence in the
  quasiparticle density of states on far-infrared absorption in the pseudogap
  state}}, Physical Review B 72 (2005), p. 134506.

\bibitem[256]{Homes1995}
C. Homes, T. Timusk, D. Bonn, R. Liang, and W. Hardy, {\itshape Optical
  properties along the $c$-axis of YBa$_2$Cu$_3$O$_{6+x}$, for x=0.50
  $\rightarrow$ 0.95 evolution of the pseudogap}, Physica C: Superconductivity
  254 (1995), pp. {265--280}.

\bibitem[257]{Okimoto1999}
Y. Okimoto, Y. Tomioka, Y. Onose, Y. Otsuka, and Y. Tokura, {\itshape Optical
  study of ${\mathrm{Pr}}_{1-x}{\mathrm{Ca}}_{x}{\mathrm{MnO}}_{3}(x=0.4)$ in a
  magnetic field: Variation of electronic structure with charge ordering and
  disordering phase transitions}, Phys. Rev. B 59 (1999), pp. 7401--7408.

\bibitem[258]{Katsufuji1996}
T. Katsufuji, T. Tanabe, T. Ishikawa, Y. Fukuda, T. Arima, and Y. Tokura,
  {\itshape {Optical spectroscopy of the charge-ordering transition in
  La$_{1.67}$Sr$_{0.33}$NiO$_{4}$}}, Phys. Rev. B 54 (1996), pp.
  R14230--R14233.

\bibitem[259]{Rini2007}
M. Rini, R. Tobey, N. Dean, J. Itatani, Y. Tomioka, Y. Tokura, R.W. Schoenlein,
  and A. Cavalleri, {\itshape {Control of the electronic phase of a manganite
  by mode-selective vibrational excitation}}, Nature 449 (2007), pp. 72--74.

\bibitem[260]{Kubler2007}
C. K\"ubler, H. Ehrke, R. Huber, R. Lopez, A. Halabica, R.F. Haglund, and A.
  Leitenstorfer, {\itshape Coherent Structural Dynamics and Electronic
  Correlations during an Ultrafast Insulator-to-Metal Phase Transition in
  VO$_2$}, Phys. Rev. Lett. 99 (2007), p. 116401.

\bibitem[261]{Carbone2008}
F. Carbone, D.S. Yang, E. Giannini, and A.H. Zewail, {\itshape Direct role of
  structural dynamics in electron-lattice coupling of superconducting
  cuprates}, Proceedings of the National Academy of Sciences 105 (2008), pp.
  20161--20166.

\bibitem[262]{Pashkin2010}
A. Pashkin et~al., {\itshape {Femtosecond Response of Quasiparticles and
  Phonons in Superconducting YBa$_2$Cu$_3$O$_{7-\delta}$ Studied by Wideband
  Terahertz Spectroscopy}}, Physical Review Letters 105 (2010), p. 067001.

\bibitem[263]{Hu2014}
W. Hu, S. Kaiser, D. Nicoletti, C.R. Hunt, I. Gierz, M.C. Hoffmann, M. {Le
  Tacon}, T. Loew, B. Keimer, and A. Cavalleri, {\itshape {Optically enhanced
  coherent transport in YBa$_2$Cu$_3$O$_{6.5}$ by ultrafast redistribution of
  interlayer coupling}}, Nat. Mater. 13 (2014), pp. 705--711.

\bibitem[264]{Kaiser2014}
S. Kaiser et~al., {\itshape Optically induced coherent transport far above
  $T_{c}$ in underdoped YBa$_2$Cu$_3$O$_{6+\delta}$}, Phys. Rev. B 89 (2014),
  p. 184516.

\bibitem[265]{Fano1961}
U. Fano, {\itshape Effects of Configuration Interaction on Intensities and
  Phase Shifts}, Phys. Rev. 124 (1961), pp. 1866--1878.

\bibitem[266]{Munzar1999}
D. Munzar, C. Bernhard, A. Golnik, J. Huml???ek, and M. Cardona, {\itshape
  Anomalies of the infrared-active phonons in underdoped YBa$_2$Cu$_3$O$_y$ as
  evidence for the intra-bilayer Josephson effect}, Solid State Communications
  112 (1999), pp. {365 -- 369}.

\bibitem[267]{Zetterer1990}
T. Zetterer, M. Franz, J. Sch\"utzmann, W. Ose, H.H. Otto, and K.F. Renk,
  {\itshape {Anomalous behavior of phonons in superconducting
  Tl$_{2}$Ba$_{2}$Ba$_{2}$Ca$_{2}$Cu$_{3}$O$_{10}$ detected by far-infrared
  spectroscopy}}, Phys. Rev. B 41 (1990), pp. 9499--9501.

\bibitem[268]{Homes1997}
C.C. Homes, J.M. Tranquada, and D.J. Buttrey, {\itshape Stripe order and
  vibrational properties of La$_2$NiO$_{4+\delta}$ for $\delta$=2/15:
  Measurements and ab initio calculations}, Phys. Rev. B 75 (2007), p. 045128.

\bibitem[269]{Jung2001}
J.H. Jung, D.W. Kim, T.W. Noh, H.C. Kim, H.C. Ri, S.J. Levett, M.R. Lees, D.M.
  Paul, and G. Balakrishnan, {\itshape Optical conductivity studies of
  La$_{3/2}$Sr$_{1/2}$NiO$_4$: Lattice effect on charge ordering}, Phys. Rev. B
  64 (2001), p. 165106.

\bibitem[270]{Uchida2011}
M. Uchida et~al., {\itshape Pseudogap of Metallic Layered Nickelate
  R$_{2-x}$Sr$_x$NiO$_4$ (R=Nd, Eu) Crystals Measured Using Angle-Resolved
  Photoemission Spectroscopy}, Phys. Rev. Lett. 106 (2011), p. 027001.

\bibitem[271]{Pashkin2011}
A. Pashkin, C. K\"ubler, H. Ehrke, R. Lopez, A. Halabica, R.F. Haglund, R.
  Huber, and A. Leitenstorfer, {\itshape Ultrafast insulator-metal phase
  transition in VO$_{2}$ studied by multiterahertz spectroscopy}, Phys. Rev. B
  83 (2011), p. 195120.

\bibitem[272]{Porer2014}
M. Porer, U. Leierseder, J.M. M\'{e}nard, H. Dachraoui, L. Mouchliadis, I.E.
  Perakis, U. Heinzmann, J. Demsar, K. Rossnagel, and R. Huber, {\itshape
  {Non-thermal separation of electronic and structural orders in a persisting
  charge density wave}}, Nat. Mater. 13 (2014), pp. 857--861.

\bibitem[273]{Holy1977}
J.A. Holy, K.C. Woo, M.V. Klein, and F.C. Brown, {\itshape Raman and infrared
  studies of superlattice formation in TiSe$_2$}, Phys. Rev. B 16 (1977), pp.
  3628--3637.

\bibitem[274]{Li2007}
G. Li, W.Z. Hu, D. Qian, D. Hsieh, M.Z. Hasan, E. Morosan, R.J. Cava, and N.L.
  Wang, {\itshape Semimetal-to-Semimetal Charge Density Wave Transition in
  1$T$-TiSe$_2$}, Phys. Rev. Lett. 99 (2007), p. 027404.

\bibitem[275]{Averitt2001}
R.D. Averitt, G. Rodriguez, A.I. Lobad, J.L.W. Siders, S.A. Trugman, and A.J.
  Taylor, {\itshape {Nonequilibrium superconductivity and quasiparticle
  dynamics in YBa$_2$Cu$_3$O$_{7-\delta}$}},  63 (2001), pp. 2--5.

\bibitem[276]{Carnahan2004}
M. Carnahan, R. Kaindl, J. Orenstein, D. Chemla, S. Oh, and J. Eckstein,
  {\itshape {Nonequilibrium THz conductivity of
  Bi$_2$Sr$_2$CaCu$_2$O$_{8+\delta}$}}, Physica C: Superconductivity 408-410
  (2004), pp. {729--730} Proceedings of the International Conference on
  Materials and Mechanisms of Superconductivity. High Temperature
  Superconductors VII-M2SRIO.

\bibitem[277]{Perfetti2015}
L. Perfetti, B. Sciolla, G. Biroli, C.J. Beekvan~der , C. Piovera, M. Wolf, and
  T. Kampfrath, {\itshape Ultrafast Dynamics of Fluctuations in
  High-Temperature Superconductors Far from Equilibrium}, Phys. Rev. Lett. 114
  (2015), p. 067003.

\bibitem[278]{Beck2011}
M. Beck, M. Klammer, S. Lang, P. Leiderer, V.V. Kabanov, G.N. {Gol'tsman}, and
  J. Demsar, {\itshape {Energy-Gap Dynamics of Superconducting NbN Thin Films
  Studied by Time-Resolved Terahertz Spectroscopy}}, Physical Review Letters
  107 (2011), p. 177007.

\bibitem[279]{Beyer2011}
M. Beyer, D. St\"{a}dter, M. Beck, H. Sch\"{a}fer, V.V. Kabanov, G. Logvenov,
  I. Bozovic, G. Koren, and J. Demsar, {\itshape {Photoinduced melting of
  superconductivity in the high-$T_c$ superconductor La$_{2-x}$Sr$_x$CuO$_4$
  probed by time-resolved optical and terahertz techniques}}, Physical Review B
  83 (2011), p. 214515.

\bibitem[280]{Beck2013}
M. Beck, I. Rousseau, M. Klammer, P. Leiderer, M. Mittendorff, S. Winnerl, M.
  Helm, G.N. Gol'tsman, and J. Demsar, {\itshape Transient Increase of the
  Energy Gap of Superconducting NbN Thin Films Excited by Resonant Narrow-Band
  Terahertz Pulses}, Phys. Rev. Lett. 110 (2013), p. 267003.

\bibitem[281]{vanderMarel1996}
D. Marelvan~der  and A. Tsvetkov, {\itshape Transverse optical plasmons in
  layered superconductors}, Czechoslovak Journal of Physics 46 (1996), pp.
  3165--3168.

\bibitem[282]{vanderMarel2001}
D. Marelvan~der  and A.A. Tsvetkov, {\itshape Transverse-optical Josephson
  plasmons: Equations of motion}, Phys. Rev. B 64 (2001), p. 024530.

\bibitem[283]{Urbach1953}
F. Urbach, {\itshape The Long-Wavelength Edge of Photographic Sensitivity and
  of the Electronic Absorption of Solids}, Phys. Rev. 92 (1953), pp.
  1324--1324.

\bibitem[284]{Wall2010}
S. Wall et~al., {\itshape {Quantum interference between charge excitation paths
  in a solid-state Mott insulator}}, Nature Physics 7 (2010), pp. 114--118.

\bibitem[285]{Fan1951}
H.Y. Fan, {\itshape Temperature Dependence of the Energy Gap in
  Semiconductors}, Phys. Rev. 82 (1951), pp. 900--905.

\bibitem[286]{Falck1992}
J.P. Falck, A. Levy, M.A. Kastner, and R.J. Birgeneau, {\itshape
  Charge-transfer spectrum and its temperature dependence in La$_2$CuO$_4$},
  Phys. Rev. Lett. 69 (1992), pp. 1109--1112.

\bibitem[287]{Lovenich2001}
R. L\"ovenich, A.B. Schumacher, J.S. Dodge, D.S. Chemla, and L.L. Miller,
  {\itshape Evidence of phonon-mediated coupling between charge transfer and
  ligand field excitons in Sr$_{2}$CuO$_{2}$Cl$_{2}$}, Phys. Rev. B 63 (2001),
  p. 235104.

\bibitem[288]{Okamoto2010}
H. Okamoto, T. Miyagoe, K. Kobayashi, H. Uemura, H. Nishioka, H. Matsuzaki, A.
  Sawa, and Y. Tokura, {\itshape {Ultrafast charge dynamics in photoexcited
  Nd$_2$CuO$_4$ and La$_2$CuO$_4$ cuprate compounds investigated by femtosecond
  absorption spectroscopy}}, Phys. Rev. B 82 (2010), p. {060513(R)}.

\bibitem[289]{Khurana1990}
A. Khurana, {\itshape Electrical conductivity in the infinite-dimensional
  Hubbard model}, Phys. Rev. Lett. 64 (1990), pp. 1990--1990.

\bibitem[290]{Tomczak2009}
J.M. Tomczak and S. Biermann, {\itshape Optical properties of correlated
  materials: Generalized Peierls approach and its application to
  ${\text{VO}}_{2}$}, Phys. Rev. B 80 (2009), p. 085117.

\bibitem[291]{Pruschke1993}
T. Pruschke, D.L. Cox, and M. Jarrell, {\itshape Hubbard model at infinite
  dimensions: Thermodynamic and transport properties}, Phys. Rev. B 47 (1993),
  pp. 3553--3565.

\bibitem[292]{Jarrell1995}
M. Jarrell, J.K. Freericks, and T. Pruschke, {\itshape Optical conductivity of
  the infinite-dimensional Hubbard model}, Phys. Rev. B 51 (1995), pp.
  11704--11711.

\bibitem[293]{Rozenberg1995}
M.J. Rozenberg, G. Kotliar, H. Kajueter, G.A. Thomas, D.H. Rapkine, J.M. Honig,
  and P. Metcalf, {\itshape Optical Conductivity in Mott-Hubbard Systems},
  Phys. Rev. Lett. 75 (1995), pp. 105--108.

\bibitem[294]{Rozenberg1996}
M.J. Rozenberg, G. Kotliar, and H. Kajueter, {\itshape Transfer of spectral
  weight in spectroscopies of correlated electron systems}, Phys. Rev. B 54
  (1996), pp. 8452--8468.

\bibitem[295]{Maier2005a}
T. Maier, M. Jarrell, T. Pruschke, and M.H. Hettler, {\itshape Quantum cluster
  theories}, Rev. Mod. Phys. 77 (2005), pp. 1027--1080.

\bibitem[296]{Kotliar2001}
G. Kotliar, S.Y. Savrasov, G. P\'alsson, and G. Biroli, {\itshape Cellular
  Dynamical Mean Field Approach to Strongly Correlated Systems}, Phys. Rev.
  Lett. 87 (2001), p. 186401.

\bibitem[297]{Lichtenstein2000}
A.I. Lichtenstein and M.I. Katsnelson, {\itshape Antiferromagnetism and
  \textbf{ \textit{d} } -wave superconductivity in cuprates: A cluster
  dynamical mean-field theory}, Phys. Rev. B 62 (2000), pp. R9283--R9286.

\bibitem[298]{Ferrero2008}
M. Ferrero, P.S. Cornaglia, L.D. Leo, O. Parcollet, G. Kotliar, and A. Georges,
  {\itshape Valence bond dynamical mean-field theory of doped Mott insulators
  with nodal/antinodal differentiation}, EPL (Europhysics Letters) 85 (2009),
  p. 57009.

\bibitem[299]{Ferrero2009}
M. Ferrero, P.S. Cornaglia, L. De~Leo, O. Parcollet, G. Kotliar, and A.
  Georges, {\itshape Pseudogap opening and formation of Fermi arcs as an
  orbital-selective Mott transition in momentum space}, Phys. Rev. B 80 (2009),
  p. 064501.

\bibitem[300]{Civelli2005}
M. Civelli, M. Capone, S.S. Kancharla, O. Parcollet, and G. Kotliar, {\itshape
  Dynamical Breakup of the Fermi Surface in a Doped Mott Insulator}, Phys. Rev.
  Lett. 95 (2005), p. 106402.

\bibitem[301]{Capone2006}
M. Capone and G. Kotliar, {\itshape Competition between $d$-wave
  superconductivity and antiferromagnetism in the two-dimensional Hubbard
  model}, Phys. Rev. B 74 (2006), p. 054513.

\bibitem[302]{Haule2007}
K. Haule and G. Kotliar, {\itshape Strongly correlated superconductivity: A
  plaquette dynamical mean-field theory study}, Phys. Rev. B 76 (2007), p.
  104509.

\bibitem[303]{Gull2008}
E. Gull, P. Werner, X. Wang, M. Troyer, and A.J. Millis, {\itshape Local order
  and the gapped phase of the Hubbard model: A plaquette dynamical mean-field
  investigation}, EPL (Europhysics Letters) 84 (2008), p. 37009.

\bibitem[304]{Civelli2008}
M. Civelli, M. Capone, A. Georges, K. Haule, O. Parcollet, T.D. Stanescu, and
  G. Kotliar, {\itshape Nodal-Antinodal Dichotomy and the Two Gaps of a
  Superconducting Doped Mott Insulator}, Phys. Rev. Lett. 100 (2008), p.
  046402.

\bibitem[305]{Kancharla2008}
S.S. Kancharla, B. Kyung, D. S\'en\'echal, M. Civelli, M. Capone, G. Kotliar,
  and A.M.S. Tremblay, {\itshape Anomalous superconductivity and its
  competition with antiferromagnetism in doped Mott insulators}, Phys. Rev. B
  77 (2008), p. 184516.

\bibitem[306]{Civelli2009}
M. Civelli, {\itshape Evolution of the Dynamical Pairing across the Phase
  Diagram of a Strongly Correlated High-Temperature Superconductor}, Phys. Rev.
  Lett. 103 (2009), p. 136402.

\bibitem[307]{Sakai2009}
S. Sakai, Y. Motome, and M. Imada, {\itshape Evolution of Electronic Structure
  of Doped Mott Insulators: Reconstruction of Poles and Zeros of Green's
  Function}, Phys. Rev. Lett. 102 (2009), p. 056404.

\bibitem[308]{Sakai2012}
S. Sakai, G. Sangiovanni, M. Civelli, Y. Motome, K. Held, and M. Imada,
  {\itshape Cluster-size dependence in cellular dynamical mean-field theory},
  Phys. Rev. B 85 (2012), p. 035102.

\bibitem[309]{Sordi2012}
G. Sordi, P. S\'emon, K. Haule, and A.M.S. Tremblay, {\itshape Strong Coupling
  Superconductivity, Pseudogap, and Mott Transition}, Phys. Rev. Lett. 108
  (2012), p. 216401.

\bibitem[310]{Sordi2013b}
---{}---{}---, {\itshape Pseudogap temperature as a Widom line in doped Mott
  insulators}, Scientific Reports 2 (2012), p. 457.

\bibitem[311]{Maier2005b}
T.A. Maier, M. Jarrell, T.C. Schulthess, P.R.C. Kent, and J.B. White, {\itshape
  Systematic Study of $d$-Wave Superconductivity in the 2D Repulsive Hubbard
  Model}, Phys. Rev. Lett. 95 (2005), p. 237001.

\bibitem[312]{Gull2010}
E. Gull, M. Ferrero, O. Parcollet, A. Georges, and A.J. Millis, {\itshape
  Momentum-space anisotropy and pseudogaps: A comparative cluster dynamical
  mean-field analysis of the doping-driven metal-insulator transition in the
  two-dimensional Hubbard model}, Phys. Rev. B 82 (2010), p. 155101.

\bibitem[313]{Gull2012}
E. Gull and A.J. Millis, {\itshape Energetics of superconductivity in the
  two-dimensional Hubbard model}, Phys. Rev. B 86 (2012), p. 241106.

\bibitem[314]{Gull2013}
E. Gull, O. Parcollet, and A.J. Millis, {\itshape Superconductivity and the
  Pseudogap in the Two-Dimensional Hubbard Model}, Phys. Rev. Lett. 110 (2013),
  p. 216405.

\bibitem[315]{Chen2015}
X. Chen, J.P.F. LeBlanc, and E. Gull, {\itshape Superconducting Fluctuations in
  the Normal State of the Two-Dimensional Hubbard Model}, Phys. Rev. Lett. 115
  (2015), p. 116402.

\bibitem[316]{Comanac2008}
A. Comanac, L. {de' Medici}, M. Capone, and A.J. Millis, {\itshape Optical
  conductivity and the correlation strength of high-temperature copper-oxide
  superconductors}, Nat Phys 4 (2008), pp. 287--290.

\bibitem[317]{Toschi2008}
A. Toschi and M. Capone, {\itshape Optical sum rule anomalies in the cuprates:
  Interplay between strong correlation and electronic band structure}, Phys.
  Rev. B 77 (2008), p. 014518.

\bibitem[318]{Nicoletti2010}
D. Nicoletti et~al., {\itshape High-Temperature Optical Spectral Weight and
  Fermi-liquid Renormalization in Bi-Based Cuprate Superconductors}, Phys. Rev.
  Lett. 105 (2010), p. 077002.

\bibitem[319]{Guerrisi1975}
M. Guerrisi, R. Rosei, and P. Winsemius, {\itshape Splitting of the interband
  absorption edge in Au}, Phys. Rev. B 12 (1975), pp. 557--563.

\bibitem[320]{Baranov2014}
V.V. Baranov and V.V. Kabanov, {\itshape Theory of electronic relaxation in a
  metal excited by an ultrashort optical pump}, Phys. Rev. B 89 (2014), p.
  125102.

\bibitem[321]{Neshat2012}
M. Neshat and N.P. Armitage, {\itshape Terahertz time-domain spectroscopic
  ellipsometry: instrumentation and calibration}, Opt. Express 20 (2012), pp.
  29063--29075.

\bibitem[322]{Morris2012}
C.M. Morris, R.V. Aguilar, A.V. Stier, and N.P. Armitage, {\itshape
  Polarization modulation time-domain terahertz polarimetry}, Opt. Express 20
  (2012), pp. 12303--12317.

\bibitem[323]{Roeser2003}
 {Roeser, C. A. D. and Kim, A. M.-T. and Callan, J. P. and Huang, L. and
  Glezer, E. N. and Siegal, Y. and Mazur, E.}, {\itshape Femtosecond
  time-resolved dielectric function measurements by dual-angle reflectometry},
  Review of Scientific 74 (2003), pp. {3413--3422}.

\bibitem[324]{Mattis1958}
D.C. Mattis and J. Bardeen, {\itshape Theory of the Anomalous Skin Effect in
  Normal and Superconducting Metals}, Phys. Rev. 111 (1958), pp. 412--417.

\bibitem[325]{Mukamel}
S. Mukamel {\itshape {Principles of Nonlinear Optical Spectroscopy}},    Oxfero
  University Press, 1995.

\bibitem[326]{Mansart2013b}
B. Mansart, J. Lorenzana, A. Mann, A. Odeh, M. Scarongella, M. Chergui, and F.
  Carbone, {\itshape Coupling of a high-energy excitation to superconducting
  quasiparticles in a cuprate from coherent charge fluctuation spectroscopy},
  Proceedings of the National Academy of Sciences 110 (2013), pp. 4539--4544.

\bibitem[327]{Toda:2014ga}
Y. Toda, F. Kawanokami, T. Kurosawa, M. Oda, I. Madan, T. Mertelj, V.V.
  Kabanov, and D. Mihailovic, {\itshape Rotational symmetry breaking in
  Bi$_2$Sr$_2$CaCu$_2$O$_{8+\delta}$ probed by polarized femtosecond
  spectroscopy}, Phys. Rev. B 90 (2014), p. 094513.

\bibitem[328]{Rossi2002}
F. Rossi and T. Kuhn, {\itshape Theory of ultrafast phenomena in photoexcited
  semiconductors}, Rev. Mod. Phys. 74 (2002), pp. 895--950.

\bibitem[329]{Okamoto2011}
H. Okamoto, T. Miyagoe, K. Kobayashi, H. Uemura, H. Nishioka, H. Matsuzaki, a.
  Sawa, and Y. Tokura, {\itshape {Photoinduced transition from Mott insulator
  to metal in the undoped cuprates Nd$_{2}$CuO$_{4}$ and La$_{2}$CuO$_{4}$}},
  Physical Review B 83 (2011), p. 125102.

\bibitem[330]{Devereaux:2007tg}
T.P. Devereaux and R. Hackl, {\itshape {Inelastic light scattering from
  correlated electrons}}, Reviews of Modern Physics 79 (2007), p. 175.

\bibitem[331]{Sugai:1989jya}
S. Sugai, {\itshape {Phonon Raman scattering in (La$_{1-x}$Sr$_x$)$_2$CuO$_4$
  single crystals}}, Physical Review B 39 (1989), pp. 4306--4315.

\bibitem[332]{Nemetschek:1997er}
R. Nemetschek, M. Opel, C. Hoffmann, P.F. M{\"u}ller, R. Hackl, H. Berger, L.
  Forro, A. Erb, and E. Walker, {\itshape {Pseudogap and Superconducting Gap in
  the Electronic Raman Spectra of Underdoped Cuprates}}, Phys. Rev. Lett. 78
  (1997), pp. 4837--4840.

\bibitem[333]{Sakai:22tRGgJS}
S. Sakai et~al., {\itshape {Raman-Scattering Measurements and Theory of the
  Energy-Momentum Spectrum for Underdoped Bi$_2$Sr$_2$CaCuO$_8$
  Superconductors: Evidence of an $s$-Wave Structure for the Pseudogap}}, Phys.
  Rev. Lett. 111 (2013), p. 107001.

\bibitem[334]{BussmannHolder:2007dpa}
A. Bussmann-Holder, R. Khasanov, A. Shengelaya, A. Maisuradze, F.L. Mattina, H.
  Keller, and K.A. M{\"u}ller, {\itshape {Mixed order parameter symmetries in
  cuprate superconductors}}, Europhysics Letters (EPL) 77 (2007), p. 27002.

\bibitem[335]{Han1990}
S.G. Han, Z.V. Vardeny, K.S. Wong, O.G. Symko, and G. Koren, {\itshape
  Femtosecond optical detection of quasiparticle dynamics in high-$T_c$
  YBa$_2$Cu$_3$O$_{7-\delta}$ superconducting thin films}, Phys. Rev. Lett. 65
  (1990), pp. 2708--2711.

\bibitem[336]{Thomas:1996wc}
T.N. Thomas, C.J. Stevens, A. Choudhary, J.F. Ryan, D. Mihailovic, T. Mertelj,
  L. Forro, G. Wagner, and J.E. Evetts, {\itshape {Photoexcited carrier
  relaxation and localization in Bi$_2$Sr$_2$Ca$_{1-y}$Y$_y$Cu$_2$O$_8$ and
  YBa$_2$Cu$_3$O$_{7-\delta}$: A study by femtosecond time-resolved
  spectroscopy}}, Physical Review B 53 (1996), pp. 12436--12440.

\bibitem[337]{Demsar:1999wh}
J. Demsar, B. Podobnik, J. Evetts, G. Wagner, and D. Mihailovic, {\itshape
  {Evidence for crossover from a Bose-Einstein condensate to a BCS-like
  superconductor with doping in YBa$_2$Cu$_3$O$_{7-\delta}$ from quasiparticle
  relaxation dynamics experiments}}, Europhysics Letters 45 (1999), pp.
  381--386.

\bibitem[338]{Dvorsek:2002p82}
D. Dvorsek, V. Kabanov, J. Demsar, S. Kazakov, J. Karpinski, and D. Mihailovic,
  {\itshape {Femtosecond quasiparticle relaxation dynamics and probe
  polarization anisotropy in YSr$_{x}$Ba$_{2-x}$Cu$_{4}$O$_{8}$(x=0, 0.4)}},
  Physical Review B 66 (2002), p. 020510.

\bibitem[339]{Gedik:2003p2289}
N. Gedik, J. Orenstein, R. Liang, D. Bonn, and W. Hardy, {\itshape {Diffusion
  of nonequilibrium quasi-particles in a cuprate superconductor}}, Science 300
  (2003), pp. 1410--1412.

\bibitem[340]{Gedik2004}
N. Gedik, P. Blake, R. Spitzer, J. Orenstein, R. Liang, D. Bonn, and W. Hardy,
  {\itshape {Single-quasiparticle stability and quasiparticle-pair decay in
  YBa$_2$Cu$_3$O$_{6.5}$}}, Physical Review B 70 (2004), p. 014504.

\bibitem[341]{Luo2003}
C.W. Luo, M.H. Chen, S.P. Chen, K.H. Wu, J.Y. Juang, J. Lin, T.M. Uen, and Y.S.
  Gou, {\itshape {Spatial symmetry of the superconducting gap of
  YBa$_2$Cu$_3$O$_{7+\delta}$ obtained from femtosecond spectroscopy}}, Phys.
  Rev. B 68 (2003), p. {220508(R)}.

\bibitem[342]{Luo2006}
C.W. Luo, C.C. Hsieh, Y.J. Chen, P.T. Shih, M.H. Chen, K.H. Wu, J.Y. Juang,
  J.Y. Lin, T.M. Uen, and Y.S. Gou, {\itshape Spatial dichotomy of
  quasiparticle dynamics in underdoped thin-film YBa$_2$Cu$_3$O$_{7-\delta}$
  superconductors}, Phys. Rev. B 74 (2006), p. 184525.

\bibitem[343]{Bianchi:2003kj}
G. Bianchi, {\itshape {Ultrafast spectroscopy of La$_{2-x}$Sr$_x$CuO$_4$ single
  crystals}}, Journal Of Low Temperature Physics 131 (2003), pp. 755--759.

\bibitem[344]{Schneider2003}
M.L. Schneider, S. Rast, M. Onellion, J. Demsar, A.J. Taylor, Y. Glinka, N.H.
  Tolk, and Y.H. Ren, {\itshape {Carrier relaxation time divergence in single
  and double layer cuprates}}, The European Physical Journal B 334 (2003), pp.
  327--334.

\bibitem[345]{Bianchi:2005p3988}
G. Bianchi, C. Chen, M. Nohara, H. Takagi, and J. Ryan, {\itshape
  {Nonequilibrium quasiparticle relaxation in the vortex state of
  La$_{2-x}$Sr$_x$CuO$_4$}}, Physical Review Letters 94 (2005), p. 107004.

\bibitem[346]{Bianchi:2005p3989}
---{}---{}---, {\itshape {Competing phases in the superconducting state of
  La$_{2-x}$Sr$_x$CuO$_4$: Temperature- and magnetic-field-dependent
  quasiparticle relaxation measurements}}, Physical Review B 72 (2005), p.
  094516.

\bibitem[347]{Kusar:2005p4778}
P. Kusar, J. Demsar, D. Mihailovic, and S. Sugai, {\itshape {A systematic study
  of femtosecond quasiparticle relaxation processes in
  La$_{2-x}$Sr$_x$CuO$_4$}}, Physical Review B 72 (2005), p. 014544.

\bibitem[348]{Kusar:2010hz}
P. Kusar, V.V. Kabanov, S. Sugai, J. Demsar, T. Mertelj, and D. Mihailovic,
  {\itshape {Dynamical Structural Instabilities in La$_{1.9}$Sr$_{0.1}$CuO$_4$
  Under Intense Laser Photoexcitation}}, Journal Of Superconductivity And Novel
  Magnetism 24 (2010), pp. 421--425.

\bibitem[349]{Smith2000}
D. Smith, C.J. {Gay, PStevens}, C. Chen, G. Yang, S.J. Abell, D.Z. Wang, J.H.
  Wang, and J.F. Ryan, {\itshape {Femtosecond Dynamics of BSCCO-2212 in the
  weak and strong excitation limits}}, Physica C 348 (2000), pp. 2221--2222.

\bibitem[350]{Murakami2002}
H. Murakami, T. Uchiyama, I. Iguchi, and Z. Wang, {\itshape {Relaxation
  dynamics of optically excited quasi-particles in BSCCO}}, Physica C 381
  (2002), pp. 320--323.

\bibitem[351]{Liu:2008p4917}
Y. Liu, Y. Toda, K. Shimatake, N. Momono, and M. Oda, {\itshape {Direct
  Observation of the Coexistence of the Pseudogap and Superconducting
  Quasiparticles in Bi$_2$Sr$_2$CaCu$_2$O$_{8+y}$ by Time-Resolved Optical
  Spectroscopy}}, Physical Review Letters 101 (2008), p. 137003.

\bibitem[352]{Chia2013}
E.E.M. Chia et~al., {\itshape {Doping dependence of the electron-phonon and
  electron-spin fluctuation interactions in the high-$T_c$ superconductor
  Bi$_2$Sr$_2$CaCu$_2$O$_{8+\delta}$}}, New Journal of Physics 15 (2013), p.
  103027.

\bibitem[353]{Smith:1999p4056}
D. Smith, P. Gay, C. Stevens, D. Wang, J. Wang, Z. Ren, and J. Ryan, {\itshape
  {Ultrafast optical response of Tl$_2$Ba$_2$CuO$_{6+\delta}$}}, Journal Of Low
  Temperature Physics 117 (1999), pp. 1059--1063.

\bibitem[354]{Easley1990}
G.L. Eesley, J. Heremans, D.A. Meyer, G.L. Doll, and S.H. Liou, {\itshape
  {Relaxation time of the order parameter in a high-temperature
  superconductor}}, Physical Review Letters 65 (1990), pp. 3445--3448.

\bibitem[355]{Chia:2007p1669}
E.E.M. Chia, J.X. Zhu, D. Talbayev, R.D. Averitt, A.J. Taylor, K.H. OH, I.S.
  Jo, and S.I. Lee, {\itshape {Observation of competing order in a High-$T_c$
  superconductor using femtosecond optical pulses}}, Physical Review Letters 99
  (2007), p. 147008.

\bibitem[356]{Demsar2000}
J. Demsar, R. Hudej, J. Karpinski, V. Kabanov, and D. Mihailovic, {\itshape
  Quasiparticle relaxation dynamics in Hg-1223 studied by femtosecond
  time-resolved optical spectroscopy}, Physica C: Superconductivity {341-348,
  Part 2} (2000), pp. {925--926}.

\bibitem[357]{Hinton2013b}
J.P. Hinton, J.D. Koralek, G. Yu, E.M. Motoyama, Y.M. Lu, A. Vishwanath, M.
  Greven, and J. Orenstein, {\itshape {Time-Resolved Optical Reflectivity of
  the Electron-Doped Nd$_{2-x}$Ce$_x$CuO$_{4+\delta}$ Cuprate Superconductor:
  Evidence for an Interplay between Competing Orders}}, Phys. Rev. Lett. 110
  (2013), p. 217002.

\bibitem[358]{Mihailovic:1999p4786}
D. Mihailovic, V. Kabanov, K. Zagar, and J. Demsar, {\itshape {Distinct charge
  and spin gaps in underdoped YBa$_2$Cu$_3$O$_{7-\delta}$ from analysis of NMR,
  neutron scattering, tunneling, and quasiparticle relaxation experiments}},
  Physical Review B 60 (1999), pp. R6995--R6997.

\bibitem[359]{Deutscher1999}
G. Deutscher, {\itshape {Coherence and single-particle excitations in the
  high-temperature superconductors}}, Nature 397 (1999), pp. 410--412.

\bibitem[360]{Kaindl2000}
R. Kaindl, M. Woerner, T. Elsaesser, D. Smith, and J. Ryan, {\itshape
  {Ultrafast Mid-Infrared Response of YBa$_2$Cu$_3$O$_{7-\delta}$}}, Science
  287 (2000), pp. 470--473.

\bibitem[361]{Gedik2003}
N. Gedik, J. Orenstein, R. Liang, D.A. Bonn, and W.N. Hardy, {\itshape
  {Diffusion of Nonequilibrium Quasi-Particles in a Cuprate}}, Science 300
  (2003), pp. 1410--1412.

\bibitem[362]{Bilbro2011b}
L.S. Bilbro, R. Vald\'es~Aguilar, G. Logvenov, I. Bozovic, and N.P. Armitage,
  {\itshape {On the possibility of fast vortices in the cuprates: A vortex
  plasma model analysis of THz conductivity and diamagnetism in
  La$_{2-x}$Sr$_{x}$CuO$_{4}$}}, Phys. Rev. B 84 (2011), p. 100511.

\bibitem[363]{Sau2011}
J.D. Sau and S. Tewari, {\itshape Diamagnetic Susceptibility Obtained from the
  Six-Vertex Model and Its Implications for the High-Temperature Diamagnetic
  State of Cuprate Superconductors}, Phys. Rev. Lett. 107 (2011), p. 177006.

\bibitem[364]{Corson:1999bo}
J. Corson, R. Mallozzi, J. Orenstein, J.N. Eckstein, and I. Bozovic, {\itshape
  {Vanishing of phase coherence in underdoped
  Bi$_2$Sr$_2$CaCu$_2$O$_{8+\delta}$}}, Nature 398 (1999), pp. 221--223.

\bibitem[365]{Bilbro2011}
L. Bilbro, R.V. Aguilar, G. Logvenov, O. Pelleg, I. Bozovic, and N.P. Armitage,
  {\itshape {Temporal correlations of superconductivity above the transition
  temperature in La$_{2-x}$Sr$_x$CuO$_4$ probed by terahertz spectroscopy}},
  Nature Physics 7 (2011), pp. 298--302.

\bibitem[366]{Kurosawa2010}
T. Kurosawa et~al., {\itshape Large pseudogap and nodal superconducting gap in
  Bi$_{2}$Sr$_{2-x}$La$_{x}$CuO$_{6+\delta}$ and
  Bi$_{2}$Sr$_{2}$CaC$_{2}$CuO$_{8+\delta}$: Scanning tunneling microscopy and
  spectroscopy}, Phys. Rev. B 81 (2010), p. 094519.

\bibitem[367]{Stojchevska:2011fz}
L. Stojchevska, P. Kusar, T. Mertelj, V.V. Kabanov, Y. Toda, X. Yao, and D.
  Mihailovic, {\itshape {Mechanisms of nonthermal destruction of the
  superconducting state and melting of the charge-density-wave state by
  femtosecond laser pulses}}, Physical Review B 84 (2011), p. 180507(R).

\bibitem[368]{Giannetti2009}
C. Giannetti, G. Zgrablic, C. Consani, A. Crepaldi, D. Nardi, G. Ferrini, G.
  Dhalenne, a. Revcolevschi, and F. Parmigiani, {\itshape {Disentangling
  thermal and nonthermal excited states in a charge-transfer insulator by time-
  and frequency-resolved pump-probe spectroscopy}}, Physical Review B 80
  (2009), p. 235129.

\bibitem[369]{Zhang2013}
W. Zhang, C.L. Smallwood, C. Jozwiak, T.L. Miller, Y. Yoshida, H. Eisaki, D.H.
  Lee, and A. Lanzara, {\itshape Signatures of superconductivity and pseudogap
  formation in nonequilibrium nodal quasiparticles revealed by ultrafast
  angle-resolved photoemission}, Phys. Rev. B 88 (2013), p. 245132.

\bibitem[370]{Smallwood2014}
C.L. Smallwood, W. Zhang, T.L. Miller, C. Jozwiak, H. Eisaki, D.H. Lee, and A.
  Lanzara, {\itshape {Time- and momentum-resolved gap dynamics in
  Bi$_2$Sr$_2$CaCu$_2$O$_{8+\delta}$}}, Phys. Rev. B 89 (2014), p. 115126.

\bibitem[371]{Rameau2014}
J.D. Rameau et~al., {\itshape Photoinduced changes in the cuprate electronic
  structure revealed by femtosecond time- and angle-resolved photoemission},
  Phys. Rev. B 89 (2014), p. 115115.

\bibitem[372]{Piovera2015}
C. Piovera, Z. Zhang, M. {d'Astuto}, A. {Taleb-Ibrahimi}, E. Papalazarou, M.
  Marsi, Z. Li, H. Raffy, and L. Perfetti, {\itshape {Quasiparticles dynamics
  in high-temperature superconductors far from equilibrium: an indication of
  pairing amplitude without phase coherence}}, Phys. Rev. B 91 (2015), p.
  224509.

\bibitem[373]{Coslovich2013b}
G. Coslovich et~al., {\itshape {Competition Between the Pseudogap and
  Superconducting States of
  Bi$_2$Sr$_2$Ca$_{0.92}$Y$_{0.08}$Cu$_2$O$_{8+\delta}$ Single Crystals
  Revealed by Ultrafast Broadband Optical Reflectivity}}, Phys. Rev. Lett. 110
  (2013), p. 107003.

\bibitem[374]{Mihailovic:1998p4077}
D. Mihailovic, B. Podobnik, J. Demsar, and G. Wagner, {\itshape {Divergence of
  the quasiparticle lifetime with doping and evidence for pre-formed pairs
  below $T^*$ in YBa$_2$Cu$_3$O$_{7-\delta}$: direct measurements by
  femtosecond time-resolved spectroscopy.}}, Journal Of Physics And Chemistry
  Of Solids  (1998).

\bibitem[375]{Long:2006gp}
Y.B. Long et~al., {\itshape {Femtosecond optical response of electron-doped
  superconductor La$_{2-x}$Ce$_x$CuO$_4$}}, Physica C: Superconductivity 436
  (2006), pp. 59--61.

\bibitem[376]{daSilvaNeto2015}
E.H. {da Silva Neto}, R. Comin, F. He, R. Sutarto, Y. Jiang, R.L. Greene, G.A.
  Sawatzky, and A. Damascelli, {\itshape {Charge ordering in the electron-doped
  superconductor Nd$_{2-x}$Ce$_x$CuO$_4$}}, Science 347 (2015), pp. 282--285.

\bibitem[377]{Stojchevska:2012ur}
L. Stojchevska, T. Mertelj, J.H. Chu, I.R. Fisher, and D. Mihailovic, {\itshape
  {Doping dependence of femtosecond quasiparticle relaxation dynamics in
  Ba(Fe,Co)$_{2}$As$_{2}$ single crystals: Evidence for normal-state nematic
  fluctuations}}, Physical Review B 86 (2012), p. 024519.

\bibitem[378]{Chia2010}
E.E.M. Chia et~al., {\itshape {Ultrafast Pump-Probe Study of Phase Separation
  and Competing Orders in the Underdoped (Ba,K)Fe$_2$As$_2$ Superconductor}},
  Physical Review Letters 104 (2010), p. 027003.

\bibitem[379]{Mansart2010}
B. Mansart, D. Boschetto, A. Savoia, F. Rullier-Albenque, F. Bouquet, E.
  Papalazarou, A. Forget, D. Colson, A. Rousse, and M. Marsi, {\itshape
  Ultrafast transient response and electron-phonon coupling in the
  iron-pnictide superconductor Ba{Fe$_{1-x}$Co$_x$}$_2$As$_2$}, Phys. Rev. B 82
  (2010), p. 024513.

\bibitem[380]{Rettig2015}
L. Rettig et~al., {\itshape {Ultrafast Structural Dynamics of the Fe-Pnictide
  Parent Compound BaFe$_2$As$_2$}}, Phys. Rev. Lett. 114 (2015), p. 067402.

\bibitem[381]{Kim:2012kh}
K.W. Kim, A. Pashkin, H. Sch{\"a}fer, M. Beyer, M. Porer, T. Wolf, C. Bernhard,
  J. Demsar, R. Huber, and A. Leitenstorfer, {\itshape {Ultrafast transient
  generation of spin-density-wave order in the normal state of BaFe$_2$As$_2$
  driven by coherent lattice vibrations}}, Nature Materials 11 (2012), pp.
  497--501.

\bibitem[382]{Schmitt:2011um}
F. Schmitt, P.S. Kirchmann, U. Bovensiepen, R.G. Moore, J.H. Chu, D.H. Lu, L.
  Rettig, M. Wolf, I.R. Fisher, and Z.X. Shen, {\itshape Ultrafast electron
  dynamics in the charge density wave material TbTe 3}, New Journal of Physics
  13 (2011), p. 063022.

\bibitem[383]{Haufe1975}
U. Haufe, G. Kerker, and K. Bennemann, {\itshape Calculation of the
  superconducting transition temperature $T_c$ for metals with
  phonon-anomalies}, Solid State Communications 17 (1975), pp. 321 -- 325.

\bibitem[384]{Iwai:2006vr}
S. Iwai and H. Okamoto, {\itshape {Ultrafast Phase Control in One-Dimensional
  Correlated Electron Systems (< Special Topics> Photo-Induced Phase
  Transitions and their Dynamics)}}, Journal of the Physical Society of Japan
  75 (2006), p. 011007.

\bibitem[385]{Toda:2011gf}
Y. Toda, {\itshape {Femtosecond Carrier Relaxation Dynamics and Photoinduced
  Phase Separation in $k$-(BEDT-TTF)$_2$Cu[N(CN$_2$])$X$ ($X$=Br,Cl)}},
  Physical Review Letters 107 (2011), p. 227002.

\bibitem[386]{Kawakami:2009p11264}
Y. Kawakami, S. Iwai, T. Fukatsu, M. Miura, N. Yoneyama, T. Sasaki, and N.
  Kobayashi, {\itshape {Optical Modulation of Effective On-Site Coulomb Energy
  for the Mott Transition in an Organic Dimer Insulator}}, Physical Review
  Letters 103 (2009), p. 066403.

\bibitem[387]{Lenarcic2012b}
Z. Lenar\v{c}i\v{c} and P. Prelov\v{s}ek, {\itshape Ultrafast Charge
  Recombination in a Photoexcited Mott-Hubbard Insulator}, Phys. Rev. Lett. 111
  (2013), p. 016401.

\bibitem[388]{Lenarcic2014}
 {Lenar\v{c}i\v{c}, Z. and Prelov\v{s}ek, P.}, {\itshape Charge recombination
  in undoped cuprates}, Phys. Rev. B 90 (2014), p. 235136.

\bibitem[389]{Matsuzaki2015}
H. Matsuzaki, H. Nishioka, H. Uemura, A. Sawa, S. Sota, T. Tohyama, and H.
  Okamoto, {\itshape Ultrafast charge and lattice dynamics in one-dimensional
  Mott insulator of CuO-chain compound Ca$_2$CuO$_3$ investigated by
  femtosecond absorption spectroscopy}, Phys. Rev. B 91 (2015), p. 081114.

\bibitem[390]{Okamoto2007}
S. Okamoto, {\itshape {Nonequilibrium transport and optical properties of model
  metal--Mott-insulator--metal heterostructures}}, Phys. Rev. B 76 (2007), p.
  035105.

\bibitem[391]{Lenarcic2014b}
Z. Lenar\v{c}i\v{c}, D. Gole\v{z}, J. Bon\v{c}a, and P. Prelov\v{s}ek,
  {\itshape Optical response of highly excited particles in a strongly
  correlated system}, Phys. Rev. B 89 (2014), p. 125123.

\bibitem[392]{Golez2012}
D. Gole\v{z}, J. Bon\v{c}a, L. Vidmar, and S.A. Trugman, {\itshape Relaxation
  Dynamics of the Holstein Polaron}, Phys. Rev. Lett. 109 (2012), p. 236402.

\bibitem[393]{Kogoj2014}
J. Kogoj, Z. Lenar\v{c}i\v{c}, D. Gole\v{z}, M. Mierzejewski, P. Prelov\v{s}ek,
  and J. Bon\v{c}a, {\itshape Multistage dynamics of the spin-lattice polaron
  formation}, Phys. Rev. B 90 (2014), p. 125104.

\bibitem[394]{Defilippis2012}
G. De~Filippis, V. Cataudella, E.A. Nowadnick, T.P. Devereaux, A.S. Mishchenko,
  and N. Nagaosa, {\itshape Quantum Dynamics of the Hubbard-Holstein Model in
  Equilibrium and Nonequilibrium: Application to Pump-Probe Phenomena}, Phys.
  Rev. Lett. 109 (2012), p. 176402.

\bibitem[395]{Eckstein2008b}
M. Eckstein and M. Kollar, {\itshape Theory of time-resolved optical
  spectroscopy on correlated electron systems}, Phys. Rev. B 78 (2008), p.
  205119.

\bibitem[396]{Eckstein2014a}
M. Eckstein and P. Werner, {\itshape Ultrafast Separation of Photodoped
  Carriers in Mott Antiferromagnets}, Phys. Rev. Lett. 113 (2014), p. 076405.

\bibitem[397]{DalConte2015}
S. {Dal Conte} et~al., {\itshape {Snapshots of the retarded interaction of
  charge carriers with ultrafast fluctuations in cuprates}}, Nature Physics 11
  (2015), pp. 421--426.

\bibitem[398]{Peli2015}
 S. Peli et al. \textit{The room temperature prodrome of charge-order in copper
  oxides.} arXiv:1508.03075 (2015).

\bibitem[399]{Li2014}
W. Li, C. Zhang, X. Wang, J. Chakhalian, and M. Xiao, {\itshape Ultrafast
  spectroscopy of quasiparticle dynamics in cuprate superconductors}, Journal
  of Magnetism and Magnetic Materials  (2014), pp. {29--39}.

\bibitem[400]{Kresin2009}
V. Kresin and S. Wolf, {\itshape {Colloquium: Electron-lattice interaction and
  its impact on high Tc superconductivity}}, Reviews of Modern Physics 81
  (2009), pp. 481--501.

\bibitem[401]{Maksimov2010}
E. Maksimov, M. Kulic, and O. Dolgov, {\itshape {Bosonic Spectral Function and
  The Electron-Phonon Interaction in HTSC Cuprates}}, Advances in Condensed
  Matter Physics  (2010), p. 423725.

\bibitem[402]{Yin2009}
W.G. Yin and W. Ku, {\itshape {Tuning the in-plane electron behavior in high-Tc
  cuprate superconductors via apical atoms: A first-principles Wannier-states
  analysis}}, Physical Review B 79 (2009), p. 214512.

\bibitem[403]{Reznik2006}
D. Reznik, L. Pintschovius, M. Ito, S. Iikubo, M. Sato, H. Goka, M. Fujita, K.
  Yamada, G.D. Gu, and J.M. Tranquada, {\itshape {Electron-phonon coupling
  reflecting dynamic charge inhomogeneity in copper oxide superconductors.}},
  Nature 440 (2006), pp. 1170--3.

\bibitem[404]{Mansart2013}
B. Mansart et~al., {\itshape Temperature-dependent electron-phonon coupling in
  La$_{2-x}$Sr$_x$CuO$_4$ probed by femtosecond x-ray diffraction}, Phys. Rev.
  B 88 (2013), p. 054507.

\bibitem[405]{Gadermaier2014}
C. Gadermaier et~al., {\itshape Strain-Induced Enhancement of the Electron
  Energy Relaxation in Strongly Correlated Superconductors}, Phys. Rev. X 4
  (2014), p. 011056.

\bibitem[406]{Rettig2012}
L. Rettig, R. Cort\'es, S. Thirupathaiah, P. Gegenwart, H.S. Jeevan, M. Wolf,
  J. Fink, and U. Bovensiepen, {\itshape {Ultrafast Momentum-Dependent Response
  of Electrons in Antiferromagnetic EuFe$_2$As$_2$ Driven by Optical
  Excitation}}, Physical Review Letters 108 (2012), p. 097002.

\bibitem[407]{Rettig2013}
L. Rettig, R. Cortés, H.S. Jeevan, P. Gegenwart, T. Wolf, J. Fink, and U.
  Bovensiepen, {\itshape Electron-phonon coupling in 122 Fe pnictides analyzed
  by femtosecond time-resolved photoemission}, New Journal of Physics 15
  (2013), p. 083023.

\bibitem[408]{Avigo2013}
I. Avigo et~al., {\itshape {Coherent excitations and electron-phonon coupling
  in Ba/EuFe$_2$As$_2$ compounds investigated by femtosecond time- and
  angle-resolved photoemission spectroscopy}}, Journal of Physics: Condensed
  Matter 25 (2013), p. 094003.

\bibitem[409]{Stojchevska2010}
L. Stojchevska, P. Kusar, T. Mertelj, V.V. Kabanov, X. Lin, G.H. Cao, Z.A. Xu,
  and D. Mihailovic, {\itshape {Electron-phonon coupling and the charge gap of
  spin-density wave iron-pnictide materials from quasiparticle relaxation
  dynamics}}, Phys. Rev. B  (2010), pp. 1--4.

\bibitem[410]{Mertelj2010}
T. Mertelj, P. Kusar, V.V. Kabanov, L. Stojchevska, N.D. Zhigadlo, S. Katrych,
  Z. Bukowski, and J. Karpinski, {\itshape {Quasiparticle relaxation dynamics
  in spin-density-wave and superconducting SmFeAsO$_{1- x}$F$_x$ single
  crystals}}, Phys. Rev. B  (2010), pp. 1--9.

\bibitem[411]{Ulstrup2014}
S. Ulstrup, J.C. Johannsen, M. Grioni, and P. Hofmann, {\itshape Extracting the
  temperature of hot carriers in time- and angle-resolved photoemission},
  Review of Scientific Instruments 85 (2014), 013907, pp.~{--}.

\bibitem[412]{Golez2014}
D. Gole\v{z}, J. Bon\v{c}a, M. Mierzejewski, and L. Vidmar, {\itshape
  {Mechanism of Ultrafast Relaxation of a Photo-Carrier in Antiferromagnetic
  Spin Background}}, Physical Review B 89 (2014), p. 165118.

\bibitem[413]{Eckstein2014}
 Eckstein, M. and Werner, P. \textit{Ultra-fast photo-carrier relaxation in
  Mott insulators with short- range spin correlations}. arXiv:1410.3956 (2014).

\bibitem[414]{Dorfner2015}
F. Dorfner, L. Vidmar, C. Brockt, E. Jeckelmann, and F. Heidrich-Meisner,
  {\itshape Real-time decay of a highly excited charge carrier in the
  one-dimensional Holstein model}, Phys. Rev. B 91 (2015), p. 104302.

\bibitem[415]{Gull2014}
E. Gull and A.J. Millis, {\itshape Pairing glue in the two-dimensional Hubbard
  model}, Phys. Rev. B 90 (2014), p. 041110.

\bibitem[416]{Kogoj2015}
 J. Kogoj et al. \textit{Thermalization after photoexcitation from the
  perspective of optical spectroscopy.} arXiv:1509.08431 (2015).

\bibitem[417]{Sentef2013}
M. Sentef, A.F. Kemper, B. Moritz, J.K. Freericks, Z.X. Shen, and T.P.
  Devereaux, {\itshape Examining Electron-Boson Coupling Using Time-Resolved
  Spectroscopy}, Phys. Rev. X 3 (2013), p. 041033.

\bibitem[418]{Cilento2014}
F. Cilento et~al., {\itshape {Photo-enhanced antinodal conductivity in the
  pseudogap state of high-Tc cuprates}}, Nature Communications 5 (2014), p.
  4353.

\bibitem[419]{He2011}
R.H. He et~al., {\itshape From a Single-Band Metal to a High-Temperature
  Superconductor via Two Thermal Phase Transitions}, Science 331 (2011), pp.
  1579--1583.

\bibitem[420]{Madan2015}
I. Madan, T. Kurosawa, Y. Toda, M. Oda, T. Mertelj, and D. Mihailovic,
  {\itshape {Evidence for carrier localization in the pseudogap state of
  cuprate superconductors from coherent quench experiments}}, Nature
  Communications 6 (2015), p. 6958.

\bibitem[421]{Saichu2008}
R.P. Saichu et~al., {\itshape {Two-Component Dynamics of the Order Parameter of
  High Temperature Bi$_2$Sr$_2$CaCu$_2$O$_{8+\delta}$ Superconductors Revealed
  by Time-Resolved Raman Scattering}}, Phys. Rev. Lett. 102 (2009), p. 177004.

\bibitem[422]{Dresselhaus1962}
G. Dresselhaus and M.S. Dresselhaus, {\itshape Interband Transitions in
  Superconductors}, Phys. Rev. 125 (1962), pp. 1212--1214.

\bibitem[423]{Molegraaf2002}
H.J.A. Molegraaf, C. Presura, D. Marelvan~der , P.H. Kes, and M. Li, {\itshape
  {Superconductivity-Induced Transfer of In-Plane Spectral Weight in
  Bi$_2$Sr$_2$CaCu$_2$O$_{8+\delta}$}}, Science 295 (2002), pp. 2239--2241.

\bibitem[424]{Deutscher2005}
G. Deutscher, A.F. Santander-Syro, and N. Bontemps, {\itshape Kinetic energy
  change with doping upon superfluid condensation in high-temperature
  superconductors}, Phys. Rev. B 72 (2005), p. 092504.

\bibitem[425]{Carbone2006}
 {Carbone, F. and Kuzmenko, A. B. and Molegraaf, H. J. A. and {van Heumen}, E.
  and Lukovac, V. and Marsiglio, F. and {van der Marel}, D. and Haule, K. and
  Kotliar, G. and Berger, H. and Courjault, S. and Kes, P. H. and Li, M.},
  {\itshape {Doping dependence of the redistribution of optical spectral weight
  in Bi$_2$Sr$_2$CaCu$_2$O$_{8+\delta}$}}, Phys. Rev. B 74 (2006), p. 064510.

\bibitem[426]{Norman2002}
M.R. Norman and C. P\'epin, {\itshape Quasiparticle formation and optical sum
  rule violation in cuprate superconductors}, Phys. Rev. B 66 (2002), p.
  100506.

\bibitem[427]{Fausti2014}
 D. Fausti et al. \textit{Dynamical coupling between off-plane phonons and
  in-plane electronic excitations in superconducting YBCO}. arXiv:1408.0888
  (2014).

\bibitem[428]{Graf2011}
J. Graf, C. Jozwiak, C.L. Smallwood, H. Eisaki, R.A. Kaindl, D.H. Lee, and A.
  Lanzara, {\itshape {Nodal quasiparticle meltdown in ultrahigh-resolution
  pump-probe angle-resolved photoemission}}, Nat Phys 7 (2011), pp. 805--809.

\bibitem[429]{Rameau2015}
 Rameau, J.D. et al. \textit{Time-resolved Boson Emission in the Excitation
  Spectrum of Bi$_2$Sr$_2$CaCu$_2$O$_{8+\delta}$}. arXiv:1505.07055 (2015).

\bibitem[430]{Yang2015}
S.L. Yang, J.A. Sobota, D. Leuenberger, Y. He, M. Hashimoto, D.H. Lu, H.
  Eisaki, P.S. Kirchmann, and Z.X. Shen, {\itshape Inequivalence of
  Single-Particle and Population Lifetimes in a Cuprate Superconductor}, Phys.
  Rev. Lett. 114 (2015), p. 247001.

\bibitem[431]{Zhang2014}
W. Zhang et~al., {\itshape {Ultrafast quenching of electron-boson interaction
  and superconducting gap in a cuprate superconductor}}, Nat Commun 5 (2014).

\bibitem[432]{Madan2014}
I. Madan, T. Kurosawa, Y. Toda, M. Oda, T. Mertelj, P. Kusar, and D.
  Mihailovic, {\itshape {Separating pairing from quantum phase coherence
  dynamics above the superconducting transition by femtosecond spectroscopy}},
  Sci. Rep. 4 (2014), p. 5656.

\bibitem[433]{Forst2011}
M. Forst, C. Manzoni, S. Kaiser, Y. Tomioka, Y. Tokura, R. Merlin, and A.
  Cavalleri, {\itshape {Nonlinear phononics as an ultrafast route to lattice
  control}}, Nat. Phys. 7 (2011), pp. 854--856.

\bibitem[434]{Chia2007}
E. Chia, J.X. Zhu, D. Talbayev, R. Averitt, a. Taylor, K.H. Oh, I.S. Jo, and
  S.I. Lee, {\itshape {Observation of Competing Order in a High-$T_c$
  Superconductor Using Femtosecond Optical Pulses}}, Physical Review Letters 99
  (2007), p. 147008.

\bibitem[435]{Mansart2010b}
B. Mansart, D. Boschetto, S. Sauvage, A. Rousse, and M. Marsi, {\itshape Mott
  transition in Cr-doped V$_2$O$_3$f studied by ultrafast reflectivity:
  Electron correlation effects on the transient response}, EPL (Europhysics
  Letters) 92 (2010), p. 37007.

\bibitem[436]{Torchinsky2013}
D.H. Torchinsky, F. Mahmood, A.T. Bollinger, I. Bo\v{z}ovi\'{c}, and N. Gedik,
  {\itshape {Fluctuating charge-density waves in a cuprate superconductor}},
  Nat Mater 12 (2013), pp. 387--391.

\bibitem[437]{Matsunaga2014}
R. Matsunaga, N. Tsuji, H. Fujita, A. Sugioka, K. Makise, Y. Uzawa, H. Terai,
  Z. Wang, H. Aoki, and R. Shimano, {\itshape Light-induced collective
  pseudospin precession resonating with Higgs mode in a superconductor},
  Science 345 (2014), pp. 1145--1149.

\bibitem[438]{Mazin1994}
I.I. Mazin, A.I. Liechtenstein, O. Jepsen, O.K. Andersen, and C.O. Rodriguez,
  {\itshape {Displacive excitation of coherent phonons in
  YBa$_2$Cu$_3$O$_{7-x}$}}, Phys. Rev. B 49 (1994), pp. 9210--9213.

\bibitem[439]{Yang2014}
X. Yang L. et~al., {\itshape Ultrafast Modulation of the Chemical Potential in
  ${\mathrm{BaFe}}_{2}{\mathrm{As}}_{2}$ by Coherent Phonons}, Phys. Rev. Lett.
  112 (2014), p. 207001.

\bibitem[440]{Mansart2012}
B. Mansart, M.J.G. Cottet, T.J. Penfold, S.B. Dugdale, R. Tediosi, M. Chergui,
  and F. Carbone, {\itshape {Evidence for a Peierls phase-transition in a
  three-dimensional multiple charge-density waves solid.}}, Proceedings of the
  National Academy of Sciences of the United States of America 109 (2012), pp.
  5603--8.

\bibitem[441]{Hellmann2012}
S. Hellmann et~al., {\itshape {Time-domain classification of
  charge-density-wave insulators}}, Nature Communications 3 (2012), p. 1069.

\bibitem[442]{Eichberger:2010p12340}
M. Eichberger, H. Sch{\"a}fer, M. Krumova, and M. Beyer, {\itshape {Snapshots
  of cooperative atomic motions in the optical suppression of charge density
  waves}}, Nature 468 (2010), pp. 799{\textendash}802--802.

\bibitem[443]{Schafer2010}
H. Sch\"afer, V.V. Kabanov, M. Beyer, K. Biljakovic, and J. Demsar, {\itshape
  Disentanglement of the Electronic and Lattice Parts of the Order Parameter in
  a 1D Charge Density Wave System Probed by Femtosecond Spectroscopy}, Phys.
  Rev. Lett. 105 (2010), p. 066402.

\bibitem[444]{Schafer2014}
H. Sch\"afer, V.V. Kabanov, and J. Demsar, {\itshape Collective modes in
  quasi-one-dimensional charge-density wave systems probed by femtosecond
  time-resolved optical studies}, Phys. Rev. B 89 (2014), p. 045106.

\bibitem[445]{Rettig2014}
L. Rettig, J.H. Chu, I.R. Fisher, U. Bovensiepen, and M. Wolf, {\itshape
  Coherent dynamics of the charge density wave gap in tritellurides}, Faraday
  Discuss. 171 (2014), pp. 299--310.

\bibitem[446]{Huber2014}
T. Huber et~al., {\itshape Coherent Structural Dynamics of a Prototypical
  Charge-Density-Wave-to-Metal Transition}, Phys. Rev. Lett. 113 (2014), p.
  026401.

\bibitem[447]{Shen2014}
W. Shen, T.P. Devereaux, and J.K. Freericks, {\itshape Exact solution for Bloch
  oscillations of a simple charge-density-wave insulator}, Phys. Rev. B 89
  (2014), p. 235129.

\bibitem[448]{Shen2014b}
W. Shen, Y. Ge, Y. Liu A.\, R. Krishnamurthy H.\, P. Devereaux T.\, and K.
  Freericks J.\, {\itshape Nonequilibrium ``Melting'' of a Charge Density Wave
  Insulator via an Ultrafast Laser Pulse}, Phys. Rev. Lett. 112 (2014), p.
  176404.

\bibitem[449]{Achkar2012}
A.J. Achkar et~al., {\itshape {Distinct Charge Orders in the Planes and Chains
  of Ortho-III-Ordered YBa$_2$Cu$_3$O$_{6+\delta}$ Superconductors Identified
  by Resonant Elastic X-ray Scattering}}, Phys. Rev. Lett. 109 (2012), p.
  167001.

\bibitem[450]{daSilvaNeto2014}
E.H. {da Silva Neto} et~al., {\itshape {Ubiquitous interplay between charge
  ordering and high-temperature superconductivity in cuprates.}}, Science 343
  (2014), pp. 393--6.

\bibitem[451]{BlancoCanosa2014}
S. Blanco-Canosa, A. Frano, E. Schierle, J. Porras, T. Loew, M. Minola, M.
  Bluschke, E. Weschke, B. Keimer, and M. Le~Tacon, {\itshape {Resonant x-ray
  scattering study of charge-density wave correlations in
  YBa$_2$Cu$_3$O$_{6+x}$}}, Phys. Rev. B 90 (2014), p. 054513.

\bibitem[452]{Tabis2014}
W. Tabis et~al., {\itshape {Charge order and its connection with Fermi-liquid
  charge transport in a pristine high-Tc cuprate}}, Nat. Commun. 5 (2014), p.
  5875.

\bibitem[453]{Comin2015}
R. Comin, R. Sutarto, E. {da Silva Neto}, L. Chauviere, R. Liang, W. Hardy, D.
  Bonn, F. He, G. Sawatzky, and A. Damascelli, {\itshape {Broken translational
  and rotational symmetry via charge stripe order in underdoped
  YBa$_2$Cu$_3$O$_{6+y}$}}, Science 347 (2015), pp. 1335--9.

\bibitem[454]{Anderson2015}
P.W. Anderson, {\itshape {Superconductivity: Higgs, Anderson and all that}},
  Nat Phys 11 (2015), p.~93.

\bibitem[455]{Varma2001}
 Varma, C.M. \textit{Higgs Boson in Superconductors}. arXiv:0109409 (2001).

\bibitem[456]{Littlewood1982}
P.B. Littlewood and C.M. Varma, {\itshape Amplitude collective modes in
  superconductors and their coupling to charge-density waves}, Phys. Rev. B 26
  (1982), pp. 4883--4893.

\bibitem[457]{Cea2014}
T. Cea and L. Benfatto, {\itshape Nature and Raman signatures of the Higgs
  amplitude mode in the coexisting superconducting and charge-density-wave
  state}, Phys. Rev. B 90 (2014), p. 224515.

\bibitem[458]{Measson2014}
M.A. M\'easson, Y. Gallais, M. Cazayous, B. Clair, P. Rodi\`ere, L. Cario, and
  A. Sacuto, {\itshape {Amplitude Higgs mode in the 2H-NbSe$_2$
  superconductor}}, Phys. Rev. B 89 (2014), p. 060503.

\bibitem[459]{Kemper2015}
A. Kemper, M. Sentef, B. Moritz, J. Freericks, and T. Devereaux, {\itshape
  {Direct observation of Higgs mode oscillations in the pump-probe
  photoemission spectra of electron-phonon mediated superconductors}}, Phys.
  Rev. B 92 (2015), p. 224517.

\bibitem[460]{Matsunaga2012}
R. Matsunaga and R. Shimano, {\itshape Nonequilibrium BCS State Dynamics
  Induced by Intense Terahertz Pulses in a Superconducting NbN Film}, Phys.
  Rev. Lett. 109 (2012), p. 187002.

\bibitem[461]{Krull2015}
 H. Krull et al. \textit{Coupling of Higgs and Leggett modes in nonequilibrium
  superconductors.} arXiv:1512.08121 (2015).

\bibitem[462]{Krull2014}
H. Krull, D. Manske, G.S. Uhrig, and A.P. Schnyder, {\itshape Signatures of
  nonadiabatic BCS state dynamics in pump-probe conductivity}, Phys. Rev. B 90
  (2014), p. 014515.

\bibitem[463]{Matsunaga2013}
R. Matsunaga, Y.I. Hamada, K. Makise, Y. Uzawa, H. Terai, Z. Wang, and R.
  Shimano, {\itshape {Higgs Amplitude Mode in the BCS Superconductors
  Nb$_{1-x}$Ti$_{x}$N Induced by Terahertz Pulse Excitation}}, Physical Review
  Letters 111 (2013), p. 057002.

\bibitem[464]{Cea2015}
 T. Cea et al. \textit{Non-linear optical effects and third-harmonic generation
  in superconductors: Cooper-pairs vs Higgs mode contribution.}
  arXiv:1512.02544 (2015).

\bibitem[465]{Barlas2013}
Y. Barlas and C.M. Varma, {\itshape {Amplitude or Higgs modes in $d$-wave
  superconductors}}, Physical Review B 87 (2013), p. 054503.

\bibitem[466]{Schnyder2011}
A.P. Schnyder, D. Manske, and A. Avella, {\itshape Resonant generation of
  coherent phonons in a superconductor by ultrafast optical pump pulses}, Phys.
  Rev. B 84 (2011), p. 214513.

\bibitem[467]{PIPT1-5}
Chapter title. {\itshape Photo-induced phase transition conference series}
  (2014), .

\bibitem[468]{Nasu2004}
K. Nasu {\itshape Photo-indiced phase transition},    World Scientific Pub Co
  Inc, 2004.

\bibitem[469]{Chanduri2010}
S. Chaudhuri, N.K. Pandey, S. Saini, and R.C. Budhani, {\itshape {Dynamics of a
  robust photo-induced insulator-metal transition driven by coherent and
  broad-band light in epitaxial films of
  La$_{0.625-y}$Pr$_y$Ca$_{0.375}$MnO$_3$}}, Journal of Physics: Condensed
  Matter 22 (2010), p. 275502.

\bibitem[470]{Rini2009}
M. Rini, R. Tobey, N. Dean, S. Wall, H. Ehrke, Y. Zhu, Y. Tomioka, Y. Tokura,
  R.W. Schoenlein, and A. Cavalleri, {\itshape Time-resolved studies of phase
  transition dynamics in strongly correlated manganites}, Journal of Physics:
  Conference Series 148 (2009), p. 012013.

\bibitem[471]{Takubo2008}
N. Takubo, I. Onishi, K. Takubo, T. Mizokawa, and K. Miyano, {\itshape
  Photoinduced Metal-to-Insulator Transition in a Manganite Thin Film}, Phys.
  Rev. Lett. 101 (2008), p. 177403.

\bibitem[472]{Ichikawa2011}
H. Ichikawa et~al., {\itshape Transient photoinduced hidden phase in a
  manganite}, Nat. Materials 10 (2011), p. 101.

\bibitem[473]{Cavalleri2004}
A. Cavalleri, T. Dekorsy, H.H.W. Chong, J.C. Kieffer, and R.W. Schoenlein,
  {\itshape {Evidence for a structurally-driven insulator-to-metal transition
  in VO$_2$: A view from the ultrafast timescale}}, Phys. Rev. B 70 (2004), p.
  161102.

\bibitem[474]{Liu2011}
M.K. Liu, B. Pardo, J. Zhang, M.M. Qazilbash, S.J. Yun, Z. Fei, J.H. Shin, H.T.
  Kim, D.N. Basov, and R.D. Averitt, {\itshape {Photoinduced Phase Transitions
  by Time-Resolved Far-Infrared Spectroscopy in V$_2$O$_3$}}, Phys. Rev. Lett.
  107 (2011), p. 066403.

\bibitem[475]{Hada2010}
M. Hada, K. Okimura, and J. Matsuo, {\itshape {Characterization of structural
  dynamics of VO$_2$ thin film on $c$-Al$_2$O$_3$ using in-air time-resolved
  x-ray diffraction}}, Phys. Rev. B 82 (2010), p. 153401.

\bibitem[476]{Baum2007}
P. Baum, D.S. Yang, and A.H. Zewail, {\itshape 4D Visualization of Transitional
  Structures in Phase Transformations by Electron Diffraction}, Science 318
  (2007), pp. 788--792.

\bibitem[477]{Morrison2014}
V.R. Morrison, R.P. Chatelain, K.L. Tiwari, A. Hendaoui, A. Bruhacs, M. Chaker,
  and B.J. Siwick, {\itshape A photoinduced metal-like phase of monoclinic
  VO$_2$ revealed by ultrafast electron diffraction}, Science 346 (2014), pp.
  445--448.

\bibitem[478]{Hilton2007}
D.J. Hilton, R.P. Prasankumar, S. Fourmaux, A. Cavalleri, D. Brassard, M.A.
  El~Khakani, J.C. Kieffer, A.J. Taylor, and R.D. Averitt, {\itshape Enhanced
  Photosusceptibility near ${T}_{c}$ for the Light-Induced Insulator-to-Metal
  Phase Transition in Vanadium Dioxide}, Phys. Rev. Lett. 99 (2007), p. 226401.

\bibitem[479]{Wall2012}
S. Wall, D. Wegkamp, L. Foglia, K. Appavoo, J. Nag, R. Haglund, J. St\"ahler,
  and M. Wolf, {\itshape Ultrafast changes in lattice symmetry probed by
  coherent phonons}, Nat. Comm. 3 (2012), p. 721.

\bibitem[480]{Cocker2012}
T.L. Cocker, L.V. Titova, S. Fourmaux, G. Holloway, H.C. Bandulet, D. Brassard,
  J.C. Kieffer, M.A. El~Khakani, and F. Hegmann, {\itshape Phase diagram of the
  ultrafast photoinduced insulator-metal transition in vanadium dioxide}, Phys.
  Rev. B 85 (2012), p. 155120.

\bibitem[481]{Hsieh2014}
W.P. Hsieh, M. Trigo, D.A. Reis, G. Andrea~Artioli, L. Malavasi, and W.L. Mao,
  {\itshape Evidence for photo-induced monoclinic metallic VO$_2$ under high
  pressure}, Applied Physics Letters 104 (2014), 021917.

\bibitem[482]{Wegkamp2014}
D. Wegkamp et~al., {\itshape Instantaneous Band Gap Collapse in Photoexcited
  Monoclinic VO$_2$ due to Photocarrier Doping}, Phys. Rev. Lett. 113 (2014),
  p. 216401.

\bibitem[483]{Biermann2005}
S. Biermann, A. Poteryaev, A.I. Lichtenstein, and A. Georges, {\itshape
  Dynamical Singlets and Correlation-Assisted Peierls Transition in VO$_2$},
  Phys. Rev. Lett. 94 (2005), p. 026404.

\bibitem[484]{Rini2008}
M. Rini et~al., {\itshape Optical switching in VO$_2$ films by below-gap
  excitation}, Applied Physics Letters 92 (2008), 181904, p. 181904.

\bibitem[485]{Yoshida2014}
R. Yoshida et~al., {\itshape Ultrafast photoinduced transition of an insulating
  ${\mathrm{VO}}_{2}$ thin film into a nonrutile metallic state}, Phys. Rev. B
  89 (2014), p. 205114.

\bibitem[486]{Stojchevska2014}
L. Stojchevska, I. Vaskivskyi, T. Mertelj, P. Kusar, D. Svetin, S. Brazovskii,
  and D. Mihailovic, {\itshape Ultrafast Switching to a Stable Hidden Quantum
  State in an Electronic Crystal}, Science 344 (2014), pp. 177--180.

\bibitem[487]{Testardi:1971p1592}
L. Testardi, {\itshape {Destruction of Superconductivity by Laser Light}},
  Physical Review B 4 (1971), p. 2189.

\bibitem[488]{Demsar2003}
J. Demsar, R.D. Averitt, A.J. Taylor, V.V. Kabanov, W.N. Kang, H.J. Kim, E.M.
  Choi, and S.I. Lee, {\itshape {Pair-Breaking and Superconducting State
  Recovery Dynamics in MgB$_2$}},  91 (2003), p. 267002.

\bibitem[489]{Kumar1968}
N. Kumar and K. Sinha, {\itshape {Possibility of Photoinduced
  Superconductivity}}, Physical Review  (1968).

\bibitem[490]{Mihailovic:1990p1474}
D. Mihailovic and A.J. Heeger, {\itshape {Pyroelectric and piezoelectric
  effects in single crystals of YBa$_{2}$Cu$_{3}$O$_{7-\delta}$}}, Solid State
  Communications 75 (1990), pp. 319--323.

\bibitem[491]{Fugol1990}
I.A. Fugol, V.N. Samovarov, I.I. Rybalko, and V.M. Zhuravlev, {\itshape
  {Luminescence of charge-sensitive oxygen centers in high-temperature
  superconductors - The effect of superconducting transition}}, Fizika Nizkikh
  Temperatur 16 (1990), pp. 580--583.

\bibitem[492]{Tanabe:1994kp}
K. Tanabe, F.H. Teherani, S. Kubo, H. Asano, and M. Suzuki, {\itshape {Effects
  of photoinduced hole doping on transport properties of YBa$_2$Cu$_3$O$_y$
  grain boundary junctions}}, Journal Of Applied Physics 76 (1994), pp.
  3679--3683.

\bibitem[493]{Hayes2004}
W. Hayes and A.M. Stoneham {\itshape {Defects and Defect Processes in
  Nonmetallic Solids}},  , 2004.

\bibitem[494]{Bridges1990}
F. Bridges, G. Davies, J. Robertson, and A.M. Stoneham, {\itshape {The
  spectroscopy of crystal defects: a compendium of defect nomenclature}},
  Journal Of Physics-Condensed Matter 2 (1990), pp. 2875--2928.

\bibitem[495]{Kudinov1993}
V.I. Kudinov, I.L. Chaplygin, A.I. Kirilyuk, N.M. Kreines, R. Laiho, E.
  L{\"a}hderanta, and C. Ayache, {\itshape {Persistent photoconductivity in
  YBa$_ {2}$ Cu$_ {3} $O$_ {6+ x}$ films as a method of photodoping toward
  metallic and superconducting phases}}, Physical Review B 47 (1993), p. 9017.

\bibitem[496]{Hoffmann1997}
A. Hoffmann, D. Reznik, and I.K. Schuller, {\itshape {Persistent photoinduced
  effects in high-Te superconductors}}, Advanced Materials 9 (1997), pp.
  271--273.

\bibitem[497]{Federici1995}
J.F. Federici, D. Chew, B. Welker, W. Savin, J. Gutierrez-Solana, T. Fink, and
  W. Wilber, {\itshape Defect mechanism of photoinduced superconductivity in
  YBa$_2$Cu$_3$O$_{6+x}$}, Phys. Rev. B 52 (1995), pp. 15592--15597.

\bibitem[498]{Bahrs2004}
S. Bahrs, A. Go{\~n}i, C. Thomsen, B. Maiorov, G. Nieva, and A. Fainstein,
  {\itshape {Light-induced oxygen-ordering dynamics in
  (Y,Pr)Ba$_{2}$Cu$_{3}$O$_{6.7}$: A Raman spectroscopy and Monte Carlo
  study}}, Physical Review B 70 (2004), p. 014512.

\bibitem[499]{Bahrs2005}
S. Bahrs, J. Guimpel, A.R. Go{\~n}i, B. Maiorov, A. Fainstein, G. Nieva, and C.
  Thomsen, {\itshape {Persistent photo-excitation in
  GdBa$_{2}$Cu$_{3}$O$_{6.5}$ in a simultaneous Raman and electrical-transport
  experiment}}, Physical Review B 72 (2005), p. 144501.

\bibitem[500]{Bruchhausen:2004hda}
A. Bruchhausen, S. Bahrs, K. Fleischer, A. Go{\~n}i, A. Fainstein, G. Nieva, A.
  Aligia, W. Richter, and C. Thomsen, {\itshape {Photoinduced chain-oxygen
  ordering in detwinned YBa$_2$Cu$_3$O$_{6.7}$ single crystals studied by
  reflectance-anisotropy spectroscopy}}, Physical Review B 69 (2004), p.
  224508.

\bibitem[501]{Pena2006}
V. Pe{\~n}a, T. Gredig, J. Santamaria, and I. Schuller, {\itshape
  {Interfacially Controlled Transient Photoinduced Superconductivity}},
  Physical Review Letters 97 (2006), p. 177005.

\bibitem[502]{Yu1991}
G. Yu, C.H. Lee, A. Heeger, N. Herron, and E. Mccarron, {\itshape {Transient
  Photoinduced Conductivity in Single-Crystals of
  YBa$_{2}$Cu$_{3}$O$_{7-\delta}$ - Photodoping to the Metallic State}},
  Physical Review Letters 67 (1991), pp. 2581--2584.

\bibitem[503]{Gedik2007}
N. Gedik, D.S. Yang, G. Logvenov, I. Bozovic, and A.H. Zewail, {\itshape
  {Nonequilibrium phase transitions in cuprates observed by ultrafast electron
  crystallography}}, Science (New York, N.Y.) 316 (2007), pp. 425--9.

\bibitem[504]{Fausti2011}
D. Fausti, R.I. Tobey, N. Dean, S. Kaiser, A. Dienst, M.C. Hoffmann, S. Pyon,
  T. Takayama, H. Takagi, and A. Cavalleri, {\itshape Light-Induced
  Superconductivity in a Stripe-Ordered Cuprate}, Science 331 (2011), pp.
  189--191.

\bibitem[505]{Nicoletti2014}
D. Nicoletti, E. Casandruc, Y. Laplace, V. Khanna, C.R. Hunt, S. Kaiser, S.S.
  Dhesi, G.D. Gu, J.P. Hill, and A. Cavalleri, {\itshape Optically induced
  superconductivity in striped
  ${\mathrm{La}}_{2\ensuremath{-}x}{\mathrm{Ba}}_{x}{\mathrm{CuO}}_{4}$ by
  polarization-selective excitation in the near infrared}, Phys. Rev. B 90
  (2014), p. 100503.

\bibitem[506]{Tamasaku1992}
K. Tamasaku, Y. Nakamura, and S. Uchida, {\itshape Charge dynamics across the
  ${\mathrm{CuO}}_{2}$ planes in
  ${\mathrm{La}}_{2\mathrm{-}\mathit{x}}$${\mathrm{Sr}}_{\mathit{x}}$${\mathrm{CuO}}_{4}$},
  Phys. Rev. Lett. 69 (1992), pp. 1455--1458.

\bibitem[507]{Subedi2014}
A. Subedi, A. Cavalleri, and A. Georges, {\itshape Theory of nonlinear
  phononics for coherent light control of solids}, Phys. Rev. B 89 (2014), p.
  220301.

\bibitem[508]{Denny2015}
S.J. Denny, S.R. Clark, Y. Laplace, A. Cavalleri, and D. Jaksch, {\itshape
  Proposed Parametric Cooling of Bilayer Cuprate Superconductors by Terahertz
  Excitation}, Phys. Rev. Lett. 114 (2015), p. 137001.

\bibitem[509]{Hoppner2015}
R. H\"oppner, B. Zhu, T. Rexin, A. Cavalleri, and L. Mathey, {\itshape
  Redistribution of phase fluctuations in a periodically driven cuprate
  superconductor}, Phys. Rev. B 91 (2015), p. 104507.

\bibitem[510]{Baskaran2012}
 Baskaran, G. \textit{Superradiant superconductivity}. arxiv.org/abs/1211.4567
  (2012).

\bibitem[511]{Orenstein2015}
 Orenstein, J. and Dodge J. S. \textit{Terahertz time-domain spectroscopy of
  transient metallic and superconducting states}. arxiv.org/abs/1506.06758
  (2015).

\bibitem[512]{Nicoletti2015}
 {Nicoletti, D. and Mitrano, M. and Cantaluppi, A. and Cavalleri A.
  \textit{Comment on "Terahertz time-domain spectroscopy of transient metallic
  and superconducting states"}. arxiv.org/abs/1506.07846 (2015)}.

\bibitem[513]{Dienst2011}
 DienstA., H. C.,  FaustiD., P. C.,  PyonS.,  TakayamaT.,  TakagiH., and
  CavalleriA., {\itshape Bi-directional ultrafast electric-field gating of
  interlayer charge transport in a cuprate superconductor}, Nat Photon 5
  (2011), pp. 485--488.

\bibitem[514]{Dienst2013}
A. Dienst et~al., {\itshape Optical excitation of Josephson plasma solitons in
  a cuprate superconductor}, Nat Mater 12 (2013), pp. 535--541 Article.

\bibitem[515]{Glossner2012}
 A. Glossner et al. \textit{Cooper Pair Breakup in YBCO under Strong Terahertz
  Fields}. arXiv:1205.1684 (2012).

\bibitem[516]{Dakovski2015}
G.L. Dakovski, W.S. Lee, D.G. Hawthorn, N. Garner, D. Bonn, W. Hardy, R. Liang,
  M.C. Hoffmann, and J.J. Turner, {\itshape Enhanced coherent oscillations in
  the superconducting state of underdoped YBa$_2$Cu$_3$O$_{6+x}$ induced via
  ultrafast terahertz excitation}, Phys. Rev. B 91 (2015), p. 220506.

\bibitem[517]{Zhang1988}
F.C. Zhang and T.M. Rice, {\itshape Effective Hamiltonian for the
  superconducting Cu oxides}, Phys. Rev. B 37 (1988), pp. 3759--3761.

\bibitem[518]{Chu2002}
S. Chu, {\itshape Cold atoms and quantum control}, Nature 416 (2002), p. 206.

\bibitem[519]{Gutzwiller1964}
M.C. Gutzwiller, {\itshape Effect of Correlation on the Ferromagnetism of
  Transition Metals}, Phys. Rev. 134 (1964), p. A923.

\bibitem[520]{Vidal2004}
G. Vidal, {\itshape Efficient Simulation of One-Dimensional Quantum Many-Body
  Systems}, Phys. Rev. Lett. 93 (2004), p. 040502.

\bibitem[521]{Feiguin2004}
S.R. White and A.E. Feiguin, {\itshape Real-Time Evolution Using the Density
  Matrix Renormalization Group}, Phys. Rev. Lett. 93 (2004), p. 076401.

\bibitem[522]{Breuer2002}
H. Breuer and F. Petruccione {\itshape The Theory of Open Quantum Systems},
  Oxford University Press, 2002.

\bibitem[523]{Rammer1998}
J. Rammer {\itshape Quantum Transport Theory},  Frontiers in physics  Perseus
  Books, 1998.

\bibitem[524]{McLachlan1964}
A. McLachlan, {\itshape A variational solution of the time-dependent
  Schrodinger equation}, Molecular Physics 8 (1964), pp. 39--44.

\bibitem[525]{Goth2012}
F. Goth and F.F. Assaad, {\itshape Time and spatially resolved quench of the
  fermionic Hubbard model showing restricted equilibration}, Phys. Rev. B 85
  (2012), p. 085129.

\bibitem[526]{Aoki2014}
H. Aoki, N. Tsuji, M. Eckstein, M. Kollar, T. Oka, and P. Werner, {\itshape
  Nonequilibrium dynamical mean-field theory and its applications}, Rev. Mod.
  Phys. 86 (2014), pp. 779--837.

\bibitem[527]{Schiro2010}
M. Schir{\'o} and M. Fabrizio, {\itshape {Time-Dependent Mean Field Theory for
  Quench Dynamics in Correlated Electron Systems}}, Phys. Rev. Lett 105 (2010),
  p. 076401.

\bibitem[528]{Schiro2011}
 {Schir\'o, M. and Fabrizio, M.}, {\itshape {Quantum quenches in the Hubbard
  model: Time-dependent mean-field theory and the role of quantum
  fluctuations}}, Phys. Rev. B 83 (2011), p. 165105.

\bibitem[529]{Sorella2005}
S. Sorella, {\itshape {Wave function optimization in the variational Monte
  Carlo method}}, Phys. Rev. B 71 (2005), p. 241103.

\bibitem[530]{Fabrizio2007}
M. Fabrizio, {\itshape Gutzwiller description of non-magnetic Mott insulators:
  Dimer lattice model}, Phys. Rev. B 76 (2007), p. 165110.

\bibitem[531]{Fabrizio2013}
 {Fabrizio, Michele}, {\itshape The Out-of-Equilibrium Time-Dependent
  Gutzwiller Approximation}, in  {\itshape New Materials for Thermoelectric
  Applications: Theory and Experiment}, , in {\itshape New Materials for
  Thermoelectric Applications: Theory and Experiment}, ed. {Zlatic, Veljko and
  Hewson, Alex}{Zlatic, Veljko and Hewson, Alex} ed.,    Springer Netherlands,
  2013, pp. 247--273.

\bibitem[532]{Sandri2012}
M. Sandri, M. Schir{\'o}, and M. Fabrizio, {\itshape {Linear ramps of
  interaction in the fermionic Hubbard model}}, Phys. Rev. B 86 (2012), p.
  075122.

\bibitem[533]{Sandri2013b}
M. Sandri and M. Fabrizio, {\itshape Nonequilibrium dynamics in the
  antiferromagnetic Hubbard model}, Phys. Rev. B 88 (2013), p. 165113.

\bibitem[534]{Kotliar1986}
G. Kotliar and A.E. Ruckenstein, {\itshape New Functional Integral Approach to
  Strongly Correlated Fermi Systems: The Gutzwiller Approximation as a Saddle
  Point}, Phys. Rev. Lett. 57 (1986), pp. 1362--1365.

\bibitem[535]{Lechermann2007}
F. Lechermann, A. Georges, G. Kotliar, and O. Parcollet, {\itshape Rotationally
  invariant slave-boson formalism and momentum dependence of the quasiparticle
  weight}, Phys. Rev. B 76 (2007), p. 155102.

\bibitem[536]{Anderson1987}
P.W. Anderson, {\itshape {The Resonating Valence Bond State in La$_2$CuO$_4$
  and Superconductivity}}, Science 235 (1987), pp. 1196--1198.

\bibitem[537]{Seibold2001}
G. Seibold and J. Lorenzana, {\itshape {Time-Dependent Gutzwiller Approximation
  for the Hubbard Model}}, Phys. Rev. Lett. 86 (2001), p. 2605.

\bibitem[538]{Metzner1989}
W. Metzner and D. Vollhardt, {\itshape Correlated Lattice Fermions in
  $d=\ensuremath{\infty}$ Dimensions}, Phys. Rev. Lett. 62 (1989), pp.
  324--327.

\bibitem[539]{Zhang1993}
X.Y. Zhang, M.J. Rozenberg, and G. Kotliar, {\itshape Mott transition in the
  \textit{d} =\ensuremath{\infty} Hubbard model at zero temperature}, Phys.
  Rev. Lett. 70 (1993), pp. 1666--1669.

\bibitem[540]{Maier2005}
T. Maier, M. Jarrell, T. Pruschke, and M.H. Hettler, {\itshape Quantum cluster
  theories}, Rev. Mod. Phys. 77 (2005), pp. 1027--1080.

\bibitem[541]{Freericks2006}
 {Freericks, J. K. and Turkowski, V. M. and Zlatic, V.}, {\itshape
  {Nonequilibrium Dynamical Mean-Field Theory}}, Phys. Rev. Lett. 97 (2006), p.
  266408.

\bibitem[542]{Toschi2005a}
A. Toschi, P. Barone, M. Capone, and C. Castellani, {\itshape {Pairing and
  superconductivity from weak to strong coupling in the attractive Hubbard
  model}}, New J. Phys. 7 (2005), pp. 7--7.

\bibitem[543]{Toschi2005b}
A. Toschi, M. Capone, and C. Castellani, {\itshape {Energetic balance of the
  superconducting transition across the BCS---Bose Einstein crossover in the
  attractive Hubbard model}}, Phys. Rev. B 72 (2005), p. 235118.

\bibitem[544]{Schwinger1961}
J. Schwinger, {\itshape {Brownian Motion of a Quantum Oscillator}}, Journal of
  Mathematical Physics 2 (1961), pp. 407--432.

\bibitem[545]{Kadanoff1962}
L. Kadanoff and G. Baym {\itshape Quantum statistical mechanics: Green's
  function methods in equilibrium and nonequilibrium problems},  Frontiers in
  physics  W.A. Benjamin, 1962.

\bibitem[546]{Keldysh1964}
L.V. Keldysh, {\itshape {}}, Sov. Phys. JETP 20 (1964), p. 1018.

\bibitem[547]{Kamenev2011}
A. Kamenev {\itshape Field Theory of Non-Equilibrium Systems},    Cambridfe
  University Press, Cambridge, England, 2011.

\bibitem[548]{Wagner1991}
M. Wagner, {\itshape Expansions of nonequilibrium Green's functions}, Phys.
  Rev. B 44 (1991), pp. 6104--6117.

\bibitem[549]{Bloch1929}
F. Bloch, {\itshape Über die Quantenmechanik der Elektronen in
  Kristallgittern}, Zeitschrift für Physik 52 (1929), pp. 555--600.

\bibitem[550]{Turkowski2007}
V. Turkowski and J.K. Freericks, {\itshape Nonequilibrium perturbation theory
  of the spinless Falicov-Kimball model: Second-order truncated expansion in
  $U$}, Phys. Rev. B 75 (2007), p. 125110.

\bibitem[551]{Freericks2008}
J.K. Freericks, {\itshape Quenching Bloch oscillations in a strongly correlated
  material: Nonequilibrium dynamical mean-field theory}, Phys. Rev. B 77
  (2008), p. 075109.

\bibitem[552]{Eckstein2011b}
M. Eckstein and P. Werner, {\itshape Damping of Bloch Oscillations in the
  Hubbard Model}, Phys. Rev. Lett. 107 (2011), p. 186406.

\bibitem[553]{Amaricci2012}
A. Amaricci, C. Weber, M. Capone, and G. Kotliar, {\itshape {Approach to a
  stationary state in a driven Hubbard model coupled to a thermostat}}, Pyhs.
  Rev. B 86 (2012), p. 85110.

\bibitem[554]{Tsuji2008}
N. Tsuji, T. Oka, and H. Aoki, {\itshape Correlated electron systems
  periodically driven out of equilibrium: $\text{Floquet}+\text{DMFT}$
  formalism}, Phys. Rev. B 78 (2008), p. 235124.

\bibitem[555]{Joura2008}
A.V. Joura, J.K. Freericks, and T. Pruschke, {\itshape {Steady-State
  Nonequilibrium Density of States of Driven Strongly Correlated Lattice Models
  in Infinite Dimensions}}, Phys. Rev. Lett. 101 (2008), p. 196401.

\bibitem[556]{Aron2012a}
C. Aron, G. Kotliar, and C. Weber, {\itshape Dimensional Crossover Driven by an
  Electric Field}, Phys. Rev. Lett. 108 (2012), p. 086401.

\bibitem[557]{Mierzejewski2011a}
M. Mierzejewski, L. Vidmar, J. Bon\v{c}a, and P. Prelov\v{s}ek, {\itshape
  Nonequilibrium Quantum Dynamics of a Charge Carrier Doped into a Mott
  Insulator}, Phys. Rev. Lett. 106 (2011), p. 196401.

\bibitem[558]{Eckstein2013c}
M. Eckstein and P. Werner, {\itshape Dielectric breakdown of Mott insulators
  – doublon production and doublon heating}, Journal of Physics: Conference
  Series 427 (2013), p. 012005.

\bibitem[559]{Eckstein2013a}
---{}---{}---, {\itshape Photoinduced States in a Mott Insulator}, Phys. Rev.
  Lett. 110 (2013), p. 126401.

\bibitem[560]{Kemper2013}
A.F. Kemper, M. Sentef, B. Moritz, C.C. Kao, Z.X. Shen, J.K. Freericks, and
  T.P. Devereaux, {\itshape Mapping of unoccupied states and relevant bosonic
  modes via the time-dependent momentum distribution}, Phys. Rev. B 87 (2013),
  p. 235139.

\bibitem[561]{Han2013a}
J.E. Han, {\itshape Solution of electric-field-driven tight-binding lattice
  coupled to fermion reservoirs}, Phys. Rev. B 87 (2013), p. 085119.

\bibitem[562]{Han2013b}
J.E. Han and J. Li, {\itshape Energy dissipation in a dc-field-driven electron
  lattice coupled to fermion baths}, Phys. Rev. B 88 (2013), p. 075113.

\bibitem[563]{Shirley1965}
J.H. Shirley, {\itshape Solution of the Schr\"odinger Equation with a
  Hamiltonian Periodic in Time}, Phys. Rev. 138 (1965), pp. B979--B987.

\bibitem[564]{Zeldovich1967}
Y. Zel'dovich, {\itshape The Quasienergy of a Quantum-mechanical System
  Subjected to a Periodic Action}, JETP 24 (1967), p. 1006.

\bibitem[565]{Floquet1883}
 Gaston Floquet, \textit{Sur les \'equations diff\'erentielles lin\'eaires \`a
  coefficients p\'eriodiques}. Annales de l'\'Ecole Normale Sup\'erieure 12:
  47-88 (1883).

\bibitem[566]{Schmidt2002}
 Schmidt, P. and Monien, H. \textit{Nonequilibrium dynamical mean-field theory
  of a strongly correlated system}. arXiv:cond-mat/0202046 (2002).

\bibitem[567]{Freericks2008p}
J.K. Freericks and A.V. Joura, {\itshape Nonequilibrium density of states and
  distribution functions for strongly correlated materials across the Mott
  transition}, in  {\itshape Electron transport in nanosystems}, J.~Bonca and
  S.~Kuchinin,  eds.,    Springer, Berlin, 2008, pp. 219--236.

\bibitem[568]{Lubatsch2009}
A. Lubatsch and J. Kroha, {\itshape Optically driven Mott-Hubbard systems out
  of thermodynamic equilibrium}, Annalen der Physik 18 (2009), pp. 863--867.

\bibitem[569]{Tsuji2009}
N. Tsuji, T. Oka, and H. Aoki, {\itshape Nonequilibrium Steady State of
  Photoexcited Correlated Electrons in the Presence of Dissipation}, Phys. Rev.
  Lett. 103 (2009), p. 047403.

\bibitem[570]{Tsuji2011}
N. Tsuji, T. Oka, P. Werner, and H. Aoki, {\itshape Dynamical Band Flipping in
  Fermionic Lattice Systems: An ac-Field-Driven Change of the Interaction from
  Repulsive to Attractive}, Phys. Rev. Lett. 106 (2011), p. 236401.

\bibitem[571]{Tsuji2012}
N. Tsuji, T. Oka, H. Aoki, and P. Werner, {\itshape {Repulsion-to-attraction
  transition in correlated electron systems triggered by a monocycle pulse}},
  Phys. Rev. B 85 (2012), p. 155124.

\bibitem[572]{Okamoto2008}
S. Okamoto, {\itshape {Nonlinear Transport through Strongly Correlated
  Two-Terminal Heterostructures : A Dynamical Mean-Field Approach}}, Phys. Rev.
  Lett. 116807 (2008), pp. 1--4.

\bibitem[573]{Heary2009}
R.J. Heary and J.E. Han, {\itshape Perturbation study of nonequilibrium
  quasiparticle spectra in an infinite-dimensional Hubbard lattice}, Phys. Rev.
  B 80 (2009), p. 035102.

\bibitem[574]{Li2015}
J. Li, C. Aron, G. Kotliar, and J.E. Han, {\itshape Electric-Field-Driven
  Resistive Switching in the Dissipative Hubbard Model}, Phys. Rev. Lett. 114
  (2015), p. 226403.

\bibitem[575]{Amaricci2014}
 A. Amaricci and M. Capone. \textit{Dynamical Mean-Field Theory description of
  the voltage induced transition in a non-equilibrium superconductor}.
  arXiv:1411.23476 (2014).

\bibitem[576]{Keller2001}
M. Keller, W. Metzner, and U. Schollw\"ock, {\itshape {Dynamical Mean-Field
  Theory for Pairing and Spin Gap in the Attractive Hubbard Model}}, Phys. Rev.
  Lett. 86 (2001), pp. 4612--4615.

\bibitem[577]{Capone2002}
M. Capone, C. Castellani, and M. Grilli, {\itshape {First-Order Pairing
  Transition and Single-Particle Spectral Function in the Attractive Hubbard
  Model}}, Phys. Rev. Lett. 88 (2002), p. 126403.

\bibitem[578]{Eckstein2013b}
M. Eckstein and P. Werner, {\itshape Nonequilibrium dynamical mean-field
  simulation of inhomogeneous systems}, Phys. Rev. B 88 (2013), p. 075135.

\bibitem[579]{Landau1932}
L. Landau, {\itshape {}}, Phys. Z. Sowjetunion 2 (1932), p.~46.

\bibitem[580]{Zener1932}
C. Zener, {\itshape Non-Adiabatic Crossing of Energy Levels}, Proceedings of
  the Royal Society of London A: Mathematical, Physical and Engineering
  Sciences 137 (1932), pp. 696--702.

\bibitem[581]{Cario2010}
L. Cario, C. Vaju, B. Corraze, V. Guiot, and E. Janod, {\itshape
  Electric-Field-Induced Resistive Switching in a Family of Mott Insulators:
  Towards a New Class of RRAM Memories}, Advanced Materials 22 (2010), pp.
  5193--5197.

\bibitem[582]{Cario2013a}
P. Stoliar, L. Cario, E. Janod, B. Corraze, C. Guillot-Deudon, S.
  Salmon-Bourmand, V. Guiot, J. Tranchant, and M. Rozenberg, {\itshape
  Universal Electric-Field-Driven Resistive Transition in Narrow-Gap Mott
  Insulators}, Advanced Materials 25 (2013), pp. 3222--3226.

\bibitem[583]{Cario2013b}
V. Guiot, L. Cario, E. Janod, B. Corraze, V. {Ta Phuoc}, M. Rozenberg, P.
  Stoliar, T. Cren, and D. Roditchev, {\itshape {Avalanche breakdown in
  GaTa$_4$Se$_8$narrow-gap Mott insulators}}, Nat. Commun. 4 (2013), p. 1722.

\bibitem[584]{Oka2003}
T. Oka and H. Aoki, {\itshape Ground-State Decay Rate for the Zener Breakdown
  in Band and Mott Insulators}, Phys. Rev. Lett. 95 (2005), p. 137601.

\bibitem[585]{Oka2010}
 {Oka, T. and Aoki, H.}, {\itshape {Dielectric breakdown in a Mott insulator:
  Many-body Schwinger-Landau-Zener mechanism studied with a generalized Bethe
  ansatz}}, Phys. Rev. B 81 (2010), p. 033103.

\bibitem[586]{Oka2012}
T. Oka, {\itshape Nonlinear doublon production in a Mott insulator:
  Landau-Dykhne method applied to an integrable model}, Phys. Rev. B 86 (2012),
  p. 075148.

\bibitem[587]{Lenarcic2012a}
Z. Lenar\v{c}i\v{c} and P. Prelov\v{s}ek, {\itshape Dielectric Breakdown in
  Spin-Polarized Mott Insulator}, Phys. Rev. Lett. 108 (2012), p. 196401.

\bibitem[588]{Eckstein2010b}
M. Eckstein, T. Oka, and P. Werner, {\itshape {Dielectric Breakdown of Mott
  Insulators in Dynamical Mean-Field Theory}}, Phys. Rev. Lett. 105 (2010), p.
  146404.

\bibitem[589]{Aron2012b}
C. Aron, {\itshape Dielectric breakdown of a Mott insulator}, Phys. Rev. B 86
  (2012), p. 085127.

\bibitem[590]{Mazza2015}
G. Mazza, A. Amaricci, M. Capone, and M. Fabrizio, {\itshape Electronic
  transport and dynamics in correlated heterostructures}, Phys. Rev. B 91
  (2015), p. 195124.

\bibitem[591]{Fotso2014}
H. Fotso, K. Mikelsons, and J.K. Freericks, {\itshape Thermalization of field
  driven quantum systems}, Scientific Reports 4 (2014), p. 4699.

\bibitem[592]{Mazza2015b}
 G. Mazza et al. \textit{Gap Collapse and Novel Mechanism for Dielectric
  Breakdown of Mott Insulators}. arXiv:1510.xxxxx (2015).

\bibitem[593]{Camjayi2014}
A. Camjayi, C. Acha, R. Weht, G. Rodriguez M.\, B. Corraze, E. Janod, L. Cario,
  and J. Rozenberg M.\, {\itshape First-Order Insulator-to-Metal Mott
  Transition in the Paramagnetic 3D System
  ${\mathrm{GaTa}}_{4}{\mathrm{Se}}_{8}$}, Phys. Rev. Lett. 113 (2014), p.
  086404.

\bibitem[594]{Lee2014}
W.R. Lee and K. Park, {\itshape Dielectric breakdown via emergent
  nonequilibrium steady states of the electric-field-driven Mott insulator},
  Phys. Rev. B 89 (2014), p. 205126.

\bibitem[595]{Eckstein2011a}
M. Eckstein and P. Werner, {\itshape Thermalization of a pump-excited Mott
  insulator}, Phys. Rev. B 84 (2011), p. 035122.

\bibitem[596]{Moritz2010}
B. Moritz, T.P. Devereaux, and J.K. Freericks, {\itshape Time-resolved
  photoemission of correlated electrons driven out of equilibrium}, Phys. Rev.
  B 81 (2010), p. 165112.

\bibitem[597]{Moritz2013}
B. Moritz, A.F. Kemper, M. Sentef, T.P. Devereaux, and J.K. Freericks,
  {\itshape Electron-Mediated Relaxation Following Ultrafast Pumping of
  Strongly Correlated Materials: Model Evidence of a Correlation-Tuned
  Crossover between Thermal and Nonthermal States}, Phys. Rev. Lett. 111
  (2013), p. 077401.

\bibitem[598]{Freericks2009}
J.K. Freericks, H.R. Krishnamurthy, and T. Pruschke, {\itshape Theoretical
  Description of Time-Resolved Photoemission Spectroscopy: Application to
  Pump-Probe Experiments}, Phys. Rev. Lett. 102 (2009), p. 136401.

\bibitem[599]{Heyl2013}
M. Heyl, A. Polkovnikov, and S. Kehrein, {\itshape Dynamical Quantum Phase
  Transitions in the Transverse-Field Ising Model}, Phys. Rev. Lett. 110
  (2013), p. 135704.

\bibitem[600]{Tsuji2013}
N. Tsuji, M. Eckstein, and P. Werner, {\itshape Nonthermal Antiferromagnetic
  Order and Nonequilibrium Criticality in the Hubbard Model}, Phys. Rev. Lett.
  110 (2013), p. 136404.

\bibitem[601]{Werner2014}
P. Werner, K. Held, and M. Eckstein, {\itshape Role of impact ionization in the
  thermalization of photoexcited Mott insulators}, Phys. Rev. B 90 (2014), p.
  235102.

\bibitem[602]{Golez2015}
Dynamics of screening in photo-doped Mott insulators; .

\bibitem[603]{Polkovnikov2011}
A. Polkovnikov, K. Sengupta, A. Silva, and M. Vengalattore, {\itshape
  {\textit{Colloquium}: Nonequilibrium dynamics of closed interacting quantum
  systems}}, Rev. Mod. Phys. 83 (2011), pp. 863--883.

\bibitem[604]{Eckstein2009}
M. Eckstein, M. Kollar, and P. Werner, {\itshape {Thermalization after an
  Interaction Quench in the {Hubbard} Model}}, Phys. Rev. Lett. 103 (2009), p.
  056403.

\bibitem[605]{Eckstein2010a}
 {Eckstein, M. and Kollar, M. and Werner, P.}, {\itshape {Interaction quench in
  the Hubbard model: Relaxation of the spectral function and the optical
  conductivity}}, Phys. Rev. B 81 (2010), p. 115131.

\bibitem[606]{Andre2012}
P. Andr\'e, M. Schir\'o, and M. Fabrizio, {\itshape Lattice and surface effects
  in the out-of-equilibrium dynamics of the Hubbard model}, Phys. Rev. B 85
  (2012), p. 205118.

\bibitem[607]{Sandri2013a}
M. Sandri, M. Capone, and M. Fabrizio, {\itshape Finite-temperature Gutzwiller
  approximation and the phase diagram of a toy model for V${}_{2}$O${}_{3}$},
  Phys. Rev. B 87 (2013), p. 205108.

\bibitem[608]{Balzer2015}
K. Balzer, F.A. Wolf, I.P. McCulloch, P. Werner, and M. Eckstein, {\itshape
  Nonthermal Melting of N\'eel Order in the Hubbard Model}, Phys. Rev. X 5
  (2015), p. 031039.

\bibitem[609]{Werner2012}
P. Werner, N. Tsuji, and M. Eckstein, {\itshape Nonthermal symmetry-broken
  states in the strongly interacting Hubbard model}, Phys. Rev. B 86 (2012), p.
  205101.

\bibitem[610]{Mentink2014}
J.H. Mentink and M. Eckstein, {\itshape Ultrafast Quenching of the Exchange
  Interaction in a Mott Insulator}, Phys. Rev. Lett. 113 (2014), p. 057201.

\bibitem[611]{Mentink2015}
J.H. Mentink, K. Balzer, and M. Eckstein, {\itshape Ultrafast and reversible
  control of the exchange interaction in Mott insulators}, Nat. Commun. 6
  (2015), p. 6708.

\bibitem[612]{Werner2007}
P. Werner and A.J. Millis, {\itshape {High-Spin to Low-Spin and Orbital
  Polarization Transitions in Multiorbital Mott Systems}}, Phys. Rev. Lett. 99
  (2007), p. 126405.

\bibitem[613]{Werner2009}
P. Werner, E. Gull, and A.J. Millis, {\itshape Metal-insulator phase diagram
  and orbital selectivity in three-orbital models with rotationally invariant
  Hund coupling}, Phys. Rev. B 79 (2009), p. 115119.

\bibitem[614]{de'Medici2011a}
L. {de' Medici}, {\itshape Hund's coupling and its key role in tuning
  multiorbital correlations}, Phys. Rev. B 83 (2011), p. 205112.

\bibitem[615]{de'Medici2011b}
L. {de' Medici}, J. Mravlje, and A. Georges, {\itshape Janus-Faced Influence of
  Hund's Rule Coupling in Strongly Correlated Materials}, Phys. Rev. Lett. 107
  (2011), p. 256401.

\bibitem[616]{Georges2013}
A. Georges, L. {de' Medici}, and J. Mravlje, {\itshape Strong Correlations from
  Hund's Coupling}, Annual Review of Condensed Matter Physics 4 (2013), pp.
  137--178.

\bibitem[617]{de'Medici2014}
L. {de' Medici}, G. Giovannetti, and M. Capone, {\itshape Selective Mott
  Physics as a Key to Iron Superconductors}, Phys. Rev. Lett. 112 (2014), p.
  177001.

\bibitem[618]{Poteryaev2008}
A.I. Poteryaev, M. Ferrero, A. Georges, and O. Parcollet, {\itshape Effect of
  crystal-field splitting and interband hybridization on the metal-insulator
  transitions of strongly correlated systems}, Phys. Rev. B 78 (2008), p.
  045115.

\bibitem[619]{Sandri2014}
M. Sandri and M. Fabrizio, {\itshape Non-equilibrium gap-collapse near a
  first-order Mott transition},  (2014),  arXiv:14104442.

\bibitem[620]{Lantz2014}
 G. Lantz et al. \textit{Surface effects on the Mott-Hubbard transition in
  archetypal V$_2$O$_3$}. arXiv:1507.02116 (2015).

\bibitem[621]{Ciuchi1997}
S. Ciuchi, F. Pasqualede~, S. Fratini, and D. Feinberg, {\itshape Dynamical
  mean-field theory of the small polaron}, Phys. Rev. B 56 (1997), pp.
  4494--4512.

\bibitem[622]{Capone1997}
M. Capone, W. Stephan, and M. Grilli, {\itshape Small-polaron formation and
  optical absorption in Su-Schrieffer-Heeger and Holstein models}, Phys. Rev. B
  56 (1997), pp. 4484--4493.

\bibitem[623]{Bonca1999}
J. Bon\v{c}a, S.A. Trugman, and I. Batisti\'{c}, {\itshape Holstein polaron},
  Phys. Rev. B 60 (1999), pp. 1633--1642.

\bibitem[624]{Capone2003p}
M. Capone and S. Ciuchi, {\itshape Polaron Crossover and Bipolaronic
  Metal-Insulator Transition in the Half-Filled Holstein Model}, Phys. Rev.
  Lett. 91 (2003), p. 186405.

\bibitem[625]{Capone2006p}
M. Capone, P. Carta, and S. Ciuchi, {\itshape Dynamical mean field theory of
  polarons and bipolarons in the half-filled Holstein model}, Phys. Rev. B 74
  (2006), p. 045106.

\bibitem[626]{Golez2012b}
D. Gole\v{z}, J. Bon\v{c}a, and L. Vidmar, {\itshape Dissociation of a
  Hubbard-Holstein bipolaron driven away from equilibrium by a constant
  electric field}, Phys. Rev. B 85 (2012), p. 144304.

\bibitem[627]{Kemper2014}
A.F. Kemper, M.A. Sentef, B. Moritz, J.K. Freericks, and T.P. Devereaux,
  {\itshape Effect of dynamical spectral weight redistribution on effective
  interactions in time-resolved spectroscopy}, Phys. Rev. B 90 (2014), p.
  075126.

\bibitem[628]{Murakami2015}
Y. Murakami, P. Werner, N. Tsuji, and H. Aoki, {\itshape Interaction quench in
  the Holstein model: Thermalization crossover from electron- to
  phonon-dominated relaxation}, Phys. Rev. B 91 (2015), p. 045128.

\bibitem[629]{Sayyad2015}
S. Sayyad and M. Eckstein, {\itshape Coexistence of excited polarons and
  metastable delocalized states in photoinduced metals}, Phys. Rev. B 91
  (2015), p. 104301.

\bibitem[630]{Sangiovanni2005}
G. Sangiovanni, M. Capone, C. Castellani, and M. Grilli, {\itshape
  Electron-Phonon Interaction Close to a Mott Transition}, Phys. Rev. Lett. 94
  (2005), p. 026401.

\bibitem[631]{Sangiovanni2006}
G. Sangiovanni, M. Capone, and C. Castellani, {\itshape Relevance of phonon
  dynamics in strongly correlated systems coupled to phonons: Dynamical
  mean-field theory analysis}, Phys. Rev. B 73 (2006), p. 165123.

\bibitem[632]{Yonemitsu2009}
K. Yonemitsu and N. Maeshima, {\itshape Coupling-dependent rate of energy
  transfer from photoexcited Mott insulators to lattice vibrations}, Phys. Rev.
  B 79 (2009), p. 125118.

\bibitem[633]{Werner2013final}
P. Werner and M. Eckstein, {\itshape Phonon-enhanced relaxation and excitation
  in the Holstein-Hubbard model}, Phys. Rev. B 88 (2013), p. 165108.

\bibitem[634]{Werner2015final}
---{}---{}---, {\itshape Field-induced polaron formation in the
  Holstein-Hubbard model}, EPL (Europhysics Letters) 109 (2015), p. 37002.

\bibitem[635]{Sansone2011}
G. Sansone, L. Poletto, and M. Nisoli, {\itshape {High-energy attosecond light
  sources}}, Nat. Photon. 5 (2011), pp. 655--663.

\bibitem[636]{Cundiff2013}
S.T. Cundiff and S. Mukamel, {\itshape Optical multidimensional coherent
  spectroscopy}, Physics Today 66 (2013), pp. 44--49.

\bibitem[637]{Esposito2015}
M. Esposito, K. Titimbo, K. Zimmermann, F. Giusti, F. Randi, D. Boschetto, F.
  Parmigiani, R. Floreanini, F. Benatti, and D. Fausti, {\itshape Photon number
  statistics uncover the fluctuations in non-equilibrium lattice dynamics},
  Nat. Commun. 6 (2015), p. 10249.

\end{thebibliography}
\label{lastpage}

\end{document}